\documentclass[12pt,twoside,openright]{report}        
\usepackage{latexsym}       
\usepackage{astron}         
\usepackage{url}            
\usepackage{a4wide}         
\usepackage{graphicx}       
\usepackage{wrapfig}        
\usepackage{longtable}      
\usepackage{color}          
\makeatletter               %
\renewcommand{\fnum@figure}[1]{\small \textbf{\figurename~\thefigure:} \sffamily}
\renewcommand{\fnum@table}[1]{\small \textbf{\tablename~\thetable:} \sffamily}

\makeatother                %
\definecolor{red}{rgb}{1.0,0.0,0.0}     
\definecolor{blue}{rgb}{0.0,0.0,1.0}
\def\microns{$\textrm{$\mu$m}$}                             
\def\orderof{\mathcal{O}}                                   

\def\sun{\hbox{$_\odot$}}                                    
\def\farcm{\hbox{$.\mkern-4mu^\prime$}}                     
\def\farcs{\hbox{$.\!\!^{\prime\prime}$}}                   
\def\arcmin{\hbox{$^\prime$}}                               
\def\arcsec{\hbox{$^{\prime\prime}$}}                       
\def\degr{\hbox{$^\circ$}}                                  

\def\ie{\textit{i.e.}}                                      
\def\eg{\textit{e.g.}}                                      
\def\etal{\textit{et al.}}                                  

\def\la{\mathrel{\mathchoice {\vcenter{\offinterlineskip\halign{\hfil
$\displaystyle##$\hfil\cr<\cr\sim\cr}}}
{\vcenter{\offinterlineskip\halign{\hfil$\textstyle##$\hfil\cr
<\cr\sim\cr}}}
{\vcenter{\offinterlineskip\halign{\hfil$\scriptstyle##$\hfil\cr
<\cr\sim\cr}}}
{\vcenter{\offinterlineskip\halign{\hfil$\scriptscriptstyle##$\hfil\cr
<\cr\sim\cr}}}}}                                            
\def\ga{\mathrel{\mathchoice {\vcenter{\offinterlineskip\halign{\hfil
$\displaystyle##$\hfil\cr>\cr\sim\cr}}}
{\vcenter{\offinterlineskip\halign{\hfil$\textstyle##$\hfil\cr
>\cr\sim\cr}}}
{\vcenter{\offinterlineskip\halign{\hfil$\scriptstyle##$\hfil\cr
>\cr\sim\cr}}}
{\vcenter{\offinterlineskip\halign{\hfil$\scriptscriptstyle##$\hfil\cr
>\cr\sim\cr}}}}}                                            
\def\num{\hbox{N$^{\underline{o}}$}}                        

\def\zg1{$z\!>\!1$}                                         
\def\zga1{$z\!\ga\!1$}                                      
\def\zsim1{$z\!\sim\!1$}                                    
\def\flux{erg\,s$^{-1}$cm$^{-2}$}                          
\def\reds{red-sequence }                               
\def\h70{$h_{70}$}                                          
\def\hinv70{$h^{-1}_{70}$}                                  

\begin{document}
\pagestyle{empty}

\begin{center} \bfseries \vspace{1cm}














\Large


\LARGE Studying Cosmic Evolution with the \\
XMM-Newton Distant Cluster Project: \\
\vspace{0.8ex}
\Large X-ray Luminous Galaxy Clusters at $\mathbf{z\!\ga\!1}$ \\
and their Galaxy Populations



\vspace{4ex}

\begin{figure}[h!]
\centering
\includegraphics[angle=0,clip,width=0.4\textwidth]{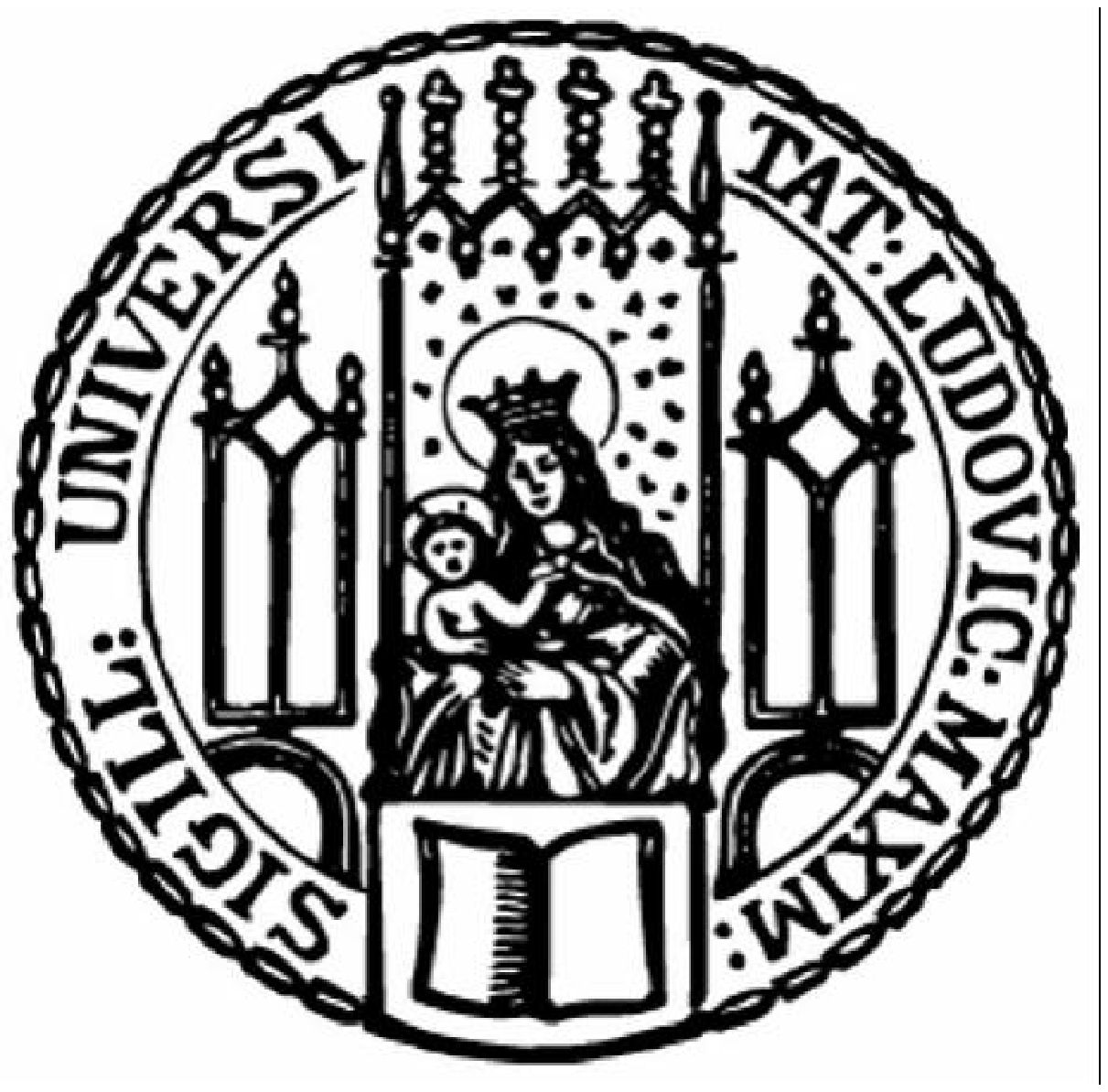}
\end{figure}

\vfill
\LARGE 
\mdseries 
\vspace{1ex}
Dissertation der Fakult\"at f\"ur Physik \\
der Ludwig-Maximilians-Universit\"at M\"unchen \\

\Large

\vspace{6ex}
vorgelegt von \\
\vspace{3ex}

{\bf \LARGE Rene Fassbender} \\


\vspace{1ex}
aus Diez/Lahn\\

\vspace{8ex}
\vfill
M\"unchen, Oktober 2007 \\

\normalsize \mdseries
\end{center} 
\vfill


\newpage
\thispagestyle{empty}
\mbox{} \vfill
{\Large
1. Gutachter: Prof. Dr. Hans B\"ohringer

2. Gutachter: Prof. Dr. Andreas Burkert

\vspace{5ex}
Tag der m\"undlichen Pr\"ufung: 29. November 2007}



\newpage
\thispagestyle{empty}
\mbox{} \vfill
\begin{center}
\Large
Dedicated to the memory of \\
\vspace{1ex}

{\LARGE Peter Schuecker} \\

\vspace{0.5ex}
(1959--2006)

\vspace{1ex}
\end{center}
\vfill \mbox{}


\cleardoublepage
\newpage
\thispagestyle{empty}

\framebox[1\hsize]{\parbox{15.8cm} {\large 
\bf 
\begin{center}


Beobachtung Kosmischer Entwicklung mit dem \\
XMM-Newton Distant Cluster Project: \\
\vspace{0.5ex}
\normalsize
R\"ontgenhelle Galaxienhaufen bei $\mathbf{z\!\ga\!1}$ und ihre Galaxienpopulationen

\end{center}


Zusammenfassung \quad \rm 

\small 
R\"ontgenhelle Galaxienhaufen bei hoher Rotverschiebung (\zga1) erm\"oglichen einzigartige, aber herausfordernde,  Untersuchungen der kosmischen Entwicklung dieser gr\"o\ss ten Objekte im Universum, der baryonischen Materie in Form des hei\ss en Haufengases (Intracluster Medium, ICM), ihrer Galaxienpopulationen und des Einflusses der  mysteri\"osen  Dunklen Energie. 
Das Hauptziel der vorliegenden Doktorarbeit ist das Erstellen der Datengrundlage f\"ur das  XMM-Newton Distant Cluster Project (XDCP). Diese neue Himmelsdurchmusterung im R\"ontgenlicht konzentriert sich  
auf die systematische Suche und Identifizierung von Galaxienhaufen bei den h\"ochsten  Rotverschiebungen von \zga1 mittels einer gezielten Vorauswahl von ausgedehnten R\"ontgenquellen, Nachbeobachtungen in zwei photometrischen B\"andern zum Nachweis und zur Entfernungssch\"atzung der Objekte sowie der abschlie\ss enden spektroskopischen   Best\"atigung. \newline
Als ersten Schritt habe ich insgesamt 80 Quadratgrad  (469 Felder) langbelichteter R\"ontgenbeobachtungen aus dem XMM-Newton Datenarchiv 
mit einem neu entwickelten automatisierten Datenreduktionssystem analysiert. 
Damit konnte ich ann\"ahernd 1\,000 
ausgedehnte Quellen als Galaxienhaufen-Kandidaten nachweisen, von denen sich  75\% durch 
vorhandene optische Daten als Haufen und Gruppen bei     
$z\!\la\!0.6$ identifizieren lie\ss en.
Die verbleibenden ca. 250 Kandidaten mit typischen  R\"ontgenfl\"ussen von $\simeq\!10^{-14}$\,\flux \ (0.5--2.0\,keV)
erfordern weitere 
Schritte, um als entfernte Galaxienhaufen best\"atigt werden zu k\"onnen. \newline
Daf\"ur entwickelte ich eine neue Nahinfrarot-Beobachtungsstrategie, mit der sich  die Identifizierung der ausgedehnten R\"ontgenquellen
mittels 
Z- und H-Band Aufnahmen effizient durchf\"uhren l\"asst und die zur Rotverschiebungsabsch\"atzung anhand der Z--H Farbe der `Red-Sequence' der Fr\"uhtypgalaxien dient.
Um das Potential dieser Methode voll auszusch\"opfen, konzipierte ich ein neues Nahinfrarot-Datenverarbeitungsprogramm, mit dem die Nachbeobachtungen zweier Kampagnen mit dem 3.5\,m Teleskop des Calar Alto Observatoriums   f\"ur 25\% der 250 entfernten Galaxienhaufen-Kandidaten ausgewertet wurden.
Anhand dieser Daten konnte ich als erstes Hauptergebnis insgesamt mehr als 20 photometrisch best\"atigte Neuendeckungen von entfernten r\"ontgenhellen  Galaxienhaufen bei $z\!\ga\!0.9$ nachweisen. \newline
Desweiteren hat es die neue Z--H Beobachtungsmethode erm\"oglicht, erstmalig eine Galaxienhaufen-Stichprobe \"uber die gesamte   
Rotverschiebungsspanne von $0.2\!\la\!z\!\la\!1.5$ zu untersuchen.
Durch den Vergleich der beobachteten Farbentwicklung der `Red-Sequence'  mit Entwicklungsmodellen konnte ich die Entstehungsepoche 
der Sternpopulationen der Fr\"uhtypgalaxien auf 
$z_{\mathrm{f}}\!=\!4.2\pm 1.1$ einschr\"anken und damit ihr hohes Alter  
best\"atigen.  
Eine weitere Vorstudie 
\"uber die H-Band Leuchtkraftentwicklung von 63 hellsten Haufengalaxien (Brightest Cluster Galaxies, BCGs) \"uber das gleiche Rotverschiebungsintervall 
erm\"oglichte es, 
erstmals  direkte Beobachtungshinweise daf\"ur zu liefern, dass sich die Masse der gr\"o\ss ten Gala\-xien im lokalen Universum  seit $z\!\simeq\!1.5$ mindestens verdoppelt hat.
Mein vorl\"aufiges Ergebnis, dass nahe BCGs alte Sternpopulationen besitzen, ihre Masse aber erst zum Gro\ss teil \"uber die letzten 9\,Milliarden Jahre angesammelt wurde, stimmt qualitativ 
mit den 
Vorhersagen der mo\-dernsten Simulationen zur hierarchischen Galaxienentstehung und -entwicklung 
innerhalb des theoretischen Standardmodells \"uberein. 
Die Best\"atigung und Verfeinerung dieser vorl\"aufigen Ergebnisse werden zur Entwicklung  einer widerspruchsfreien Beschreibung 
kosmischer Entwicklung von Galaxienpopulationen und der gro\ss r\"aumigen Struktur beitragen.

\vspace{1ex}}}

\vfill

\cleardoublepage
\framebox[1\hsize]{\parbox{15.8cm}{\large \bf
\begin{center}
\large Studying Cosmic Evolution with the \\
XMM-Newton Distant Cluster Project: \\
\vspace{0.5ex}
\normalsize
X-ray Luminous Galaxy Clusters at $\mathbf{z\!\ga\!1}$ and their Galaxy Populations


\end{center}


Abstract \quad \rm 

\small
Investigating X-ray luminous galaxy clusters at high redshift
(\zga1) provides a challenging but fundamental constraint
on evolutionary studies of the largest virialized structures
in the Universe, the baryonic matter component in form of
the hot intracluster medium (ICM), their galaxy populations, 
and the effects of the  mysterious Dark Energy.
The main aim of this thesis work is to establish
the  observational foundation for
the XMM-Newton Distant Cluster Project (XDCP).
This new generation serendipitous X-ray survey is focused on 
the most distant galaxy clusters at \zga1,
based on the selection of extended X-ray sources, their  identification as clusters and redshift estimation via two-band imaging, and  their final spectroscopic confirmation. \newline
As a first step, I have analyzed 80\,deg$^2$ (469 fields) of deep XMM-Newton archival X-ray data with a new pipeline processing system and selected almost 1\,000 extended sources as galaxy cluster candidates,
75\% of which 
could be identified as clusters or groups at 
$z\!\la\!0.6$ using available optical data.
This left about  250  candidates with typical 0.5--2.0\,keV X-ray fluxes of $\sim\!10^{-14}$\,\flux \,  
in need of confirmation as distant cluster sources. \newline
Therefore, I have adopted 
a new strategy to efficiently establish the 
nature of these extended X-ray sources and estimate their redshifts,
based on medium deep Z- and H-band photometry and the observed Z--H  `red-sequence'  color of early-type cluster galaxies.
To fully exploit this technique, I have designed a new near-infrared  data reduction code, which was 
applied to the data collected for 25\% of the 250  distant cluster candidates in two imaging campaigns
at the 3.5m telescope at the Calar Alto Observatory.
As a first main result, more than 20 X-ray luminous clusters
were discovered to lie at a photometric redshift
of $z\!\ga\!0.9$. \newline
Furthermore, the new Z--H  \reds method has allowed  a cluster sample study over an unprecedented redshift baseline of $0.2\!\la\!z\!\la\!1.5$.
From a comparison of the observed color evolution
of the cluster red-sequence galaxies with model predictions, I could constrain the formation epoch
of the bulk of their stellar populations as $z_{\mathrm{f}}\!=\!4.2 \pm 1.1$.
This confirms the well-established old age
of the stellar populations of early-type galaxies in clusters.
The preliminary investigation of the H-band luminosity evolution
of 63 brightest cluster galaxies (BCGs) over the same redshift range
provides for the first time direct observational indications that 
the most massive cluster galaxies in the local Universe have doubled their stellar mass 
since $z\!\simeq\!1.5$.
My tentative finding that nearby BCGs have old, passively evolving
stellar populations and were assembled in the last 9\,Gyr
is in qualitative agreement with predictions
from the latest numerical simulations based on the standard
cold dark matter scenario of galaxy formation and evolution
via hierarchical merging.
The confirmation and refinement of these preliminary results will 
contribute to the 
development of a consistent picture of the cosmic evolution of galaxy populations and the
large-scale structure.





\vspace{1ex}}}

\cleardoublepage
\newpage
\thispagestyle{empty}










\newpage
\pagenumbering{roman} 
\tableofcontents 

\chapter{Introduction}
\pagestyle{headings} \pagenumbering{arabic}
\label{c1_intro}


\noindent 
X-ray studies of clusters of galaxies have come a long way over the past four decades since the first (spurious) detection of X-ray emission from  Coma  and the speculations of its thermal origin \cite{Felten1966a}.
After the first pioneering imaging X-ray telescope EINSTEIN in the late seventies and the ground-breaking success of the ROSAT mission in the nineties, we are currently in the era 
of the third generation of X-ray observatories led by XMM-Newton and Chandra.

Early survey results (Gioia \etal, 1990; Henry \etal, 1992) 
\nocite{Gioia1990a} \nocite{Henry1992a}
were suggestive that the population of X-ray luminous galaxy clusters exhibits a rather rapid evolution, which gave rise to 
the widespread belief that these objects were absent by redshifts of $z\!\sim\!1$.
This was also predicted by theory for a $\Omega_{\mathrm{m}}\!=\!1$ Universe, the standard cosmological model at the time.
Only one decade later, the situation had been completely reversed based on the results of the second generation cluster surveys. After discovering the first very distant systems, 
Rosati \cite*{Rosati2001a} summarized the status for X-ray clusters in the following
 statement that 
has remained valid until the present:
{\it
``ROSAT deep surveys have failed to show any dramatic change in the space density of clusters out to $z\!\sim\!1$, with the exception of the most luminous, massive systems.''}


Fundamental questions within the cosmic evolution scenario have so far lacked definite observational answers:
(i) What is the formation epoch of the first X-ray clusters? (ii) How does the intracluster medium (ICM) form and when does it thermalize? (iii) When and how is the ICM polluted with metals and is it pre-heated? 
(iv) At what epoch do  cluster galaxies form? (v) What drives their evolution in clusters? (vi) How and when do the most massive galaxies in cluster centers form?

Great advances in observational cosmology over the last years have profoundly impacted the standard world model and have led to the establishment of a new {\em `concordance' cosmology}. Galaxy clusters in this respect have provided early evidence of a low matter density Universe and have proven to be sensitive cosmological probes.
Meanwhile, the properties and physical foundations of the
{\em Dark Energy\/}  component have emerged as the most fundamental open questions of the concordance model. 
One of the most promising approaches for constraining {\em Dark Energy\/}  is provided by {\em distant\/} galaxy clusters.

These open and central questions of cosmic evolution gave rise to the motivation to initiate the {\bf XMM-Newton Distant Cluster Project (XDCP)} as a {\em third generation\/} X-ray cluster survey with special focus on \zg1 systems.
Starting about four years ago, the XDCP is a long-term, medium-size project with the following main goals:







\vspace{-0.5ex}
\begin{enumerate}
    \item Providing \zga1 galaxy clusters at various masses as  {\bf individual astrophysical laboratories} for detailed cluster and galaxy evolution studies (Chaps.\,\ref{c2_cluster_theory} \& \ref{c10_HizClusterStudies});
\vspace{-0.5ex}
    \item Compiling {\bf pathfinder samples} to characterize high-redshift clusters as cosmological probes for major upcoming SZE and X-ray surveys (Chaps.\,\ref{c2_cluster_theory}, \ref{c3_cosmo_theory}, \ref{c9_SurveyResults} \& \ref{c11_Outlook});
\vspace{-0.5ex}
    \item Establishing a statistically complete {\bf cosmological distant cluster sample} selected from a sufficiently large search volume  for cosmological studies (Chaps.\,\ref{c3_cosmo_theory} \& \ref{c9_SurveyResults}).
\end{enumerate}    
\vspace{-0.5ex}

\noindent
These different survey scopes are associated with increasing time lines. Whereas the current XDCP results mainly contribute to the first category, the project focus will shift towards the pathfinder role 
 over the next 2--3 years and 
cosmological studies 
within about 4--5 years.  
The main scope for this work is to {\em establish the observational foundation} of the XMM-Newton Distant Cluster Project.
The main agenda for the thesis can be subdivided into four parts:






\vspace{-0.5ex}
\begin{description}
    \item[Part I:] 
 The {\em theoretical background and state-of-the-art\/} in distant cluster studies is discussed in Chaps.\,\ref{c2_cluster_theory}--\ref{c4_Xsurveys}.   
Chapter\,\ref{c2_cluster_theory} introduces the basic properties of clusters, Chap.\,\ref{c3_cosmo_theory}  establishes the cosmological framework  and discusses clusters as cosmological probes, and  Chap.\,\ref{c4_Xsurveys} provides an overview of X-ray surveys. Owing to the rapid developments in the field over the last decade, the introductory concept aims at a broad overview of the current status in galaxy cluster research. 

\vspace{-0.5ex}    
    \item[Part II:] 
 The {\em observational strategies and methods} are introduced in Chaps.\,\ref{c5_XDCP}--\ref{c8_SpecAnalysis}. 
Chapter\,\ref{c5_XDCP} discusses the general XDCP survey strategy, Chap.\,\ref{c6_XrayAnalysis} summarizes the X-ray analysis techniques,  Chap.\,\ref{c7_NIRanalysis} reviews follow-up imaging methods, and Chap.\,\ref{c8_SpecAnalysis} provides a brief overview of optical spectroscopy. The treatise of these technical chapters is aimed at providing a profound insight to 
the different observational methods to those without deep prior knowledge of the applied techniques. 


\vspace{-0.5ex}
    \item[Part III:] 
    The {\em survey status and first results} are summarized in Chaps.\,\ref{c9_SurveyResults}--\ref{c10b_science_outlook}.
    Chapter\,\ref{c9_SurveyResults} evaluates the current status of the different survey stages and discusses open survey tasks.   Chapter\,\ref{c10_HizClusterStudies} presents  first science applications 
based on spectroscopically confirmed clusters.
Preliminary results of an extended photometrically characterized cluster sample are shown in  Chap.\,\ref{c10b_science_outlook}.
    
    
\vspace{-0.5ex}    
    \item[Part IV:] 
    The thesis closes with an {\em outlook on future developments} in Chap.\,\ref{c11_Outlook}.
    
\end{description}    
\vspace{-0.5ex}







\chapter{Galaxy Clusters as Astrophysical Laboratories}
\label{c2_cluster_theory}

\noindent  This chapter will review selected aspects of the physical foundation of galaxy clusters.
The main focus will be placed on the X-ray properties of the intracluster medium (ICM) and optical characteristics of the (distant) cluster galaxy population.
The discussions on  X-ray emission mechanisms, ICM structure and formation, and cool cores provide the background for understanding the high-redshift cluster detection procedure. X-ray scaling relations are critical for the mass calibration, whereas the \reds of cluster galaxies is used for the initial redshift estimation. The sections on galaxy formation, evolution, and environmental effects set the stage for the high-redshift cluster studies in Chaps.\,\ref{c10_HizClusterStudies}\,\&\,\ref{c10b_science_outlook}.      
More details can be found in the classic work of Sarazin \cite*{Sarazin1986a}, the review of Voit \cite*{Voit2004a}, or Biviano \cite*{Biviano2000a} for a historical treatise of clusters.


\section{General Properties}
\label{s2_general_properties}


\noindent
From a modern point of view, galaxy clusters can be considered as {\em dark matter halos\/}, whose gravitational potential wells are filled with a dilute hot gas, the {\em intracluster medium\/}, and which confine the bound {\em cluster galaxy populations}.  
The mass fractions of these three dominant cluster components are approximately 80--85\% for the dark matter (DM), 12--15\% for the intracluster medium, and only  2--5\% for the total mass of the galaxies. 
 

Galaxy clusters occupy the {\em top level of the cosmic object hierarchy\/} and 
are extraordinary objects in many respects. 
With typical masses of $10^{14}$--\,$10^{15}$$M_{\sun}$ and sizes of a few Mega parsecs (Mpc), they are by far the largest and most massive virialized objects in the Universe. 
Galaxy clusters also lead the object classes in terms of their {\em dark matter domination}, which is manifested in a mass-to-light ratio of $M/L\!\simeq\!200\,(M_{\sun}/L_{\sun})$.
On the other hand, clusters are unique for studying  the {\em baryonic matter component}, since they  define the only large volumes in the Universe from which the majority of baryons emit detectable radiation.




Clusters constitute important {\em astrophysical laboratories} for various fields in physics and are observed and studied over the full electromagnetic spectrum. 
The {\em dark matter component} of the matter density is the main ingredient for cosmological structure formation theories in which the largest collapsed dark matter halos are identified with galaxy clusters (see Sect.\,\ref{s3_struct_formation}). The theoretical dark matter distribution of clusters (Sect.\,\ref{s2_DM_halos}) has been studied in detail with elaborate numerical simulation experiments for about three decades. New observational techniques using the {\em weak gravitational lensing} effect caused by the {\em total} mass concentration of a cluster have recently enabled direct observations of the DM structure of clusters.
 

The {\em hot intracluster medium} (Sect.\,\ref{s2_ICM_properties}) can be considered as a giant plasma physics laboratory filling several Mpc$^3$ 
 of volume at very low electron densities of  $n_{\mathrm{e}}\!\simeq\!10^2$--10$^5$\,m$^{-3}$, an environment which is not accessible for table top experiments. Owing to the plasma temperatures of 
$k_{\mathrm{B}}\,T\!\simeq\!2$--10\,keV, corresponding to $T\!\simeq\!(20$--$100)\!\times\!10^6$\,K, the thermal ICM emission gives rise to  X-ray luminosities of  $L_{\mathrm{X}}\!\simeq\!10^{43}$--\,$3\!\times\!10^{45}$\,erg\,s$^{-1}$, manifesting clusters as the most X-ray luminous objects in the Universe next to Active Galactic Nuclei (AGN).
X-ray observations of the intracluster medium of nearby clusters  nowadays allow precision measurements of the ICM structure, its thermodynamic state, and elemental abundances.  
Furthermore, the ICM enables studies of hydrodynamical and plasma physical processes on the largest scales such as shock fronts, contact discontinuities, propagation of sound waves, turbulence, heat conduction, and diffusion processes. 






Last but not least, the {\em galaxy populations\/} of clusters (Sect.\,\ref{s2_galaxy_populations}), 
accessible through optical and infrared observations, 
can be investigated in well defined environments for detailed studies on
 galaxy formation, evolution, and transformation processes.
We will now have a closer look at the three major individual matter components of galaxy clusters, with a focus on important properties for high-redshift cluster studies.




\section{Clusters as Dark Matter Halos}
\label{s2_DM_halos}

\begin{figure}[t]
\centering
\includegraphics[angle=0,clip,width=0.7\textwidth]{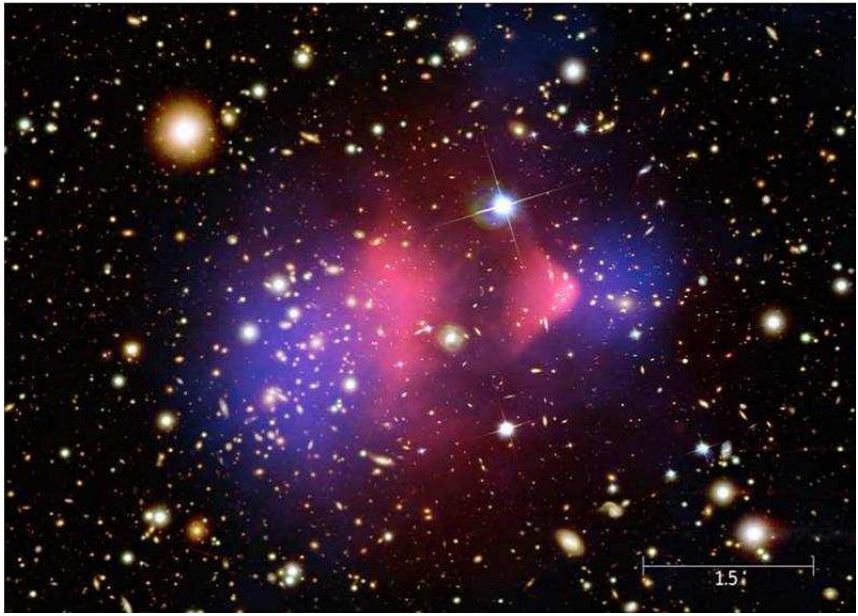}

\caption[Bullet Cluster]{The `{\em Bullet Cluster}' 1E\,0657-56 at $z\!=\!0.296$ provides the best empirical evidence for the existence of dark matter \cite{Clowe2006a}. The cluster was formed through a major merger event which spatially separated the cluster constituents. The X-ray emission of the ICM, as the dominant baryonic component, is shown in pink; the blue shade indicates the weak lensing reconstructed mass concentrations, which are  coincident with the galaxy overdensities. The fluid-like hot ICM gas trails the  collisionless dark matter and galaxy components resulting in spatially distinct centers of the dominant visible matter (ICM) and the major mass constituent (dark matter). Image from press release\footnotemark.  }
\label{f2_BulletCluster}       
\end{figure}

\noindent
The {\em missing mass problem\/} was established  70 years ago in a seminal paper by Zwicky  \cite*{Zwicky1937a}, who noted  that the visible mass in the Coma cluster was not consistent with total mass estimates based on the {\em Virial theorem}.
Besides these very early indications for the need of a dark matter component, the best `direct' empirical evidence for the existence of dark matter to date is also provided by galaxy clusters. Recent detailed studies of the `Bullet Cluster'  have shown that the center of mass of the luminous matter and the total mass distribution have been spatially separated as the consequence of a major merger event \cite{Clowe2006a}. 
Figure\,\ref{f2_BulletCluster} shows the observed X-ray emission of the cluster in pink and the weak lensing reconstructed total mass distribution in blue. The bullet-shaped ICM emission in the right part is trailing the total mass of the infalling  subcluster after the initial collision. Being the dominant {\em baryonic \/} mass component, the ICM center of mass can only deviate from the total mass center if {\em non-baryonic dark matter \/} is the main cluster constituent.






\addtocounter{footnote}{+1} \footnotetext{The image can be found at \url{http://chandra.harvard.edu/photo/2006/1e0657/1e0657_scale.jpg}.}
\addtocounter{footnote}{-2}

The properties of dark matter halos, \ie \ DM aggregates that form a quasi equilibrium configuration, have been investigated in detail by means of 
high-resolution numerical simulations. 
Navarro, Frenk, and White (NFW) \cite*{Navarro1997a} found a universal DM halo density profile  for virialized objects
for the now standard {\em cold dark matter}\footnote{{\em Cold \/} means that the kinetic energy of the particles is much smaller than their rest mass energy.} (CDM) paradigm.  
The widely used NFW fitting formula for the density profile $\rho_{\mathrm{DM}}(r)$ has the form


\begin{equation}\label{e2_NFW_profile}
    \rho_{\mathrm{DM}}(r)  = \frac{\rho_{\mathrm{s}}}{\left(\frac{r}{r_{\mathrm{s}}} \right)\left(1+\frac{ r}{r_{\mathrm{s}}} \right)^2} \ ,
\end{equation}


\noindent
where $r_{\mathrm{s}}$ is a characteristic scale length and $\rho_{\mathrm{s}}\!=\!\delta_{\mathrm{c}}\,\rho_{\mathrm{cr}}(z)$
the central density (see Equ.\,\ref{e3_crit_density}), which is linked to the cosmological model (see Sect.\,\ref{s3_struct_formation}).
At small radii $r\!\ll\!r_{\mathrm{s}}$, the NFW profile exhibits a `cusp' associated with the inner slope $\rho_{\mathrm{DM}}(r)\!\propto\!r^{-1}$, whereas at cluster-centric distances of  $r\!\gg\!r_{\mathrm{s}}$  the density follows $\rho_{\mathrm{DM}}(r)\!\propto\!r^{-3}$.

The characteristic central `cuspy' shape of the NFW profile is a fundamental prediction of the CDM paradigm, quasi independent   
of halo mass and cosmological parameters. 
X-ray observations of galaxy clusters have meanwhile confirmed the predicted universal CDM profile (\eg \ Pratt \& Arnaud, 2002; Pointecouteau \etal, 2005), as shown in Fig.\,\ref{f2_mass_profiles}. 
\nocite{Pratt2002a}
\nocite{Pointec2005a}
Concerning clusters at high redshift, the NFW profile predicts denser cores compared to recently formed objects as a consequence of the increased critical density  $\rho_{\mathrm{cr}}(z)$ at collapse time (see Sect.\,\ref{s3_struct_formation}).












\begin{figure}[t]
\centering
\includegraphics[angle=0,clip,width=\textwidth]{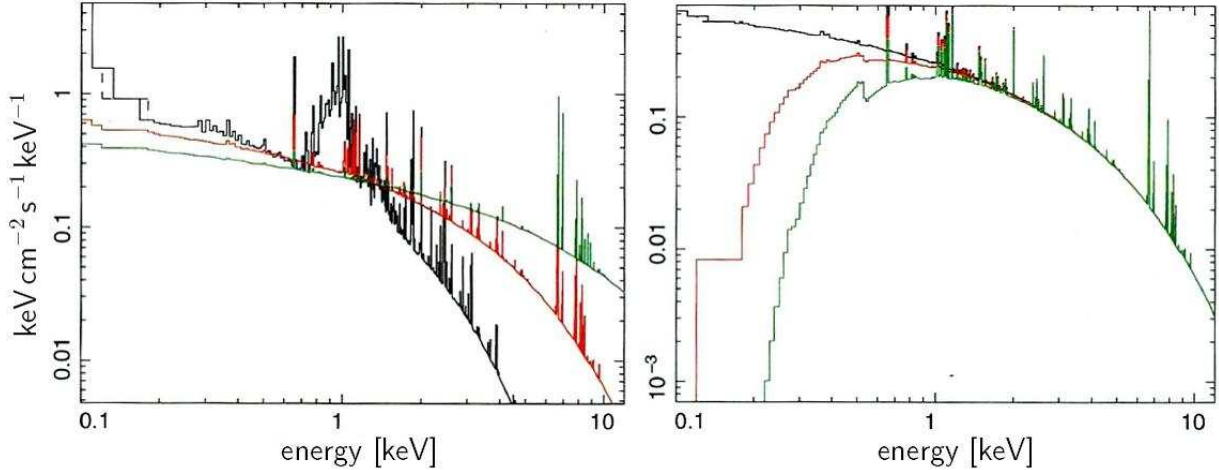}

\caption[Thermal X-ray Spectra]{X-ray emission of the ICM. 
{\em Left:} Unabsorbed ICM model spectra of the bremsstrahlung continuum and line emission for plasmas with 0.4 solar metallicity and  temperatures of 1\,keV (black), 3\,keV (red), and 9\,keV (green).   
{\em Right:} Absorbed model spectra of $T\!=\!3$\,keV observed through galactic  hydrogen columns of $10^{21}$\,cm$^{-2}$ (green),   $3\!\times\!10^{20}$\,cm$^{-2}$ (red), and for the unabsorbed case (black).  
Plots from Schneider \cite*{Schneider2006a}.}
\label{f2_Thermal_Spectra}       
\end{figure}

\section{X-ray Properties of the Intracluster Medium}
\label{s2_ICM_properties}
\noindent
``{\it Clusters of galaxies are X-ray sources because galaxy formation is inefficient}'' \cite{Voit2004a}.
About 90\% of all baryons in the Universe reside in intergalactic space, but X-ray emitting gas temperatures are only reached in the deep potential wells of galaxy clusters.
The  main X-ray properties of the hot intracluster medium will be summarized in this section.


\subsection{Emission mechanisms}
\label{s2_emission_mechan}

\noindent
The dominant X-ray emission mechanism for X-ray clusters at temperatures of $T\!\ga\!10^7$\,K ($k_{\mathrm{B}}T\!\ga\!1$\,keV) is {\em thermal bremsstrahlung} (free-free radiation).
The intracluster medium is a collisionally ionized, optically thin plasma, which  implies that essentially every emitted photon will escape from the cluster volume.  
The bremsstrahlung emissivity $\epsilon^{\mathrm{ff}}_{\nu}$ for a plasma at temperature $T$, \ie \ the luminosity per unit volume and frequency interval, is given by




\begin{equation}\label{e2_bremstr_emissivity}
\epsilon^{\mathrm{ff}}_{\nu} \approx 6.8\!\times\!10^{-38}\,Z^2_{\mathrm{i}}\,n_{\mathrm{e}}\,n_{\mathrm{i}}
\,g^{\mathrm{ff}}(\nu, T)
\ \frac{e^{- \frac{h_{\mathrm{P}}\,\nu}{k_{\mathrm{B}}\,T}}}{\sqrt{T}} \ \mathrm{erg\,s}^{-1}\mathrm{cm}^{-3}\mathrm{Hz}^{-1} \ ,
\end{equation}

\noindent
where n$_{\mathrm{e}}$ and $n_{\mathrm{i}}$ denote\footnote{To avoid confusion with the Hubble constant parameter $h$ and wavenumbers $k$, the Boltzmann constant is denoted as $k_{\mathrm{B}}$ and the Planck constant as $h_{\mathrm{P}}$ throughout this thesis.} the number densities of electrons and ions, $Z_{\mathrm{i}}$ the ion charge, and $g^{\mathrm{ff}}(\nu, T)\!\propto\!\ln (9k_{\mathrm{B}}T/(4h_{\mathrm{P}}\nu))$ is the Gaunt factor, which is of order unity 
and corrects for quantum mechanical effects and distant collisions. Equation\,\ref{e2_bremstr_emissivity} implies that thermal bremsstrahlung spectra are essentially constant (in frequency space) for $h_{\mathrm{P}}\nu\!\ll\!k_{\mathrm{B}}\,T$ and exhibit a temperature dependent steep exponential cut-off at high frequencies of $h_{\mathrm{P}}\nu\!\ga\!k_{\mathrm{B}}\,T$. 
One of the key features of X-ray observations is the proportionality of the bremsstrahlung emissivity (and all other radiation processes induced by particle collisions)   to the {\em square} of the gas density
$\epsilon^{\mathrm{ff}}_{\nu}\!\propto\!n_{\mathrm{e}}\,n_{\mathrm{i}}\!\approx\!n^2_{\mathrm{e}}$, which results in a peaked, high contrast signature of the central cluster regions. The total integrated (bolometric) volume emissivity of a thermal plasma with solar metal abundances  is then

\begin{equation}\label{e2_bremstr_bol}
\epsilon_{\mathrm{bol}}^{\mathrm{ff}}  = \int_0^{\infty}d\nu\,\epsilon^{\mathrm{ff}}_{\nu} \approx
 9.5\!\times\!10^{-32}\,\sqrt{\frac{T}{1\,\mathrm{keV}}}\,\left(\frac{n_{\mathrm{e}}}{10^{-3}\,\mathrm{cm}^{-3}} \right)^2
 \mathrm{erg\,s}^{-1}\mathrm{cm}^{-3} \propto \sqrt{T}\ .
\end{equation}



\noindent
At temperatures of $T\!\la\!2$\,keV, recombination radiation (free-bound) and line emission radiation (bound-bound) become important\footnote{Modern ICM radiation codes account for {\em all\/} emission mechanisms at  {\em all\/} temperatures.} as the ionization levels of the ICM metals decrease.
For ions with an ionization potential $\chi_{\mathrm{i}}$ of state $i$, the free-bound emissivity has the form


\begin{equation}\label{e2_recomb_emissivity}
\epsilon^{\mathrm{fb}}_{\nu} \propto \,g^{\mathrm{fb}}(\nu, T)
\ \frac{e^{- \frac{h_{\mathrm{P}}\,\nu - \chi_{\mathrm{i}}}{k_{\mathrm{B}}\,T}}}{T^{\frac{3}{2}}} \ . 
\end{equation}


\noindent
The most important X-ray line feature for massive clusters is  the K-shell line complex of hydrogen-like iron Fe\,{\small XXVI} around 6.7\,keV, with slightly shifted energies for other ionization states. At lower temperatures, additional important line features originate from the Fe\,L line complex at $\sim\!1$\,keV and  ions of O, Mg, Si, S, Ar, Ca, and Ne.  



Figure\,\ref{f2_Thermal_Spectra} illustrates in the left panel three thermal ICM model spectra for plasma temperatures of 9\,keV (green), 3\,keV (red), and 1\,keV (black) including the bremsstrahlung continuum, recombination radiation, and line emission.  
The right panel shows the effects of the interstellar medium (ISM) absorption on an observed 3\,keV cluster spectrum for  hydrogen column densities of $10^{21}$\,cm$^{-2}$ (green), $3\!\times\!10^{20}$\,cm$^{-2}$ (red), and in the unabsorbed case (black). The photoionization cross-section of neutral hydrogen, and similarly for the heavier elements of the ISM, exhibits 
approximately the photon energy dependence 
$a_{E}\!\propto\! E^{-3}$ \cite{allen2000}, \ie \ the extinction  increases rapidly with decreasing  energy (above the absorption edge).  
As can be seen from Fig.\,\ref{f2_Thermal_Spectra}, the galactic ISM extinction introduces an effective cut-off in the observed spectrum at energies of about $\la\!0.3$\,keV for the  typical hydrogen columns of $2$--$5\!\times\!10^{20}$\,cm$^{-2}$  in extragalactic survey fields. 


The total luminosity of a cluster is  obtained from the emissivity by integrating over the plasma volume and frequency
$L_{\mathrm{X}}\!=\!\int_{\nu}\int_V \epsilon_{\nu}\,dV\,d\nu\!\propto\!\mathrm{EM}$, where
 $\mathrm{EM}\!=\!\int_V n^2_{\mathrm{e}}\,dV$  is the emission measure.
Observed along the line-of sight at projected radius $r\!=\!d_{\mathrm{ang}}(z)\,\Theta$ (see Equ.\,\ref{e3_ang_dist}), the emission measure
$\mathrm{EM}\!=\!\int n^2_{\mathrm{e}}\,dl\!=\!4\,\pi\,(1\!+\!z)^4\,S(\Theta)/\Lambda(T,z)$ 
gives direct observational access to the local electron density  $n_{\mathrm{e}}$,
deduced from the measured surface brightness profile $S(\Theta)$ as a function of cluster-centric angle $\Theta$  in the soft X-ray band (see Equ.\,\ref{e2_beta_model}), and the redshifted and absorption corrected emissivity $\Lambda(T,z)$.
The gas temperature $T_{\mathrm{X}}$, as the second local X-ray observable, is obtained from a model fit to the redshifted X-ray spectrum, which achieves the highest temperature sensitivity in the hard band region.








\subsection{ICM structure}
\label{s2_ICM_struct}

\noindent
The angle-averaged global spatial structure of the intracluster medium  
is commonly described  by the {\em King model\/}  for a self-gravitating, isothermal sphere \cite{King1966a}.
If one allows for an additional scaling of the ICM density profile with respect to the underlying dark matter profile 
of the form $\rho_{\mathrm{gas}}\!\propto\!\rho^{\beta}_{\mathrm{DM}}$, the three dimensional {\em $\beta$-model\/} for the radial ICM profile $\rho_{\mathrm{gas}}(r)$ is obtained \cite{Cavaliere1976a}

\begin{equation}\label{e2_3D_beta_model}
\rho_{\mathrm{gas}}(r) = \frac{ \rho_{\mathrm{gas,0}}}{ \left[ 1 +  \left(\frac{r}{r_{\mathrm{c}}}\right)^2 \right]^{\frac{3 \beta}{2}}} \ .
\end{equation}

\noindent
The core radius $r_{\mathrm{c}}$ determines the characteristic extent scale of the source, whereas the $\beta$ parameter modulates the overall steepness of the profile. 
The original {\em King model\/} for the underlying dark matter potential is recovered by setting $\beta\!=\!1$.
Note that the central gas density  $\rho_{\mathrm{gas,0}}$ defines the peak of the {\em core} of the profile, in contrast to the rising {\em cusp} of the NFW profile of Equ.\,\ref{e2_NFW_profile}.

The {\em $\beta$-model\/} for the global ICM structure assumes a spherically symmetric, isothermal gas in hydrostatic equilibrium. 
For $\beta$ parameters deviating from unity, the galaxies, as dynamic tracers of the dark matter potential, exhibit a different velocity dispersion than the ICM gas according to  $\sigma^2_\mathrm{gal}\!=\!\beta\,\sigma^2_\mathrm{gas}$. 
Typical values of $\beta\!\simeq\!2/3\!\pm\!0.2$ \cite{Jones1984a} indicate that the galaxy velocity dispersion in clusters is lower than for the gas. In effect, $\beta$ is the ratio between the kinetic energy of tracers of the gravitational potential (\eg \ galaxies) and the thermal energy of the gas
$\beta\!=\!\mu\,m_{\mathrm{p}}\,\sigma^2_{\mathrm{r}}/(k_{\mathrm{B}}T_{\mathrm{gas}})$, where $\sigma_{\mathrm{r}}$ denotes the radial velocity dispersion, $m_{\mathrm{p}}$ the proton mass, and $\mu$ the mean molecular weight of the gas ($\mu\!\simeq\!0.6$ for a primordial gas composition). The empirically determined $\beta$ values imply that the mean energy per unit mass is higher for the intracluster medium component than for the galaxies.
 

 





The observed two dimensional X-ray surface brightness profile $S(\Theta)$ of the {\em $\beta$-model\/} is obtained by relating the gas density profile $\rho_{\mathrm{gas}}(r)$ to the X-ray emissivity (Equ.\,\ref{e2_bremstr_emissivity}) and projecting the distribution on the plane of the sky. The resulting radially symmetric profile

\begin{equation}\label{e2_beta_model}
S(\Theta) = \frac{ S_0}{ \left[ 1 + \left(\frac{\Theta}{\Theta_\mathrm{c}}\right)^2\right]^{3 \cdot \beta - \frac{1}{2}}} = \frac{ S_0 }{\left[ 1 + \left(\frac{\Theta}{\Theta_\mathrm{c}}\right)^2\right]^{\frac{3}{2}}} \ \ \ \ \ \ \mathrm{for} \ \beta = \frac{2}{3} \\
\end{equation}

\noindent
contains the fit parameters $\Theta_{\mathrm{c}}$ for the angular core radius size, the central surface brightness $S_0$, and the $\beta$ value.
In the last step, the slope parameter is fixed to $\beta\!=\!2/3$ in order to reduce the number of degrees of freedom for model fits of weak sources. This constrained  surface brightness profile serves as a reference model for the detection of serendipitous extended X-ray  sources (\eg \ clusters) in  Sect.\,\ref{s6_source_detection}, which contain typically a few hundred X-ray photons.
Detailed investigations of the morphological properties of nearby clusters (\eg \ B\"ohringer \etal, 2007b)  
\nocite{HxB2007a} require more elaborated structural parameters that account for multiple components and deviations from spherical symmetry.







\subsection{ICM formation}
\label{s2_ICM_form}

\noindent
Studies of the ICM formation process will be one of the important future applications of the XDCP survey; 
some basic considerations and models are now discussed.
Scenarios for the intracluster medium formation need to account for two essential observational constraints: (i) the {\em total ICM mass} of $M_{\mathrm{ICM}}\!\sim\!10^{14}\,M_{\sun}$ for massive clusters, and (ii) the {\em metallicity enrichment} 
of the intracluster gas with values of $Z_{\mathrm{ICM}}\!\sim\!0.4\,Z_{\sun}$.


We will first take a look at  {\em infall models} for the ICM, which were proposed shortly after the first X-ray observations of clusters became available (\eg \ Gott\,\&\,Gunn, 1971;  Gunn\,\&\,Gott, 1972).  
\nocite{Gott1971a} 
\nocite{Gunn1972a} 
Let us consider a primordial, cool gas initially at rest at large distance that falls into the potential well of an existing massive dark matter halo. Applying the {\em Virial theorem\/} for the final equilibrium state of the gas settled to the bottom of the potential, we obtain 

\begin{equation}\label{e2_virial_theorem}
    2\,\bar{E}_{\mathrm{kin}} + \bar{E}_{\mathrm{pot}} = 0 \simeq \langle v^2 \rangle - \frac{G\,M}{R_{\mathrm{g}}} \simeq  \frac{3\,k_{\mathrm{B}}T_{\mathrm{gas}}}{\mu\,m_{\mathrm{p}}} - \frac{G\,M}{R_{\mathrm{g}}} \ ,
\end{equation}

\noindent 
where $\langle v^2 \rangle\!\equiv\!\sigma^2\!=\!3\,\sigma_{\mathrm{r}}^2$ is the squared (radial) velocity dispersion of particles in the potential and $R_{\mathrm{g}}\!\equiv\!G\,M^2/|\bar{E}_{\mathrm{pot}}|$ is the gravitational radius of the cluster. In the last step, the kinetic particle energy was equated with the thermal energy of a monoatomic gas (\ie \ $\beta\!\simeq\!1$).
This first order approximation yields ICM gas temperature estimates expressed in terms of the velocity dispersion of the potential of

\begin{equation}\label{e2_ICM_temp}
     T_{\mathrm{gas}} \simeq  \frac{\mu\,m_{\mathrm{p}}\,\sigma^2_{\mathrm{r}}}{k_{\mathrm{B}}} 
     \simeq 7.3 \times 10^7  \ \mathrm{K} \cdot
     \left( \frac{\sigma_{\mathrm{r}}}{10^3\,\mathrm{km\,s}^{-1}} \right)^2  \simeq 6.3 \ \mathrm{keV} \cdot \left( \frac{\sigma_{\mathrm{r}}}{10^3\,\mathrm{km\,s}^{-1}} \right)^2 
\end{equation}  

\noindent
in reasonably good agreement with observations. The {\em hot ICM\/} temperatures can hence be explained as a natural consequence of gas infall to the dark matter halo. 
The {\em total ICM mass\/} can be accounted for when considering that clusters form via collapse of cosmic regions of $R_{\mathrm{col}}\!\sim\!\orderof$(10\,Mpc) with an associated approximate baryon reservoir of $M_{\mathrm{bar}}\!\sim\!(4/3)\,\pi\,R^3_{\mathrm{col}}\cdot \rho_{\mathrm{cr}}\,\Omega_{\mathrm{b}}\,(1\!+\!\delta)\!\sim\!10^{14}\,M_{\sun}$ (see Sect.\,\ref{s3_struct_formation} for explanation).

Several assumptions for the derived temperature estimate might not be well justified and could modify the gas temperatures: (i)   
the ICM infall {\em after \/} the DM halo formed, (ii) initially only {\em marginally bound\/} gas coming from a large distance, (iii) a {\em cold\/} medium before infall, and  (iv) no radiative energy losses during collapse. 

 

The purely gravitationally driven thermalization process of the ICM can be qualitatively described with a spherically symmetric smooth accretion model: (i) incoming cold gas enters the cluster in concentric shells, (ii) the gas is adiabatically compressed as it reaches the denser central regions, (iii) accretion shocks form when the gas motions become supersonic and the shocks propagate outwards, (iv) subsequent infalling shells pass through the accretion shocks of the inner shells and are shock-heated to the virial temperature of the cluster within about one sound crossing time $t_{\mathrm{sound}}$.  
For hierarchical merger scenarios, accretion shocks are not well defined but form a complex network of shocks as different infalling subclusters mix with the ICM of the main halo.
After shocks have passed, the ICM gas is in a nearly hydrostatic state and further evolution proceeds quasi-static.
Since the cluster and ICM formation physics contains no characteristic scale, a self-similarity of the dark matter and 
gas internal structure is expected, \ie \ the main object properties should only be scaled versions of each other.

The characteristic cluster formation time scale is given by the sound crossing time $t_{\mathrm{sound}}$ which can be expressed as

\begin{equation}\label{e2_soundcrossing_time}
    t_{\mathrm{sound}} \simeq \frac{R}{c_{\mathrm{sound}}}  \simeq 0.66  \, \mathrm{Gyr} \cdot \left(\frac{R}{1\,\mathrm{Mpc}}\right)  \left(\frac{T_{\mathrm{gas}}}{10^8 \,\mathrm{K}} \right)^{-\frac{1}{2}}
\end{equation}

\noindent
derived from the intracluster medium sound speed of

\begin{equation}\label{e2_soundspeed}
    c_{\mathrm{sound}} \simeq 1\,480  \ \mathrm{km\,s}^{-1} \cdot \left(\frac{T_{\mathrm{gas}}}{10^8 \,\mathrm{K}} \right)^{\frac{1}{2}} .
\end{equation}

\noindent
The sound crossing timescale is hence short compared to the ages of low-redshift clusters, implying that the  central 
cluster gas will reach a quasi-hydrostatic equilibrium about 1\,Gyr after formation or major merger events. 
The situation for clusters at redshift $z\!\sim\!1.5$ with a cosmic age of $\sim\!4$\,Gyrs (see Fig.\,\ref{f3_cosmic_time}) is already quite different and one would expect to find a large fraction of systems that have not yet fully virialized.
Equation\,\ref{e2_soundcrossing_time} additionally implies that 
cosmic regions with dimensions of $\ga\!10$\,Mpc will not be able to virialize within a Hubble time and hence no relaxed objects larger than the cluster scale exist at the current epoch.


The ICM sound speed (Equ.\,\ref{e2_soundspeed}) is comparable to the velocities of the cluster galaxies. 
The galaxy crossing time scale for a passage through the cluster $t_{\mathrm{cross}}$


\begin{equation}\label{e2_crossing_time}
    t_{\mathrm{cross}} \simeq \frac{R}{\sigma_v} \simeq 1 \, \mathrm{Gyr} \cdot \left(\frac{R}{1\,\mathrm{Mpc}}\right)  \left(\frac{\sigma_v}{10^3 \,\mathrm{km}\,\mathrm{s}^{-1}}\right)^{-1} 
\end{equation}

\noindent
is hence similar to $t_{\mathrm{sound}}$ implying that the same arguments apply for the relaxation time scale of the galaxy population.







ICM {\em infall models} can reproduce basic properties of the intracluster medium, but do not account for 
the observed heavy element abundances   of the initially metal free primordial gas. 
The detection of the 6.7\,keV iron line \cite{Mitchell1977a} provided early evidence that at least some of the gas must have been processed by stars and later been ejected from the galaxies.
Heavy element abundances and their distribution in low redshift clusters can now be consistently attributed to supernova explosion yields, which also provide sufficient energy for ejecting the metals out of the galaxies into the ICM (\eg \ B\"ohringer, 2003).
\nocite{HxB2003a}

As a last point concerning the formation process, we can now address the question of what determines the density of the intracluster medium, which is closely linked to the entropy of the ICM.
The cluster entropy\footnote{$K$ is the constant of proportionality in the equation-of-state for an adiabatic monoatomic gas $P\!=\!K\rho_{\mathrm{gas}}^{5/3}$ and is related to the standard thermodynamic entropy per particle through $s\!=\!k_{\mathrm{B}}\,\ln K^{3/2}\!+\!s_0$. An alternative, widely used definition for the cluster entropy is 
$S\!=\!K_{\mathrm{e}}\!=\!k_{\mathrm{B}}\,T / n_{\mathrm{e}}^{2/3}$.} $K\!\equiv\!k_{\mathrm{B}}\,T / (\mu\,m_{\mathrm{p}}\,\rho_{\mathrm{gas}}^{2/3})$ and the shape of the dark matter potential well are the main determinants for the ICM X-ray structure \cite{Voit2004a}. As a second fundamental key feature, the cluster entropy contains a record of the   
thermodynamic history of the ICM gas.



 


The ICM structure of clusters is influenced by $K$ because, in a simple picture, 
{\it high entropy gas floats and low entropy gas sinks}.
The ICM gas thus convects\footnote{Observations deviate from the expected polytropic index of $\gamma\!=\!5/3$ for a convectively established equilibrium, suggesting that also other processes are significant.} until the equilibrium condition $dK/dr\!\geq\!0$ is fulfilled throughout the cluster, \ie \
a quasi-static configuration is achieved when the isentropic surfaces coincide with the equipotential surfaces of the dark matter potential.



For a purely gravitationally driven ICM collapse with initially cold gas, the accretion shocks are the only source of entropy for the ICM gas, with the consequence that the self-similar entropy profile $K(r)\!\propto\!r^{1.1}$ depends entirely on the mass accretion history $M(t)$.
In the case that the accreted gas was not cold, an isentropic central entropy core is expected with a level similar to the pre-shock entropy. 
This so called {\em pre-heating\/} of ICM gas before infall makes gas harder to compress and thus lowers the density in the cluster center, where entropy is smallest. This mechanism  can give rise to a breaking of the cluster  self-similarity and is hence of great interest. Observations seem to be consistent with a minimum entropy floor\footnote{This picture has recently been adjusted in terms of a gradual change of the entropy level as a function of ICM temperature.} level of about 100--200\,keV\,cm$^2$,  but the simple pre-heating model is likely to be complemented by entropy changing radiative cooling and heating processes.

Many fundamental questions concerning the intracluster medium formation have remained unanswered and are awaiting investigations using appropriate high-redshift X-ray cluster samples.
(i) At what epoch does the  thermal ICM  X-ray emission start?
(ii) Are non-gravitational processes  important during the formation process? 
(iii) Is the gas pre-heated before falling into the cluster potential well? 
(iv) What is the source of (extra) entropy?
(v) What is the early metal enrichment history and thermal evolution of the ICM?

\subsection{Scaling relations}
\label{s2_scaling_relations}

\begin{figure}[t]
\centering
\includegraphics[angle=0,clip,width=0.486\textwidth]{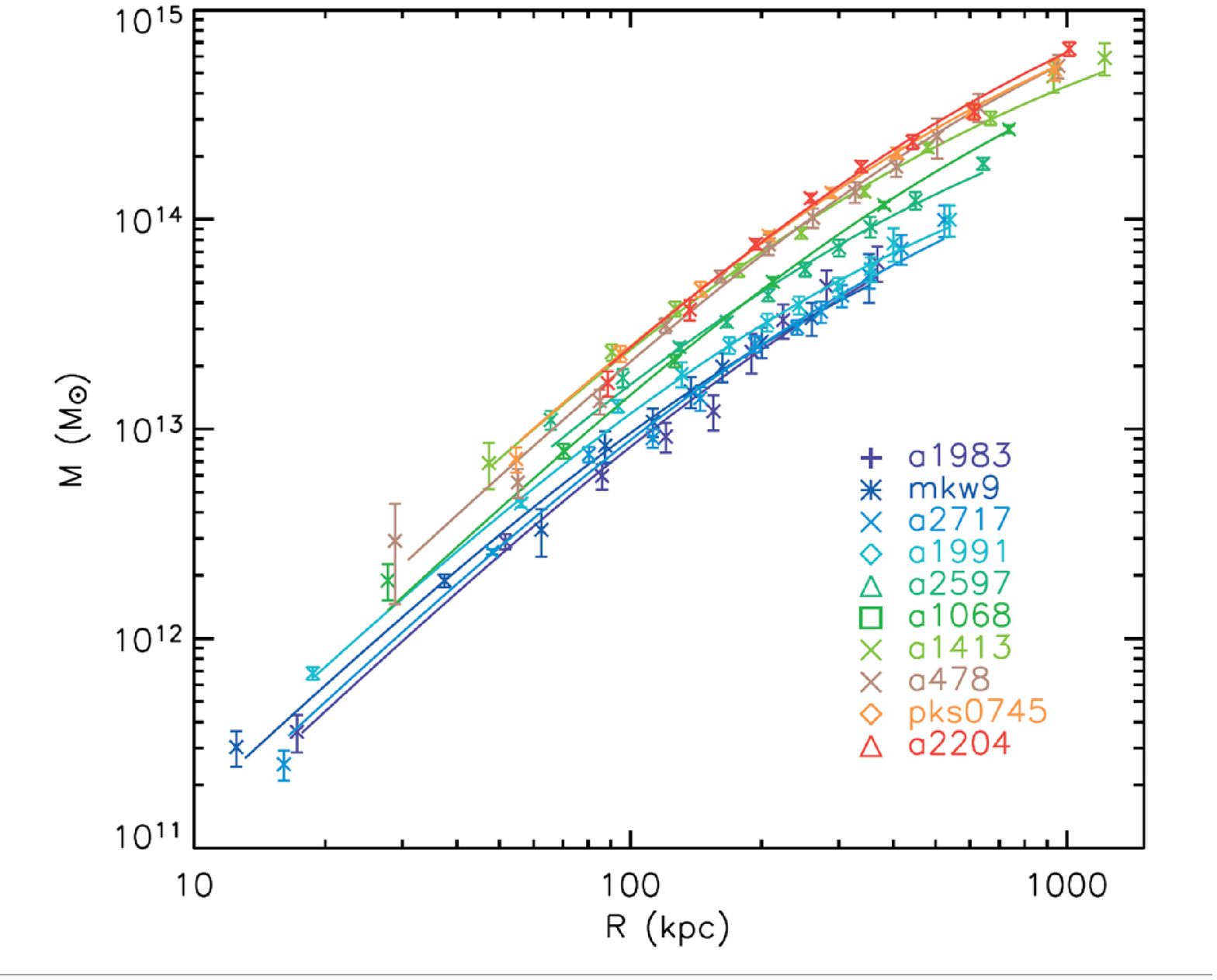}
\includegraphics[angle=0,clip,width=0.505\textwidth]{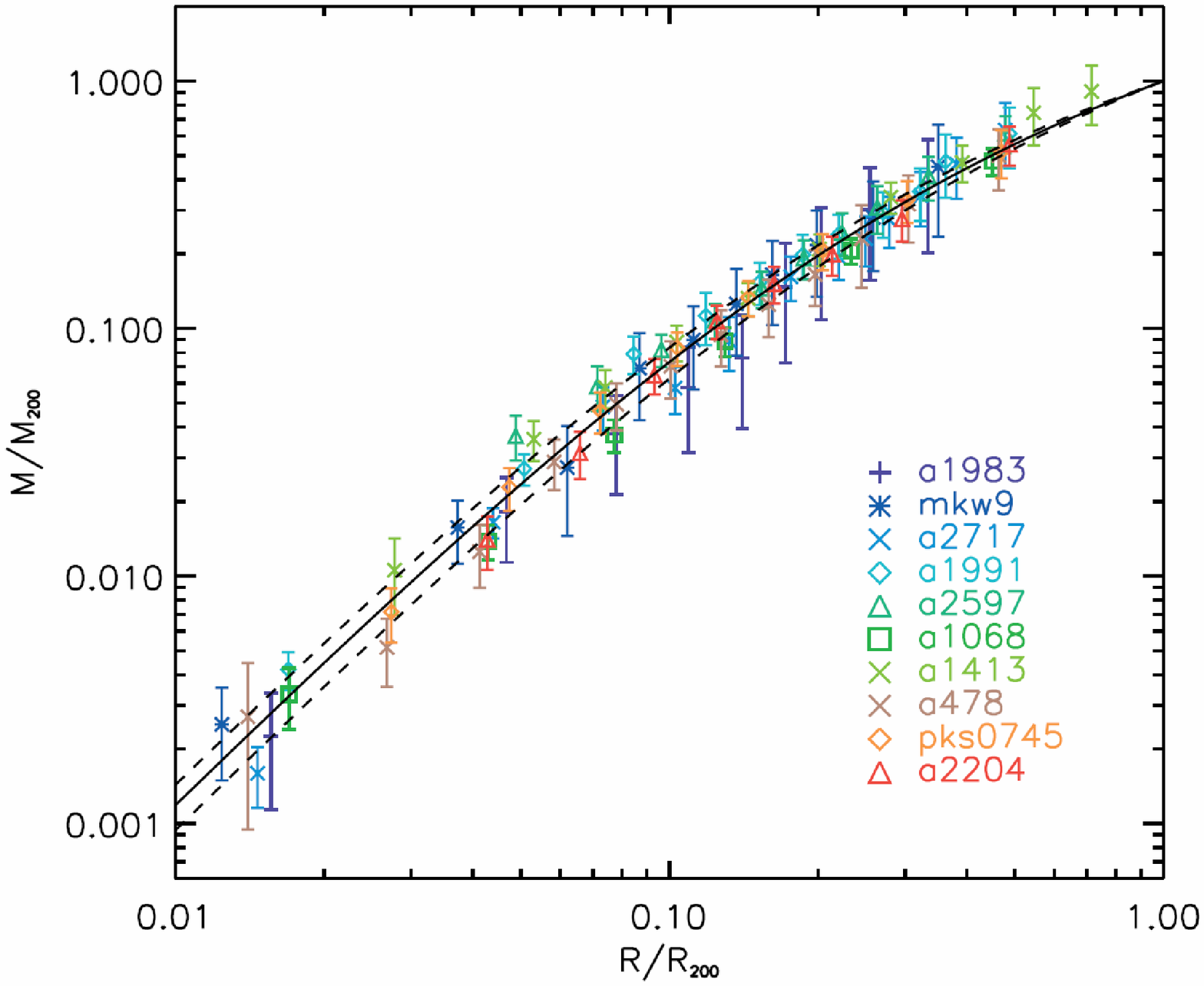}
\caption[Cluster Mass Profiles]{Galaxy cluster mass profiles for nearby clusters.
{\em Left:} Integrated total mass profiles as a function of  physical radius (kpc) for ten clusters with temperatures in the range 2--9\,keV. Solid lines represent the best fitting NFW profiles.
{\em Right:} Scaled mass profiles for the same clusters with masses scaled to $M_{200}$ and radii to $R_{200}$.  The black line gives the  mean scaled NFW model. Plots from Pointecouteau \etal \  \cite*{Pointec2005a}.}
\label{f2_mass_profiles}       
\end{figure}

\begin{figure}[t]
\parbox{0.58\textwidth}{
\includegraphics[angle=0,clip,width=0.6\textwidth]{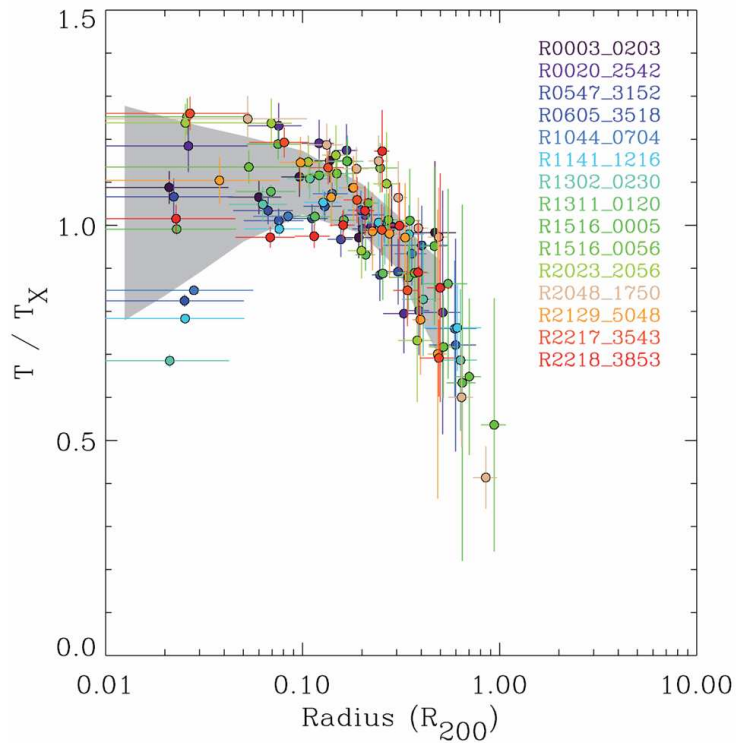}
}
\hfill
\parbox{0.39\textwidth}{
\caption[Cluster Temperature Profiles]{Scaled temperature profiles of 15 nearby clusters. The grey shaded area indicates the region enclosing the 1\,$\sigma$ deviations around the mean profile. Temperatures are scaled to the averaged temperature in the outer cluster parts. Plot from Pratt \etal \ \cite*{Pratt2007a}.
}\vfill
}
\label{f2_temperature_profiles}       
\end{figure}

\noindent
The expected (nearly) self-similar cluster formation process gives rise to general scaling relations of physical cluster properties. The comparison of observed relations with the scaling behavior of the self-similar predictions yield important clues on additional contributing physical processes. Scaling relations for distant cluster applications provide the crucial {\em mass proxy} in absence of high quality X-ray data for a detailed analysis, which requires $\ga\!1\,000$ X-ray photons. We will first introduce the standard method for deriving total cluster masses from X-ray data and will then discuss some important scaling relations.      

With XMM-Newton and Chandra, the following physical cluster properties can be directly derived from sufficiently deep X-ray data:
(i) the X-ray luminosity $L_{\mathrm{X}}$ from the source flux $f_{\mathrm{X}}$ (and redshift),  (ii) the radial gas density profile $\rho_{\mathrm{gas}}(r)$ from the local emission measure $\mathrm{EM}(r)$
(iii) the total gas mass $M_{\mathrm{gas}}$ from the integrated density profile,
(iv) the global mean temperature $T_{\mathrm{X}}$ from the observed spectrum,
and (v) the temperature profile $T_{\mathrm{X}}(r)$ from spatially resolved spectra.
Using the spatially resolved density and temperature distributions, the (vi) entropy profile $K(r)$ and (vii) the total integrated mass profile $M_{\mathrm{tot}}(<r)$ can be obtained.
 




Under the assumptions of spherical symmetry and hydrostatic  equilibrium, the integrated total mass profile $M_{\mathrm{tot}}(<r)$
can be derived from the ideal gas equation $p_{\mathrm{gas}}\!=\!\rho_{\mathrm{gas}}\,k_{\mathrm{B}}\,T_{\mathrm{X}}/(\mu\,m_{\mathrm{p}})$ and the 
hydrostatic equation for pressure equilibrium


\begin{equation}\label{e2_hydrostatic_equilibr}
    \frac{dp_{\mathrm{gas}}(r)}{dr} = - \rho_{\mathrm{gas}}(r)\frac{G\,M_{\mathrm{tot}}(<\!r)}{r^2} =  \frac{k_{\mathrm{B}}}{\mu\,m_{\mathrm{p}}}\frac{d[\rho_{\mathrm{gas}}(r)\,T_{\mathrm{X}}(r)]}{dr} \ .
\end{equation}

\noindent
Solving for $M_{\mathrm{tot}}(<\!r)$ yields the final important relationship for the total gravitating mass within radius $r$

\begin{equation}\label{e2_cluster_mass}
    M_{\mathrm{tot}}(<\!r) = - \frac{k_{\mathrm{B}}\,T(r) \ r}{G\,\mu\,m_{\mathrm{p}}} \left( \frac{d\ln \rho_{\mathrm{gas}}}{d\ln  r} + \frac{d\ln T}{d\ln  r} \right) \ .
\end{equation}

\noindent
The integrated mass at radius $r$ is hence determined by the local temperature $T(r)$ and the sum of the logarithmic slopes of the density and temperature profiles. Since galaxy clusters do not have a well-defined boundary, the total virialized mass $M_{\mathrm{vir}}\!\equiv\!M_{200}$ is commonly defined within the radius $R_{\mathrm{vir}}\!\equiv\!R_{200}$, where the average cluster density equals a specified density contrast $\delta\!\sim\!200$ with respect to the critical density $\rho_{\mathrm{cr}}$ at the cluster redshift $z$ (see Chap.\,\ref{c3_cosmo_theory}) 

\begin{equation}\label{e2_cluster_radius}
    \frac{M_{200}}{\frac{4\,\pi}{3}\,R^3_{200}} = \langle \rho \rangle_{\mathrm{vir}} = \delta\,\rho_{\mathrm{cr}}(z) \ \ \ , \ \ \ \delta \sim 200 \ .
\end{equation}





\noindent
The left panel of Fig.\,\ref{f2_mass_profiles} shows the determined integrated total mass profiles of ten nearby clusters within the  temperature range of 2--9\,keV \cite{Pointec2005a} with the best-fitting NFW dark matter profiles (Equ.\,\ref{e2_NFW_profile}) indicated by the solid lines. Once the mass profiles are scaled to units of $R_{200}$ and $M_{200}$  
(right panel) the clusters reveal their self-similar property and are all consistent with a single universal NFW mass profile illustrated by the solid line. A similar universal scaling behavior can be observed for the outer regions of the temperature profiles as displayed in Fig.\,2.4 
 \cite{Pratt2007a}. The inner cluster regions  at $R\!\la\!0.1\,R_{200}$  exhibit an increased temperature scatter due to additional non-gravitational physics in cluster centers (Sect.\,\ref{s2_CCC}).

As for the scaled radial profiles, a self-similar 
behavior is expected for global properties \cite{Kaiser1986a}. 
Under the assumptions that for systems in virial equilibrium (i) the bremsstrahlung emissivity scales with temperature as  $\epsilon_{\mathrm{bol}}^{\mathrm{ff}}\!\propto\!\sqrt{T_{\mathrm{X}}}$ (Equ.\,\ref{e2_bremstr_bol}), (ii) the mass scales with a characteristic radius $R_*$ as $M_{\mathrm{tot}}\!\propto\!R_*^3$, (iii) a constant gas mass fraction $f_{\mathrm{gas}}\!=\!M_{\mathrm{gas}}/M_{\mathrm{tot}}\!\approx\!\mathrm{const}$, and (iv) self-similar dark matter halos (Equ.\,\ref{e2_NFW_profile}), global relations for purely gravitationally driven cluster physics can be derived.  
Since the NFW dark matter halos form a two parameter family, where the parameters are the total halo mass $M_{\mathrm{tot}}$ and the formation epoch $z_{\mathrm{form}}$, the relations are expected to hold for all clusters that formed at the same redshift.
The proper redshift dependence of the scaling relations can additionally be accounted for by considering the redshift evolution of the collapse threshold \cite{Kitayama1996a}, which  predominantly  depends on the critical density at a given redshift
$\rho_{\mathrm{cr}}(z)\!=\!E^2(z)\,\rho_{\mathrm{cr}}(0)$, parameterized by the evolution function $E(z)$ (see Equs.\,\,\ref{e3_crit_density}\,\&\,\ref{e3_evolution_funct}),
and to a lesser extent on the cosmological model parameters (see Sect.\,\ref{s3_struct_formation}).
As an example for the full redshift dependence, we consider the evolution  of the cluster size scaling relation $R_{200}(z)\!\propto\!M^{1/3}_{200}\,E^{-2/3}(z)$, which indicates\footnote{For $E(z)\!\geq\!1$, \eg \ for a concordance model.} that objects at high redshift have smaller dimensions and are hence more compact.

We will now return to the discussion of four important global scaling relations for clusters of the same formation epoch. 
A summary of these relations is listed in Tab.\,\ref{t2_scaling_relations}. For a more detailed discussion of empirical parameters see Zhang \cite*{Zhang2005a}.

\begin{description}
    \item [$\mathbf{M}$--$\mathbf{T_{\mathrm{\bf X}}}$:] Equation\,\ref{e2_virial_theorem} yields $T_{\mathrm{X}}\!\propto\!\sigma^2_{\mathrm{r}}\!\propto\!M/R_*\!\propto\!M^{2/3}$ and consequently  
$M\!\propto\!T^{3/2}_{\mathrm{X}}$. Slopes of empirical $M$--$T_{\mathrm{X}}$  relations  are consistent with the self-similar expectation (\eg \ Arnaud \etal, 2005). The $M$--$T_{\mathrm{X}}$ relation is one of the fundamental scaling laws because of the direct connection of mass and gas temperature through the {\em Virial theorem\/}.  
With the assumed constant gas mass fraction $M_{\mathrm{gas}}\!=\!f_{\mathrm{gas}}M$, the same self-similar scaling behavior is obtained for the  $M_{\mathrm{gas}}$--$T_{\mathrm{X}}$ relation.

    \item [$\mathbf{L^{\mathrm{\bf bol}}_{\mathrm{\bf X}}}$--$\mathbf{T_{\mathrm{\bf X}}}$:]  
An averaged integration of the emissivity over the cluster volume yields 
$L^{\mathrm{bol}}_{\mathrm{X}}\!\propto\!\rho^2_{\mathrm{gas}}\,R^3_{*}\,T^{1/2}_{\mathrm{X}}\!\propto\!
(f_{\mathrm{gas}}M/R^3_{*})\,R^3_{*}\,T^{1/2}_{\mathrm{X}}\!\propto\!M\,T^{1/2}_{\mathrm{X}}\!\propto\!T^{2}_{\mathrm{X}}$, where in the last step the result for the $M$--$T_{\mathrm{X}}$ relation was used.
The measured scaling behavior of the 
$L^{\mathrm{bol}}_{\mathrm{X}}$--$T_{\mathrm{X}}$ relation reveals a systematically steeper logarithmic slope than expected. 


    \item [$\mathbf{L^{\mathrm{\bf bol}}_{\mathrm{\bf X}}}$--$\mathbf{M}$:] Combining the first two relations, we obtain
    $L^{\mathrm{bol}}_{\mathrm{X}}\!\propto\!T^{2}_{\mathrm{X}}\!\propto\!M^{4/3}$.
    The $L^{\mathrm{bol}}_{\mathrm{X}}$--$M$ correlation is of fundamental importance for distant cluster cosmology since it     provides the easiest means to determine masses for
faint X-ray clusters.  Reiprich\,\&\,B\"ohringer \cite*{Reiprich2002a} 
empirically calibrated the relation and obtained a slightly steeper slope than the self-similar expectation (see Fig.\,\ref{f4_LX_M_relation}). The approximate full empirical relationship between X-ray luminosity and total cluster mass is given by

\begin{equation}\label{e2_LX_M}
    L_{\mathrm{X}} \simeq 10^{45.0 \pm 0.3} \, h^{-2}_{70}\, \mathrm{erg\,s}^{-1}  \left( \frac{M_{200}}{10^{15} h^{-1}_{70} M_{\sun}}\right)^{1.8} .
\end{equation}

    \item [$\mathbf{M}$--$\mathbf{Y_{\mathrm{\bf X}}}$:]
The use of the combined parameter $Y_{\mathrm{X}}\!=\!M_{\mathrm{gas}}\!\cdot\!T_{\mathrm{X}}$ as a robust, low-scatter mass proxy has been motivated by studies of the {\em Sunyaev-Zeldovich effect} (see Sect.\,\ref{s2_SZE}) and has recently been calibrated with X-ray observations by Arnaud \etal \ \cite*{Arnaud2007a}. The self-similar expectation is obtained from the first relation as $Y_{\mathrm{X}}\!\propto\!M_{\mathrm{gas}}T_{\mathrm{X}}\!\propto\!f_{\mathrm{gas}}M\,M^{2/3}\!\propto\!M^{5/3}$ and hence $M\!\propto\!Y^{3/5}_{\mathrm{X}}$, consistent with the observed slopes.

\end{description}


\begin{table}[t]    
\begin{center}

\begin{tabular}{|c|c|c|c|c|}
\hline

\bf{Relation} & \bf{Form} & \bf{Self-Similar $\mathbf{\alpha}$} & \bf{Observed $\mathbf{\alpha}$} & \bf{Reference}  \\

\hline\hline

$M_{500}$--$T_{\mathrm{X}}$  & $M_{500}\!\propto\!T^{\alpha}_{\mathrm{X}}$  & 3/2 &  $1.6 \pm0.1$ &  Arnaud \etal \ \cite*{Arnaud2005a}  \\

$L^{\mathrm{bol}}_{\mathrm{X}}$--$T_{\mathrm{X}}$  & $L^{\mathrm{bol}}_{\mathrm{X}}\!\propto\!T^{\alpha}_{\mathrm{X}}$  &  2 &  $2.9\pm0.15$ & Arnaud\,\&\,Evrard \cite*{Arnaud1999a}  \\


$L^{\mathrm{bol}}_{\mathrm{X}}$--$M_{\mathrm{200}}$  & $L^{\mathrm{bol}}_{\mathrm{X}}\!\propto\!M^{\alpha}_{200}$  & 4/3  &  $1.8 \pm 0.1$ & Reiprich\,\&\,B\"ohringer \cite*{Reiprich2002a} \\

$M_{500}$--$Y_{\mathrm{X}}$  & $M_{500}\!\propto\!Y^{\alpha}_{\mathrm{X}}$  & 3/5   &  $0.55 \pm 0.05$ &  Arnaud \etal \ \cite*{Arnaud2007a}  \\

\hline
\end{tabular}

\caption[Cluster Scaling Relation]{Selected galaxy cluster scaling relations. The table lists the relation, the expected self-similar scaling behavior, and approximate observed values.} \label{t2_scaling_relations}
\end{center}
\end{table}

\noindent
The introduced scaling relations are of fundamental importance for  the XDCP survey in two ways. First, 
a selected subset of newly discovered {\em individual\/} distant systems can be used to study and calibrate the evolution of the mass-observable relations at high  redshift, possibly with deep X-ray follow-up observations 
in order to improve the achievable accuracies.
Second, the calibrated scaling relations can then  provide the mass proxy for  future cosmological applications for the {\em full\/} survey sample.    

Deviations from the expected gravitational self-similar scaling relations are commonly attributed to 
non-gravitational effects in the intracluster medium.
ICM pre-heating and central cluster cooling and heating processes have been mentioned as  possible 
mechanisms able to break
self-similarity.
The next section will briefly summarize some of the complex baryonic physics phenomena observed in cluster centers.

\subsection{Cool core clusters}
\label{s2_CCC}

\noindent
Cooling and heating in the centers of galaxy clusters is the subject of very active current research since the observed physical processes affect the  evolution of both the cluster {\em and\/} the galaxy populations. 
A compilation of the latest results is available in the conference proceedings of B\"ohringer \etal \ \cite*{Cool06}. 
{\em Cool core clusters\/} (CCC) could also be important for high-redshift studies for two reasons. 
(i) The abundance and properties of high 
redshift cool core clusters are essentially unexplored (\eg \ Vikhlinin \etal, 2007; Santos \etal, submitted),
\nocite{Vik2007a}
\nocite{Santos2007a}
 and (ii)  CCC systems at \zg1, if they exist, could introduce survey selection biases owing to their highly peaked X-ray emission and the associated  difficulty to discriminate them from point sources. 



The radiative cooling time scale of the  ICM can be approximated from the ratio of the specific enthalpy\footnote{The specific enthalpy $h$ is to be used for an {\em isobaric\/} cooling of the gas.} and the total volume emissivity and is inversely proportional to the gas density $n$

\begin{equation}\label{e2_cooling_time}
    t_{\mathrm{cool}} \simeq \frac{\frac{5}{2}n\,k_{\mathrm{B}}\,T }{\epsilon_{\mathrm{bol}}^{\mathrm{ff}}} \simeq 8.5 \times 10^{10} \ \mathrm{yr} \cdot \left(\frac{n}{10^{-3}\,\mathrm{cm}^{-3}}\right)^{-1} \left(\frac{T}{10^8\,\mathrm{K}}\right)^{1/2} \propto \frac{1}{n} \ .
\end{equation}

\noindent
Throughout most regions of typical clusters, the  cooling time scale is much longer than the Hubble time $t_{\mathrm{H}}\!=\!H_0^{-1}$,  justifying the assumption of hydrostatic equilibrium. However, in dense cluster cores,  $t_{\mathrm{cool}}$ can become less than the age of the Universe and radiative cooling effects gain significance.
The {\em cooling surface\/} is defined as the radius $r_{\mathrm{cool}}$, where the cooling time equals the age of the cluster.
Inside $r_{\mathrm{cool}}$ radiative cooling losses are  significant and can cause a decline of the ICM temperature towards the center, as visible in some of the temperature profiles of Fig.\,2.4.  
Such {\em cool core clusters\/}
exhibit very peaked surface brightness profiles with central emission exceeding 
the  $\beta$-model expectations of Equ.\,\ref{e2_beta_model}.
Contributing about half to the {\em local\/} cluster population, current data is suggestive that the fraction of CCC is rapidly declining at  $z\!\ga\!0.5$ \cite{Vik2007a}. 


Early models of {\em cool core clusters\/} (\eg \ Silk, 1976; Fabian\,\&\,Nulsen, 1977)
\nocite{Silk1976a}
\nocite{Fabian1977a} 
predicted that the central cooled ICM  should gradually condense and be replaced by surrounding gas giving rise to a constant inward flow  with mass deposition rates in the cluster centers of $\orderof (100\,M_{\sun})$ per year. However, observations with XMM-Newton showed that 
this {\em cooling flow \/} picture does not hold \cite{Peterson2001a} and that the central gas temperatures do not decrease below a few keV. These results emphasized the need of a central heating source, which is sufficiently energetic and self-regulated 
to balance the radiative cooling losses and establish a stable ICM configuration. The current standard paradigm (\eg \ B\"ohringer \etal, 2002b)
\nocite{HxB2002b}
attributes the heat source to a central AGN, which exhibits duty-cycles of increased activity and intermediate quiescent periods.
This paradigm is consistent with observations of complex hydrodynamic phenomena in cluster centers such as X-ray cavities and shock fronts (\eg \ McNamara \etal, 2005), buoyantly rising bubbles and propagating sound waves (\eg \ Churazov \etal, 2001; Fabian \etal, 2003), or AGN jet interactions (\eg \ McNamara \etal, 2007).   
\nocite{Fabian2003a}
\nocite{McNamara2007a}
\nocite{McNamara2005a}
\nocite{Churazov2001a}






For cooling time scales of $\orderof$(1\,Gyr), characteristic features of CCC might already be observable at  \zg1. 
The associated peaked X-ray surface brightness profile  and possible additional central AGN activity have to be adequately considered for the cluster candidate selection process (Sect.\,\ref{s6_Source_Screening}) and the computation of the selection function (Sect.\,\ref{s9_Selection_Funct}).

\subsection{Sunyaev-Zeldovich effect}
\label{s2_SZE}

\noindent
Measurements of the {\em Sunyaev-Zeldovich effect} (SZE) are expected to have a significant impact for future distant cluster studies and are therefore briefly summarized here. The hot electrons of the intracluster medium not only emit X-ray photons, they also lead to a characteristic distortion of the cosmic microwave background (CMB) radiation through {\em inverse Compton scattering}. 

Low energy CMB photons propagating through the hot cluster ICM are on average `upscattered' by the electrons, \ie \ the collective effect of many scattering processes gives rise to a shift of the CMB spectrum towards higher temperatures.
The SZE \cite{Sun1972a} introduces spectral changes of the CMB intensity $I$ of order   $\Delta I / I_0\!\sim\!10^{-4}$ resulting in an increased CMB intensity in the higher energy {\em Wien\/} tail of the spectrum and a {\em CMB decrement} in the 
{\em Rayleigh-Jeans\/} tail. The transition from a  CMB decrement to an increment occurs at a wavelength of $\lambda_{\mathrm{null}}\!\simeq\!0.14$\,cm ($\sim\!218$\,GHz). The SZE hence imprints a characteristic CMB spectral modulation of a decreased CMB intensity at $\lambda\!>\!\lambda_{\mathrm{null}}$ ($\nu\!<\!218$\,GHz), a null effect at $\lambda_{\mathrm{null}}$, and an increment at $\lambda\!<\!\lambda_{\mathrm{null}}$ ($\nu\!>\!218$\,GHz).


The CMB decrement in the radio regime can be quantified as
$\Delta I(\nu)\!=\!-2\,y\,I(\nu)$ by defining the {\em Compton-$y$ parameter}


\begin{equation}\label{e2_SZE_y}
    y \equiv \frac{\sigma_{\mathrm{T}}\,k_{\mathrm{B}}}{m_{\mathrm{e}}\,c^2} \int T_{\mathrm{e}}\,n_{\mathrm{e}}\,dl \ ,
\end{equation}

\noindent 
where $\sigma_{\mathrm{T}}$ is the Thomson electron cross-section,  $m_{\mathrm{e}}$ the electron mass, and the integration runs along the line-of-sight path element $dl$. To first order, the distorted CMB spectrum depends on the single $y$ parameter, which is proportional to the probability that a photon will Compton scatter and the amount of energy the photon gains on average. 
A comparison to the ideal gas law $p\!=\!n\,k_{\mathrm{B}}\,T$ reveals that the $y$ parameter is proportional to the integrated pressure along the line-of-sight  $y\!\propto\!\bar{p}$. The linear dependence on the gas density $y\!\propto\!n_{\mathrm{e}}$ implies that cluster outskirts can be probed and that the SZE signal is less sensitive to ICM inhomogeneities compared to X-ray measurements.




In the case that the SZE signal of the cluster is not spatially resolved, an integrated version  of the distortion parameter $Y$ over the projected surface area is measured, 
$Y\!=\!\int y\,dA\!\propto\!n_{\mathrm{e}}\,T\,dV\!\propto\!E_{\mathrm{thermal}}$, which is proportional to the total thermal energy of the electrons. Since the $Y$ parameter is directly linked to the cluster gas mass and the temperature, a scaling relation of the form $M_{\mathrm{tot}}\!\propto\!Y^{\alpha}$ promises to yield a new robust mass proxy once calibrated.



The key feature of the SZE concerning distant cluster studies is its (almost) redshift independence (for resolved systems). 
The cosmological surface brightness dimming $S_{\mathrm{bol}}\!\propto\!(1\!+\!z)^{-4}$ (see Equ.\,\ref{e3_SB_dim})
of the SZE signal is exactly compensated by the increased CMB intensity at the given redshift $\Delta I\!\propto\!I\!\propto\!T^4_{\mathrm{CMB}}\!\propto\!(1\!+\!z)^{4}$. 
For this reason, future SZE cluster surveys (see Sect.\,\ref{s11_SZE}) bear great hope for a profound impact on cosmological studies.








\section{The Cluster Galaxy Population}
\label{s2_galaxy_populations}

\noindent
Despite the small contribution of only a few percent to the total cluster mass, optical observations of the galaxies have  long been the only means for cluster studies. Abell \cite*{Abell1958a} compiled the first large optically selected cluster catalog based on galaxy overdensities within a projected radius of $\Theta_{\mathrm{A}}\!=\!1\farcm 7/z$, corresponding to a physical scale of about 2\,Mpc. Redshift estimates for the 1\,682 identified clusters containing more than 50 (projected) galaxies within two magnitudes of the third brightest member $m_3$ were obtained by using the tenth brightest cluster member $m_{10}$ as a `standard candle' (see Equ.\,\ref{e3_hubble_diagram}).

Galaxy clusters play an important role for observational studies of galaxy evolution for several reasons: (i) they provide a large number of galaxies with the same general environmental parameters and redshift, (ii) evolutionary processes are accelerated in  high-density environments, and (iii) some classes of galaxies are only found in clusters.
By going to high redshift, galaxy evolution effects are directly accessible to observations. Future studies of the evolutionary effects on the  cluster galaxy populations will explore galaxy properties as a function of redshift {\em and\/} cluster mass over an increased redshift baseline. Whereas the cluster redshift fixes the cosmic age at the time of observation, the mass of the system determines the environmental properties. Disentangling the influence of different cluster specific effects on galaxies will thus require also a wide mass range at high redshift.
The following sections illustrate selected aspects of the cluster galaxy populations that are of importance for distant cluster surveys.










\subsection{Cluster versus field galaxies}
\label{s2_transformation_mechanisms}

\noindent
The galaxy population is commonly divided into {\em cluster galaxies\/} and {\em field galaxies\/} depending on the environment they reside in. In the light of hierarchical structure formation scenarios, this distinction is not well defined over cosmic time as galaxies from the `field' can enter cluster or group environments and groups subsequently merge with larger clusters.  
As a working definition, we can consider as {\em cluster galaxies\/} all objects which are located within the  virial radius of a massive dark matter halo filled with an X-ray emitting ICM of minimum luminosity $L_{\mathrm{X}}\!\ga\!10^{43}$\,erg\,s$^{-1}$. 
Based on this `cluster definition', three main differences distinguish  the cluster  environments from the `field':
\begin{itemize}
    \item the {\bf confining dark matter potential well} associated with the DM density profile of Equ.\,\ref{e2_NFW_profile}; 
    \item the {\bf dilute hot intracluster medium}  permeating  the potential well with increasing densities towards the center;
    \item the {\bf highly increased galaxy density} following a similar radial profile to the ICM. 
\end{itemize}    

\noindent 
All items of this special environment, as known from {\em existing\/} clusters, can introduce {\em nurture\/} effects on the galaxy population, \ie \ object properties are transformed as a {\em reaction\/} to the environment.  On the other hand, differences in the cluster versus field galaxy populations could originate from their different {\em nature\/}, \ie \ they are inherited from the structure formation process covered in Sect.\,\ref{s3_struct_formation}. In the following, some major consequences of the different physical cluster setting on the galaxy population are discussed.

\begin{description}

    \item[Violent relaxation:] 
    A fluctuating gravitational potential $\Phi$, as expected during the cluster collapse process, changes the energy of particles (galaxies or DM) according to $\dot{E}\!=\!m\,\dot{\Phi}$ and can produce a thermal equilibrium configuration on short timescales even for collisionless systems \cite{LyndenBell1967a}. The initial cluster galaxy population is hence expected to be virialized shortly  after cluster formation due to the collective gravitational effects of {\em violent relaxation}.


    \item[Dynamical friction and cannibalism:]
A massive galaxy moving at velocity  $v$ through a homogeneous background medium of lighter (dark matter) particles with mass density $\rho$ suffers 
a drag force named {\em dynamical friction\/} \cite{Chandra1943a}. 
Along the trajectory of the massive object, the background medium is `gravitationally polarized' by the attractive force of the galaxy, leaving an overdense concentration of background particles in the wake of the galaxy.  
The collective effect of this friction force results in  a slowing of the galaxy according to 
$\dot{v}\!\propto\!\rho\,M\,v^{-2}$. The drag force is independent of the mass of the individual background particles, but proportional to the galaxy mass $M$, which leads to a mass segregation of the largest  galaxies  in clusters on the {\em dynamical friction\/} time scale $t_{\mathrm{DF}}$

\begin{equation}\label{e2_dynamic_friction}
   t_{\mathrm{DF}} \simeq 2\,\mathrm{Gyr} \, \left( \frac{M}{10^{12} M_{\sun}}  \right)^{-1} \left( \frac{v}{1\,000\,\mathrm{km\,s}^{-1}}  \right)^{3}  \left( \frac{R}{100\,\mathrm{kpc}}  \right)^{2} \left( \frac{\sigma_{\mathrm{r}}}{300\,\mathrm{km\,s}^{-1}}  \right)^{-2}  ,
\end{equation}

\noindent
where $R$ is the distance from the cluster center, and $\sigma_{\mathrm{r}}$ the radial velocity dispersion of the dark matter halo potential well. Massive cluster galaxies can hence spiral towards the center on time scales of a few Giga years and form a giant object via a process named  {\em galactic cannibalism\/}, the growing of a single galaxy by consuming its neighbors \cite{Hausman1978a}.

 


    

    \item[Galaxy merging:]    
    
The key parameter to determine the probability that two approaching galaxies will merge is their relative velocity $V$, which needs to be smaller than the relative escape velocity of the system of $\orderof$(200\,km\,s$^{-1}$) for a merging event to occur.
If a significant fraction of the orbital energy is transferred  to the stellar systems by tidal forces in a close encounter, a merger is likely.   
For two equal mass, spherically symmetric galaxies with a mass-weighted mean-square radius $\langle r^2 \rangle$ and relative velocity $V$, merging occurs if the bodies have an impact parameter $b$ smaller than a critical value 
$b_{\mathrm{crit}}\!=\!(32\,G^2 M^2 \langle r^2 \rangle /3)^{1/4}/V\!\propto\!\sqrt{M}/V$ \cite{Peacock1999a}.
The expected merger rate $\Gamma$ can then be approximated as the product of the merging cross section $\sigma_{\mathrm{merge}}\!=\!\pi \,b^2_{\mathrm{crit}}$ and the particle flux $j_{\mathrm{gal}}\!=\!n_{\mathrm{gal}}V$ 

\begin{equation}\label{e2_merger_rate}
   \Gamma =  \sigma_{\mathrm{merge}}\,j_{\mathrm{gal}} \simeq \pi \, b^2_{\mathrm{crit}} n_{\mathrm{gal}}V \propto \frac{1}{V} \ .
\end{equation}

\noindent
The inverse proportionality to the velocity generally favors merger events in lower mass groups rather than in clusters, where velocity dispersions are higher. 
 
 

    \item[Galaxy harassment:]  
Even if galaxies do not merge, the cumulative effects of the tidal forces of many weaker encounters can significantly influence cluster galaxies, in particular objects with lower masses. This {\em galaxy harassment\/}  can lead to stripping of the outer galactic halos, disk deformations in spirals and S0 galaxies, or trigger star formation activity 
\cite{Richstone1976a}.
    
 
    \item[Ram pressure stripping:]
Objects moving through a gaseous medium with density $\rho_{\mathrm{gas}}$ at velocity $v$ are exerted to a {\em ram pressure\/} 
$p_{\mathrm{ram}}\!=\!\rho_{\mathrm{gas}}\,v^2_{\mathrm{gal}}$.
A galaxy passing through the dense central region of the ICM can this way be efficiently stripped from its internal gas on time scales of $10^7$--$10^8$ years. A single passage  of a  galaxy through a cluster core can hence almost completely remove the  interstellar medium from this galaxy  and quench its star formation.    




\end{description}

\noindent
One could expect that these  additional physical effects taking place in clusters modify or re-shape the galaxy luminosity functions (LF) of the cluster galaxy populations. However, the 
{\em individual\/} galaxy classes (\eg \ E, S0, S) maintain a universal  luminosity function  
with the same general {\em Schechter\/} function form as field galaxies (\eg \ Andreon 1998; Popesso \etal, 2005) 
\nocite{Andreon1998a} 
\nocite{Popesso2005a}

\begin{equation}\label{e2_Schechter_LF}
    \phi(L)\,dL = \phi_0 \left(\frac{L}{L*}\right)^{\alpha} \exp\left(\frac{L}{L*}\right) \, \frac{dL}{L*} \ ,
\end{equation}

\noindent 
where $L*$ is the characteristic luminosity,  $\phi_0$ the space density normalization, and $\alpha$ the faint-end slope.
This universal form of the luminosity function has been confirmed out to redshifts of $z\!\simeq\!1.3$ with typical slope parameters of $\alpha\!\simeq\!-1\!\pm\!0.3$ for the bright part of the composite LF and characteristic rest-frame absolute (AB) magnitudes of $\mathrm{Ks*}\!\simeq\!-23.5\!\pm\!0.5$ \cite{Strazzullo2006a}.
Note that the frequently used notation m*+1 refers to the apparent (observed) magnitude of a galaxy with a luminosity that is one magnitude, \ie \ a factor of 2.51, {\em fainter\/} than the characteristic luminosity L* at the given redshift.
Similarly, M*-2 denotes galaxies which are two magnitudes {\em brighter\/} in absolute luminosity, \ie \ a factor of 6.3, compared to a characteristic L* object of the given galaxy class.







\subsection{Formation of ellipticals and the cluster red-sequence}
\label{s2_red_sequence}

\noindent
Early-type galaxies, \ie \ elliptical (E) and lenticular (S0) galaxies, play a distinct role in galaxy clusters.
Not only they constitute the dominant galaxy class in the core regions of clusters, but their properties exhibit a high degree of homogeneity, which allows to place strong constraints on their formation history.

One of the key features of {\em cluster\/} galaxy populations is the existence of a tight sequence of red galaxies in the color-magnitude diagram (CMD), the  so-called {\em cluster red-sequence}. This color-magnitude relation (CMR) in clusters has first been noticed by Baum \cite*{Baum1959a}, but the full cosmological implications have only later been realized by Sandage \& Visvanathan \cite*{Sandage1978a}.
Bower, Lucey \& Ellis \cite*{Bower1992b,Bower1992a} confirmed the universality of the CMR in Virgo and Coma
and established the \reds as distance indicator for clusters. The precision photometric study of these two clusters showed that the color-magnitude relations are intrinsically identical and that the observed color difference originates from the  effects of the distance modulus (see Equ.\,\ref{e3_dist_modulus}).  
This study additionally established that the relation for the S0 population is less tight than for ellipticals, which exhibit
a measured rms scatter along the \reds of 0.05\,mag, 
mostly contributed by observational errors.
The characteristic \reds  with its minimal intrinsic scatter in low redshift clusters is of
key importance for the understanding of galaxy evolution and can furthermore be used for an optical search and identification of new clusters (\eg \ Gladders\,\&\,Yee, 2005). 
\nocite{Gladders2005a}

The \reds in the CMD  of the photometric bands X and Y can be characterized by (i) the color normalization (X$-$Y)$_{\mathrm{CMR}}$(Y*) at the location of L* galaxies, (ii) the slope of the sequence d(X$-$Y$)_{\mathrm{CMR}}$/dY, and (iii) the (intrinsic) scatter $\sigma_{\mathrm{CMR}}$.
The color of the sequence is related to the luminosity-weighted age of the stellar populations in galaxies and can be compared to galaxy evolution models for the derivation of a formation redshift estimate. The intrinsic scatter $\sigma_{\mathrm{CMR}}$ provides information on the star formation histories of the \reds galaxies as a tight relation should be only observable if these objects formed their stars over a period 
which is short compared to the age of the galaxies. 

The rising slope towards redder colors for more luminous objects can now be attributed to a {\em metallicity sequence\/} \cite{Kodama1997a}.
The required systematic increase towards higher metallicities and hence redder stellar populations\footnote{This reddening with increasing metallicity is primarily caused by a lower temperature of the metal rich stars at the main sequence turnoff (\eg \ Renzini, 2006). \nocite{Renzini2006a} } 
can be quantitatively explained with supernova-driven wind models of ejected heated gas from the system. 
The deeper potential wells of massive galaxies, with  increased  gravitational binding energy per unit mass compared to smaller systems, can retain more metals  and additionally trigger a more rapid metal production at early cosmic epochs. 


The explanation of the observed local CMR  with stellar population evolutionary synthesis models 
was not straightforward due to a general {\em age-metallicity degeneracy\/}  \cite{Worthey1994a}. Galaxy colors become redder as a consequence of increasing metallicity {\em or \/} increasing age of the stellar population (in absence of dust). This degeneracy could be broken by moving to higher redshifts, where the two scenarios predict different redshift evolutions of the CMR.
Models for an evolving metallicity sequence predict the observed (almost) redshift independent slope. 
An age sequence, on the other hand,  
would feature a rapid steepening and a CMR truncation  at higher redshifts, since the formation epoch is approached earlier for the faint younger blue galaxies than for the luminous older red ones \cite{Kodama1998a}.



\begin{figure}[t]
\begin{center}
\includegraphics[angle=0,clip,width=0.75\textwidth]{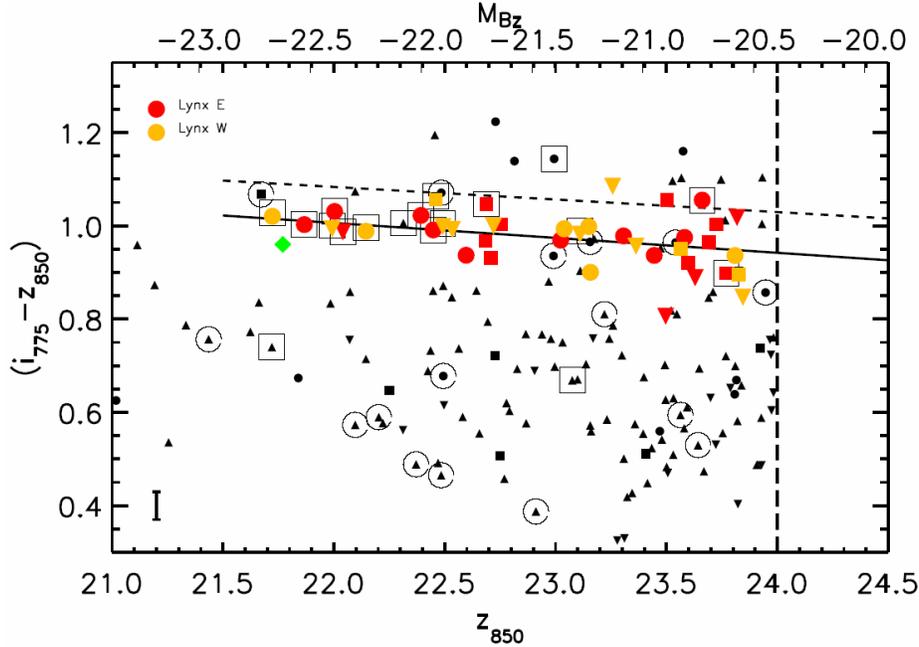}

\end{center}
\vspace{-3ex}
\caption[Color-Magnitude Diagram for the Lynx Supercluster]{Color-magnitude diagram for the Lynx supercluster at $z\!=\!1.26$, including the two most distant clusters known from the ROSAT era. Orange and red symbols indicate the individual clusters Lynx East and Lynx West. Elliptical galaxies are indicated by circles, S0s by squares, and spirals by triangles. Square boxes mark spectroscopically confirmed cluster members, open circles show interlopers, \ie \ foreground objects. Plot from Mei \etal \ \cite*{Mei2006a}.} \label{f2_Lynx_CMD}
\end{figure}

The tightness of the \reds  is expected to break down when 
 the lookback time at the observed redshift $z$  approaches
the main epoch of star formation activity in (massive) early-type galaxies, \ie \ when the stellar populations are `young'. 
However, all current known spectroscopically confirmed high-redshift galaxy clusters out to $z\!\sim\!1.3$ show a pronounced \reds with a surprisingly small scatter. Figure\,\ref{f2_Lynx_CMD} displays a recent Hubble Space Telescope (HST) color-magnitude diagram of the  Lynx supercluster. Its two high-redshift  clusters, the most distant ones discovered with ROSAT, exhibit intrinsic \reds scatters of merely $0.025\!\pm\!0.015$\,mag  for the elliptical population  \cite{Mei2006a}.   
From these observations, the formation redshift of the \reds galaxies can be estimated as $z_{\mathrm{f}}\!>\!2.5$, yielding strong constraints for evolution models.

Two competing formation paradigms for elliptical galaxies have been contrasted 
in the last few decades: the {\em monolithic collapse \/} model and the {\em hierarchical formation\/} scenario.
The {\em monolithic collapse\/} model \cite{Eggen1962a} 
assumes that ellipticals and bulges formed at high redshift, when gas clouds collapse rapidly 
and turn {\em instantaneously} into stars. After this violent star formation burst,
 no further generations of stars are formed, implying that the stellar population properties of a galaxy follow a passive evolution, \ie \ the luminosity changes only due to the aging of stars. 
Within the {\em monolithic collapse\/} model, elliptical galaxies have not changed their overall appearance, mass, or identity since the formation redshift $z_{\mathrm{f}}$, except for the ageing stellar population and the associated reddening of the spectral energy distribution (SED) with cosmic time. Note that essentially all observed cluster red-sequence colors are consistent with this simple galaxy formation model for  bursts taking place at $z_\mathrm{f}\! \ga \!3$ (see Sect.\,\ref{s7_imaging_strategy}).    

Conversely, the {\em hierarchical formation\/} scenario (\eg \ White\,\&\,Reese, 1978; Kauffmann, 1996; Kauffmann\,\&\,Charlot, 1998) 
\nocite{White1978a}
\nocite{Kauffmann1996a}
\nocite{Kauffmann1998b}
 rests upon the physically motivated standard structure formation and evolution paradigm discussed in Sect.\,\ref{s3_struct_formation}. {\em Hierarchical formation\/} in this respect means that small galaxy-size dark matter halos form first and massive galaxies originate from subsequent merging of smaller units.
Major merging events can significantly alter the overall galaxy properties and lead to morphological  transformations, trigger star burst events, or start nuclear activity in the galaxy centers.  
The main difference to the {\em monolithic collapse\/} model, however, is the expected late  appearance of massive spheroids, as they  are assembled over cosmic time. Consequently, a rapid decline of the space density of the most massive systems with redshift is expected.
Luminous ellipticals are relatively young objects in this scenario, mostly appearing at $z<\!1\!$, which has long been at odds with discoveries of massive spheroids at $z\!>\!1.5$ (\eg \ Dunlop \etal, 1996; Cimatti \etal, 2004). 
\nocite{Dunlop1996a}
\nocite{Cimatti2004a}

An important aspect of {\em hierarchical galaxy formation\/}  is the fact that the appearance and identity of objects can change over cosmic time due to merging and evolutionary effects. This implies that the cluster elliptical population observed locally does not necessarily correspond to the ellipticals in high-redshift clusters, a consequence which has been named {\em progenitor-bias\/} \cite{VanDokkum1996a}.
A key for the consolidation of {\em hierarchical models\/} with observations is the strict distinction between the {\em formation epoch\/} of the stellar population in ellipticals and their {\em mass assembly history\/} (see Fig.\,\ref{f2_BCG_assembly}).   
If star formation after merger events is absent ({\em dry merging}), massive ellipticals can exhibit old, passively evolving stellar populations, while their mass and luminosity  growth  
might still continue to the present.  

Recent progress in numerical simulations and semi-analytic modelling techniques have led to detailed predictions from the {\em hierarchical formation scenario\/}  for the properties of the elliptical galaxy populations in clusters, now being  largely consistent with  observations \cite{DeLucia2006a}. 
The star formation activity for the largest local model spheroids peaks at $z\!\sim\!5$, but only half of the systems possess progenitors  with at least 50\% of the final mass at $z\!\sim\!1.5$,  the lag between star formation epoch and assembly time increasing for more massive ellipticals.
Hence the leverage for an observational distinction between {\em monolithic collapse\/} and {\em hierarchical formation\/} scenarios is highest for clusters at \zg1.

\subsection{Evolution and environmental effects}
\label{s2_environmental_effects}

\noindent
Environmental effects and galaxy evolution are closely interlinked since the discussed external physical mechanisms at work in galaxy clusters can drastically change the galaxy properties over cosmic time. 
One of the key features of the cluster environment is the existence of a pronounced morphology-density relation \cite{Dressler1980a}, \ie \
the observed correlation between the frequency of various Hubble types and the local projected  galaxy number density.
In the local Universe, the fraction of spiral galaxies decreases from 60--70\% in field environments to about 10\% in dense cluster cores. The elliptical population, on the other hand, increases its relative contribution from approximately 10\% in the field to roughly 50\% in cluster centers. The fraction of  lenticular (S0) galaxies rises more moderately from $\sim$30\% in low density environments to about 40\% in high density ones.


\begin{figure}[t]
\centering
\includegraphics[angle=0,clip,width=\textwidth]{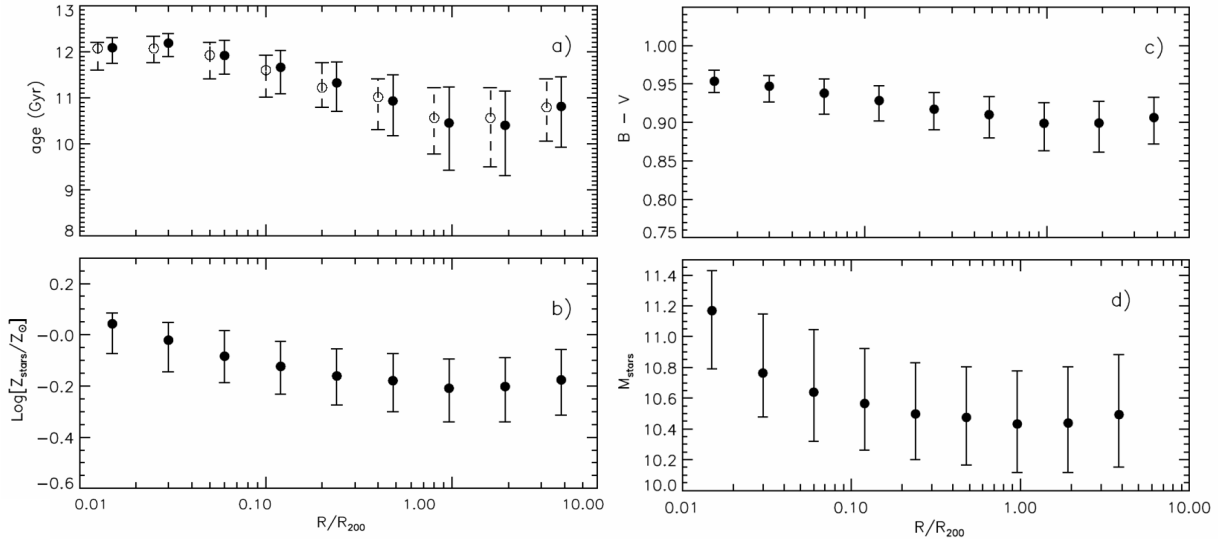}

\caption[Simulated Cluster Ellipticals]{Variations of the properties of the elliptical galaxy population  with cluster-centric distance as predicted at low redshift by the {\em Millennium Simulation}. Galaxy properties are extracted from massive dark matter halos with $M_{\mathrm{tot}}\!\ga\!8\!\times\!10^{14}$\,M$_{\sun}$. Going from the cluster center towards the outskirts, the  age (lookback time) of the stellar populations of ellipticals is predicted to decrease ({\em top left}), the metallicity to decrease ({\em bottom left}), and the typical galaxy colors to become bluer ({\em top right}). These trends are partly driven by mass segregation, \ie \ the more massive galaxies are found close to the center ({\em bottom right}). From De Lucia \etal \ \cite*{DeLucia2006a}.}
\label{f2_SimulatedEllipticals}       
\end{figure}

%







Significant changes of the morphological galaxy mix  can already be observed at redshifts of $z\!\sim\!0.5$ \cite{Dressler1997a}.
From $z\!=\!0$ to $z\!=\!0.5$, the S0 galaxy fraction in cluster centers has dropped by a factor of 2--3 to 15--20\%, whereas the spiral fraction has proportionally increased to about 30--35\%. 
Recent studies of the morphology-density relation at \zsim1 confirm this trend and measure a combined early type fraction of $\langle f_{\mathrm{E+S0}}\rangle\!\simeq\!0.72\pm0.10$, which seems to be correlated with the bolometric X-ray luminosity of the cluster through the relation $f_{\mathrm{E+S0}}\!\propto\!L^{0.33\pm 0.09}_{\mathrm{X,bol}}$  \cite{Postman2005a}.
The rising fraction of  ellipticals  towards higher densities remains essentially unchanged, implying that the population of elliptical galaxies has been essentially in place since \zsim1.
Conversely, the decreasing S0 fraction with redshift suggests that an environment induced morphological transformation from spirals to S0s occurs at redshifts of $z\!\la\!1$ and continues until the present epoch.






This picture is qualitatively consistent with the {\em Butcher-Oemler effect\/}, which provided the first 
direct observational evidence for the evolution of the cluster galaxy population \cite{Butcher1978a}.
The two-band photometry of clusters at $z\!\sim\!0.4$ revealed an excess of blue galaxies compared to local rich clusters.  
A more recent study of 295 Abell clusters at $z\!<\!0.38$ confirmed the original findings and quantified the evolution of the blue galaxy fraction as $f_{\mathrm{B}}\!=\!(1.34\pm 0.11)\cdot z - 0.03$, where objects more than 0.2\,mag below the \reds were defined as {\em blue\/} \cite{Margoniner2001a}. Although the trend of such a relation is well established, the current form will obviously not hold at higher redshifts as it `predicts' a vanishing of the \reds by $z\!\simeq\!0.8$, at odds with observations (see Fig.\,\ref{f2_Lynx_CMD}). 







\enlargethispage{4ex}

The galaxy colors are closely linked to their star formation rates (SFR), which can be probed as a function of galaxy density with the Sloan Digital Sky Survey (SDSS) or the 2dF survey for low redshift cluster environments (Lewis \etal, 2002; G{\'o}mez \etal, 2003). 
\nocite{Lewis2002a}
\nocite{Gomez2003a}
These studies showed the existence of a sharp transition in the density--SFR relation starting from field-like star formation rates at local projected galaxy densities of
$\Sigma\!\simeq\!1$\,Mpc$^{-2}$   and converging  to the 
 suppressed star formation activity of cluster cores at $\Sigma\!\ga\!7$\,Mpc$^{-2}$. The transition from field to central cluster  SFR properties is hence completed over a range of about seven in the local galaxy density and occurs roughly at the cluster virial radius. 
 This large radius for the onset of  star formation suppression, and the fact that a similar density--SFR relation is found for group environments, give rise to a scenario where the transition from star forming to passive cluster objects already occurs during the infall along filaments towards the cluster \cite{Bower2004a}.



Using the high-resolution {\em Millennium Run\/} simulation \cite{Springel2005a}, detailed predictions for the changes of galaxy properties as a function of  cluster-centric distance can be obtained. The results for the elliptical cluster galaxy population at low redshift are displayed in Fig.\,\ref{f2_SimulatedEllipticals} \cite{DeLucia2006a}.
Moving outwards from the high-density cluster centers, the simulations predict (i)  decreasing  ages, (ii) decreasing galaxy metallicities, (iii) bluer object colors, and (iv) decreasing stellar masses for the ellipticals. In other words, elliptical galaxies in cluster centers are expected to be older, more massive, redder, and more metal rich than their counterparts in lower density environments. These trends  are partly driven by the dynamic friction induced mass segregation.


Moving towards higher redshifts, a study of A\,851 at $z\!=\!0.41$ yielded the first indication for a sharp radial transition in the properties of  faint ($<$0.1\,L*) galaxies \cite{Kodama2001a}. Using deep photometric data, an abrupt change from a predominantly red galaxy population towards  bluer galaxies  was observed at projected densities of  $\Sigma\!\simeq\!100$\,Mpc$^{-2}$, corresponding to the location of the  cluster subclumps in the filaments of the surrounding large-scale structure (LSS).
At medium distant redshifts, investigations of the LSS around a 
 $z\!=\!0.73$ cluster in the COSMOS field (see Sect.\,\ref{s4_distant_cl2004}) show a weak trend towards redder colors with increasing local density for early-type galaxies  \cite{Cassata2007a}.
For RXJ\,J0910+5422 at $z\!=\!1.11$., Mei \etal \ \cite*{Mei2006b} observe an S0 population that is systematically bluer than the elliptical \reds by about 0.1\,mag, suggesting that this population is still evolving towards the redder CMR colors of the ellipticals.
However, in the Lynx clusters of Fig.\,\ref{f2_Lynx_CMD}, ellipticals and S0 galaxies lie on the same red-sequence, but the scatter increases with distance from the center \cite{Mei2006a}.

Recently Cucciati \etal \ \cite*{Cucciati2006a} and Cooper \etal \ \cite*{Cooper2007a} have investigated the general color--density relation out to $z\!\sim\!1.5$ based on large spectroscopic galaxy surveys. These studies show
that the fraction of red galaxies in  dense environments gradually decreases with redshift, 
with a slower trend for the brightest objects. By redshifts of $z\!\sim\!1.5$, the segregation of red and blue galaxies in dense regions has completely disappeared. 
These new results suggest that the maximal star formation activity shifts  to galaxies with lower luminosities  {\em and\/} in lower density environments with decreasing redshifts. This is possibly governed by a combination of biased galaxy formation in high density regions and environmentally induced star formation quenching.

\enlargethispage{6ex}

The observation that the star formation activity in the most massive galaxies ceases at earlier epochs 
is a manifestation of the so-called `downsizing' scenario (Faber \etal, 1995; Cowie \etal, 1996). 
\nocite{Faber1995a}
\nocite{Cowie1996a}
A direct consequence of  `downsizing' in high-redshift clusters would be the `truncation' of the \reds at a given redshift          dependent magnitude, \ie \ an increasing deficiency of faint red  galaxies with increasing lookback time. Galaxies move towards the  \reds in the color-magnitude diagram after star formation activity has ceased. If this occurs at earlier epochs for the more luminous galaxies, then the \reds will be progressively populated over cosmic time starting from the bright end. Observations of the resulting  deficit of faint red galaxies on the \reds have been claimed by De Lucia \etal \ \cite*{DeLucia2004a}  for clusters at $z\!\sim\!0.8$ and by Tanaka \etal \ \cite*{Tanaka2007a} for a cluster at redshift $z\!=\!1.24$.

\pagebreak

\subsection{Brightest cluster galaxies}
\label{s2_BCGs}

\noindent
For distant cluster searches, the brightest cluster galaxies (BCGs) are of particular importance  in the cluster identification procedure. In fact,  (i) BCGs often mark the center of the associated galaxy overdensity, (ii) they usually have the best photometry and color information available among all cluster galaxies, (iii) the BCG location in the color-magnitude diagram usually marks the bright end of the cluster \reds and is hence important for the redshift estimate, and (iv) BCGs are prime targets for spectroscopic follow-up.


The brightest cluster galaxies constitute a very special  category of galaxies  that still lacks a detailed understanding of the formation history. 
Compared to other galaxy classes,  low redshift BCGs  are more massive, have larger spatial extent, and are ultra-luminous with typical luminosities of $L_{\mathrm{BCG}}\!\sim\!10\,L*$ \cite{Schombert1986a} that correlate with cluster mass in the form  $L_{\mathrm{BCG}}\!\propto\!M^{0.26\pm 0.04}_{200}$ \cite{Lin2004a}. In fact, the observed luminosity gap of the BCGs to the second ranked  galaxies of massive clusters cannot be accounted for by a statistical sampling of the bright end of the Schechter LF (Equ.\,\ref{e2_Schechter_LF}), indicating the need for a special formation process  \cite{Bhav1985a}.  
In the local Universe, the first ranked cluster galaxies are  predominantly located very close to the cluster's DM potential well minimum; 75\%  are found within the central 3\% of the virial radius.  
They are often classified as radio-loud objects and are associated with the cooling core phenomenon (Sect.\,\ref{s2_CCC}) in a complex way.
BCGs are also found to have a higher fraction of dark matter compared to similar massive objects resulting in larger velocity dispersions for the stellar component \cite{vonLinder2007a}.



More than half  of all BCGs in local X-ray clusters are classified as cD galaxies, a galaxy type  exclusively found in the centers of clusters. 
cD galaxies were originally defined by Matthews, Morgan \& Schmidt \cite*{Matthews1964a} as supergiant galaxies with 
an additional extended low surface brightness envelope in excess of the radial 
light distribution  of a normal elliptical galaxy at large radii. The outer cD halo has no clear edge and can extend far into 
 the cluster \cite{Schombert1988a}. 
 The most prominent formation scenario for these centrally dominating supergiant galaxies is the {\em galactic cannibalism\/} model of Ostriker\,\&\,Tremaine \cite*{Ostriker1975a}. In this picture, cD galaxies  grow through cannibalism of their neighbors as the most massive galaxies sink to the cluster center driven by dynamic friction. The formation scenario for giant ellipticals forming via merging in cluster centers has been confirmed by numerical simulations (\eg \ White, 1976),
\nocite{White1976a}  
but the reproduction of the characteristic extended halo has remained challenging until the present (\eg \ Nipoti \etal, 2003). 
\nocite{Nipoti2003a} 
It seems likely, however, that the origin of the cD halo starlight is connected to tidally stripped  material  from the outer parts of galaxies or their tidal disruption (\eg \ Zibetti \etal, 2005).
\nocite{Zibetti2005a}

\enlargethispage{6ex}

Despite their possibly `violent' formation history, the brightest cluster galaxies constitute a surprisingly  homogeneous class of objects with a low dispersion in their absolute magnitudes of $\langle M_{\mathrm{V}} \rangle_{\mathrm{BCG}}\!\simeq\!-24.5 \pm 0.3$. 
Local BCGs have been characterized as good `standard candles' exhibiting a Gaussian scatter of the absolute magnitudes around the mean value \cite{Postman1995a}.  
Because of this characteristic,  BCGs of massive clusters at redshifts $z\!\la\!1$ have been used for cosmological tests by applying the Hubble diagram of Sect.\,\ref{s3_cosmo_tests} (\eg \ Aragon-Salamanca \etal, 1998; Collins\,\&\,Mann, 1998).  
\nocite{Aragon1998a}
\nocite{Collins1998a}
Over the  redshift range covered, no signs of a significant BCG evolution have been found, other than the passive luminosity evolution \cite{Burke2000a}.

\pagebreak



\begin{figure}[t]
\centering
\includegraphics[angle=0,clip,width=\textwidth]{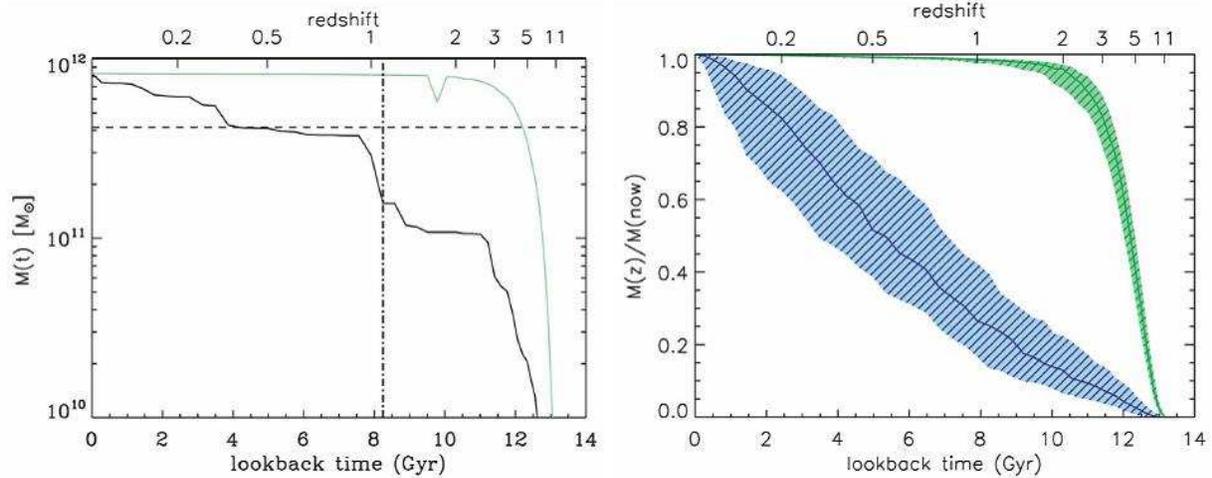}
\vspace{-4ex}
\caption[Mass Assembly History of BCGs]{Mass assembly histories and formation times of BCGs as seen from the {\em Millennium Simulation}. {\em Left:} Stellar mass of the main BCG progenitor (lower black line) and the sum of all progenitors (top line) for a {\em single\/} case study BCG. The horizontal dashed line represents half of the final BCG mass. {\em Right:} Mass assembly history (blue) and stellar formation epoch (green) for the full BCG sample.  Plots from De Lucia\,\&\,Blaizot \cite*{DeLucia2007a}.}
\label{f2_BCG_assembly}       
\end{figure}

\begin{figure}[t]
\centering
\includegraphics[angle=0,clip,width=0.615\textwidth]{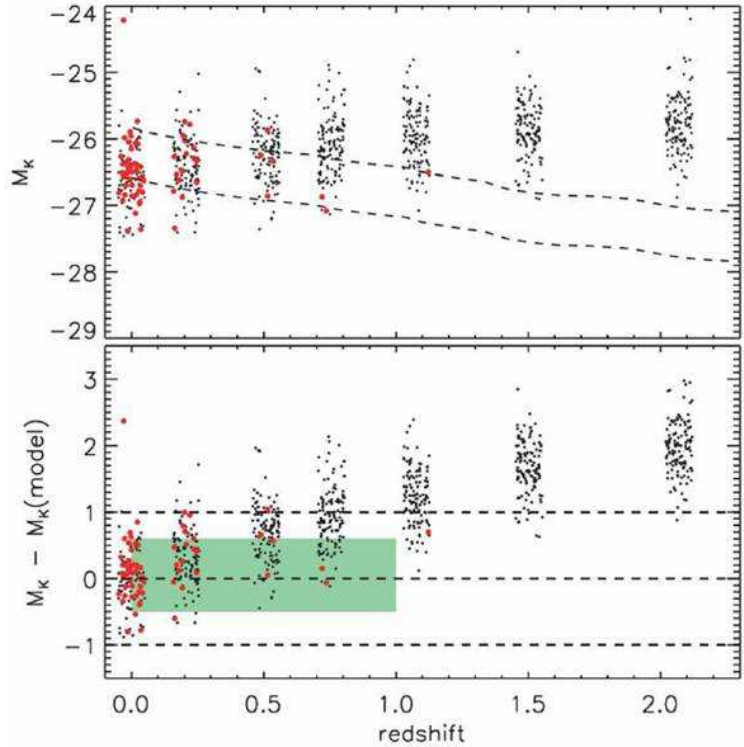}
\vspace{-1.5ex}
\caption[BCG  Model Evolution]{{\em Top:} Evolution of the absolute BCG K-band rest-frame magnitude as a function of redshift.  Black dots represent model BCGs, red symbols indicate objects in halos more massive than $10^{15}M_{\sun}$. The dashed lines show burst model predictions for a formation redshift of $z_{\mathrm{f}}\!=\!5$ and fixed mass. {\em Bottom:} Residuals after subtracting the burst model. The green shaded area indicates the redshift range which has been probed observationally. Plot from De Lucia\,\&\,Blaizot \cite*{DeLucia2007a}.}
\label{f2_sim_BCG_HD}       
\end{figure}

\begin{figure}[b]
\centering
\includegraphics[angle=0,clip,width=0.7\textwidth]{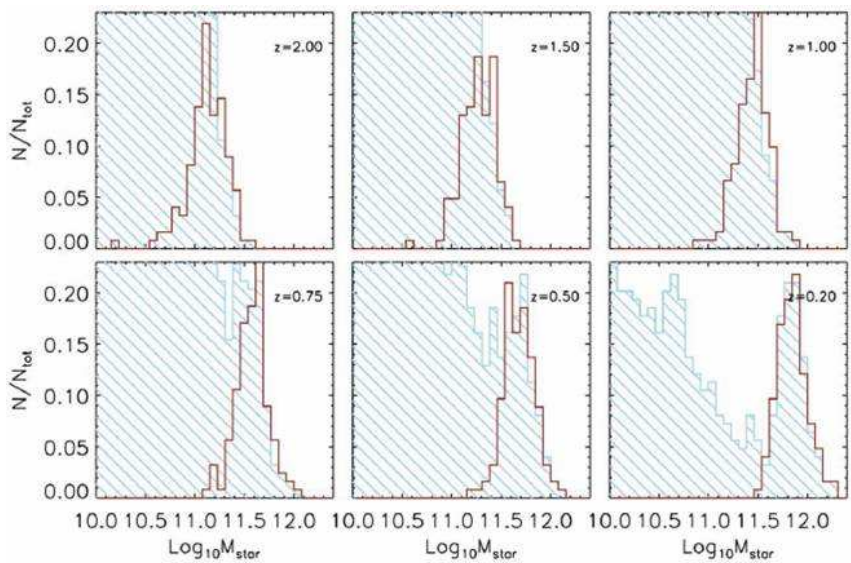}
\vspace{-1.5ex}
\caption[BCG Stellar Mass Distribution]{Evolution of the stellar mass distribution of high-redshift model BCGs (red histograms) and of progenitors of local BCGs (blue histograms). Plot from De Lucia\,\&\,Blaizot \cite*{DeLucia2007a}.}
\label{f2_BCG_masses}       
\end{figure}

To complete this introductory section on the cluster galaxy population, the predictions of the state-of-the-art  {\em Millennium Run Simulation\/} concerning the cosmic evolution of BCGs are presented, which have been recently discussed by De~Lucia\,\&\,Blaizot \cite*{DeLucia2007a}. 
Figures\,\ref{f2_BCG_assembly}--\ref{f2_BCG_masses} establish the simulated reference model for comparison with observations.
 The 125 traced model BCGs were selected from local cluster halos more massive than $7\!\times\!10^{14}\,M_{\sun}$ and have  absolute K-band magnitudes of $\langle M_{\mathrm{K}}\rangle\!=\!-26.58$\,mag  with a dispersion of 0.2\,mag.   
The galaxy {\em assembly time\/}  is defined in the analysis as the epoch when the {\em single \/} main BCG progenitor contains half of the final ($z\!=\!0$) stellar mass.  At the  {\em formation time\/}, on the other hand, half of the final total stellar mass is present in the sum of {\em all\/}  progenitor objects.

\enlargethispage{10ex}

The predicted extreme hierarchical nature of the brightest cluster galaxies is illustrated in Fig.\,\ref{f2_BCG_assembly}, which shows the mass evolution of a single case study BCG in the left panel and  the averaged assembly history of the full sample in the right panel. The upper curves in both plots indicate that most of the BCG stars form at high redshift, 50\% by $z\!\sim\!5$ and 80\% by  $z\!\sim\!3$, implying old, evolved, and red stellar populations as observed at the accessible redshifts. The mass assembly history of the main BCG progenitor is displayed by the lower lines showing the discrete merging events for the single galaxy on the left and the average sample evolution on the right, with a predicted mean mass growth by a factor of three from $z\!=\!1$ to $z\!=\!0$. 
Figure\,\ref{f2_sim_BCG_HD} provides testable predictions (see Sect.\,\ref{s10_bcg_assembly}) for the evolution of the absolute rest-frame K-band  luminosities (top panel) and after subtracting a passive luminosity evolution model with formation redshift $z_{\mathrm{f}}\!=\!5$ (lower panel). 
The distribution of BCG stellar masses at  different cosmic 
epochs is illustrated by the red histograms in Fig.\,\ref{f2_BCG_masses}.

Observationally, the big questions {\em when\/} and {\em how\/}  cD  galaxies formed are yet to be explored.
Between the 
{\em spiderweb galaxy\/}  \cite{Miley2006a},  a possible cD precursor in a protocluster environment at redshift $z\!=\!2.2$, and the well studied end products of the BCG evolution at $z\!\la\!0.8$, there is still a lot of redshift space to fill in.  

\pagebreak



\clearpage 
 
\chapter{Galaxy Clusters as Cosmological Probes}
\label{c3_cosmo_theory}
\noindent
Besides their use as astrophysical laboratories, galaxy clusters are also sensitive cosmological probes of increasing importance.
The NASA Dark Energy Task Force report\footnote{\url{http://www.science.doe.gov/hep/DETF-FinalRptJune30,2006.pdf}}, for example, recommends the study of cluster evolution as one of the four most promising pillars of future research in fundamental cosmology.

The distant galaxy clusters under focus for this thesis are observed at an epoch which corresponds to the second quarter of the lifetime of the Universe. Their general observational properties have thus to be 
interpreted within the standard (homogeneous) cosmological model and modern structure formation scenarios.  
More complete and detailed general introductions to cosmology can be found in the textbooks by \eg \ Schneider \cite*{Schneider2006a} and Coles \& Lucchin \cite*{Coles2002a}. 
This chapter pursues the following main motivations and aims:
\begin{enumerate}
    \item Provide the general cosmological framework and tools for the consistent description of \zg1 clusters within the concordance\footnote{$\Lambda$CDM means that the dominant energy components are cold dark matter (CDM) and {\em Dark Energy\/}.} $\Lambda$CDM cosmology (Sect.\,\ref{s3_cosmo_framework}); 
    \item Introduce the main aspects of modern hierarchical structure formation scenarios with the focus on cluster formation and evolution (Sect.\,\ref{s3_struct_formation});
    \item Compile a summary of cosmological tests using (distant) galaxy clusters which are feasible within the next decade(s) (Sect.\,\ref{s3_cosmo_tests}).
\end{enumerate}

    




\section{Cosmological Framework}
\label{s3_cosmo_framework}


\subsection{Dynamics of the expanding Universe}

\noindent
The dynamics of the scale factor $R(t)$ of standard relativistic world models is governed by the Friedman--Lem\^{a}itre equations, which are derived from the field equations of general relativity (GR) for the case of a perfect homogeneous fluid

\begin{eqnarray}\label{e3_friedmann}
  \ddot{R} = -\frac{4}{3} \pi G R \left(\rho + \frac{3p}{c^2}\right) + \frac{1}{3} \Lambda  R \\ \label{e3_friedmann2}
  \dot{R}^2 =  \frac{8 }{3} \pi G \rho R^2 + \frac{1}{3} \Lambda R^2 - kc^2 \ .
\end{eqnarray}  
\noindent
Here dots represent time derivatives of the scale factor $R(t)$, $\rho$ is the matter density, $p$ the pressure, $\Lambda$ Einstein's cosmological constant, and $k$ the space time curvature constant. 
The scale factor $R(t)$ with the dimension of a length can be replaced by the dimensionless parameter $a(t)=R(t)/R_0$, which is normalized to the  scale factor of the current epoch $R_0$, \ie \ $a(t_0) \equiv 1$. 

The two equations for the expansion velocity $\dot{R}$ (or $\dot{a}$) and acceleration rate $\ddot{R}$ (or $\ddot{a}$) are not independent, but are linked\footnote{To see the general relationship, the time derivative of Equ.\,\ref{e3_friedmann2} is taken and solved for $ \ddot{a}$. Reordering the right-hand side to match Equ.\,\ref{e3_friedmann} leaves extra terms that can be identified with Equ.\,\ref{e3_adiabatic_equat}.} through the adiabatic expansion equation (energy conservation in thermodynamics: $dU=-p\,dV$)

\begin{equation}\label{e3_adiabatic_equat}
    \frac{d}{dt}(a^3 \rho c^2) = - p \, \frac{d}{dt}(a^3) \ .
\end{equation}

\noindent
From Equ.\,\ref{e3_adiabatic_equat} the density evolution during the expansion of the Universe can be derived for the different relevant energy components with their respective equations of state (EoS) $p=w \rho c^2$. In $\Lambda$CDM cosmologies, the major energy components with their density scaling relations are: 


\begin{eqnarray}\label{e3_density_evol_m}
 \mathrm{matter \ (CDM \ and \ baryonic):} \ \ \ \ \ \ w=0  \ \ \ \ \ \ \ \ \ \ \ \ \ \ \ \ \rho_{\mathrm{m}} \propto a^{-3} \\ \label{e3_density_evol_r}
\mathrm{radiation:} \ \ \ \ \ \ w=1/3  \ \ \ \ \ \ \ \ \ \ \ \ \ \ \rho_{\mathrm{r}} \propto a^{-4} \\ \label{e3_density_evol_DE}
\mathrm{Dark \ Energy:} \ \ \ \ \ w\!<\!-1/3  \ \ \ \ \ \rho_{\mathrm{DE}} \propto a^{-3(1+w)}
\end{eqnarray}

\noindent
The pressureless matter density (Equ.\,\ref{e3_density_evol_m}) is just diluted as space expands, the radiation component (and relativistic matter) experiences an additional dilution factor of $a^{-1}$ due to the cosmological redshift (see Equ.\,\ref{e3_redshifts}) and thus its energy density decreases 
faster than matter. In the last step, the {\em Dark Energy\/} component was generalized to account for its unknown nature. For the special case of the cosmological constant $\Lambda$ ($w\!=\!-1$), Equ.\,\ref{e3_density_evol_DE} implies a constant energy density $\rho_{\Lambda}\!\propto\!a^0\!=\!\mathrm{const} $.

The ratio of the expansion velocity and the scale factor defines the Hubble parameter $H(a)$, \ie \ the fractional increase of the Universe per unit time

\begin{equation}\label{e3_hubble_para}
    H(a) \equiv \frac{\dot{a}}{a} \ .
\end{equation}


\noindent
Its present value $H_0\!\equiv\!H(a_0)$ is the Hubble constant with $H_0 \approx 70\,\mathrm{km}\,\mathrm{s}^{-1}\,\mathrm{Mpc}^{-1} \approx 2.3 \times 10^{-18}\,\mathrm{s}^{-1}$ 
(\eg \ Freedman \etal, 2001). \nocite{Freedman2001a}
The lowest order approximation of Equ.\,\ref{e3_hubble_para} yields the famous Hubble law, which relates the distance $D$ to the cosmological recession velocity $v_{\mathrm{recession}}$ (for small redshifts) via $D\!=\!v_{\mathrm{recession}}/H_0$. Hence, at constant velocity (or redshift) all cosmological distances scale with the inverse of the value of the Hubble constant $H_0^{-1}$, \ie \ a faster expansion rate results in a smaller (comoving) distance at a given redshift. This distance scaling property can be parameterized with the dimensionless constant $h\!=\!H_0 / (100\,\mathrm{km}\,\mathrm{s}^{-1}\,\mathrm{Mpc}^{-1})$, or for a concordance cosmology as 
$h_{70}\!=\!H_0 / (70\,\mathrm{km}\,\mathrm{s}^{-1}\,\mathrm{Mpc}^{-1})$. The dependency on the exact numerical value of the Hubble constant can then be accounted for by using the (parameterized) distance units $[D]\!=\!h^{-1}\,\mathrm{Mpc}$. Analogously all secondary derived quantities that are linked to cosmological distances scale with the appropriate powers of $h$: volumes as $[V]\!=\!h^{-3}\,\mathrm{Mpc}^3$, total cluster masses\footnote{The distance scaling  depends on the mass determination method and is referring to Equ\,\ref{e2_cluster_mass}. The X-ray determined cluster gas mass scales as $M_{\mathrm{gas}}\!\propto\!h^{-5/2}$ (\eg \ Sasaki, 1996). \nocite{Sasaki1996a}} as $[M]\!=\!h^{-1}\,\mathrm{M}_{\sun}$, densities as $[\rho]\!=\!h^{2}\,\mathrm{g}\,\mathrm{cm}^{-3}$, and luminosities as $[L]\!=\!h^{-2}\,\mathrm{erg}\,\mathrm{s}^{-1}$.     


Since all distance scales are stretched by the expansion of the Universe, this also applies to the wavelength $\lambda$ of light propagating through space. This leads to the cosmological redshift $z$, which is given by the ratio of the scale factors, \ie \ the stretching factor,  at the time of observation $a_0$ and  emission of the light $a(t)$

\begin{equation}\label{e3_redshifts}
    1+z \equiv \frac{\lambda_{\mathrm{obs}}}{\lambda_{\mathrm{em}}} = \frac{a_0}{a} \ .
\end{equation}

\noindent
The redshift $z$ is a direct observable and is hence of particular importance for observational cosmology. Through Equ.\,\ref{e3_redshifts} redshifts are uniquely linked to the expansion state of the Universe and other quantities such as cosmic time and distances through the underlying cosmological model. 
Note that the cosmological redshift should not be interpreted as a recession velocity $v_{\mathrm{rec}}$ as in special relativity theory (SRT), although for small redshifts ($z\!\la\!0.3$) the results are consistent 
as is the case for the Hubble law.
At high redshift ($z\!\ga\!1$), however, it can be shown empirically \cite{Davis2004a}  that the SRT interpretation fails and that super-luminal recession velocities are indeed allowed in the general relativistic treatment. 
In particular the {\em Hubble radius} $d_{\mathrm{H}}$, defined as the sphere that presently recedes at the velocity of light 

\begin{equation}\label{e3_hubble_radius}
d_{\mathrm{H}} =  \frac{c}{H_0}  = 2\,997.9\,h^{-1}\,\mathrm{Mpc} \simeq  4\,280\,h_{70}^{-1}\,\mathrm{Mpc} \ ,   
\end{equation}


\noindent
corresponds to a redshift of $z\!=\!1.46$ in concordance cosmology. 
This implies that objects at higher redshift recede  super-luminally at the current epoch, and the furthest spectroscopically confirmed galaxies at $z\!\sim\!6$ have done so at all times\footnote{It is still possible to observe them, since the Hubble sphere is receding even  more rapidly and can `overtake' the emitted photons, see Davis and Lineweaver \cite*{Davis2004a} for more details.}. 


As the next step, all energy densities can be transformed into dimensionless parameters by stating them 
in units of the 
critical energy density $\rho_{\mathrm{cr}}$. $\rho_{\mathrm{cr}}$ is defined as the total energy density that results in a flat global geometry of the Universe, \ie \ $k\!\equiv\!0$ in Equ.\,\ref{e3_friedmann2} (for $\Lambda\!=\!0$), and is thus given by

\begin{equation}\label{e3_crit_density}
\rho_{\mathrm{cr}} \equiv  \frac{3 H_0^2}{8\pi G} = 1.879 \times 10^{-29} \, h^2 \, \mathrm{g\,cm}^{-3} = 2.775 \times 10^{11} \, h^2 M_{\sun} \mathrm{Mpc}^{-3} \ .
\end{equation}

\noindent
The dimensionless energy components as measured today are then defined as

\begin{equation}\label{e3_cosmoparam}
\Omega_\mathrm{m} = \frac{\rho_\mathrm{m}}{\rho_{\mathrm{cr}}} \ ; \ \ \ \
\Omega_{\mathrm{DE}} = \frac{\rho_{\Lambda}}{\rho_{\mathrm{cr}}} = \frac{\Lambda c^2}{3H_0^2} \ ; \ \ \ \
\Omega_\mathrm{r} = \frac{\rho_\mathrm{r}}{\rho_{\mathrm{cr}}} \ ; \ \ \ \
\Omega_\mathrm{k} = 1 - \Omega_\mathrm{m} - \Omega_{\Lambda} - \Omega_\mathrm{r} \ . 
\end{equation}

\noindent
The total energy density is  the sum of all contributing components
\begin{equation}\label{e3_tot_energy}
\Omega_{\mathrm{tot}} = \Omega_\mathrm{m} + \Omega_{\mathrm{DE}} + \Omega_\mathrm{r}  \ .
\end{equation}

\noindent
The first two terms for the matter density $\Omega_\mathrm{m}$ and the {\em Dark Energy\/} density $\Omega_{\mathrm{DE}}$ are the dynamically dominant contributions today. In concordance cosmology, their sum (see Tab.\,\ref{t3_cosmo_param}) is very close to the critical density $\Omega_\mathrm{m} + \Omega_{\mathrm{DE}} \approx \Omega_{\mathrm{tot}} \approx 1$., which implies a flat (or very close to flat) Universe with $\Omega_\mathrm{k}\!\approx\!0$ with percent level accuracy \cite{Spergel2007a}.
The radiation term $\Omega_\mathrm{r}\!\sim\!\orderof (10^{-4})$ is negligible today but has been of significant importance at early epochs of $z \ga 1000$ (see Sect.\,\ref{s3_struct_formation}), which can be seen from the density scaling relations in Equ.\,\ref{e3_density_evol_r}.


When combining equations \ref{e3_friedmann2}, \ref{e3_hubble_para}, and \ref{e3_cosmoparam}, the general Hubble expansion history is obtained as a function of the normalized scale factor

\begin{equation}\label{e3_HubbleExpansion_a}
H^2(a) = H_0^2 \cdot \left[\Omega_\mathrm{m}\, a^{-3} + \Omega_{\mathrm{DE}} \, a^{-3(w+1)} + \Omega_\mathrm{r} \, a^{-4} -(\Omega_{\mathrm{tot}}-1) \, a^{-2}  \right] \ .
\end{equation}


\noindent
Here the {\em Dark Energy\/} equation-of-state parameter $w$ is left as a free parameter as in Equ.\,\ref{e3_density_evol_DE}. However,   
the most general form of {\em Dark Energy\/} also allows for a time evolution of the equation-of-state $w(z)$. 
In its simplest form the evolution can be parameterized as $w(z)\!=\!w_0 + w_1\!\cdot\!z$ with a constant part $w_0$ and an evolving contribution $w_1$. 
For an arbitrary time variation $w(z)$, the {\em Dark Energy\/} redshift 
scaling factor is integrated according to

\begin{equation}\label{e3_general_DE}
\int_0^{z} d\left[(1\!+\!z')^{3[w(z')\!+\!1]}\right] = \exp \left(\int_0^{z} d\,\ln\left[(1\!+\!z')^{3[w(z')\!+\!1]}\right] \right) =
\exp \left(3\!\int_0^z \frac{1\!+\!w(z')}{1\!+\!z'} \, dz' \right). 
\end{equation}

\noindent
For expressing all relevant relations in terms of the observable redshift $z$, the scale factors 
are substituted according to $a\!\rightarrow\!(1+z)^{-1}$  \  and \  $da\!\rightarrow\!-dz(1+z)^{-2}$. 
As a final step, we can now define the cosmic evolution function $E(z)$, with which the final general form of the expansion history of the Universe $H(z)$ can be expressed in terms of $H_0$

\begin{eqnarray}
\label{e3_evolution_funct}
    E^2(z) = \Omega_\mathrm{m} (1+z)^{3} + \Omega_{\mathrm{DE}}\cdot e^{3\int_0^z \frac{1 + w(z')}{1+z'} dz'}  + \Omega_\mathrm{r} (1+z)^{4} +(1-\Omega_{\mathrm{tot}})(1+z)^{2} \\ \label{e3_expansion_history_z}
H^2(z) = H^2_0 \cdot E^2(z) \ . \hspace{26ex}
\end{eqnarray}

\noindent
Detailed measurements of the Hubble expansion history $H(z)$ are the core of many cosmological tests, in particular for future {\em Dark Energy\/} studies (see Sect.\,\ref{s3_cosmo_tests}).



\subsection{Distance measures}

\noindent
A meaningful general definition of distances in the Universe requires the knowledge of the metric for the global geometry.
For the isotropic and homogeneous Universe (maximally symmetric space time) the general relativistic solution is the Robertson-Walker-Metric (RWM). The space time line element in spherical coordinates with curvature constant $k$ and scale factor $R(t)$ can be written as

\begin{equation}\label{e3_RWM}
    ds^2 = c^2 dt^2 - R^{2}(t) \left[ \frac{dr^2}{1-kr^2} + r^2\left(\sin^2\!\theta \, d\phi^2 + d\theta^2\right) \right] 
\end{equation}

\noindent
with proper time $t$, radial comoving coordinate $r$, and angular coordinates $\phi$ and $\theta$.
As the global space time metric, the RWM is valid for all distances. For many applications the light propagation along radial coordinates is of prime interest. Since light travels on null geodesics, \ie \ $ds^2\!=\!0$, Equ.\,\ref{e3_RWM} reduces to the simple form

\begin{equation}\label{e3_radial_light}
    c^2 dt^2 - R^{2}(t) \frac{dr^2}{1-kr^2} = 0 \ .
\end{equation}

\noindent
For the following discussion of distance measures, the specialization for a flat Universe ($k\!\equiv\!0$) is assumed, which is valid for concordance cosmology.
Small deviations from a flat geometry could still be possible, however,  the additional correction functions $\sin(R_0 r)$ for $k\!=\!1$ and $\sinh(R_0 r)$ for $k\!=\!-1$ for non-flat space times would 
inhibit the readability and clarity for this discussion and are therefore omitted.




Integrating  Equ.\,\ref{e3_radial_light} (for $k\!=\!0$) yields the fundamental
comoving distance $D$, \ie \ the distance 
an object at redshift $z$ 
has today,

\begin{equation}\label{e3_comoving_distance}
D(z) = R_0 \cdot r(z) = R_0 \int_0^r dr' = c  \int_t^{t_0} \frac{dt'}{a(t')} = c \int_0^z \frac{dz'}{H(z')} = \frac{c}{H_0} \int_0^z \frac{dz'}{E(z')} \ .
\end{equation}


\noindent
The second last step took advantage of Equ.\,\ref{e3_redshifts} and its time derivative $dt\!=\!-dz(1\!+\!z)^{-1}H(z)^{-1}$ to relate time $t$ and redshift $z$, and the final step made use of Equ.\,\ref{e3_expansion_history_z}.
From the observational point of view, two other alternative cosmological distance measures are of prime importance. Firstly,
the luminosity distance $d_{\mathrm{lum}}$ relates the luminosity $L$ of an object to its observed flux $f$ in a way to reproduce the inverse square law of  flat Euclidean space $f\!=\!L/(4 \pi  d_{\mathrm{lum}}^2)$.
Secondly, the angular diameter distance $d_{\mathrm{ang}}$ is defined to yield the Euclidean relation $dl = \theta \cdot d_{\mathrm{ang}}$ for the apparent angular size $\theta$ of an object with physical size $dl$. Both observationally motivated distance measures can be expressed in terms of the comoving distance $D$ as

\begin{eqnarray}
\label{e3_lum_dist}
d_{\mathrm{lum}} = D \cdot (1+z) \  \\ \label{e3_ang_dist}
d_{\mathrm{ang}} = \frac{D}{(1+z)} \ .  \ \ \ 
\end{eqnarray}



\begin{figure}
\centering
\includegraphics[angle=0,clip,width=0.49\textwidth]{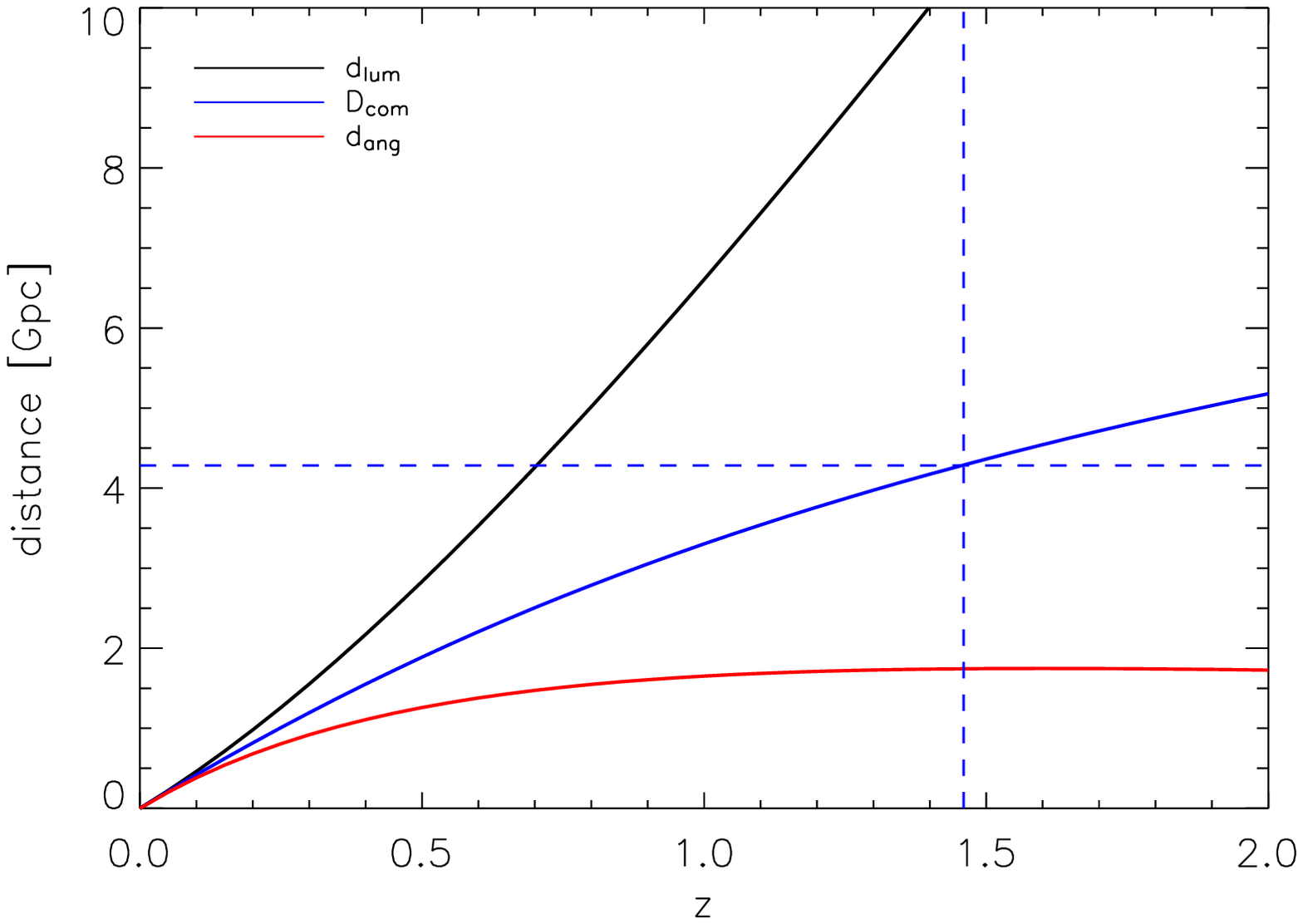}
\includegraphics[angle=0,clip,width=0.49\textwidth]{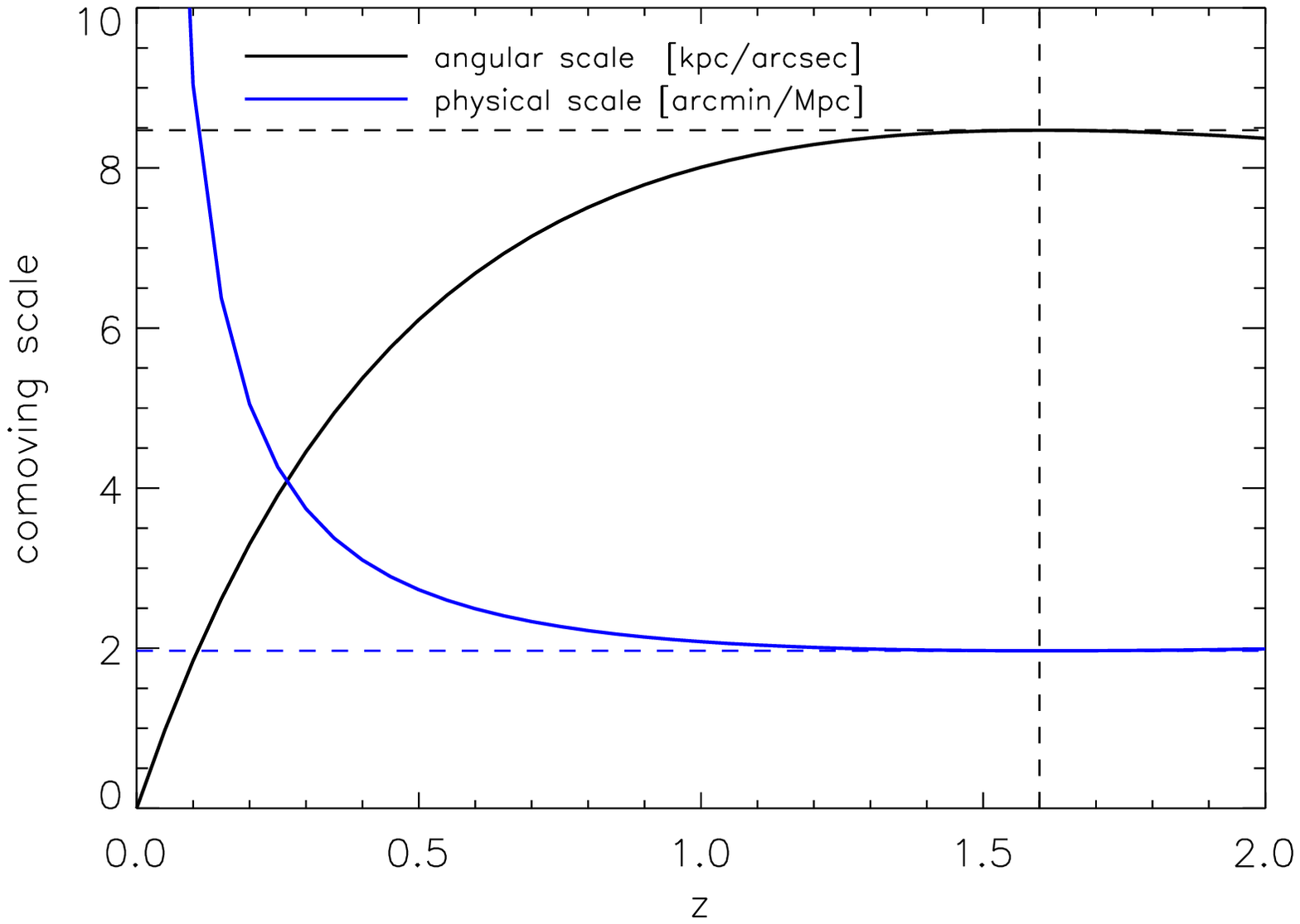}
\caption[Cosmological Distances]{Cosmological distances and scales for objects at redshift $z$ using concordance model parameters. {\em Left:} Comoving distance $D$ 
 ({\em blue solid line}), luminosity distance $d_{\mathrm{lum}}$ ({\em black}), and angular size distance $d_{\mathrm{ang}}$ ({\em red}) as a function of redshift. The Hubble radius $d_{\mathrm{H}}$ 
is indicated by the horizontal dashed line and corresponds to the comoving distance at redshift $z\!=\!1.46$. {\em Right:} Physical and angular scale evolution with redshift. The black line depicts the physical comoving size of a fixed angular scale of 1\arcsec \ in units [kpc/arcsec], the blue line traces the apparent angular size of a fixed physical scale of 1\,Mpc in units [arcmin/Mpc]. Due to the global maximum of $d_{\mathrm{ang}}$ at $z\!\approx\!1.6$, both functions assume their extrema at this redshift. However, at \zg1 the scales are almost redshift independent at approximately 8.4\,kpc\,arcsec$^{-1}$ and correspondingly 2.0\,arcmin\,Mpc$^{-1}$.  }
\label{f3_cosmological_distances}       
\end{figure}

\begin{figure}
\centering
\includegraphics[angle=0,clip,width=0.49\textwidth]{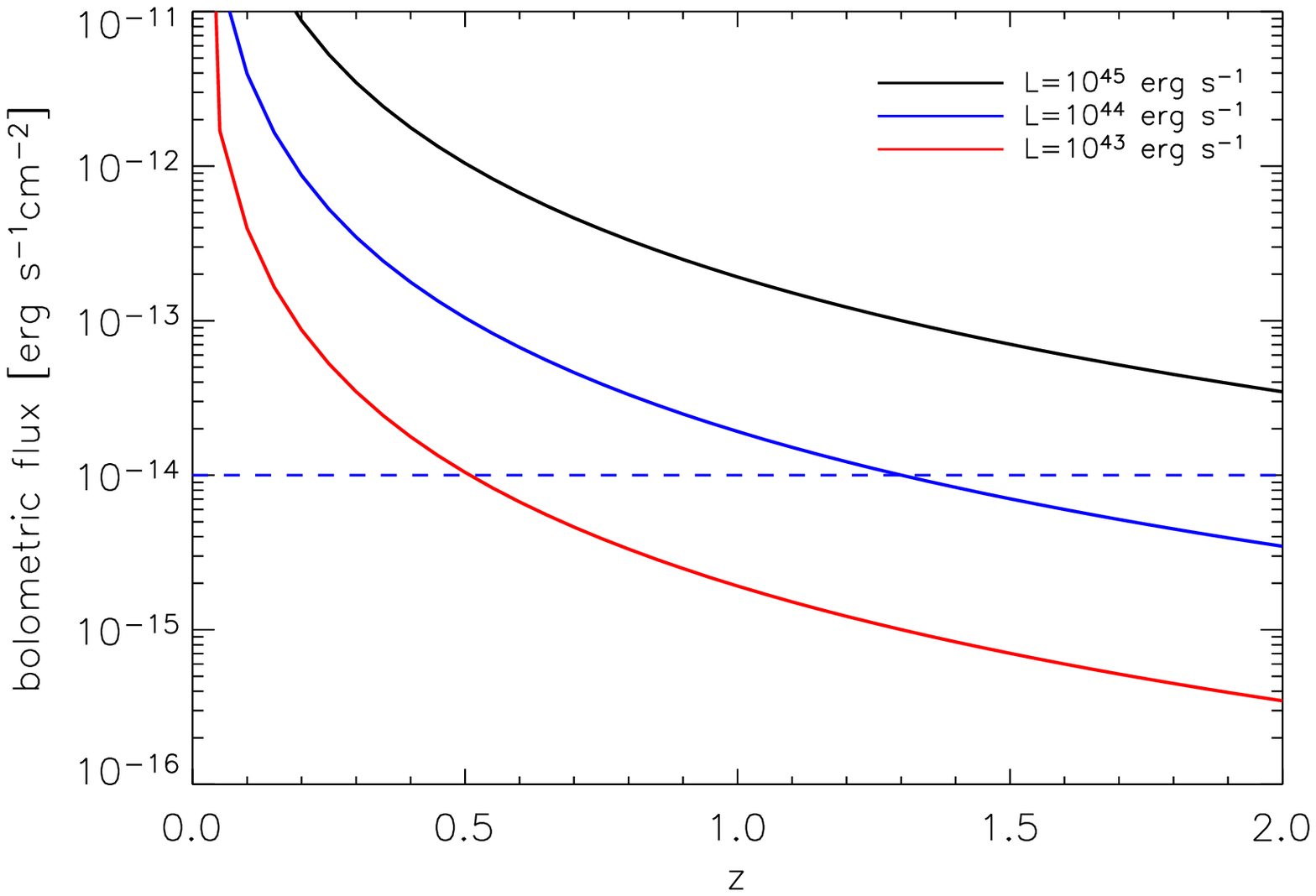}
\includegraphics[angle=0,clip,width=0.49\textwidth]{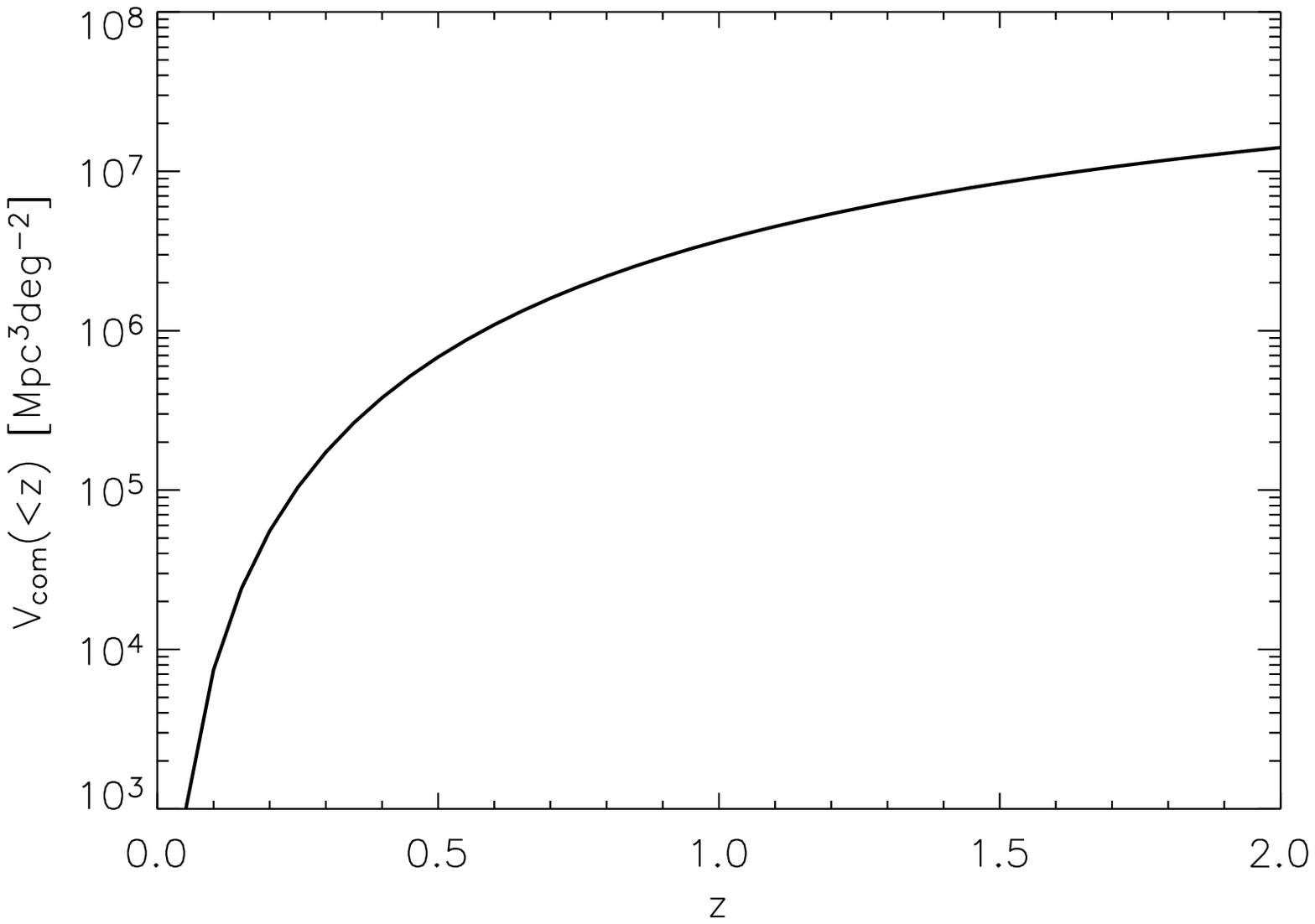}
\caption[Fluxes and Comoving Volumes]{Bolometric flux evolution and comoving volumes as a function of redshift. {\em Left:} Received bolometric flux for three fiducial objects at redshift $z$ with luminosities $10^{43}$\,erg\,s$^{-1}$ ({\em red line}), $10^{44}$\,erg\,s$^{-1}$ ({\em blue line}), and $10^{45}$\,erg\,s$^{-1}$ ({\em black line}). A limited detection bandwidth will shift the curves down with an offset that depends on the flux fraction outside the filter band. 
At a given flux limit (\eg \ $10^{-14}$\,erg\,s$^{-1}$cm$^{-2}$ for the dashed horizontal line) objects are detectable out to a maximum redshift $z_{\mathrm{max}}$, which is increasing with luminosity. {\em Right:} Comoving volume $V_{\mathrm{com}}(<\!z)$ per square degree sky coverage from redshift 0 to $z$. The search volume increases roughly by a factor of 5 from redshift 0.5 to 1.0, and doubles again out to $z\!=\!1.5$. } 
\label{f3_flux_evolution}       
\end{figure}

\begin{figure}[t]
\centering
\includegraphics[angle=0,clip,height=5.5cm]{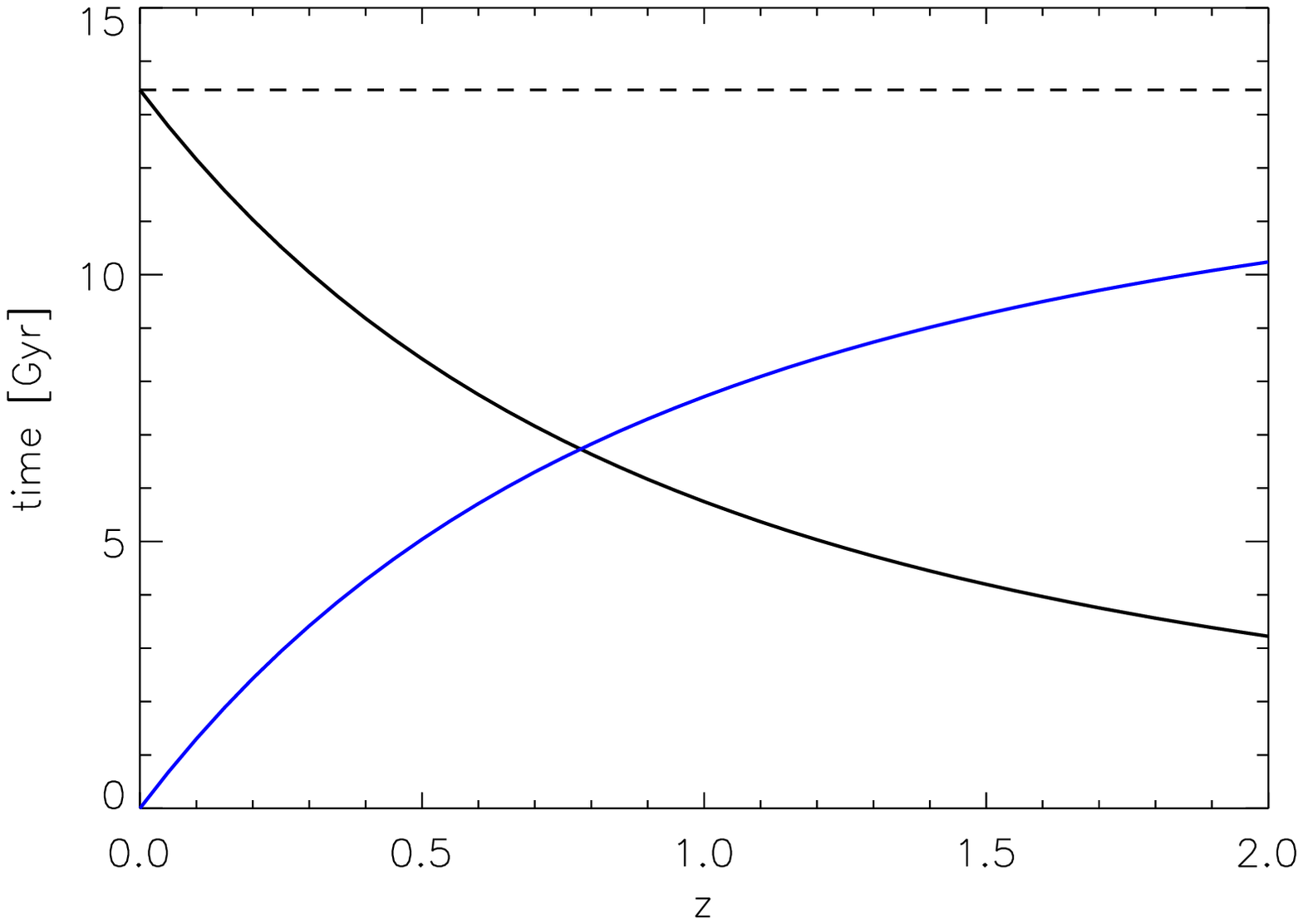}
\includegraphics[angle=0,clip,height=5.5cm]{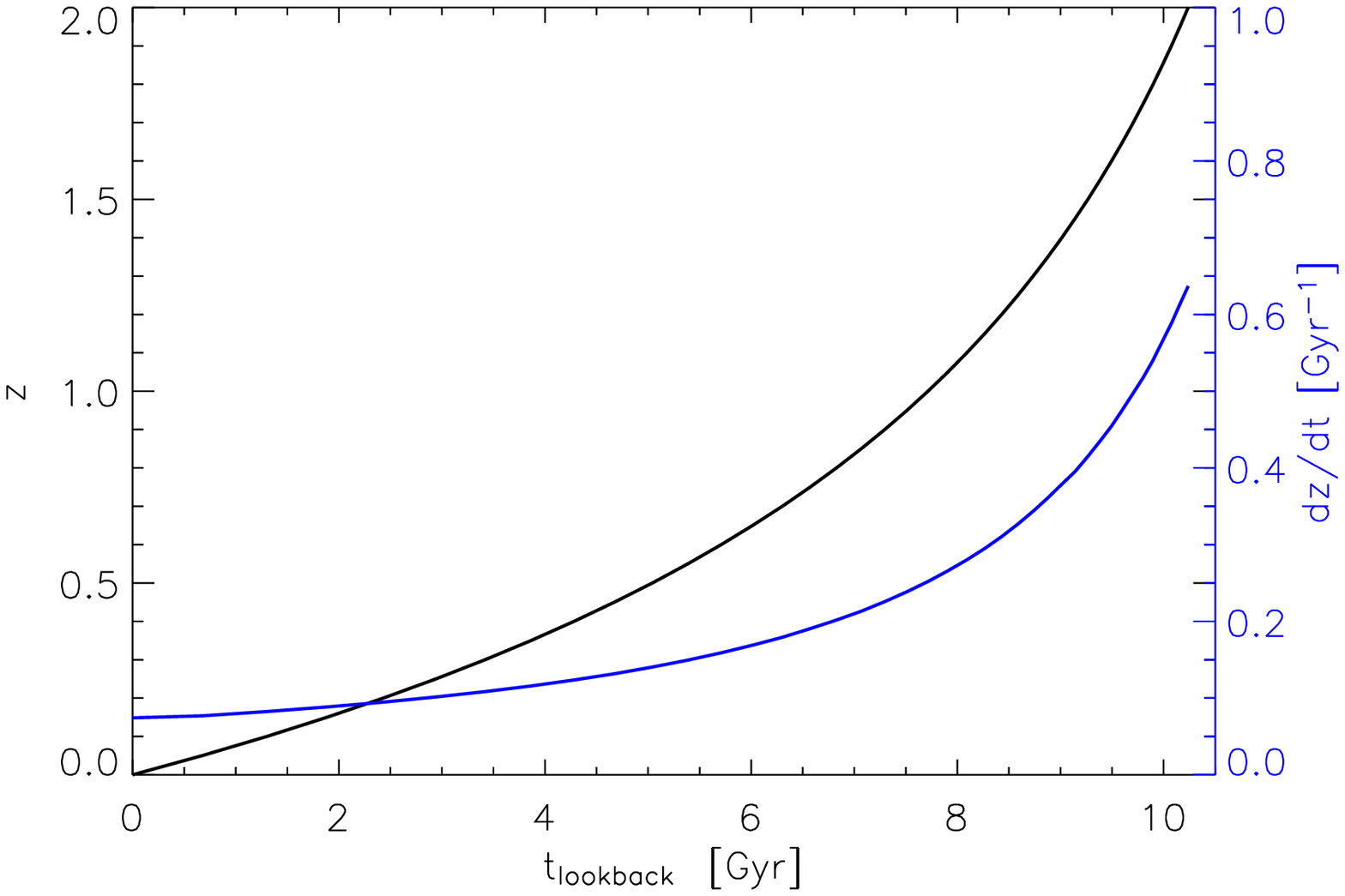}
\caption[Cosmic Time]{Cosmic time as a function of redshift for concordance model parameters. {\em Left:} The black solid line indicates the age of the Universe $t_{\mathrm{age}}$ at redshift $z$ with the total age $t_0\!=\!13.46$\,Gyr as represented by the horizontal dashed line. The blue line illustrates the lookback-time $t_{\mathrm{lookback}}$ from the current epoch to redshift $z$. The second quarter of the lifetime of the Universe ($6.8\,\mathrm{Gyr} \la t_{\mathrm{lookback}} \la 10.1\,\mathrm{Gyr}$) is spanned by the redshift range $0.8\!\la\!z\!\la\!1.9$; from redshift $z\!=\!1$ to $z\!=\!1.5$ the lookback-time increases from 7.7\,Gyr to 9.3\,Gyr. {\em Right:} Redshift $z$ versus lookback time $t_{\mathrm{lookback}}$ (black line with left scale). Many physical processes related to structure formation have characteristic timescales of $\sim\!1$\,Gyr, which translate into redshift in a non-linear way. The blue  line (right scale) illustrates the changing redshift interval per Giga year lookback time $dz/dt$. Out to $z\!\la\!0.6$ the redshift intervals are almost constant at $\sim\!0.1$\,Gyr$^{-1}$, \ie \ redshift and time are linear to first order. By $z\!=\!1$ the redshift interval per Giga year has increased to $0.25$\,Gyr$^{-1}$, at $z\!=\!1.5$ to $0.4$\,Gyr$^{-1}$, and at $z\!=\!2$ it is $0.6$\,Gyr$^{-1}$.
}
\label{f3_cosmic_time}       
\end{figure}

\noindent
The three main cosmological distance measures $D$, $d_{\mathrm{lum}}$, and $d_{\mathrm{ang}}$ are illustrated in the left panel of Fig.\,\ref{f3_cosmological_distances}\footnote{The models for Figs.\,\ref{f3_cosmological_distances}, \ref{f3_flux_evolution}, and \ref{f3_cosmic_time} were computed with the cosmological calculators at 
\url{http://www.astro.ucla.edu/}{\tt \%}\url{7Ewright/CosmoCalc.html} and \url{http://faraday.uwyo.edu/}{\tt \%}\url{7Echip/misc/Cosmo2/cosmo.cgi}.} 
for concordance model parameters. The fact that the angular size distance $d_{\mathrm{ang}}$ has a maximum at $z\!\approx\!1.6$ has the important observational consequence that the apparent angular size of objects is practically constant at \zg1,  as shown in the right panel of Fig.\,\ref{f3_cosmological_distances}. This implies that high-redshift galaxies with typical physical sizes of 10\,kpc are observed at angular scales of 1.2\arcsec, 100\,kpc cluster cores at 12\arcsec, and 1\,Mpc scale clusters of galaxies at 2\arcmin. 

Combining the relation $d_{\mathrm{lum}}=d_{\mathrm{ang}}\cdot (1+z)^{2}$ and the fact that the solid angle $d\Omega$ covered by an object follows $d\Omega \propto d^2_{\mathrm{ang}}$ leads to the cosmological surface brightness dimming ({\em Tolman's Law})

\begin{equation}\label{e3_SB_dim}
    I_{\mathrm{bol}}(\mathrm{observed}) = \frac{I_{\mathrm{bol}}(\mathrm{emitted})}{(1+z)^4} \ .
\end{equation}

\noindent
For monochromatic flux measurements, the surface brightness dimming reduces to  $I_{\mathrm{mono}}(\mathrm{observed}) \propto (1+z)^{-3}$, since the  factor $(1+z)^{-1}$ for the effective narrowing of the observed rest-frame bandwidth is recovered.


If the rest-frame spectral energy distribution (SED) has a non-zero slope at the spectral interval of interest, then the observed flux will not only depend on the luminosity distance $d_{\mathrm{lum}}$ but also on the actually observed part of the rest-frame spectrum 
redshifted by the factor $(1+z)$. This is taken into account with the so-called $K$-correction, which can be defined as an additive term $K_{\mathrm{QR}}$ for the observed apparent magnitude $m_{\mathrm{R}}$ in band-pass $R$


 \begin{equation}\label{e3_k_correction}
    m_{\mathrm{R}} = M_{\mathrm{Q}} + DM + K_{\mathrm{QR}} \ ,
\end{equation}

\noindent
where $M_{\mathrm{Q}}$ is the absolute magnitude of the source in the rest-frame band-pass $Q$ and $DM$ is the distance modulus

\begin{equation}\label{e3_dist_modulus}
DM=m\!-\!M=25\!+\!5\log \left(d_{\mathrm{lum}}\, [\mathrm{Mpc}]\right)\!-\!5\log h_{70} \ .
\end{equation}

\enlargethispage{4ex}

\noindent
This last relation follows directly from the definition of absolute magnitudes transformed to \hinv70\,Mpc units\footnote{The last term vanishes for concordance cosmology, but reflects the constant offset when changing the value of the Hubble constant $H_0$}. For concordance cosmology, the last term is zero, and $d_{\mathrm{lum}}$ can be read off 
from Fig.\,\ref{f3_cosmological_distances}.
The $K$-correction term $K_{\mathrm{QR}}$ depends on the filter band, the redshift, and the spectral properties of the object under investigation (see Fig.\,\ref{f10_bcg_kcorrection}).
Positive correction terms imply a resulting $K$-dimming due to a decreasing SED with increasing frequency; negative terms result in $K$-brightening, \ie \ objects actually appear brighter than expected at higher redshift. For efficient observations of the high-redshift Universe, the $K$-correction is an important consideration for the observing strategy as discussed in Sect.\,\ref{s7_NIR_strategy}.

With the distance measures at hand, it is now possible to define comoving volume elements $dV_{\mathrm{com}}$ as

\begin{equation}\label{e3_volume_elements}
    dV_{\mathrm{com}} = D^2 d\Omega \, dr = \frac{c D^2}{H(z)}  \, d\Omega \, dz \ .
\end{equation}

\noindent
Here $D$ is the comoving distance, $d\Omega$ the solid angle, and $dr$ the radial thickness\footnote{Strictly speaking $d(R_0 r)$, as the radial coordinate $r$ is dimensionless.} of the volume element. In the last step, the radial component was expressed in terms of a redshift interval $dz$ as in Equ.\,\ref{e3_comoving_distance}. Figure\,\ref{f3_flux_evolution} (right panel) displays the enclosed comoving volume out to redshift $z$ per square degree for concordance cosmology parameters. The left panel depicts the observed (bolometric) flux for a selection of fiducial object luminosities as a function of redshift and illustrates the luminosity dependent search volumes for flux-limited observations.

When integrating 
a homogeneously distributed population of sources in Euclidean space over all luminosities (with arbitrary luminosity function), one obtains a universal relation for the
cumulative number counts $N(>S)$ above flux $S$\footnote{Same meaning  as flux $f$, but $S$ is typically used in the context of the $\log N$--$\log S$ relation.} 

\begin{equation}\label{e3_number_counts}
    N(>\!S) \propto S^{-\frac{3}{2}} \ .
\end{equation}

\noindent
The $\log N(>\!S)$--$\log S$ plot thus shows a constant slope of $-3/2$ down to the flux limit of the observation, where the counting becomes incomplete (see also Equ.\,\ref{e3_cluster_NumberCounts}). The slope is due to the fact, that a lower flux reaches out further in distance (for any luminosity interval) $d_{\mathrm{lum}} \propto S^{-\frac{1}{2}}$, and the enclosed volume grows proportional to the third power of the distance. Since this is only valid for static flat Euclidean space, the relation is expected to intrinsically flatten at a certain flux level when the effects of the expanding Universe and evolving populations become significant (see Sects\,\ref{s3_cosmo_tests}\,\&\,\ref{s4_DeepSurveys}).


For a population of objects with a constant comoving number density $n_0\!=\!dN/dV_{\mathrm{com}}$, 
the redshift distribution $dN/dz$ is obtained by using Equ.\,\ref{e3_volume_elements}

\begin{equation}\label{e3_redshift_distib}
    \frac{dN}{dz} = n_0 \frac{dV_{\mathrm{com}}}{dz} =  n_0 \,  \frac{c \, D^2 }{H(z)} \, d\Omega \ .
\end{equation}

\noindent
The unique link between redshift $z$ and cosmic distance measures can also be generalized for cosmic time $t$. The time derivative of Equ.\,\ref{e3_redshifts} yields $\dot{a}\!=\!-(1\!+\!z)^{-2}\,dz/dt$. Dividing by $a(t)$ and substituting with relation\,\ref{e3_hubble_para} results in $H(z)\!=\!-(1\!+\!z)^{-1}\,dz/dt$, which can be readily integrated by separation of variables to arrive at the lookback time $t_{\mathrm{lookback}}$

\begin{equation}\label{e3_lookback_time}
    t_{\mathrm{lookback}}(z)= \int_{t_0}^{t} dt'= \frac{1}{H_0} \int_0^{z} \frac{dz'}{(1+z')\cdot E(z')} \ ,
\end{equation}

\enlargethispage{4ex}

\noindent
\ie \ the time light from an object at redshift $z$ has been travelling (light travelling time). Integrating to $z=\infty$ returns the age of the Universe $t_0$, and the difference yields the age of the Universe at a given redshift $t_{\mathrm{age}}\!=t_0-t_{\mathrm{lookback}}(z)$. Fig.\,\ref{f3_cosmic_time} 
illustrates $t_{\mathrm{age}}$ and $t_{\mathrm{lookback}}$ for a concordance cosmology with $t_0\!=\!13.46$\,Gyr.

Table\,\ref{t3_cosmo_param} summarizes the main cosmological parameters of the concordance model following Spergel \etal \ \cite*{Spergel2007a}. These parameters will be used throughout this thesis for any background cosmological model unless otherwise stated.

\pagebreak



\begin{table}[h]    
\begin{center}

\begin{tabular}{|c|c||c|c|}
\hline

\bf{Parameters} & \bf{Value} & \bf{Parameters} & \bf{Value} \\
 
\hline\hline

matter density  & $\Omega_{\mathrm{m}}=0.3$  &   total energy density    & $\Omega_0=1$  \\ 
power spectrum normalization  & $\sigma_{8}=0.8$         & primordial PS slope  & $n_{\mathrm{S}}=1$  \\                   
Dark Energy density   & $\Omega_{\Lambda}=0.7$ & age of Universe [Gyr] &  $t_0=13.46$ \\
Dark Energy EoS   & $w=-1$                     & recombination redshift  & $z_\mathrm{rec} \sim 1100$ \\
Hubble constant [km\,s$^{-1}$\,Mpc$^{-1}$ ]  &  $H_0=70$   & matter-radiation equality  &  $z_\mathrm{equ} \sim 3500$   \\
baryon density  & $\Omega_{\mathrm{b}}=0.045$ &  &  \\

\hline
\end{tabular}

\caption[Cosmological Concordance Parameters]{Cosmological concordance model parameters as used throughout this thesis. \emph{Left:} Parameters for which galaxy cluster samples are particularly  sensitive to. \emph{Right:} Additional concordance model parameters that will be assumed for the background cosmology. The parameters $\sigma_{8}$ and $n_{\mathrm{S}}$ describe the properties of the initial power spectrum of density perturbations, all other parameters are linked to the global geometry of the Universe, which is assumed to be flat ($k\!=\!0$) in concordance cosmology. The chosen parameter values follow Spergel \etal \ \cite*{Spergel2007a}.} \label{t3_cosmo_param}
\end{center}
\end{table}

\clearpage

\section{Cosmic Structure Formation}
\label{s3_struct_formation}

\noindent
Whereas Sect.\,\ref{s3_cosmo_framework} set the framework for the {\em homogeneous Universe}, this part will introduce some relevant aspects of the {\em inhomogeneous Universe}, \ie \ the formation and evolution of cosmic structure.
In the hierarchical structure formation scenario, galaxy clusters play a distinct role as
the largest and most recently virialized    
 objects to form and mark the important transition regime between linear and non-linear gravitational dynamics.

The standard paradigm for the growth of density perturbations rests upon the following qualitative picture:  
(i) Quantum fluctuations in the very early Universe serve as initial density fluctuations. (ii) These fluctuations are stretched to macroscopic scales by an early inflationary phase which also imprints a characteristic primordial fluctuation power spectrum.  (iii) After inflation, density perturbations grow only through gravitational instability within the expanding background cosmology. From this point on, the structure evolution can be well traced with large numerical simulations using cold dark matter (CDM) as the main ingredient. 
(iv) The structure evolution in the non-linear regime and the increasing importance of complex  non-gravitational physics (`{\em ga(s-a)strophysics}') on small scales is only accessible with elaborate numerical experiments. This regime includes all relevant  galaxy scales and hence turns the detailed understanding of galaxy formation and evolution into a very difficult and unsolved task. However, down to the galaxy cluster scale of $\sim$10\,Mpc analytic approximations and models can still provide the basic structure formation framework and valuable physical insights. 

In preparation for the cosmological tests presented  in the next section and to provide a basic understanding of the formation process of galaxy clusters, we will  
 summarize four main aspects of structure formation theory with a focus on galaxy clusters: (i) the general growth of density perturbations, (ii) the top-hat collapse model, (iii) the Press-Schechter mass function, and (iv) the spatial distribution 
of clusters.

\subsection{Growth and collapse of density perturbations}


\subsubsection{Perturbation Growth}

\noindent
We will first have a closer look at how the primordial density fluctuations seeded by the inflationary epoch grow in general. 
This growth is mainly driven by the cold dark matter density component $\rho_{\mathrm{CDM}}$ which only interacts gravitationally. In the early Universe, however, the strong coupling of photons and baryons (the {\em photon-baryon fluid}) has important consequences on the structure formation. The homogeneous {\em Dark Energy\/} density  $\rho_{\mathrm{DE}}$, as the fourth player of perturbation growth, has an additional indirect influence by governing the dynamics of the expanding background cosmology.


The hydrodynamic equations are the starting point for the description of the gravitational instability of density fluctuations. 
Besides {\em Euler's Equation} (momentum conservation), the continuity equation (mass conservation) and {\em Poissions's Equation} are needed to close the system. These equations are transformed to comoving coordinates and the matter density field $\rho(\mathbf{x},t)=\bar{\rho}_{\mathrm{m}}(t) \cdot [1+\delta(\mathbf{x},t)]$ is expressed 
through the dimensionless density contrast $\delta(\mathbf{x},t)$

\begin{equation}\label{e3_density_contrast}
\delta(\mathbf{x},t) = \frac{\rho(\mathbf{x},t)-  \bar{\rho}_{\mathrm{m}}(t)}{ \bar{\rho}_{\mathrm{m}}(t)} \ .
\end{equation}
 
\noindent
The system can then be linearized for small fluctuation amplitudes $|\delta \ll 1|$ on top of  the mean matter density field $\bar{\rho}_{\mathrm{m}}$ and separated into the homogeneous background solution for the Hubble expansion (Equ.\,\ref{e3_density_evol_m}) and the inhomogeneous solutions for the evolution of the density contrast $\delta$.
The result of the perturbation analysis is a second order differential equation that governs the gravitational amplification of the fractional density contrast $\delta$, which can be expressed in its general form as a damped wave equation

\begin{equation}\label{e3_StructureGrowth}
\ddot{\delta} + 2 \frac{\dot{a}}{a}\dot{\delta} = \delta \left[4 \pi G \bar{\rho}_{\mathrm{m}} - \frac{c_{\mathrm{s}}^2 k^2}{a^2} \right] \ .
\end{equation}

\noindent
The left hand side contains the 
acceleration and the damping term, where the Hubble expansion $H\!=\!\dot{a}/a$ acts as a friction force or a {\em Hubble drag}.
The right hand side of Equ.\,\ref{e3_StructureGrowth} shows the gravitational force due to density fluctuation and an additional 
term originating from {\em Euler's Equation} when considering fluids with pressure, as is the case for the {\em photon-baryon fluid} of the early Universe. The speed of sound for this relativistic fluid with equation-of-state according to Equ.\,\ref{e3_density_evol_r} is then given by  $c_{\mathrm{s}}^{2}\!=\!\partial p / \partial \rho\!\simeq\!c^2/3$.
For sufficiently large perturbation wave numbers $k\!=\!2\!\pi / L$, corresponding to small wavelength scales $L$, the right hand side is negative implying an oscillating solution. 

At this point, we will focus on the growing 
 solution, which is obtained for the large scale perturbations with $k\!\ll\!(a/c_{\mathrm{s}})\sqrt{4 \pi G \bar{\rho}_{\mathrm{m}}}$ in the early Universe or as the general perturbation growth mode after matter-radiation decoupling, when the baryonic pressure vanishes. For a negligible oscillation term on the right hand side, Equ.\,\ref{e3_StructureGrowth} does not contain any spatial derivatives implying that the time evolution 
for the perturbation growth is separable from the initial spatial perturbation field $\delta_{i+}(\mathbf{x},t_i)$ according to 

\begin{equation}\label{e3_time_separation}
\delta(\mathbf{x},t) = D_{+}(t) \cdot \delta_{i+}(\mathbf{x},t_i)  + D_{-}(t) \cdot \delta_{i-}(\mathbf{x},t_i) \ .
\end{equation}
 
\noindent
The decaying solution $D_{-}(t)$ is of little physical interest and will not be considered. The general expression for the growing perturbation solution $D_{+}(t)$ (\eg \ Borgani, 2006 \nocite{Borgani2006a}) in linear approximation as a function of redshift $z$ is given by




\begin{equation}\label{e3_Growth_Factor}
D_{+}(z) = \frac{5}{2} \Omega_{\mathrm{m}}  E(z) \cdot \int_{z}^{\infty} \frac{1+z'}{E(z')^3}dz' \ .
\end{equation}



\noindent
In an Einstein-de\,Sitter Universe ($\Omega_{\mathrm{m}}\!=\!1$) the perturbation amplitudes grow proportional to the scale factor $D_{+}\!\propto\!a\!\propto\!(1\!+\!z)^{-1}\!\propto\!(t/t_i)^{2/3}$. For an empty Universe ($\Omega_{\mathrm{tot}}\!\equiv\!0$), (virtual) perturbations are frozen and do not grow at all. In a low density Universe (\eg \ $\Omega_{\mathrm{m}}\!\sim\!0.3$, $\Omega_{\mathrm{DE}}\!=\!0$) the driving gravitational force on the right hand side of Equ.\,\ref{e3_StructureGrowth} is decreased resulting in a slower growth rate at low redshifts. At early epochs, however, the matter density of such models is still close to critical, \ie \ to the Einstein-de\,Sitter case, with an analogous fast early growth phase which slows
once the matter density drops significantly below the critical value. When looking backwards in time from the present perspective, the effect of this slowed structure evolution is an increased number of objects, \eg \ clusters, at high redshift relative to a critical matter density Universe. For a concordance cosmological model ($\Omega_{\mathrm{m}}\!\sim\!0.3$, $\Omega_{\mathrm{DE}}\!=\!0.7$), the addition of a {\em Dark Energy\/} component results in an intermediate degree of evolution, since the Hubble expansion must have been slower in the past, relative to an open Universe, leading to a smaller friction term in Equ.\,\ref{e3_StructureGrowth}. Note that the {\em `Hubble drag'}  term depends on the cosmological parameters, which implicitly influence the perturbation growth in addition to the explicit dependance on the mean matter density $\bar{\rho}_{\mathrm{m}}$.

\subsubsection{Top-hat collapse}

An important approach to tackle the non-linear structure evolution regime is provided by {\em Birkhoff's theorem}, which states that a closed sphere within a homogeneous Universe evolves independent of its surroundings, \ie \ as if no external forces are exerted on the sphere\footnote{In reality residual tidal forces are expected which can be neglected for this discussion.}. This implies that any overdense region of the Universe can be conceived as a homogeneous mini-Universe with the evolution driven by the local density parameters of the region under consideration.

The spherical top-hat model is the only system for which the collapse of an overdense region of the Universe can be treated analytically. For this we consider a spherical region with radius $R$ and a homogeneous overdensity $\delta_{\mathrm{TH}}$ embedded in a background field with constant density $\bar{\rho}_{\mathrm{m}}$. According to {\em Birkhoff's theorem} the evolution of this region is independent of the background field and can be described by the Newtonian approximation of the first Friedmann equation\,\ref{e3_friedmann}, \ie \ with vanishing pressure and zero cosmological constant term. In analogy to a closed Universe, this finite region will expand up to a maximum radius $R_{\mathrm{max}}$ at turn-around time $t_{\mathrm{turn}}$ and then re-collapse 
in a time-symmetric fashion at time $t_{\mathrm{col}}\!=\!2\!\cdot\!t_{\mathrm{turn}}$. In practice, the overdense region will not collapse to a point but stabilize in a bound dynamic equilibrium state at radius $R_{\mathrm{vir}}\!\simeq\!R_{\mathrm{max}}/2$ once the {\em Virial\/} condition $2\!\cdot\!\bar{E}_{\mathrm{kin}}\!+\!\bar{E}_{\mathrm{pot}}\!=\!0$ is fulfilled. The gravitationally bound, virialized object with total energy\footnote{A factor of 3/5 arises for the potential energy of a sphere with uniform density.} $E_{\mathrm{tot}}\!=\!\bar{E}_{\mathrm{pot}}/2\!=\!-3GM^2/(5\cdot 2R_{\mathrm{vir}})$ is now detached from the Hubble flow, \ie \ it does not take part in the overall expansion of the background cosmology. 

In the case of an Einstein-de\,Sitter cosmology with $\Omega_{\mathrm{m}}\!=\!1$ some universal characteristics of the collapse process can be specified, with only small variations for  
 other cosmologies. 
 The density contrast at turn-around time $t_{\mathrm{turn}}$, \ie \ at the time of decoupling from the Hubble flow, is about $\delta_{\mathrm{turn}}\!\simeq\!4.55$. 
 At collapse time  $t_{\mathrm{col}}$ the contrast has increased to its final (universal) equilibrium value\footnote{This density contrast is roughly obtained from the turn-around density by considering the compression factor during collapse and the evolution of the critical density during the formation process.} of $\delta_{\mathrm{vir}}\!\simeq\!177$, which gives rise to the widely used definition of $R_{200}$ as the cluster radius with an average density  of 200 times the critical density $\rho_{\mathrm{cr}}$. In comparison, the linear extrapolation of the density contrast according to Equ.\,\ref{e3_Growth_Factor} at $t_{\mathrm{col}}$ yields 
$\delta_{\mathrm{vir}}(\mathrm{linear})\!\simeq\!1.69$, and at turn-around $\delta_{\mathrm{turn}}(\mathrm{linear})\!\simeq\!1.06$,  emphasizing the 
onset of the non-linear regime and the break-down of the linear approximation during object collapse. Note that a virialized cluster with final radius $R_{\mathrm{vir}}\!\simeq1.5$\,\hinv70\,Mpc originates from a collapsed region of size $\sim\!6\!\cdot\!R_{\mathrm{vir}}\!\simeq\!10$\,\hinv70\,Mpc.

These considerations imply that any spherical overdensities reaching the critical density contrast $\delta_{\mathrm{turn}}$ detach from the Hubble flow and form a virialized object at $t_{\mathrm{vir}}\!=\!t_{\mathrm{col}}\!\simeq\!2\!\cdot\!t_{\mathrm{turn}}$.  
For the cold dark matter standard model with a scale invariant initial perturbation spectrum,
sub-galactic halos are the first to decouple from the Hubble flow, since perturbations on small length scales have greater amplitudes (see Equ.\,\ref{e3_filtered_variance}). Galaxy clusters on the other hand  have formed relatively recently and hence appear late on the cosmic stage, marking the largest mass scales that have had sufficient time to virialize.

 
The top-hat collapse model enables 
important physical insights into the cluster formation process. However, it rests upon many simplifications (spherical symmetry, homogeneous mass distribution, no external tidal forces, only gravitational physics) which are at best only an approximate description of the cluster formation conditions. The {\em Millennium Run} \ simulation of Springel \etal \ \cite*{Springel2005a} provides the most detailed and realistic cluster formation scenarios currently available. Figure\,\ref{f3_ClusterFormation} shows simulation snap-shots  of the formation of a rich galaxy cluster over the last 10\,Giga years of cosmic time 
($2\!\geq\!z\!\geq\!0$). Besides the detailed dark matter distribution in the left panels, these simulations also trace the intracluster medium gas density (center panels) and the ICM temperature (right panels), which are directly accessible through  X-ray observations. Note that the ICM gas density is a very good tracer of the underlying dark matter distribution, and that the substructure and  asymmetry increases drastically beyond \zga1. 

\subsection{Galaxy cluster statistics }

After discussing the principal formation process of individual objects of the large-scale structure, we will now turn to the statistical properties of the  overall galaxy cluster population. This section will focus on two essential properties of clusters which provide a sensitive link to cosmological models: (i) the distribution of cluster masses (mass function), and (ii) the spatial clustering properties (power spectrum).

\subsubsection{Cluster mass function}

The basic framework to quantitatively approximate the comoving number density of (relaxed) dark matter halos as a function of mass and redshift $n(M,z)$ is provided by the 
{\em Press-Schechter} formalism \cite{Press1974a}. The main idea of this procedure is the statistical evaluation of the   
Gaussian fluctuation field $\delta(\mathbf{x},t)$ and the identification of density peaks above the critical collapse value
after applying a spatial filter function corresponding to a minimum object mass $M_{\mathrm{min}}$.

\addtocounter{footnote}{1} 
\footnotetext{The simulation movie, from which the snap-shot images are taken, can be found at \url{http://www.mpa-garching.mpg.de/galform/data_vis/index.shtml}.}
\addtocounter{footnote}{-2} 

\begin{figure}[h]
\centering
\includegraphics[angle=0,clip,width=0.76\textwidth]{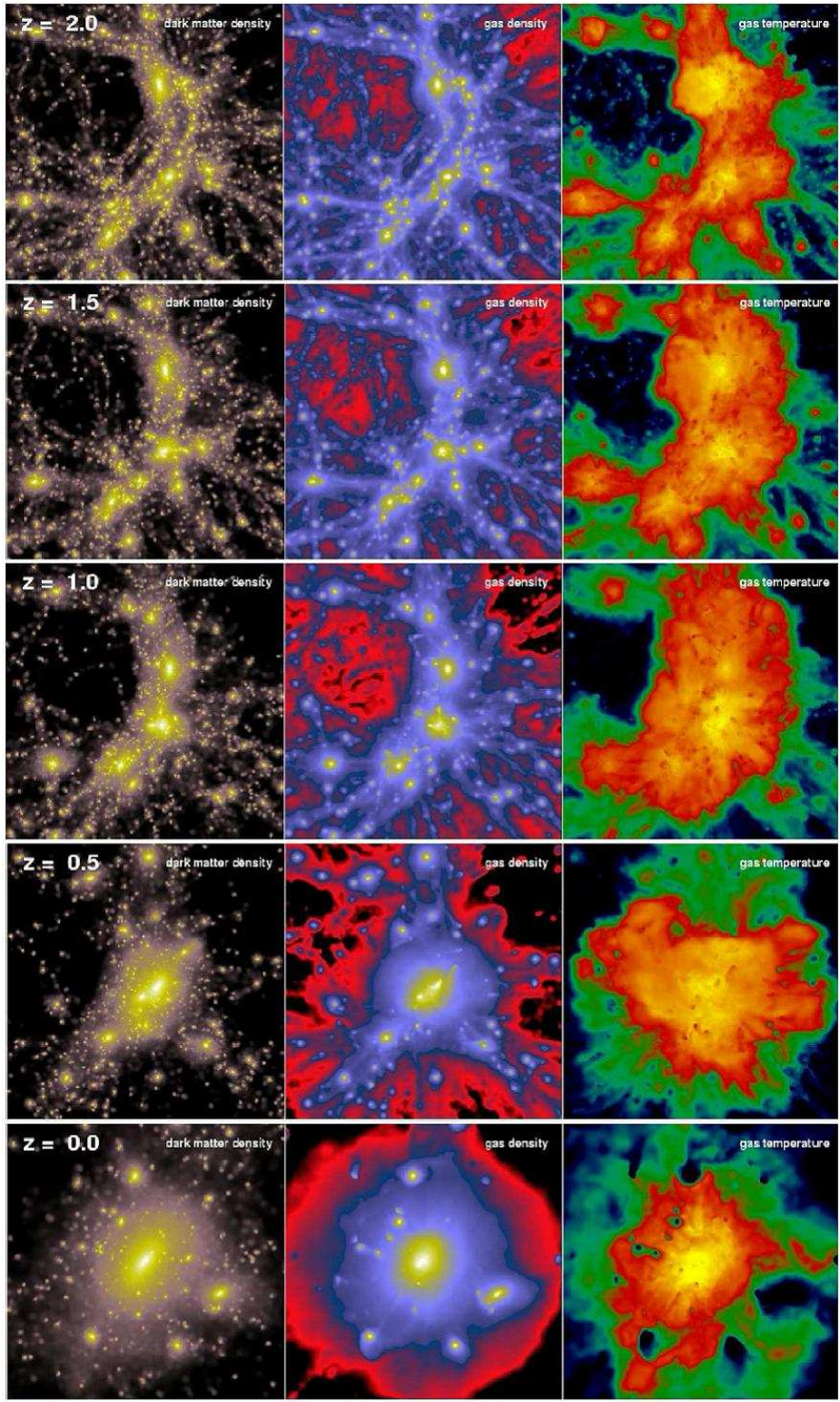}
\vspace{-1ex}
\caption[Cluster Formation Simulations]{High resolution simulation of the formation of a rich cluster of galaxies from the {\em Millennium Run}. From {\em top} to {\em bottom} snapshots at redshifts of 2.0, 1.5, 1.0, 0.5, and $z\!=\!0$ are shown. The {\em left panels} illustrate the dark matter distribution, the {\em center panels} the gas density, and the {\em right panels} the temperature structure at the corresponding redshift. Note the significant increase of substructure and asymmetry with redshift. The gas density and temperature are in principle accessible cluster observables by means of X-ray observations. Credits: V. Springel.\footnotemark}
\label{f3_ClusterFormation}       
\end{figure}

\clearpage


Since the filtering concept will be of central importance for this section, we will first have a closer look at the properties of smoothed density fields. Let us denote $W_{\mathrm{R}}(\mathbf{x})$ as a (3-D) spatial window function with smoothing length scale $R$. In the simplest case of a top-hat filter this will just be a sphere with radius $R$, which has a zero value outside 
 and a normalized constant value  $W_{\mathrm{R}}(\mathbf{|x|}\!\leq\!R)=(4 \pi R^3 / 3)^{-1}$ inside  the sphere.
A filtered density field will hence contain the average value within this sphere surrounding any given point $\mathbf{x}$. 
Another widely used window function is a Gaussian filter, but more complicated functional forms are conceivable. Mathematically spatial filtering is equivalent to a convolution of the density field $\delta(\mathbf{x})$ with the window function $W_{\mathrm{R}}$

\begin{equation}\label{e3_spatial_filter}
    \delta_{\mathrm{R}}(\mathbf{x}) = \delta_{\mathrm{M}}(\mathbf{x}) = \int \delta(\mathbf{y}) \, W_{\mathrm{R}}(\mathbf{|x-y|}) \, d\mathbf{y} \ ,
\end{equation}

\noindent
where $\delta_{\mathrm{R}}(\mathbf{x})$ denotes the filtered field. In the case of the spherical top-hat filter, the average mass $M$ contained inside the window function is 
$\langle M \rangle\!\simeq\!4 \pi R^3 \bar{\rho}_{\mathrm{m}}/ 3$, 
providing a one-to-one correspondence between the comoving smoothing length scale $R$ and the mass scale $M$.
Since a spatial convolution translates into a simple multiplication in $\mathbf{k}$-space, working with the  
Fourier representation of the density contrast is often advantageous 

\begin{equation}\label{e3_Fourier_contrast}
\tilde{\delta}({\mathbf k}) = \frac{1}{(2 \pi)^{3/2}} \int d{\mathbf x} \, \delta({\mathbf x}) \,  e^{i\mathbf{kx}} \ .
\end{equation}

\noindent
The filtered density contrast in Fourier space is then $\tilde{\delta}_{\mathrm{R}}({\mathbf k})\!\propto\!
 \tilde{\delta}({\mathbf k}) \cdot \tilde{W}_{\mathrm{R}}({\mathbf k})$, with $\tilde{W}_{\mathrm{R}}({\mathbf k})$ being the Fourier transform of the window function.
The averaged squared density fluctuations in Fourier space define the important {\em power spectrum} $P(k)$, or more accurately, the power spectral density function

\begin{equation}\label{e3_Power_Spec}
P(k) \equiv \langle |\tilde{\delta}({\mathbf k})|^2 \rangle = \frac{1}{2 \pi^2}  \int dr \, r^2 \, \xi(r) \, \frac{\sin(kr)}{kr} \ .
\end{equation}

\noindent
The right hand side shows that the power spectrum  $P(k)$ is in effect the Fourier transform of the  
2-point correlation function $\xi(r)$, which states the probability to find two objects at separation $r$ in excess to a random distribution. The power spectrum expresses the amount of cosmic structure as function of wavenumber $k$ corresponding to the length scale $L\,\simeq\,2\pi/k$.
As a second order statistic, the power spectrum provides a complete statistical  characterization   of the underlying Gaussian\footnote{Gaussianity is one of the underlying assumptions for the {\em Press-Schechter} theory.} fluctuation field, in analogy to the mean and standard deviation as parameters for the one dimensional Gaussian distribution.


Combining the discussions on spatial filters and Fourier representations, we can introduce the variance of the smoothed fluctuation field $\delta_{\mathrm{R}}(\mathbf{x})$ as 


\begin{equation}\label{e3_filtered_variance}
\sigma^2_{\mathrm{R}} = \sigma^2_{\mathrm{M}} = \langle \, \delta^2_{\mathrm{R}} \, \rangle =  \langle \, \tilde{\delta}^2_{\mathrm{R}} \, \rangle =   
 \frac{1}{2 \pi^2} \int dk \, k^2 \, P(k) \, \tilde{W}^2_{\mathrm{R}}(k) \ .
\end{equation}

\noindent
Here {\em Parseval's theorem}\footnote{The norm of a function is invariant under Fourier transformation, \ie \ $\int |f({\mathbf x})|^2 d{\mathbf x} = \int |\tilde{f}({\mathbf k})|^2 d{\mathbf k}$\,.} was applied and in the last step, Equ.\,\ref{e3_Power_Spec} and the isotropy assumption were used, implying that the final result can only depend on the norm of the wave vector ${\mathbf k}$ but not its direction. As discussed, the smoothing scale $R$ is directly related to the mass scale $M$ and thus to the mass variance $ \sigma^2_{\mathrm{M}}\!\equiv\!\langle |\delta M/M|^2  \rangle$, with the exact functional form depending on the shape of the used filter.

We can now return to the {\em Press-Schechter} theory and outline the procedure to obtain an approximate dark matter halo mass function  $n(M,z)$, which can be interpreted as galaxy cluster mass function. In addition to the spatial dependence of the fluctuation field, the time evolution, or equivalently the redshift dependence, has to be considered according to the linear theory evolution of Equs.\,\ref{e3_time_separation}\,\&\,\ref{e3_Growth_Factor}. As pointed out, the collapsed objects reach a virialized state once a critical linearly extrapolated density contrast $\delta_{\mathrm{c}}\!=\!\delta_{\mathrm{vir}}(\mathrm{linear})\!\simeq\!1.69$ is reached, with a weak dependance of the exact numeric value on the cosmological model and the redshift. For a linear extrapolation to the present time $t_0$ or $z\!=\!0$,  density contrast peaks with amplitude $\delta_{\mathrm{c}}$ can be identified with objects currently reaching the virial equilibrium state, \ie \  presently formed systems. On the other hand, objects that have already collapsed and formed at higher redshift $z_{\mathrm{form}}$ will have a higher density contrast today $\delta_{\mathrm{0,c}}\!=\!\delta_{{\mathrm{c}}}/D_{+}(z_{\mathrm{form}})$ (see  Equ\,\ref{e3_time_separation}). The linear growth factor $D_{+}(z)$ with normalization $D_{+}(z\!=\!0)\!\equiv\!1$ is hence the main factor for the redshift evolution of collapsed objects. For an Einstein-de\,Sitter Universe the last  expression reduces to $\delta_{\mathrm{0,c}}(\mathrm{EdS})\!=\!1.69\cdot (1+z_{\mathrm{form}})$, but generally both the critical overdensity $\delta_{\mathrm{c}}$ and the linear growth factor $D_{+}(z)$ are cosmology dependent.


In order to obtain the number of collapsed objects at redshift $z$ with mass $M$ the following procedure is applied: (i) 
Determine the 
smoothing length scale $R$ corresponding to the mass scale $M$. (ii) Filter the present fluctuation field with the window function $W_{\mathrm{R}}$ to arrive at  $\delta_{\mathrm{R}}(\mathbf{x}, t_0)$. (iii) Count the smoothed density peaks for amplitudes above $\delta_{\mathrm{c}}$  at redshift $z$, \ie \ for the minimum present amplitude $\delta_{\mathrm{min}}\!>\!\delta_{\mathrm{0,c}}\!=\!\delta_{{\mathrm{c}}}/D_{+}(z)$. These peaks are interpreted as collapsed objects of mass $\geq\!M$ at redshift $z$. 
The probability $p(\delta_{\mathrm{0,c}})$ to find amplitudes above the collapse threshold $\delta_{\mathrm{0,c}}$ follows from Gaussian statistics and is given by


\begin{equation}\label{e3_fill_factor}
    p(\delta_{{\mathrm{0,c}}}, R) = \int_{\delta_{{\mathrm{0,c}}}}^{\infty} p(\delta, R) \, d\delta = \frac{1}{2} \mathrm{erfc} \left(\frac{\delta_{{\mathrm{0,c}}}}{\sqrt{2} \cdot \sigma_{R} } \right) \ ,
\end{equation}

\noindent
where erfc$(x)$ is the error function\footnote{The error function is defined as $\mathrm{erfc}(x)\!=\!\frac{2}{\sqrt{x}} \int_x^{\infty} e^{-t^2} dt$.} and $\sigma_{R}\!=\!\sigma_{M}$ the smoothed fluctuation amplitude of Equ.\,\ref{e3_filtered_variance}. 
This probability can be interpreted as spatial filling factor or mass fraction of collapsed objects above the threshold mass.

The final {\em Press-Schechter} \ mass function $n(M,z)$ (\eg \ Borgani, 2006 \nocite{Borgani2006a}), \ie \ the redshift evolution of the dark matter halo number density, is derived from the above filling factor of collapsed objects at redshift $z$ in the  mass interval $[M,M\!+\!dM]$ as


\begin{equation}\label{e3_PS_massfct}
  \frac{dn_{\mathrm{PS}}(M,z)}{dM} = -\frac{\partial p}{\partial M} \, \frac{2}{V_M} =
  \sqrt{ \frac{2}{\pi}}  \cdot   \frac{\bar{\rho}_{\mathrm{m}}}{M^2}  \cdot  \frac{\delta_{{\mathrm{c}}}}{\sigma_{M}(z)}  \cdot     \left| \frac{d\log \sigma_M(z)}{d\log M } \right| \cdot  \exp\left({-\frac{\delta_{\mathrm{c}}^{2}}{2 \, \sigma_M ^2(z)}}\right) .
\end{equation}

\noindent
To go from a filling factor per unit volume to a cluster number, one has to divide by the occupied volume of the object $V_M\!\simeq\!M/\bar{\rho}_{\mathrm{m}}$. The additional factor of two corrects for a shortcoming of the {\em Press-Schechter} \ theory namely that the integral of Equ.\,\ref{e3_fill_factor} over all mass scales only accounts for half the total mass (the `cloud-in-cloud' problem). The mass dependence of the  probability $p(\delta_{\mathrm{0,c}})$ is implicitly contained in the mass-scale smoothed amplitude  $\sigma_{M}$ of the fluctuation field. This important factor also contains the dependence on cosmological parameters through the linear growth factor  $D_{+}(z)$ governing the $\sigma_{M}$ evolution with redshift.
This implies that the  high-mass end of the cluster mass function,  where the last term dominates,
is `exponentially sensitive' to variations of cosmological parameters.



The {\em Press-Schechter\/}  theory in the form presented here rests upon several strong assumptions, \eg \ a spherically symmetric collapse without substructure, and some {\em ad hoc} procedures such as the correct filter form or the cloud-in-cloud problem.
Meanwhile {\em extended Press-Schechter} formalisms provide more rigorous derivations of the mass function (\eg \ Bond \etal , 2001; \nocite{Bond1991a} Bower, 1991) \nocite{Bower1991a}. 
Sheth \& Tormen \cite*{Sheth2001a} have  
 generalized the mass function for the
more realistic assumption of an ellipsoidal collapse.
A  particularly well tested and widely used fitting formula for the mass range $10^{12}$--$10^{15} M_{\sun}$ has been given by Jenkins \etal \ \cite*{Jenkins2001a} based on large numerical simulations. A comparison to the more precise {\em Jenkins} mass function 


\begin{equation}\label{e3_Jenkins_MassFct}
  \frac{d n_{\mathrm{J}}(M,z)}{dM} = 0.315 \cdot \frac{\bar{\rho}_{\mathrm{m}}}{M^2} \cdot  \frac{d\ln \sigma_M^{-1}}{d\ln M} \cdot \exp\left( -| \ln \sigma_M ^{-1} + 0.61 |^{3.8}\right) 
\end{equation}

\noindent
reveals that the {\em Press-Schechter}  mass function underestimates the number density of massive clusters and overestimates the abundance of low-mass objects.







\begin{figure}[t]
\centering
\includegraphics[angle=0,clip,width=0.8\textwidth]{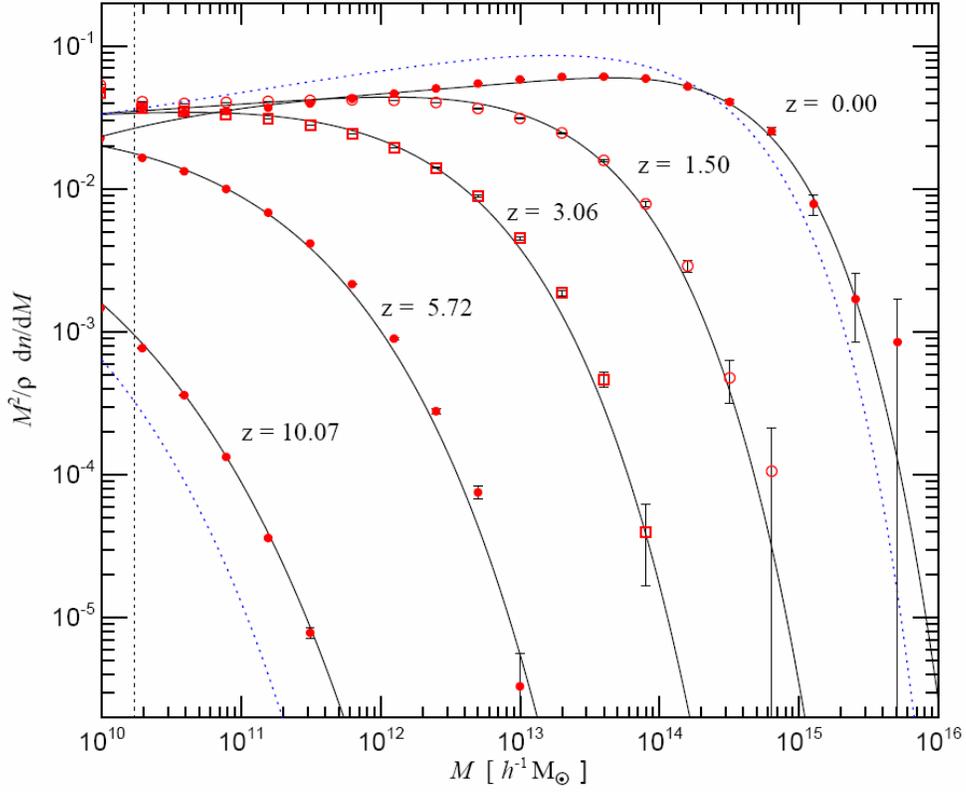}

\caption[Halo Number Density]{Differential dark matter halo number density as a function of mass and epoch as obtained with the {\it Millennium Run} simulation. Shown is the halo multiplicity function ($M^2/\bar{\rho}_{\mathrm{m}}) \, dn/dM$ against the halo mass for different redshifts. Of particular interest are the observationally accessible curves at redshift $z\!=\!0$ and $z\!=\!1.5$. Solid lines show predictions using the {\em Jenkins}  fitting formula (Equ.\,\ref{e3_Jenkins_MassFct}), dotted lines show {\em Press-Schechter}  model predictions of Equ.\,\ref{e3_PS_massfct}. Plot from Springel \etal \ \cite*{Springel2005a}.}
\label{f3_Halo_Number_Density}       
\end{figure}

Figure\,\ref{f3_Halo_Number_Density} illustrates the results of the {\it Millennium Run} simulation 
and the very good agreement with the predictions of the {\em Jenkins} mass function. Shown is dark matter halo multiplicity function ($M^2/\bar{\rho}_{\mathrm{m}}) \, dn/dM = N\!\cdot\!(M/dM)\!\cdot\!(V_M/V)$, \ie \ the 
mass fraction carried by objects in a unit mass bin. Note that the density from $z\!=\!0$ to $z\!=\!1.5$ drops by three orders of magnitude for halo masses of $\sim\!10^{15}\,h^{-1}M_{\sun}$ and by one order of magnitude for masses of $\sim\!10^{14}\,h^{-1}M_{\sun}$.





\subsubsection{Cluster power spectrum}
\label{s3_cluster_PS}



\footnotetext{\url{http://www.mpa-garching.mpg.de/galform/virgo/millenium/index.shtml}}
\addtocounter{footnote}{-2}

\begin{figure}[t]
\centering
\includegraphics[angle=0,clip,width=0.9\textwidth]{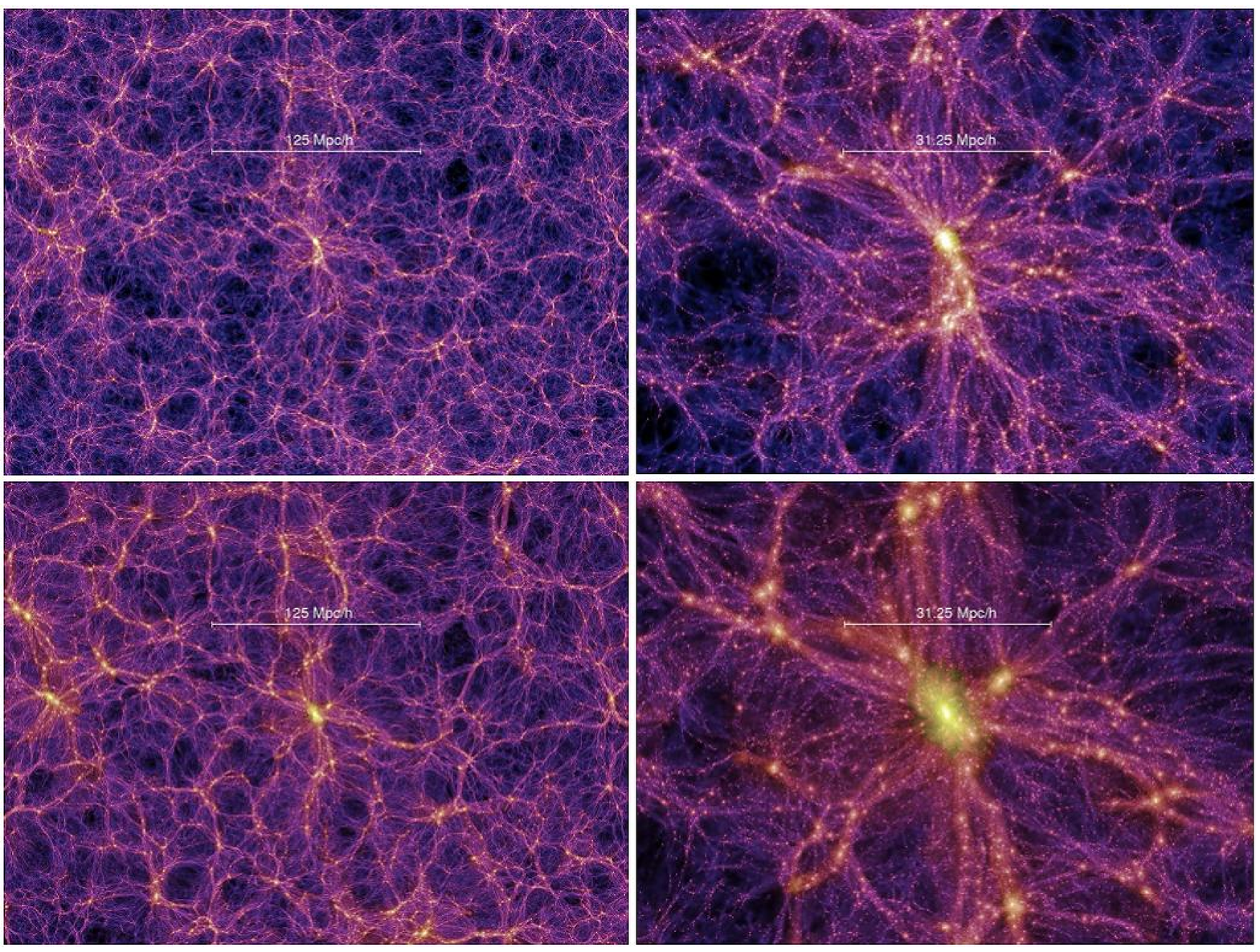}
\caption[LSS Simulations]{The cosmic large-scale structure as seen with the {\it Millennium Run} simulation. The {\em top panels} show the dark matter distribution at redshift $z\!=\!1.4$  for a comoving image scale of approximately 500\,Mpc ({\em left}) and 125\,Mpc ({\em right}) on a side. The same regions at redshift $z\!=\!0$ are illustrated in the {\em bottom panels}. Note that the cosmic web is basically in place at $z\!>\!1$, but the nodes, \ie \ the clusters, still actively accrete large amounts of matter until the present. Credits: V. Springel\footnotemark.  }
\label{f3_LSS_MilleniumRun}       
\end{figure}


As a last application of the standard structure formation scenario, we will take a closer look at the spatial distribution of galaxy clusters, \ie \ the clustering properties of clusters. Figure\,\ref{f3_LSS_MilleniumRun} shows the simulated cosmic web of the large-scale structure on a 500\,Mpc scale (left) and zoomed in by a factor of four (right). The top panels show the simulated dark matter distribution at redshift $z\!=\!1.4$ and the bottom panels the present day structure. The main aim of this section is to introduce the basic tools for a quantitative statistical description of the large-scale structure and how the observed spatial galaxy cluster distribution is connected to cosmological parameters. The tools and tests of this subsection will not have direct applications for the current XDCP survey but will be of prime importance for the next generation of galaxy cluster surveys (see Chap.\,\ref{c11_Outlook}), for which XDCP assumes a pathfinder role for high-redshift clusters.





The power spectrum $P(k)$ is the main tool for the statistical description of the large-scale structure and has been introduced in Equ.\,\ref{e3_Power_Spec}. Whereas the overall shape and features of $P(k)$ can be theoretically derived, the amplitude or normalization of the power spectrum has to be empirically determined. This can be achieved with the help of $\sigma_8$, a very important  but probably the least intuitive cosmological parameter.   

For a given filtering length scale $R$, Equ.\,\ref{e3_filtered_variance} provides the link between the mass variance $\sigma^2_R$ on this scale and the power spectrum $P(k)$, which fixes the {\em a priori} unknown absolute normalization of $P(k)$.
The chosen filter scale  $R$ should on one hand be sufficiently large to ensure the validity of linear theory and on the other hand small enough to allow  observations of a large amount of structure\footnote{The homogeneity of the Universe on very large scales decreases the mass variance with increasing filtering length $R$. For large $R$ and the use of a spherical top-hat filter one obtains $\sigma_M \simeq \delta M/M \propto M^{-2/3}$.}. 
A sensible choice for fixing the filtering scale is $R \equiv 8\,h^{-1}$Mpc, which was motivated by  results of early galaxy surveys finding $ \delta_{\mathrm{gal}}(R\!=\!8\,h^{-1}\mathrm{Mpc}) \simeq \delta N_{\mathrm{gal}}/N_{\mathrm{gal}} \simeq \delta M/M\!\simeq\!1$, \ie \ the variance  of galaxy counts in spatial bins of radius 8\,$h^{-1}$Mpc is similar to the mean value. 
The mass scale of this filtering length is $M\!\simeq\!5\!\times\!10^{14} M_{\sun}$ and corresponds to the collapsed region of a massive galaxy cluster. The mass variance of order unity on the $8\,h^{-1}\mathrm{Mpc}$ scale also marks the transition region  from the linear regime with $\delta\!\la\!1$, \ie \ $L\!\ga\!8\,h^{-1}$Mpc, to the non-linear regime at $L\!\ll\!8\,h^{-1}$Mpc. 




In short, $\sigma_8$ is the average fluctuation amplitude\footnote{Meaning the root-mean-square (rms) or standard deviation of the Gaussian field.} of the linearly evolved (Equs.\,\ref{e3_time_separation}\,\&\,\ref{e3_Growth_Factor}), present day dark matter density contrast field $\delta^{\mathrm{DM,lin}}_{\mathrm{8}}(\mathbf{x}, t_0)$ smoothed with  a top-hat filter of comoving radius  $R\!=\!8\,h^{-1}$Mpc.
The formal definition and its connection to the power spectrum is then given by 


\begin{equation}\label{e3_sigma_eight}
\sigma^2_{8}  = \langle \delta^2_{8\,h^{-1}\mathrm{Mpc}} \rangle_0 = 4 \pi \int \frac{dk \, k^2}{(2\pi)^3} \, P^{\mathrm{DM}}_0(k) \, \tilde{W}_{8\,h^{-1}\mathrm{Mpc}}^2(k) \ .
\end{equation}

\noindent
The single dimensionless number $\sigma_8$ hence determines the height of density peaks and consequently the object abundance. The discussed galaxy cluster mass function (Equs.\,\ref{e3_PS_massfct}\,\&\,\ref{e3_Jenkins_MassFct}) provides a direct measure of $\sigma_8$. (Nevertheless, $\sigma_{8}$ is probably the least securely determined prime cosmological parameter.)



The second important parameter related to the large-scale structure in the Universe is the power law index $n_{\mathrm{S}}$ of the  primordial (\ie \ initial) power spectrum $P_{\mathrm{i}}(k)\!\propto\!k^{n_{\mathrm{S}}}$.
This index is observationally confirmed \cite{Spergel2007a} to be very close to the `scale-free' {\em Harrison-Zeldovich spectrum}
$P^{\mathrm{HZ}}_{\mathrm{i}}(k)\!\propto\!k^{1}$ (\ie \ $n_{\mathrm{S}}\!=\!1$), which is also  predicted by inflationary theories. 





From the known evolution of the density fluctuation field $\delta (t)\!\propto\!D_+(t)$ (Equ.\,\ref{e3_time_separation}) and the definition of the power spectrum (Equ.\,\ref{e3_Power_Spec}), the time evolution of $P(k,t)$ starting with the initial fluctuation spectrum seeded after inflation is then given by

\begin{equation}\label{e3_linear_PS_growth}
P(k,t) = D_{+}^2(t) \cdot A \, k^1 \cdot T^2(k,t) \ .
\end{equation}


\noindent
The {\em a priori} unknown amplitude $A$ of the  power spectrum (today) is directly linked to $\sigma_8$ as an alternative normalization method (cluster normalization). $T(k)$ is the so-called {\em transfer function} which is an extra term  containing all evolutionary effects that alter the original linear form of the primordial seed power spectrum. In particular, the {\em transfer function}
incorporates important scale-imprinting effects on the initially scale-invariant power spectrum.

The first imprinted physical scale with prime influence on the present day overall shape of $P(k)$ is the (particle) horizon at the epoch of matter-radiation equality  $D^{\mathrm{H}}_{\mathrm{equ}}$ at redshift  $z_{\mathrm{equ}}$.
This characteristic length scale represents the distance over which causal influence can propagate during the radiation dominated era of the Universe and is given by

\begin{equation}\label{e3_matterrad_equality}
D^{\mathrm{H}}_{\mathrm{equ}} = \frac{c}{\sqrt{2} \, H_0} \, \frac{1}{\sqrt{(1+z_{\mathrm{eq}}) \, \Omega_{\mathrm{m}}}}
\simeq 16\,(\Omega_{\mathrm{m}} h^2)^{-1}\,\mathrm{Mpc} \sim 100\,\mathrm{Mpc} \ .
\end{equation}

\noindent
During the radiation dominated early epoch with a coupled {\em photon-baryon fluid\/}, the pressure term in Equ.\,\ref{e3_StructureGrowth} becomes important and causes an oscillatory behavior of the fluctuations {\em inside\/} the horizon $D^{\mathrm{H}}_{\mathrm{equ}}$. Fluctuation growth on scales $L\!\la\!D^{\mathrm{H}}_{\mathrm{equ}}$ is therefore effectively suppressed, and the structure growth can only proceed on super-horizon scales  $L\!>\!D^{\mathrm{H}}_{\mathrm{equ}}$ beyond the causal influence of the radiation pressure. 



For the following arguments, we only consider the (cold) dark matter component as the dominant contribution to the matter density.
The decisive criterion for the growth of perturbation modes  with wavenumber $k\!=\!2 \pi/L$ during the rapid expansion phase of the radiation-dominated early epoch  is determined by  whether or not the length scale of the mode is contained within the expanding horizon $D^{\mathrm{H}}(z)$ at redshift $z$. Modes that are outside this region of causal influence  $k_{\mathrm{grow}}(z)\!<\!2 \pi/D^{\mathrm{H}}(z)$ do not `feel' the 
radiation pressure 
 implying that the modes grow according to the linear expectations of Equ.\,\ref{e3_linear_PS_growth} with $T(k)\!=\!1$. Once the expanding causality horizon encompasses perturbation modes $k_{\mathrm{suppr}}(z)\!>\!2 \pi/D^{\mathrm{H}}(z)$, the growth suppression sets in and  the perturbation amplitudes essentially remain constant.    

For the growth evolution in the radiation dominated phase starting after inflation and ending with the epoch of matter radiation equality at $z_\mathrm{equ}\!\sim\!3500$, we obtain the following scenario: (i) Initially the primordial {\em Harrison-Zeldovich spectrum} evolves homogeneously as the modes on all length scales grow proportional to $D_+^2(t)$, \ie \ only the amplitude of the linear power spectrum increases with time. (ii) Once the length scale of a perturbation mode enters the  particle horizon at $z_{\mathrm{enter}}$, \ie \ $k_{\mathrm{enter}}(z_{\mathrm{enter}})\!=\!2 \pi/D^{\mathrm{H}}(z_{\mathrm{enter}})$, further growth is suppressed and the amplitude of the mode halts at the value of horizon entering. (iii) This process proceeds from the smallest scales, \ie \ large wavenumbers $k$, to larger scales as the horizon expands with time, which implies that the structure growth suppression first influences small regions and over time gradually halts the growth for modes of increasing  scales. Upon horizon entering, the spectral power density of the corresponding perturbation mode $P(k_{\mathrm{enter}})$ will be held constant, while the modes of smaller wavenumbers continue to grow. The overall shape of the power spectrum $P(k,t)$ will hence bend over towards lower amplitudes for modes with $k\!\ga\!k_{\mathrm{enter}}(z)$.
(iv) After matter-radiation equality $z_{\mathrm{eq}}$, the matter density dominates the  dynamics of the Universe. 
The {\em photon-baryon fluid\/} decouples, resulting in a vanishing pressure for the baryons and
 an associated resumed structure growth on all scales. The largest length scale which experiences a growth suppression is thus the introduced horizon scale at equality $D^{\mathrm{H}}_{\mathrm{eq}}$, which defines the break point $k_{\mathrm{eq}}$ of the present power spectrum. 

This horizon-crossing growth suppression mechanism in the early Universe completely fixes the overall shape of the present day cold dark matter power spectrum $P_{\mathrm{CDM}}(k,t_0)$ imprinting the characteristic length scale  $D^{\mathrm{H}}_{\mathrm{eq}}$ as a break point. For small wavenumbers  $k\!\ll\! 2 \pi /D^{\mathrm{H}}_{\mathrm{eq}}$ the form of the primordial power spectrum is conserved with $P_{\mathrm{CDM}}(k\!\ll\!k_{\mathrm{eq}},t_0)\!=\!A\,k$, \ie \ $T(k)\!\approx\!1$. 
Large wavenumbers with  $k\!\gg\!k_{\mathrm{eq}}$ have experienced an extended early period of growth suppression resulting in a transfer function  $T(k)\!\approx\!(D^{\mathrm{H}}_{\mathrm{eq}}\,k)^{-2}$ and consequently a power spectrum of the form  
$P_{\mathrm{CDM}}(k\!\gg\!k_{\mathrm{eq}},t_0)\!\propto\,k^{-3}$. The break point $k_{\mathrm{eq}}\!=\! 2 \pi /D^{\mathrm{H}}_{\mathrm{eq}}\!\propto\!(\Omega_{\mathrm{m}} h^2)\,\mathrm{Mpc}^{-1}$, \ie \ the maximum of the power spectrum, is sensitive to the matter density $\Omega_{\mathrm{m}}$ and moves towards higher wavenumbers with increasing matter content,  reflecting the earlier epoch of matter-radiation equality.

In principle, this result compresses the statistical information present in the simulated large-scale dark matter distribution shown in Fig.\,\ref{f3_LSS_MilleniumRun}. The locally observed galaxy cluster power spectrum is shown in the next Chapter in Fig.\,\ref{f4_REFLEX_PS} (upper curve in right panel). The broad maximum of the power spectrum is visible but will only be well constrained with the extended sample of local X-ray cluster surveys currently in preparation.
Note that the well-fit $P(k)\!\propto\,k^{-3}$ behavior at large wavenumbers confirms the cold dark matter paradigm. Hot (HDM) or warm (WDM) forms of dark matter would exhibit a different power spectrum shape, as small scale perturbations are additionally damped by diffusion processes of the fast particles (`free streaming').

The final ingredient to fully understand the observed cluster power spectrum of Fig.\,\ref{f4_REFLEX_PS} and its vertical offset to the galaxy power spectrum is the concept of {\em biasing} \cite{Kaiser1984a}. So far we have mostly related the discussion to the cold dark matter power spectrum $P_{\mathrm{CDM}}(k)$, which is not directly accessible to observations. 
Different classes of observable objects could have systematic offsets of their density field with respect to the dark matter density contrast $\delta_{\mathrm{object}}\!=\!b\cdot \delta_{\mathrm{CDM}}$. The pre-factor $b$ is the {\em bias} of the object class which can be determined from first principles. The origin of galaxy clusters from the highest peaks in the initial density contrast field results in a large bias factor of $b\!\simeq\!2.6$, \ie \ clusters have an amplified\footnote{The biasing factor depends on the typical cluster mass in a survey. The given numeric values are taken from  the REFLEX survey (see Sect.\,\ref{s4_LocalSurveys}).} power spectrum $P_{\mathrm{cluster}}(k)\!=\!b^2\cdot P_{\mathrm{CDM}}(k)\!\simeq\!6.8\!\cdot\!P_{\mathrm{CDM}}(k)\!\sim\!6.8\!\cdot\! P_{\mathrm{galaxy}}(k) $.  
This greatly enhanced clustering strength can be understood in terms of the underlying long-wavelength fluctuation modes of the dark matter density field. The probability for a cluster-scale density perturbation to reach the critical collapse threshold value $\delta_{\mathrm{c}}$ is highly increased if this region is found on top of a larger perturbation mode with small wavenumber. On the other hand, the probability is suppressed if the cluster-scale perturbation coincides with a minimum of the  underlying long-wavelength mode. The {\em biasing} of different object classes is hence expected to increase with object mass. Clusters in this respect are amplified probes and can thus yield valuable results with a lower number of objects compared to  galaxies. 





The second important imprinted physical scale in the transfer function $T(k)$ is the sound horizon at matter-radiation decoupling.
We are now considering the baryon component of the matter density which implies that the expected second  order  baryonic effects are about an order of magnitude smaller than those due to the dark matter.
Nevertheless, the theory is well understood (\eg \ Hu\,\&\,Sugiyama, 1996; Eisenstein\,\&\,Hu, 1998),  
\nocite{Hu1996a} 
\nocite{Eisenstein1998a} 
and observations of the so-called {\em baryonic acoustic oscillations\/} (BAOs) bear great promise to yield valuable results 
 for the {\em Dark Energy\/} studies over the next two decades.


The inclusion of baryons during the strongly coupled baryon-photon fluid  phase  before recombination at redshift $z_{\mathrm{rec}}$ requires the consideration of the last oscillation term in Equ.\,\ref{e3_StructureGrowth}. 
For small scales inside the sound horizon (SH) $D^{\mathrm{SH}}(z)$  the relativistic photon-baryon system behaves like a damped, forced oscillator where the self-gravitation of matter overdensities acts as driving force and the photon induced pressure as counter action.
In this relativistic fluid sound waves, \ie \ pressure and density perturbations, can propagate at the 
 sound speed of $c_{\mathrm{s}}\!=\!\sqrt{\partial p / \partial \rho}\!\simeq\!c/\sqrt{3}$.
At the epoch of hydrogen recombination at $z_{\mathrm{rec}}\!\approx\!1\,100$, photons and baryons decouple, which implies that 
 pressure waves cease to expand and are frozen into the (baryonic) matter distribution.
The longest {\em baryonic acoustic oscillation\/} wavelength, \ie \ the first harmonic, is hence the comoving distance a sound wave could have propagated before recombination and is given by the sound horizon 
 at  $z_{\mathrm{rec}}$  \cite{Hu2004a}

\begin{equation}\label{e3_acoustic_horizon}
D^{\mathrm{SH}}_{\mathrm{rec}} \simeq 147 \cdot (\Omega_\mathrm{m} h^2 / 0.13)^{-0.25} (\Omega_\mathrm{b} h^2 / 0.023)^{-0.08} \, \mathrm{Mpc} = 147.8 \pm 2.6\,\mathrm{Mpc} \ .
\end{equation}


\noindent
The numeric value was measured with WMAP \cite{Spergel2007a} with about 2\% accuracy and is expected to be determined on the 
percent level within the next few years. The first harmonic acoustic wave hence imprints an overdensity in the matter distribution which will be detectable as a peak, implying enhanced structure, in the 2-point correlation function $\xi(r)$ at  scale $D^{\mathrm{SH}}_{\mathrm{rec}}$ or as a modulation peak 
in the power spectrum at the corresponding wavenumber. Higher order harmonics give rise to additional acoustic peaks at shorter wavelengths resulting in a characteristic {\em baryonic acoustic oscillation\/} pattern in the power spectrum.  








The imprinted sound horizon of $D^{\mathrm{SH}}_{\mathrm{rec}}\!\simeq\!148$\,Mpc acts as a high-precision comoving `standard ruler'. In principle, any object class can be used to probe the {\em baryonic acoustic oscillations\/} and to determine their locations in the power spectrum as a function of redshift. However, as second order effect, BAO measurements are extremely difficult and require high-precision measurements of $P(k)$. It is also expected that the oscillations are damped or washed out at low redshift due to the onset of non-linear effects. 

The detection of the first acoustic peak in the correlation function of luminous red SDSS galaxies at $z\!\sim\!0.35$ has been  reported \cite{Eisenstein2005a}. The measurement of {\em baryonic acoustic oscillations\/} in power spectra of galaxy clusters provides a potentially powerful cosmological tool for the future (\eg \ Angulo \etal, 2005) \nocite{Angulo2005a} but requires cluster samples of the order of ten thousand.




\section{Cosmological Tests with Galaxy Clusters }
\label{s3_cosmo_tests}


\noindent
In the last two sections some important aspects of the close interplay between the observable properties of galaxy clusters and the underlying global geometry and structure formation process in the Universe were introduced.
This section will shortly summarize a number of cosmological tests that use galaxy clusters as probes and have been  successfully applied in the past or will become feasible within the next decade. 

For the dawning era of precision cosmology it will be crucial to measure cosmological parameters with tests sensitive to the global geometry of the {\em homogeneous Universe} {\bf and} methods probing the structure evolution of the {\em inhomogeneous Universe}.
The very different physical nature of the methods and their underlying theoretical frameworks will allow important consistency cross-checks of the emerging cosmological concordance  model.   
Galaxy clusters in this respect enable several independent cosmological tests for both fundamental aspects 
which qualifies the cluster population as one of the most versatile cosmological probes.





The majority of cosmological tests can be reduced to the measurement of effectively four cosmic evolution functions for which the dependence on the underlying cosmological parameters have been discussed: (i) the cosmic expansion history $H(z)$ (Equ.\,\ref{e3_expansion_history_z}), (ii) cosmic distances $D(z)$ (Equs.\,\ref{e3_comoving_distance},\,\ref{e3_lum_dist},\,\ref{e3_ang_dist}), (iii) cosmic age $t(z)$  (Equ.\,\ref{e3_lookback_time}), and (iv) the linear growth factor  $D_+(z)$ (Equ.\,\ref{e3_Growth_Factor}).
The first three methods are linked to the global geometry and probe the {\em homogeneous Universe}, the last one traces the structure  growth of the {\em inhomogeneous Universe}.






\begin{description}
    \item[(1) Cluster Mass Function:] 
The cluster abundance of the local Universe mainly depends on the matter density $\Omega_\mathrm{m}$ and the amplitude of the density fluctuation field $\sigma_8$. The mass function is related to the statistics of rare high amplitude peaks in the primordial matter distribution.  
 For a direct comparison to the observed cluster population, the predicted mass function $n(M)$ is 
converted to the measured X-ray luminosity function (XLF) via
$ \phi (L_{\mathrm{X}}) = n[M(L_{\mathrm{X}})] \cdot dM/dL_{\mathrm{X}}$\,.
The X-ray luminosity function at $z\!\sim\!0$ has been well established \cite{HxB2002a} and serves as local reference
for evolution measurements.
The local XLF is degenerate concerning the parameter combination $\sigma_8\!\cdot\!\Omega_{\mathrm{m}}^{\sim 0.5}$, \ie \ the effect of a larger fluctuation amplitude can be compensated by a lower matter density. The degeneracy can be broken in combination with the power spectrum or by measuring the redshift evolution in the next test.

   




\item[(2) Number Density Evolution:]
The redshift evolution of the cluster abundance out to $z\!\sim\!1.5$ will be  
the main cosmological test of the final XDCP sample.
The evolution of the cluster mass function $n(M,z)$ combines two tests simultaneously: 
(i) the comoving volume elements $dV_{\mathrm{com}}\!\propto\!D^2/H(z)$, governed by the expansion history, and
(ii) the growth of the mass fluctuation spectrum driven by  $D_+(z)$.
The cosmological test also becomes increasingly sensitive to variations of the {\em Dark Energy\/} parameters $\Omega_\mathrm{DE}$ and $w$
at redshifts above $z\!\ga\!0.8$. 
Up to now, the changing demographics of the entire cluster population has been investigated out to intermediate redshifts of $z\!\sim\!0.8$ by Mullis \etal \ \cite*{Mullis2004a} and Borgani \etal \cite*{Borgani2001a} (see Fig.\,\ref{f4_XLF}), but only weak evolutionary effects have been found. The redshift at which the cluster number density decreases significantly is yet to be determined and will be one of the goals for XDCP.

    \item[(3) Amplitude and Shape of Power Spectrum:] 
The clustering properties and the general shape of the cluster power spectrum depend on the matter density $\Omega_{\mathrm{m}}$
and the amplitude of the dark matter fluctuation spectrum $\sigma_8$ as discussed in Sect.\,\ref{s3_cluster_PS}.
Reliable measurements require large area contiguous surveys such as the ROSAT All-Sky Survey. The currently best measured cluster power spectrum has been obtained with the REFLEX survey (see Fig.\,\ref{f4_REFLEX_PS}) which will soon be improved upon with the extended NORAS\,2 and REFLEX\,2 samples (see Sect.\,\ref{s4_LocalSurveys}). Future X-ray surveys such as eROSITA (see Sect.\,\ref{s11_eROSITA}) will be able to measure the power spectrum as a function of redshift and will thus directly probe the evolution of the growth function $D_+(z)$.


    
 

    \item[(4) Baryonic Acoustic Oscillations:] 
Measurements of the locations of the acoustic peaks in the cluster power spectrum as a function of redshift will be one of the main aims and challenges of the eROSITA mission (Sect.\,\ref{s11_eROSITA}) and will require a total of 50\,000--100\,000 clusters with approximate redshifts. This cosmological test, however, is one of the cleanest methods to probe {\em Dark Energy\/} since it is least affected by systematic uncertainties. It is based on the well-understood `standard ruler' imprinted by the sound horizon at the epoch of last scattering $D^{\mathrm{SH}}_{\mathrm{rec}}$. Measurements of the transverse `baryonic wiggles' probe the angular size distance $d_{\mathrm{ang}}(z)$ (Equ.\,\ref{e3_ang_dist}) while measurements in the radial direction trace the expansion history $H(z)$ through Equ.\,\ref{e3_comoving_distance}.




    \item[(5) Cluster Baryon Mass Fraction:] 
This conceptually simple and attractive test rests upon the assumption that clusters, as objects that have collapsed from a region of ~10\,$h^{-1}$\,Mpc, represent a fair accounting of the relative matter content in the Universe, \ie \ that their baryon mass fraction $f_{\mathrm{b}}$ is representative. The cluster gas mass fraction 
$f_{\mathrm{gas}}$ can be well measured with X-ray observations. With corrections for the galaxy mass in clusters, one obtains the universal total cluster baryon fraction  which is  asymptotically approaching values of $f_{\mathrm{b}}\!\sim\!0.12$--0.14\,\h70$^{3/2}$ towards the cluster outskirts. If the  baryon content $\Omega_{\mathrm{b}}$ of the Universe is taken as prior from  cosmic nucleosynthesis calculations, the total matter density can be determined by $\Omega_{\mathrm{m}}\!=\!\Omega_{\mathrm{b}}/f_{\mathrm{b}}$ (\eg \ Allen \etal, 2003). \nocite{Allen2003a}
    


    \item[(6) Evolving Gas Mass Fraction:] 
    This method is a generalization of test\,(5) and uses the cluster gas mass fraction $f_{\mathrm{gas}}$ as `standard ruler' 
to probe the global geometry of the Universe \cite{Sasaki1996a}. The additional assumption is that the gas mass fraction is universal at all redshifts $f_{\mathrm{gas}}(z)\!\approx\!\mathrm{const}$. Since measurements of the gas mass ($M_{\mathrm{gas}}\!\propto\!h^{-5/2}\,d_{\mathrm{ang}}^{5/2}$)
and total mass ($M_{\mathrm{tot}}\!\propto\!h^{-1}\,d_{\mathrm{ang}}$) obey different scaling relations with distance, the resulting observed gas mass fraction scales with $f_{\mathrm{gas}}(z)\!\propto\!d_{\mathrm{ang}}^{3/2}(z)$. The measured gas mass fraction hence depends on the assumed angular diameter distance to the cluster, and will deviate from the (assumed) universal value $\bar{f}_{\mathrm{gas}}(z\!=\!0)$, if the  reference cosmology is incorrect. This geometric test can constrain the parameters $\Omega_{\mathrm{m}}$, $\Omega_{\mathrm{DE}}$, and $w$ and has been applied by Allen \etal \ \cite*{Allen2004a} based on detailed X-ray observations of 26 dynamically relaxed clusters at $z\!\la\!0.9$.

    \item[(7) Absolute Distance Measurements:] 
When X-ray and SZE observations are combined, the absolute distance to a cluster can be determined,  
 allowing the measurement of the Hubble constant $H_0$.
The SZ effect (see Sect.\,\ref{s2_SZE}) yields the observed temperature decrement
 $\Delta T\!\propto\!\int n_{\mathrm{e}}\,T_{\mathrm{e}}\,dl\!\propto\!n_0\,T_0\,R_{\mathrm{cl}}$\,, and the observation of the thermal X-ray emission (see Sect.\,\ref{s2_emission_mechan}) results in the flux  $S_{\mathrm{X}}\!\propto\!\int n_{\mathrm{e}}^2\, \Lambda(T_{\mathrm{e}})\,dl\!\propto\!n_0^2\,R_{\mathrm{cl}}$\,, where $R_{\mathrm{cl}}$ is the clusters radius, $n_0$  the gas density, and  $T_0$ the gas temperature. 
Elimination of $n_0$ results in an expression for the physical cluster radius $R_{\mathrm{cl}}$, which can be related to the observed angular diameter $\theta_{\mathrm{cl}}$ to yield the angular diameter distance via 

\begin{equation}\label{e3_cluster_H0}
d_{\mathrm{ang}}(z) = \frac{R_{\mathrm{cl}}}{\theta_{\mathrm{cl}}} =  \frac{\Delta T^2}{S_{\mathrm{X}}} \frac{1}{T_0^2 \, \theta_{\mathrm{cl}}} \cdot {\mathrm{const}} \approx \frac{c\,z}{H_0}\,.
\end{equation}

\noindent 
The last step is a low redshift approximation for $H_0$, which can be generalized for more distant clusters to $H(z)$.
Achievable accuracies for $H_0$ based on individual cluster measurements are about  $\pm 20$\,km\,s$^{-1}$Mpc$^{-1}$.





    \item[(8) Cluster Number Counts:]
Cumulative cluster number counts provide
a model independent test since only X-ray flux measurements are needed.
Generalizing Equ.\,\ref{e3_number_counts} for the cluster population yields 
the number of clusters per steradian  brighter than flux $S$ (\eg \ Borgani, 2006) \nocite{Borgani2006a} as  

\begin{equation}\label{e3_cluster_NumberCounts}
    N(>\!S) = \int_0^{\infty} dz \frac{c D^2(z)}{H(z)} \cdot \int_S^{\infty} dS \ n[M(S,z);z] \, \frac{dM}{dS} \ .
\end{equation}

\noindent
Here the first integral accounts for the comoving volume per steradian (Equ.\,\ref{e3_volume_elements}) and the second term yields the observable clusters in this volume at flux $S$.
Deviation of the Euclidean slope of  Equ.\,\ref{e3_number_counts} are hence related to (i) the flux limit, (ii) evolution effects of the cluster population, and (iii) cosmological distances. 
A compilation of different $\log\,N$--$\log\,S$ functions is shown in Fig.\,\ref{f4_ROSAT_logSlogN}. The observed slope on the bright end starts with the expected Euclidian value 1.5 and then flattens to a value of $\simeq\,1$ towards faint fluxes \cite{Rosati2002a}. Cluster number counts  are nowadays mainly used as a consistency check with other surveys
and for an estimation of the survey flux limit.







    \item[(9) BCG Hubble Diagram:]
The brightest cluster galaxies exhibit a remarkably low scatter of about $\pm\!0.3$\,mag in their average absolute magnitude $\bar{M}_{\mathrm{BCG}}$, which qualifies them as decent `standard candles', similar to supernovae type\,Ia.
With the observed apparent BCG magnitudes $m_{\mathrm{BCG}}$ of a given filter, we arrive at the $m_{\mathrm{BCG}}$--$\log (z)$  Hubble diagram by combining Equ.\,\ref{e3_k_correction}\,\&\,\ref{e3_dist_modulus} 

\begin{eqnarray}\label{e3_hubble_diagram} \nonumber
    m_{\mathrm{BCG}}(z) =  M_{\mathrm{BCG}} + K_{\mathrm{BCG}}(z) + 25 + 5\log \left(d_{\mathrm{lum}}\, [\mathrm{Mpc}]\right)  - 5\log h_{70} \\ \nonumber
= M_{\mathrm{BCG}}\!+\!K_{\mathrm{BCG}}(z)\!+\!25\!+\!5\log \left(\frac{c\,h^{-1}_{70}}{H_0} [\mathrm{Mpc}]\right)\!+\!5\log \left((1\!+\!z)\int_0^z\!\frac{dz'}{E(z')} \right) \\ 
= M_{\mathrm{BCG}} + K_{\mathrm{BCG}}(z) - 5\log h_{70} + 43.16 +   5\log (1\!+\!z) + 5\log \left(\int_0^z \frac{dz'}{E(z')} \right)  
\end{eqnarray}

\noindent
Here $K_{\mathrm{BCG}}(z)$ is the redshift dependent $K$-correction and in the second line the luminosity distance\footnote{Assuming a flat geometry, \ie \ $k\!=\!0$.} $d_{\mathrm{lum}}$ was substituted with Equ.\,\ref{e3_lum_dist}.
Assuming a concordance model cosmology and considering redshift dependent terms up to an order of $\orderof (z^2)$ in Equ\,\ref{e3_hubble_diagram}, we obtain the low-redshift approximation

\begin{equation}\label{e3_approx_HD}
m_{\mathrm{BCG}}(z) \approx  M_{\mathrm{BCG}} + K_{\mathrm{BCG}}(z) + 43.16 +  5\log z + 1.68\!\cdot\!z \ ,
\end{equation}

\noindent
with the expected linear $m(z)\!\propto\!\log z$ behavior for the local approximation $z\!\ll\!1$. 
Variations of the Hubble constant $H_0$ result in a constant vertical offset, whereas 
the sensitivity on the cosmological parameters  $\Omega_{\mathrm{m}}$, $\Omega_{\mathrm{DE}}$, and $w$
change the form of the Hubble relation   at higher redshifts  (last term in Equ\,\ref{e3_hubble_diagram}).
Based on this effect, SNe\,Ia measurements have established the existence of a non-vanishing {\em Dark Energy\/} component \cite{Perlmutter1999a}. 
The BCG Hubble diagram has been applied for cosmological studies by Collins\,\&\,Mann \cite*{Collins1998a} and by Arag{\'o}n-Salamanca, Baugh\,\&\,Kauffmann \cite*{Aragon1998a} using clusters at $z\!<\!1$.
With the concordance model in place, the BCG Hubble diagram is nowadays more suitable to probe evolutionary effects in the brightest cluster galaxy population as will be shown in Sect.\,\ref{s10_bcg_assembly}.






\end{description}

\noindent
The  cosmological tests using X-ray luminous galaxy clusters as probes are summarized in Tab.\,\ref{t3_cosmo_tests}.
The methods can be loosely classified into two categories: (A) tests {\bf 1--6} having the potential to yield competitive constraints on cosmological parameters, and (B)  tests {\bf 7--9} which are more to be considered as consistency checks, either for the underlying model ({\bf 7}), survey cross-comparisons ({\bf 8}), or to disentangle evolutionary effects ({\bf 9}).    

For distant cluster cosmology, the uncertainties in the mass calibration derived from observed properties 
and the redshift evolution of the scaling relations pose the
largest current limitations. Improving these uncertainties and characterizing \zg1 clusters as cosmological probes are important goals for the XDCP survey. The main cosmological application of the XDCP sample will be test {\bf 2}, which is expected to yield   competitive parameter constraints 4--5\,years from now. For tests {\bf 6} \&  {\bf 7}, XDCP can provide additional high-$z$ clusters in order to increase the redshift leverage. Methods {\bf 8} \&  {\bf 9} are applicable even for an uncompleted survey and are     
presented in Sect.\,\ref{s9_logS_logN}\,\&\,\ref{s10_bcg_assembly}.

\begin{table}[t]    
\begin{center}

\begin{tabular}{|c|c|c|c|c|}
\hline

 & \bf{Cosmological Test} & \bf{Effective Test } & \bf{\num} & \bf{Parameters} \\ 
 
\hline\hline

1 & cluster mass function $n(M)$ & $D_+$ & tens  & $\Omega_\mathrm{m}$, $\sigma_8$  \\ 

2 & number density evolution $n(M,z)$ & $D_+(z)$, $V_\mathrm{com}(z)$  & tens  & 
$\Omega_\mathrm{m}$,$\sigma_8$,$\Omega_\mathrm{\Lambda}$,$w$   \\ 

3 & cluster power spectrum $P(k)$ & $D_+$  & hundreds  & $\Omega_\mathrm{m}$, $\sigma_8$  \\ 

4 & baryonic acoustic oscillations & $d_\mathrm{ang}(z)$, $H(z)$  & ten thousands  & $\Omega_\mathrm{m}$, $\Omega_\mathrm{\Lambda}$, $w$ \\

5 & cluster  baryon fraction $f_\mathrm{b}$ &  $\Omega_\mathrm{b} / \Omega_\mathrm{m}$ & few  &  $\Omega_\mathrm{m}$  \\ 

6 & evolving gas mass fraction $f_\mathrm{gas}(z)$& $d_\mathrm{ang}(z)$ & tens  & $\Omega_\mathrm{m}$, $\Omega_\mathrm{\Lambda}$, $w$  \\ 

\hline

7 & absolute distances &  $d_\mathrm{ang}(z)$ & few  & $H_0$    \\ 

8 & cluster number counts $N(>S)$ &  $D_+$  & hundreds  &  $\Omega_\mathrm{m}$, $\sigma_8$  \\ 

9 & BCG Hubble diagram $m_{\mathrm{BCG}}(z)$ & $d_\mathrm{lum}(z)$  & tens  & $\Omega_\mathrm{m}$, $\Omega_\mathrm{\Lambda}$ \\ 

\hline
\end{tabular}

\caption[Cosmological Tests with Galaxy Clusters]{Cosmological tests using X-ray galaxy clusters as probes. The table summarizes the discussed methods, the effectively measured cosmic property, lower limits on the number of objects required, and the cosmological parameters for which the method is most sensitive to. The first six tests have the potential to provide competitive parameter constraints, while the last three are to be seen as consistency checks without strong constraining power.} \label{t3_cosmo_tests}
\end{center}
\end{table}

\chapter{X-ray Cluster Surveys}
\label{c4_Xsurveys}
\noindent
This last introductory chapter will shortly summarize some of the major achievements 
of galaxy cluster X-ray surveys. In addition, the state-of-the-art of distant X-ray luminous clusters at the  start of this thesis  will be discussed. 


\section{X-ray Selection}

\noindent
Cosmological evolution studies of the galaxy cluster population impose three essential requirements on any suitable survey sample (\eg \ Rosati, Borgani \& Norman, 2002)\nocite{Rosati2002a}:

\begin{enumerate}
    \item An efficient method to {\bf find clusters} over a wide redshift range; 
    \item An observable {\bf estimator of cluster mass};
    \item A method to {\bf evaluate the survey volume} within which clusters are found.
\end{enumerate}    

\noindent
Finding distant clusters is the main aim of this thesis and will be discussed in detail in the next four chapters.
A suitable observable proxy for the cluster mass has to be calibrated with local surveys and is shown in the next section. 
The survey volume $V_{\mathrm{max}}$ or equivalently the selection function $S_{\mathrm{sky}}[f(L,z)]$\footnote{Here the flux is denoted with $f$ and the sky coverage in units of steradian with $S_{\mathrm{sky}}$, not to be confused with the flux $S$ often used as a notation for number counts.}, \ie \ the effective sky coverage as a function of limiting flux, are essential to determine the comoving density of clusters and hence to relate the observations to model predictions.

For a strictly flux-limited X-ray survey with limit $f_{\mathrm{lim}}$ and corresponding sky coverage  $S[f_{\mathrm{lim}}]$, the comoving maximum search volume $V_{\mathrm{max}}(L)$ \ for objects with luminosity $L$ is straightforward to calculate using Equ.\,\ref{e3_volume_elements}. The maximum redshift $z_{\mathrm{max}}$ is obtained from  the luminosity distance (Equ.\,\ref{e3_lum_dist}) by solving $d_{\mathrm{lum}}(z_{\mathrm{max}})\!\equiv\!\sqrt{L/(4 \pi\!f_{\mathrm{lim}})}$ for $z_{\mathrm{max}}$. The search volume (see Fig.\,\ref{f3_flux_evolution}) is then given by

\begin{equation}\label{e4_survey_volume}
    V_{\mathrm{max}}(L) = V_{\mathrm{com}}[z_{\mathrm{max}}(f_{\mathrm{lim}},L)] = \int_0^{z_{\mathrm{max}}} S_{\mathrm{sky}}[f_{\mathrm{lim}}] \left( \frac{d_{\mathrm{lum}}(z)}{1+z}\right)^2 \frac{c \, dz}{H(z)} \ .
\end{equation}


\noindent
For an inhomogeneous sensitivity distribution, as typical for serendipitous surveys, the sky coverage\footnote{Additional correction terms might be necessary, \eg \ to take the detection efficiency as a function of cluster core radius or off-axis angle into account (see Sect.\,\ref{s9_Selection_Funct}).}
rises gradually over an extended flux range from the most sensitive regions with limit  $f_{\mathrm{low}}$ to the flux $f_{\mathrm{high}}$ covering the full survey area. In this case, the search volume is obtained by a second 
 integration over the flux limit and proper accounting of the additional survey area $dS/df$ at higher fluxes

\begin{equation}\label{e4_survey_volume_serendip}
    V_{\mathrm{max}}(L)\!=\!\int_{f_{\mathrm{low}}}^{f_{\mathrm{high}}} \frac{dV_{\mathrm{com}}[z_{\mathrm{max}}(f,L)]}{df}\,df\!=\!\int_{f_{\mathrm{low}}}^{f_{\mathrm{high}}}
\int_0^{z_{\mathrm{max}}(f)} \frac{dS_{\mathrm{sky}}[f]}{df} \left( \frac{d_{\mathrm{lum}}(z)}{1+z}\right)^2 \frac{c \, dz }{H(z)} \, df\,.
\end{equation}



\noindent
In principle, galaxy clusters can be systematically selected in several independent ways: (i) Optical/NIR selection based on the cluster signature of a spatial galaxy overdensity (and clustering in color or redshift space), (ii) X-ray detection of the extended ICM  emission, and for future surveys (iii) selection based on the SZE, or (iv) the weak gravitational lensing signature of clusters.     
The main important advantages of X-ray galaxy cluster surveys can be summarized as follows:

\begin{itemize}
    \item Bright thermal X-ray emission is only observed for {\bf well evolved clusters} with a deep gravitational well (Sect.\,\ref{s2_ICM_properties});
    \item The X-ray emission is highly peaked which {\bf minimizes projection effects} (Sect.\,\ref{s2_ICM_properties});
    \item The X-ray luminosity is tightly correlated with the gravitational mass and is providing a {\bf reliable mass proxy} based on the survey data (Sect.\,\ref{s4_LocalSurveys}); 
    \item The {\bf X-ray selection} can be accurately assessed allowing a simple evaluation of the survey volume (this section);
    \item Cosmological applications with X-ray clusters can build upon {\bf several decades of survey experience} (Chaps.\,\ref{c1_intro}\,\&\,\ref{c4_Xsurveys}). 
\end{itemize}



\section{Local Cluster Surveys}
\label{s4_LocalSurveys}

\enlargethispage{4ex}

\noindent
The success story of X-ray surveys for clusters of galaxies is largely built upon the German-American ROSAT (ROentgenSATellit) mission from 1990--1999.   
The  X-ray imaging capabilities of its main instrument, the Position Sensitive Proportional Counter (PSPC), allowed for the first time a cluster selection based on source extent, in particular for the mission phase of pointed observations  after the completion of
the ROSAT All-Sky Survey (RASS). The central  0.2\,deg$^2$ of the PSPC field-of-view achieved a point spread function with a FWHM resolution of 30\arcsec--45\arcsec \ enabling the resolved detection of extended emission of clusters with fiducial core radii of 250\,$h^{-1}$\,kpc out to  \zsim1 (\eg \ Rosati, Borgani \& Norman, 2002\nocite{Rosati2002a}). 
Considering that the extragalactic AGN point source population outnumbers clusters by approximately a factor of $10\!:\!1$ at faint flux levels ($f_{\mathrm{X}}\!\sim\!10^{-14}$\,\flux), the selection based on the extent criterion greatly enhances the cluster survey efficiencies.

\begin{figure}[t]
\begin{center}
\includegraphics[angle=0,clip,width=0.8\textwidth]{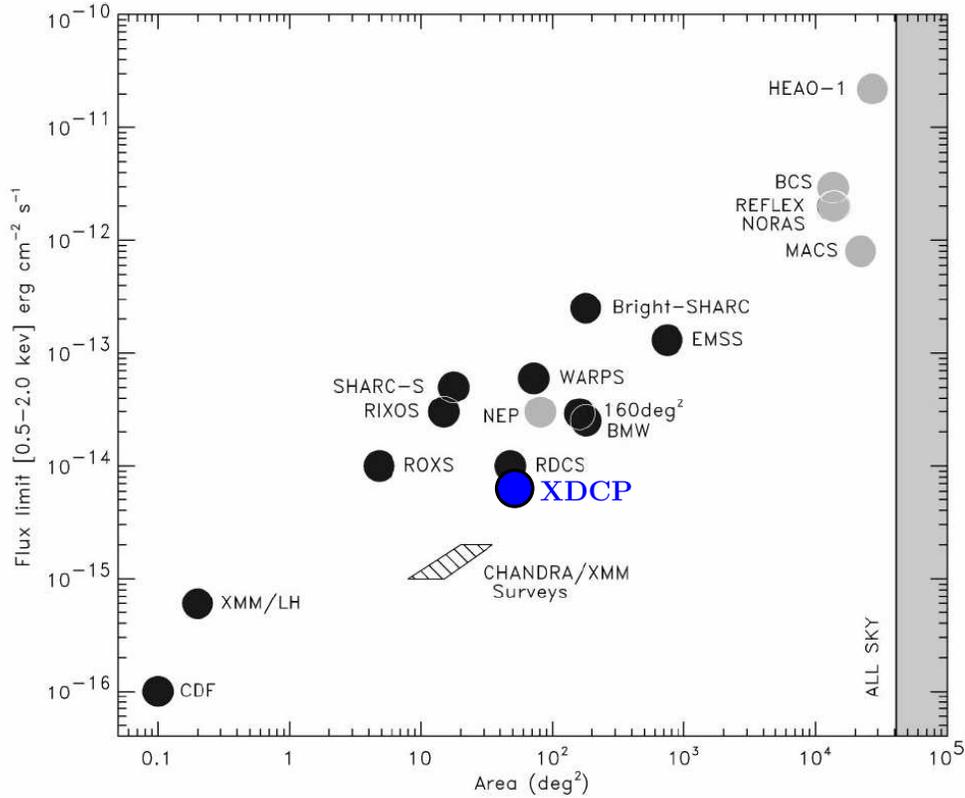}
\vspace{-3ex}
\end{center}
\caption[X-ray Cluster Survey Comparison]{Comparison of X-ray cluster surveys over the past two decades. The approximate survey flux limits and solid angles are plotted for different contiguous area surveys (light shaded circles) and serendipitous surveys (dark circles). The position of the XDCP survey has been added. Plot adapted from Rosati, Borgani \& Norman \cite*{Rosati2002a}.} \label{f4_Xray_Surveys}
\end{figure}



Two basic types of X-ray surveys can be distinguished. (i) {\em Contiguous sky area surveys} covering a single connected region, and (ii) {\em serendipitous surveys} making use of the ensemble of sky patches of numerous individual pointed observations.
Both survey types have their merits and strengths for different science applications. Contiguous surveys, in particular with all-sky coverage, span large solid angles with typically relatively bright flux limits. They are well-suited to detect the rarest, most luminous systems in the large survey volume and to investigate their clustering properties. Serendipitous surveys, on the other hand, are typically sensitive to significantly fainter flux limits, thus providing complementary information on more common systems with lower luminosities and more distant objects. Samples compiled from this survey type are particularly fit for cluster evolution studies.     




\enlargethispage{4ex}

Figure\,\ref{f4_Xray_Surveys} provides an overview of the major X-ray cluster projects of the last two decades. Contiguous area surveys are indicated by the light shaded symbols, whereas serendipitous surveys are represented by dark circles. The approximate location of the XMM-Newton Distant Cluster Project in the solid angle versus flux limit plane has been added in blue. 
First, we will have a short look at the major all-sky survey samples providing a local cluster census and reference system.     


\newpage

\begin{figure}[t]
\begin{center}
\includegraphics[angle=0,clip,width=\textwidth]{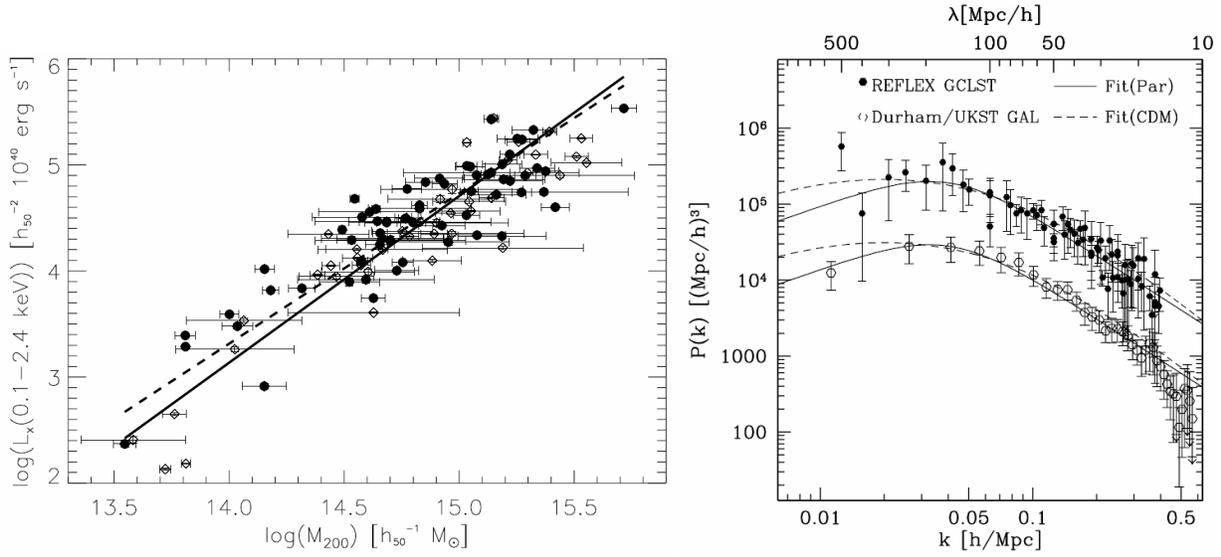}
\end{center}
\vspace{-2ex}
\caption[Important Local Survey Results]{Important results of local cluster surveys. {\em Left:} The empirical $L_{\mathrm{X}}$--$M$ relation of Reiprich \& B\"ohringer \cite*{Reiprich2002a} based on the HIFLUGCS sample of the 63 brightest galaxy clusters in the sky ({\em solid line}) and an extended sample of 106 objects ({\em dashed line}). {\em Right:}
Galaxy cluster power spectrum from the REFLEX survey ({\em top filled symbols}) in comparison to a galaxy power spectrum ({\em bottom open symbols}). The power spectra amplitude of the two object classes differ by the {\em biasing factor} $b^2\!\simeq\!6.8$. Plot from Schuecker \etal \ \cite*{Schuecker2001b}.} \label{f4_REFLEX_PS} \label{f4_LX_M_relation}
\end{figure}

\vspace{8ex}

\begin{figure}[t]
\begin{center}
\includegraphics[angle=0,clip,width=\textwidth]{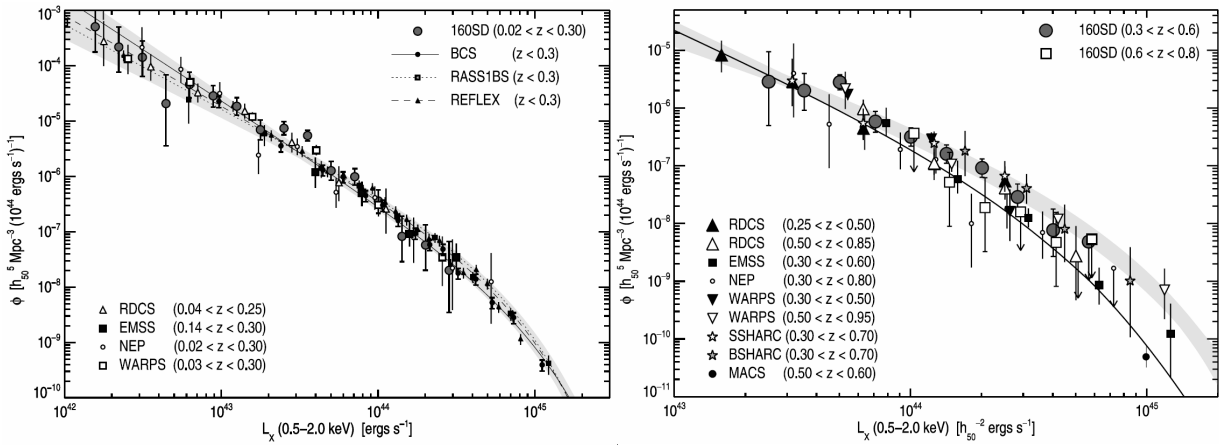}
\end{center}
\vspace{-2ex}
\caption[X-ray Cluster Luminosity Function]{X-ray luminosity functions (XLF) as determined by different cluster surveys. {\em Left:} The local ($z\!<\!0.3$) X-ray luminosity functions of eight flux-limited surveys show excellent agreement and serve as the no-evolution baseline for comparisons with distant cluster samples.
{\em Right:} XLFs derived from cluster samples of intermediate redshifts ($0.25\!<\!z\!<\!0.95$). The shaded regions indicates  the local XLF of the left panel. Significant evolution effects have only been observed for the highest luminosity clusters manifested in a lower observed space density at $z\!\sim\!0.8$. 
Plots from Mullis \etal \ \cite*{Mullis2004a}. } \label{f4_XLF}
\end{figure}

\begin{figure}[t]
\begin{center}
\parbox{0.45\textwidth}{
\includegraphics[angle=0,clip,width=0.45\textwidth]{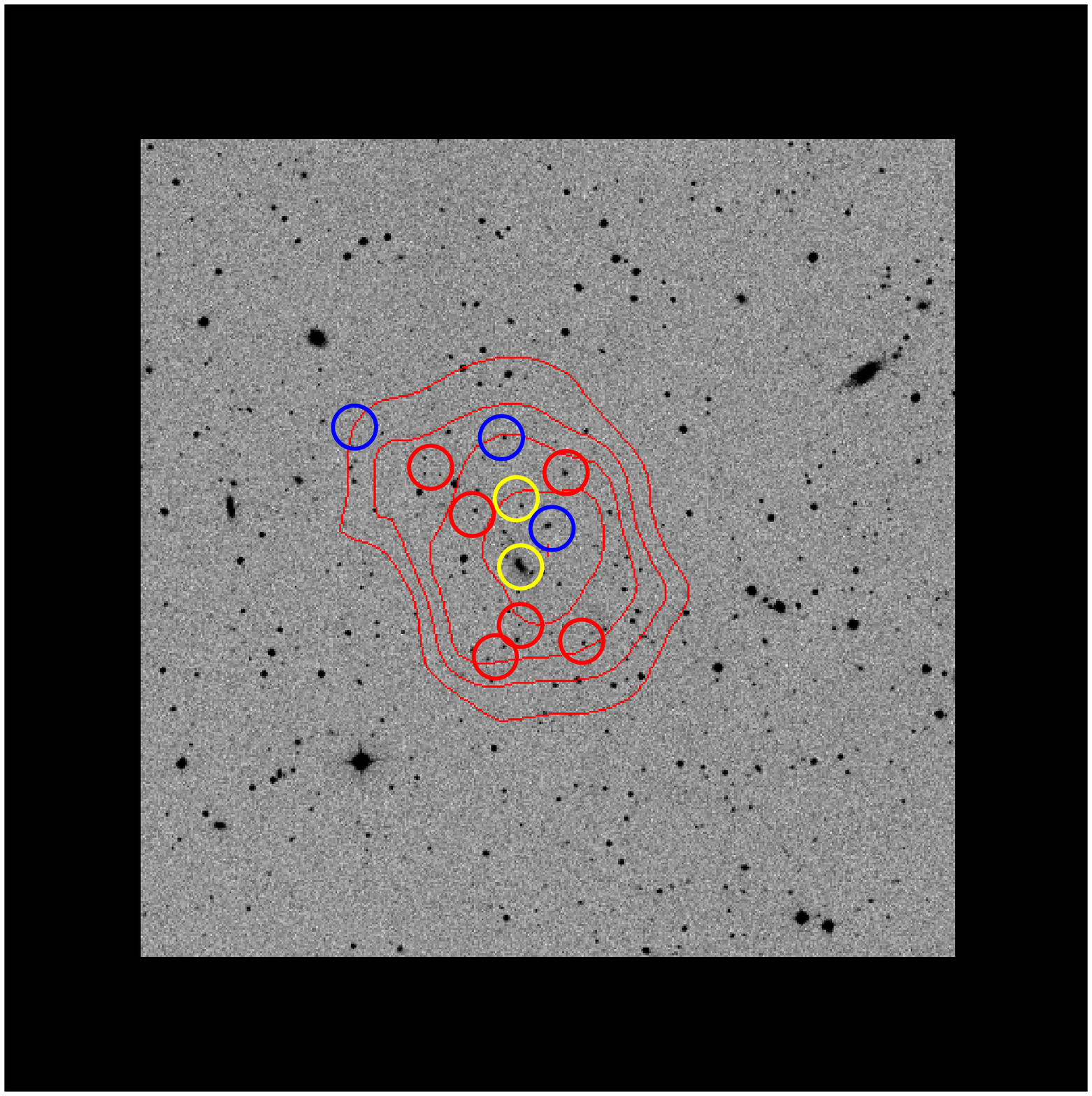}
}
\hfill
\parbox{0.45\textwidth}{
\caption[NORAS\,2 Galaxy Cluster]{The NORAS\,2 cluster of galaxies RXC\,J0834.9+5534 at $z\!=\!0.239$ as a typical example of a low-redshift object detected in the ROSAT All-Sky Survey. The red X-ray contours are shown on top of the $12\arcmin\!\times\!12\arcmin$ \ optical DSS image. The circles indicate the locations of spectroscopically confirmed cluster galaxies of early type (red), late type (blue), and Seyfert type\,1 (yellow).} \label{f4_SampleCl_NORAS2}
}
\end{center}
\end{figure}

\begin{figure}[b]
\begin{center}
\includegraphics[angle=0,clip,width=0.955\textwidth]{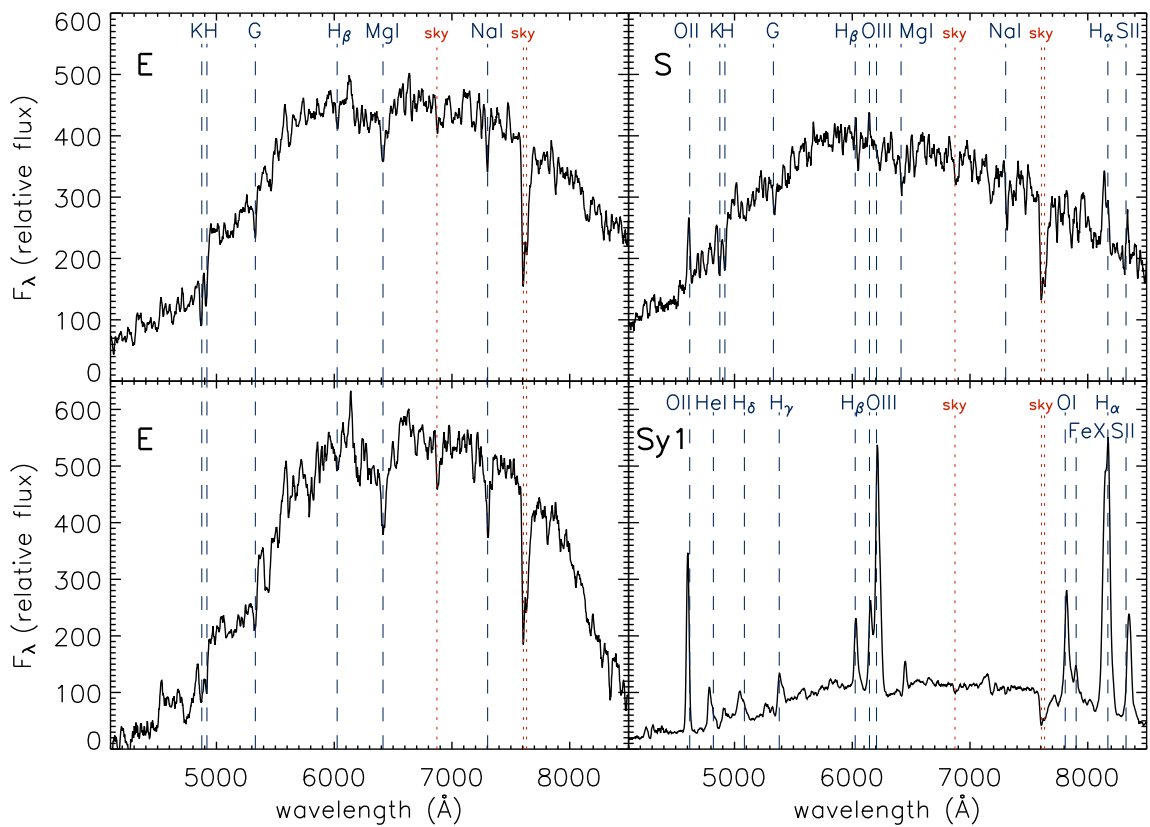}
\end{center}
\vspace{-4ex}
\caption[NORAS\,2 Galaxy Cluster Spectra]{Typical spectra of galaxy members of the cluster RXC\,J0834.9+5534 at $z\!=\!0.239$  shown above.  
Important spectral features are indicated by dashed blue lines with corresponding labels, atmospheric features by red dotted 
lines. {\em Left:} Early type galaxy spectra with their characteristic D4000 break redshifted to about 5\,000\,\AA \ and other labelled absorption features. {\em Right:} Late-type galaxy spectrum ({\em top}) and a Seyfert type\,1 spectrum ({\em bottom}) with indicated emission lines.} \label{f4_Noras2_spectra}
\end{figure}

\clearpage


\begin{description}

    \item[HIghest X-ray FLUx Galaxy Cluster Sample (HIFLUGCS):] 
    This sample includes the 63 brightest X-ray clusters in the (extragalactic) sky and has a flux limit of $f_{\mathrm{X}}\!=\!2\!\times\!10^{-11}$\,\flux \ in the ROSAT 0.1--2.4\,keV band \cite{Reiprich2001a}.
    The major achievement with this brightest cluster sample has been the empirical calibration of the (local) 
    $L_{\mathrm{X}}$--$M$ relation as shown in the left panel of Fig.\,\ref{f4_LX_M_relation} \cite{Reiprich2002a}.
    This relation establishes 
     the crucial mass proxy and hence the link to cosmological applications for large X-ray survey samples.    
        
    \item[ROSAT-ESO Flux-Limited X-ray cluster survey (REFLEX):] 
    The REFLEX sample \cite{HxB2004a} 
    is currently the largest statistically complete published compilation of X-ray clusters. It contains 447 objects in the Southern sky ($\delta\!<\!+2.5\degr$)  down to the flux limit of $f_{\mathrm{X}}\!=\!3\!\times\!10^{-12}$\,\flux \ [0.1--2.4\,keV].
  REFLEX provides the local reference of the luminosity function (left panel of Fig.\,\ref{f4_XLF}) \cite{HxB2002a} 
  and has been used for a wide variety of cosmological studies (\eg \ Schuecker \etal, 2003a, Schuecker \etal, 2003b). 
  \nocite{Schuecker2003b} 
  \nocite{Schuecker2003a}
  A major achievement has been the measurement of the galaxy cluster power spectrum on scales of  15--800\,$h^{-1}$\,Mpc as shown in the right panel of Fig.\,\ref{f4_LX_M_relation} \cite{Schuecker2001b}.
    An extension of the survey, REFLEX\,2, is in preparation and will roughly double the number of objects by  lowering the flux limit to  $f_{\mathrm{X}}\!=\!1.8\!\times\!10^{-12}$\,\flux. 
  

    \item[NOrthern ROSAT All-Sky survey (NORAS):] 
    NORAS \cite{HxB2000a} 
    is the counterpart of REFLEX in the Northern hemisphere containing 378 clusters out to $z\!\la\!0.3$.
    The survey extension NORAS\,2 is  in preparation. The combined extended samples NORAS\,2 and REFLEX\,2 with a total of         approximately 1\,800 X-ray clusters will provide the next milestone for a local power spectrum analysis. 
    This thesis work also included the spectroscopic analysis of 35 clusters for the extended NORAS\,2 survey.   
    Figures\,\ref{f4_SampleCl_NORAS2}\,\&\,\ref{f4_Noras2_spectra} 
    show one example of a low-redshift cluster at  $z\!=\!0.239$ and some typical cluster galaxy spectra as local reference for the discussions on the spectral energy distribution (SED) in Chap.\,\ref{c7_NIRanalysis} and spectroscopic features (Chap.\,\ref{c8_SpecAnalysis}) of distant cluster galaxies. 
        
    
\end{description}




\section{Deep Surveys}
\label{s4_DeepSurveys}

\enlargethispage{4ex}

\noindent
Three of the major deep ROSAT  surveys are introduced and the corresponding cluster samples used as the basis for discussing evolution effects of the  X-ray luminosity function.


\begin{description}
    \item[North Ecliptic Pole survey (NEP):] 
    The North ecliptic pole is the area with the deepest effective exposure time of the ROSAT All-Sky Survey owing to the survey scan pattern along great circles  perpendicular to the ecliptic. 81 square degrees with the most sensitive coverage around the North pole have been used as the basis for a complete spectroscopic identification of all X-ray sources. The North Ecliptic Pole survey \cite{Henry2006a} 
with a median sensitivity of $f_{\mathrm{X}}\!\sim\!7\!\times\!10^{-14}$\,\flux \
includes 62 galaxy clusters out to a maximum redshift of $z\!=\!0.81$ and is the only deep ROSAT survey with a contiguous area coverage.
    
    \item[160\,deg$^{2}$ large area survey (160\,deg$^{2}$):] 
    The 160\,deg$^{2}$ serendipitious survey \cite{Vik1998a} with a median flux limit of $1.2\!\times\!10^{-13}$\flux \ [0.5--2.0\,keV] and deepest coverage of  $4\!\times\!10^{-14}$\flux \
    contains 201 clusters. 
    The 22 objects at  $z\!>\!0.5$ allow an assessment of evolution effects out to redshifts $z\!\sim\!0.8$.
    This survey has recently been extended to a sky coverage of 400\,deg$^2$ \cite{Burenin2007a}.
    

    

    \item[ROSAT Deep Cluster Survey (RDCS):] 
    The RDCS  \cite{Rosati1998a} is the deepest ROSAT survey available and has compiled a sample of 103 clusters out to $z\!=\!1.27$, which marks the `ROSAT redshift horizon'. Based on deep PSPC pointed observations, the serendipitous survey covers a sky area of about 50 square degree down to a flux limit of $f_{\mathrm{X}}\!=\!3\!\times\!10^{-14}$\,\flux \ [0.5-2.0\,keV] yielding 26 systems at $0.5\!<\!z\!<\!0.85$. 
Only the high-redshift extension of RDCS with an effective survey area of about 5 square degrees and a sensitivity down to $f_{\mathrm{X}}\!=\!1\!\times\!10^{-14}$\,\flux \ allowed the detection of the first four \zg1 X-ray clusters.



\end{description}



\begin{figure}[t]
\begin{center}
\includegraphics[angle=0,clip,width=0.85\textwidth]{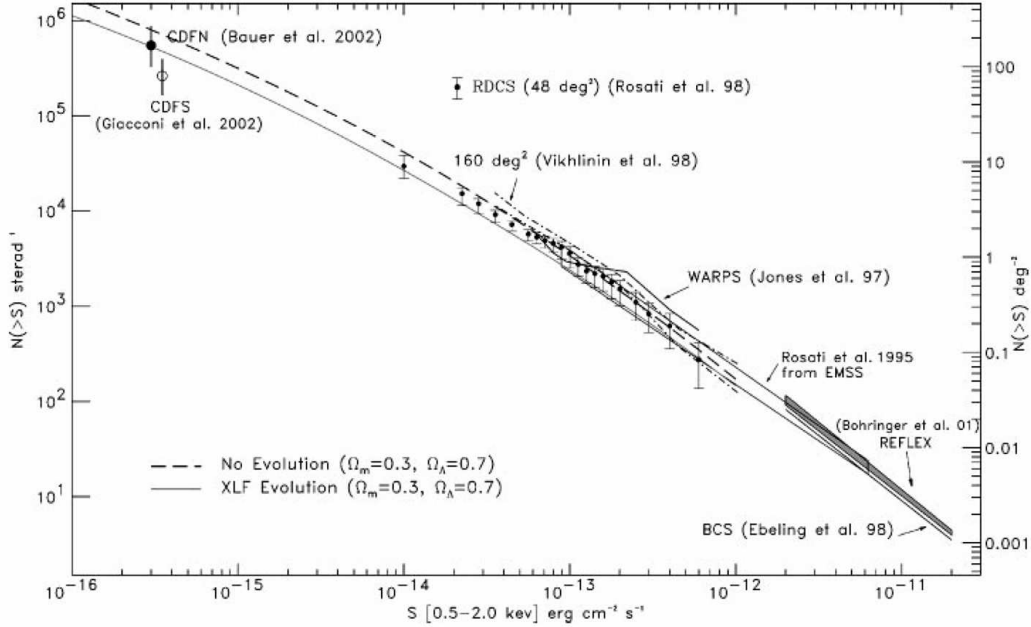}
\vspace{-3ex}
\end{center}
\caption[ROSAT Cluster Number Counts]{Compilation of cumulative cluster number counts as a function of X-ray flux for different ROSAT surveys. The log\,$N$--log\,$S$ has been measured over four orders of magnitude in flux. The right-hand scale indicates the total cluster surface density per square degree at a given flux. Plot from Rosati, Borgani, \& Norman \cite*{Rosati2002a}.} \label{f4_ROSAT_logSlogN}
\end{figure}

\noindent
Figure\,\ref{f4_ROSAT_logSlogN} shows a compilation of the cumulative cluster number counts (log\,$N$--log\,$S$) derived for different local and deep surveys. From the deepest flux levels probed by ROSAT, a surface density estimate of about 10 galaxy clusters per square degree is expected at $f_{\mathrm{X}}\!\sim\!1\!\times\!10^{-14}$\,\flux \ [0.5-2.0\,keV].
On the opposite end, \ie \ for the HIFLUGCS  selection criterion, only one cluster per 400 square degree sky area exists. 

We can now assess the evidence for an evolution of the galaxy cluster X-ray luminosity function (XLF) as established by ROSAT-based surveys. The compilation of Mullis \etal \ \cite*{Mullis2004a} in Fig.\,\ref{f4_XLF} summarizes the measurements of the local cluster luminosity function (left panel) and the distant XLF out to $z\!\sim\!0.8$ (right panel).
The differential luminosity function $\phi(L_{\mathrm{X}},z)$ is commonly parameterized in the form


\begin{equation}\label{e4_XLF_SchechterFct}
\phi(L_{\mathrm{X}},z)\,dL_{\mathrm{X}}= \frac{d^2 N}{dV\,dL_{\mathrm{X}}}(L_{\mathrm{X},z}) = \phi^{*}\,\left( \frac{L_{\mathrm{X}}}{L^{*}_{\mathrm{X}}}\right)^{-\alpha}
\exp \left(-\frac{L_{\mathrm{X}}}{L^{*}_{\mathrm{X}}}\right)\,\frac{dL_{\mathrm{X}}}{L^{*}_{\mathrm{X}}} \ ,
\end{equation}

\noindent 
where $N$ denotes the number of clusters of luminosity $L_{\mathrm{X}}$ in volume $V$ at redshift $z$.
The {\em Schechter\/} function  on the right hand side is parameterized by the characteristic luminosity $L^{*}_{\mathrm{X}}$, the faint-end slope $\alpha$, and the  normalization  $\phi^{*}$ (in units $h^3$\,Mpc$^{-3}$), which is directly related to the space density of clusters brighter than a minimum luminosity $L_{\mathrm{min}}$ via $n_0\!=\!\int_{L_{\mathrm{min}}}^{\infty}\phi(L)\,dL$.  

The local X-ray luminosity function shows a remarkable overall agreement between the surveys considering the independent selection techniques and data sets and is hence a robustly determined reference measure for evolution studies. The parameters for the local XLF can be summarized as follows (\eg \ B\"ohringer \etal, 2001; Rosati, Borgani \& Norman, 2002): \nocite{HxB2001a} \nocite{Rosati2002a}
$L^{*}_{\mathrm{X}}\!\simeq\!4\!\times\!10^{44}$\,erg\,s$^{-1}$ [0.5--2.0\,keV], $\alpha\!\simeq\!1.8$ (15\% variation), and  
$\phi^{*}\!\simeq\!2.7\times\!10^{-7}\,h_{70}^{3}$\,Mpc$^{-3}$ (50\% variation).






\enlargethispage{4ex}

A meaningful parametrization of  evolution effects of the XLF 
\cite{Rosati2000a} 
is given by    

\begin{equation}\label{e4_evolving_XLF}
\phi(L,z) = \phi_{0}\,(1+z)^{A} \cdot L^{-\alpha}\exp \left( -\frac{L}{L^{*}(z)} \right)  ,
\end{equation}

\noindent
with the free parameters $A$ for the evolution of the local space density normalization  $\phi_{0}$, and $B$ for an evolving characteristic luminosity $L^{*}(z)\!=\!L^{*}_0\,(1+z)^{B}$. 
The current data out to $z\!\sim\!0.8$ show no detectable evolution effects for galaxy clusters with luminosities $L_{\mathrm{X}}\!\la\!10^{44}$\,erg\,s$^{-1}$, \ie \ the measurements are consistent with $A\!=\!B\!=\!0$ (right panel of Fig.\,\ref{f4_XLF}). Only at the highest luminosities at $L_{\mathrm{X}}\!>\!10^{44}$\,erg\,s$^{-1}$ \ a mild but significant deficit in the volume density compared to the local value is confirmed at a confidence level of $>\!3\sigma$ \cite{Mullis2004a}. 

In summary, a moderate negative evolution, \ie \ objects were rarer at high redshift,  has been established 
for the most luminous, massive clusters out to $z\!\sim\!0.8$, but the bulk of the cluster population is approximately constant for at least  the second half of cosmic time. This means that the epoch of a significant number density change of the overall cluster population is yet to be determined.






\section{Distant Galaxy Clusters in 2004}
\label{s4_distant_cl2004}

\begin{figure}[!t]
\begin{center}
\includegraphics[angle=0,clip,height=6.5cm]{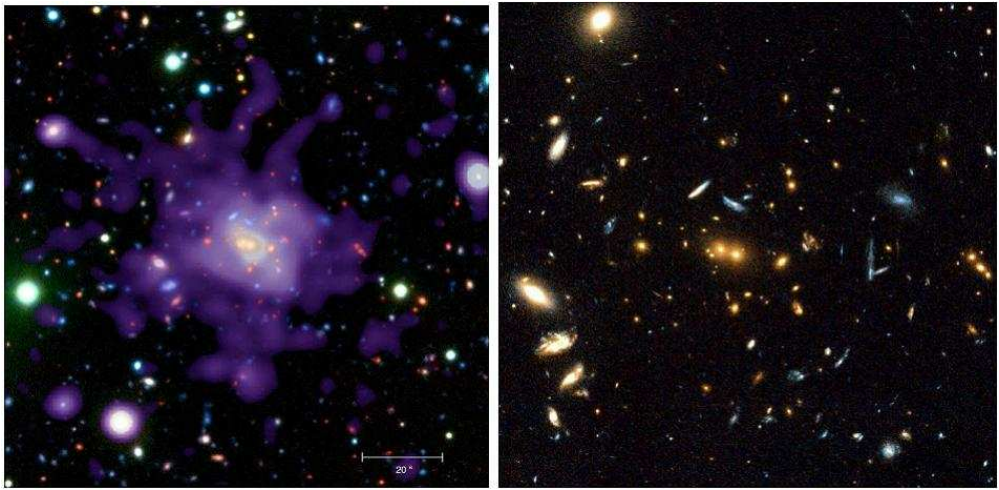}
\end{center}
\vspace{-2ex}
\caption[Distant Cluster RDCS\,1252.9-2927]{Multi-wavelength view of the distant cluster of galaxies RDCS\,J1252.9-2927 at redshift $z\!=\!1.237$, the most massive and best studied \zg1 \ cluster discovered with ROSAT. 
{\em Left:} $2\arcmin\!\times\!2\arcmin$ \ color composite image of the cluster X-ray  emission (purple) overlaid on the optical/near-infrared data showing the red galaxy population. 
{\em Right:} $1\arcmin\!\times\!1\arcmin$ \ high resolution Hubble Space Telescope image of the cluster core. 
Images from press release\footnotemark.
} \label{f4_RDCS1252}
\end{figure}

\noindent
The ROSAT era has led to the discovery of five X-ray luminous galaxy clusters at \zg1 (see Tab.\,\ref{t10_Xray_clusters}), four of them detected within the RDCS survey, 
and one in the deep Mega-second observation of the  Lockman Hole (LH) field \cite{Hashimoto2005a}.
The ongoing detailed multi-band  follow-up observations of these rare systems have revealed rather advanced  evolutionary states despite the large lookback time, showing a metal-rich, hot ICM and a red and passively evolving galaxy population.
Figure\,\ref{f4_RDCS1252} displays a multi-wavelength overlay (left panel) and a high-resolution HST image (right panel) of 
RDCS\,J1252.9-2927  at $z\!=\!1.237$, the most massive and best studied \zg1 \ galaxy cluster in 2004. 

The situation of distant cluster studies at the start of this thesis is summarized in the cone diagram of Fig.\,4.8, 
 illustrating the distribution of  known X-ray clusters in redshift space.
Out to redshifts of $z\!\sim\!0.7$ (red circles), the cluster population is well-sampled by the local and deep ROSAT surveys. Orange circles represent the small sample of all known X-ray clusters at $z\!>\!0.7$, about ten in the redshift range $0.7\!<\!z\!\la\!1$, and the mere five test objects at $1\!<\!z\!\leq\!1.27$. The yellow star indicates the position of the cluster XMMU\,J2235.3-2557 in this diagram, which will be presented in detail in Sect.\,\ref{s5_xmmu_2235}.



The highly active phase of galaxy cluster  surveys  has  
continued into the XMM-Newton era. Here we summarize some of the major 
projects, which will have a significant impact on the study of distant X-ray clusters.


\begin{description}
    \item[Cosmic Evolution Survey (COSMOS):] 
    The 2\,deg$^2$ field of the   COSMOS survey \cite{Scoville2006a} 
    has obtained deep multi-wavelength coverage with basically all major astronomical facilities. 
    The project is designed to probe the correlated evolution of galaxies, star formation, AGN, dark matter, and the large-scale structure. More than 2\,Msec of XMM-Newton data have been obtained \cite{Hasinger2006a} 
    allowing the detection of extended sources down to flux levels of $f_{\mathrm{X}}\!\sim\!10^{-15}$\,\flux.
    An early set of the COSMOS X-ray data has been used to test the XDCP selection procedure and efficiency (Sect.\,\ref{s6_COSMOS_tests}).
    The COSMOS survey has the ability to detect and study dozens of high-redshift groups and clusters \cite{Alexis2006a}.  
    However, the limited solid angle restricts the survey to the inclusion of  a single 
    structure of order $2\!\times \!10^{14}\,M_{\sun}$, implying that the high mass end of the XLF cannot be sufficiently probed. 
    

    \item[XMM Large-Scale Structure survey (XMM-LSS):] 
    The XMM-LSS project \cite{Pierre2006a} 
    is designed as a designated multi-wavelength galaxy cluster survey over a fairly large contiguous sky area. However, the large demand of XMM resources has limited the currently available X-ray data to roughly 9\,deg$^2$ with a typical depth of 10\,ksec.
    The XDCP sample hence provides a wider and deeper coverage.  
    It has an overlap with XMM-LSS of about 0.8\,deg$^2$, which can be used for a survey comparison study.
    
    \item[XMM Cluster Survey (XCS):] 
    The ambitious XCS \cite{Romer2001a} 
    serendipitous survey aims at the detection of {\em all} galaxy clusters in {\em all} XMM archive fields. In principle, XCS fully overlaps with the XDCP survey. However, the  XCS project is centered on fairly bright cluster sources with a sufficient number of photons for an accurate temperature determination and is not focussed on very distant clusters with their typically low flux and count levels.

\end{description}    



\addtocounter{footnote}{1} 
\footnotetext{The images can be found at \url{http://chandra.harvard.edu/photo/2004/rdcs1252/rdcs1252_xray_opt.jpg} and \url{http://chandra.harvard.edu/photo/2004/rdcs1252/rdcs1252_hst_optical.jpg}.}
\addtocounter{footnote}{-2} 

\newpage

\begin{figure}[!b]
\begin{center}
\parbox{0.45\textwidth}{
\includegraphics[angle=0,clip,width=0.45\textwidth]{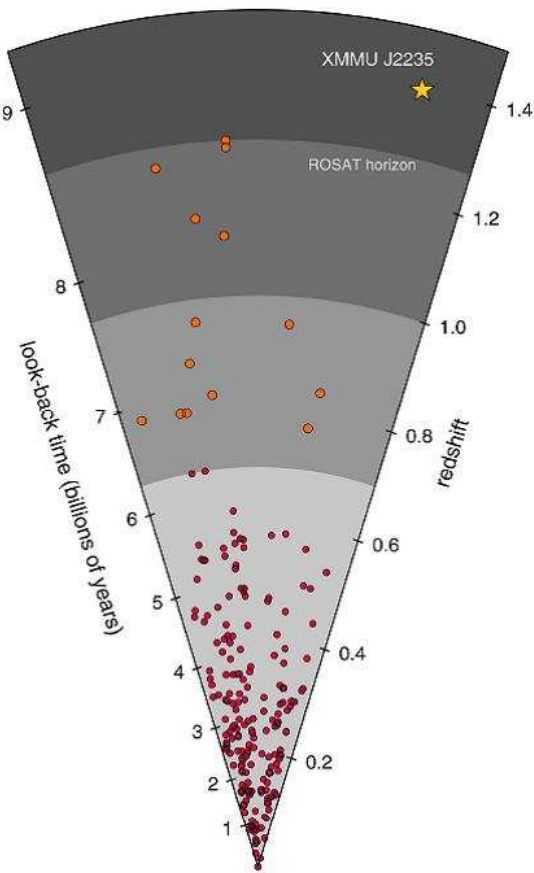}
}
\hfill
\parbox{0.45\textwidth}{
\caption[Cone Diagram of Known Galaxy Clusters]{Schematic cone diagram of known X-ray luminous galaxy clusters in 2004. The redshift increases from bottom to top, left to right illustrates the right ascension distribution. The red symbols at $z\!<\!0.7$ represent a small selection of known low-redshift clusters, orange circles at $0.7\!\leq\!z\!\la\!1.3$ indicate all known ROSAT X-ray clusters in the high-redshift regime. The XDCP cluster XMMU\,J2235.3-2557 at $z\!\sim\!1.4$ (see Sect.\,\ref{s5_xmmu_2235}) was the first XMM selected system that could break the `ROSAT redshift horizon'.
Plot from C. Mullis.}
}
\end{center}
\label{f4_cone_diagram}
\end{figure}
\newpage





\chapter{The XMM-Newton Distant Cluster Project}
\label{c5_XDCP}

\noindent This chapter will provide an overview of the XMM-Newton Distant Cluster Project. The principal survey strategy is introduced followed by a discussion of early results and gained experiences of a pilot study. 


\section{Motivation}
\label{s5_motivation}

\noindent
The main {\em scientific motivations\/} for starting an extensive new serendipitous distant cluster survey have been outlined in the introductory chapters \ref{c2_cluster_theory}--\ref{c4_Xsurveys}. Among the vast number of astrophysical and cosmological  applications of distant X-ray luminous galaxy clusters, the accurate determination of the cluster number density evolution out to $z\!\sim\!1.5$ is the prime goal and science driver of the XDCP.
The time window for a `new generation' survey apt for this ambitious task has only opened 
in the year 2003 owing to the rapid {\em advance of technology\/} over the last few years. The European X-ray satellite XMM-Newton has been providing an ever increasing amount of archive data with unprecedented sensitivity to a depth that will be unrivalled for at least another decade (see Sect.\,\ref{s11_eROSITA}). These advances in X-ray technology go hand-in-hand with the availability of ground-based 8--10 meter class telescopes, most notably the European Very Large Telescope (VLT), which now allow the spectroscopic confirmation and further detailed studies of cluster galaxies at \zg1. 

As discussed in Chap.\,\ref{c1_intro}, the cluster samples associated with the scientific goals of the XDCP can be grouped into three different categories with increasing time line and scope: (1) provide \zg1 galaxy clusters as \emph{individual astrophysical laboratories} for detailed studies, (2) 
compile \emph{pathfinder samples} for the characterization of high-redshift clusters as cosmological probes for large upcoming surveys, and (3) establish a statistically complete \emph{cosmological sample} for various cosmological tests.
The fundamental pre-requisite for compiling these cluster samples is an efficient survey strategy 
for finding and confirming the rare distant X-ray luminous systems; the 
 basic steps are introduced in the following section.


\newpage
\section{XDCP Survey Strategy}
\label{s5_survey_strategy}

\begin{figure}[t]
\begin{center}
\includegraphics[angle=0,clip,width=11.8cm]{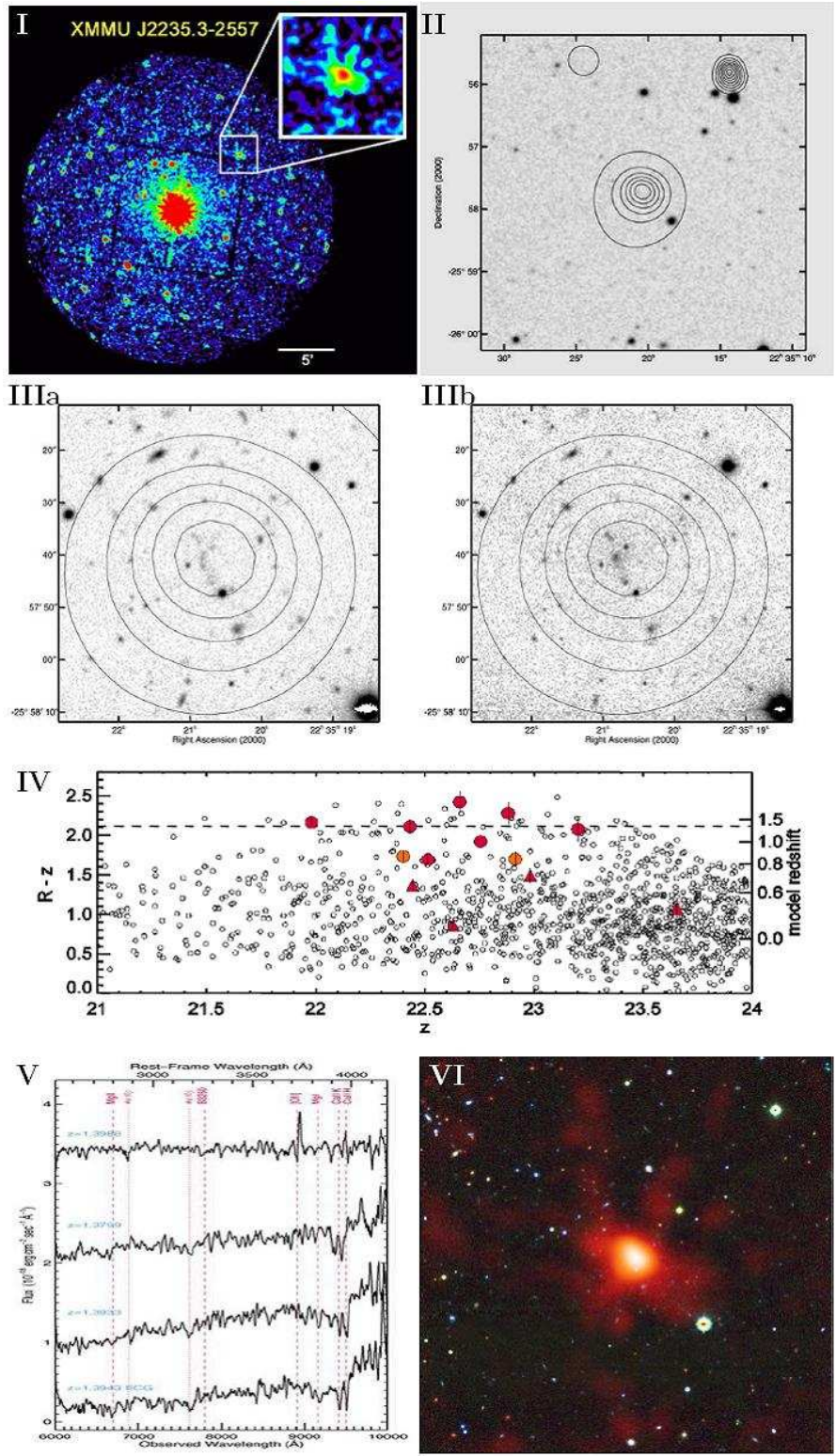}
\end{center}
\vspace{-4ex}
\caption[XDCP Survey Strategy]{Schematic steps of the XDCP survey strategy. {\em I:} Detection of serendipitous extended X-ray sources in suitable archival XMM-Newton fields. {\em II:} X-ray source screening using available optical and NIR data (DSS). {\em III:} Two-band follow-up imaging to verify the presence of a cluster as overdensity of red galaxies (IIIa R-band, IIIb Z-band). {\em IV:} Photometric redshift estimation using the color of the \reds \ as distance indicator. {\em V:} Spectroscopic confirmation of the cluster redshift. {\em VI:} Detailed multi-wavelength follow-up of the most interesting objects. Plots provided by C. Mullis.} \label{f5_survey_strategy}
\end{figure}

\noindent
The XMM-Newton Distant Cluster Project \cite{HxB2005a} is based on the following six-step strategy aiming at the efficient selection and confirmation of X-ray luminous \zga1 clusters of galaxies. The intermediate results of the different survey steps are illustrated in Fig.\,\ref{f5_survey_strategy} using the galaxy cluster XMMU\,J2235.3-2557 as an example case.

\begin{description}
    \item[Step\,1: Detection of serendipitous extended X-ray sources.] 
    Deep  XMM-Newton archive fields ($\ga$10\,ksec) at high galactic latitude ($|b|\!\geq\!20$\degr) are analyzed for serendipitous extended X-ray sources in the field-of-view. 
    \item[Step\,2: Source screening and selection of candidates.] The detected extended X-ray sources are visually inspected and their sky positions cross-correlated with available optical data and extragalactic database information. X-ray sources without an identified optical counterpart are selected as {\em distant\/} cluster candidate sources.
    \item[Step\,3: Two-band follow-up imaging:] For the candidate targets, follow-up imaging data is obtained in (at least) two filter bands to sufficient depth, which allows a preliminary cluster identification based on a red overdense galaxy population coincident with the extended X-ray source.
    \item[Step\,4: Photometric redshift estimate.] The full photometric analysis yields a cluster redshift estimate based on the color of the \reds \ which allows a refined selection of photometrically confirmed $z\!\ga\!0.9$ candidates. 
    \item[Step\,5: Spectroscopic confirmation.] Promising high-redshift cluster candidates are spectroscopically observed for the final cluster confirmation and redshift determination.
    \item[Step\,6: Detailed follow-up studies.] For the most interesting \zga1 systems,  additional detailed imaging and spectroscopic follow-up observations in X-ray, optical, NIR, IR, and the radio regime will be obtained to disclose the high-redshift cluster properties and physics.
\end{description}    

\noindent
It should be emphasized at this point that the vast majority of {\em extended\/} X-ray sources (steps\,1\,\&\,2) in high galactic latitude fields are expected to originate from galaxy clusters (see \eg \ the NEP survey in Sect.\,\ref{s4_DeepSurveys}).  
Due to the sequential dependency of the different survey steps and the increasing  demands on observatory resources,  the completion rates   
at an intermediate survey stage will drastically decrease from top to bottom. The first two steps are discussed in detail in Chap.\,\ref{c6_XrayAnalysis} and are based on publicly available data only, hence providing a 100\% coverage for the current survey phase. Steps\,3 \& 4 are presented in Chap.\,\ref{c7_NIRanalysis} achieving a survey coverage of about 40\% for the thesis data. The observational bottleneck is the spectroscopy of step\,5, which is introduced in Chap.\,\ref{c8_SpecAnalysis}, and which is currently available only for few systems. The final step\,6 will not be discussed in this thesis beyond the overview provided in  Sect.\,\ref{s5_xmmu_2235} since detailed follow-up data has only  been obtained for the cluster XMMU\,J2235.3-2557.

\clearpage

\section{Pilot Study Results}
\label{s5_pilot_study}

\noindent
The XMM-Newton Distant Cluster Project started in the year 2003 with a pilot study to test the feasibility and efficiency of the survey strategy. Since this feasibility study and most of the related work began prior to this thesis, the results and main lessons of this initial project phase are summarized without going into the details.

120 archival XMM-Newton pilot study fields were analyzed for serendipitous extended X-ray sources ({\bf step 1}). From these about 60 distant cluster candidate sources were selected ({\bf step 2}) and observed with VLT FORS\,2 R and Z-band snap-shot imaging ({\bf step 3}). Based on the R--Z \reds \ analysis, a dozen promising high-redshift candidates with photometric redshift estimates $z\!\ga\!0.9$ could be identified ({\bf step 4}) and have been proposed for spectroscopic confirmation ({\bf step 5}). For the first discovered high-$z$ cluster (see Sect.\,\ref{s5_xmmu_2235}) detailed multi-wavelength follow-up observations have been proposed and obtained ({\bf step 6}).

The most important XDCP pilot study conclusion was the confirmation that high-redshift galaxy clusters can indeed be efficiently selected based on their extended X-ray signature in conjunction with two-band imaging data. The 
strategy described in Sect.\,\ref{s5_survey_strategy} could thus be established as the basis for a large systematic distant galaxy cluster archival survey. The early results have been very encouraging and have pointed to the great scientific potential achievable with such a project.  

However, besides the positive overall observational feedback, the XDCP pilot study has also revealed 
optimization potentials for most of the survey chain, in particular for the first four process steps. 

\begin{itemize}
    \item {\bf Step 1:} The pilot study X-ray selection of distant cluster candidates was in effect based on sources with at least 200 detected X-ray photons. For a 10\,ksec net XMM exposure, this implies an approximate detection limit of $f_{\mathrm{lim}}\simeq 2 \times 10^{-14}$\,\flux (0.5--2.0\,keV), 
\ie \ comparable to the deepest ROSAT surveys (see Fig.\,\ref{f4_Xray_Surveys}). In order to achieve significant sensitivity improvements compared to existing surveys, the detection of extended sources with less than 100 X-ray photons is a desirable aim. Consequently, the first optimization goal is an improved sensitivity for detecting extended X-ray sources.    
    \item {\bf Step 2:} The pilot study returned a large fraction of intermediate-redshift systems at $z\simeq 0.3$--0.6, which contributed almost 40\% to the identified cluster sample but are not of particular interest for the prime XDCP science objectives. An improved screening could significantly decrease the number of such intermediate-redshift objects and thus increase the \zg1 \ cluster fraction in the candidate sample. A more efficient low-redshift cluster identification and rejection is the second optimization goal.  
    \item {\bf Step 3:} For several confirmed high-redshift candidates and clusters between $z\!\simeq\!1.1$--1.45 the VLT FORS\,2 `snap-shot imaging' with net exposure times 
    of 480\,s in Z (Z$_{\mathrm{lim}}(\mathrm{AB})\!\sim\!23.0$\,mag) and 960\,s in R     (R$_{\mathrm{lim}}(\mathrm{AB})\!\sim\!24.5$\,mag) did not allow a secure cluster identification, \ie \ the data did not have sufficient depth. The third optimization goal is thus a more complete  two-band imaging cluster confirmation at the high redshift end.
    \item {\bf Step 4:} The R--Z color exhibits large intrinsic redshift uncertainties at \zg1 . The final optimization goals are thus improved and more secure redshift estimates based on two-band imaging data. 
\end{itemize}




\noindent
The first two optimization items dealing with the X-ray selection process will be discussed and implemented in Chap.\,\ref{c6_XrayAnalysis}. The latter two concerning the two-band follow-up imaging will be treated in detail in Chap.\,\ref{c7_NIRanalysis}. As the highlight result of the XDCP pilot study, the 
galaxy cluster XMMU\,J2235.3-2557 is now introduced.


\section{XMMU\,J2235.3-2557}
\label{s5_xmmu_2235}

\noindent
The XDCP discovery of XMMU\,J2235.3-2557 was announced\footnote{Press release articles can be found under \url{http://www.eso.org/public/outreach/press-rel/pr-2005/pr-04-05.html} or \url{http://www.mpg.de/bilderBerichteDokumente/dokumentation/pressemitteilungen/2005/pressemitteilung20050228/presselogin}.} in March 2005 with the results published in the accompanying  paper of Mullis \etal \ \cite*{Mullis2005a}.
At the time of publication,  XMMU\,J2235.3-2557, at a redshift of $z\!=\!1.393$, was the most distant galaxy cluster known, and the first XMM-Newton selected system at \zg1. It was also the first XDCP object for which spectroscopic follow-up observations were obtained.

The selection procedure for XMMU\,J2235.3-2557 is shown in Fig.\,\ref{f5_survey_strategy} as an illustration of the survey strategy steps. The cluster's extended X-ray emission was detected in a 44\,ksec XMM-Newton observation of the Seyfert galaxy NGC\,7314 at an off-axis angle of 7.7\arcmin \ (panel I). The corresponding sky position on the DSS plates is devoid of objects, which classified the source as a distant cluster candidate (panel II). The subsequent R (panel IIIa) and Z-band (panel IIIb) follow-up images reveal a galaxy cluster signature of red galaxies with a significant central overdensity. The color-magnitude diagram (panel IV)  exhibits a cluster \reds \ with a color of R--Z\,$\sim$\,2.1 indicating a high-redshift cluster candidate at \zg1. Deep VLT-FORS\,2 spectroscopy with an exposure time of 4\,hours confirmed the cluster at a redshift of  $z\!=\!1.393$ based on 12 identified spectroscopic cluster member galaxies. The detailed multi-band follow-up program (panel VI) is discussed below.

\begin{figure}[t]
\begin{center}
\includegraphics[angle=0,clip,width=0.95\textwidth]{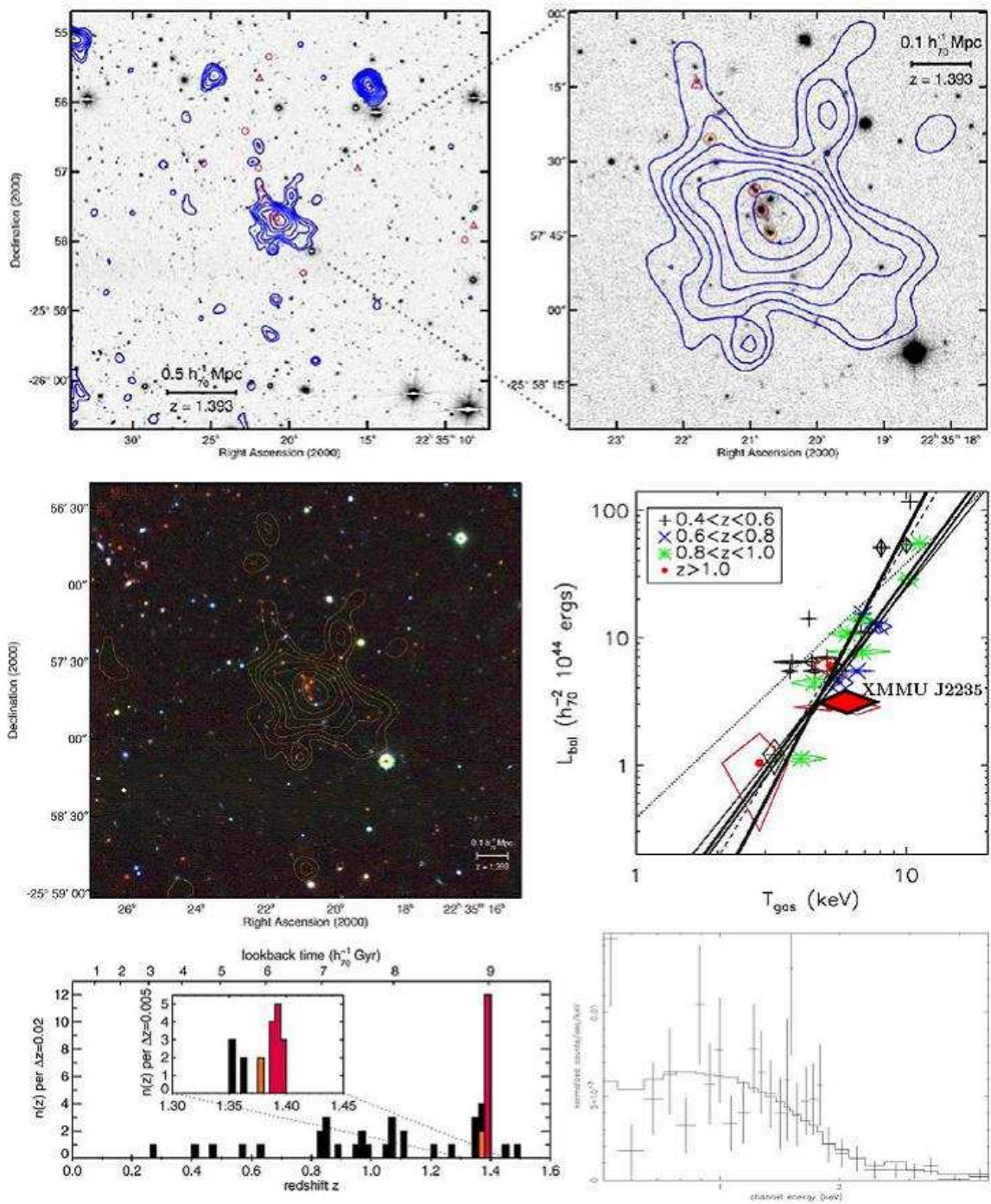}
\end{center}
\vspace{-4ex}
\caption[XMMU~J2235 Properties]{A detailed view of  galaxy cluster XMMU~J2235.3-2557 at $z\!=\!1.393$. {\em Top left:} VLT-FORS\,2 R-band image (1\,140\,s) with X-ray contours overlaid in blue. {\em Top right:} 1.5\arcmin$\times$1.5\arcmin \ zoom on a 3\,600\,s VLT-Isaac Ks-band image. The colored objects correspond to the red and orange peaks of the redshift histogram in the {\em bottom left panel}.
{\em Center left:} R+Z+Ks pseudo-color composite with field size  2.5\arcmin$\times$2.5\arcmin. {\em Center right:} XMMU~J2235.3-2557 (red filled diamond) in comparison to $L_{\mathrm{X}}$-$T$ relations as derived for different surveys \cite{Ettori2004a}. {\em Bottom right:} Spectral X-ray energy distribution of XMMU~J2235.3-2557 (vertical bars) in the 0.5--4.0\,keV range and the thermal model fit (solid line) based on the 280 X-ray photons in the archival XMM-Newton field. Plots from Mullis \etal \ \cite*{Mullis2005a} and H.~B\"ohringer.} \label{f5_xmmu2235}
\end{figure}

The cluster properties, as they have been derived from the survey data set (\ie \ steps 1--5), are summarized in Tab.\,\ref{t5_XMMU2235} and a more detailed view of XMMU\,J2235.3-2557 is presented in Fig.\,\ref{f5_xmmu2235}.
With  280 net source counts in the 0.3--4.5\,keV energy range, XMMU\,J2235.3-2557 is a fairly bright X-ray source which had been serendipitously detected, without optical identification, in a ROSAT pointed observation. The source flux of $(2.6 \pm 0.2)\!\times\!10^{-14}$\,\flux \ in the 0.5--2.0\,keV band with a corresponding total unabsorbed flux of $(3.6 \pm 0.3)\!\times\!10^{-14}$\,\flux \ yields a rest-frame luminosity estimate of 
 $(3.0 \pm 0.2)\!\times\!10^{44}$\,$h_{70}^{-2}$\,erg\,s$^{-1}$ (0.5--2.0\,keV). Assuming a thermal (MEKAL) emission model with a metallicity of $0.3\,Z_{\sun}$ and a galactic hydrogen absorption column of $n_{\mathrm{H}}\!=\!1.47\!\times\!10^{20}$\,cm$^{-2}$, a spectral temperature fit (lower right panel of Fig.\,\ref{f5_xmmu2235}) with  $kT_\mathrm{X} = 6.0^{+2.5}_{-1.8}$\,keV is obtained. With a mass estimate of $\sim\!3\!\times\!10^{14}$\,M$_{\sun}$ based on the $L_{\mathrm{X}}$--$M$ relation (see Fig.\,\ref{f4_LX_M_relation}),  XMMU\,J2235.3-2557 is likely the most massive galaxy cluster known at redshifts \zg1. The derived X-ray properties for the cluster are consistent, within errors, with the determined  $L_{\mathrm{X}}$--$T$  relations of other surveys as shown in the right central panel of Fig.\,\ref{f5_xmmu2235}.   

The follow-up imaging observations (top panels and central left panel of Fig.\,\ref{f5_xmmu2235}) of XMMU\,J2235.3-2557 revealed an old and evolved galaxy population in the cluster center already in place at a lookback time of 9\,Gyr. The brightest cluster galaxy exhibits an extended surface brightness profile typical for massive cluster cDs and has a spatial offset to the X-ray centroid of 3\farcs 7, corresponding to 31\,kpc at $z\!=\!1.393$, which might hint at a recent subcluster merger in the Northeast-Southwest direction. 
The spectroscopic redshift distribution in the bottom left panel of Fig.\,\ref{f5_xmmu2235} shows the clear redshift peak at $z\!\sim\!1.4$ with 12 confirmed cluster members (red) and two additional galaxies at slightly lower redhsift (orange). The preliminary velocity dispersion estimate based on this data is $762 \pm 262$\,km\,s$^{-1}$.


Meanwhile, XMMU\,J2235.3-2557 has been the subject of an extensive multi-wavelength follow-up program ({\bf step 6}). The above discussion has been purposely limited to the discovery data set\footnote{All preliminary cluster properties have been confirmed (within the stated errors) by the deeper follow-up observations.} since this will be typical in data quality and quantity for routine detections within the XDCP core survey program ({\bf step 1--5}). As an outlook of the follow-up science for the most interesting XDCP clusters, we will summarize the obtained follow-up observations and the principal science objectives for XMMU\,J2235.3-2557.     

In X-rays, the cluster has been re-observed with XMM-Newton for a nominal total exposure time of 80\,ksec complemented by a 200\,ksec observation with Chandra. 
The main science objectives for the high spatial resolution Chandra data are (i) the reliable measurement of the structural parameters free from the potential impact of point sources for an accurate mass determination, (ii) probing small-scale features 
(\eg \ substructure) 
to provide further insight into the global state of the system, and (iii) to test for the presence of a cool core. The enhanced spectroscopic performance of XMM-Newton in addition allows (iv) an ICM temperature determination to better than 10\%, (v) precise measurements of the X-ray surface brightness, gas mass, and gravitational mass profiles, (vi) the first ICM metallicity evaluation and detection of the iron line at this redshift, and (vii) test the predicted evolving scaling relation (\eg \ $L_{\mathrm{X}}$--$T$, $M$--$T$) and self-similarity of the surface brightness  and entropy structure. 
With all this at hand, XMMU\,J2235.3-2557 can be (viii) established as a `standard candle' at $z \sim 1.4$ and used for the cosmological baryon fraction test of Sect.\,\ref{s3_cosmo_tests}.

 


With the additional VLT FORS\,2 spectroscopy (ix) more cluster member galaxies can be identified 
for an improved velocity dispersion measurement and reference data set for photometric studies.
The imaging data baseline has been broadened to the R, I, Z, J, H, Ks, 3.6\,\microns, and 4.5\,\microns-bands, \ie \ from 0.6--4.5\,\microns \ corresponding to a rest-frame photometric coverage from the UV to the NIR. Deep I, Z, and H-band coverage has been obtained with the Hubble Space Telescope ACS and NICMOS cameras ($>$30\,hours in total), J (2\,700\,s) and  Ks-band (3\,600\,s) data with VLT ISAAC, and the mid-infrared 3.6\,\microns \ and 4.5\,\microns-bands with the Spitzer observatory (4\,hours in total).
This eight-band data set allows (x) detailed SED fitting of the spectroscopic members to yield formation ages, crude star formation histories and stellar masses, (xi) an accurate morphology mix determination based on the high-resolution HST data to probe whether the galaxy Hubble sequence is already in place, and (xii) a comparison of the ages and stellar masses of early-type cluster galaxies  with the field elliptical population at similar redshifts to probe the expected accelerated galaxy evolution in high-density environments. (xiii) The accurate measurement of the normalization, scatter, and slope of the Z--Ks \reds \ will accurately constrain the formation epoch and can probe evolutionary trends compared to lower redshift studies.
(xiv) The construction of the rest-frame NIR luminosity function out to M*+3 can shed light on the role of merging and stellar mass assembly of galaxies, since the faint-end slope is expected to steepen  at high redshift as the cluster galaxies in hierarchical models break up into their progenitors. Additional science objectives mentioned here are (xv) fundamental plane (FP) studies of the cluster ellipticals and (xvi) an investigation of the   
 cluster mass-to-light ratio in combination with the available X-ray data.

In summary, the XDCP discovery of the massive galaxy cluster XMMU\,J2235.3-2557 at $z\!=\!1.393$ has triggered an extensive follow-up program to address a number of open questions within the cluster and galaxy evolution scenarios to be discussed in detail in forthcoming publications.
At this point it should be emphasized again that the search for such rare massive high-redshift systems is best performed within the scope of large-area serendipitious surveys covering several dozen square degrees. 
For the discussions in the upcoming chapters one should keep in mind that less than one out of a hundred serendipitous extended X-ray sources will be of similar cluster quality and scientific impact. XMMU\,J2235.3-2557 has thus marked a very positive XDCP start and has encouraged the extended systematic survey for this thesis work.

\begin{table}[t]    
\begin{center}

\begin{tabular}{|c|c||c|c|}
\hline

\multicolumn{2}{|c||}{\bf{X-ray Properties}} & \multicolumn{2}{c|}{\bf{Optical Properties}} \\
 
\hline\hline

$f_\mathrm{X}$ \ [\flux] & $(2.6\!\pm\!0.2)\!\times\!10^{-14}$  & spectroscopic members  &  12 \\

$L_\mathrm{X}$ \ [$h_{70}^{-2}$\,erg\,s$^{-1}$] &  $(3.0\!\pm\!0.2)\!\times\!10^{44}$  & $\bar{z}$ &  1.393 \\
$kT_\mathrm{X}$  & $6.0^{+2.5}_{-1.8}$\,keV   &  velocity dispersion & $762 \pm 262$\,km\,s$^{-1}$\\
$M_\mathrm{X}$  &  $\sim\!3\!\times\!10^{14}$\,M$_{\sun}$  & color of \reds  & R--Z $\sim$ 2.1 \\
X-ray photons & $\sim$\,280  &   BCG type  &  cD-like    \\
core radius   &   $\sim$\,140\,kpc & BCG X-ray centroid offset  & 31\,kpc  \\

\hline
\end{tabular}

\caption[Properties of XMMU~J2235.3-2557]{X-ray and optical properties of galaxy cluster XMMU~J2235.3-2557 as derived from the initial data set. The unabsorbed flux $f_\mathrm{X}$ and the rest-frame X-ray luminosity $L_\mathrm{X}$ are given for the 0.5--2.0\,keV band.} \label{t5_XMMU2235}
\end{center}
\end{table}

\clearpage

\section{Thesis Data Overview}
\label{s5_pilot_study}

\noindent
Before turning to the detailed discussion of the individual survey steps in the next three chapters, a data overview is provided in Tab.\,\ref{t5_thesis_data}.
One of the prime goals for the thesis is a systematic XMM-Newton archival survey for serendipitous distant cluster sources with subsequent follow-up observations following the survey steps 1--5 of Sect.\,\ref{s5_survey_strategy}.
In total, the following chapters are based on more than 1\,TB of processed data products.

The XDCP survey area for the thesis project phase is defined by the XMM archival status as of November 2004 and consists of approximately 550 X-ray fields. The sky coverage is thus more than quadrupled compared to the pilot study and requires the processing of about 280\,GB of raw X-ray data, which will be presented in the next Chap.\,\ref{c6_XrayAnalysis}.

The second large amount of data included here has been acquired through several follow-up near-infrared imaging campaigns as discussed in Chap.\,\ref{c7_NIRanalysis}. These imaging observations cover about 40\% of all identified distant cluster candidates. 

The spectroscopic confirmation of the distant cluster candidates is currently gaining momentum but is still at an early stage with only few XDCP designated observations available. Achieving a spectroscopic survey completion will be the most challenging part and will realistically require another five years. The thesis work included the spectroscopic analysis of 35 low-redshift clusters for the NORAS\,2 survey (see Sect.\,\ref{s4_LocalSurveys}) which is not further discussed here. In order to maintain the balance for the current project status, Chap.\,\ref{c8_SpecAnalysis} covering the spectroscopy is kept short and is focussed on the main aspects.


\begin{table}[t]    
\begin{center}

\begin{tabular}{|l|c|c|c|c|}
\hline

\bf{Data} & \bf{\num Fields} & \bf{Raw Data} & \bf{Reduced Data} & \bf{Survey Fraction} \\
 
\hline\hline

XMM X-ray archive & 546 & 280\,GB & 600\,GB & 100\%\\
NIR imaging & 56 & 112\,GB & 480\,GB & 40\% \\
optical spectroscopy & 1 (+35) & 3\,GB & 8\,GB & 2\% \\

\hline
\end{tabular}

\caption[Thesis Data Overview]{Thesis data overview. The data type, number of different fields, and data amounts are indicated. The last column gives an estimate of the corresponding data fraction for the ongoing XDCP survey phase.} \label{t5_thesis_data}
\end{center}
\end{table}



\chapter{X-ray Analysis of XMM-Newton Archival Data}
\label{c6_XrayAnalysis}

\noindent 
In this chapter, we will discuss in detail the X-ray selection of distant cluster candidates. 
The first three sections will introduce the X-ray data, survey fields, and source detection procedures (survey {\bf step 1}). Section four will review the source classification and candidate selection (survey {\bf step 2}), and the final two sections will present test results and general diagnostic plots. 

The X-ray analysis constitutes the backbone of the XDCP survey and is crucial for the overall success of the project. 
The final goal of this X-ray part is the compilation of a well-selected sample of distant cluster candidates which should 
(i) exploit the maximum  sensitivity achievable with XMM archive data, (ii) minimize  the low-redshift cluster fraction, and (iii) include only an acceptable fraction of false positive sources while remaining highly complete for true high-redshift clusters.

The assessment of the extended nature of  X-ray sources is of prime importance for the cluster selection process. Estimations of the source extent and the X-ray flux are the main accessible observables  at this point of  the survey chain. In combination with a qualitative evaluation of the X-ray morphology, first clues to the dynamical state of the cluster sources can be obtained. After completion of survey {\bf steps 3--5}, the X-ray source properties can be transformed to physical object quantities such as luminosity, temperature (for brighter sources), and  a mass estimate.







\section{XMM-Newton in a Nutshell}
\label{s6_XMM}

\noindent
The XMM-Newton data archive constitutes the basis and starting point of the XDCP survey. A detailed understanding of the data properties,  capabilities, and limitations is hence the pre-requisite for a proper analysis and for the survey application. 
In this introductory section, the main XMM-Newton characteristics related to distant cluster searches are summarized and evaluated.  

\newpage


\begin{figure}[t]
\centering
\includegraphics[angle=0,clip,width=\textwidth]{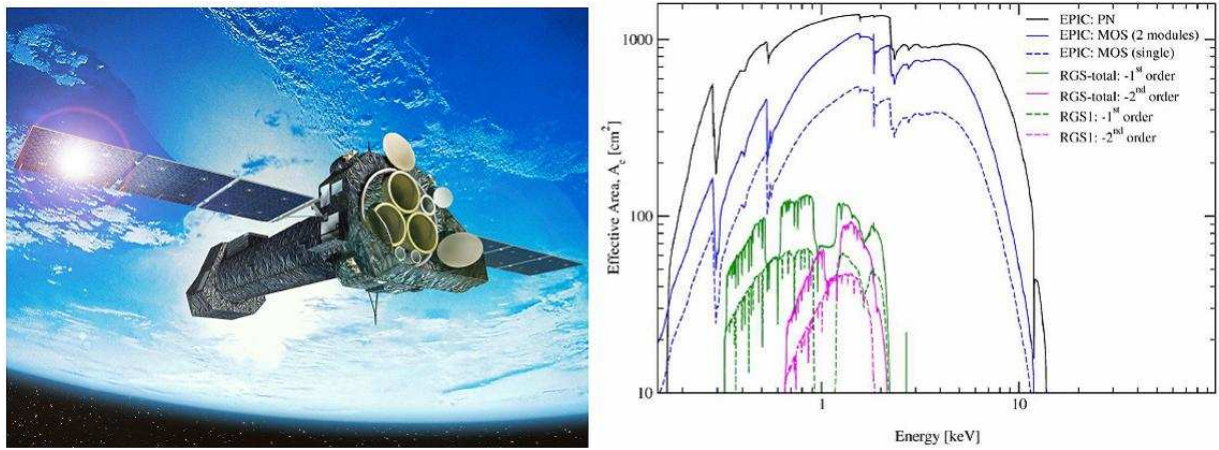}
\caption[XMM-Newton]{{\em Left:} The XMM-Newton observatory with its three X-ray telescopes. {\em Right:} Effective area as function of energy. The total XMM-Newton effective area for X-ray imaging consists of the sum of the two upper curves. Plots from Ehle \etal \ \cite*{xmm_manual}.
}
\label{f6_XMM_Newton}       
\end{figure}


XMM-Newton, the Xray Multi-Mirror Mission, was launched on  10 December 1999 as the second of ESAs four cornerstone missions of the {\em Horizon\,2000} program. 
With a length of more  than  10\,m and a weight of almost 4\,tons, XMM-Newton is the largest  scientific satellite ever built in Europe (see Fig.\,\ref{f6_XMM_Newton}).  
The observatory was launched 
into a highly elliptical 48\,hour orbit with a perigee of 6\,000\,km and an apogee\footnote{An orbital manoeuvre in February 2003 has changed the orbital parameters to the current perigee and apogee values of 20\,000\,km and 101\,000\,km \cite{xmm_manual}.} of 115\,000\,km   
  allowing about 130\,ksec of uninterrupted science observations per revolution outside Earth's radiation belts\footnote{Science observations can typically be conducted at elevations above 46\,000\,km.} .

The main scientific payload consists of three co-aligned X-ray mirror systems made up of 58 nested mirror shells each. This large number of Wolter type X-ray mirrors provide an unprecedented effective geometric\footnote{The combined area, \ie \ the sum of all thin shell-like regions from where (on-axis) photons are focussed onto the detector, is equivalent to the area of a circle with a diameter of 44.4\,cm.} area of 1\,550\,cm$^2$ per module. The total gathering power is shared by five simultaneously operating focal plane X-ray instruments, two Reflection Grating Spectrometers (RGS) and three European Photon Imaging Cameras (EPIC). 

The XMM-Newton EPIC imaging instruments use two different kinds of X-ray CCD technologies. The two identical  MOS (Metal Oxide Semiconductors) cameras use seven front-illuminated $600\!\times\!600$ pixel CCDs with a frame store region as data buffer. Each MOS camera shares a mirror module with one of the RGS spectrometers resulting in a reduced effective area for imaging of 44\%. The more sensitive PN camera,\footnote{PN is derived from the pn-junction of the silicon semiconductor technology used in the detectors. For easier readability,  the capitalized form PN, instead of pn, is used throughout the thesis.} on the other hand, receives the full reflected light of one of the three X-ray telescopes. The twelve  backside-illuminated CCD chips with a total of $376\!\times\!384$ pixels are fabricated from a single Silicon wafer. They have a higher quantum efficiency (QE$_{\mathrm{PN}}\!>\!90\%$) than the MOS detectors (QE$_{\mathrm{MOS}}\!\sim$40--85\%), and allow higher frame rates through the integration of readout nodes for each individual pixel column. The PN chips lack frame store buffers, which in practice has the consequence of  so-called out-of-time events (OoT). Since the pixels still register incoming X-ray events during the column readout phase, which lasts a few percent  of the full integration cycle,\footnote{The out-of-time fraction is 6.3\% for full frame mode and 2.3\% for the extended full frame mode.} bright sources will imprint a smeared event streak 
over the full pixel column (Y direction) due to falsely identified  photon positions during  readout.      

For survey applications, we are mainly  interested in full-frame imaging mode observations of the instruments.\footnote{Various other timing, burst, or small window modes are available for different science applications.} In contrast to optical or NIR pixel arrays (see Chaps.\,\ref{c7_NIRanalysis}\,\&\,\ref{c8_SpecAnalysis}), X-ray CCDs are operated in single photon counting mode enabling non-dispersive low resolution spectroscopy. Incident X-ray photons produce a number of recorded electrons inside the detector pixel via an internal photoelectric effect which scales with the X-ray energy. If the CCD readout is sufficiently fast to  ensure at most a single X-ray photon interaction per pixel neighborhood\footnote{X-ray photons can deposit their energy  over several connected pixels, a pixel cluster, which have to be isolated spatially and temporally for an accurate energy reconstruction. Typically, only single and double pixel events are considered for the science analysis.} and clocking cycle, the  energy resolution 
is achieved by matching the measured pixel signal to a calibrated energy.\footnote{Pile-up events, \ie \ several photons  recorded in the same pixel, can be identified and discarded during the data processing.} 



Table\,\ref{t6_XMM_properties} summarizes the most important XMM-Newton imaging mode characteristics. After the general overview, we will now discuss the main instrumental factors for distant galaxy cluster searches ordered with decreasing priority.   

\begin{figure}[t]
\centering
\includegraphics[angle=0,clip,width=0.495\textwidth]{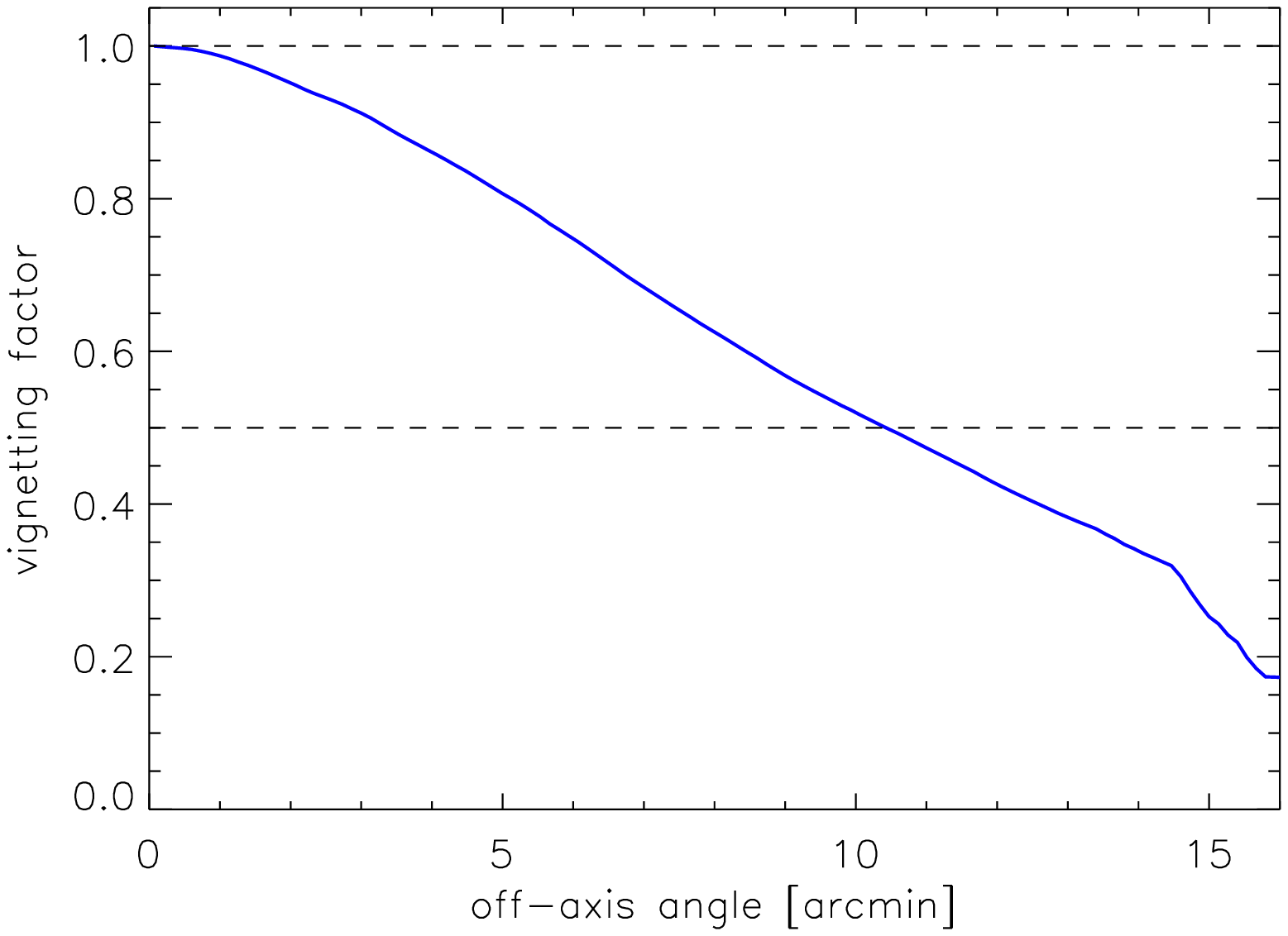}
\includegraphics[angle=0,clip,width=0.495\textwidth]{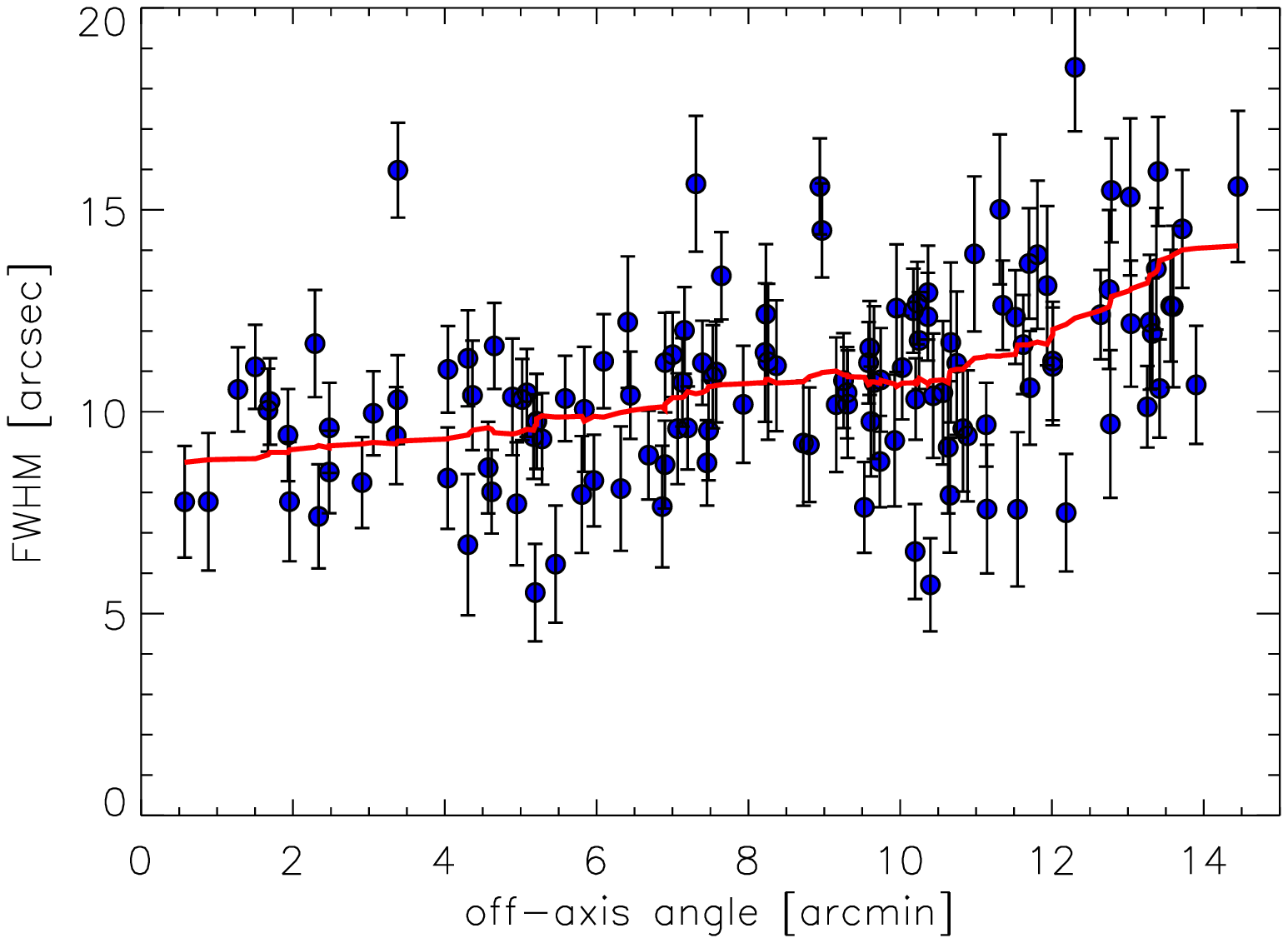}
\vspace{-4ex}
\caption[XMM Vignetting and PSF Functions]{XMM vignetting and PSF properties as function of off-axis angle. {\em Left:} Average vignetting function of all three XMM X-ray telescope systems, with the PN optical axis used as central reference point. The horizontal dashed lines indicate the 100\% and 50\% effective area levels, the latter is reached at an off-axis angle of about 10.5\,arcmin. The sharp drop starting at $\sim$\,14.5\,arcmin off-axis angle indicates the onset of the FoV edge of some detectors.  {\em Right:} Gaussian FWHM of sources as determined in three deep fields. The measurements were performed in the combined image of all three detectors. 
The error bar length is proportional to the ratio of the FWHM in the X and Y direction, \ie \ to the amount of PSF asymmetry. The red solid line illustrates the smoothed trend of the average FWHM between about 8\arcsec \ in the center and 15\arcsec \ at the FoV edge. }

\label{f6_vignetting}       
\end{figure}

\begin{figure}[b]
\centering
\includegraphics[angle=0,clip,width=0.80\textwidth]{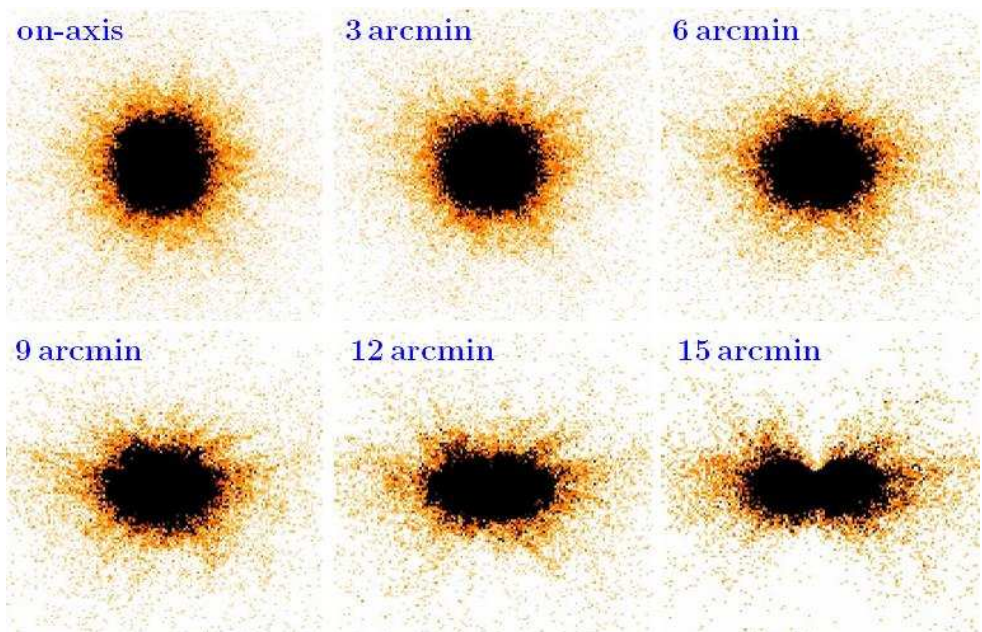}
\vspace{-1ex}
\caption[XMM PSF Shape]{XMM PSF shape variations as a function of off-axis angle. Shown are the 2-D PSF images of the PN mirror module in off-axis angle steps of 3\,arcmin  as provided by the XMM calibration data base. The increasing asymmetry towards the FoV edge poses a bigger problem to the detection of extended sources than the moderate overall broadening of the PSF core.
}

\label{f6_xmm_PSF}       
\end{figure}

\begin{description}
    \item[Effective Area:] 
    The effective area $A_{\mathrm{e}}$ is the prime instrument characteristic related to the achievable sensitivity per unit time. The effective {\em instrumental} photon gathering power  is determined as the product of the geometric area of the mirror modules, their reflectivity,  and the quantum efficiency of the detectors, shown in the right panel of Fig.\,\ref{f6_XMM_Newton} for all XMM-Newton camera systems. The solid black  line illustrates  $A_{\mathrm{e}}$ as a function of photon energy for the PN camera as the most sensitive imaging instrument. The two MOS\footnote{The PN camera is effectively a factor of 3.3 more sensitive than the individual MOS cameras at 1\,keV. A factor of 2.3 is attributed to the different geometric  mirror area, the remaining factor of 1.5 is due to the QE differences.} detectors provide lower quantum efficiencies and are only exposed to 44\% of the geometric mirror area resulting in the  dotted blue line for the individual instrument and the solid blue line for the sum of both modules.    
The total effective area for all three XMM imaging instruments adds up to $A_{\mathrm{e}}\!\simeq\!2\,500$\,cm$^2$ at 1\,keV. 
XMM-Newton achieves a  photon gathering power 
 which outperforms the Chandra observatory sensitivity by about a factor of four, and roughly by a factor of ten in comparison to ROSAT (\eg \ Romer \etal, 2001). \nocite{Romer2001a}

The grazing-incidence reflection efficiency of X-ray photons on the mirror shells is very sensitive to the incident angle with respect to the optical axis of the module. This gives rise to the {\em vignetting} \ effect, \ie \ the decreasing effective area with increasing off-axis angle from the optical axis, which is shown in the left panel of Fig.\,\ref{f6_vignetting}.
At $\Theta\!\simeq\!10.5$\,arcmin the photon gathering power has dropped to 50\% of the central value and continues to fall towards the outskirts of the FoV. An integration of the vignetting function over the detector area out to a given off-axis angle yields the average vignetting factor $\bar{V}_\mathrm{vig}(\leq\!\Theta)$ as
$\bar{V}_\mathrm{vig}(\leq\!10\arcmin)\!\simeq\!0.72$, $\bar{V}_\mathrm{vig}(\leq\!12\arcmin)\!\simeq\!0.65$, and $\bar{V}_\mathrm{vig}(\leq\!15\arcmin)\!\simeq\!0.55$.


    \item[Field-of-View:] 
    The total XMM-Newton field-of-view is about 30\,arcmin in diameter, corresponding to $\sim$\,0.2\,deg$^2$.
    The relevant instrument characteristic for survey applications is the product of the FoV solid angle and the effective area, the  so-called {\em grasp} $g\!\equiv\!A_{\mathrm{e}}\!\cdot\!\Omega$, which determines the required time to cover a unit sky area to a specified depth. Taking the average vignetting factors into account, we obtain (at 1.5\,keV) $g_{15}\!\simeq\!270$\,cm$^2$\,deg$^2$ for a maximum off-axis angle of 15\,arcmin, and $g_{12}\!\simeq\!204$\,cm$^2$\,deg$^2$ and $g_{10}\!\simeq\!157$\,cm$^2$\,deg$^2$ when restricting to FoV to off-axis angles of 12\,arcmin or 10\,arcmin, respectively. The motivation to constrain the field-of-view for a cluster survey arises from the PSF properties discussed below.
    
\enlargethispage{4ex} 

    \item[Spatial Resolution:] 
    The first two items determine the number of photons per source and unit time and the total sky area per observation.
    The most crucial point for any serendipitous cluster survey is the assessment of the source extent properties, which serve as  primary selection criteria. Simply speaking, a source can be spatially resolved if its extent is larger than the instrumental point-spread-function (PSF). This implies that an accurate knowledge of the PSF shape at the source position is the pre-requisite for determining a spatial extent of the source. As was the case for the vignetting factor, the PSF  varies strongly as a function of off-axis angle. Figure\,\ref{f6_xmm_PSF} illustrates the changing shape of the XMM point-spread-function in 3\,arcmin steps from the center. The increasing asymmetry and complexity towards the FoV outskirts  is obvious, and is the main reason for prevailing uncertainties in the off-axis PSF calibration even after more than seven years of XMM operations.

The central PSF is well described by a King profile of the form

\begin{equation}\label{e6_psf_king}
    PSF(r) = \frac{A}{\left[1 + \left( \frac{r}{r_{\mathrm{c}}} \right)^2  \right]^{\alpha}} \ ,
\end{equation}

with core radius $r_{\mathrm{c}}$, index $\alpha$, and normalization $A$.  All fitting parameters are 
functions of off-axis angle and energy. The on-axis PSF core radii at 1.5\,keV have been determined\footnote{PSF calibration documentation EPIC-MCT-TN-012, 2002 by S. Ghizzardi. See \url{http://xmm.vilspa.esa.es/external/xmm_sw_cal/calib/documentation/index.shtml}. }  as $r_{\mathrm{c}}(\mathrm{MOS1})\!\simeq\!4.7\arcsec$, $r_{\mathrm{c}}(\mathrm{MOS2})\!\simeq\!4.5\arcsec$, and $r_{\mathrm{c}}(\mathrm{PN})\!\simeq\!6.2\arcsec$. 
The instrumental XMM on-axis resolution limit is hence set at cluster core radii of $r_{\mathrm{c}}\!\simeq$\,5\arcsec--6\arcsec, corresponding to a physical resolved cluster core scale at \zg1 (see Fig.\,\ref{f3_cosmological_distances}) of $r_{\mathrm{c}}^\mathrm{lim}\!\sim\!50$\,kpc.
However, the on-axis PSF sets a lower ideal resolution limit, which is not applicable outside the very center of the FoV. From the lower row of PSF images in Fig.\,\ref{f6_xmm_PSF} it becomes obvious that the azimuthally symmetric King profile is not a good approximation to the increasing asymmetry of the core. Meaningful calibration data characterizing the off-axis PSF is sparse, and a formal application of the King profile fit even returns a decreasing core radius at increasing off-axis angles. 
A more practical illustration of the off-axis dependence on the resolution limit is given in the right panel of  Fig.\,\ref{f6_vignetting}, which shows Gaussian full-width at half-maximum (FWHM) measurements of sources in three deep fields. 
The combined photon images of all three detectors were used, thus returning the properties of the average XMM PSF. The red line in Fig.\,\ref{f6_vignetting} indicates the averaged results and is consistent with the expected off-axis trend of a FWHM broadening from about 8\arcsec \ in the center to roughly 14\arcsec--15\arcsec \ in the detector periphery. 
The uncertainty in the off-axis PSF calibration is one of the main reasons for the detection of spurious extended sources and has to be adequately taken into account for the source classification.  

\enlargethispage{4ex} 
    
    \item[Background Characteristics:]
The X-ray background is the second critical issue for the detection of extended sources, as uncertainties in the
 model background can also result in spurious extended sources.
For low surface brightness objects such as clusters, the X-ray background is the prime limitation of the achievable sensitivity.
Concerning the background properties, XMM-Newton is not in an ideal orbit (see Sect.\,\ref{s6_field_ statistics}) as observations are contaminated from frequent background flaring events and exhibit rather high quiescent levels. 
This has the consequence that cluster studies require significant efforts for the suppression and proper treatment of the X-ray background (see Sect.\,\ref{s6_Xray_data_reduction}).
    
    \item[Energy Resolution:] 
The XMM-Newton energy resolution in imaging mode of $\sim\!80$\,eV at 1\,keV is the least critical among the discussed items for distant cluster searches. However, it enables (i) the {\em a posteriori} definition of band-passes, (ii) the assessment of hardness ratios for all sources, and (iii) more detailed spectroscopic studies of confirmed clusters, \eg \ for a temperature determination of sufficiently bright sources.     
  
\end{description}      
    



\begin{table}[t]    
\begin{center}

\begin{tabular}{|c|c|}
\hline

\bf{Characteristic} & \bf{XMM Performance}   \\
 
\hline\hline

number of telescopes and imaging detectors & 3 \\
total effective area at 1keV &  $\sim$2\,500\,cm$^{2}$ \\
field-of-view & 30\arcmin \ diameter $\simeq$ 0.2\,deg$^2$ \\
spatial resolution (FWHM) & 5\arcsec--15\arcsec \\
half-energy width (HEW)  &  14\arcsec--20\arcsec        \\  
pixel scale PN / MOS     &  4\farcs 1 / 1\farcs 1 \ per pixel \\
energy resolution at 1keV & $\sim\!80$\,eV \\
 point source sensitivity in 10\,ksec (all cameras) & $\sim\!5\!\times\!10^{15}$\,\flux  \\

\hline
\end{tabular}

\caption[XMM-Newton in a Nutshell]{XMM-Newton imaging-mode characteristics. } \label{t6_XMM_properties}
\end{center}
\end{table}

\clearpage


\vfill

\clearpage
\section{X-ray Data Archive and Survey Definition}
\label{s6_survey_definition}

\noindent
All five X-ray instruments of the XMM-Newton observatory are operating simultaneously during targeted observations. Even for science programs that are not focussed on XMM's imaging capabilities, any idle PN or MOS camera proceed with the data acquisition. 
All observations are collected in the online XMM data archive\footnote{The XMM-Newton data archive can be accessed through \url{http://xmm.vilspa.esa.es/external/xmm_data_acc/xsa/index.shtml}.}, where they become publicly available after a proprietary period of typically twelve months. The XMM archive is hence constantly growing with now more than seven years of successful operations.

\begin{figure}[t]
\centering
\includegraphics[angle=0,clip,width=0.82\textwidth]{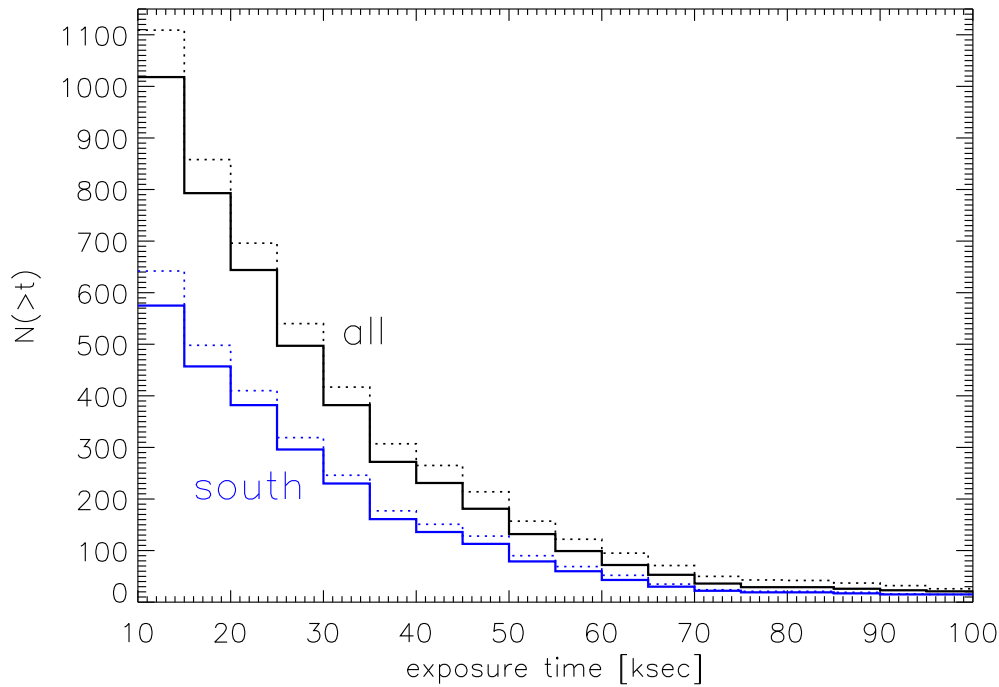}
\vspace{-1ex}
\caption[XMM-Newton Archival Fields]{Cumulative distribution of XMM-Newton archival fields suitable for a serendipitous survey as of status November 2004. The blue solid line shows the Southern sample (DEC$<\!+20$\degr) of 575 XMM-Newton observations used for the analysis in this thesis. The black solid line depicts the full sample including the Northern sky, dashed lines include in addition observations of designated deep surveys (\eg \ XMM-LSS, COSMOS, see Sect.\,\ref{s4_distant_cl2004}).}
\label{f6_ArchiveFields}       
\end{figure}

\begin{figure}[b]
\centering
\includegraphics[angle=0,clip,width=0.73\textwidth]{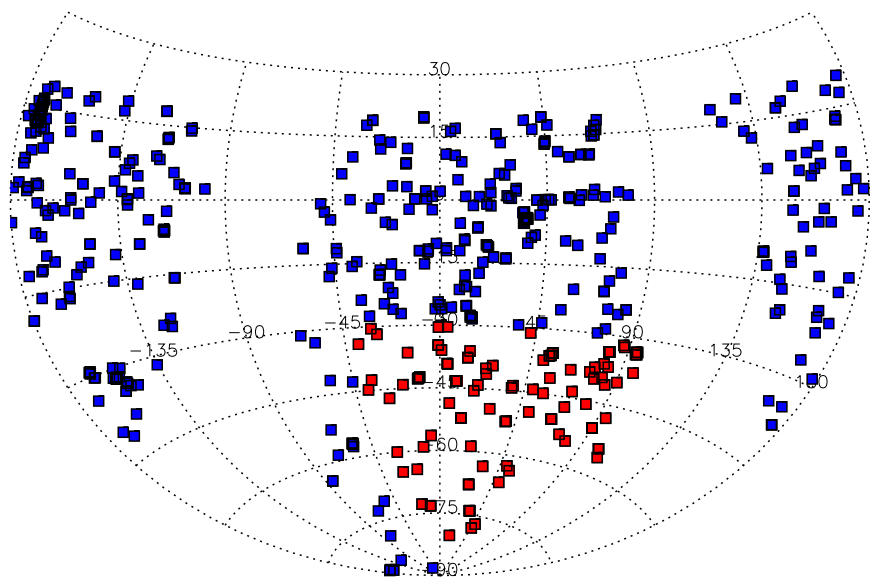}
\vspace{-1ex}
\caption[XMM-Newton Sky Coverage]{Sky distribution of the 575 Southern XMM-Newton fields suitable for a survey ({\em raw XDCP sample}). The red squares indicate fields which will be covered by SZE observations with the South Pole Telescope. Square symbols are not to scale.}
\label{f6_Sky_Coverage}       
\end{figure}

In order to obtain a well controlled sample for the current project phase, the general XDCP survey field definition has been frozen at the archive status of {\bf 2 November 2004}. After almost five years of in-orbit operations, the XMM data archive   
exhibited the following statistical properties\footnote{The software for processing XMM archive lists and for automated data download was originally written by Chris Mullis.}:

\begin{itemize}
    \item 2960 fields with expired proprietary period  were available in total with a total nominal exposure time sum of $\Sigma\!\simeq\!72.3$\,Msec;
    \item 2016 of these observations had at least one camera operating in full frame imaging mode;
    \item 1910 remained after rejecting sky regions `contaminated' by the Large and Small Magellanic Clouds and by the Andromeda galaxy M31; 
    \item 1109 fields were located at sufficiently high galactic latitudes with $|b|\!\geq\!20\degr$ and had nominal exposure times of more than 10\,ksec (see dotted black line in Fig.\,\ref{f6_ArchiveFields} for exposure time distribution);
    \item 1018 observations were not attributed to designated survey programs with separate source identification programs, \eg \ XMM-LSS, COSMOS, Chandra Deep Fields (black solid line in  Fig.\,\ref{f6_ArchiveFields});
    \item 575 fields were VLT accessible at declinations DEC$\leq\!+20$\degr \ (blue solid line) defining the {\em XDCP raw survey sample} with the sky distribution shown in Fig.\,\ref{f6_Sky_Coverage};
    \item 546 data sets (see field list in Tab.\,\ref{tA_field_list}) passed a first quicklook check and were  processed with the XDCP X-ray reduction pipeline ($\Sigma\!\simeq\!17.5$\,Msec); 
    \item  {\bf 469} fields were fully analyzed and are in principle  suitable for distant cluster searches,  defining the {\bf XDCP core survey sample} \ ($\Sigma\!\simeq\!15.2$\,Msec);
    \item 106 analyzed observations are located within the South Pole Telescope SZE survey region at DEC$<\!-30\deg$ and  $21\,\mathrm{h}\!<\!\mathrm{RA}\!<\!7\,\mathrm{h}$ indicated by red symbols in Fig.\,\ref{f6_Sky_Coverage} ($\Sigma\!\simeq\!3.3$\,Msec).
\end{itemize}









\clearpage

\section{Development of an X-ray Reduction and Analysis Pipeline}
\label{s6_Xray_Pipeline}


\noindent
In this section, the basic concepts of the X-ray data reduction and source detection processes are introduced.
The handling of hundreds of XDCP survey field data sets required the development of a designated, distant cluster-optimized  X-ray reduction pipeline. 
The driver and final goal of the X-ray pipeline development is 
the maximally achievable sensitivity for faint extended cluster sources in the archival data. This development is based upon the available tools of the XMM Science Analysis Software\footnote{The SAS homepage can be found at \url{http://xmm.vilspa.esa.es/sas}.} ({\tt SAS}), the mission software package to reduce and analyze XMM data.\footnote{{\tt SAS} tasks such as {\tt eboxdetect} are set in typewriter font for easy identification.} In particular for the 
 critical source detection procedure, the performance of the {\tt SAS} tasks impose a `boundary condition' for the survey outcome.  

The overall reduction and analysis approach  aims at the compilation of a versatile {\em raw} survey masterlist, obtained by homogeneously  applying the same procedure to all fields of the predefined XDCP sample. These {\em raw} results should comprise a super-set of the finalized survey sample in terms of number of analyzed fields, detector area, significance thresholds, and source characterization and should enable cross-checks and performance evaluations. 
The final XDCP {\em science survey sample} can then be defined {\em a posteriori} based on optimized and more restrictive cuts concerning (i) the flux limit, (ii) significance thresholds, (iii) usable detector area, and (iv) survey field selection. 
Since the XDCP is pushing the limits of   
what is achievable  with XMM  for distant cluster detection,
this versatile approach promises the most flexibility for selection adjustments and the final survey characterizations once additional observational  feedback is available.





\subsection{Search strategy}
\label{s6_search_strategy}

\enlargethispage{4ex}

\noindent
For faint object detections in any waveband, the crucial 
 characteristics is always the signal-to-noise ratio (SNR), \ie \ the 
source signal in a given aperture divided by the total noise in the same area.   
A first possible distant cluster optimization is hence related to the 
energy range of the applied detection bands.
As discussed in Sect.\,\ref{s2_ICM_properties}, the observed thermal bremsstrahlung spectra of plasmas at temperature $T$  (Equ.\,\ref{e2_bremstr_emissivity}) are characterized by an exponential cut-off 
 beyond energies of $E_{\mathrm{cut}}\!\simeq\!k_{\mathrm{B}}T$ at the high-energy end and a galactic hydrogen absorption suppression at the low-energy end of  $E_{\mathrm{abs}}\!\la\!0.3$\,keV.
While the latter spectral boundary is redshift-independent, the first one scales as $E_{\mathrm{cut}}(z)\!\simeq\!k_{\mathrm{B}}T/(1\!+\!z)$. A fiducial $T\!\sim\!5$\,keV cluster at \zsim1 \  will hence exhibit the spectral break at about 2.5\,keV, which will shift towards lower energies at even higher redshifts.    
These simple considerations imply that distant cluster searches should be focussed on the soft bands of the XMM sensitivity range.

\begin{figure}[t]
\begin{center}
\includegraphics[angle=0,clip,width=0.73\textwidth]{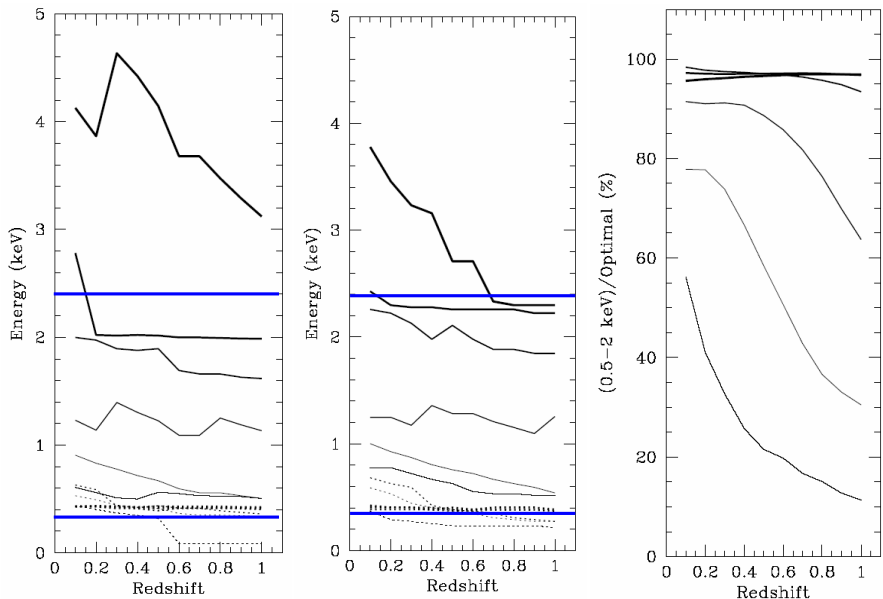}
\end{center}
\vspace{-4ex}
\caption[Optimized Detection Bands]{Optimized X-ray source detection bands for galaxy clusters and groups as derived by Scharf \cite*{Scharf2002a}. {\em Left:} Upper (black solid lines) and lower (black dotted lines) detection band-pass versus object redshift for the XMM MOS detectors for systems with different temperatures. From top to bottom the assumed cluster/group temperature is 6, 4, 2, 1, 0.5, 0.2\,keV. The blue solid lines illustrate the XDCP detection band with an energy range of 0.35--2.4\,keV, which was chosen to yield an optimized high-redshift detection performance over a wide range of temperatures.  {\em Center:} Same plot for the XMM PN detector. {\em Right:} Expected PN signal-to-noise ratio of the standard 0.5--2.0\,keV band divided by the SNR of the optimized bands (black lines) in the center panel. The SNR gain is most notable for lower temperature systems at $T\!\la\!2$\,keV. Plots adapted from Scharf \cite*{Scharf2002a}.} \label{f6_Optimal_Xbands}
\end{figure}
\begin{figure}[b]
\begin{center}
\includegraphics[angle=0,clip,width=0.60\textwidth]{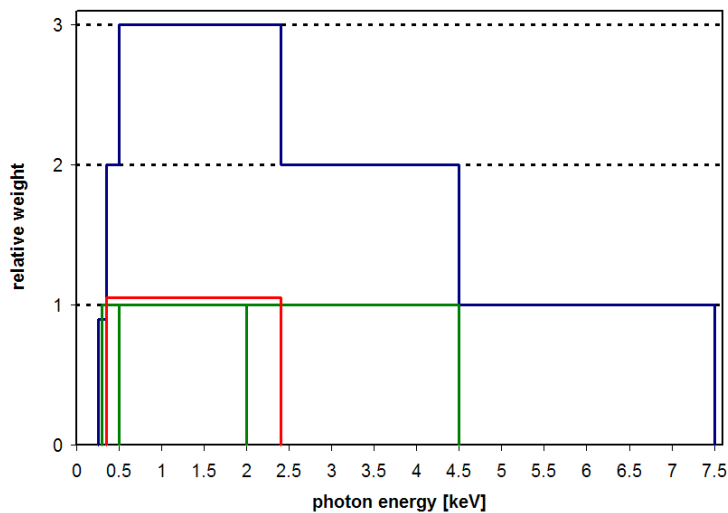}
\end{center}
\vspace{-4ex}
\caption[Detection Band Schemes]{Detection schemes. The relative photon weight factor is plotted against the energy for the unweighted {\em single band scheme} (red line), the three-band {\em standard scheme} (green), and the {\em spectral matched filter scheme} (blue) featuring different photon weights across the spectral range. Small offsets are applied for easier distinction of band boundaries.} \label{f6_Xdetection_schemes}
\end{figure}

Scharf \cite*{Scharf2002a} has determined the optimal XMM detection bands for yielding the highest SNR for galaxy clusters and groups with different temperatures and redshifts. Figure\,\ref{f6_Optimal_Xbands} summarizes these results for the MOS detectors (left panel), the PN instrument (center panel), and the  SNR improvement for the PN (right panel).
The XDCP prime target objects are \zga1 \ clusters with temperatures $2\,\mathrm{keV}\!\la\!T\!\la\!8\,\mathrm{keV}$, \ie \ we are mainly interested in the extrapolation of the upper three black lines for 6\,keV, 4\,keV, and 2\,keV model clusters in  Fig.\,\ref{f6_Optimal_Xbands}. Based on this study, a single optimized XDCP distant cluster detection band with an energy range of 0.35--2.4\,keV (blue lines) was defined, which promises an enhanced performance 
for distant objects over a wide range of temperatures. 
The lower limit of this optimized  band is essentially representing the average of the case study model clusters (dotted lines in Fig.\,\ref{f6_Optimal_Xbands}), the upper limit is motivated by the high-$z$ trends of hot systems (upper solid lines of left and central panel). Two main aspects determining the efficacy of an optimized single detection band are (i) the SNR of the sources as prime parameter for the {\em detection}, and (ii) the number of source photons for the source {\em characterization}, \ie \ in particular the extent measurement (see Fig.\,\ref{f6_EXTML_counts}).
Whereas the achieved improvements of the SNR for hot systems are moderate compared to a standard 0.5--2.0\,keV band-pass (see right panel of Fig.\,\ref{f6_Optimal_Xbands}), the total number of photons for distant cluster sources  can be significantly increased ($\sim\!30$--50\%) for the optimized, wider 0.35--2.4\,keV band, which in turn yields more accurate source parameters.
In the following, three different band-pass schemes are defined, which aim at a detection optimization in different ways:






\begin{description}

\item[(1) Single Band Scheme:] 
The conceptually simplest search strategy for distant clusters is the use of a single detection band.
As discussed, 
the 0.35-2.4\,keV band is expected to
optimize the target signal-to-noise ratio {\em and\/} the total number of source photons over a wide range of the expected source characteristics.
This band-pass is used as the primary XDCP detection scheme
and almost coincides with the ROSAT band (0.1--2.4\,keV). 
A single defined energy band will simplify the survey calibration and the determination of the selection function through extensive simulations (see Sect.\,\ref{s9_Selection_Funct}).   


\item[(2) Spectral Matched Filter Scheme (SMF):] 
The first detection scheme does not 
allow the determination of spectral hardness ratio properties and additionally makes use of justified but fairly strong priors on the expected spectral source properties. It is conceivable that very hot systems at $T\!\ga\!10$\,keV exist in the distant Universe, \eg \ as the result of a recent major merger event. In order to broaden the energy baseline and hence increase the sensitivity for hotter objects, the second scheme covers the full 0.3--7.5\,keV range, but assigns different weights to parts of the spectral range, as shown by the blue line in Fig.\,\ref{f6_Xdetection_schemes}.        
By enhancing the softer energy regions, the shape of the weight function roughly follows the (expected) continuum form of hot distant clusters (see lower right panel of Fig.\,\ref{f5_xmmu2235}) and is consequently termed {\em spectral matched filter scheme} (SMF). The effective weight function is achieved in practice by using five  (partially) overlapping\footnote{For the source characterization and the associated likelihoods, this implementation is  statistically equivalent to  the use of a single broad band (0.3--7.5\,keV) with different assigned weights over the spectral range.}  bands (0.3--0.5, 0.35-2.4, 0.5--2.0, 2.0--4.5, 0.5--7.5\,keV). The SMF scheme is to be considered as an {\em experimental\/} approach, which has yielded promising performance results in early comparison tests. The empirically determined likelihood offset factors for the significance evaluation and the comparison to the other schemes will be discussed in Sect.\,\ref{s6_extendedsource_ statistics}.


\item[(3) Standard Scheme (STD):] 
The third applied scheme makes use of the combination of three XMM standard bands in the range 0.3--0.5\,keV, 0.5--2.0\,keV, and 2.0--4.5\,keV, \ie \ it provides continuous coverage from 0.3--4.5\,keV. This approach is widely used and can hence be considered as the {\em standard} scheme.

\end{description}

\begin{samepage}
\enlargethispage{4ex}

\noindent
Figure\,\ref{f6_Xdetection_schemes} summarizes all three detection schemes and their energy coverage.
All methods offer some advantages and complementarity. The first detection scheme is attractive for its simplicity, the second for its additional sensitivity to `exotic systems', and the third for its standardized 
 source parameters.    
Three main reasons motivated the application of a complementary multi-scheme detection method:

\begin{itemize}

    \item achieving the best possible survey completeness of detected extended sources;
    \item enabling a cross-comparison of methods to test for systematic selection effects;
    \item increasing the reliability of the source classification close to the detection threshold by 
    evaluating the stability of important source parameter solutions.
\end{itemize}      

\end{samepage}

\subsection{X-ray data reduction}
\label{s6_Xray_data_reduction}

\begin{figure}[t]
\begin{center}
\includegraphics[angle=0,clip,height=21cm]{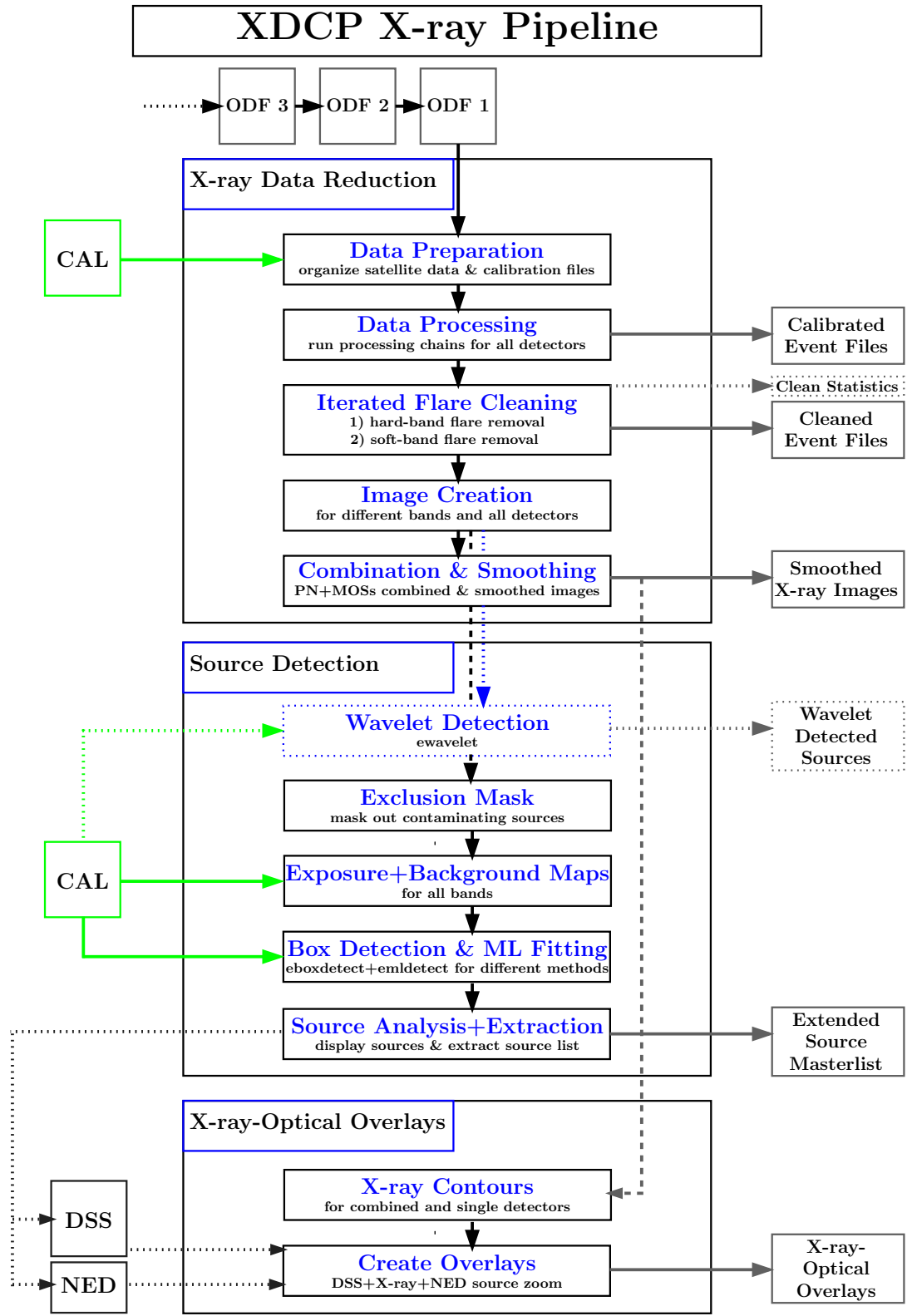}
\end{center}
\caption[X-ray Pipeline Flow Chart]{X-ray reduction and analysis pipeline flow chart.} \label{f6_Xray_pipeline}
\end{figure}

\noindent
The general outline of the developed XDCP X-ray reduction pipeline is illustrated in Fig.\,\ref{f6_Xray_pipeline}. 
Raw XMM-Newton data for each observation are organized in a so-called Observation Data File (ODF), a directory with a collection of $\sim\!200$ files containing the uncalibrated  data of all instruments, satellite attitude files, and calibration information. 
ODFs have unique identifications numbers (OBSID)  and can be considered as the basic starting data set from which all science products can be derived. 
An arbitrarily long list of ODFs serves as the input for the X-ray pipeline and 
is passed on to the 
top-level script, which organizes the ODF handling and the calls of the individual reduction modules. In practice, 1--3 modules 
are applied to typically 100 data sets at a time, followed by quality and completion checks of the intermediate data products. All data sets have been homogenously\footnote{An early reduction run of 120 XMM observations using {\tt SAS\,6.1} has been repeated with {\tt SAS\,6.5}, released in August 2005.} reduced with the XMM Science Analysis Software
version {\tt SAS\,6.5}.

The full XDCP pipeline can be subdivided into three main parts: (i) The actual {\em X-ray data reduction} modules, which create calibrated science images from the raw satellite data, (ii) the {\em source detection} and characterization procedures  resulting in lists of extended sources, and (iii) the {\em interactive\/} individual  detailed source analysis and classification using {\em X-ray-optical image overlays}. In this sub-section, we will have a closer look at the data reduction modules\footnote{The individual X-ray reduction modules are adapted and extended versions of codes originally written by Hans B\"ohringer.} of the first part.



\clearpage





\subsubsection{XMM data setup}

\noindent
The first task is to transform the XMM sets of raw satellite data, each ODF containing typically $\sim$500\,MB of data, 
into a suitable format for further processing.
This is achieved with few straightforward {\tt SAS} procedures which keep the complexity of the underlying raw data hidden from the user. The task {\tt cifbuild} matches the input data with the appropriate calibration files, which are  accessible through the so-called Calibration Access Layer (CAL). {\tt odfingest} extracts and organizes the satellite housekeeping data necessary to accurately reconstruct the observations. The instrument specific processing chains {\tt epchain} for the PN  and {\tt emchain} for the MOS cameras generate calibrated photon event lists for the complete observation with reconstructed sky positions, energies, arrival times, and event characteristics for each individual X-ray photon.





In Sect.\,\ref{s6_survey_definition}, one of the survey field selection criteria was a minimal {\em nominal} exposure time of 10\,ksec, which is the period XMM was pointing at the target location. However, the {\em nominal} on-target times are not equivalent to the effective {\em science-usable clean} exposure times, which can be significantly lower for two main reasons.
(i) Instrument calibration overheads at the beginning of an observation require about 0.5--1.5\,h for the PN, and 10\,min for the MOS instruments in order to compute the {\em  optical loading} of the CCD chips, \ie \ the signal fraction attributed to optical light rather than X-ray photons. The more sensitive PN camera  hence achieves on average about 1\,h less on-target time than the MOS instruments. (ii) In addition to the constant hardware overheads, the more or less unpredictable {\em space weather} can contaminate observations resulting in extended periods of lost time. The {\em space weather} effects and the different XMM background components are discussed in the following.



\subsubsection{X-ray backgrounds and flare cleaning}

\noindent
The XMM background is the prime constraint for the achievable sensitivity limit of low surface brightness extended sources.
For its importance, the main background components\footnote{All XMM background components are summarized by A. Read at \url{http://www.star.le.ac.uk/}{\tt \%}\url{7Eamr30/BG/BGTable.html}.}
 are shortly introduced followed by a discussion of the implemented flare cleaning procedure.  


\begin{description}

\item[Cosmic X-ray Background (CXB):]
The cosmic X-ray background can be separated into two main contributions. (i) The {\em hard component} of the CXB dominates at energies $\ga$1\,keV and is mostly attributed to the integrated light of faint, unresolved extragalactic X-ray sources, predominantly Active Galactic Nuclei (AGN). Extremely deep 1\,Msec 
pencil-beam observations with Chandra have resolved the largest fraction of the hard CXB into individual point sources 
(\eg \ Mushotzky \etal, 2000; Giacconi \etal, 2001).   
\nocite{Mushotzky2000a}
\nocite{Giacconi2001a}
The total CXB flux in the 2--10\,keV hard band reaches an approximate surface brightness of $5 \times 10^{-15}$\flux arcmin$^{-2}$.
(ii) The  {\em soft component}  of the CXB, dominating below 1\,keV, can be attributed to thermal diffuse emission of galactic origin. It originates from plasma of the Local Hot Bubble, the Galactic Disk, and the Galactic Halo
and is consistent with thermal emission of about $T\!\sim\!0.1$--0.2\,keV.
The CXB is constant in time, but the {\em soft component} varies spatially across the sky.



\item[Flaring Particle Background:]

This background component was severely underestimated before the launch  of XMM and is responsible for most of the lost science time.
The  flaring background  has strong rapid variability and is attributed to soft protons of solar origin with energies of a few 100\,keV. These protons, which seem to be organized in `clouds' populating the Earth's magnetosphere, are funnelled by the X-ray mirrors towards the detectors, where they typically deposit several keV of their energy in the CCD pixels. The occurrence of the soft proton flares is unpredictable but is related to the solar activity  and the satellite position with respect to the magnetosphere. Flare intensities can reach 1\,000 times the quiescent background level implying that contaminated time intervals have to be carefully cleaned and removed from the science data. Since the soft protons pass the mirror system, they imprint a (partially) vignetted background component onto the detectors. 





\item[Quiescent Particle Induced Background: ]

This second particle induced component dominates the quiescent background  level. 
It is attributed to high-energy cosmic ray events  of some 100\,MeV, where a charged particle interacts with the detector      material and the surrounding structures to produce fluorescence emission. This more stable internal background component  typically varies only on the $\pm 15$\% level during observations, but can be an order of magnitude enhanced in high radiation periods.
The cosmic rays induce a background component originating from the detector material itself, \ie \ the quiescent particle background is not vignetted, but reflects the spatial structure of the surrounding detector material. 
The most prominent instrumental fluorescence lines are due to the  Al-K$\alpha$ transition at 1.5\,keV for all instruments and  additional Si-K$\alpha$ emission at 1.7\,keV  for the MOS detectors. Another intense line complex of several elements (\eg \ Cu, Ni, Zn) arises around 8\,keV for the PN and is hence beyond the considered bands for the survey.









\item[Instrumental Noise:]

The last considered background contribution is electronic detector noise. This component is negligible in the harder bands  but becomes significant at low energies of $\la\!0.3$\,keV. The low energy band limit for the survey of 0.3\,keV is thus motivated by the onset of increased instrumental noise in conjunction with  enhanced galactic hydrogen absorption.



\end{description}

\begin{figure}[h]
\begin{center}
\includegraphics[angle=-90,clip,width=0.49\textwidth]{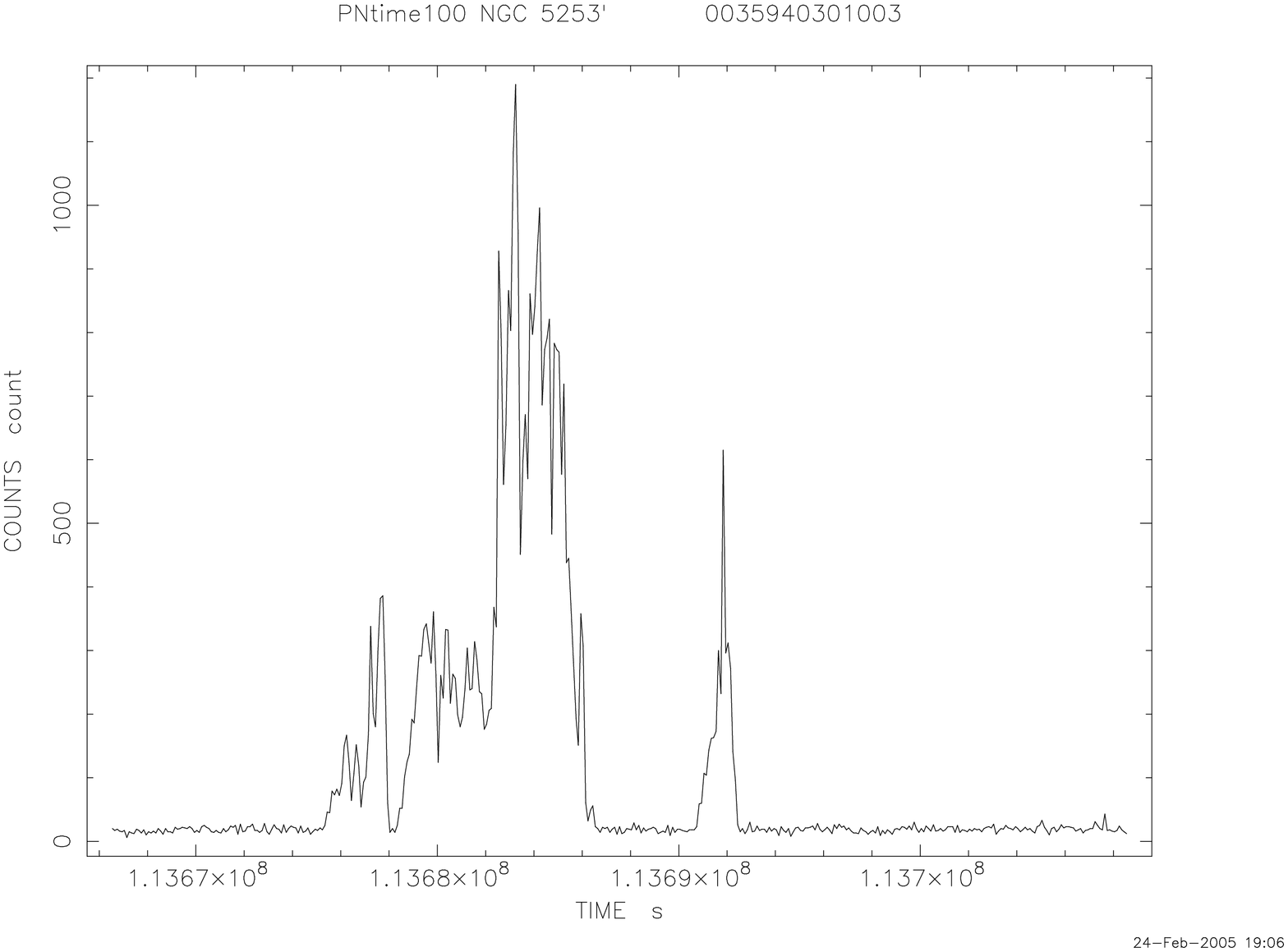}
\includegraphics[angle=-90,clip,width=0.49\textwidth]{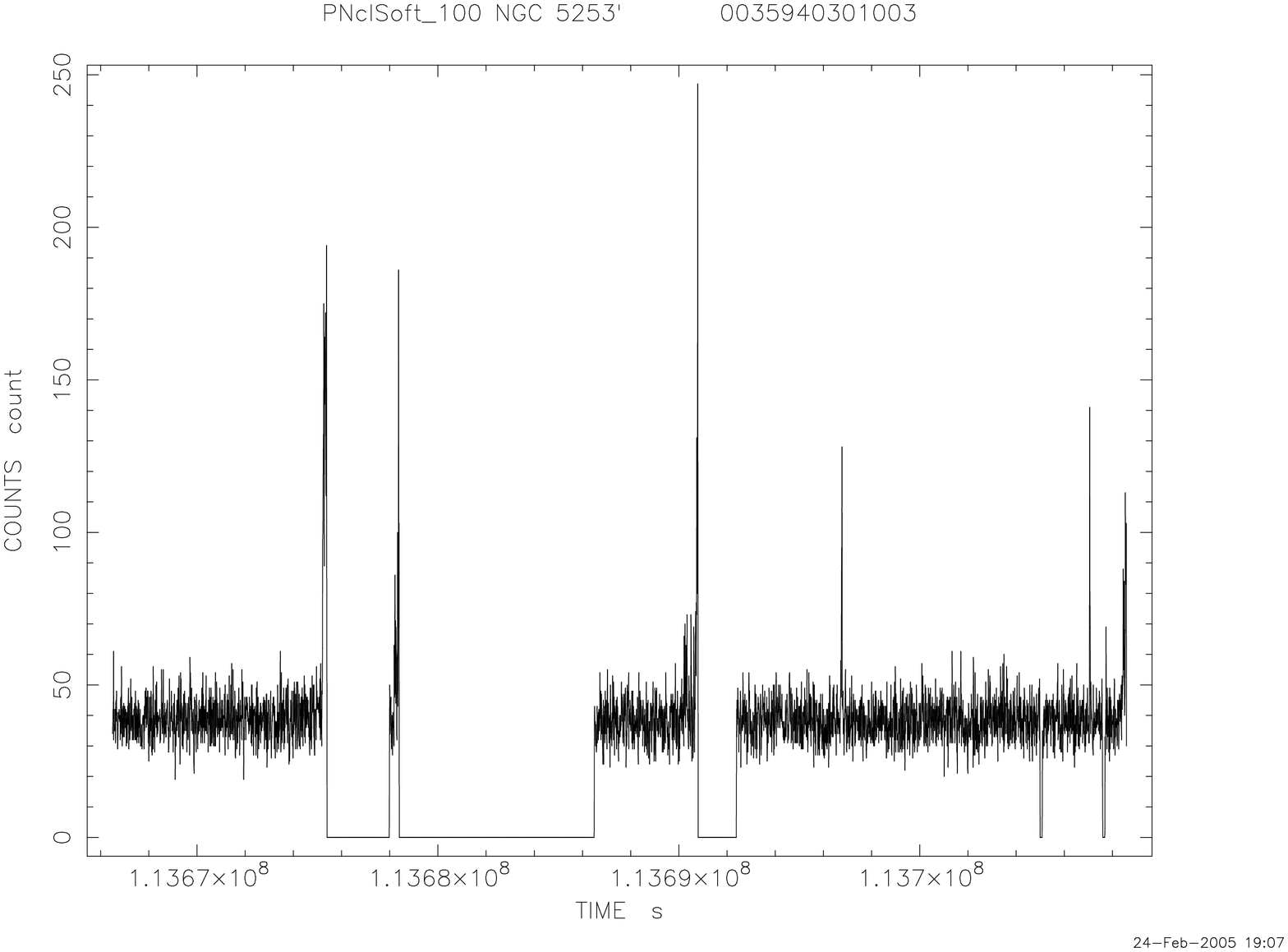}
\includegraphics[angle=0,clip,width=0.495\textwidth]{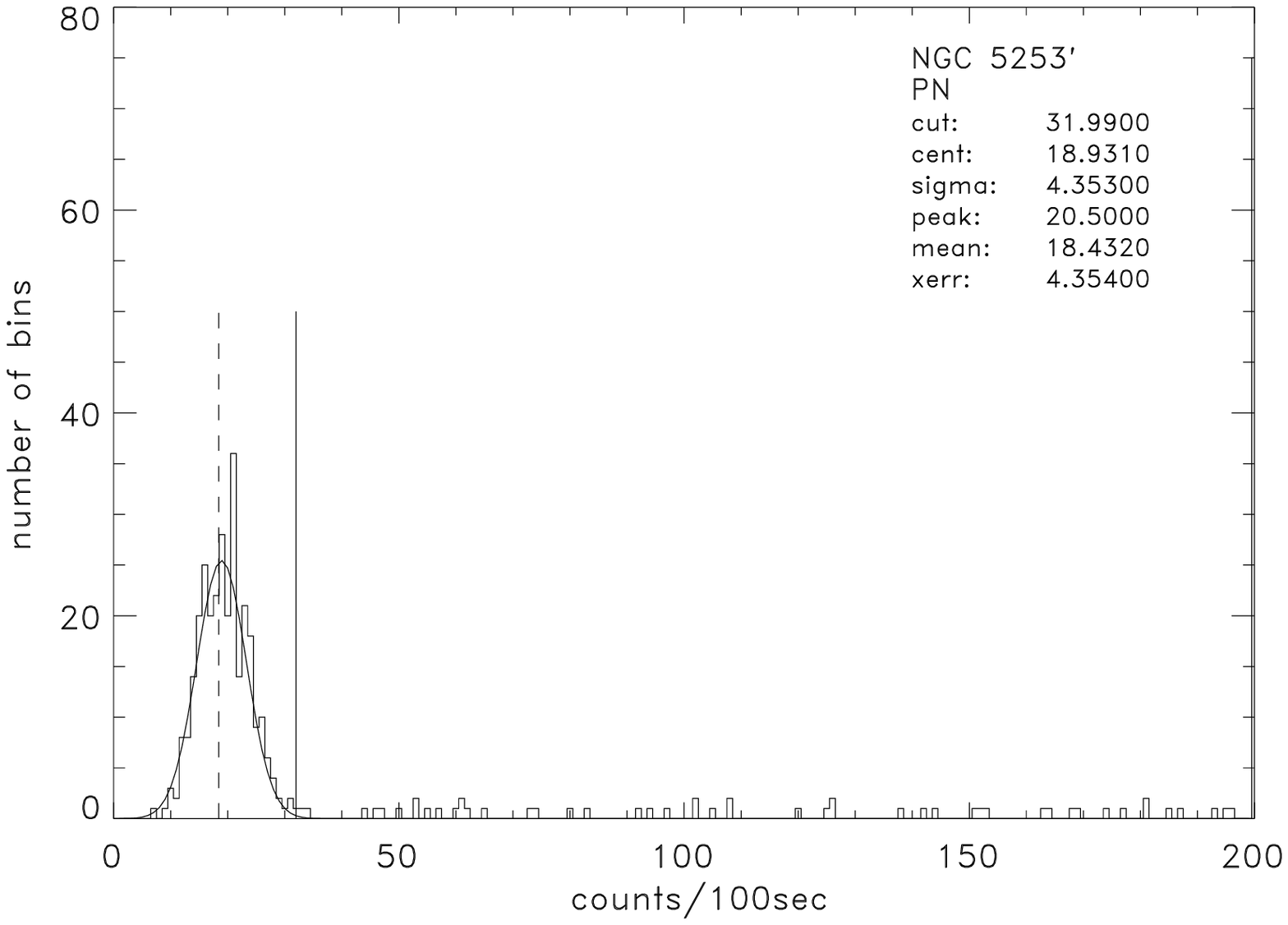}
\includegraphics[angle=0,clip,width=0.495\textwidth]{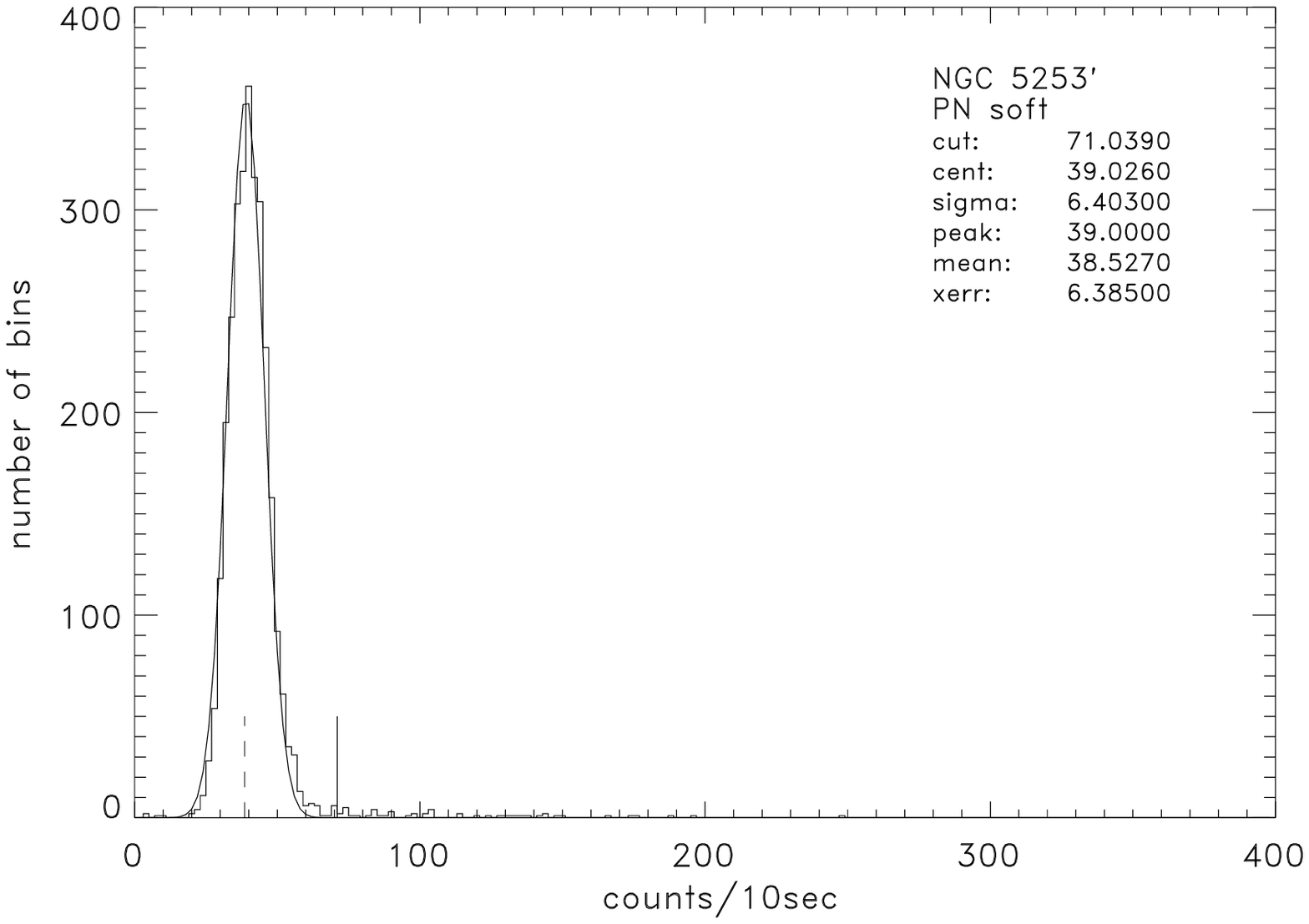}
\includegraphics[angle=-90,clip,width=0.49\textwidth]{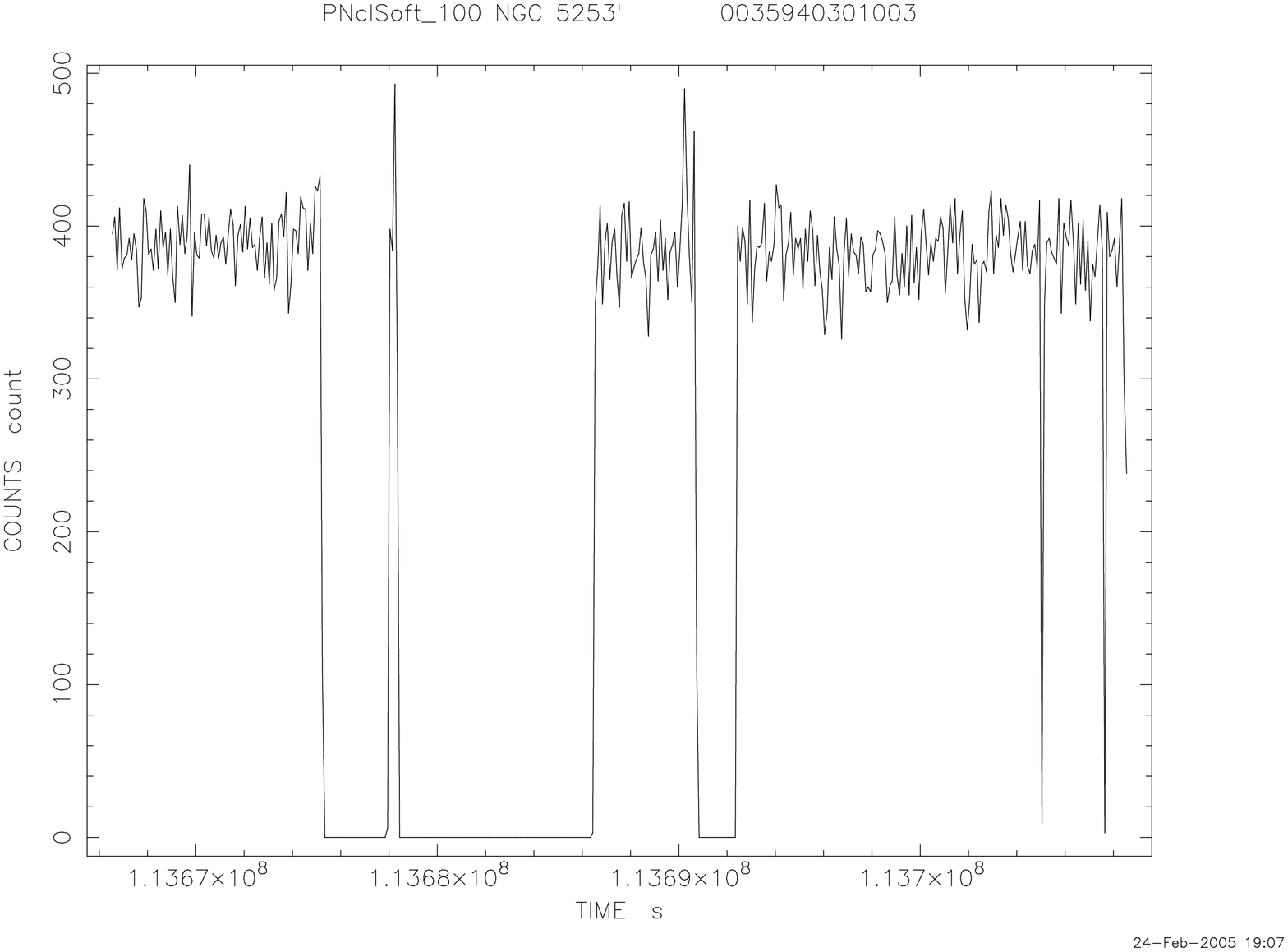}
\includegraphics[angle=-90,clip,width=0.49\textwidth]{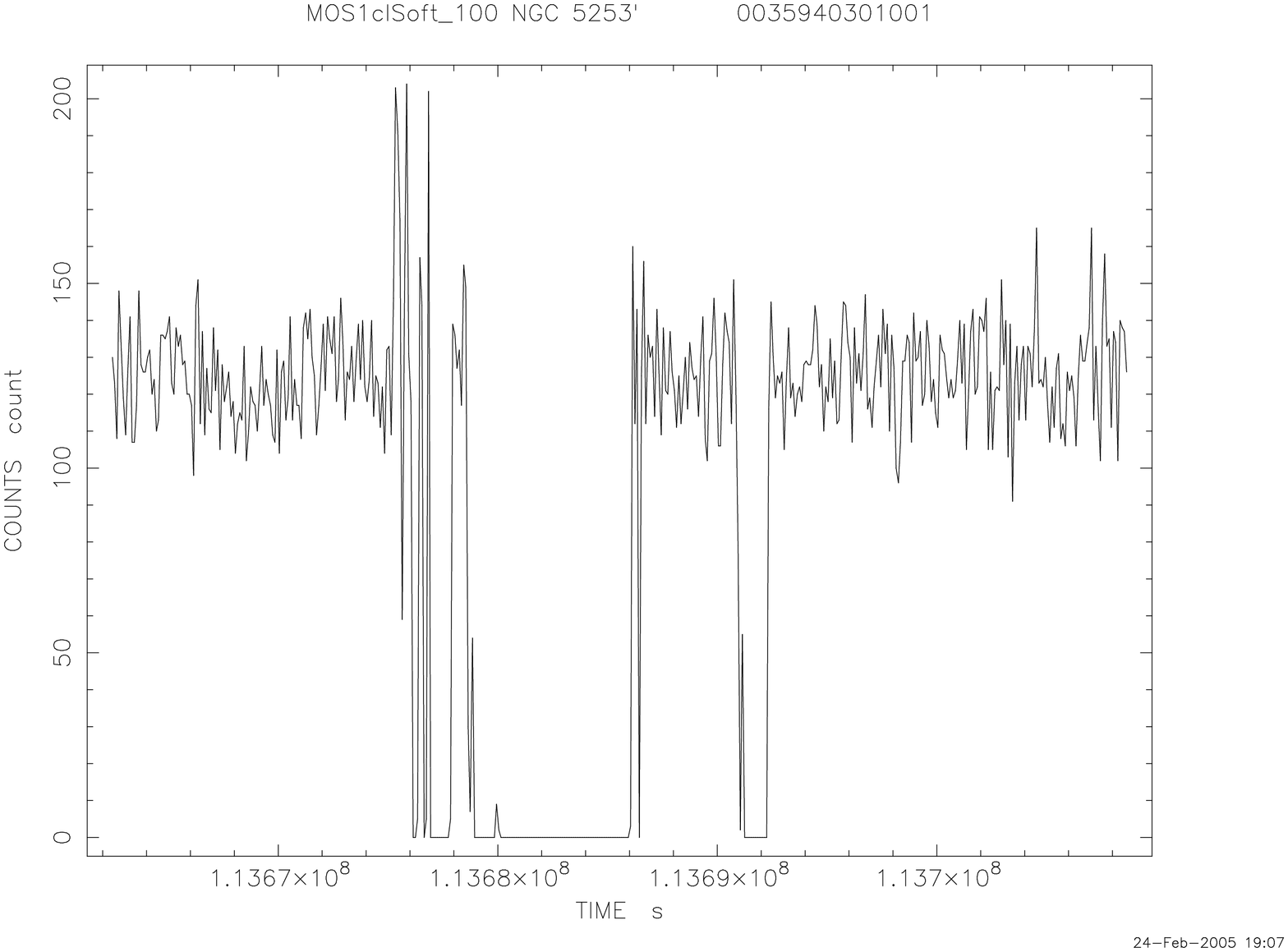}
\end{center}
\caption[X-ray Flare Cleaning]{X-ray flare cleaning procedure. {\em Top row left:} 100\,sec-binned PN light curve of the complete field in the hard X-ray band at 12--14\,keV. The flared periods are easily visible and are removed during the first cleaning stage. {\em Right:} 
10\,sec-binned PN `soft' band light curve at 0.3--10\,keV. The missed rising flanks close to the onset of flare periods and short flares are identified during the second cleaning stage. {\em Center row:} Automatic good time interval analysis based on the identification of the quiescent level and specified $\kappa\sigma$-clipping cut-levels for cleaning stage one ({\em left}) and stage two ({\em right}). {\em Bottom row:} Final light curve of the identified good time intervals for the PN ({\em left}) and MOS ({\em right}) detectors.} \label{f6_flare_cleaning}
\end{figure}

\noindent
The top left panel of Fig.\,\ref{f6_flare_cleaning} shows a typical 100\,sec-binned light curve in the PN 12-14\,keV hard band of a  47\,ksec pointing. The soft proton flare during the central part of the observation is obvious and can be particularly well identified in the hardest band due to the flat nature of the flare spectrum. 
The pipeline flare removal procedure of the {\em first cleaning stage} automatically identifies the stable quiescent level  based on a Gaussian peak-fit in the count-level histogram of the time series (center left panel of Fig.\,\ref{f6_flare_cleaning}). 
All observation periods with background levels less than three standard deviations above the determined quiescent level are accepted as  good time intervals (GTI) for the initial hard band cleaning.

\clearpage

This {\em first cleaning stage} efficiently removes the largest fraction of flare-contaminated data. However, 
at lower energies the flares can exhibit a different time behavior than the main hard component, which is currently attributed to 
lower energy particles at the edges of the encountered proton `clouds'. The upper right panel of Fig.\,\ref{f6_flare_cleaning} 
displays the 10\,sec-binned light curve in the full 0.3--10\,keV soft X-ray band after the initial flare removal. The residual flare peaks are identified during the {\em second soft-band cleaning stage}\footnote{The double flare cleaning procedure was originally proposed and applied by Pratt and Arnaud \cite*{Pratt2003a}.} and are removed from the remaining good time intervals via 5-$\sigma$ clipping of the soft energy histogram (center right panel of Fig.\,\ref{f6_flare_cleaning}).

The lower panels of  Fig.\,\ref{f6_flare_cleaning} show the resulting light curves (100\,sec-binned) after the iterated flare cleaning procedure for the PN instrument (left) and one of the MOS cameras (right). The contaminated time intervals are completely removed from the final cleaned photon event lists resulting in an effective science-usable exposure time of 29.6\,ksec or 63\% of the nominal time for this field.

The flare removal procedure is applied to all instruments independently since the higher low energy sensitivity of the PN camera results in a different temporal response of the system.
For about 80\% of all XMM fields, the {\em automatic} double flare cleaning procedures yields good and stable results in identifying the good time intervals. If more than half of the observation is contaminated, the automatic determination of the quiescent background level might fail and require a manual definition of the GTI cuts.








\enlargethispage{6ex}

\subsubsection{X-ray images}

\noindent
With the cleaned photon event lists at hand, we can now create  X-ray images for all subsequent analysis tasks. 
The energy information of the X-ray photon events allows the arbitrary definition of customized band-passes at this point. As discussed in Sect.\,\ref{s6_search_strategy}, five different energy bands are considered for the XDCP survey.
Photons of the 0.35--2.4\,keV energy range are selected for the optimized single detection band, 
the ranges 0.3--0.5\,keV, 0.5--2.0\,keV, and  2.0--4.5\,keV for the standard band method, and an additional 0.5--7.5\,keV broad band for the spectral matched filter scheme, which uses all five bands. 


The X-ray images are reconstructed with 4\,arcsec pixels from the cleaned photon event lists by applying the appropriate energy cuts and conservatively selecting only well calibrated events as characterized by event quality flags. The absolute astrometry, \ie \ the sky coordinate information, is typically accurately determined to within approximately one arcsecond.  
Images are produced for each instrument individually resulting in a total of 15 band-detector combinations.

At this stage, the correction for  out-of-time events of the PN detector is performed.
Based on the observed count rate in each pixel, the smeared streaks of photons with mismatched Y position information can be statistically reconstructed in an OoT image\footnote{A reconstructed OoT event list can be obtained with the task {\tt epchain}.}. This OoT frame is created for each band and subtracted from the corresponding raw PN data to yield  a first order correction of the out-of-time image artifacts. Although the fraction of smeared photons is only 2.3\% or 6.3\% (depending on the imaging mode), the trails of very bright central target objects can mimic apparently extended sources and should therefore be corrected.

\pagebreak

Observations that have been interrupted, \eg \ due to technical problems or strong flares, are often split up into several shorter data blocks per instrument. The separate event lists are individually reduced, cleaned, and transformed into X-ray images.  
These shorter exposures are co-added for each band-pass and instrument at this point in order to restore the full on-target integration time of the observation. 

As a last step of the data reduction part, images of the same band are combined for the three detectors to yield the full XMM imaging information in the specified energy range. In order to enhance the contrast of weak sources for visual inspection, the combined images are additionally smoothed with a 4\,arcsec Gaussian filter\footnote{The filtering should average over several pixels to reduce the noise (pixel scale: 4\arcsec /pixel), but should not exceed the PSF FWHM to allow a distinction between point sources and extended objects.}. Figure\,\ref{f6_Xray_products} displays the raw (center left panel) and smoothed (upper left panel) combined 0.35--2.4\,keV images of one of the deep XMM  survey fields. 
The first part of the XDCP pipeline is  now finished and we can proceed with the source detection module.






\subsection{Detection of extended X-ray sources}
\label{s6_source_detection}

\begin{figure}[t]
\begin{center}
\includegraphics[angle=0,clip,width=\textwidth]{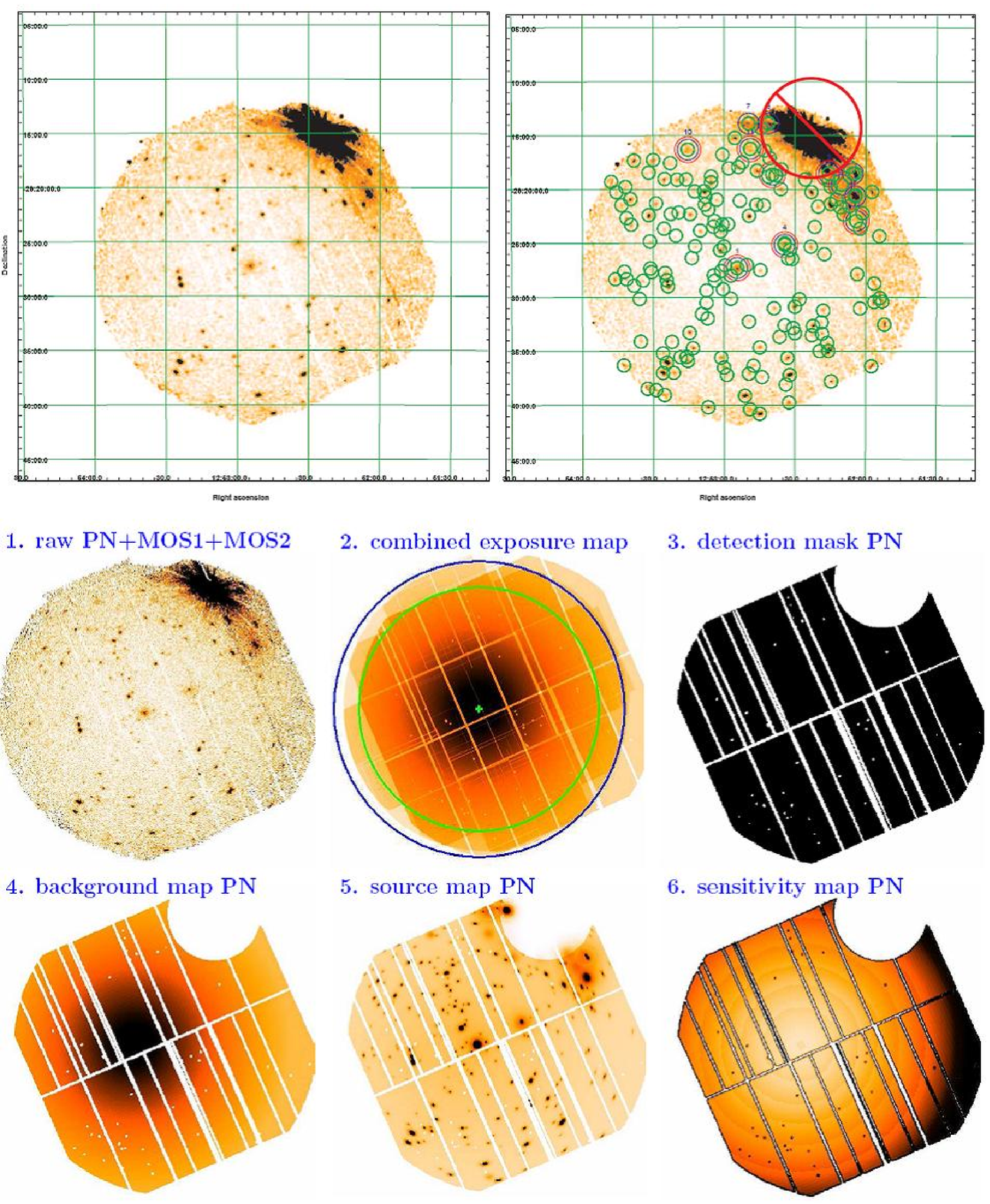}
\end{center}
\vspace{-3ex}
\caption[X-ray Source Detection Products]{X-ray source detection products. {\em Top row left:} Smoothed X-ray image in the 0.35--2.4\,keV detection band of the field with the distant cluster RDCS\,J1252-29 at $z\!=\!1.24$ in the center. {\em Right:} Source detection results in the same field with point sources marked in green, extended sources indicated by larger circles, and the bright source in the upper right corner excluded from the detection area. {\em Center and bottom rows:} Labelled,  intermediate image products of the source detection procedure. The green circle in the exposure map marks the 12\,arcmin off-axis radius, blue indicates the 14.5\,arcmin circle coinciding with the FoV edge of the MOS detectors.} \label{f6_Xray_products}
\end{figure}

\noindent
The source detection procedure\footnote{The box detection procedure follows in part the scheme of Georg Lamer.} as shown in Fig.\,\ref{f6_Xray_pipeline} requires several additional preparatory steps and intermediate data products.
Figure\,\ref{f6_Xray_products} illustrates the X-ray image of the current reduction status (top left panel), the display of the final X-ray detection results (upper right panel), and a selection of 
 intermediate data products in the numbered center and bottom panels.

The first important calibration frames besides the raw photon count images (panel\,1) are the {\em exposure maps} (panel\,2), which correct for the varying sensitivity across the XMM field-of-view, similar to {\em flatfields} in optical and NIR imaging (see Sect.\,\ref{s7_NIR_reduction_steps}). Dividing a raw photon count image by its associated exposure map results in a 
sensitivity-corrected count rate image with units [counts$\cdot$s$^{-1}$], normalized to the central, on-axis values. 
These calibrated count rates  differ from the final physical flux units in each pixel only by a multiplicative energy-dependent constant, the (on-axis) count-rate-to-flux conversion factor. 

In effect, the exposure maps contain the effective, local integration times associated with each detector pixel, scaled to the on-axis exposure.
The radial shape of the exposure maps  is mainly due to the 
vignetting function\footnote{As an alternative method,  vignetting effects can be accounted for by applying a weighting factor for each pixel.}
  (see Fig.\,\ref{f6_vignetting}), which accounts for the decreasing off-axis sensitivity.  The task {\tt eexpmap} also includes 
the calibration information on the energy dependent detector quantum efficiency, chip gaps, dead columns, the transmission of the optical blocking filters, and the detector field-of-view. Panel\,2 shows the combined exposure map of all three instruments with dark regions indicating the most sensitive parts and lighter colors the areas with a reduced effective exposure time. 
The position of the optical axis,  also named  {\em boresight}, is indicated by the green cross.
The circles illustrate the 12\arcmin \ (green)  and 14.5\arcmin \ off-axis radius, with the latter one roughly coinciding with the FoV boundaries of the two MOS cameras.


\clearpage

The next step is the generation of detection masks with {\tt emask}, as shown in panel\,3 for the PN detector. The masks define the detector regions over which the source detection is performed (black areas), excluding chip gaps and dead columns (white).
In addition, {\em manually defined\/}  regions with bright contaminating sources are also cut out and excluded from the detection area. In the example case in Fig.\,\ref{f6_Xray_products}, the bright off-axis  source in the upper right corner has been removed (manually), but more typically the central target objects are cut out to ease serendipitous detections. Excluding bright and large area sources helps in two ways: (i) the occupied region does not contribute to the effective area of the survey and is this way automatically accounted for, and (ii) the background model is improved.     



The crucial {\em background maps} for each detector and band  are determined in the final preparatory step.
The following procedure is applied. (i) A first sliding box detection run is performed with {\tt eboxdetect} in {\em local mode}, \ie \ the background is approximated locally around the detection cell. This yields a preliminary source list which is used to identify the positions of all sources and excise them from the field. (ii) The resulting `cheese image' only contains X-ray background contributions and is used as input for the background 
modelling process with {\tt esplinemap}.\footnote{The task {\tt esplinemap} can be used for spline fits or two-component model fits.} The task performs a two-component model fit for the linear combination of a  spatially constant background contribution (quiescent particle induced background and instrumental noise) and a vignetted component (CXB and residuals of the soft proton particle background). The final background model result is shown in panel\,4 of Fig.\,\ref{f6_Xray_products}. The applied reconstruction method is physically motivated by the different contributing background components and has the main advantage of a smooth radial profile, but the disadvantage that the global fit over the 30\arcmin \ FoV does not account for local variations. In any case, this model is preferable to the alternative spline fitting method, which accounts for local 
variations, but is also prone to producing numerous spurious extended sources as a consequence of local minima of the fitted splines.




\subsubsection{Sliding box detection and maximum likelihood fitting}

The actual source detection is performed in two steps. (i) A list of source candidates is obtained with the sliding box method, followed by (ii) the subsequent detailed analysis and source characterization via maximum likelihood (ML) fitting. 

In order to achieve the best sensitivity for the candidate source list,
the task {\tt eboxdetect} is run in {\em map mode}, \ie \ the improved reconstructed global background maps are used as input rather than the local approximation. In addition to the photon images and  background maps, the exposure maps and the detection masks are supplied to the task. {\tt eboxdetect} then takes a 5$\times$5 pixel (20\arcsec$\times$20\arcsec) detection cell and slides it over the input images. At any image point, the PSF-weighted, integrated cell counts are compared to the expected background level and the likelihood for a random background fluctuation of this level is calculated. The sensitivity to extended sources is achieved, by iterating the procedure three times in conjunction with  doubling the detection cell size to $10\!\times\!10$, $20\!\times\!20$, and $40\!\times\!40$ pixel boxes. All candidate sources above a specified combined likelihood are saved to an input file for the detailed source characterization.
 
The significance for the detection or the extent of an X-ray source is usually expressed in terms of the  likelihood 
$L\!=\!-\ln p_{\mathrm{Pois}}$ \cite{Cruddace1988a}, where $p_{\mathrm{Pois}}$ is the probability of a Poissonian random background fluctuation of counts in the detection cell, which would result in at least the number of observed counts.
The $L$ likelihood values are linear, \ie \ the total likelihood of a source determined from the full band width is equivalent to the sum of the $L_i$ values of the individual input bands.
The probability of a source to be real  $p_{\mathrm{real}}$ and {\em not}  originate from a statistical fluctuation is hence given by $p_{\mathrm{real}}\!=\!1\!-\!p_{\mathrm{Pois}}\!=\!1\!-\!\exp(-L)$. For a selection of  likelihood values $L$ the following statistical probabilities for a real physical source are obtained: $p_{\mathrm{real}}(L\!=\!1)\!\simeq\!0.632$,  $p_{\mathrm{real}}(L\!=\!2)\!\simeq\!0.865$, $p_{\mathrm{real}}(L\!=\!3)\!\simeq\!0.950$, $p_{\mathrm{real}}(L\!=\!4)\!\simeq\!0.982$, $p_{\mathrm{real}}(L\!=\!5)\!\simeq\!0.993$,  $p_{\mathrm{real}}(L\!=\!6)\!\simeq\!0.998$, $p_{\mathrm{real}}(L\!=\!8)\!\simeq\!0.9997$, and $p_{\mathrm{real}}(L\!=\!10)\!\simeq\!0.99995$.       
The likelihoods are roughly related to positive outliers of Gaussian statistics with $n\,\sigma$ significance via $n\!\sim\!L/2$, \ie \ $L\!=\!4$ corresponds approximately to 2-$\sigma$ significance  and $L\!=\!6$ to a 3-$\sigma$ detection.


The task {\tt emldetect} performs the final characterization of the candidate sources using the same input frames and the additional candidate source list of {\tt eboxdetect}. A maximum likelihood PSF fit in the photon images is applied simultaneously for all input bands. Each candidate source is fit with free parameters for the source location (RA \& DEC), the source extent, and the source count rate in each band. Before a source is classified as extended, {\tt emldetect} attempts to fit two overlapping point sources to the photon distribution. Only if a single extended source fit yields a significant likelihood improvement in comparison to the 2-PSF fit of a double point source, the extended nature of the source is established.    

{\tt emldetect} characterizes extended sources based on a King profile fit (Equ.\,\ref{e2_beta_model}) with a fixed standard beta value of $\beta\!=\!2/3$ and the core radius $\Theta_{\mathrm{c}}$ (or equivalently  $r_{\mathrm{c}}$) as free fit parameter.
The observed photon distribution is to be compared to the  King profile of the extended source convolved  with the two-dimensional PSF shape at the specific detector location, which is a function of off-axis angle (Fig.\,\ref{f6_xmm_PSF}), azimuthal angle (PSF rotation), and energy. If the maximum likelihood for the source extent, determined simultaneously for all input bands, exceeds a specified minimum threshold value the measured parameters for the {\em extended} 
source are saved to the final {\tt emldetect} source list. In the case that the extent likelihood falls below the threshold, the corresponding {\em point source} parameters are determined and saved.

For the XDCP search strategy of Sect.\,\ref{s6_search_strategy}, the {\tt emldetect} maximum likelihood fitting procedure is performed once for each of the three 
different detection schemes. The minimum {\em detection} likelihood for a source 
is fixed at  $L\!=\!6$ ($\sim\!3\,\sigma$), whereas the {\em extent} likelihood is successively lowered for the different schemes.
For the {\em single band scheme} with highest XDCP priority the minimum {\em extent} significance is set to $\sim\!2.5\,\sigma$ ($L\!=\!5$), for the  {\em spectral matched filter scheme}\footnote{This is an effective value. The corresponding {\tt emldetect} parameters is set to $L_{\mathrm{min}}\!=\!10$ but the enhanced weight of the central spectral region enters the total likelihood accordingly through the sum of overlapping bands.} to $\sim\!2\,\sigma$ ($L\!\simeq\!4$), and for the {\em standard scheme\/} to  $\sim\!1.5\,\sigma$ ($L\!=\!3$). The decreasing extent threshold is to ensure that the majority of sources are detected as a resolved objects for several detection schemes in order to allow a parameter cross-comparison. 
The aim for the raw extended source list as input for the screening procedure is a minimum extent significance of 2--2.5\,$\sigma$ as will be discussed in Sect.\,\ref{s6_Source_Screening}.   

Panel\,5 of Fig.\,\ref{f6_Xray_products} displays the final reconstructed source image for the PN detector. Note the increasing tangential elongation of the PSF as a function of off-axis angle. Detected sources are also marked in the smoothed combined photon image in the upper right panel, where green circles symbolize point source detections and additional blue or red circles identify the extended sources.








The last panel\,6 of Fig.\,\ref{f6_Xray_products} shows the PN sensitivity map of the observation, where light colors indicate the most sensitive central areas.
The task {\tt esensmap} approximates the local sensitivity limits for {\em point source} detections of a specified significance based on the 
exposure maps, the background maps, and the detection masks using Poissonian count statistics.
Similar maps for the {\em extended source} sensitivity have to be calculated for the final XDCP survey assessment. The procedure for this more complex but crucial task is briefly discussed in Sect.\,\ref{s9_Selection_Funct}.




\subsubsection{Wavelet detection}

Besides the  traditional sliding box  method,  wavelet techniques have become 
a popular alternative approach  for the detection of extended X-ray sources (\eg \ Rosati \etal, 1995; Vikhlinin \etal, 1995; Herranz \etal, 2002).
\nocite{Rosati1995a} 
\nocite{Vik1995a} 
\nocite{Herranz2002a} 
For the XDCP source detection procedure, the  {\tt SAS} task 
{\tt ewavelet}\footnote{See the {\tt ewavelet} manual at \url{http://xmm.vilspa.esa.es/sas/7.1.0/doc/ewavelet/index.html} for more information.}
is used as a complementary source detection method to provide additional
extent information, in particular for the evaluation of faint sources.
The basic concepts of the wavelet technique are briefly summarized.



Wavelet functions allow a multi-scale analysis of images  and enable a significant enhancement of the contrast of sources of different sizes against a non-uniform background. Wavelets can be considered as an extension of  Fourier transformations with the difference that they are (i) localized in space, and have (ii) a zero normalization.   
Wavelets additionally satisfy  (iii) translational invariance and (iv) scaling properties according to $W_{b,\sigma}(x)\!=\!(1/\sigma)W((x\!-\!b)/\sigma)$, where $W(x)$ is the wavelet function, $b$ the translation, 
 and $\sigma$ the scaling parameter. 


The most popular wavelet function for X-ray source detection is the so-called {\em Mexican hat} (MH) wavelet, which is related to the second derivative of a Gaussian function and can be written in its two dimensional form as 

\begin{equation}\label{e6_mexican_hat}
    g_{\mathrm{MH}}\left( \frac{r}{a}\right) = \left( 2 - \frac{r^2}{a^2}\right) e^{-\frac{r^2}{2\,a^2}}  \ ,
\end{equation}

\noindent
with the wavelet scale $a$ and $r^2\!=\!x^2+y^2$. The MH has a Gaussian core with positive values out to radius $r\!=\!\sqrt{2}\,a$  and negative values outside.  


\begin{figure}[t]
\begin{center}
\includegraphics[angle=0,clip,height=5.2cm]{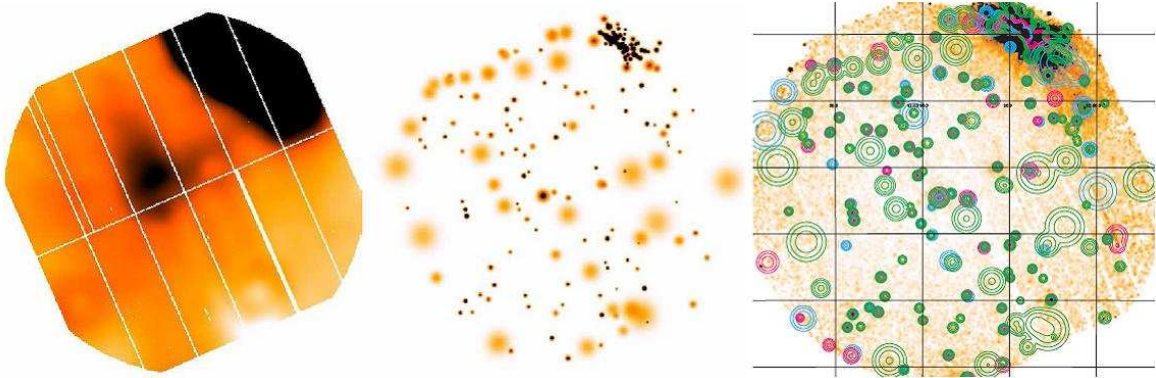}
\end{center}
\vspace{-2ex}
\caption[Wavelet Source Detection]{Wavelet source detection products of the same field as in Fig.\,\ref{f6_Xray_products}. {\em Left:} Wavelet-reconstructed background map for the PN detector. {\em Center:} Reconstructed PN source map. {\em Right:} Wavelet detection results for the PN (green) and the MOS (blue and pink) detectors. This example illustrates for the PN camera how a poorly determined background leads to numerous spurious extended sources in the center and right panel. The {\tt SAS ewavelet} detection results are hence only used as complementary information for the source evaluation.} \label{f6_Xray_wavedetect}
\end{figure}

\enlargethispage{2ex}

The basic principle of the {\tt ewavelet}  detection method works as follows: (i)
The X-ray image is convolved with the {\em Mexican hat} wavelet (analogous to Equ.\,\ref{e3_spatial_filter}). 
In effect, the MH acts like a sliding cell, where the positive part resembles the source cell of scale $a$, and the negative part the background area.
The normalization ensures that the wavelet transform of a homogeneous background is zero. Local maxima in the convolved image correspond to source candidates of wavelet scale $a$.
(ii) If the significance of the local maximum in the wavelet transform is above a specified threshold, the source location and peak height at scale $a$ is saved. (iii) The convolution procedure is repeated for various wavelength scales $a$, starting with the PSF size, and successively increasing the scale by factors of $\sqrt{2}$ up to a given fraction of the image size.
(iv) The final  scale of the identified objects  corresponds to the highest peak in all wavelet transforms at the source position.   

The wavelet technique does not require a supplied external background, but rather reconstructs it from the data. 
The {\tt ewavelet} determined PN background for the same XMM field as before is shown in the left panel of Fig.\,\ref{f6_Xray_wavedetect}. In contrast to the box detection case, the wavelet reconstructed background traces local variations in addition to the off-axis dependence. The reconstructed PN source map in the central panel displays the effect of a poorly determined local background, in this case probably induced by the bright source, which cannot be excised for the {\tt ewavelet} task. Numerous spurious extended sources are detected at the positions of local minima (lighter regions in left panel) in the background.
The right panel illustrates the wavelet detection results for all detectors with sources and their scales overlaid on the smoothed photon image. Since the {\tt ewavelet} source characterization is known to be less reliable than the 
maximum likelihood fitting method, the results of the wavelet detection, as visualized in the latter image, are used as an additional qualitative 
extent indicator for the cross-comparison with the sliding box detected extended sources.








\subsubsection{X-ray source characterization}

As the final step of the source detection procedure, the {\em extended} sources for each XMM field are extracted from the final  {\tt emldetect} source lists associated with the different detection schemes. These field lists are then merged for all XMM observations 
resulting in a  {\em global extended source masterlist}, which is the starting point for the screening procedure of Sect.\,\ref{s6_Source_Screening}.  

Listed in decreasing priority, the relevant {\tt emldetect\/} parameters for the characterization of extended sources are 
(i) the maximum likelihood of the source extent ({\tt EXT\,ML\/}), 
(ii) the detection  maximum likelihood ({\tt DET\,ML\/}),
(iii) the source extent, \ie \ the determined core radius ($r_{\mathrm{c}}$),  
 (iv) the total number of detected counts per band, (v) the derived flux ($f_{\mathrm{X}}$), (vi) the off-axis angle of the source ($\Theta$),  (vii) the hardness ratios for the multi-band schemes (HR), and (viii) the total effective exposure at the source location.


{\tt emldetect} measures the total count rate of a source in a given energy band. 
Physical X-ray fluxes are obtained from the count rate values by multiplying with the Energy Conversion Factors (ECF), which   
depend on the instrument response in the considered energy band, the optical blocking filter, the initial photon selection cuts, the galactic hydrogen column density, and the spectral source properties. For galaxy cluster sources, this implies that the true ECFs are a function of cluster temperature, ICM metallicity, and redshift, which are {\em a priori} not known and additionally differ from object to object. 
Final accurate flux measurements for individual clusters hence require the full source characterization, which is generally not available prior to the redshift confirmation (\ie \ survey step 5).
For the global flux characterization of the source population in early survey stages, {\em fiducial\/} source fluxes are obtained by using standard, fixed Energy Conversion Factors.\footnote{Following the XMM-Newton calibration document SSC-LUX-TN-0059 (2003) at \url{http://isar.star.le.ac.uk/pubdocs/docindex.shtml}. A power law spectrum with a photon index of 1.7 and an absorbing column of $3\!\times\!10^{20}$\,cm$^{-2}$ is assumed.} 
These ECFs are similar to the values for a fiducial distant cluster ($z\!\sim\!1$, $T\!\sim\!5$\,keV) and have been confirmed for several sources to typically yield results with 10--20\% accuracy in the standard 0.5--2.0\,keV band.








As the final point of the X-ray data discussion, we consider the main reasons for {\em spurious} detections of extended sources using the sliding box detection and maximum likelihood fitting method.
Two types of  spurious extended sources are to be distinguished: (i)  {\em spurious detections} (SD), \ie \ the X-ray source is not real, and (ii) {\em spurious extents} (SE) implying that the actual source is point-like or a blend of several point sources. It should be emphasized, that the determined detection and extent likelihoods are purely of statistical nature and cannot account for systematic errors of the supplied calibration data. 
In decreasing order of occurrence, the dominant reasons for false positive detections beyond the statistical expectations are: 

\begin{enumerate}
    \item secondary detections in wings of large extended sources (SD);
\vspace{-0.5ex}
    \item PSF residuals due to uncertainties in the off-axis calibration (SE+SD);
\vspace{-0.5ex}
    \item detections at detector or chip boundaries (SE+SD);
\vspace{-0.5ex}
    \item blends of three or more point sources (SE);
\vspace{-0.5ex}
    \item background residuals due to a poor local determination (SD+SE);
\vspace{-0.5ex}
    \item out-of-time trail residuals (SD);
\vspace{-0.5ex}
    \item optical loading residuals caused by bright optical sources (SD).
\end{enumerate}    

\noindent
The brackets indicate, wether the artifacts result mostly in spurious extents (SE), spurious detections (SD), or both (SE+SD).
The upper right panel of Fig.\,\ref{f6_Xray_products} illustrates as example several spurious sources in the wings of the excised bright object in the upper right corner.





Valtchanov, Pierre \& Gastaud \cite*{Valtchanov2001a}  
have benchmarked half a dozen techniques for the detection of extended X-ray sources in XMM-Newton data.
The main conclusion is that there is no existing single detection technique without any drawbacks.
Concerning the performance of {\tt eboxdetect} in combination with {\tt emldetect}, they confirm a very good detection rate and decent parameter reconstruction at the expense of numerous spurious sources, in particular related to the splitting of larger objects. {\tt ewavelet} also shows a good sensitivity, but exhibits poor photometric parameter determinations.      




In any case, the detection of faint extended X-ray sources is a very challenging task.
The available techniques and methods are still  evolving, in particular for their application to ongoing missions such as XMM-Newton.  This section has provided an overview of the principal methods and has discussed some of the 
 complexities and issues  arising therefrom. 

At this point, it should be emphasized again that one of the main objectives of the  XDCP approach is to {\em push the X-ray source detection to the very limit} and fully exploit the achievable XMM sensitivity. 
The inevitable price for maintaining a high completeness level of real cluster sources at very faint flux levels is an increased  contamination fraction at early stages of the survey confirmation process (steps 1--4), \ie \ before the observationally expensive spectroscopy (see Chap.\,\ref{c8_SpecAnalysis}).  
The largest fraction of detected spurious extended sources  can be removed during the screening procedure described in Sect.\,\ref{s6_Source_Screening} (step 2); the remaining non-cluster sources will be identified during the follow-up imaging stage (steps 3\,\&\,4) discussed in Chap.\,\ref{c7_NIRanalysis}. For the challenging aim of finding and confirming the most distant X-ray clusters at \zg1, using less than half of the (relatively inexpensive) imaging time for the identification of the remaining false positive sources seems to be an adequate and tolerable observational price to pay.



\subsection{X-ray-optical overlays}
\label{s6_overlays}

\noindent
As a preparatory step for the source screening procedure of Sect.\,\ref{s6_Source_Screening},
the final part of the XDCP pipeline handles the generation of X-ray-optical overlays for the individual detected extended sources. These overlays provide essential information for the final assessment of (i) the X-ray quality of the source, and (ii) possible optical counterparts. 

\newpage


\begin{figure}[t]
\begin{center}
\includegraphics[angle=0,clip,width=0.66\textwidth]{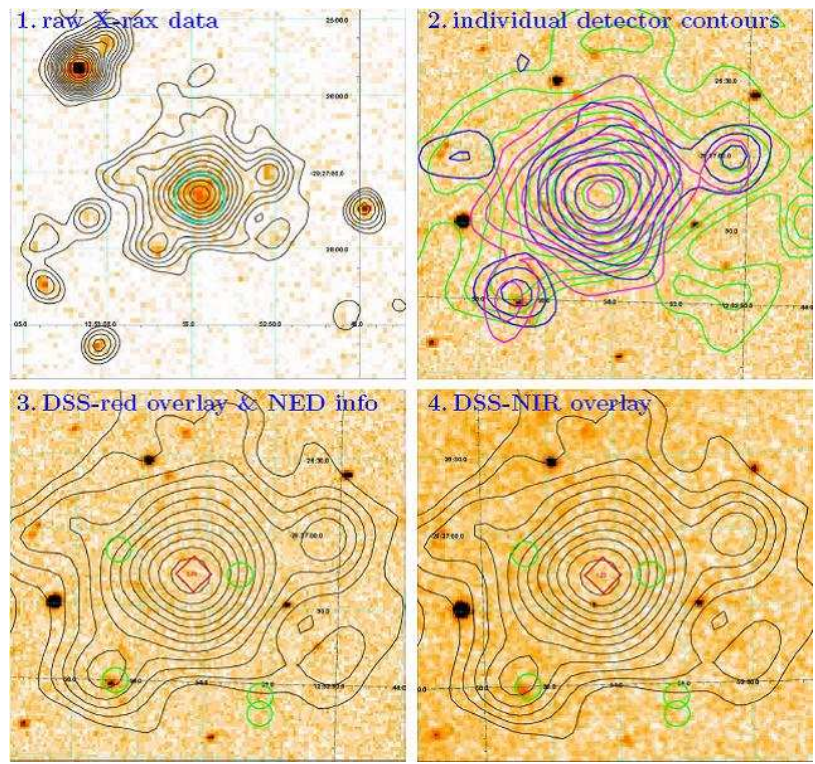}
\end{center}
\vspace{-2ex}
\caption[X-ray-Optical Overlays]{X-ray-optical overlays for the source classification procedure as complementary information to the {\tt emldetect} source parameters.  {\em Top left:} Unfiltered 5\arcmin$\times$5\arcmin \ photon image in the 0.35--2.4\,keV band with the smoothed contours overlaid and the core radius indicated by the blue circle. {\em Top right:} Individual detector contours for the PN (green) and the MOS instruments (blue and pink) as a means for cross-checking instrumental artifacts. {\em Bottom row left:}
DSS-red image (2.5\arcmin$\times$2.5\arcmin) with smoothed X-ray contours and NED object information overlaid. The source is  identified as the known cluster RDCS\,J1252-29 at $z\!=\!1.24$ (red diamond), green circles indicate NED sources without redshift information. {\em Right:} Same material overlaid on the DSS NIR image.} \label{f6_DSS_Overlay}
\end{figure}


\enlargethispage{2ex}

Combining data of several different  wavebands into a single image can be achieved through the use of contours, which are overlaid onto a second image. The first step for the overlay production is hence the generation of X-ray contours, which are obtained from the smoothed single detection band images at 0.35--2.4\,keV. The contours are computed for logarithmically-spaced, pre-defined flux levels for the combined image and the three detector frames individually.  

For the assessment of the X-ray quality of the source, the combined contours are overlaid onto the raw X-ray photon image with a FoV size of  5\arcmin$\times$5\arcmin \ centered on the source. In addition the determined King profile core radius of the extended source is overplotted. Panel\,1 of Fig.\,\ref{f6_DSS_Overlay} illustrates this raw data zoom for the example source of the distant cluster  RDCS\,J1252-29 at $z\!=\!1.24$. Note the point source in the upper left corner for comparison.
As a means for checking artifacts introduced by one of the detectors, the contours for the individual instruments are placed on top of each other in the 2.5\arcmin$\times$2.5\arcmin \ image shown in panel\,2.


The optical source classification is based on image data provided through the Second
Digitized Sky Survey\footnote{The ESO DSS archive can be accessed at \url{http://archive.eso.org/dss}.} 
(DSS-2). Of particular interest for distant cluster applications are the red band (DSS-red) and the NIR data (DSS-NIR), which reach almost all-sky coverage and are online accessible. 
The imaging data is complemented by a near-position query around the source location to the NASA Extragalactic Data Base\footnote{The NED homepage is found under \url{http://nedwww.ipac.caltech.edu}.} (NED), which contains information on all published extragalactic objects such as source type, spectral properties, and redshift, if available.
The returned object list is filtered and transformed into region files, which can be overlaid onto the image data in addition to the X-ray contours. 

Panels\,3\,\&\,4 of Fig.\,\ref{f6_DSS_Overlay} display the resulting 2.5\arcmin$\times$2.5\arcmin \ image zooms for the DSS-red data (left) and the  DSS-NIR (right). 
Additional 5\arcmin$\times$5\arcmin \ overlay images of the source environment allow  an inspection on a larger scale. 
Green circles indicate NED objects without redshift information, whereas red diamonds 
display the location of objects with the NED specified redshift. In the example of Fig.\,\ref{f6_DSS_Overlay}, the indicated central redshift is associated with the distant ROSAT-discovered target cluster of the observation.







\subsection{Distant cluster optimizations}
\label{s6_optimizations}

The largely automated XDCP X-ray pipeline processing, as outlined in Fig.\,\ref{f6_Xray_pipeline}, is now completed.
The two newly tested non-standard detection schemes (single band and spectral matched filter), for a  complementary use to the standard source detection approach, can be considered as key characteristics.  
Numerous additional 
 features of the XDCP pipeline   aim at an  
optimization of the distant cluster detection and source classification efficiency. 
The efficacy of the newly implemented XDCP pipeline characteristics has been tested in initial performance comparisons to `standard method' results and cross-checked with early observational feedback from follow-up imaging campaigns.   
In general, four categories can be distinguished for which 
improvement effects compared to a standard detection procedure are expected:
 (i) the survey usability of additional XMM fields (F), (ii) an improved  extended source sensitivity (S), (iii) a higher  extent reliability (R), and (iv) an optimized  low-redshift screening efficiency (Z).
The main high-$z$ optimization features are summarized in the following, effect categories (F, S, R, Z) are indicated in brackets after the items: 




\vspace{-1ex}

\begin{itemize}
    \item a strict double flare cleaning procedure (F+S);
\vspace{-0.5ex}
    \item correction for out-of-time events (R);
\vspace{-0.5ex}
    \item use of detection masks with excised bright sources (F+S+R);
\vspace{-0.5ex}
    \item use of an optimized distant cluster band for the  detection and visual inspection (S);
\vspace{-0.5ex}
    \item use of a smooth background model fit instead of spline interpolations (R+S);
\vspace{-0.5ex}
    \item  four box detection iterations  with increasing cell size for an improved extended source sensitivity (S);
\vspace{-0.5ex}
    \item a maximum extent radius of 20 pixels (80\arcsec) to constrain the possible extended source parameter space and prevent very extended spurious sources (R);
\vspace{-0.5ex}
    \item maximum likelihood tests for a possible double point source blend  (R);
\vspace{-0.5ex}
    \item use of detection parameter information  of three different schemes for cross-comparison (S+R);
\vspace{-0.5ex}
    \item use of the wavelet detection method for complementary extent information (R);
\vspace{-0.5ex}
    \item individual detector artefact assessment using separate X-ray source contours  (R);
\vspace{-0.5ex}
    \item additional use of DSS-NIR data for an object cross-correlation with DSS-red images  (Z);
\vspace{-0.5ex}
    \item automated NED object projections for source identifications  (Z).

\end{itemize}

\section{Source Screening and Candidate Selection}
\label{s6_Source_Screening}

\noindent
The XDCP X-ray pipeline 
 has been applied to the 546 XMM-Newton archive fields that have passed the survey definition selection (see Sect.\,\ref{s6_survey_definition}) and a first visual screening\footnote{Some (contaminated) XMM fields can be discarded from the survey sample based on the inspection of available quicklook data images.}. These fields are listed in Tab.\,\ref{tA_field_list} in the appendix with their positions, nominal and clean exposure times, galactic hydrogen column density, and the survey status.
The total nominal exposure time of this data set of 17.5\,Msec corresponds to 6.7 months of continuous observations, and the full X-ray reduction required about two months of CPU time. 

77 of the data sets have been discarded from survey use for 
three possible reasons. (i) The observation was completely contaminated by soft proton flares, (ii) the central target covered most of the field-of-view, or (iii) data corruption preventing a proper processing.   
The remaining 469 XMM fields with a total nominal exposure time of 15.2\,Msec define the {\bf XDCP core survey sample}, \ie \ the data sets that are in principle usable for serendipitous distant cluster searches.

More than {\bf 2\,000 extended X-ray sources} have been detected in these survey fields.
In this section, we will discuss the screening and classification procedure for these sources with the final goal of compiling a well-selected  XDCP {\em distant cluster candidate list} as a starting point for further follow-up observations. 

\enlargethispage{4ex}

The challenges associated with the detection of extended sources at faint flux levels, in particular the various reasons for 
the expected significant fraction of {\em spurious\/} non-cluster sources, have been discussed in  Sect.\,\ref{s6_source_detection}.
The 
final classification of the detected sources is 
too complex (see \eg \ Fig.\,\ref{f6_Xray_products})  to be reliably implemented  in an automated and objective fashion. The screening process of the individual extended X-ray sources hence requires human intervention for a due assessment of possible problems and for achieving the best results.
A certain level of subjectivity is inevitable at this point. However, after an initial `training phase' with observational feedback (see \eg \ Sect.\,\ref{s6_COSMOS_tests}), a relatively homogeneous classification of the sources is possible.



\begin{figure}[h]
\begin{center}
\includegraphics[angle=0,clip,height=21cm]{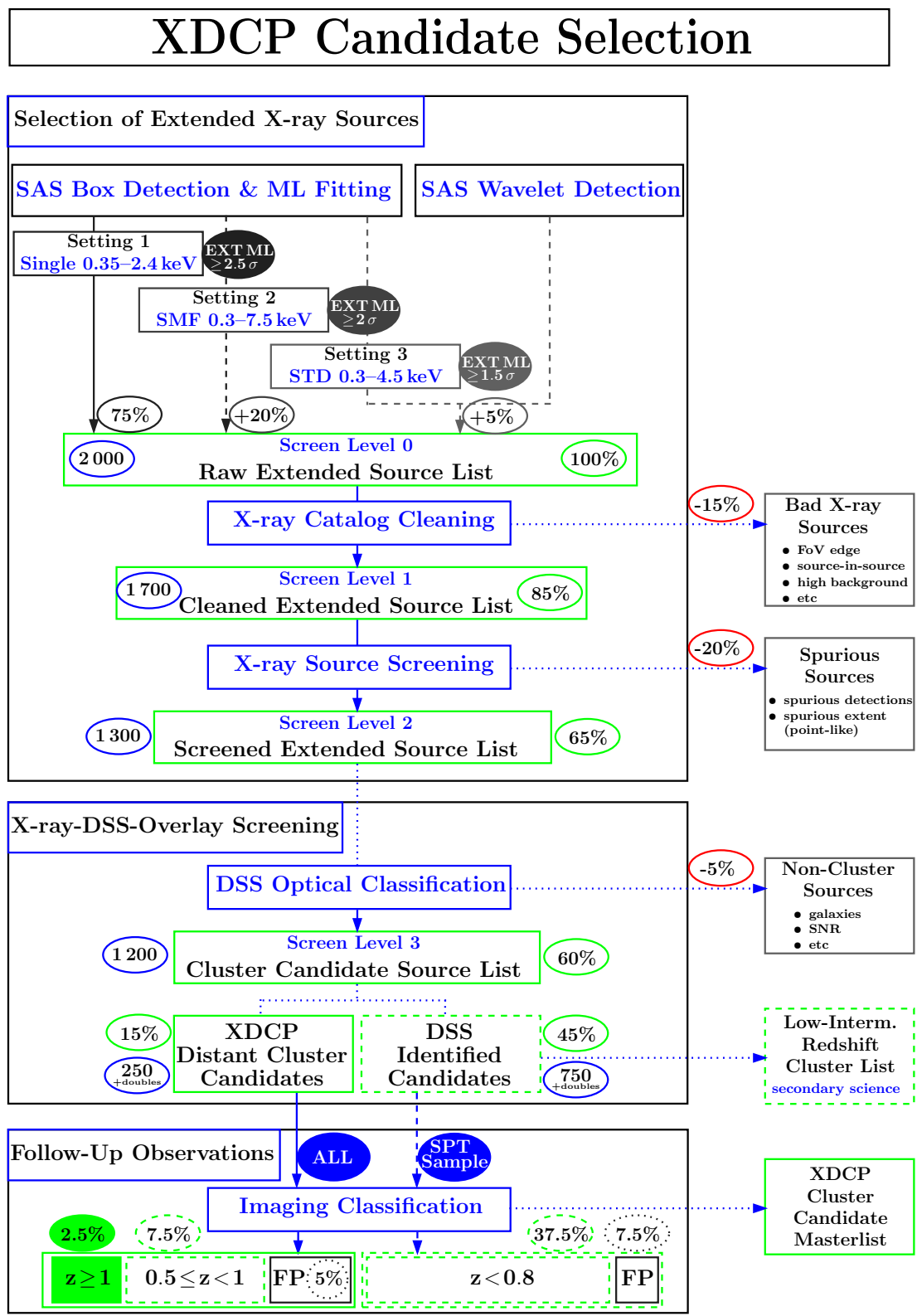}
\end{center}
\vspace{-3ex}
\caption[Source Screening Flow Chart]{Source screening flow chart. Used abbreviations are EXT\,ML: Extent Maximum Likelihood, SMF: Spectral Matched Filter, STD: Standard Method, DSS: Digitized Sky Survey, SPT: South Pole Telescope, FP: False Positives. Stated percentages refer to the {\em Level 0} starting sample, absolute numbers are circled in blue.} \label{f6_source_screening}
\end{figure}

\clearpage

\subsection{Screening steps}
\noindent
The complete XDCP screening and classification procedure is illustrated in the flow chart of Fig.\,\ref{f6_source_screening}.
The selection process can be structured into the two independent parts: (i) the X-ray quality assessment of the source and (ii) the optical counterpart identification.  
The third part in Fig.\,\ref{f6_source_screening} displays some rounded preliminary success rates 
of follow-up observations, covered in Chap.\,\ref{c7_NIRanalysis}, in order to complete the start-to-end source accounting. Numbers circled in blue refer to approximate absolute extended source counts at the given screening stage, the green circled percentages are relative to the starting sample at {\em screen level 0}.

The {\em sample completeness} is a prime factor for the XDCP science applications. Consequently care has to be taken not to lose any real clusters during the screening procedure. Ambiguous sources at a given screening stage are therefore passed to the next level for a re-evaluation before they are rejected. In the end, the survey completeness and  the follow-up efficiency have to be balanced, \ie \ the false positive fraction should be acceptable for a given completeness level. The source flagging and classification scheme should thus be flexible enough to allow re-adjustments of the selection cuts as a consequence of observational follow-up feedback.


\subsubsection{X-ray selection}


\noindent
The starting point for the selection process is the {\em raw extended source list} or {\em screen level~0}, which is compiled in a  automated and objective way. The first aim of this starting source sample is the complete coverage of the {\em single band scheme} detections on the 2.5\,$\sigma$ extent significance level, which contribute about 75\% (see  Fig.\,\ref{f6_source_screening}). The second aim is to test for possible selection effects introduced by a given detection scheme.  We thus include the extended sources detected with the {\em spectral matched filter scheme} on the 2.0\,$\sigma$
significance level which are not on the list and comprise about  20\% of the initial sample. Below an extent significance of 2.0\,$\sigma$ most additional  detected sources  are very uncertain. For the {\em standard scheme} results at the 1.5\,$\sigma$ extent level, only those new sources are added to the list, which are also confirmed as extended by the wavelet method. With these last 5\% of sources detected at lower significance, the {\em raw extended source list} with a total of just over 2\,000 extended sources is completed. From this point on, all subsequent classification steps  ({\em screen levels~1--3\/}) rely on human intervention. 

As a first screening level, the raw X-ray source catalog is cleaned of obvious spurious detections based on the summary plots as shown in the upper right panel of Fig.\,\ref{f6_Xray_products}. At this time, the majority of the source-in-source detections, FoV boundary sources, wing detections of bright point sources,  and pronounced out-of-time residual detections  are removed. With about 15\% of the X-ray sources flagged as `artifact detections', the cleaned {\em screen level~1} sources are passed on two the next level.

The detailed X-ray source screening of the second level makes use of the zoomed X-ray data overlays as shown in panels\,1\,\&\,2 of Fig.\,\ref{f6_DSS_Overlay} and (up to) three sets of returned source parameters of the maximum likelihood fitting procedure. 
At this stage, sources with more subtle reasons for a spurious {\em extent} or a spurious {\em detection} are identified. The most common causes are misclassified point sources, point source blends, high noise environments, poorly determined backgrounds, and detections at high off-axis angle due to PSF residuals. Besides the visual confirmation of the {\em existence} and {\em extent} using the overlay panels\,1\,\&\,2, the source quality is evaluated based on (i) the wavelet detection result, (ii) 
the number of extent detections and their significance within the three schemes, (iii) the stability of the extent determination, and (iv) the hardness ratio of the source.
Spurious {\em extents} of point sources can often be identified based on cross-checks with items (i), (ii), and (iv); spurious    {\em detections} on the other hand typically do not exhibit stable extent solutions of item (iii) or are not re-detected at all for tests (i) and (ii).

An additional $\sim$20\% of the initial source sample are identified as spurious sources in this way and flagged accordingly.
Extended sources that passed the X-ray quality screening but are close to the cut threshold are marked as sources with lower X-ray significance and retained in the object list.
The {\em screen level~2} source list has been cleaned of about 35\% of the initial sample. This large fraction of spurious sources is to  be predominantly attributed to systematic effects inherent to the available {\tt SAS} source detection and calibration tools and {\em not} to the expectation rate of false positives of the statistical likelihood.   
The main reason for the built-in redundancies in the XDCP source detection procedure is the improved  identification and removal of this spurious source fraction while retaining the full sensitivity.   
The 
{\em screen level~2} extended source list 
completes the X-ray object selection and characterization and
should now closely reflect the statistical expectations of a sample with a 2--2.5\,$\sigma$ significance  threshold.






\subsubsection{Optical classification}

\enlargethispage{2ex}

\noindent
The screened extended source list is the input for the optical classification, which constitutes the second part of the candidate selection process. 
The main tools for the optical counterpart search are the X-ray-DSS overlays for each source as shown in panels\,3\,\&\,4 of Fig.\,\ref{f6_DSS_Overlay}. 
As a first  step of the optical classification, the non-cluster sources (\eg \ single nearby galaxies) are visually identified using the DSS imaging data and removed from the sample.
The remaining 60\% percent of the input sample comprise the {\em screen level~3} cluster candidate source list of roughly 1\,200 objects, still including multiple detections of the same source in different fields.
  
In the last step, these sources are visually classified either as {\em DSS-identified cluster candidates}  or as {\em XDCP distant cluster candidates} based on their optical counterpart or the lack thereof. The classification scheme will be discussed in the following Sect.\,\ref{s6_classification_scheme}. 
About three quarters of the {\em screen level~3} cluster candidates (45\% of the initial sample) 
can be identified with an optical cluster counterpart signature on the X-ray-DSS overlays. These approximately 750 distinct objects are flagged as low--intermediate redshift cluster candidates and are saved for secondary science projects beyond the XDCP scope.

The $\sim$\,250 extended X-ray sources without an DSS-identified  optical counterpart enter the {\em distant cluster candidate} master list, which constitutes the basis of the subsequent follow-up imaging program. 
In order to provide a full accounting of the extended X-ray source population at the XDCP sensitivity level, approximate cluster yields are stated in Fig.\,\ref{f6_source_screening} as derived from the follow-up imaging programs of Chap.\,\ref{c7_NIRanalysis}. The numbers related to the DSS-identified candidates are based on a preliminary evaluation of the wide-field follow-up program in the South Pole Telescope (SPT) survey region (see Sect.\,\ref{s9_imaging_status}\,\&\,\ref{s11_SZE}). The first important finding from these data is the  confirmation that the sources flagged as {\em spurious} are indeed {\em not} associated with optical clusters out the data limit of $z\!\sim\!0.9$.  

The final cluster accounting establishes the following approximate cluster yields for the distant cluster candidates: About 
one out of six 
candidates is a true \zga1 \ system, half of the sources are associated with $z\!<\!1$ clusters, and roughly a third are false positives (FP), \ie \ spurious sources.  For the DSS-identified candidate sources, about 5/6 are $z\!\la\!0.8$ clusters or groups with a false positive rate of 1/6.  

The XDCP candidate selection task (including spurious sources) can hence be globally summarized as follows (see Fig.\,\ref{f6_source_screening}).
(i) About 1/40 of all extended X-ray sources are associated with the targeted population of \zga1 galaxy clusters.
(ii) $\sim$1/10 of all extended sources are clusters beyond the DSS identification limit.
(iii) $\sim$1/2 of all extended sources are galaxy clusters.
(iv) $\sim$3/4 of all  cluster sources can be identified with an optical DSS counterpart.

Concerning the required cleaning of spurious sources, the following picture is obtained.
(v) For each distant cluster candidate more than two spurious sources were removed from the initial sample in early cleaning stages.
(vi) $\sim$3/4 of all spurious sources can be identified as such  within the screening procedure  without follow-up observations.
(vii) $\sim$1/10 of all spurious sources pass the screening procedure and need to be identified as {\em false positive} sources in the follow-up data of the distant cluster candidate sample. (viii) After removing all spurious sources from the object catalog, $\ga\!90$\% of the  {\em non-spurious\/} extended sources in the high galactic latitude survey fields are associated with galaxy clusters.  
 Additional source diagnostic plots will be presented in Sect.\,\ref{s6_diagnostic_plots}.






\subsection{Classification scheme}
\label{s6_classification_scheme}


\enlargethispage{2ex}

\noindent
After the  general overview of the screening procedure, we will now have a closer look at the cluster candidate classification scheme. As discussed, the selection process is comprised of the two independent parts of the X-ray quality assessment of the source, followed by the subsequent evaluation of possible optical counterparts on DSS imaging data. 

\begin{figure}[t]
\centering
\includegraphics[angle=0,clip,width=0.45\textwidth]{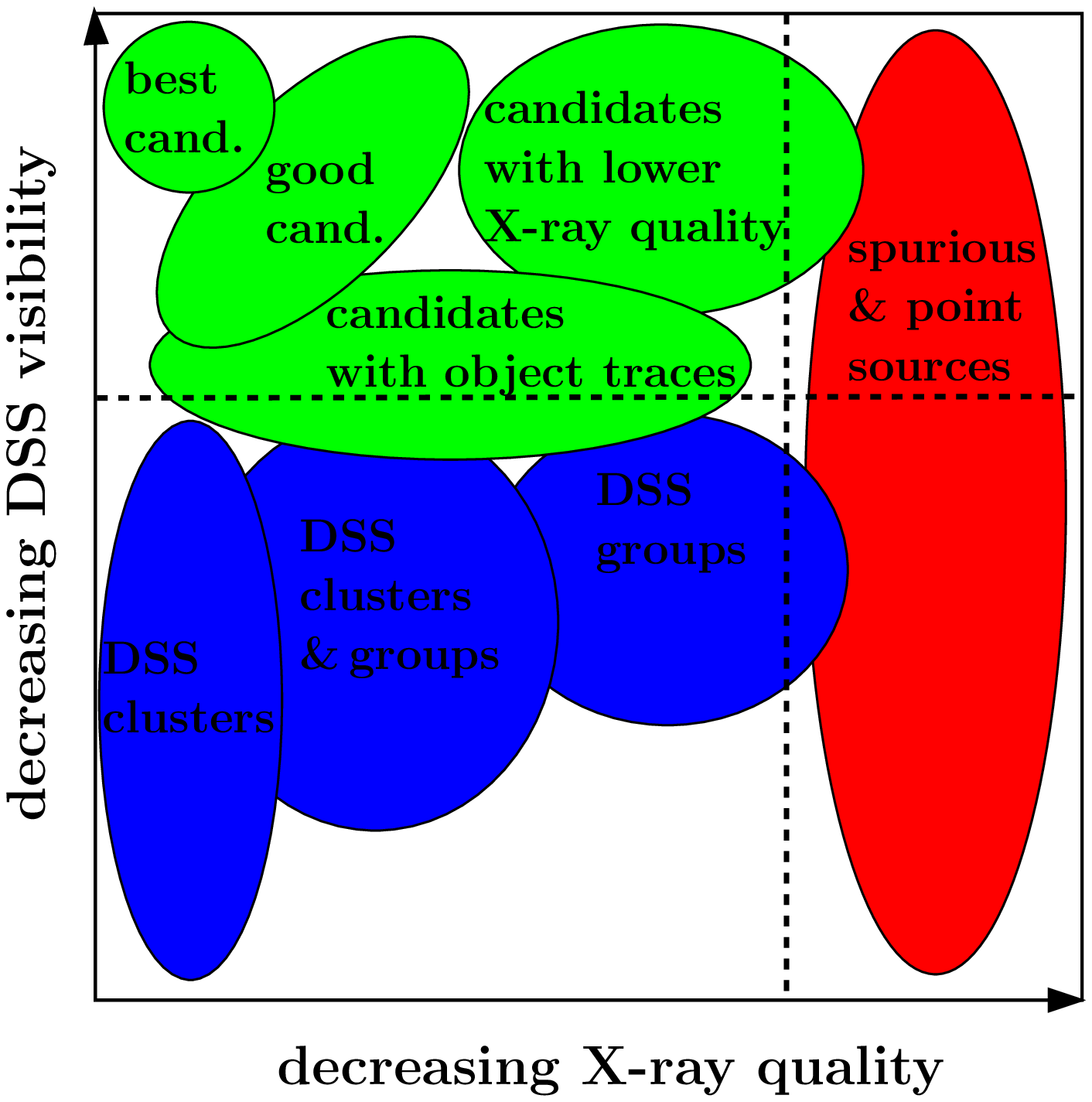}
\includegraphics[angle=0,clip,width=0.45\textwidth]{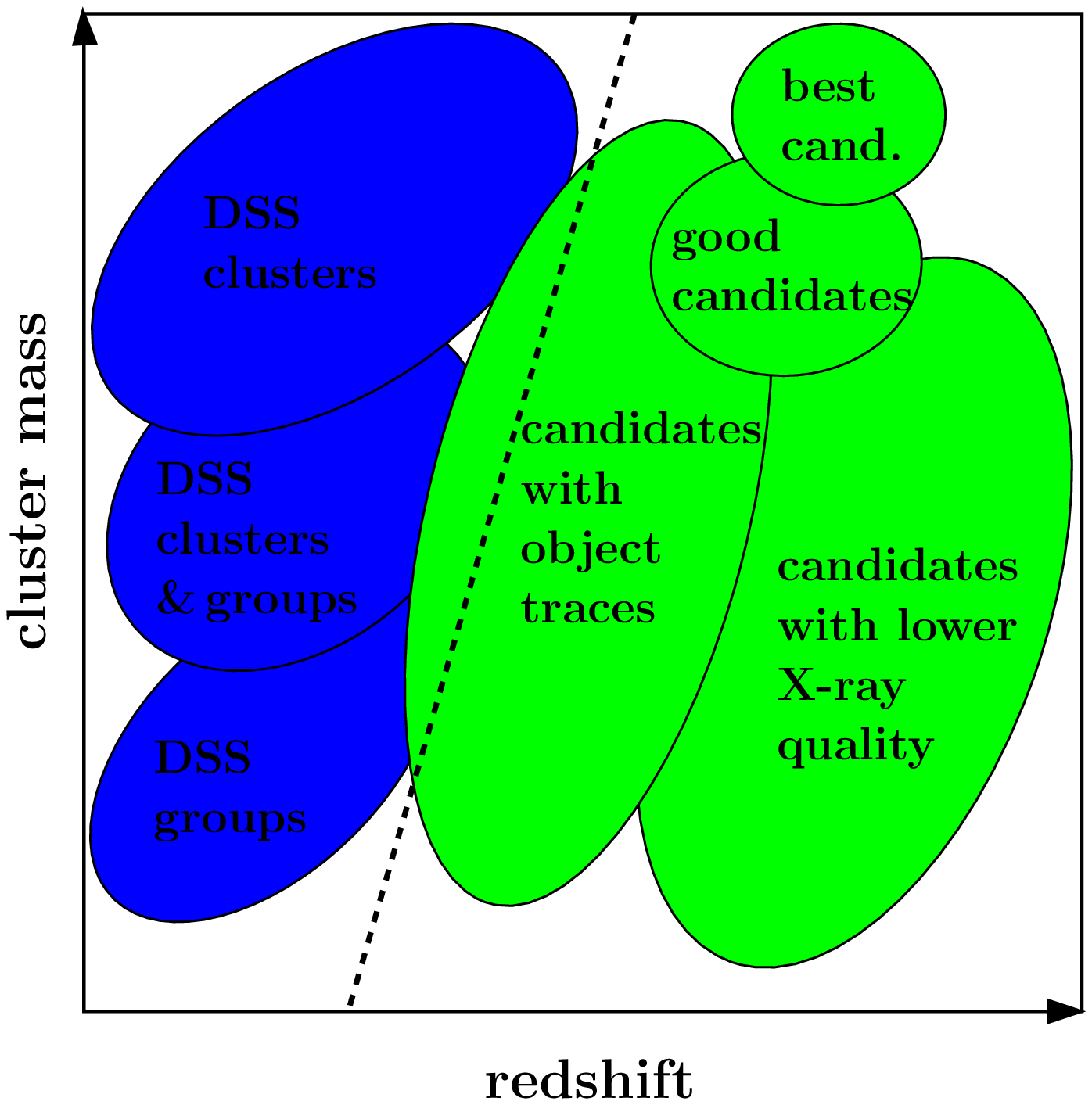}
\vspace{-1ex}
\caption[Flagging Scheme]{Qualitative X-ray source flagging scheme. {\em Left}: Practical implementation in the DSS-object-visibility versus X-ray-quality plane. The dashed lines indicate qualitatively the DSS limit (horizontal) and the X-ray quality limit (vertical) beyond which sources are classified as spurious or point-like. {\em Right:}  Qualitative 
 correspondence of the flagging scheme in the cluster mass versus redshift plane for ideal clusters. The dashed line represents the DSS identification limit.}
\label{f6_FlaggingScheme}       
\end{figure}

These two screening dimensions are reflected in the classification scheme illustrated in the left panel of Fig.\,\ref{f6_FlaggingScheme}.
In a schematic way, the optical DSS visibility is plotted against the X-ray quality of the source. Beyond the X-ray significance and quality threshold (vertical dashed line), sources are flagged as spurious (red region). The DSS limit (horizontal dashed line) divides the plane into the distant cluster candidates in the upper half (green) and DSS-identified systems in the lower half (blue). In this scheme the ideal distant candidates, featuring  excellent X-ray quality in combination with a blank DSS sky region, are located in the upper left corner. Approaching the DSS limit, \ie  \ moving downwards, results in 
traces of  possible cluster galaxies at a level insufficient for a secure cluster identification and towards the right side of the plot, the X-ray significance of the extended sources is reduced.


\begin{figure}[b]
\centering
\includegraphics[angle=0,clip,width=0.78\textwidth]{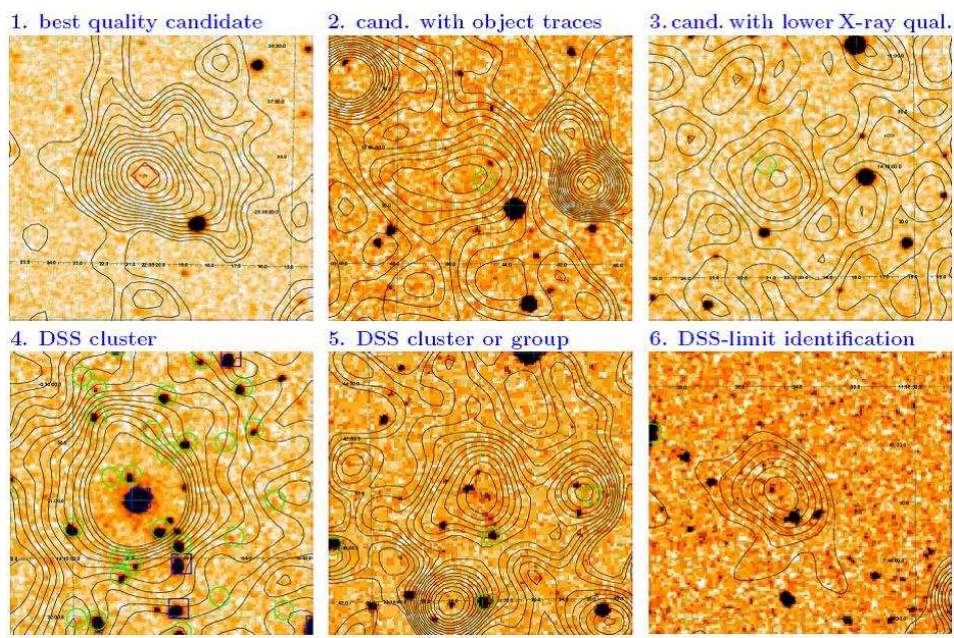}
\vspace{-1ex}
\caption[Classification Prototypes]{Candidate classification prototypes ($2.5\arcmin\!\times\!2.5\arcmin$ \ FoV) following the scheme in the left panel of Fig.\,\ref{f6_FlaggingScheme}. {\em Top row:} Distant cluster candidates of best quality ({\em left}), with slight object traces ({\em center}) and with weaker X-ray significance ({\em right}). {\em Bottom row:} DSS-identified cluster and group sources with high confidence ({\em left}), medium-high confidence ({\em center}), and a DSS-limit case ({\em right}) requiring the cross-comparison of the optical and NIR DSS data for enhancing the object likelihood of weak traces. The NED label in the upper left panel identifies XMMU\,J2235.3-2557 at $z\!=\!1.39$.}
\label{f6_prototypes}       
\end{figure}

\clearpage

Relating DSS visibility qualitatively to redshift and X-ray quality to cluster mass, the classification  scheme is transformed, in first order, to depend on the prime cluster characteristics (right panel of Fig.\,\ref{f6_FlaggingScheme}). The diagonal dashed line indicates the DSS limit which is pushed towards higher redshift for increased cluster masses. Increasing the redshift of a system or decreasing the mass results on both cases in a lower X-ray `quality' of the cluster candidate (\ie \ a lower signal).

In general, the qualitative classification dimensions DSS visibility and X-ray quality are related to the physical cluster characteristics redshift and mass in a more complex way.   
Besides the mass and redshift dependence, the observed X-ray significance and quality of the candidate is influenced by (i) the effective exposure time at the source location, (ii) the source extent, (iii) the level of the X-ray background, (iv) the galactic hydrogen column at the source location, (v) nearby contaminating sources in the field,  (vi) AGN emission in the cluster, or (vii) ongoing merging activity in the system.    
The DSS counterpart identification, on the other hand, is additionally influenced by (i) the magnitude and location of the brightest cluster galaxies, (ii) the local DSS depth, (iii) an increasing foreground object density at lower galactic latitude, (iv) contamination from optically bright nearby objects, and (v) the lack of DSS-NIR data in some sky regions.

Figure\,\ref{f6_prototypes} illustrates the classification prototypes following the discussed scheme.
The classification is applied to all cluster candidates of the {\em screen level~3} list.
The prototypes in the upper row show distant cluster candidates, whereas in the bottom part examples for DSS-identified systems are displayed.
In the next section, we will address the question of the DSS identification limit in some more detail.





\subsection{DSS identification limit}
\label{s6_identification_limit}

\noindent
The  image data of the second generation Digitized Sky Survey \cite{Reid1991a} is based on photographic plate scans of several 
 wide-field surveys using 1\,m-class Schmidt telescopes.
The typical exposure times for the DSS-red plates, corresponding roughly to the R filter band (see Tab.\,\ref{t7_XDCP_filters}), 
are 60-70\,min. The DSS-NIR plates have been exposed for 80-90\,min and with a central wavelength of $\lambda_{\mathrm{c}}\!\sim\!8\,500$\,\AA \ this band is similar to the (optical) I band.
The average limiting point source sensitivities (Vega magnitudes) are R$_{\mathrm{lim}}\!\simeq\!20.8$\,mag (DSS-red), and I$_{\mathrm{lim}}\!\simeq\!19.5$\,mag (DSS-NIR) with local variations of about $\pm$0.4\,mag for sky regions with the deepest and shallowest coverage.
The limit in the R-band for the DSS-red data corresponds to the expected apparent magnitude of an $M$* elliptical galaxy at redshift $z\!\sim\!0.5$, an $M$*$-$1 elliptical at $z\!\sim\!0.65$, or an $M$*$-$2 object at $z\!\sim\!0.8$ (see Fig.\,\ref{f7_CMD_models} and Tab.\,\ref{t7_magnitude_evolution}). For the I-band of the DSS-NIR images the corresponding redshifts are $z\!\sim\!0.45$ for an $M$* galaxy, $z\!\sim\!0.65$ for $M$*$-$1, and $z\!\sim\!0.85$ for an $M$*$-$2 object.

The brightest cluster galaxies (BCGs) are typically 1--2 magnitude brighter than the characteristic luminosity $L$* (see Sect.\,\ref{s10_bcg_assembly}). This implies that BCG signatures should be generally visible in DSS-NIR images out to redshifts of  $z\!\sim\!0.5$--0.8.
The significance of DSS sources close to the limiting magnitude can be additionally enhanced by cross-comparing the DSS-red and DSS-NIR images for spatially coincident object traces.  

\newpage


\begin{figure}[t]
\centering
\includegraphics[angle=0,clip,width=0.86\textwidth]{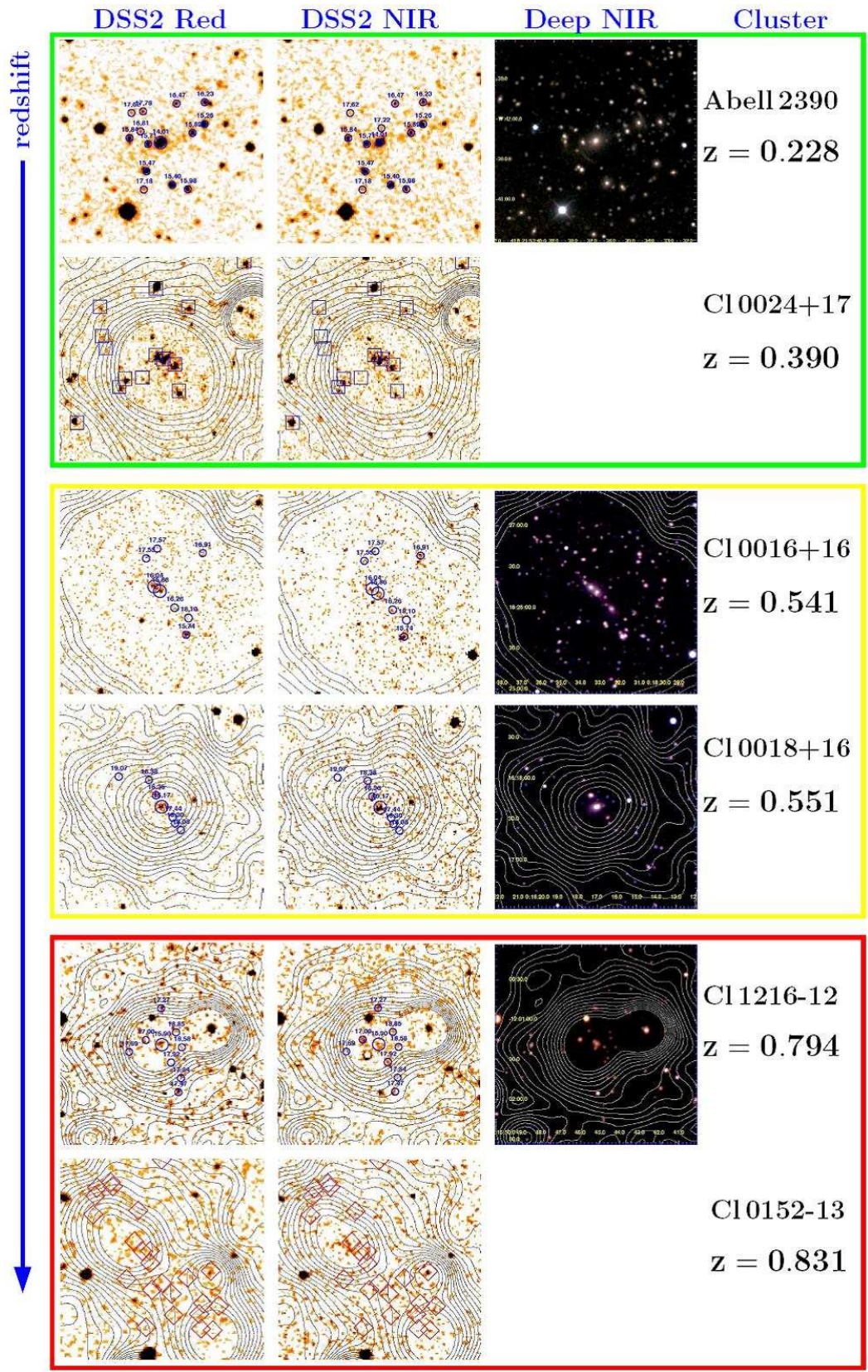}

\vspace{-1ex}
\caption[DSS Identification Limit]{DSS identification limit for rich clusters. Marked object positions are either associated with color selected cluster members (circles) with shown H-band Vega  magnitudes or spectroscopically confirmed cluster members (squares and diamonds). All images are $2.5\arcmin\!\times\!2.5\arcmin$ \ in size.}
\label{f6_DSS_limit}       
\end{figure}

\clearpage

Figure\,\ref{f6_DSS_limit} shows a compilation of six spectroscopically confirmed galaxy clusters with redshifts $0.2\!\la\!z\!\la\!0.8$. The corresponding $2.5\arcmin\!\times\!2.5\arcmin$ DSS-red (left) and DSS-NIR images (center) are displayed with high-contrast color cuts around the noise limit of the data. The right panels present NIR color composites taken, if available, from the follow-up imaging data of Chap.\,\ref{c7_NIRanalysis} to indicate the  underlying deep optical cluster signature. The overlaid blue circles on the DSS data represent color selected cluster members with associated H-band magnitudes. For the two systems without additional follow-up data, the sky positions of spectroscopically confirmed cluster members are marked.  

As can be seen from Fig.\,\ref{f6_DSS_limit}, the optical DSS-identification of clusters out to $z\!\sim\!0.4$ (green box) is straightforward and is generally possible at high confidence levels.
In the redshift range  $0.4\!\la\!z\!\la\!0.6$ (yellow box), the majority of objects can still be identified based on the BCG and the next few brightest cluster galaxies. In the case of cluster Cl0016+16 5--6 objects are visible, but for Cl0018+16 only 2--3 cluster galaxies are above the DSS noise limit.   
At $z\!>\!0.6$ the DSS-based cluster identification becomes increasingly difficult and is beyond the DSS capabilities in the general case. However, for favorable combinations of BCG properties, cluster richness, and local DSS depth, the optical counterpart can, in some cases, be identified out to  $z\!\sim\!0.8$.
Cl1216$-$12 exhibits weak traces of about half a dozen probable cluster galaxies, whereas for Cl0152$-$13 only one probable AGN counterpart is securely detected and  some weak indications of the BCG are visible in the DSS-NIR image. 

We conclude that the average DSS cluster identification limit is in the redshift range $z\!\simeq\!0.5$--0.6, if the full depth of the DSS-red {\em and} DSS-NIR images is exploited. This limit is also confirmed by early XDCP selection performance tests in the COSMOS field as presented in the next section.






\section{Performance Tests in the COSMOS Field}
\label{s6_COSMOS_tests}

\noindent
The XDCP X-ray pipeline of Sect.\,\ref{s6_Xray_Pipeline} and the candidate selection procedure of Sect.\,\ref{s6_Source_Screening}
has been tested in its early version in the COSMOS survey field (see Sect.\,\ref{s4_distant_cl2004}). The COSMOS data set\footnote{Only the first XMM data set has been used, meanwhile the available COSMOS X-ray coverage has more than doubled in exposure time.} is particularly well suited for a performance benchmark since (i) the individual XMM pointings are of similar depth as the average XDCP survey fields, and (ii) the  multi-wavelength coverage of the field enables a direct cross-comparison between detected extended X-ray sources and  galaxy overdensities with corresponding accurate photometric redshifts.  

The  XDCP reduction and selection machinery has been (blindly) applied to 20 individual XMM fields of the first COSMOS X-ray data set, corresponding to about 85\% of the observations required to cover the full 2.1\,deg$^2$ field. 
The average cleaned exposure time of 24.9\,ksec per field is about 30\% deeper than the median coverage of the XDCP fields.
The fields are overlapping to contiguously map the 1.4\degr$\times$1.4\degr area. 
In contrast to the  COSMOS  X-ray analysis of Finoguenov \etal \ \cite*{Alexis2006a}, which made use of the reconstructed mosaic image  with the full  co-added effective exposure at each position,  
the XDCP test procedure was applied to the {\em individual} XMM fields as for the real survey fields. 
Following the compilation of the final {\em screen level~3} source sample with distant cluster candidates and DSS-identified systems, the XDCP object list was cross-correlated with the proper COSMOS X-ray cluster list, which is based on the deeper mosaic data 
in conjunction with the photometric confirmation based on the optical data. 

\begin{figure}[t]
\begin{center}
\includegraphics[angle=0,clip,height=5.5cm]{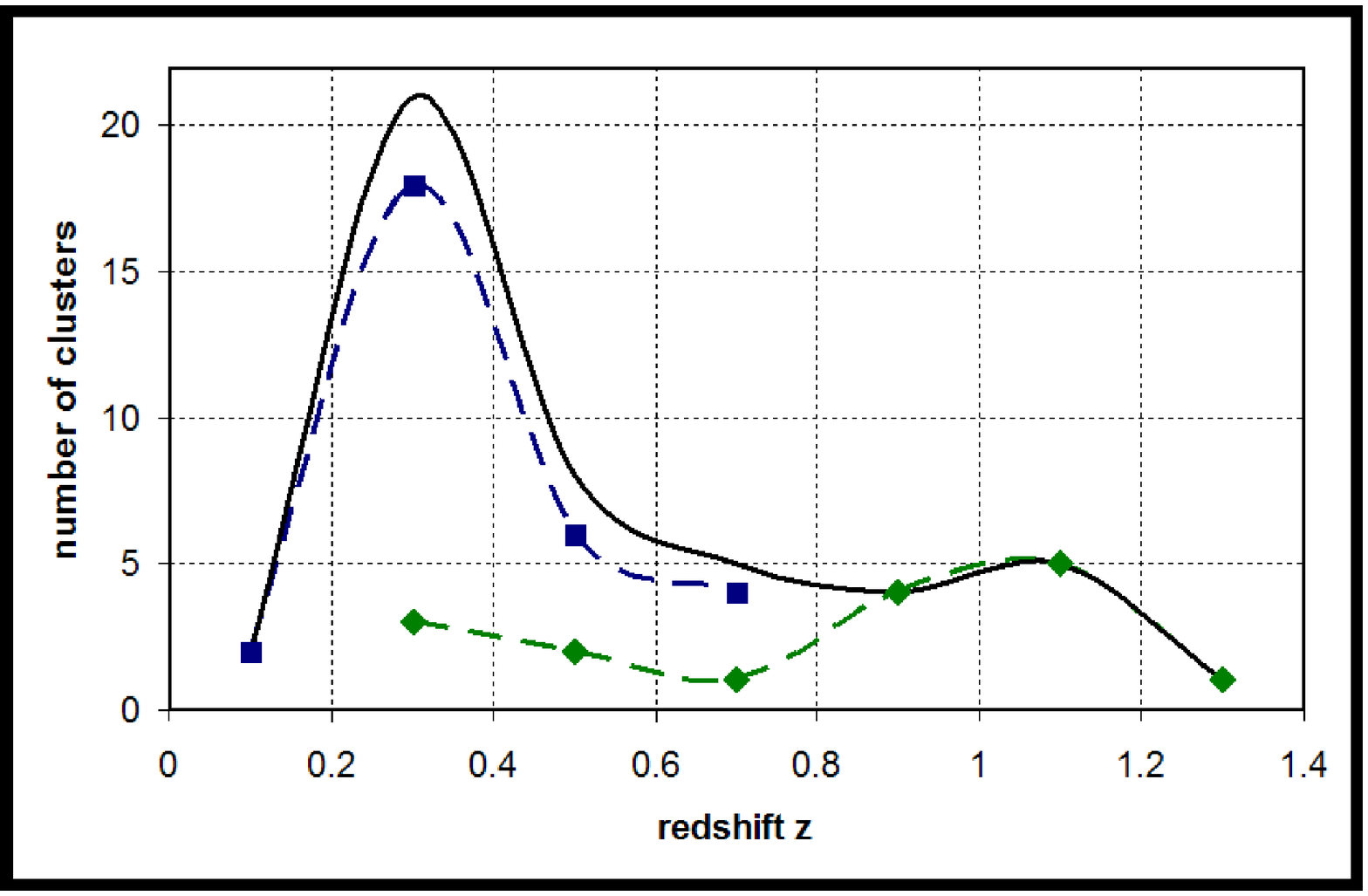}
\includegraphics[angle=0,clip,height=5.45cm]{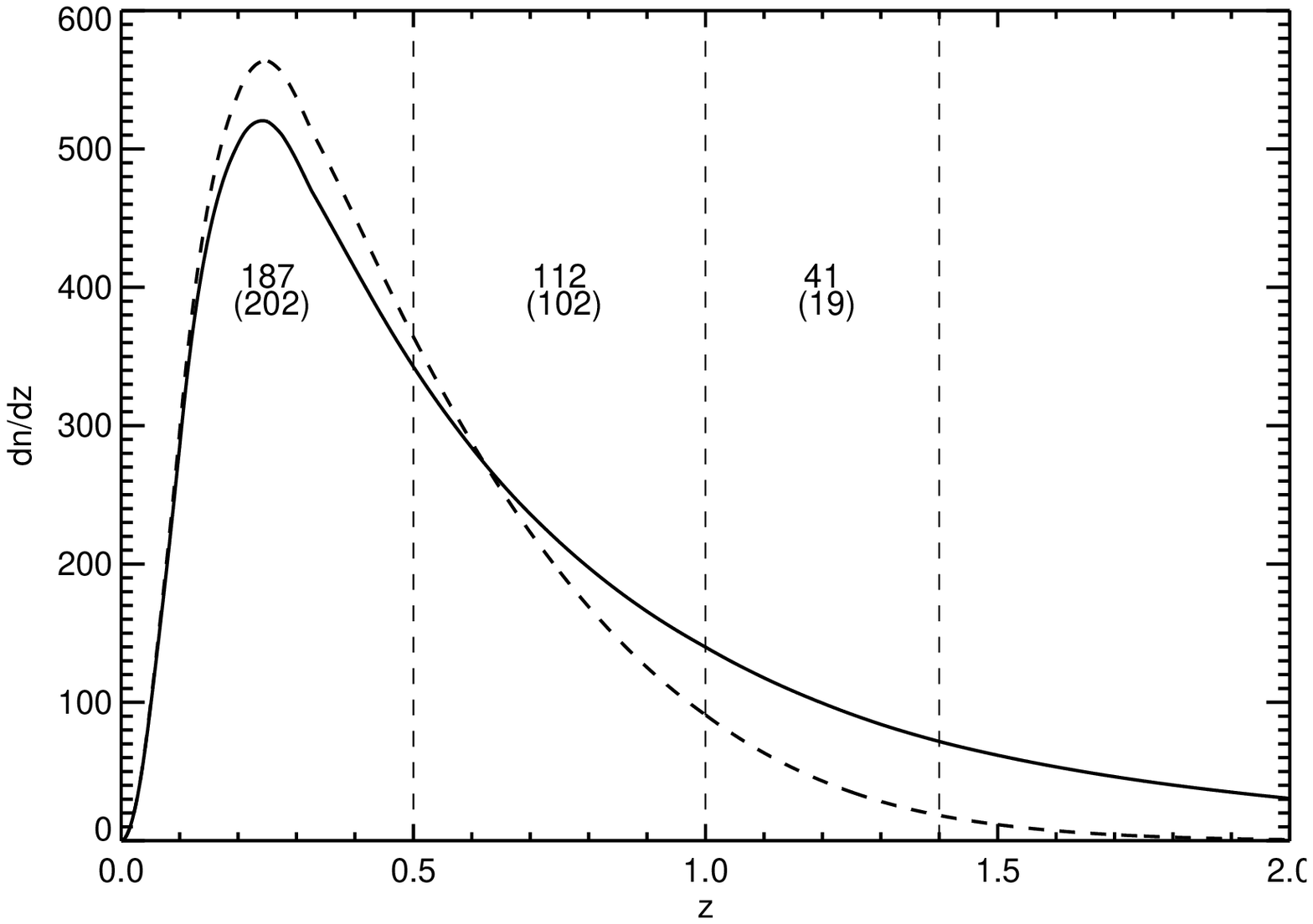}
\end{center}
\vspace{-2ex}
\caption[Redshift Distribution of Identified COSMOS Clusters]{{\em Left:} Redshift distribution of identified clusters and candidates in the COSMOS field (black line). The blue dotted line shows the distribution of DSS-identified clusters in redshift bins of $\Delta z\!=\!0.2$, the green dashed line represents the distant cluster candidates. The photometric redshift estimates are based on early internal versions of the COSMOS galaxy catalog. The bump at \zsim1 \ is also present in the published cluster catalog \cite{Alexis2006a} and is still under investigation. The five candidates at $z\!<\!0.6$ were identified as low mass groups, which are harder to identify on DSS images.
{\em Right:} The expected shape of the redshift distribution for an XDCP-like survey without cluster evolution (solid line) and with evolution (dashed line). Plot from C. Mullis.
}  \label{f6_COSMOS_clusters}
\end{figure}

The left panel of Fig.\,\ref{f6_COSMOS_clusters} illustrates the results of this test in form of the binned photometric redshift distribution of all 46 XDCP identified clusters (black line). The blue dashed line represents the {\em DSS-identified clusters}, whereas the green line indicates the selected {\em distant cluster candidates}. 
Six distant \zga1 \ clusters with  X-ray fluxes down to  $\sim\!4\!\times\!10^{-15}$\,\flux \ [0.5--2.0\,keV]
were successfully identified. 
The five 
candidates at the lower redshift end at $z\!\la\!0.6$ (green line) have been revealed as lower mass groups, mostly in the 
 range $M\!\simeq\!3$--$4\!\times\!10^{13}\,M_{\sun}$. Due to their lower optical richness, these systems are 
 harder to distinguish from possible foreground objects. The blue line for the DSS-identified clusters, on the other hand, confirms the  average DSS redshift limit of  $z\!\simeq\!0.5$--0.6.
The fraction of false positive sources in the XDCP sample turned out to be 6/22 (27\%) for the distant cluster candidates, and 2/32 (6\%) for the DSS flagged objects. 
A slightly worse performance for the real XDCP survey data can be expected due to more contaminations in serendipitous fields,
 which have not been specifically  selected for deep extragalactic survey applications.




\enlargethispage{4ex}

The right panel of Fig.\,\ref{f6_COSMOS_clusters} shows the expected shape of the cluster redshift distribution for an XDCP-like survey, derived for early sensitivity estimates. The solid line represents a model without cluster evolution, whereas the dashed line includes evolution effects as expected from the X-ray luminosity function analysis of Fig.\,\ref{f4_XLF}.

Finoguenov \etal \ \cite*{Alexis2006a} find no substantial evolution in their analysis of the initial 72 cluster COSMOS sample out to $z\!\sim\!1.3$, selected from the first 36 XMM pointings. The COSMOS cluster luminosities cover the range $8\!\times\!10^{42}$--$2\!\times\!10^{44}$\,erg\,s$^{-1}$ and hence do not include the high mass end for which mild evolution effects have been established (Sect.\,\ref{s4_DeepSurveys}).
The pronounced bump in the observed redshift distribution around \zsim1 (left panel), also present in the published COSMOS data and still under investigation, is likely attributed to either physical large-scale structure or systematic effects in the derived photometric galaxy redshifts.

In summary, the XDCP X-ray pipeline performance test in the COSMOS field has confirmed (i) the high achievable detection sensitivity, (ii) the good identification efficiency of low--intermediate redshift DSS clusters, and (iii) a high fraction of probable \zga1 \ systems in the distant cluster candidate sample,
giving additional support to the selection procedure.

\section{Field and Source Diagnostics}
\label{s6_diagnostic_plots}

\noindent
In the last section of this chapter on XMM archival X-ray data, we will have a closer look at some selected statistical properties of the survey fields and the detected extended sources.


\subsection{XMM data yield and field statistics}
\label{s6_field_ statistics}
\enlargethispage{2ex}

\begin{figure}[t]
\begin{center}
\includegraphics[angle=0,clip,width=0.85\textwidth]{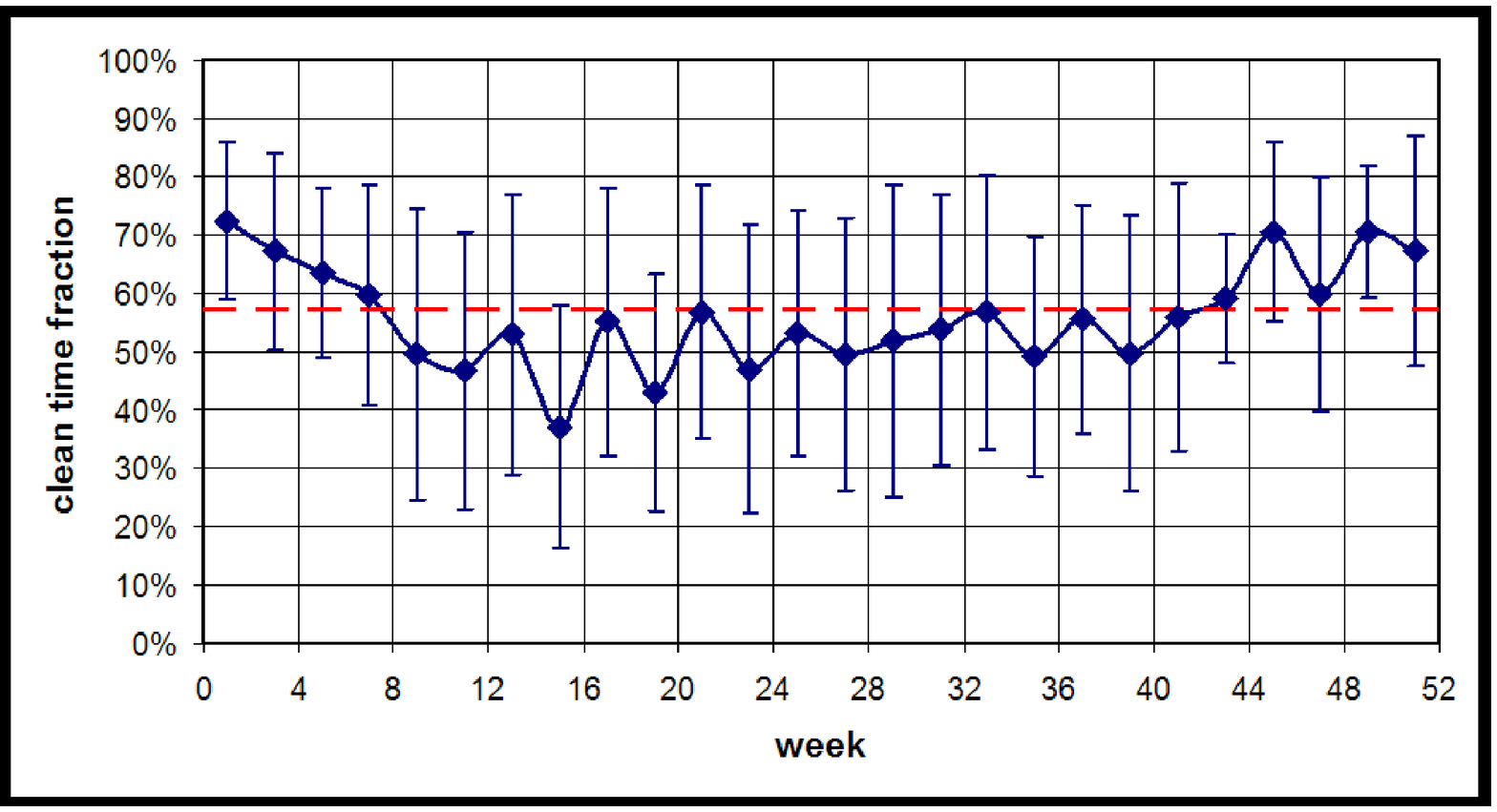}
\includegraphics[angle=0,clip,width=0.83\textwidth]{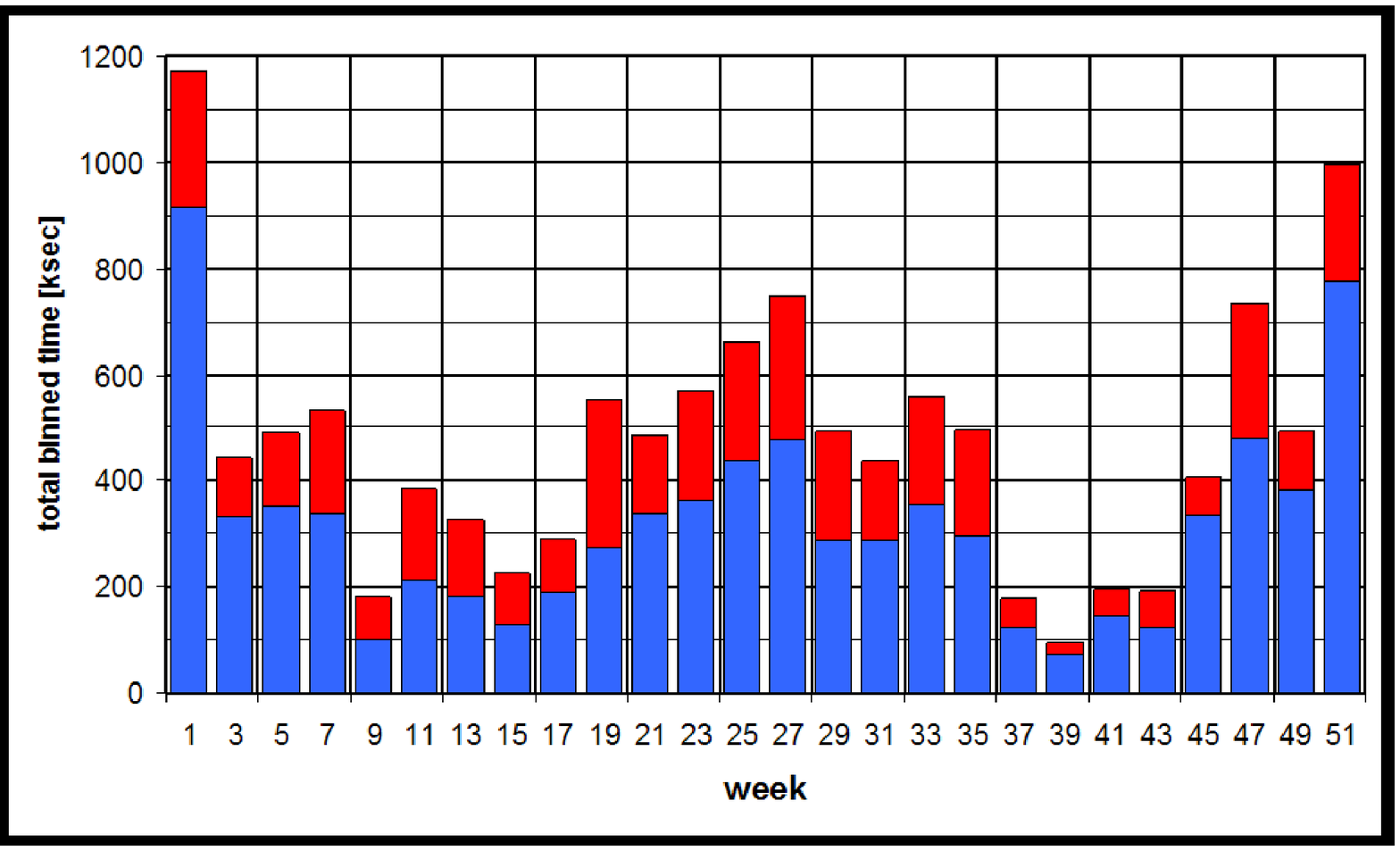}
\end{center}
\caption[Space Weather Seasons]{Seasonal changes of the clean time fraction. {\em Top:}
Average clean time fraction versus the time of the calender year in bins of two weeks. The data points are based on 540 processed XMM archive fields observed within a 4.5 year period between January 2000 and August 2004. The dashed horizontal line represents the sample average of the weighted effective clean imaging time fraction of 57\%. In contrast to most optical telescope sites, the space weather is favorable in terms of data yield in the winter months between October and February.
{\em Bottom:}
Absolute clean exposure times (blue) and flared periods (red) in two week bins for the 439 fields for which all three detectors have been operating in imaging mode. The average clean time fraction is now 67\% and is representative for the overall XMM science usable data fraction.   
} \label{f6_Saisonal_SpaceWeather}
\end{figure}

\noindent
The 540 reduced fields of the first 4.5 years of XMM operations give the possibility to trace the XMM science data yield over the course of a year and to assess the data losses due to flare periods (see Sect.\,\ref{s6_Xray_data_reduction}). Figure\,\ref{f6_Saisonal_SpaceWeather} illustrates the seasonal changes of the science-usable data fraction. The upper panel shows the instrument-weighted clean imaging time fraction with respect to the nominal exposure time for {\em all} fields that passed the flare removal procedure. In effect, this represents the total archival {\em imaging data percentage\/},
and has an average value\footnote{The 10\% difference compared to the overall XMM science data yield is due to the fact that  the full sample of 540 fields contains observations, in which some instruments were not operated in imaging mode, \ie \ science modes not usable for the XDCP survey.} of 57\%.

The lower panel displays the absolute clean and lost times for the fields with all instruments in imaging mode. The resulting mean clean time fraction of 67\% is representative for the {\em overall XMM science data yield\/}. Hence one third of the nominal pointing time is on average lost, predominantly due to soft proton flares and a  smaller fraction due to 
instrument overheads.  
        
Both representations show an increased lost data fraction during the summer months and a higher data yield during winter with differences of about 20\% of the total time. 
These pronounced seasonal {\em space weather effects} are thought\footnote{See longterm 
compilations
at \url{http://xmm.vilspa.esa.es/docs/documents/USR-TN-0014-1-0.pdf}.} to be related to the solar cycle and the XMM orbit evolution  caused by external perturbations.
The latter effect can also introduce long term changes from year to year.

\begin{figure}[t]
\begin{center}
\includegraphics[angle=0,clip,width=0.92\textwidth]{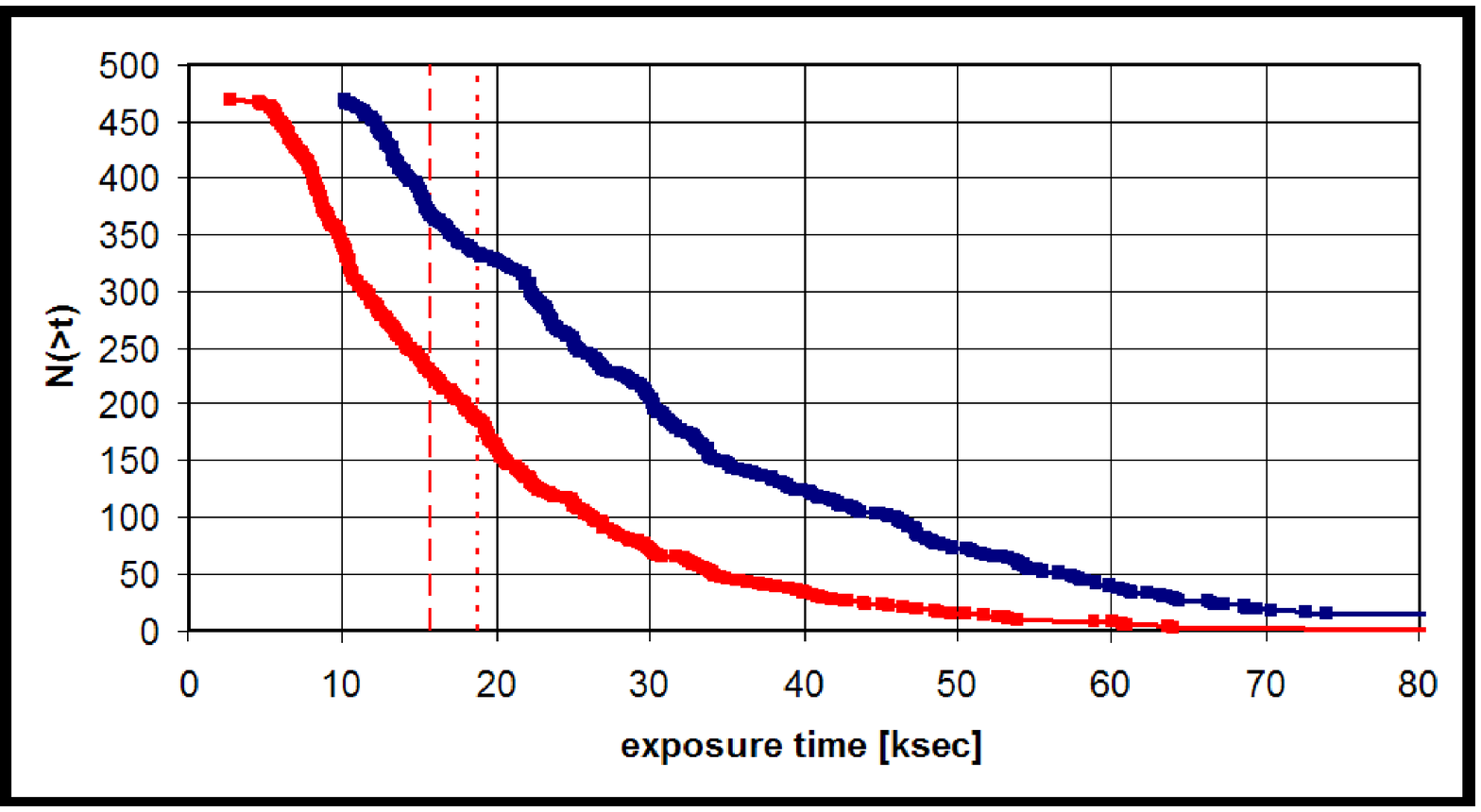}
\includegraphics[angle=0,clip,width=0.94\textwidth]{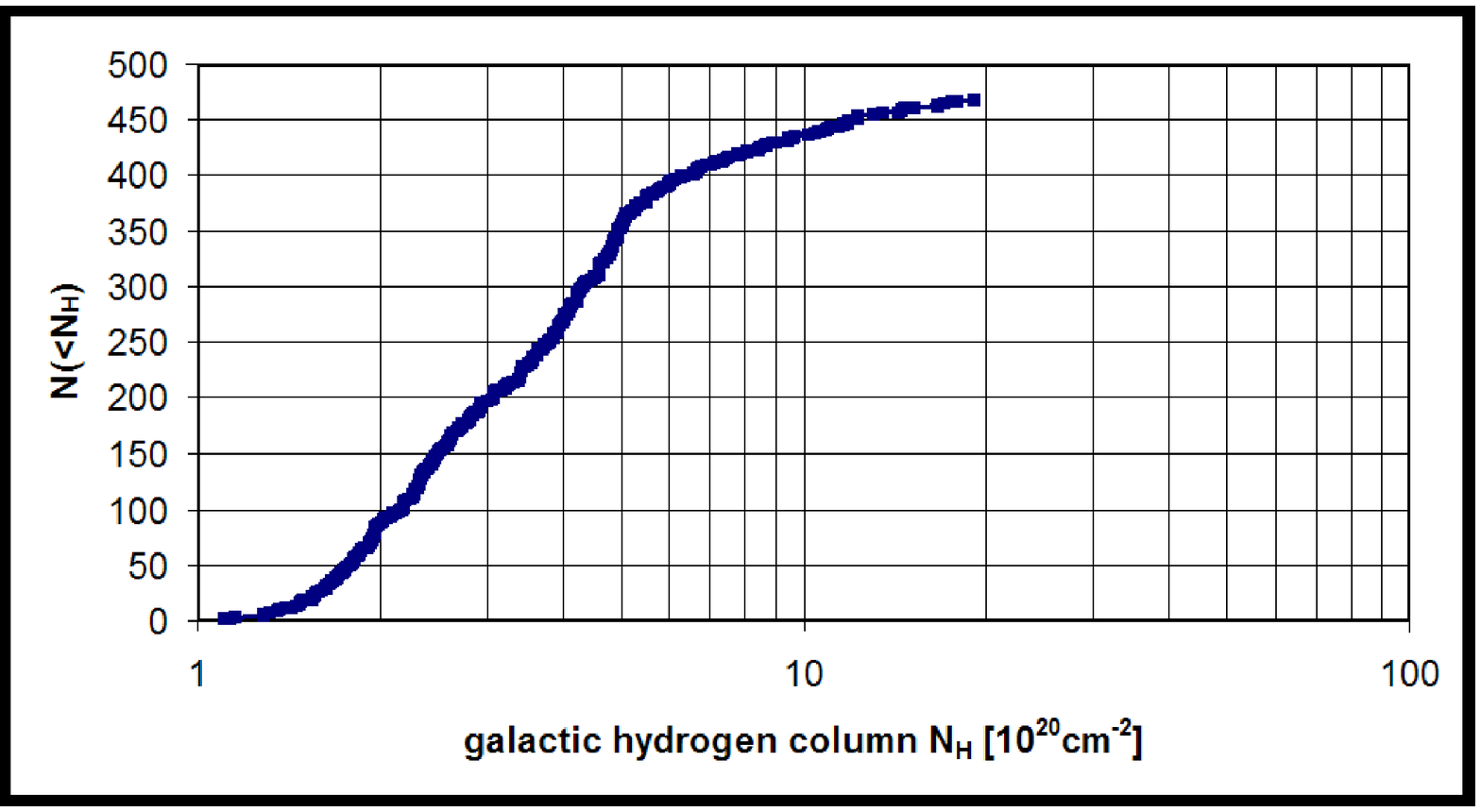}
\end{center}
\caption[XDCP Survey Field Statistics]{XDCP survey field statistics. {\em Top:} Cumulative distribution of nominal exposure times (blue line) and flare cleaned times (red line) for the 469 usable XDCP survey fields. The average clean exposure time of 18.78\,ksec is indicated by the vertical dotted line, the median of 15.71\,ksec is illustrated by the dashed line.  
{\em Bottom:} Cumulative distribution of the galactic hydrogen column for the survey fields.
} \label{f6_field_statistics}
\end{figure}


\clearpage








The statistical properties of the 469 fields of the XDCP core survey sample are summarized in Fig.\,\ref{f6_field_statistics}. 
The top panel illustrates the cumulative distribution of the resulting clean exposure times (red) in comparison to the nominal distribution (blue). The average clean XDCP field depth is 18.78\,ksec (dotted vertical line), the median is 15.71\,ksec (dashed line).   
The lower panel shows the cumulative field distribution as a function of the galactic hydrogen column $N_{\mathrm{H}}$   
with a median value of $3.55\!\times\!10^{20}$\,cm$^{-2}$.




\subsection{Extended source diagnostics}
\label{s6_extendedsource_ statistics}

\begin{figure}[t]
\centering
\includegraphics[angle=0,clip,width=0.9\textwidth]{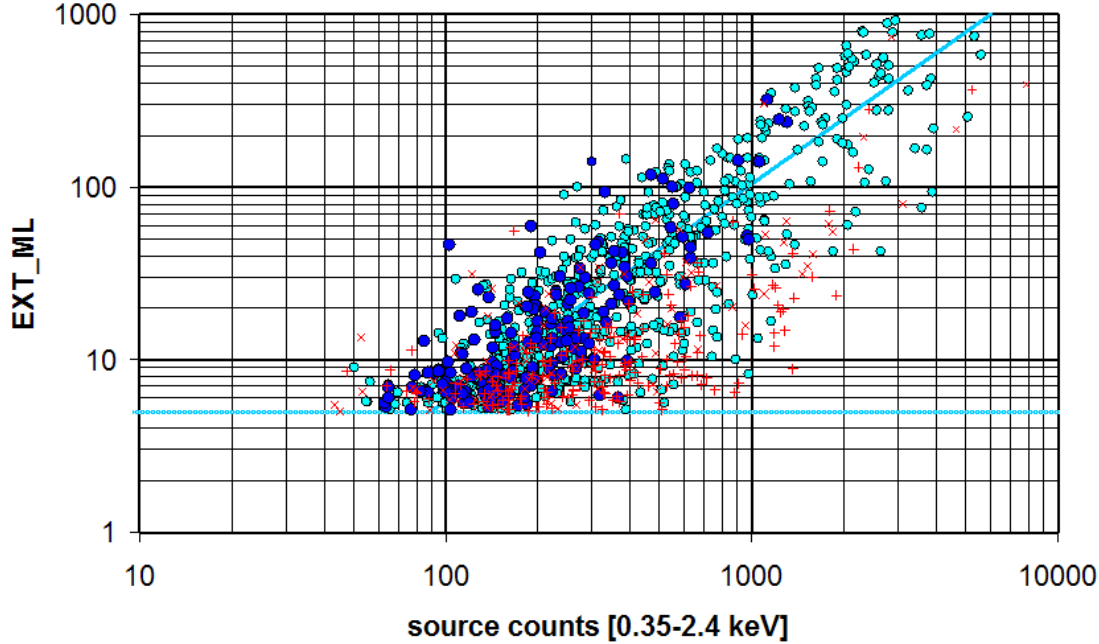}
\caption[Extent Likelihood versus X-ray Source Counts]{Extent likelihood versus X-ray source counts. Dark blue circles represent distant cluster candidates, light blue circles DSS-identified clusters, and red symbols sources that have been classified as spurious detections. The lower horizontal line indicates the minimum required extent likelihood, whereas the diagonal line shows the general correlation for the good sources. Sources that have been classified as spurious tend to have a lower extent likelihood compared to clusters candidates with the same number of counts. However, the populations are still largely overlapping which prevents an automated identification of false detection.}
\label{f6_EXTML_counts}       
\end{figure}

\noindent
We now turn to the properties of the extended X-ray sources and their detection efficiency.
Figure\,\ref{f6_EXTML_counts} provides a global summary of all extended X-ray sources detected with the {\em single band detection scheme}. Plotted are the maximum likelihoods for the extent of the source ({\tt EXT\,ML\/}) versus the total source counts in the 0.35-2.4\,keV band. The horizontal cut-off at {\tt EXT\,ML\/}$=$5 indicates the lower significance threshold for the detection run.

{\em Distant cluster candidates} are represented by dark blue circles. The median number of total source counts for this population is 193, the median  {\tt EXT\,ML\/} is 11.7, corresponding to 5--6\,$\sigma$ significance, and the median detection likelihood amounts to  36.6.
The lighter blue circles indicate the population of {\em DSS-identified cluster candidates}. Red symbols illustrate the distribution of sources classified as {\em spurious extent} (plus signs) or {\em spurious detections} (crosses).
The  extent maximum likelihood exhibits a decent correlation with the total photon number of the cluster candidates (blue symbols), with a scatter of about a factor of two. The extended sources classified as {\em spurious} show on average a lower {\tt EXT\,ML\/} for a given total number of source counts. Even though {\em spurious} sources show a systematic offset in their {\tt EXT\,ML\/}, they still occupy almost the full parameter space populated by cluster sources, in particular at the important fainter levels.


The first important conclusion from this global view on the detected extended source population is that {\em spurious} sources {\em cannot} be automatically separated from real clusters by means of appropriate parameter cuts. This is to be attributed to the discussed systematic (calibration) effects that dominate the {\em spurious} detections for the {\tt eboxdetect} and {\tt emldetect} methods.
The second preliminary conclusion 
is that the sensitivity goal for extended sources down to a total number of 100 counts and less seems to be achievable. The typical significance at which objects with $\sim$100  X-ray photons can be resolved into extended sources is about 4\,$\sigma$.
The final survey sensitivity limit has to be quantified based on detailed simulations (see Sect.\,\ref{s9_Selection_Funct}).







\begin{figure}[t]
\centering
\includegraphics[angle=0,clip,width=0.91\textwidth]{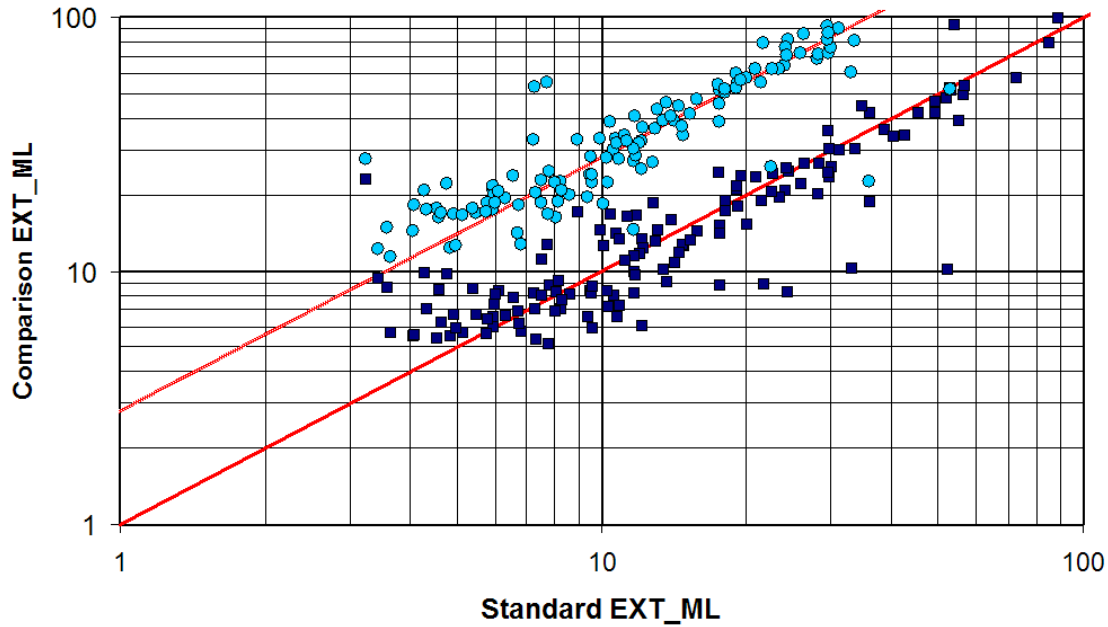}
\caption[Comparison of Detection Schemes]{Extent likelihood comparison of the three detection schemes. 
The dark squares display the
extent likelihoods of the {\em single} 0.35-2.4\,keV detection band versus the {\tt EXT\,ML\/}s of the {\em standard detection scheme} for distant cluster candidates with the lower red line indicating matching likelihoods. 
The top line and the lighter blue circles show the situation for the {\em spectral matched filter scheme} which exhibits an offset of roughly a factor of 2.8 
corresponding to the average effective weight. 
}
\label{f6_EXTML_comp}       
\end{figure}

As a next step, we can compare the extent likelihoods of the sources for the three different detection schemes, which is shown in Fig.\,\ref{f6_EXTML_comp} for the distant cluster candidate sample. The {\tt EXT\,ML\/}s of the {\em standard scheme} are plotted along the abscissa, whereas the corresponding source likelihoods for the  {\em single band scheme} (dark squares) and {\em spectral matched filter scheme} (light blue circles) are represented by the ordinate.   

The likelihoods for the {\em single band scheme} and  {\em standard scheme} scatter along the  lower red line which indicates matching values.
The corresponding comparison between the schemes with a wider spectral coverage   exhibits a slightly tighter correlation about the upper red line, which indicates the average offset factor of 2.8 and corresponds to the mean effective weight.
The effective weighting factor of the {\em spectral matched filter scheme} is not fixed {\em a priori}, but depends on the spectral properties of the source.
Below standard scheme {\tt EXT\,ML\/}s of about 8, the distant cluster candidates appear to be systematically above the average correlation which points towards an achieved enhancement of the likelihoods at faint flux levels.
The global statistics for the {\em spectral matched filter} detections confirm that the average likelihood offsets with respect to the  {\em standard scheme} are increased by about 6\% for distant cluster candidates  compared to sources classified as spurious.
However, no firm conclusion on the effectiveness of the  {\em spectral matched filter} scheme can be drawn  before 
a significant number of distant candidate clusters at lower {\tt EXT\,ML\/} levels have been spectroscopically confirmed.



Last but not least important, the optimal detector area for serendipitous surveys is investigated.  
As discussed in Sect.\,\ref{s6_XMM}, XMM-Newton's {\em grasp} as survey instrument increases with the 
maximum off-axis angle $\Theta$. On the other hand, the vignetting effect (Fig.\,\ref{f6_vignetting}) decreases the sensitivity towards the outskirts, and more importantly, the PSF shape (Fig.\,\ref{f6_xmm_PSF}) broadens and becomes increasingly ill-behaved at large   
$\Theta$, which hampers the detection of extended sources.

\newpage

\begin{figure}[b]
\begin{center}
\includegraphics[angle=0,clip,width=0.89\textwidth]{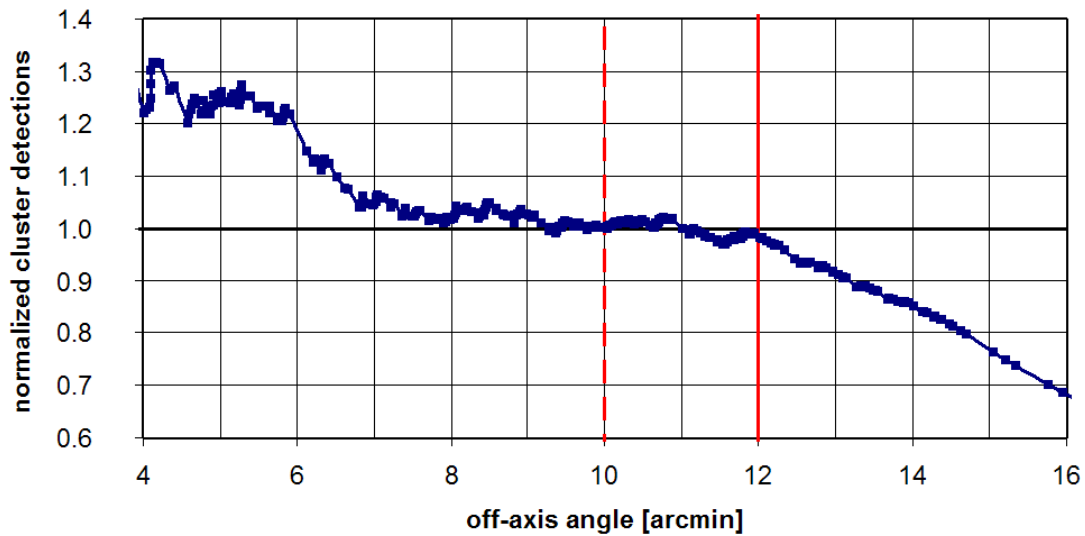}
\end{center}
\vspace{-3ex}
\caption[Cluster Detections versus Off-axis Angle]{Cluster detections per unit sky area as a function of off-axis angle normalized to the detections at 10\arcmin \ off-axis angle (dashed red line). The detection rate is almost constant between 7\arcmin \ and 12\arcmin \ from the optical axis. In the central region the detection rate rises owing to the increased effective area. Beyond 12\arcmin \ the increasing vignetting and the broadened PSF decrease the cluster detections per unit area. The inner 12\arcmin \ (solid red line) are well-behaved and can be used for the core survey.  
} \label{f6_norm_ClusterDetections}
\end{figure}

\clearpage

One way to address the issue of the maximum  off-axis angle for a strictly selected survey sample is to ask at what point the cluster detections per unit sky area start dropping significantly. 
Figure\,\ref{f6_norm_ClusterDetections} shows the results obtained from 120 fields without masked regions, \ie \ no large foreground objects, and a total of 375 cluster sources ({\em screen level~3} objects). The plot traces the normalized cluster detections per unit sky area as a function of off-axis angle, which is derived from the cumulative distribution $N(<\!\Theta$) divided by the enclosed sky area $\Omega(<\!\Theta$) and normalized to the average value within 10\arcmin \ radius (dashed red line).
From Fig.\,\ref{f6_norm_ClusterDetections} it can be seen that the cluster detections per unit area are almost constant between off-axis angles of $7\arcmin\!\la\!\Theta\!\la\!12\arcmin$. At larger radii the detection rate decreases significantly. Hence, a maximum off-axis angle of $\mathbf{\Theta_{\mathrm{max}}\!=\!12\arcmin}$ \ promises to be the best compromise between survey grasp and completeness. With this cut-off radius for the final XDCP core sample, the XMM survey grasp  is $g_{12}\!\simeq\!204$\,cm$^2$\,deg$^2$, a 30\% improvement with respect to a 10\arcmin \ constraint.

\section{Summary}

\noindent
At the end of the technical chapters\,\ref{c6_XrayAnalysis}--\ref{c8_SpecAnalysis}, the main contributions and achievements of {\em this thesis work} towards the overall XDCP survey project are briefly listed.  
Concerning the X-ray analysis of XMM-Newton archival data this work can be summarized as follows:



\begin{itemize}
    \item Definition of the current XDCP survey field sample based on a screening of the full XMM archive;
    \item Development of a distant-cluster-optimized  X-ray pipeline for the data reduction, source detection, and creation of X-ray optical overlays;
    \item Conduction of performance tests for different detection schemes;
    \item Tests of the full XDCP selection procedure in the COSMOS field;
    \item Reduction of 546 XMM observations with a complete analysis of 469 XDCP suitable survey fields;
    \item Screening and classification of about 2\,000 detected extended sources;
    \item Compilation of a final XDCP master source list with $\sim$250  distant cluster candidates and about  
    750 additional DSS-identified cluster candidates at low or intermediate redshifts.
\end{itemize}



\chapter{Near-Infrared Follow-up Imaging}
\label{c7_NIRanalysis}


\noindent
This chapter will deal with the follow-up imaging of the X-ray selected, distant cluster candidates, \ie \ with 
{\bf steps 3} and {\bf 4} of the survey strategy discussed in Sect.\,\ref{s5_survey_strategy}. Although the R--Z method used for the pilot study resulted in the discovery of XMMU\,J2235.3-2557, certain restrictions and limitations gave rise to the motivation to develop an alternative follow-up strategy. The following sections will discuss in detail the theoretical expectations of a NIR-based follow-up approach, its practical implementation, and the required software and analysis tools.

\section{A New NIR Imaging Strategy}
\label{s7_NIR_strategy}

\noindent

\subsection{Motivation}


\noindent
A new two-band imaging approach should ideally cure both the practical restrictions related to the observatory site and telescope access and the intrinsic systematic limitations of the R--Z method.
Wanted is thus an imaging strategy that:

\begin{enumerate}
    \item is applicable to {\bf 4\,m-class} telescopes and is thus not dependent on the VLT with its high over-subscription rate;
    \item is extendable to the {\bf Northern sky} for a future enlargement of the XDCP survey area;
    \item yields {\bf more accurate redshift estimates} in the main target range $1.0\!\la\!z\!\la\!1.5$;
    \item has the potential to {\bf reach beyond redshift 1.5} to allow detections of the most distant systems.
\end{enumerate}    

\noindent
The first two items are telescope related practical issues for a survey project, the last two, however, are related to the performance of the R--Z method itself. At first sight, the goal of achieving a better performance (items 3+4) with smaller telescopes (item 1) seem mutually exclusive, but a closer look reveals that this does not have to be the case.

\begin{figure}[h]
\centering
\includegraphics[angle=0,clip,width=0.49\textwidth]{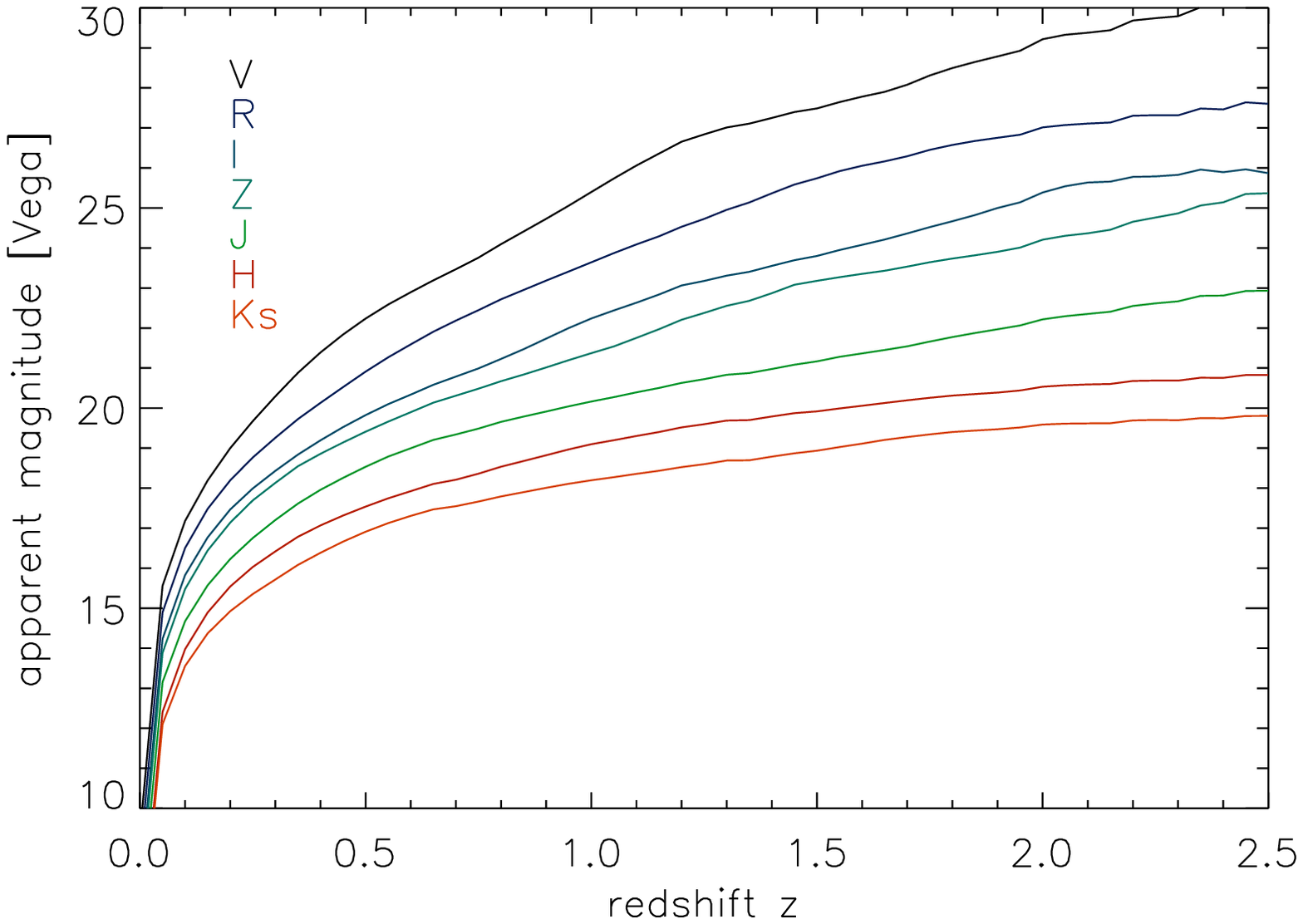}
\hfill
\includegraphics[angle=0,clip,width=0.49\textwidth]{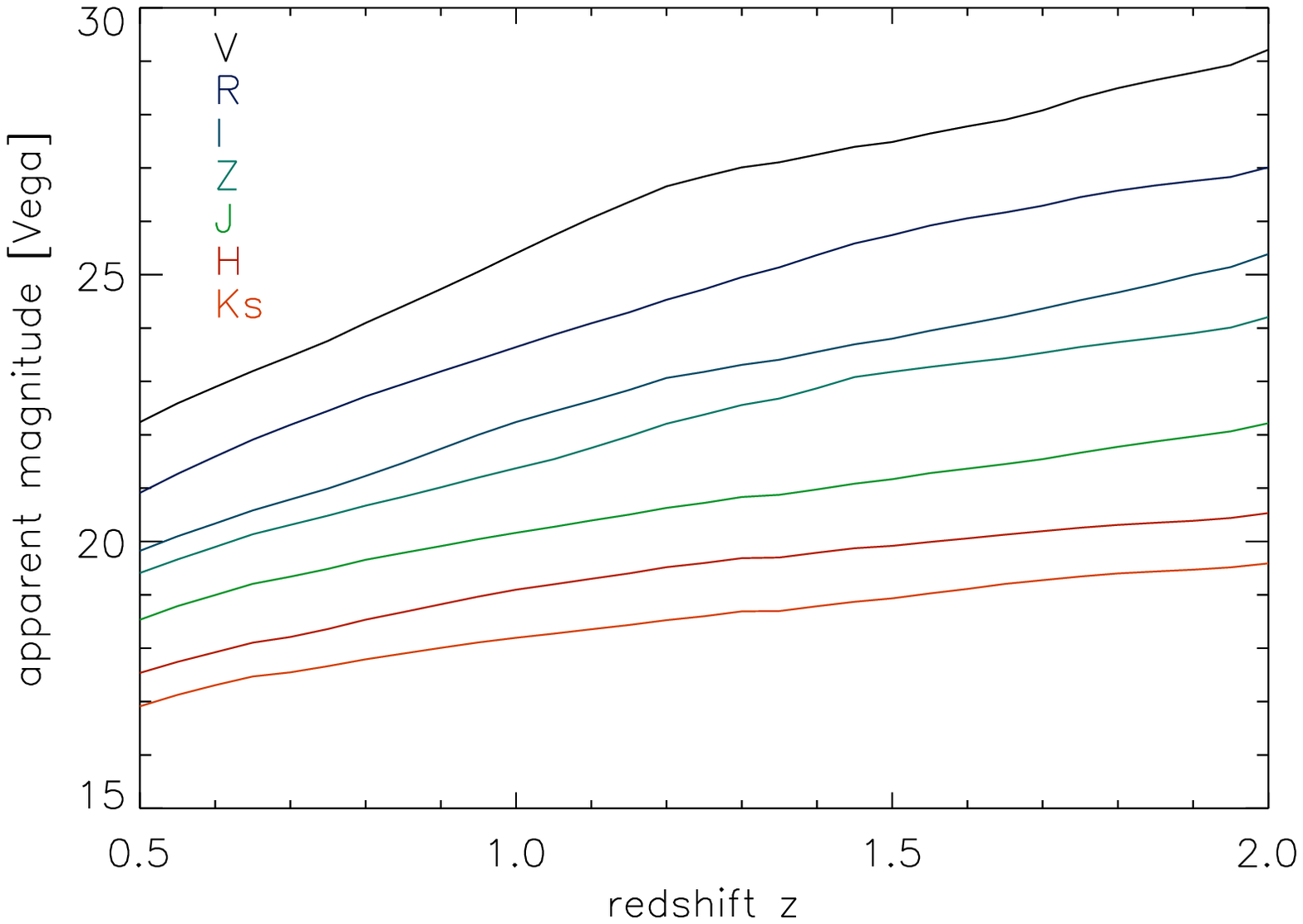}
\includegraphics[angle=0,clip,width=0.49\textwidth]{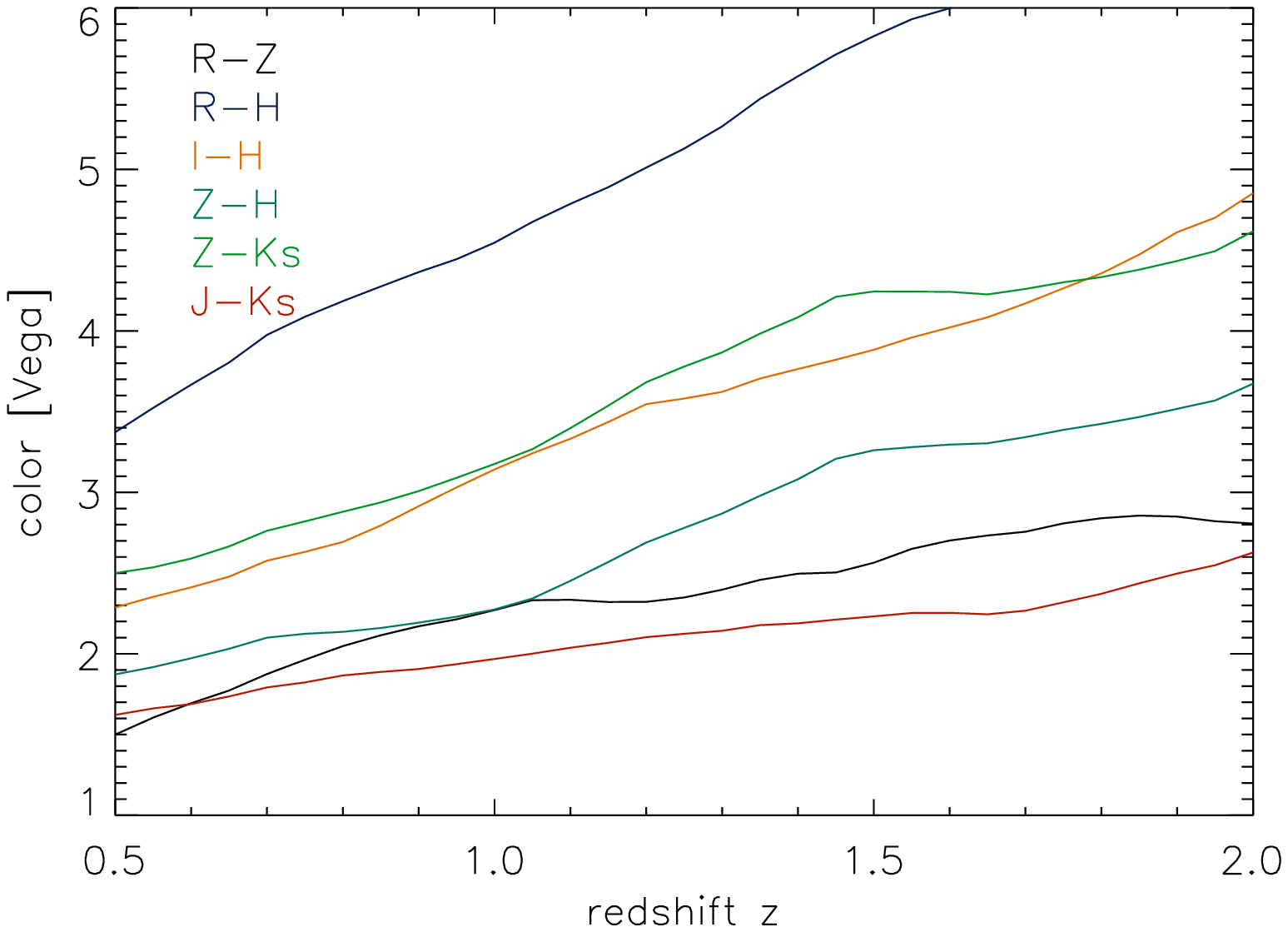}
\hfill
\includegraphics[angle=0,clip,width=0.49\textwidth]{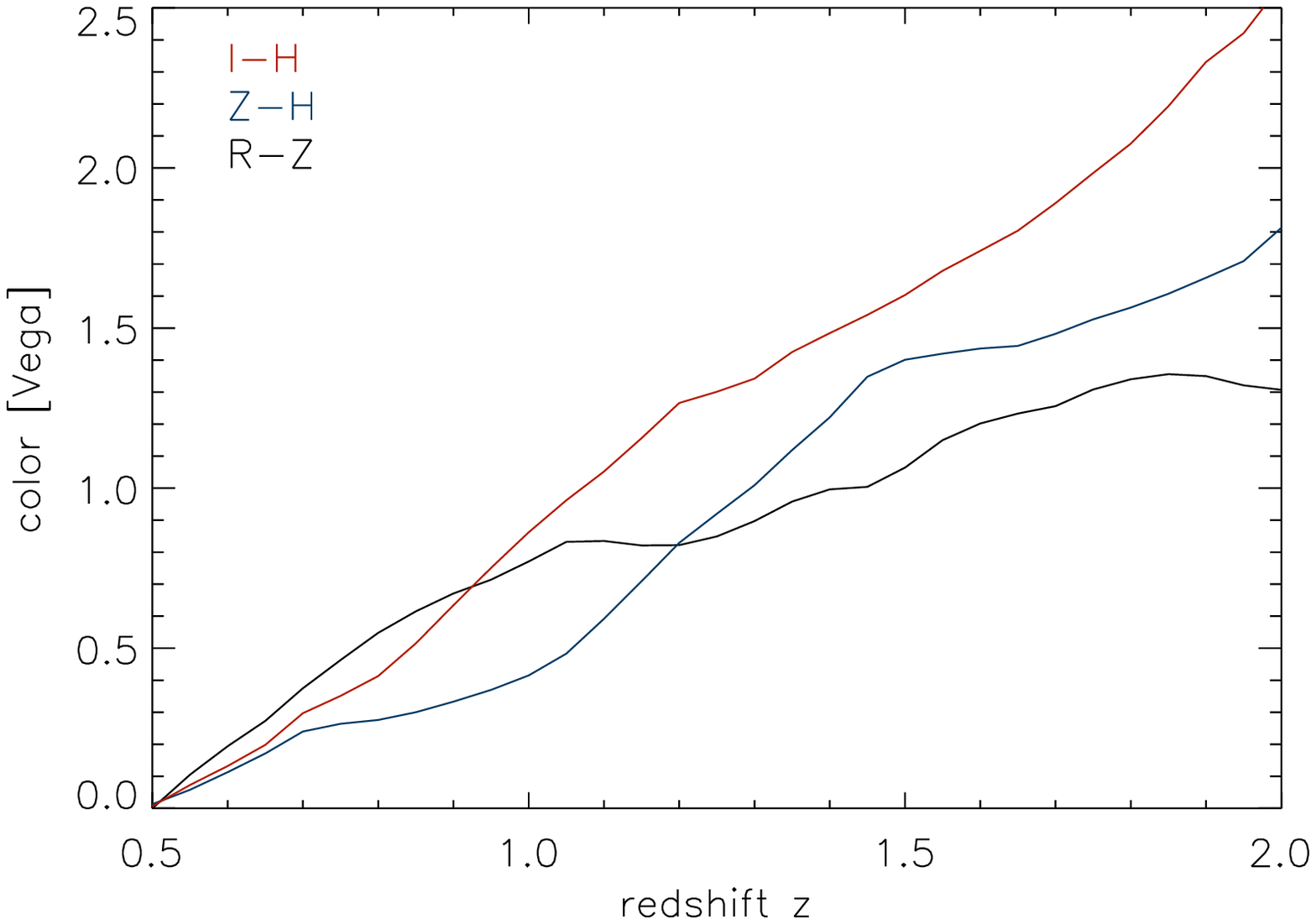}
\includegraphics[angle=0,clip,width=0.49\textwidth]{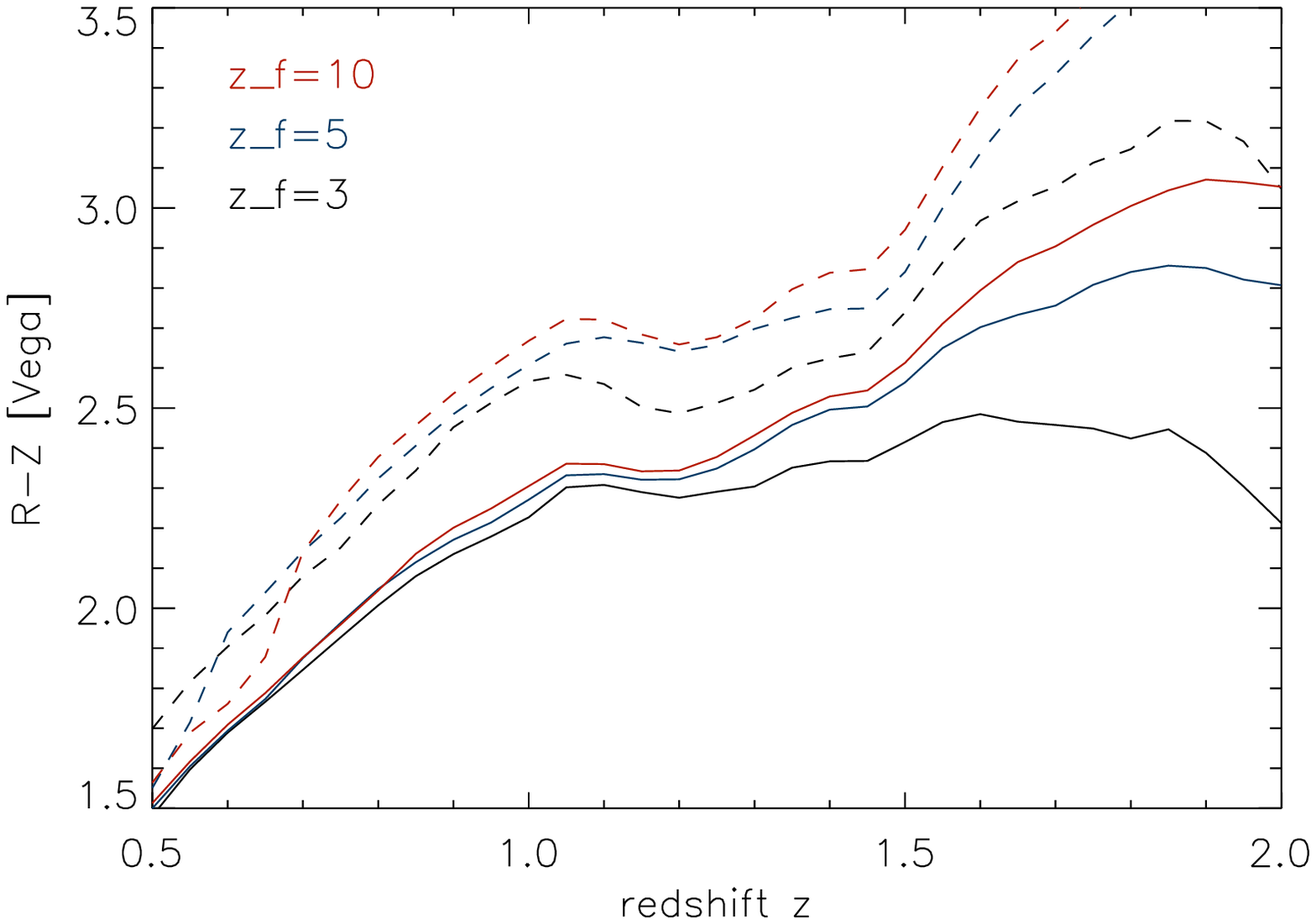}
\hfill
\includegraphics[angle=0,clip,width=0.49\textwidth]{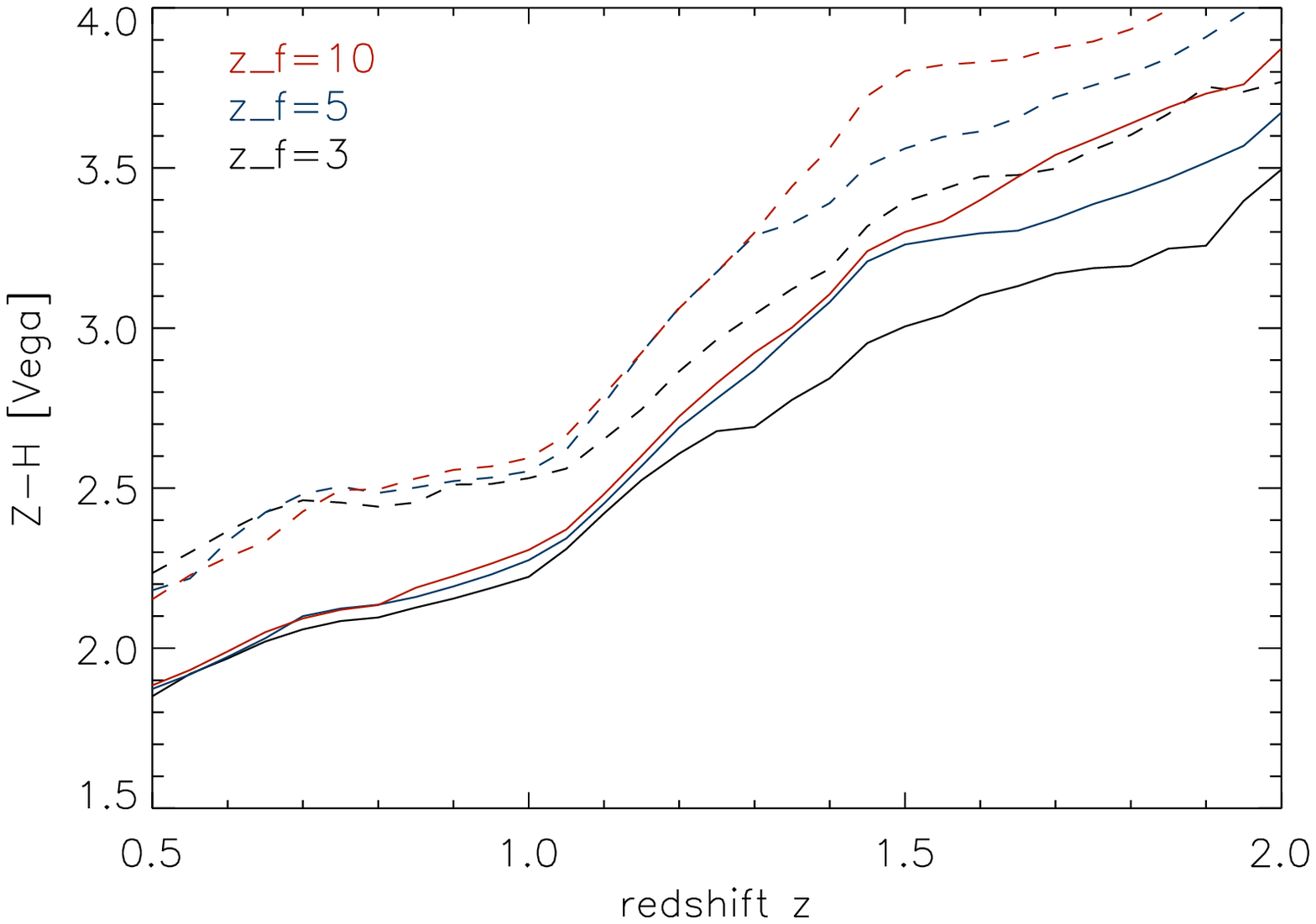}
\vspace{-2ex}
\caption[CMD Models]{Simple stellar population (SSP) models for passively evolving galaxies.{\em Top panels:} Apparent magnitudes of L* galaxies (Vega system) versus redshift diagrams for a formation redshift of $z_{\mathrm{f}}\!=\!5$ for the optical and NIR bands V, R, I, Z, J, H, and Ks. {\em Center left:} Color evolution diagram for a selection of colors ($z_{\mathrm{f}}\!=\!5$). {\em Center right:} R--Z, Z--H, and I--H color shifted to the same origin at $z$=0.5. The slope of the relations determine the redshift sensitivity in the different redshift regimes. {\em Bottom panels:} Model grid of the R--Z ({\em left}) and Z--H ({\em right}) color evolution for formation redshifts of three, five, and ten ({\em different colors}) and solar ({\em solid lines}) and three times solar metallicity ({\em dashed lines}). Models computed by Daniele Pierini using PEGASE2 \cite{Fioc1997a}.}
\label{f7_CMD_models}       
\end{figure}

\clearpage

\subsection{Imaging strategy}
\label{s7_imaging_strategy}

\begin{table}[t]    
\begin{center}

\begin{tabular}{|c|c|c|c|c|}
\hline

\bf{Filter} & \bf{Center}  & \bf{Cut-on 5\%} & \bf{Cut-off 5\%}  & \bf{m$_{\textrm{AB}}$ (Vega)}\\

 &   [\microns] &  [\microns] \ \ ($z_{\mathrm{start}}^{D4000}$)  & [\microns] \ \ ($z_{\mathrm{end}}^{D4000}$) &  \\ 
\hline\hline

R$_{\mathrm{special}}$\footnotemark  & 0.655 & 0.53 \ (0.3) & 0.75  \ (0.9) & 0.195 \\
I\footnotemark  & 0.798 & 0.70  \ (0.8) & 0.89 \  (1.2)  & 0.440 \\
\bf{Z}\footnotemark  & 0.90 & 0.81  \ (1.0)  & 0.99  \ (1.5)  & 0.521 \\
\hline
J  & 1.22 & 1.09  \ (1.7)  & 1.36  \ (2.4)  & 0.857 \\
\bf{H}  & 1.64 & 1.49  \ (2.7)  & 1.81 \  (3.5)  & 1.372 \\
Ks & 2.15 & 1.97  \ (3.9)  & 2.33  \ (4.8)  & 1.855 \\

\hline
\end{tabular}

\caption[XDCP Filters]{XDCP filters of interest. For the three optical filters R, I, and Z and the standard NIR bands J, H, and Ks the central wavelength and the 5\% transmission start and end points are listed. The values in brackets indicate the redshift when the D4000 break enters ($z_{\mathrm{start}}$) and exits ($z_{\mathrm{end}}$) the filter band. The last column shows the offsets between the standard photometric AB-system widely used in optical bands, and the classic Vega system which still prevails in NIR observations. The relation $m_{\textrm{AB}}\!=\!m_{\textrm{Vega}}\!+\!m_{\textrm{AB}}$(Vega) transforms the magnitudes.} \label{t7_XDCP_filters}
\end{center}
\end{table}

\addtocounter{footnote}{-2}
\footnotetext{For FORS\,2 R band see \url{http://www.eso.org/instruments/fors/inst/Filters}.} \stepcounter{footnote}
\footnotetext{For EMMI I band see \url{http://www.ls.eso.org/lasilla/sciops/ntt/emmi/emmiFilters.html}.} \stepcounter{footnote}
\footnotetext{For Z, J, H, Ks filters see \url{http://www.mpia.de/IRCAM/O2000/INFO/FILTERS/Filter-Status.html}.}

\noindent
The natural extension for distant cluster imaging beyond the reddest suitable optical combination of filters, R--Z, is to completely move into the near-infrared regime. Table\,\ref{t7_XDCP_filters} summarizes the characteristics of the reddest optical (standard) broad band filters (R, I, Z\footnote{In order to avoid confusion with the redshift $z$, the SDSS broad-band filter termed z is capitalized and denoted as Z throughout this thesis.}) and the three standard NIR bands J, H, and Ks\footnote{Ks is a modified `short' K filter used by the 2MASS survey to reduce the thermal background at the long wavelength end.}. Near-infrared imaging has only recently become a competitive alternative to optical observations and has been strongly boosted by great advances in the NIR imaging array technology. The latest generation of 2k$\times$2k NIR arrays have been available for about 5 years now and have driven the development of a new generation of sensitive wide-field NIR instruments.

For a first approach to the problem, we can assume that suitable instrumentation exists for all six filter bands of interest and we can thus restrict ourselves to physical arguments related to the targets and the restrictions imposed by ground-based observations in general. A discussion of the practical implementation will follow in Sect.\,\ref{s7_practical_implementation}.

Secondly, the discussion will focus on the properties of the ubiquitous elliptical galaxy population in clusters, as introduced in Sect.\,\ref{s2_galaxy_populations}. As `red and dead' objects, the spectral energy distribution (SED) of elliptical galaxies can nowadays be accurately modelled and predicted with simple stellar population (SSP models) codes such as PEGASE2 \cite{Fioc1997a} or GISSEL \cite{Bruzual2003a}.
 The SSP models under consideration assume an instantaneous star formation burst at a given formation redshift $z_{\mathrm{f}}$ and take a stellar initial mass function (IMF), the total stellar mass, and the rate of chemical enrichment as input. With this it is possible to compute the age-dependent distribution of stars in the Hertzsprung-Russell (HR) diagram and thus the properties of the stars contributing to the integrated light. Using this {\em evolutionary population synthesis} technique \cite{Tinsley1978a}, the full SED of the passively evolving galaxies can be obtained at any time $t$ after the burst by summing the contributions of the different stellar populations at this age using  a stellar spectral flux library \cite{Pickles1998a}.  

\begin{table}[t]    
\begin{center}

\begin{tabular}{|c|c|c|c|c|c|c|}
\hline

\bf{Filter} & \bf{Sky} &\bf{m*$\mathbf{(z\!=\!0.5)}$} & \bf{m*$\mathbf{(z\!=\!1.0)}$} & \bf{m*$\mathbf{(z\!=\!1.5)}$} & \bf{m*$\mathbf{(z\!=\!2.0)}$} & $\mathbf{\Delta}$\bf{m}$_{\mathbf{1\!\rightarrow\!2}}$ \\
  & [mag asec$^{-2}$] & [Vega (AB)] & [Vega (AB)] & [Vega (AB)] & [Vega (AB)] & [mag] \\ 
 
\hline\hline

R$_{\mathrm{special}}$ & 19.9--21.1 & 20.9 (21.1)& 23.6 (23.8) & 25.7 (25.9) & 27.0 (27.2) & 3.4 \\
I  & 19.2--19.9                     & 19.8 (20.2) & 22.2 (22.6) & 23.8 (24.2) & 25.4 (25.8) & 3.2 \\
\bf{Z} & 18.1-18.8                  & 19.4 (19.9) & 21.4 (21.9) & 23.3 (23.8) & 24.2 (24.7) & 2.8 \\
\hline
J  & 15.8--16.7                     & 18.5 (19.4) & 20.2 (21.1) & 21.2 (21.9) & 22.2 (23.1)  & 2.0 \\
\bf{H} & 13.8--15.0                 & 17.5 (18.9) & 19.1 (20.5) & 19.9 (21.3) & 20.5 (21.9)  & 1.4 \\
Ks & 12.7--13.0                     & 16.9 (18.8) & 18.2 (20.1) & 18.9 (20.8) & 19.6 (21.5)  & 1.4 \\


\hline
\end{tabular}

\caption[Apparent Magnitude Evolution]{Benchmark of the apparent magnitude evolution of L* passive galaxies with formation redshift $z_{\mathrm{f}}\!=\!5$ and solar metallicity for various filter bands. The second column states the typical sky surface brightness as measured for the ESO La Silla site\footnotemark. The given intervals span  bright time (first value) and dark time (second value) in the Vega system.   For faint galaxy observations with long exposure times, the  assumption of background limited observing conditions is valid for all bands of interest. Columns 3--6 show the model expectations for the apparent galaxy magnitudes at various redshifts. The last column summarizes by how many magnitudes the galaxies drop between redshift 1 and 2, indicating that for the bluer bands even the largest telescopes will start losing the objects between \zsim1--1.5. } \label{t7_magnitude_evolution}
\end{center}
\end{table}

\addtocounter{footnote}{-1}
\footnotetext{See \url{http://www.eso.org/gen-fac/pubs/astclim/lasilla/l-vanzi-poster}, \url{http://www.ls.eso.org/lasilla/Telescopes/2p2T/D1p5M/misc/SkyBrightness.html}, and \url{http://www.eso.org/observing/etc/doc/ut2/uves/helpuves.html}.}

The prediction of these simple stellar population models\footnote{PEGASE2 models computed by Daniele Pierini.} for a formation redshift $z_{\mathrm{f}}\!=\!5$ are summarized in the top panels of Fig.\,\ref{f7_CMD_models}, where the apparent magnitudes in various bands are shown for passively evolving galaxies with characteristic absolute magnitude $M*$. The derived color evolution for a variety of suitable colors are given in the left central panel. The illustrated (red) colors show mostly a well behaved monotonic functional form in the redshift range up to $z\!\sim\!2$, which qualifies them in principle as redshift estimators. The redshift sensitivity, however, depends primarily on the slope of the color evolution  $(d(\mathrm{X}\!-\!\mathrm{Y})/dz)^{-1}$, where $(\mathrm{X}\!-\!\mathrm{Y})$ denotes a given color. This sensitivity is emphasized in the right central panel for the color combinations I--H, Z--H, and R--Z, which are normalized to a common color origin at redshift $z\!=\!0.5$, since this is the fiducial XDCP starting redshift beyond which follow-up imaging is needed. Note the flattening of the R--Z color at \zsim1 \ and with a smaller effect at $z\!\sim\!1.2$ for I--H and at $z\!\sim\!1.5$ for Z--H. This feature occurs when the 4\,000\,\AA \ break is completely redshifted beyond the cut-off (see Tab.\,\ref{t7_XDCP_filters}) of the blue filter, in this case R, I, or Z. The small amount of flux in ellipticals below 4\,000\,\AA \ (see spectra in Fig.\,\ref{f4_Noras2_spectra}) is far less redshift dependent compared to the break passing through the filter.

The second important parameter influencing the slope of the color evolution is the formation redshift of the galaxy. The characteristic 4\,000\,\AA \ break of ellipticals only emerges when the  young, hot, and blue stellar populations with their dominant UV flux have disappeared from the main sequence (main sequence turn-off), \ie \ the flux break will evolve with the age of the stars. A younger stellar population at the epoch when the galaxy is observed, \ie \ a lower formation redshift, will thus shift the observed color towards the blue. This bluing effect increases with decreasing galaxy age at the time of observation and is due to the rapid evolution of young stellar populations. The influence of the formation redshift $z_{\mathrm{f}}$ on the color evolution is depicted in the bottom panels of Fig.\,\ref{f7_CMD_models} for the R--Z and Z--H colors.

The model grid also illustrates the effect of increasing the metallicity, which makes up the final critical input parameter for the predictions. It can be seen that to first order an increasing metallicity results in a redder overall color, \ie \ the color evolution curves are shifted upwards as a whole.      
Looking at epochs with a still relatively young stellar population, \ie \ at $z\!\sim\!1.5$, the age-metallicity degeneracy 
becomes obvious in this illustration as increasing age and metallicity have similar effects on the color.  
This degeneracy can be disentangled with a long redshift baseline as shown in Sect.\,\ref{s10_formation_epoch}.


The pre-requisites for measuring accurate galaxy colors are sufficiently deep exposures in each of the filter bands.
Table\,\ref{t7_magnitude_evolution} gives a benchmark summary of the predicted m* apparent magnitudes for various redshifts as shown in Fig.\,\ref{f7_CMD_models}. The second column of Tab.\,\ref{t7_magnitude_evolution} lists typical levels of the sky surface brightness in the different bands as reference. Comparing these values in the J, H, and Ks-bands to the expected magnitudes of the targeted galaxies illustrates the main challenge of ground-based NIR observations, namely that the sources of interest are between a factor of 100 ($\Delta m\!=\!5$) and 1\,000 ($\Delta m\!=\!7.5$) fainter than the sky background.

For the optical R, I, and Z-bands, observations of distant galaxies are still background-limited ($b\!\gg\!s$) .  
Under these observing conditions the 
signal-to-noise ratio (SNR) for a source  with signal $s(t_{\mathrm{exp}})$ and total background\footnote{The Poisson statistics with standard deviation $\sigma_{b}\!=\!\sqrt{b}$ is applicable to the total number of detected electrons and not directly to the digital count units. For a proper statistical treatment the signal and background have to be transformed to these physical units first by multiplying with the {\em gain} $g$, \ie \ the electrons-per-count conversion factor. Relation\,\ref{e7_background_limit} further assumes that the dark current and read noise are small compared to the background noise, which is well-justified from Tab.\,\ref{t7_omega2k_characteristics} for the NIR.} $b(t_{\mathrm{exp}})$ enclosed in the source aperture will increase with exposure time $t_{\mathrm{exp}}$ as

\begin{equation}\label{e7_background_limit}
 \mathrm{SNR} = \frac{s(t_{\mathrm{exp}})}{\sqrt{b(t_{\mathrm{exp}})}} \propto \sqrt{t_{\mathrm{exp}}} \ .
\end{equation}


\noindent
For a fixed minimal SNR requirement the flux at the detection limit thus follows 
$f_{\mathrm{lim}}(t)\!\propto\!1\!/\!\sqrt{t_{\mathrm{exp}}}$.
Expressing this relationship in magnitudes yields $\Delta m_{\mathrm{lim}}\!=\!1 \rightarrow t_{\mathrm{deep}}\!=\!6.3\cdot t_{\mathrm{shallow}}$, \ie \ going just one magnitude deeper requires more than six times the initial exposure time.
For a doubling of the exposure time, the gain at the  magnitude limit is $\Delta m_{\mathrm{lim}}\!=\!0.38$\,mag.
These simple considerations imply that the magnitude limit which can be reached within about $t_{\mathrm{exp}}\!\sim\!30$\,min
at a given telescope and filter is a solid constraint which cannot be pushed by more than $\Delta m_{\mathrm{lim}}\!\sim\!0.5$\,mag with reasonable effort for large area survey projects. The last column of Tab.\,\ref{t7_magnitude_evolution} states the galaxy dimming for the given filter from redshift 1 to 2. The rapid dimming in the optical R and I-bands of $\Delta m_{1\!\rightarrow\!2}\!=\!3.4$ compared to the moderate changes of $\Delta m_{1\!\rightarrow\!2}\!=\!1.4$ for H and Ks can be understood in terms of the $K$-correction introduced in Equ.\,\ref{e3_k_correction}. The long-wavelength filters H and Ks still probe the rest-frame  peak emission of ellipticals even at very high redshifts, whereas the optical bands are restricted to the residual rest-frame  UV-flux below the D4000 feature once the break is redshifted beyond the band. Consequently, reaching a depth to observe an $m*$ galaxy at $z\!\sim\!1.5$ in the R-band is already challenging even with the Very Large Telescope, but the task becomes drastically easier when considering the longer wavelength-bands. The very same galaxy  appears a factor of 100 brighter in flux in the Ks-band, roughly 60 times higher in the H-band, and the gain in the Z-band is still a factor of 7.     


For a high-redshift two-band imaging strategy, the choice of the I or Z-filter (instead of R) as the blue band hence results in a significant gain in terms of depth requirements at the faint end. From the galaxy properties point-of-view, the red filter selection should be at the longest feasible wavelength. However, as pointed out in Tab.\,\ref{t7_magnitude_evolution} Ks-band observations are very challenging and time consuming (long observational overheads) due to the dominating thermal background contribution. The red filter of choice for a new approach is consequently the H-band, which combines (i) good high-$z$ galaxy observability, (ii) high-redshift sensitivity in conjunction with I or Z, and (iii) improved background conditions compared to Ks. 

This discussion has shown that at the high-redshift end of the target regime around $z\!\sim\!1.5$ a modified choice of observation bands can (over-)compensate for the geometrical telescope difference between the VLT and a 4\,m-class telescope, which is of the order of 4--6 in collecting area. The next section will investigate the performance expectations concerning the redshift estimation of selected methods in more detail.

\subsection{Performance expectations}
\label{s7_performance_expectations}

\noindent
The standard R--Z method is now compared to the two alternative methods I--H and Z--H. For a quantitative performance benchmark of the redshift sensitivity some simplifying assumptions are needed. For the total error budget of a \reds \ measurement one has to consider (i) intrinsic scatter about the sequence, (ii) photometric measurement errors of the galaxies, and (iii) calibration errors of the photometric zero points. The intrinsic \reds \ scatter for all known clusters out to  $z\!\sim\!1.3$ is typically $\sigma_{\mathrm{scatter}}\la\!0.05$\,mag. Of course, studying the evolution of the \reds \ properties at the highest redshifts will be an important science goal, but for now the assumption is that a well-defined and tight cluster \reds \ exists and that the average color can be measured with good precision. 
Concerning the photometric measurement errors, the assumption of sufficiently deep observations in both filters has to be made. The discussion of the last section has shown (see Tab.\,\ref{t7_magnitude_evolution}) that this is not the case for the optical filters R, I, and Z. Depending on the telescope size, the limiting magnitude will unavoidably drop below $m*$ somewhere between redshift 1 and 2, and the bluer the band the earlier this will happen. The performance evaluation will thus assume idealized conditions, but in practice the achievable limiting magnitude of the blue filter will determine at what redshift the galaxy photometry will break down. Finally, calibration errors for the determination of the photometric zero points will add a constant color uncertainty. For well-calibrated data this offset should also be $\sigma_{\mathrm{ZP}}\la\!0.05$\,mag, which also neglects possible practical calibration issues. In any case, the total achievable color error will have a constant part, due to zero point offsets, and a redshift dependent part which is related to the increasing photometric error of fainter galaxies, fewer objects on the \reds \ due to the magnitude limit and a possibly intrinsically less populated locus of galaxies. For the following discussion a total \reds \ color error of  $\sigma_{\mathrm{color}}\approx 0.05 \cdot (1+z)$\,mag is assumed, which is a reasonable parametrization for good quality data ($\sigma_{\mathrm{color}}\!\sim\!0.1$ at $z\!=\!1$) out to $z\!\sim\!1.5$.

\begin{figure}[b]
\begin{center}
\includegraphics[angle=0,clip,width=0.87\textwidth]{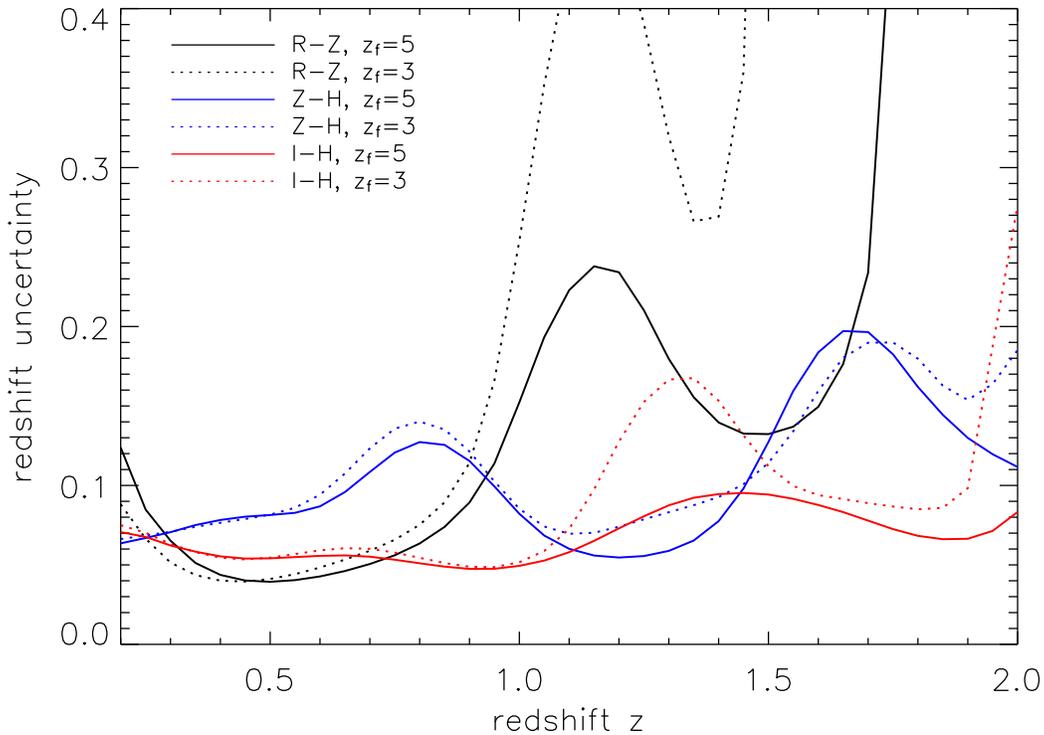}
\end{center}
\vspace{-3.5ex}
\caption[Redshift Uncertainties of Red Sequence Methods]{Absolute redshift uncertainties of different \reds \ methods for a realistic photometric color error assumption of $\sigma_{\mathrm{color}}\approx 0.05 \cdot (1+z)$\,mag. Error estimates were obtained from the derivatives of the smoothed model colors in Fig.\,\ref{f7_CMD_models} using $\sigma_{z}\approx dz/d(\mathrm{X}-\mathrm{Y}) \cdot \sigma_{\mathrm{color}}$, where $(\mathrm{X}-\mathrm{Y})$ denotes the photometric method. The black solid line illustrates the estimated redshift error for the R--Z technique under the assumption of a formation redshift of $z_{\mathrm{f}}\!=\!5$, blue shows the Z--H method, and red I--H. The dotted lines use a model formation redshift of $z_{\mathrm{f}}\!=\!3$ for the same methods. The R--Z color is an accurate redshift estimator up to about $z\!\sim\!0.9$, but at \zga1 the uncertainties grow drastically and for low formation redshift models, the $z$-estimates even become degenerate (flat slope in lower left panel of Fig.\,\ref{f7_CMD_models}). The Z--H strategy (blue) yields good redshift estimates all the way to $z\!\sim\!1.5$, has the lowest uncertainties in the targeted regime $1\!\la\!z\!\la\!1.5$ and is rather insensitive to model variations. The I--H approach (red) is another promising alternative with good performance expectations at all redshifts. Note that the highest accuracy for any method is achieved  in the redshift regime where the D4000 break passes through the blue filter (see Tab.\,\ref{t7_XDCP_filters}).} \label{f7_redshift_uncertainties}
\end{figure}

The expected redshift uncertainties $\sigma_{z}$ can then be estimated from the first derivative of the color evolution as presented in Fig.\,\ref{f7_CMD_models} as

\begin{equation}\label{e7_redshift_uncertainty}
 \sigma_{z} \approx \frac{dz}{d(\mathrm{X}\!-\!\mathrm{Y})} \cdot \sigma_{\mathrm{color}} = \left( \frac{d(\mathrm{X}\!-\!\mathrm{Y})}{dz} \right)^{-1} \cdot \sigma_{\mathrm{color}} \ ,  
\end{equation}

\noindent
where $(\mathrm{X}-\mathrm{Y})$ denotes the photometric method R--Z, I--H, or Z--H. The derivative of the color evolution was computed numerically using a boxcar-smoothed input curve in order to suppress the influence of small scale features of the coarsely sampled model on the local slope estimation. The obtained redshift uncertainty estimates are shown in Fig.\,\ref{f7_redshift_uncertainties} for solar metallicity models and formation redshifts three and five. 

As expected, the best redshift performance is obtained in the redshift regime where the D4000 break is passing through the blue filter (see Tab.\,\ref{t7_XDCP_filters}). The lowest achievable uncertainties of  $\sigma_{z}\!\approx\!0.04$--0.08 are assumed in the redshift interval [0.3--0.9] for the R--Z method, [0.3--1.2] for the I--H, and [1.0--1.4] for the Z--H approach.

The main performance characteristics of different \reds \ methods are summarized in Tab.\,\ref{t7_imaging_methods}.
Based on the former discussions and the performance expectations, the final evaluation of the different suitable \reds \ imaging strategies is as follows:

\begin{itemize}
    \item The {\bf R--Z} approach has the advantage of being an optical method, \ie \ observations and data reduction tools are standard. However, sensitive Z-band instruments are rare, which basically limits an efficient strategy to VLT FORS\,2. The method has two systematic drawbacks for high-$z$ cluster searches: (i) The redshift uncertainties at \zg1 \ increase to at least $\Delta z\!\ga\!\pm 0.2$ in the optimistic case ($z_{\mathrm{f}}\!=\!5$), and could be as much $\pm 0.4$, \ie \ basically redshift-degenerate, for a lower formation redshift model. Since both bands have to be photometrically calibrated using standard star observations, there is little room to achieve a better performance by improving the calibration errors. (ii) The depth limitation in the R-band poses a serious risk to not obtain sufficiently accurate color information on the galaxies or miss the highest redshift clusters completely. With a limiting magnitude of R$_{\mathrm{AB}}\sim$24.5--25.0 one can probe the cluster luminosity down to m*+1 at $z\!=\!1$, but by  $z\!=\!1.5$ this has dropped to m*-1. If the cluster does not contain several galaxies brighter than the characteristic magnitude m*, the identification of a \reds \ could be an issue at redshifts as low as $z\!\sim\!1.2$.
 
    \item {\bf J--Ks} has been the standard NIR approach for high-redshift cluster investigations \cite{Stanford1997a}. The color evolution (Fig.\,\ref{f7_CMD_models}, central left panel) has a constant but shallow slope, with modest uncertainties at all redshifts. The advantage of the method is that sufficient depth in the J and Ks-bands to probe galaxies out to $z\!\sim\!2$ is  fairly easy to achieve even with small telescopes. The disadvantage is the long overheads for NIR observations, in particular for Ks, which lowers the efficiency for routine follow-up. However, J--Ks can be very important to confirm very high-redshift cluster candidates that have been identified with any of the other approaches. 

 
    \item  The {\bf I--H} method shows very promising performance characteristics at all redshifts. Sufficient I-band depth to at least m* at $z\!=\!1.5$ should be achievable with 4\,m-class telescopes with reasonable exposure times, which allows this approach to be implemented at many observatories. The disadvantage, however, is the need to use different instruments for the optical and NIR observations. This results in less flexibility during the observations, more scheduling constraints, and adds extra steps to the data analysis. This method is  currently in the test phase with first data obtained with the EMMI and SOFI instruments at the New Technology Telescope (NTT) at La Silla. 
        
    \item  The two-band imaging method of choice for this thesis is the {\bf Z--H} strategy. This approach has the best performance expectations in the target redshift regime $1\!\la\!z\!\la\!1.5$ and modest uncertainties at lower and higher redshifts. The Z-filter as the blue band should allow sufficient depth with a 4\,m-class telescope. The method has the advantage that both bands can be observed with a single modern NIR instrument, \ie \ no additional optical instrument is required. The disadvantage is that Z-band observations with NIR instruments have never been tested. The practical implementation of the Z--H strategy will be discussed in the next section.  
\end{itemize}

\begin{table}[t]    
\begin{center}

\begin{tabular}{|c|c|c|c|c|c|}
\hline

\bf{Color} & \bf{Instrument} & $\mathbf{\sigma_{z}}$ \bf{in} $\mathbf{0.5\!<\!z\!<\!1}$ & $\mathbf{\sigma_{z}}$ \bf{in} $\mathbf{1\!\la\!z\!\la\!1.5}$ & $\mathbf{z\!>\!1.5}$ & \bf{Comment} \\

\hline\hline

R--Z  & optical & 0.05--0.15 & 0.15--0.4 & no & sensitive to $z_{\mathrm{f}}$ \\
I--H  & optical+NIR & 0.06 & 0.06--0.15 & possibly & instrument change \\
\bf{Z--H} & NIR (+opt.)  & 0.08--0.12 & 0.06--0.1 & yes &  \\
J--Ks  & NIR & 0.08--0.14 & 0.14--0.18 & yes & long overheads \\

\hline
\end{tabular}

\caption[Summary of Imaging Methods]{Performance summary of the different imaging methods. Columns 3 and 4 state the estimated redshift errors as discussed in Fig.\,\ref{f7_redshift_uncertainties}. Column 5 indicates whether the method has the potential to be used at redshifts $z\!>\!1.5$.} \label{t7_imaging_methods}
\end{center}
\end{table}


\clearpage

\section{Practical Implementation}
\label{s7_practical_implementation}

\noindent
The Z--H two-band imaging strategy can be ideally carried out with the NIR wide-field camera OMEGA2000 at the Calar Alto 3.5\,m telescope. This section will briefly introduce the capabilities of the instrument, the observing strategy, and the obtained data.

\subsection{OMEGA2000 in a nutshell}
\label{ss7_OMEGA2000}


\begin{figure}[t]
\begin{center}
\includegraphics[angle=0,clip,width=0.995\textwidth]{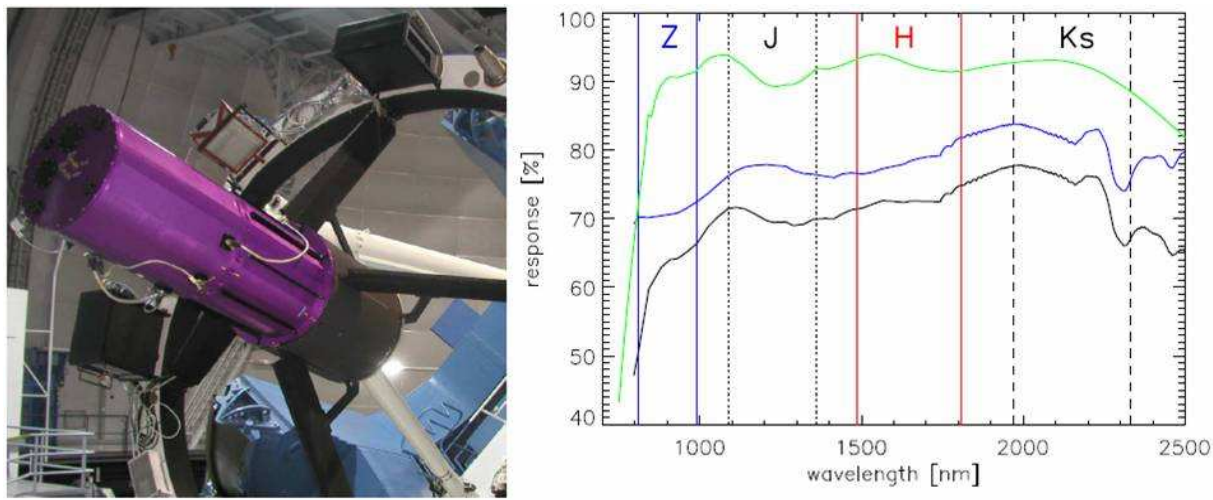}
\end{center}
\caption[OMEGA2000 Instrument Response]{{\em Left:} The NIR wide-field camera OMEGA2000 at the Calar Alto 3.5m telescope. {\em Right:} OMEGA2000 instrument response. The green solid line indicates the transmission efficiency through the optical system as a function of wavelength, the blue line shows the HAWAII-2 detector quantum efficiency, and the black solid line gives the full response of the camera system. Filter band boundaries are overplotted as vertical lines. With a quantum efficiency of $\ga\!70\%$, the Z-band sensitivity of the OMEGA2000 NIR array is comparable to or better than the best optical instruments in this critical filter band. Data from H.-J. R\"oser (private communication).} \label{f7_O2k_QE}
\end{figure}

\begin{table}[h]    
\begin{center}

\begin{tabular}{|c|c|c|c|}
\hline

\multicolumn{2}{|c|}{\bf{Instrument}}   & \multicolumn{2}{|c|}{\bf{Detector}}  \\

\hline\hline

field-of-view  & 15\farcm 4\,$\times$\,15\farcm 4 (237 arcmin$^2$) & array type  & HAWAII-2  \\
pixel scale  & 0.45\arcsec / pixel & array size  & 2\,048\,$\times$\,2\,048 pixels \\
weight & 300\,kg & pixel size  & 18\,\microns  \\
focal station  & prime focus & wavelength range  & 0.8--2.5\,\microns  \\
focal ratio  & f/2.35 & quantum efficiency  & 70--85\%  \\
telescope primary  & 3.5\,m & dark current  & $<0.03$\,e$^{-}$\,s$^{-1}$ \\
broad band filters  & Z, Y, J, H, Ks, K & read noise  & $<15$\,e$^{-}$  \\
data rate  & $\sim$10\,GB / night & good pixels  & $>$99\%  \\  
online reduction  & yes & fill factor  & 90\%  \\    
commissioned  & 2003 & full well capacity  & $\simeq$ 100\,000\,e$^{-}$   \\

\hline
\end{tabular}

\caption[OMEGA2000 Characteristics]{OMEGA2000 instrument and detector characteristics. From R\"oser \cite*{o2kmanual} and Haas \cite*{hawaii2}.} \label{t7_omega2k_characteristics}
\end{center}
\end{table}

\noindent
The left panel of Fig.\,\ref{f7_O2k_QE} shows the prime focus instrument OMEGA2000 mounted at the 3.5\,m telescope. Three main features establish this camera as the instrument of choice for testing the Z--H approach: (i) the high sensitivity from 0.8--2.5\,\microns, (ii) the large 15\farcm 4 $\times$ 15\farcm 4 field-of-view, and (iii) an online reduction system for an immediate data evaluation.

The instrument sensitivity is shown in the right panel of Fig.\,\ref{f7_O2k_QE} as a function of wavelength. The blue solid line illustrates the quantum efficiency (QE) of the HAWAII-2 NIR array, the green line follows the transmission properties of the optical system of the camera, and the black line shows the total response of OMEGA2000 as the product of the two former functions. The QE efficiency in the classical NIR band J, H, K is about 80\%, and at the short wavelength end in the Z-band one expects still a 70\% sensitivity. For comparison, the red-optimized CCDs of the VLT FORS\,2\footnote{See \url{http://www.eso.org/projects/odt/Fors2/qeu.html}.} have an average Z-band QE of 65\%, but most optical instruments provide quantum efficiencies $<$30\% in Z. Modern NIR arrays hence offer an attractive sensitive alternative at the long-wavelength end of the traditional optical regime. An additional advantage of NIR arrays with their individually isolated pixels is the absence of {\em fringing}, \ie \ interference patterns caused by strong monochromatic atmospheric emission lines in CCDs; the disadvantage on the other hand are the numerous imperfections and flaws of NIR detectors that have to be accounted for during the reduction process (see Sect.\,\ref{s7_NIR_pipeline}). 

The second prime feature of OMEGA2000 is the large field-of-view, which is still one of the largest available fields in the near-infrared. The 237 square arcminute FoV covers one quarter of the XMM-Newton field and is five times as large as the FORS\,2 FoV. This has several important advantages: (i) often several cluster candidates can be simultaneously observed, (ii) lower redshift candidates and calibration sources can be placed in the FoV at no extra observational cost, and (iii) the photometric calibration in the J, H, and Ks bands is simplified by taking advantage of the numerous 2MASS sources in the field (see Sect.\,\ref{s7_NIR_calibration}).

The available OMEAG2000 online reduction system is an additional important characteristic which helps to optimize the observational efficiency. Since the NIR background is so dominant (see Tab.\,\ref{t7_magnitude_evolution}), an evaluation of the data quality and the depth of the observation is usually difficult to obtain, in particular if the targets of interest are very faint. The online reduction pipeline can provide quicklook data products with sufficient quality while the data taking is active. This way, the reduced and co-added images of the ongoing observation are available with an approximate delay-time of 3\,min to the running exposure and thus allows an instant evaluation of the target source and the observational status.

Further OMEGA2000 instrument and detector characteristics are summarized in Tab.\,\ref{t7_omega2k_characteristics}.

\subsection{Observational strategy}
\label{s7_observational_strategy}

\noindent
As a follow-up program of a large sample of X-ray selected cluster candidates, the observational survey mode strategy of the Z--H approach is driven by the two main goals (i) source identification and (ii) efficiency. The first goal implies that the observations have to be deep enough in order to identify even the most distant clusters as local galaxy overdensities. On the other hand, the bulk of clusters at lower redshifts require much shorter exposure times for a safe identification, which increases the survey efficiency if properly accounted for. For these considerations the following observing strategy has been pursued:

\begin{enumerate}
    \item Start H-band observation of a distant cluster candidate field; 
    \item Check the online reduced co-added image products every 4--5 minutes and evaluate the appearance of the cluster candidate;
    \item Stop the H-band observation once a galaxy cluster signature is clearly visible with at least 10--15 galaxies close to the candidate X-ray center;
    \item If the target shows no cluster signature, stop after an exposure time of 50\,min, which corresponds to a limiting 5$\,\sigma$ H Vega magnitude of H$_{\mathrm{lim}}\!\ga\!21$ (m*+1 at $z\!\sim\!1.5$). The candidate source is classified as `spurious' and is not observed in any other band; 
    \item  Observe sources with cluster signatures in the Z-band. Check again the pipeline products for sufficient depth and abort observations once reached; 
    \item Go to next candidate target field;
    \item Obtain J and Ks-band observations for some of the most distant candidates to backup with a J--Ks color.
\end{enumerate}

\noindent
Starting with the redder H-band is important for the identification of the highest redshift clusters, which might be heavily dimmed in the Z-band (see Tab.\,\ref{t7_magnitude_evolution}), and correspondingly for the secure classification of a null-detection, \ie \ spurious candidates. Galaxy clusters at redshift $z\!\la\!1$ can typically be well identified with exposure times of $t_{\mathrm{exp}}\!\la$15--20 minutes, \ie \ the efficiency gain compared to the deepest observations is about a factor of three.

The photometric Z-band calibration requires additional observations of SDSS Z-band standard stars several times per night. Each targeted Z-band field also needs at least a few minutes of coverage under photometric conditions, which might have to be re-observed if the conditions during the initial exposure were not sufficient in quality.

The total exposure time $t_{\mathrm{exp}}$ for a deep observation is split up into many shorter observations $t_{\mathrm{tel}}$, after which the telescope is offset, and these again consist of the sum of several single frames of integration time $t_{\mathrm{single}}$. The elementary exposure time $t_{\mathrm{single}}$ is motivated by two boundary conditions. Firstly, the background in a given filter sets the maximum time before the detector saturates. Secondly, the difference between the saturation limit and the background level defines the effective dynamic range of the exposure, \ie \ the interval in which the flux of astronomical targets can be accurately measured. In practice this time is set in a way to include the brighter calibration objects well within the dynamic range and was fixed at $t_{\mathrm{single}}\!=\!5$\,s (10\,s) in the H-band (Z-band). These single exposures can be co-added in memory before they are saved to disk after a time $t_{\mathrm{tel}}$ during which the telescope points at the same position. The maximum time for  $t_{\mathrm{tel}}$ is limited by the  timescale of the temporal sky background variations which is a few  minutes in the NIR and is less critical in bands with a lower background level. $t_{\mathrm{tel}}$, as the exposure time of individual saved frames, is then directly linked to the data reduction strategy for the sky modelling (Sect.\,\ref{s7_NIR_pipeline}) and is set to $t_{\mathrm{tel}}\!=\!40$\,s (60\,s) in the  H-band (Z-band). After the time $t_{\mathrm{tel}}$ the telescope is {\em dithered}, \ie \ moved by a small pre-defined offset of 20\arcsec--30\arcsec \ in order to place the objects at a slightly different position on the detector for the next exposure. This procedure is repeated until the total exposure time $t_{\mathrm{exp}}\!=\!n\cdot t_{\mathrm{tel}}\!=\!n \cdot m \cdot t_{\mathrm{single}}$, with $n$ the number of telescope positions and $m$ the number of in-memory co-adds, is sufficient to reach the required depth.
The total overheads\footnote{In order to minimize the read-out overheads, it is recommended to use a large number of in-memory co-adds $m$ and short exposure times $t_{\mathrm{single}}$, since an idle dummy read is performed at the beginning of each cycle.}, \ie \ the time the camera is not observing the target, can be 50--80\% of the actual integration time and thus amount to a significant time fraction to be carefully considered and optimized within the requirements of the scientific program.


\subsection{Imaging data overview}
\label{s7_CA_data}

\noindent
Following the above observing strategy, we have conducted two runs at the Calar Alto observatory, which were complemented by subsequent service observations. Table\,\ref{t7_NIR_data_overview} summarizes the observation runs and the data yields.
The total observing project was designed as a 10-day program to follow-up XDCP candidates in the Northern part of the survey region at $-20\degr$$<$DEC$<$$+20\degr$.
The proposed program has reached a satisfactory completion of about 80\%. This was made possible only through the enormous support of the Calar Alto staff and the use of buffer time outside the scheduled observing periods.

Table\,\ref{t7_NIR_data_overview} lists an additional NIR observing run at the  Cerro Tololo Inter-American Observatory (CTIO). This run followed-up Southern cluster candidates at DEC$<$$-30\degr$ in H-band only, which complemented a multi-band optical observing program in this area. The data has been reduced and analyzed, but will not be further discussed here since the optical program is ongoing and the observing strategy followed a multi-band rather than a two-band imaging approach. 

The listed data corresponds to a total of roughly 7\,000 raw NIR images including calibration data. The data was partly taken in modest observing conditions with thin to moderate cloud coverage and seeing variations from 0.7\arcsec--3\arcsec. Additional non-ideal occurrences contained in the image data include (i) moon-light reflections from the dome, (ii) reflections and ghost images of nearby bright stars, (iii) scattered light from passing cars, (iv) airplane and satellite trails, and (v) time-varying wave patterns caused by the detector read-out electronics. All this has to be considered for the data reduction procedure which will now be discussed in detail.

\begin{table}[t]    
\begin{center}

\begin{tabular}{|c|c|c|c|c|c|}
\hline

{\bf Run}   & {\bf Date} & {\bf Good Nights} & {\bf Data Quality} & {\bf Raw Data} & {\bf Included} \\

\hline\hline

Calar Alto I   & Jan 06  & 2.5 / 10   &  ok / non-phot. &  29\,GB  &  yes  \\
CA Service A   & May 06 & 0.2   & fair / partly phot. & 4\,GB   &  yes  \\
Calar Alto II   & Nov 06  & 3.5 / 13   &  good / photometric  &  40\,GB   &  yes  \\
CA Service B   & Jan 07 & 1.0   &  fair / partly phot. & 12\,GB    &  yes  \\
CTIO I  & Sep 06  &  2.0 / 3  &  ok / non-phot.  & 27\,GB   &  no  \\

\hline
\end{tabular}

\caption[NIR Data Overview]{NIR data overview. The data for this thesis was obtained during two Calar Alto runs in 2006 complemented by two service observation blocks. Including data that was taken in challenging conditions,  the proposed observing program achieved a satisfactory completion level of about 80\%.
The results of the CTIO run  in the last row using the ISPI instrument are not included in this thesis, since the data is not self-contained (H-band only) and complements an ongoing optical imaging program.} \label{t7_NIR_data_overview}
\end{center}
\end{table}


\section{Development of a NIR Science Reduction Pipeline}
\label{s7_NIR_pipeline}

\noindent
Near-infrared wide-field imaging instruments have only emerged over the last few years and 
OMEGA2000 in this respect is still a camera with one of the widest available field-of-views. 
Additionally, the NIR data reduction process is more challenging and complex than in the optical.
Both aspects combined result in the fact that there is no generic ready-to-go data reduction software package available yet that would fulfill all demands. This gave rise to the motivation to develop a new science-grade NIR reduction pipeline optimized for the XDCP demands. 


\subsection{Requirements and specifications}

\enlargethispage{4ex}

\noindent
From the expected science target characteristics discussed in Sect.\,\ref{s7_observational_strategy} and the raw data properties of the last section, we can now define the requirements and demands on a science-grade reduction pipeline.
The distant galaxy cluster detection and characterization result in the {\bf scientific requirements} of:

\begin{itemize}
    \item the best possible performance for faint extended objects;
    \item the maximum achievable depth for the detection of the faintest objects;
    \item accurate object photometry down to the limiting magnitude;
    \item quality control via intermediate data products.
\end{itemize}    

\noindent
{\bf Practical demands} on the other hand motivate the following pipeline specifications of:

\begin{itemize}
    \item handling of thousands of raw images from dozens of pointings and several filter bands;
    \item minimal requirements of manual interaction;
    \item handling of strongly varying observing conditions and data contamination;
    \item an OMEGA2000 optimization with the flexibility to easily adjust to other NIR instruments;
    \item a portable software package for easy public distribution and use for similar survey projects.
\end{itemize}



\noindent
The science-grade NIR pipeline developments built upon the existing quicklook reduction pipeline which is discussed and characterized in detail in Fassbender \cite*{RF2003}. The following discussion will mainly focus on the upgraded  pipeline components and will only summarize the functionality of existing critical components. The reduction pipeline is implemented as a C-application program within the MIDAS environment. The MIDAS system provides the interface and tools for the image data handling, whereas all pixel-by-pixel image manipulations are based on C-coded algorithms.

\subsection{NIR reduction steps}
\label{s7_NIR_reduction_steps}

\enlargethispage{4ex}

\noindent
The implemented near-infrared data reduction scheme is illustrated in Fig.\,\ref{f7_NIR_reduction}. Data products for the different reductions steps from the raw image to the final science frame are provided in Fig.\,\ref{f7_NIR_RedSteps}.  
The full data reduction procedure can be broken up into the two independent processing blocks (i) {\em single image reduction} and (ii) {\em image summation}.
The pipeline is designed in a way to yield final image products for one or both of the processing blocks without further manual interaction. The user provides an input catalog with an arbitrary number of raw images belonging to the same telescope pointing and taken with the same filter and upon pipeline start all further image handling procedures are automated.

\subsubsection{Single image reduction}
\begin{figure}[h]
\begin{center}
\includegraphics[angle=0,clip,height=21cm]{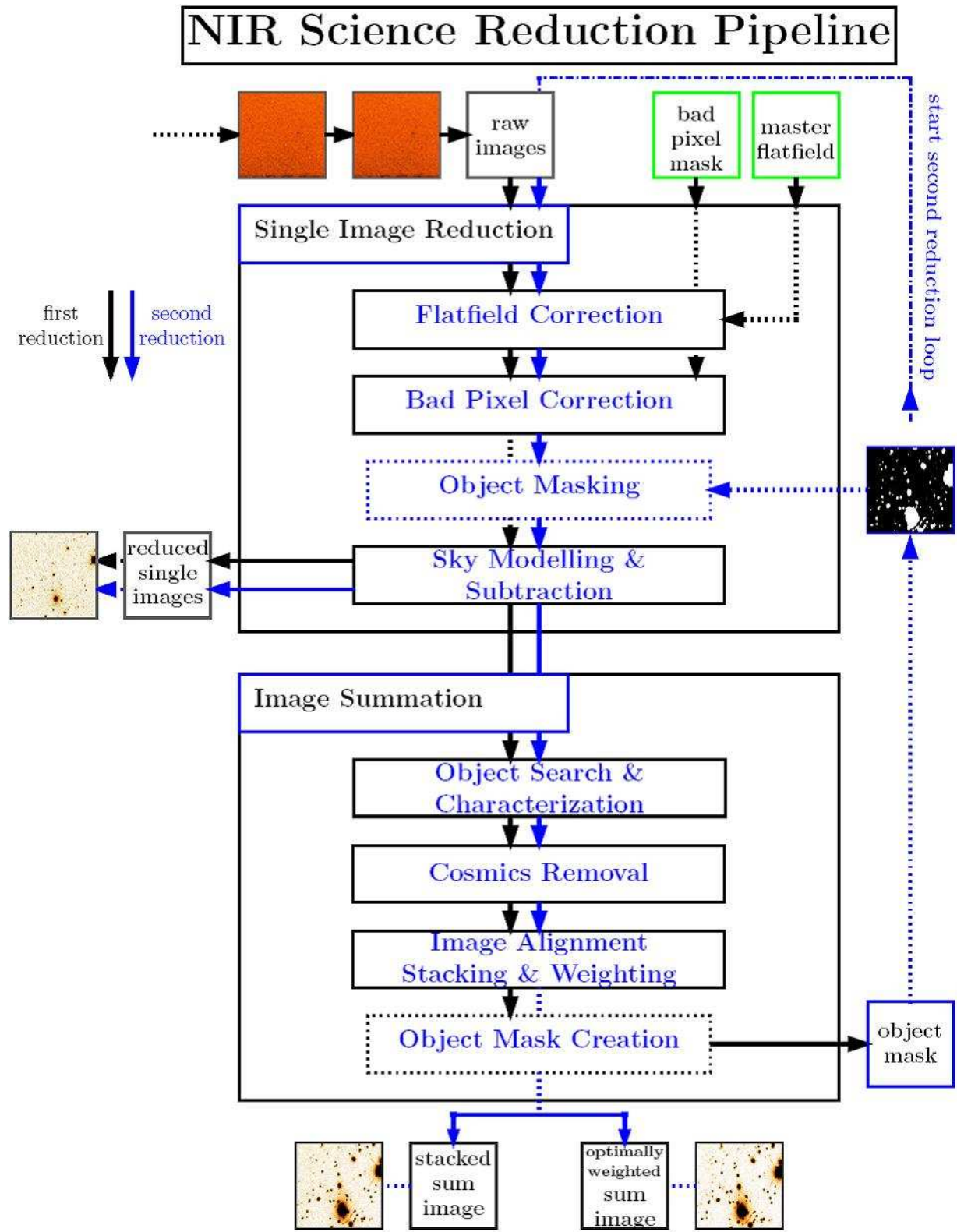}
\end{center}
\caption[NIR Science Reduction Pipeline Flow Chart]{Near-infrared science reduction pipeline flow chart.} \label{f7_NIR_reduction}
\end{figure}

\begin{figure}[h]
\begin{center}
\includegraphics[angle=0,clip,width=0.96\textwidth]{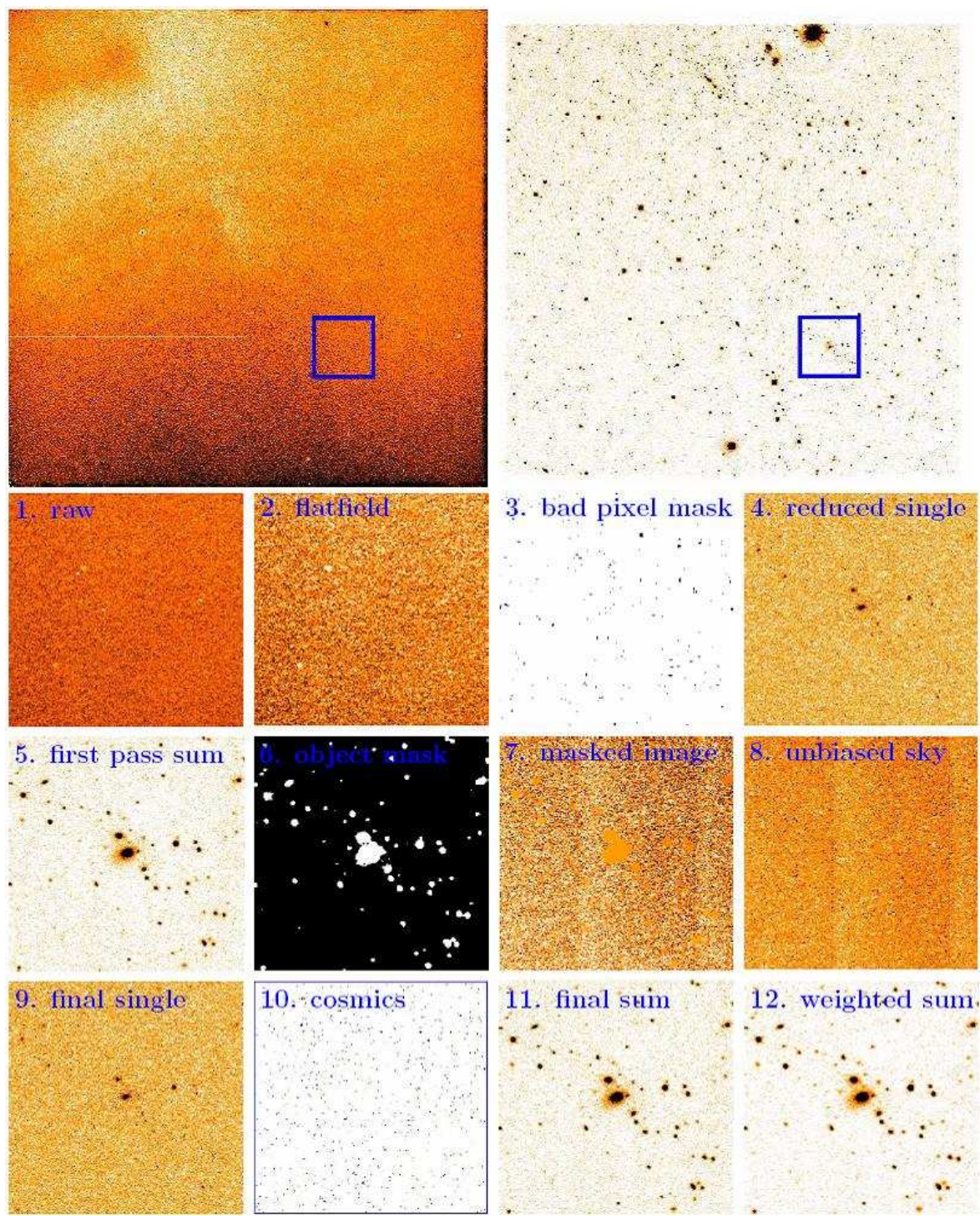}
\end{center}
\vspace{-2ex}
\caption[NIR Science Reduction Steps]{Near-infrared science reduction steps. The top panel shows the full $15.4\arcmin\!\times\!15.4\arcmin$  FoV for a raw H-band image ({\em left}) and the final reduced sum image containing 75 input frames. Note that about 30\arcsec are lost at each field edge due to the dithering ({\em right}). The blue squares indicate the $2\arcmin\!\times\!2\arcmin$ zoom field for the small panels centered on the galaxy cluster RX\,J0018.2+1617 at $z\!=\!0.55$. The small cutouts with numbers 1-12 illustrate the reduction steps of Fig.\,\ref{f7_NIR_reduction} by showing intermediate data products.
} \label{f7_NIR_RedSteps}
\end{figure}

The first processing block handles the calibration steps of the individual raw frames (upper left panel of Fig.\,\ref{f7_NIR_RedSteps} and subpanel\,1) and the subsequent sky modelling and subtraction as shown in Fig.\,\ref{f7_NIR_reduction}. 

As initial calibration step, the raw input image is {\em flatfielded\/} to correct the detector inhomogeneities.
For NIR detector arrays this is particularly critical since (i) the quantum efficiency can vary by a factor of two over the FoV (global variations), and (ii) the electronically independent NIR pixels exhibit significant pixel-to-pixel changes (local variations).
A {\em master flatfield} for each filter can be created from {\em dome flatfields} taken inside the telescope dome and {\em sky flatfields} taken during twilight in a region devoid of bright objects. Since the thermal dome emission imprints a significant amount of structure in the dome images, the {\em master flatfield} mainly relies on the detector variations determined from the sky calibration data with the {\em dome flatfields} acting only as initial correction of the former. The final {\em master flatfield} contains the information on the global and local detector variations. It is normalized to a median value of 1 and placed in a calibration frame directory, where the reduction pipeline can access it.  
By dividing the raw input frame by this  {\em master flatfield} (subpanel\,2) the intermediate image product contains  variations due to the received signal and not the detector.  

\pagebreak

\clearpage

As a second step, a {\em bad pixel correction} is applied. The OMEGA2000 detector array contains about 0.5\% pixels which are either not functioning at all (dead pixels) or exhibit significant deviations from a time-linear signal response (hot or flickering pixels). These pixels can be identified through a linearity analysis and are flagged in the {\em bad pixel mask} (subpanel\,3) of the calibration directory. The values of these flagged bad pixels in  the flatfielded image are then replaced by the interpolated pixel value of the four closest good pixels.

Note that near-infrared measurements are usually obtained by reading out the detector array at the beginning and end of the integration time and subtracting the two to determine the signal. Hence this {\em double-correlated readout} scheme does not contain a {\em bias} offset as in CCD data. The detector {\em dark current}, \ie \ the measured pixel values without an external signal, is in principle an additional (small) contribution to be corrected. However, tests have shown 
(H.-J. R\"oser, private communication) that the {\em dark current} is not stable enough to be accurately modelled and subtracted since it  depends sensitively on the exact detector temperature and its thermal history. Instead of an explicit subtraction, the  {\em dark current} is treated as part of the overall background and implicitly corrected with the  {\em sky background subtraction}.     

The most critical and important part in the near-infrared reduction process is the {\em sky background modelling}. The dominating background components are (i) atmospheric airglow (see Fig.\,\ref{f8_sky_features}) and (ii) thermal emission of the ambient structures at long wavelengths. Smaller contributions arise from (iii) scattered moonlight, (iv) Zodiacal scattered light \cite{allen2000}, and (v) the uncorrected instrumental dark current. As discussed in Sect.\,\ref{s7_NIR_strategy}, the science objects of interest can be easily a factor of 100--1\,000 fainter than the background surface brightness. Consequently this implies that the local background around the astronomical objects has to be known with uncertainties of $<\!10^{-3}$. The dominating airglow emission exhibits significant temporal changes on time scales of several minutes and spatial variations on scales of a few arcminutes. Hence the sky modelling requires a temporally and spatially local background estimation derived directly from science data. The pre-requisites to an accurate sky modelling have been implemented in the observing strategy through (i) a sufficiently short telescope pointing time $t_{\mathrm{tel}}$ that allows tracing the temporal variations and (ii) well-chosen dither offsets that enable a reconstruction of the local background without being biased by the flux  of the astronomical objects of interest.  

The following recipe yields an accurate but preliminary estimate of the sky background. (i) Take the image together with the three\footnote{Optimized pipeline parameters  for the XDCP NIR data reduction are cited for this discussion. However, all critical parameters of the software can be specified and adjusted to the actual science needs. The background for the first (last) images of an observation series are estimated from the first (last) 7 frames.} preceding and following frames, \ie \ the background is determined from a total of seven frames taken within a period of $\pm$3--5 minutes. (ii) For every detector pixel (image coordinates) select the corresponding pixel values of all seven frames. Due to the dither strategy, which places the real objects at different detector positions for each image, most pixel values will only contain a background signal, whereas a minority will additionally have an object signal component\footnote{This assumption is in general well justified for extragalactic sparse fields. However, for (low redshift) objects larger than the dither offsets of 20\arcsec--30\arcsec \ the assumption is not fulfilled, but the introduced initial error is corrected during the second iteration.}. (iii) The median of these seven values is taken as a good initial guess of the actual background level for the specified pixel. However, any object flux in one or more of the seven values systematically biases the median to a slightly higher level, \ie \ the background is overestimated due to the additional non-background component in some pixels. (iv) This background bias can be corrected in next order by applying a $\kappa\sigma$-clipping algorithm to the data values. A Poisson  estimate of the local background standard deviation $\sigma$ is obtained from the initial background estimate, the gain\footnote{The gain of a detector is the conversion factor between digital count units and physically detected electrons via the photoelectric effect.} of the detector array, and the local quantum efficiency of the detector pixel\footnote{In practice, an additional normalization factor has to be considered which is necessary to standardize the background levels in each frame.}. Any pixel values deviating more than three standard deviations from the initial background estimate are clipped, and the median is recomputed from the remaining object-cleaned values. 
(v) The $\kappa\sigma$-clipped median value is taken as the background estimate for the specified pixel. The procedure is repeated for all detector pixels resulting in a  full sky frame for the input science image. (vi) The modelled background sky is subtracted from the science image and replaced by a constant representing the average sky value in order to preserve the proper counting statistics.

The flatfielded, bad pixel corrected, and sky subtracted frame is written to disk as the final result of the first single image reduction loop (subpanel\,4). However, the applied $\kappa\sigma$-clipping for the first order correction of the local sky value could only eliminate object flux which has been detected at more than three standard deviations above the normal background level in the single frame, \ie \ for fairly bright sources. The bulk of fainter sources and the wings of brighter objects below this clipping threshold will still bias the local background model to slightly higher values. The subtracted sky  might thus still be overestimated  resulting in a systematically lower flux for an object at the given position, which can reach significant levels  for faint galaxies (see Sect.\,\ref{s7_faint_object_optim}). This lost flux can be largely  recovered with a second iterated background modelling loop as discussed below.

\subsubsection{Image summation}
The reduced single image contains the  data of integration time $t_{\mathrm{tel}}$, \ie \ typically 60\,seconds or less. 
In order to reach the required depth, the single images have to be stacked to build up a deep sum image with the full on-target exposure time $t_{\mathrm{exp}}$. To achieve this, the dither offsets have to be reversed and the images aligned in world coordinates, \ie \ the physically same objects in the individual frames have to end up on top of each other. During this stacking procedure cosmic ray events, or short {\em cosmics}, can be identified and removed.

The alignment and stacking procedure uses  the following approach: (i) The first input image is declared as master frame onto which all others are positioned and co-added. (ii) The reduced single master image is searched for typically 50--100 reference stars, whose centers-of-mass, as derived from the stellar flux distribution, are recorded. (iii) From the image header of the next input frame, the expected offsets with respect to the reference master frame are computed. Using these approximate offsets, the catalogued reference stars are searched for in the vicinity of the expected detector positions. Once matched, the exact object centers are determined resulting in an improved offset determination. (iv) The final image offset is determined from the outlier-clipped average of all matched stars. Since the OMEGA2000 optics exhibits negligible image distortions over the whole FoV, the frames can be co-added at this stage by simple XY-coordinate shifting to the nearest integer pixel offsets. 

The single images still contain the unwanted {\em cosmics} which are to be corrected for without modifying the flux of the real astronomical objects. Cosmic ray events, caused by charged particles going through the detector, can be identified from their characteristic signature of single or few isolated pixels with increased count levels without a corresponding counterpart in the next aligned image. In principle, the {\em cosmics} would be removed by a simple median process over a sufficient number of pixel-aligned images, similar to the procedure for the background determination. The complication arises that the median procedure efficiently removes everything that is not present in at least half of the images, including the outer wings of real astronomical objects in variable seeing conditions, and the central flux peak of objects. The peaks  can exhibit large gradients towards the neighboring pixels and are thus likely damped when medianing over the different object flux distributions in each frame.   
This effect can decrease the total flux of stellar sources by about 20\%, which is avoided by applying the following procedure: (v) The median image of eleven world-coordinate aligned input frames is determined and used as object reference image. This intermediate sum image is now devoid of {\em cosmics} and contains the real objects at the proper locations but with biased flux levels. (vi) Each individual frame is now checked pixel-by-pixel whether the value is more than five standard deviations above the corresponding median reference value, where the standard deviation is determined in analogy to the sky modelling procedure. (vii) If a {\em cosmics} candidate is detected, the reference median image is searched for an astronomical object within a radius of three pixels around the outlier position. In case a source is found, the initial pixel value is classified as object flux and co-added to the master image as is. If no source is matched, the outlier value is classified as {\em cosmic} and replaced by the median value of the reference image, which is in turn co-added to the master sum image. (viii) This procedure is repeated until all input images have been aligned, {\em cosmics} cleaned, and co-added to yield a final deep sum image (subpanel\,5). The cosmic ray events, as shown in subpanel\,10 of Fig.\,\ref{f7_NIR_RedSteps}, have now been removed from the field without influencing the total flux of the astronomical sources. 

\enlargethispage{8ex}
The last two subsections have provided a summary of the full first data reduction loop, which is in principle equivalent to the quicklook reduction system available at the telescope. The next two subsections will discuss further improvements towards the goal of the best achievable science-grade data reduction quality.


\subsubsection{Iterated background subtraction}

The complete science-grade reduction process requires two additional main components. Firstly, the creation of an {\em object mask} with registered positions of all sources. Secondly, an additional iterated loop through the reduction procedure but now with removed source fluxes for an unbiased second order sky background correction.   

The {\em object mask} is created from the final deep sum image of the first reduction pass (subpanel\,5) which contains the best available information on the location and size of the astronomical objects at this processing stage. The objective of the {\em object mask} is to cover all sky areas in the FoV which contain a detectable source signal and use only the pixels with an unbiased background value for the reconstruction of the sky model. In practice the mask is created using an iterative $\kappa\sigma$-clipping algorithm applied to the smoothed deep sum image. An initial smoothing is helpful for the identification of faint object halos, \ie \ the PSF wings for point sources are extended halos for galaxies. Pixel values  deviating more than about two standard deviations  from the median background level are classified as object pixels and flagged in the mask frame. This procedure is iterated to push the depth of the mask to sufficiently  faint levels. The final {\em object mask} is shown in subpanel\,6, which now distinguishes the black regions with background pixel values and the masked white areas with object flux contributions.  

With the {\em object mask} in hand, the second reduction loop can start from the very beginning (see Fig.\,\ref{f7_NIR_reduction}). The initial calibration steps {\em flatfielding} and {\em bad pixel correction} are applied as before. The {\em sky modelling} procedure on the other hand is extended by two additional steps. (i) The {\em object mask} is projected onto each individual input frame. To do this projection properly, the exact image dithering offsets as determined during the first image summation loop are used to shift the {\em object mask} to the correct world-coordinate aligned location in each input image. (ii) The pixel values of the masked object regions are replaced\footnote{Two additional replacement schemes with lower performance expectations are implemented. 1) The masked pixel is ignored for the sky determination, which leads to difficulties for very large objects. 2) The masked pixel is replaced by the global median, which does not take into account local variations.} by the unbiased median background level of the 100 nearest  unmasked values. This replacement scheme is designed to yield robust results for arbitrary large objects and to take into account local background variations across the FoV. The masked input image (subpanel\,7) with all object flux replaced is then passed on to the sky modelling procedure. The final unbiased sky model (subpanel\,8) is now the result of a second order corrected flux-removed and $\kappa\sigma$-clipped background determination. Subtraction from the original input image yields the final iterated reduced single image (subpanel\,9), and after the subsequent summation and {\em cosmics} removal (subpanel\,10) of all frames the final co-added deep sum image is available after the second processing loop (subpanel\,11).

\subsubsection{Advanced optimizations}

The final sum image is a well-reduced and science-ready data product. However, with respect to the objective of reaching the maximum possible depth, two additional optimization schemes can be applied. Firstly,  
{\em fractional pixel offsets} instead of integer values can be implemented for the stacking procedure, and secondly,  {\em optimal weighting} of the individual input frames can improve the final signal-to-noise ratio. 

The discussed image summation procedure determined the accurate XY-offsets of the frames with respect to the master frame, but then rounded the exact values to the nearest integer pixel offsets in order to match the underlying pixel grid. This way, systematic truncation errors of up to half a pixel in each spatial direction are unavoidable. For the OMEGA2000 pixel scale of 0.45\arcsec /pixel this implies systematic XY-shifting errors of 0--0.225\arcsec \ per direction, which can add up in quadrature to a maximum systematic displacement error of 0.32\arcsec \ for a stacked frame. The integer pixel shifting has thus two consequences, both gaining importance in good seeing conditions. (i) The stacked object flux is not as compact as defined by the seeing limit and (ii) the 2-dimensional flux distribution of point sources does not follow its natural radially symmetric Gaussian shape. This latter tendency of the object cores to become boxy or square-like is easily visible for the fainter objects in the left panels of Fig.\,\ref{f7_WeightSchemes}. The shape improvements after applying the {\em fractional pixel offset} scheme can be seen by eye in the right panels of the same figure.

In order to implement {\em fractional offsets} onto the underlying fixed detector grid of 2\,048$\times$2\,048 pixels, the flux values in each pixel have to be distributed over four cells of the master grid in a flux conserved way. The flux distribution matrix is calculated once for each frame from the exact offsets according to the fractional geometrical overlap of a pixel with the underlying master grid. The pixel values of the frame to be co-added are then distributed using this pre-determined matrix. In turn, each pixel of the master grid, receives fractional flux contributions from four adjacent pixel values of the single image.   

The question whether a faint object will be detected depends primarily on the signal-to-noise ratio of the source (see Equ.\,\ref{e7_background_limit}), \ie \ on its contrast  with respect to the noise properties of the background. So far, all input frames for the summation were treated as if they had the same SNR properties. However, this is in general not the case since the observing conditions for the seeing $\sigma$, the sky transparency $T$, and the background surface brightness levels $B$ can change significantly during the observations. The source signal $s$ will then vary as $s\!\propto\!T$ and the total background $b$ confined in a seeing limited aperture will follow $b\!\propto\!B\cdot \sigma^{2}$. One can then ask the question for the best way to co-add  two individual frames in order to achieve the best SNR for their sum. Following Gabasch \cite*{Gabasch2004a} for the case of faint sources ($b\!\gg\!s$), an optimal weight factor $w_{2}$ for the second frame can be derived by maximizing the resulting\footnote{For the second factor, the background noise is added in quadrature $\sigma_{tot}^{2}\!=\!\sigma_{1}^{2}\!+\!(w_2 \sigma_{2})^{2}\!=\!b_1\!+\!w_2^{2}b_2$ . Note that in this relation $\sigma$ denotes the standard deviation and not the seeing.} $\mathrm{SNR_{tot}}\!=\!(s_1 + w_{2}s_2) \cdot (b_1 + w_{2}^{2} \cdot b_2)^{-1/2} $. Using dimensionless parameters for the source signal $\tilde{s}\!\equiv\!s_{2} / s_{1}$, the transparency change $\tilde{T}\!\equiv\!T_{2} / T_{1}$, the background variation  $\tilde{B}\!\equiv\!B_{2} / B_{1}$ or $\tilde{b}\!\equiv\!b_{2} / b_{1}$, and the seeing ratio 
$\tilde{\sigma}\!\equiv\!\sigma_{2} / \sigma_{1}$ of the two observations, the optimal weight for the second frame is given by

\begin{equation}\label{e7_weight_factor}
    w_{2} = \frac{\tilde{s}}{ \tilde{b}} = \frac{\tilde{T}}{ \tilde{B} \cdot \tilde{\sigma}^{2} } = \frac{\tilde{f}}{ \tilde{B} \cdot \tilde{\sigma}^{2} } \ .
\end{equation}

\noindent
In the last step, the change in the sky transparency $\tilde{T}$ was replaced by the equivalent 
but directly observable source flux ratio of the two observations $\tilde{f}\!\equiv\!f_{2} / f_{1}\!=\!\tilde{T}$. The optimal weighting factor $w_{2}$ for a new image to be co-added to the master frame is thus proportional to the received signal ratio $\tilde{s}$ and inversely proportional to the total background ratio $\tilde{b}$ in the object aperture, which is particularly sensitive to seeing variations. This is expressed in terms of the direct observables $\tilde{f}$, $\tilde{B}$, and $\tilde{\sigma}$ that allow an automatic monitoring for each image. The background factor $\tilde{B}$ is determined from the median sky level in each frame. The flux ratio $\tilde{f}$ and the seeing ratio $\tilde{\sigma}$ can be obtained by averaging the stellar source properties as measured in the first step of the summation process, where all matched objects are characterized in terms of their 5\arcsec -aperture flux and the full-width-at-half-maximum (FWHM) of their radial source profiles.   

In practice, the {\em fractional pixel offset} and {\em optimal weighting} schemes are combined and implemented
as a separate optimized final sum image, as shown in the last subpanel\,12 of Fig.\,\ref{f7_NIR_RedSteps}. The effects of these additional optimization procedures as part of the science-grade data reduction will be investigated in the next section \ref{s7_faint_object_optim}.  

This last section has discussed the main steps and procedures to go from raw NIR data to science-ready optimized stacked images and how they are implemented as part of the newly developed NIR reduction pipeline. For the routine reduction of the data listed in Tab.\,\ref{t7_NIR_data_overview} the following approach was pursued for each science field and filter band: (i) A full automatic pipeline reduction is run on the field with standard parameters. (ii) The data quality of the intermediate and final data products is inspected. This includes checks of the individual reduced images for possible data corruption, contamination from scattered light, cloud coverage, and satellite trails. Additionally, the object mask is investigated for proper depth, the masked input images for a correct mask projection, and the sky models and cosmics frame  for unusual features. (iii) 
Identified  contaminated data frames are then excluded from the reduction and the pipeline parameters are optimized if necessary. (iv) The reduction pipeline is then re-run on the optimized input data set to yield the final science-grade image products.

\subsection{Faint object optimization}
\label{s7_faint_object_optim}

\enlargethispage{-2ex}

\begin{figure}[t]
\begin{center}
\includegraphics[angle=0,clip,width=0.65\textwidth]{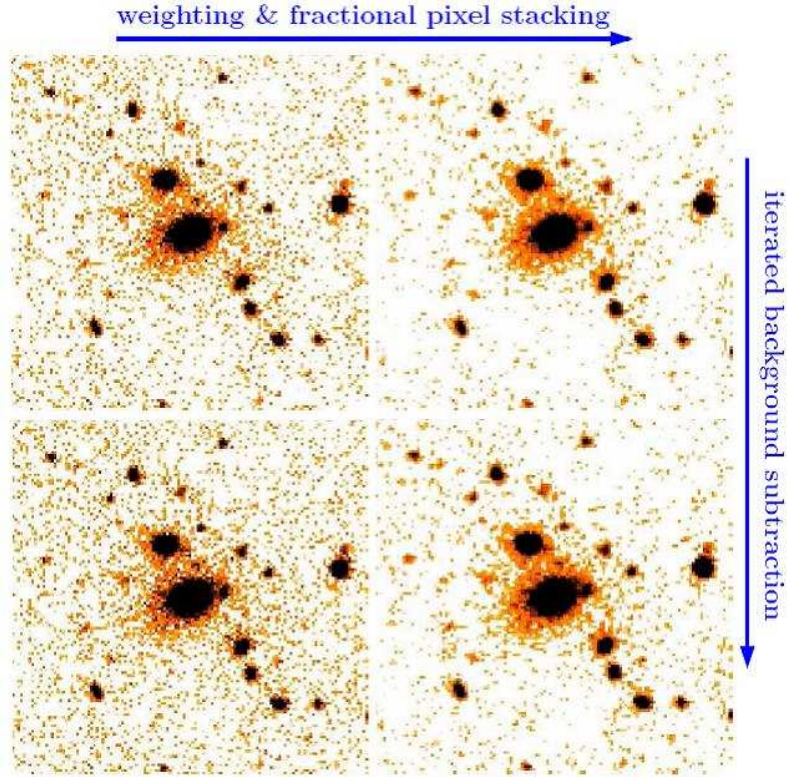}
\end{center}
\caption[Faint Object Optimization]{Faint object optimization. The {\em upper left panel} shows a high-contrast 1\arcmin$\times$1\arcmin \ zoom on the calibration cluster RX\,J0018.2+1617 at $z\!=\!0.55$ as seen in the stacked sum image after the first reduction pass (see panel 5 in Fig.\,\ref{f7_NIR_RedSteps}). Going to the right panels adds fractional pixel offsets and optimal weighting. The lower panels show the image results after the second iterated sky subtraction. The final best image including all optimizations is thus placed in the {\em lower right panel}. While the iterated sky subtraction recovers a significant amount of flux of faint and extended objects, the optimal weighting scheme increases the limiting magnitude by improving the signal-to-noise ratio of the faint sources (see Fig.\,\ref{f7_weight_performance} for quantitative results). } \label{f7_WeightSchemes}
\end{figure}

\noindent
After introducing the basic functionality of the main science pipeline components in the last section, we will now turn to a discussion of the effects of the implemented reduction optimizations. Figure\,\ref{f7_WeightSchemes} illustrates the different optimization schemes for 1\arcmin$\times$1\arcmin \ co-added image cutouts  centered on the calibration cluster RX\,J0018.2+1617 at $z\!=\!0.55$, displayed with high image contrast. The upper panels show the sum images after the first reduction loop, whereas the lower panels have the results after the {\em iterated background subtraction}. Going from left to right adds the {\em optimal weighting} scheme and the {\em fractional pixel offsets}. Hence the upper left panel displays the reduction quality achievable with a single loop (quicklook) reduction, and the lower right panel shows the same data with all optimizations incorporated requiring about the threefold CPU time. 

We will first consider the effects of the {\em iterated background subtraction}, \ie \ the difference between the upper and lower panels of Fig.\,\ref{f7_WeightSchemes}. By masking out all objects and their halos during the second reduction loop, the sky background determination is unbiased compared to the systematically slightly higher median levels when source fluxes are included. This lower background value at the object positions results in additional object flux when the sky background is subtracted. The 
{\em iterated background subtraction} hence recovers object flux and consequently makes the sources brighter. This flux recovery is expected to be more significant for extended low surface brightness objects compared to stellar point sources. 
A useful approximation to convert measured magnitude offsets into flux differences $\Delta f\!=\!f_{2}-f_{1}$ for small variations ($\Delta m\!\ll\!1$) is given by the relationship 

\begin{equation}\label{e7_flux_approximation}
\Delta m  = -2.5 \cdot \log \left(\frac{f_{2}}{f_{1}} \right) = -2.5 \cdot \log \left(1+\frac{\Delta f}{f_{1}}  \right) \approx - \frac{2.5}{\ln(10)} \cdot \frac{\Delta f}{f_{1}}  \approx - 1.1 \cdot  \frac{\Delta f}{f_{1}} \ ,
\end{equation}

\noindent
implying that small magnitude differences reflect approximately the fractional flux difference.



\begin{figure}[h]
\begin{center}
\includegraphics[angle=0,clip,width=\textwidth]{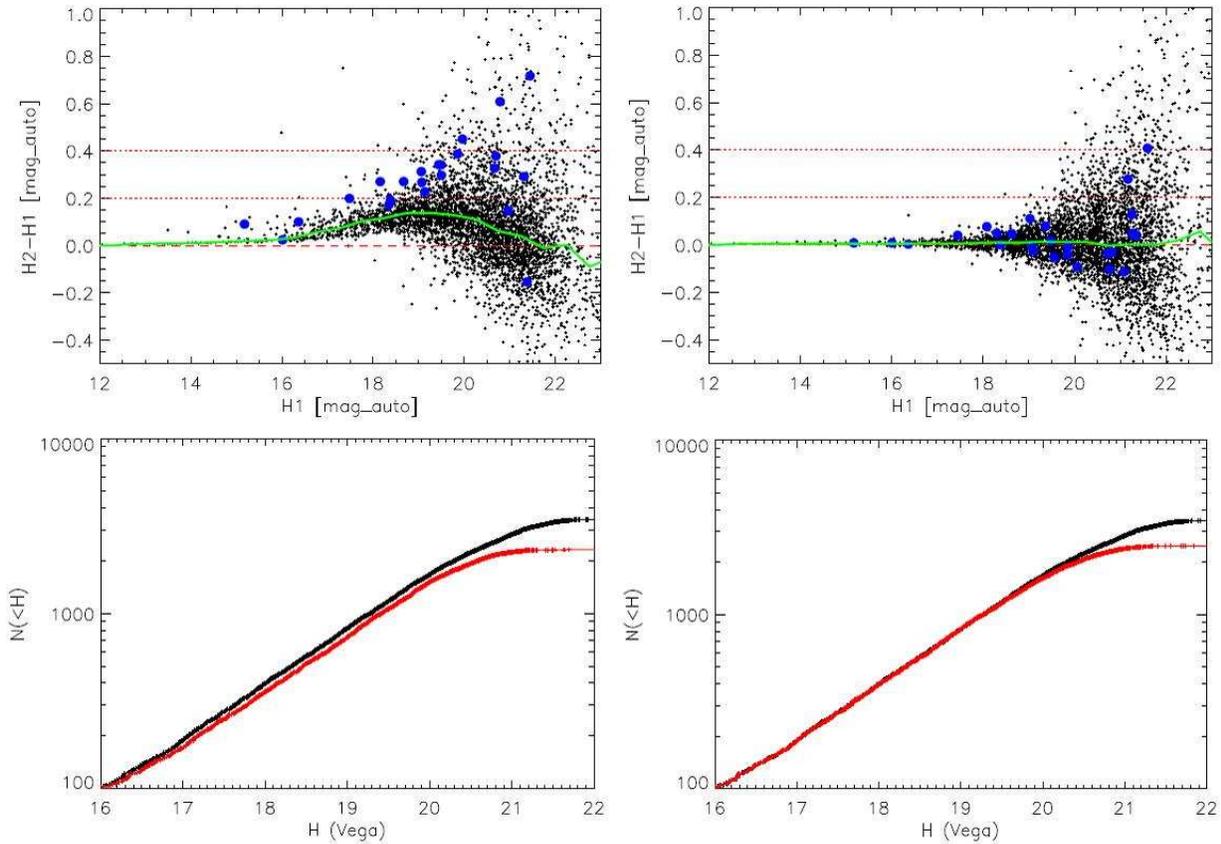}

\end{center}
\vspace{-4ex}
\caption[Advanced Reduction Schemes]{Effects of advanced reduction schemes. {\em Top panels:} Difference photometry in the H-band for various reduction stages as shown in Fig.\,\ref{f7_WeightSchemes} using the same deep detection image. 
{\em Left:} Total magnitude difference between the sum image with iterated background subtraction (here H1, lower left panel in Fig.\,\ref{f7_WeightSchemes}) and the sum image after a single reduction loop (here H2, top left panel in Fig.\,\ref{f7_WeightSchemes}). The dashed red line represents a zero offset (\ie \ the same measured magnitudes), black crosses show all objects in the FoV, big blue circles represent the galaxies seen in Fig.\,\ref{f7_WeightSchemes} (\ie \ cluster members of RX\,J0018.2+1617), and the green solid lines follows the median offset taken in 0.5\,mag bins. At H$>$16, the recovered flux from the iterated background subtraction becomes noticeable and increases to a median difference of 0.1--0.15\,mag (about 10-15\% in flux) between H magnitudes of 18--21. At H$>$21 the difference drops back to zero which reflects the depth of the object mask. Note that the recovered flux is drastically increased for the faint extended cluster galaxies (blue circles) compared to the median difference.  {\em Right:} Same plot for the difference between the final weighted image (H1, lower right panel in Fig.\,\ref{f7_WeightSchemes}) and the iterated background subtracted image without weighting (H2). The total magnitudes are now consistent, since weighting cannot change the flux, but only the signal-to-noise ratio of the sources. 
{\em Left bottom panel:}  $\log N$--$\log S$ of the final reduced image (black solid line) and the first sum image (red solid line) in a field with significant seeing and transparency variations during the observations. The systematic vertical offset is due to the iterated background subtraction. The difference in total number of detected sources on the other hand can be attributed to the optimal weighting scheme. The effect is separated in the {\em right panel}, which shows the results of the best final image (black) and the non-weighted image (red). The limiting magnitude is significantly improved by the weighting and the total number of detected sources can be boosted by as much as 30--40\% in fields with strongly varying conditions.} \label{f7_weight_performance}
\end{figure}


\clearpage

The effect of the {\em iterated background subtraction} is quantified in the upper left panel of Fig.\,\ref{f7_weight_performance} where the difference photometry with and without the second reduction loop is shown against the total object magnitudes. For both images the measurements were performed in the exact same apertures by using a common deep multi-band detection image (see Sect.\,\ref{s7_NIR_source_detection}), \ie \ any offsets are due to object flux differences. No measurable effects would thus imply zero offsets (dashed null line), and more object flux after the iterated loop will manifest itself in a positive magnitude offset (the brighter, \ie \ smaller, magnitudes are subtracted). 
The measurements for all objects are plotted as black dots, for the cluster members in Fig.\,\ref{f7_WeightSchemes} as blue circles, and the median averaged differences in 0.5\,mag bins are traced by the green solid line. Following this line  shows practically matching magnitudes out to H$\sim$15, which is expected due to the photometric calibration procedure using bright 2MASS stars as discussed in Sect.\,\ref{s7_NIR_calibration}. 
At H$>$16, the recovered flux from the iterated background subtraction becomes noticeable, \ie \ the magnitudes are systematically brighter, which increases to a median difference of 0.1--0.15\,mag (about 10--15\% in flux) between H magnitudes of 18--21. 
At H$>$21 the averaged difference drops back to zero reflecting the depth of the object mask which does not cover the faintest sources anymore. 
Note that the recovered flux is drastically increased for the cluster galaxies (blue circles) compared to the median difference.
The difference for the bright cD galaxy with its extended halo is 0.1\,mag and in the important regime between $18\!<\!H\!<\!21$ the effect grows to 0.2--0.4\,magnitudes or equivalently to approximately 20--40\% in flux. This pronounced effect for the cluster environment, \ie \ for the main science applications, has two reasons. (i) Most cluster galaxies are faint extended low surface brightness objects and (ii) the large halo of the central cD galaxy biases the background modelling in its vicinity, which is the main contribution for this particular environment.    

The effects of {\em fractional pixel offsets} and the {\em optimal weighting} are investigated in combination, since both optimizations only improve the signal-to-noise ratio of faint sources but do not change the flux of the objects. This can be seen in the right panel of  Fig.\,\ref{f7_weight_performance} where now the difference photometry is applied to the fully optimized image (lower right panel of Fig.\,\ref{f7_WeightSchemes}) and the non-weighted version (lower left panel of Fig.\,\ref{f7_WeightSchemes}). The median net flux differences is now consistent with zero and the increased scatter at the faint end can primarily be attributed to photometric measurement errors dominated by the image with lower SNR.

The main contribution of the stacking procedure using {\em fractional pixel offsets} is the improved reconstruction of the natural seeing-limited point-spread function (PSF). In Fig.\,\ref{f7_WeightSchemes} this is best seen for the stellar source at the right edge center of the panels and the fainter galaxies which change from box-like object cores in the left panels to symmetrically round PSF shapes (for the star) on the right. The recovered natural PSF shape is also the most compact flux configuration with an expected improved FWHM of approximately the  average displacement error of $\orderof$(0.1\arcsec). On the other hand, the implemented {\em fractional pixel offset} scheme introduces correlated noise, since the flux from a single independent pixel is now distributed onto four pixels of the master grid. A main concern would be spurious detections introduced from correlated single pixel noise, \eg \ an uncorrected cosmic or flickering pixel, that passed the object detection threshold of at least four adjacent pixels more than one standard deviation above the background (see Sect.\,\ref{s7_NIR_source_detection}). However, for a stacked sum of only ten images (typically 20--80) this would already require a single image pixel-noise outlier of $\sigma_{\mathrm{single}}\!\ga\!4 \sqrt{10}\!\approx\!12$ above the background, which implies that it would have been easily detected and removed by the {\em cosmics} removal procedure. Even a constantly hot pixel not included in the {\em bad pixel mask} would most likely not lead to a spurious source since (i) it would show up in the sky model and be subtracted with it, and (ii) these outliers would again be removed from the  {\em cosmics} routine since they do not align in world coordinates. In conclusion, the {\em fractional pixel offsets} procedure is not expected to cause any significant number of spurious source detections through correlated noise.

The last step towards the maximally achievable image depth is the implemented {\em optimal weighting} scheme. The measurable effects of this optimization strongly depend on the variability of the observing conditions according to Equ.\,\ref{e7_weight_factor}. However, since the execution time per field and filter is typically  30--80\,min long, changes of the external conditions are likely, leading to significant improvements if weighting is applied. Variations that modulate the weighting factors include (i) pronounced background changes at the beginning or end of the night or due to the moon setting or rising, (ii) transparency drops induced by partial cloud coverage, and most sensitively (iii) seeing variations in the atmosphere or the telescope dome, \eg \ if the primary mirror is warmer than the ambient air.     

The optimization of the signal-to-noise ratio at the faint source end  has two important effects. Firstly, objects can be characterized with higher photometric accuracy, \ie \ the measurement errors decrease. Secondly, very faint formerly undetected sources are now pushed over the detection limit requiring a minimum fixed SNR. This latter effect is of particular importance since the total number of sources increases rapidly with the flux limit as $N_{\mathrm{tot}}\!\propto\!f^{-3/2}_{\mathrm{lim}}$  (Equ.\,\ref{e3_number_counts}) and allows to probe the galaxy luminosity function in clusters to fainter levels. The number counts for a field observed under strongly varying conditions in the lower right panel of Fig.\,\ref{f7_weight_performance} illustrates the achievable gain. The black curve shows the detection results for the image with the applied  {\em optimal weighting}  scheme, the red line gives the outcome on the non-weighted image. Up to H$\la$20 the number counts are basically identical but at H$\sim$20.5 the red non-weighted curve turns over and reaches its plateau. The black weighted line on the other hand exhibits a 
significantly improved limiting magnitude, which in turn boosts the total number of detected sources for this field by almost 40\% at no extra observational cost. The depth optimization via {\em fractional pixel offsets} and  {\em optimal weighting} thus takes effect in the magnitude range 20$\la$H$\la$22, which includes about half of all detectable sources.

For comparison, the lower left panel of Fig.\,\ref{f7_weight_performance} shows the number counts in the same field but with the first loop non-weighted (quicklook) image used for the red line. The number counts now exhibit a  systematic vertical offset starting at H$>$16 due to the discussed {\em iterated background subtraction}, which is the consequence of the objects being systematically brighter, hence more sources exist at a given fixed magnitude.


In summary, the implemented optimization schemes\footnote{The faint object optimization effects in this section have been discussed using H-band data, since this is the master filter for the Z--H imaging strategy. The achievable improvements scale with the background brightness and the number of individual images to be co-added. For both cases the expected optimization performance will increase with filter wavelength, \ie \ in the order Z, J, H, Ks.} for the science-grade reduction pipeline have an important impact on the targeted science applications. The {\em iterated background subtraction} corrects the systematic biasing of the total object magnitudes, whereas the combined effects of the {\em fractional pixel offsets} and the {\em optimal weighting} scheme improve the limiting  depth of the data.
With respect to the intended distant galaxy cluster identifications the achieved improvements are of prime interest. (i) The measured object magnitudes need to be accurate in the full range 16$\la$H$\la$21 in order to derive accurate colors and thus allow a \reds \ identification and model comparisons. (ii) The increased number of detected cluster galaxies enhances the cluster signature, eases the evaluation, and increases the redshift grasp.

\subsection{Pipeline and instrument performance}
\label{s7_pipeline_performance}


\noindent
In this section we will apply the NIR reduction pipeline to real multi-band data of the XDCP follow-up program, compare the achieved depth to the requirements of Sect.\,\ref{s7_imaging_strategy}, and discuss the expected background object densities.   
The analysis is performed in the follow-up field North of the cluster of galaxies Abell\,383 with deep and a fairly homogeneous data coverage 
for the bands Z, J, H, and Ks.

\begin{figure}[t]
\centering
\includegraphics[angle=0,clip,width=\textwidth]{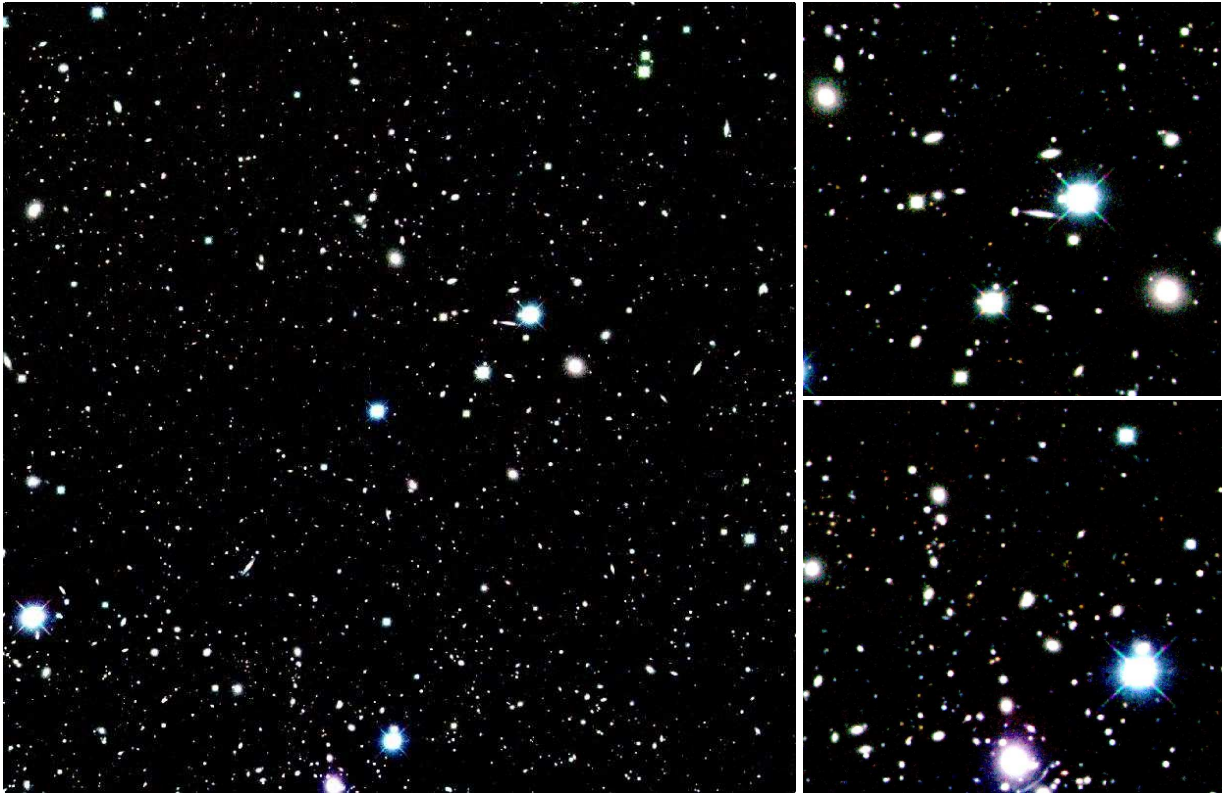}

\caption[Abell\,383 Field]{Z+J+H pseudo color image of the field North of Abell\,383 used for the number counts in Fig.\,\ref{f7_NumberCounts}. {\em Left:} The full 192 square arcminute FoV  considered for the analysis. The cD galaxy of Abell\,383 is just visible at the bottom center. {\em Right:} 4x4\,arcmin zoom on the central part of the field ({\em top}) and the Northern part of Abell 383 with its giant gravitational arc ({\em bottom}).}
\label{f7_NumberCountField}       
\end{figure}

Figure\,\ref{f7_NumberCountField} displays a Z+J+H pseudo-color composite of the study field. The two zoomed regions give an impression of the achievable image quality with OMEGA2000 under good observing conditions. The distortion-free camera optics in combination with the applied reduction schemes yield a homogeneously high image quality over the full FoV, which is demonstrated by the recovery of the subtle gravitational arc at the field edge in the lower right panel. 

The observing strategy of Sect.\,\ref{s7_observational_strategy} aimed for a maximum net exposure time of 50\,min per filter and field, in particular for the Z--H approach. The A\,383 field was covered  in all four bands to approximately this exposure time limit and is thus well suited for a cross-comparison. The data includes observations under photometric conditions for 58\,minutes in Z, 50\,minutes in J, and 50\,minutes in the Ks-band. The 60\,minute coverage of the H-band has been obtained under non-photometric conditions with partial cloud coverage, hence it provides only lower limits on the achievable depth.
We follow Harris' \cite*{Harris1990a}  definition of a limiting magnitude as the magnitude at which the completeness of detected sources drops to 50\% (50\%-completeness limit). This limit can be robustly estimated from the differential number counts
and corresponds approximately to a 5\,$\sigma$ detection significance of the sources.

\begin{figure}[h]
\centering
\includegraphics[angle=0,clip,width=\textwidth]{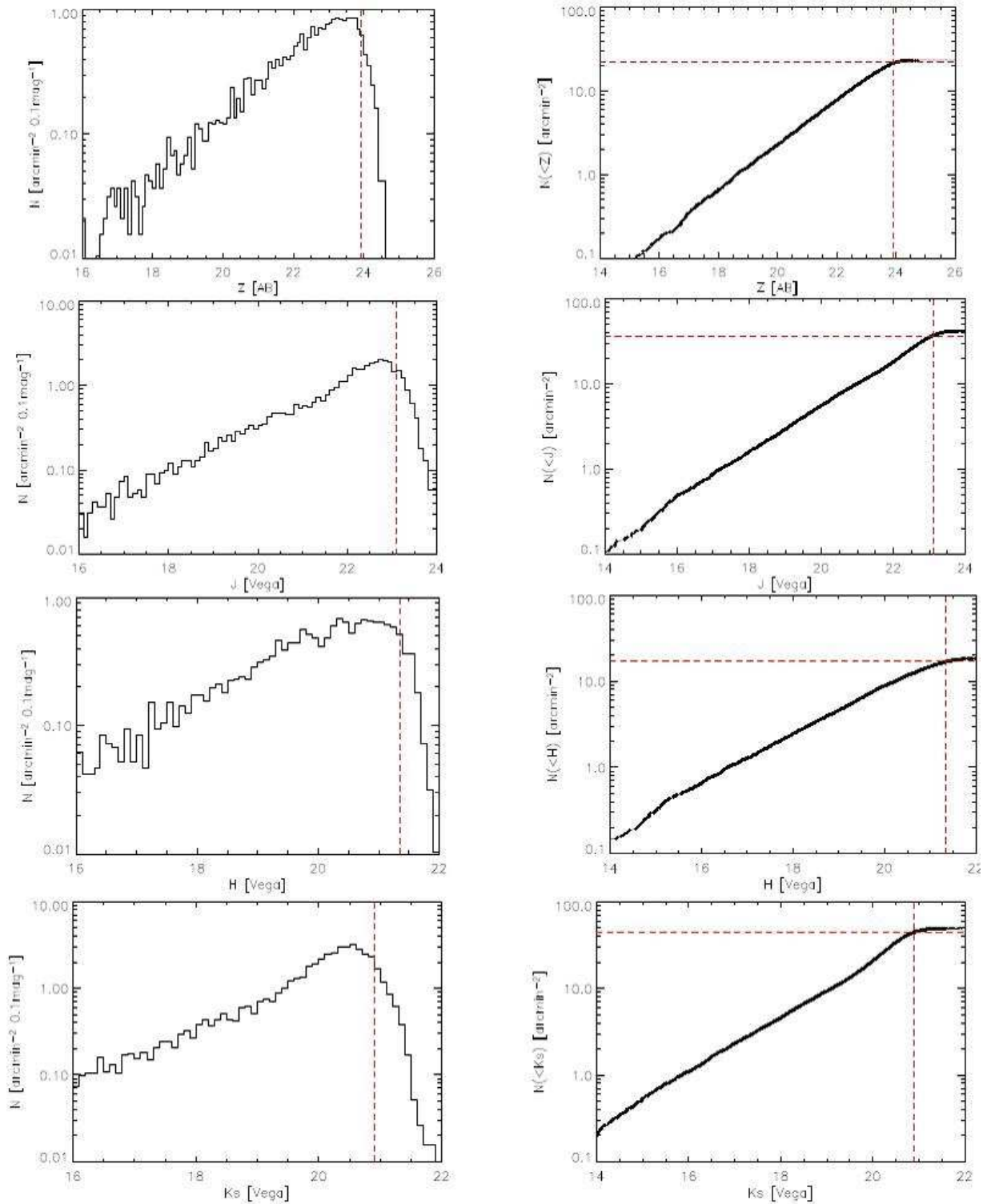}
\vspace{-2ex}
\caption[NIR Number Counts]{Z, J, H, and Ks number counts of the field displayed in Fig.\,\ref{f7_NumberCountField}. Z (58\,min), J 
(50\,min), and Ks (50\,min) were observed in photometric 1\arcsec-seeing  conditions, the H-band (60\,min) observations were not photometric and suffered from partial cloud coverage, \ie \ the limit is a lower bound. The {\em left panels} show the differential number counts in 0.1 magnitude bins. The 50\%-completeness limit is given by the dashed vertical line. The $\log N$--$\log S$ distributions are displayed in the {\em right panels}, with the horizontal dashed lines indicating the average total object density at the limiting magnitude. }
\label{f7_NumberCounts}       
\end{figure}

The cumulative and differential number counts as derived for the central 192 square arcminutes (13\farcm 8$\times$13\farcm 8) of the FoV are shown for all four filter bands in Fig.\,\ref{f7_NumberCounts}.  From the differential number counts in 0.1 magnitude bins in the left panels, the 50\%-completeness limits are estimated as indicated by the dashed vertical lines. These limiting magnitudes are also shown for the cumulative number counts in the right panels.
The normalization to square arcminute units directly  provides the surface densities at a given magnitude limit 
given by   the dashed horizontal lines for the derived limiting magnitude in each band.  


The 50\%-completeness limit for the {\bf Z-band} in the top panels of Fig.\,\ref{f7_NumberCounts}  is determined as 
Z$_{\textrm{lim}}(\textrm{AB})\!\simeq\!23.9$  corresponding to   Z$_{\textrm{lim}}(\textrm{Vega})\!\simeq\!23.4$. This limit confirms the expected high sensitivity of the NIR detector array in this last optical band.
Comparing to Tab.\,\ref{t7_magnitude_evolution} yields expectation values to what extent the Z-band galaxy luminosity function of a cluster will be probed at this maximum depth with respect to the characteristic magnitude m*. At $z\!=\!1$ this corresponds to m*+2, at $z\!=\!1.5$ to m*+0.1, and at $z\!=\!2$ to m*$-$0.8, revealing $z\!\sim\!1.5$ as the redshift where the maximum Z-band depth falls below the characteristic magnitude. As a related consequence it can be expected that a possible  \reds \ at  $z\!\ga\!1.5$ will be observable only with large scatter due to the increased photometric uncertainties in the Z-band.  

The average background surface source density  at the limiting Z-magnitude is 22\,arcmin$^{-2}$. At these densities even a {\it bona fide\/} cluster at intermediate redshifts as  RX\,J0018.2+1617 in Fig.\,\ref{f7_WeightSchemes} reaches merely an overdensity of 40\% above the background in the cluster center. For a 4\arcmin$\times$4\arcmin \ region centered on the cluster, the overdensity drops to zero and the cluster galaxies completely dissolve in the background of roughly 350 objects within this sky area. This example illustrates the importance of (i) accurate positional information (5\arcsec--10\arcsec) as provided by the X-ray selection of candidates, and (ii) reliable color information in order to enhance the density contrast and allow a cluster identification.    

The number counts in the {\bf H-band} are shown in the third row of Fig.\,\ref{f7_NumberCounts}. The 50\%-completeness limit of 
H$_{\textrm{lim}}(\textrm{Vega})\!\simeq\!21.4$ is routinely achieved in a good fraction of the XDCP fields even with shortened exposure times but better observing conditions concerning the transparency. This limiting magnitude is approximately 1\,000 times fainter than the sky surface brightness. The H-band surface density of sources at this limit is 17\,arcmin$^{-2}$ and is thus about 20\% lower than in the Z-band. At the same time the redshift grasp is greatly enhanced yielding a LF coverage to m*+2.3 at $z\!=\!1$ , m*+1.5  at $z\!=\!1.5$, and  m*+0.9 at $z\!=\!2$. This implies that the fraction of high-redshift objects is significantly increased in the H-band due to the very moderate redshift dimming of elliptical galaxies and the efficient relative suppression of blue foreground sources at these NIR wavelength. The second important conclusion is that the cluster luminosity function can be sufficiently probed out to redshifts of $z\!\sim\!2$. For the Z--H method this yields the performance expectations that a sufficient number of cluster galaxies will be detectable in the H-band even at  $z\!\ga\!1.5$, but that the quality of the color information is decreased due to the photometric uncertainties of the Z-band magnitudes.

The situations in the {\bf J-band} (second row) and {\bf Ks-band} (bottom row) are similar to H and can both probe the cluster luminosity function to about m*+1 at $z\!=\!2$. The deep J-band limit of J$_{\textrm{lim}}(\textrm{Vega})\!\simeq\!23.1$ exhibits a surface density of $\ga$30\,arcmin$^{-2}$. The Ks observations suffered from dome-scattered moonlight in a good  fraction of the FoV which manifests itself in the suspicious upturn\footnote{The highest positive spatial fluctuations of the additional scattered light surpass the detection threshold and are falsely interpreted as faint objects.} of the number counts at Ks$\ga$20. The counts are thus only trustworthy up to a number density of $\sim$20\,arcmin$^{-2}$ at Ks$\sim$20. A realistic corrected limiting Ks magnitude is approximately Ks$_{\textrm{lim}}(\textrm{Vega})\!\simeq\!20.5$. In any case, the J--Ks method is capable of providing accurate color information out to the highest redshifts and is thus very valuable in conjunction with the Z--H strategy for the identification of the most distant clusters to be found.     

In summary, we have shown that the achievable data quality and reduction performance fulfills the depth requirements and expectations of Sects.\,\ref{s7_imaging_strategy} \& \ref{s7_observational_strategy}. The applied imaging strategy
with maximal net exposure times of  $\sim$50\,min per filter and field  is sufficiently deep to provide stand-alone Z--H colors out to $z\!\sim\!1.5$ and has the potential to go significantly beyond this limit with complementary J and Ks observations. The expected surface density of background sources of typically $\sim$20\,arcmin$^{-2}$ in deep XDCP fields emphasizes the need for accurate color information. The procedures for obtaining the final photometry, object catalogs, and pseudo color images from the reduced data will be introduced in the next section.



\begin{figure}[h]
\begin{center}
\includegraphics[angle=0,clip,width=0.95\textwidth]{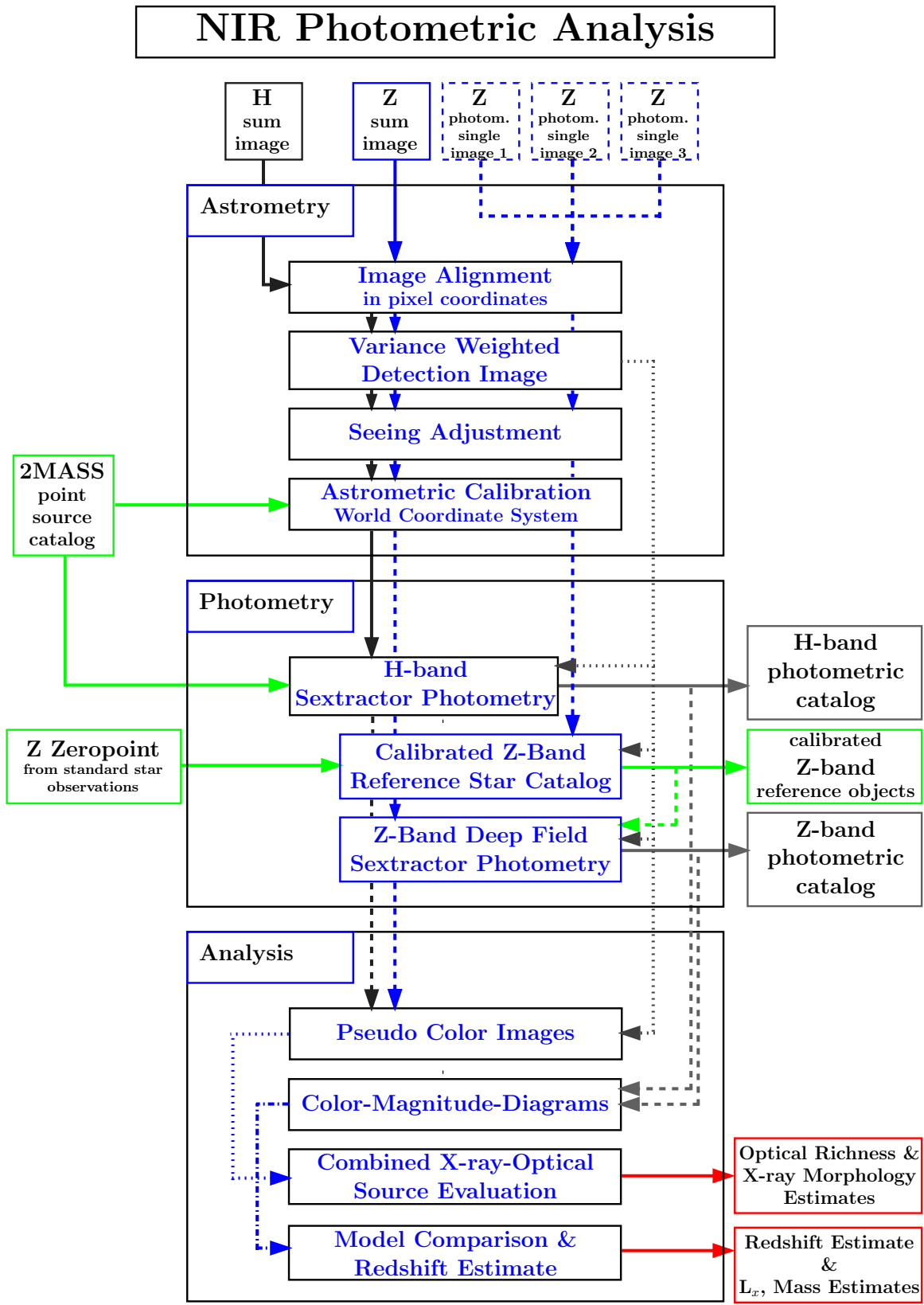}
\end{center}
\caption[NIR Photometric Analysis]{Near-infrared photometric analysis flow chart.} \label{f7_NIR_photometry}
\end{figure}

\clearpage

\section{Photometric Analysis}
\label{s7_photom_analysis}

\noindent
With the final optimized data products from the NIR reduction process in hand, we can now turn to the photometric analysis. The general objective will be the measurement of the cluster \reds \ location with an absolute color accuracy of $\sigma_{\mathrm{color}}\!\sim\!0.1$\,mag in order to achieve the redshift accuracies ($\sigma_z\!\sim\!0.1$) as discussed in Fig.\,\ref{f7_redshift_uncertainties}.

\subsection{General procedure}

\noindent
The generic photometric analysis procedure as implemented for the XDCP follow-up program is illustrated in the flow chart of Fig.\,\ref{f7_NIR_photometry}. The analysis process from the image level to final redshift estimates can be subdivided into three main modules: (i) The {\em astrometry part} during which images are aligned, combined, seeing adjusted, and finally transformed to sky coordinates, (ii) the  {\em photometry module} dealing with object detection, characterization, and calibration for all filter bands, and finally (iii) the  {\em analysis and evaluation section} based on composite color images, color-magnitude diagrams, and comparison to model predictions.  The discussion will focus on the constituents of the Z--H method, however, the  procedures for complementary J or Ks-band data are analogous to the H-band analysis.

\subsubsection{Astrometry}
\label{s7_astrometry}

\noindent
%
Accurate image alignment in pixel coordinates is a pre-requisite for the source extraction software {\em SExtractor} \cite{Bertin1996a} and is performed as a first step. Input images for the alignment procedure are the final deep summed images in the H-band and Z-band, plus additionally three single Z-band images acquired under photometric conditions for the calibration process. In practice, the H image is used as master grid, and all Z-band images are integer-pixel\footnote{Fractional pixel offsets cannot be applied in this case since the Z-band images are only aligned to the master grid, but not co-added.} aligned in sky coordinates via object matching and saved with the new coordinate grid. Since this procedure is similar to the summation process in Sect.\,\ref{s7_NIR_reduction_steps}, the NIR pipeline can be tuned to provide the accurate automatic object alignment in conjunction with a meta script organizing the proper image handling.   

During the same process a combined deep detection image DET$_{\mathrm{ZH}}$ is created from the variance-weighted sum of the Z and H-band frames as DET$_{\mathrm{ZH}} = \mathrm{Z}\cdot \sigma_{\mathrm{Z}}^{-2} + \mathrm{H}\cdot \sigma_{\mathrm{H}}^{-2}$, where $\sigma_{\mathrm{Z}}$ and $\sigma_{\mathrm{H}}$ are the measured {\em global\/} background  standard deviations in the frames. This detection frame serves the three main objectives (i) to maximize the depth and the number of detectable objects, 
 (ii)   to provide averaged object profiles for an unbiased\footnote{Due to possible positional truncation errors during the alignment, detecting sources in H and measuring the magnitude in Z bears the risk of a systematic bias towards fainter Z magnitudes.} magnitude and color determination, and (iii) to be used in the green (neutral) channel for color-composite images.  
This detection frame will be handed over to {\em SExtractor} for finding sources and determining their source apertures, while the actual measurement is performed in the individual H or Z image.
The variance-weighted image combination makes direct use of the deep frames as provided by the NIR pipeline in order to achieve a maximized detection depth, even though the frames might still exhibit different seeing levels. Since there will be no color information without an initial detection of a source, the source detection depth is given priority over the color accuracy. 

A fixed measurement aperture applied to images observed under discrepant seeing conditions probes different physical scales of the same galaxy, \eg \ a 1.5\arcsec \ aperture might contain most of the (distant) galaxy light under good conditions while under bad seeing the measured flux will be dominated by the galaxy center. Since elliptical galaxies often exhibit color gradients as a function of distance from the center, the same physical scales need to be considered for an unbiased and well-defined color measurement. The potentially discrepant seeing conditions are adjusted in the next step, where additional smoothing with a Gaussian kernel of radius $r_{\mathrm{smooth}}$ is applied to the image with better seeing $\sigma_{\mathrm{good}}$.       
Since the standard deviation of the Gaussian kernels add in quadrature, the smoothed frame will have the same effective seeing $\sigma_{\mathrm{bad}}$, measured as the full-width-at-half-maximum\footnote{The relationship between the standard deviation $\sigma$ and the full-width-at-half-maximum of a Gaussian profile is FWHM$=2.36 \cdot \sigma$.}, if the condition 
$\sigma^{2}_{\mathrm{bad}}\!=\!\sigma^{2}_{\mathrm{good}}\!+\!(2.36 \cdot r_{\mathrm{smooth}} )^{2}$
for the smoothing radius $r_{\mathrm{smooth}}$ is fulfilled. Measuring the color in these seeing adjusted images ensures a flux extraction from the same physical galaxy scale while maintaining the maximized  depth of the detection image.  




As a final preparatory step, the aligned and seeing-adjusted images are equipped with a proper equatorial world coordinate system (WCS). The transformation equations from XY image coordinates to the  right ascension ($\alpha$ or RA) and declination ($\delta$ or DEC) system  are obtained by finding the astrometric plate solution from sources in the FoV with accurately known absolute positions.    In practice this task can be performed with the {\em WCS Tools}\footnote{The free software is available at \url{http://tdc-www.harvard.edu/software/wcstools}.} software package which automatically queries the 2MASS point source catalog for reference sources, finds and matches objects, solves for the plate solution via least-square fitting, and updates the image header with the appropriate WCS keywords. Details on the exact transformations and the representation of the world coordinate system in the FITS image header can be found in Calabretta \& Greissen \cite*{Calabretta2002a}.

\subsubsection{Source detection and characterization}
\label{s7_NIR_source_detection}

\noindent
We can now turn to the actual objective of the imaging enterprise, the detection and photometry of astronomical sources with the focus on faint distant galaxies. This central task is performed with the help of the widely used source extraction and characterization software {\em SExtractor}. This versatile photometry code 
detects, de-blends, measures and classifies sources from astronomical images and saves the results in a photometric catalog for further processing.

For the science objectives of deep imaging and accurate colors we make use of {\em SExtractor's dual image mode}. In this configuration, the  object search and structural characterization takes place in the depth-optimized variance-weighted detection image, from which the final source list and the individual object extraction regions are determined. 
The actual   photometry measurements, \ie \ the final source flux determinations, are subsequently performed in the aligned individual filter band images, Z or H, using the exact object position and extraction region information of the input source list. This technique ensures that all objects in all filters have the same photometry areas and apertures  appropriate for measuring colors and eases the cross-identification of objects in different catalogs.
The pre-requisites for this method, aligned images with the same size, have been fulfilled in the astrometry module. The last requirement of coincident sky areas of the different input images is met by providing external variance maps to  {\em SExtractor} which exclude the non-overlapping dither and pointing offsets of the images from the photometric analysis. 
For the overlapping region of interest, the variance maps contain estimates of the local noise levels for the individual images. These noise maps are obtained from an empirical variance measurement in a central statistics window divided by the smoothed flatfield to account for variations due to large-scale changes of the detector quantum efficiency. The detection and measurement images, variance maps, the photometric zero point, seeing and saturation level estimates, and other data characteristics are then pipelined to {\em SExtractor} for the photometry. 



In short, {\em SExtractor} performs the following steps: (i) estimate the background and its local noise properties, (ii) determine whether pixels belong to background or objects, (iii) split up object areas into separate objects, and (iv) determine the properties of each object. A critical component for the detection performance is again the quality of the background estimation, which should be determined from a fine enough grid to account for large-scale variations  and at the same time robust enough to avoid local biasing from nearby bright objects. The second crucial set of parameters is related to the threshold for an object detection. As minimal object significance, the implemented threshold requires at least four connected pixels with values higher than one standard deviation above the background noise, which implies a 2-$\sigma$ detection for the limiting case.     
Since the main scientific interest focusses on the spatial and color distribution of cluster galaxy populations rather than individual faint galaxies, the expected spurious source rate on the percent level is an acceptable trade-off for a deeper coverage. The second shortcoming of the implemented approach is a less accurate determination of the structural object parameters\footnote{The expected median image alignment offset of about 0.1\arcsec \ between the H and Z frames introduces a slight object `smearing', which decreases the quality of the stellarity index measurement, \ie \ the point source identification.}  from the combined, and thus averaged, detection image, which is again outweighed by the optimized depth and color information. 
Finally, the photometric precision and error estimates are compromised somewhat by performing the measurements in the deep sum images, instead of the individual single images, which would allow the object properties to be determined from an ensemble of measurements. The latter  approach enables 
\eg \ the elimination of cosmic ray events inside the object apertures, but the expected marginal gain does not justify the significant extra effort for red-sequence-based methods.    





\enlargethispage*{4ex}

The last consideration concerning {\em SExtractor} photometry deals with relating the measurements to the galaxy model predictions. The models specify total apparent object magnitudes for the different filters, which are best matched by the 
{\tt MAG\,AUTO\/} magnitude, for which {\em SExtractor} applies automatic flexible elliptical apertures to each object in order to estimate the total flux.  
For an  accurate object color determination, however, the use of a smaller aperture containing only pixels with a sufficiently high signal-to-noise ratio is advantageous over  measurements including the low SNR object periphery. This can be achieved in practice through {\tt MAG\,ISO\/} magnitudes for the Z--H color determination. These isophotal magnitudes are measured from all connected object pixels above the detection threshold. 



 






\subsubsection{Analysis and redshift estimation}
\label{s7_analysis_redshifts}

\noindent
The analysis and candidate evaluation represents the last module of the NIR imaging procedure. For this all available information on the cluster candidate are combined into final data products.  

For the visual cluster identification, a pseudo color composite image is created from all filter bands. In the common case with only Z and H-band data available, the detection image is used for the third color channel. Color images in general are obtained by overlaying a red, a green, and a blue data channel (RGB colors). Applied to our case, the Z-data is loaded into the blue channel, the detection image into the (neutral) green channel, and the H-band is represented in red. The combined imaging data can be complemented by overlaying the matching X-ray contours and the extracted photometric information for each object.   

As the second important diagnostic tool, color-magnitude diagrams are created from the photometric Z and H-band catalogs for the quantitative analysis. The pre-matched object catalogs are read, the Z-magnitudes transformed to the Vega system, and the Z--H isophotal object colors determined. These are then plotted against the total H-magnitude for all objects, and sources within 30\arcsec \ and 1\arcmin-apertures around the centers of cluster candidates are highlighted. As an additional diagnostic the object overdensity with respect to the average background density is computed for the given apertures. No special efforts are taken to separate galaxies from stellar sources since (i) stars are typically easily recognizable in the color composites and can be appropriately taken into account if located close to a cluster, (ii) {\em SExtractor's} stellarity classification is less reliable when applied to the variance-weighted detection image, and (iii) the automatic classification breaks down at faint magnitudes anyway.   


If a \reds \ is identified for a cluster candidate, the best-fit color estimate can be directly transformed into a redshift estimate based on a formation redshift $z_{\mathrm{f}}\!=\!5$ model with solar metallicity (see Sect.\,\ref{s10_formation_epoch}) as shown in the lower right panel of Fig.\,\ref{f7_CMD_models}. Since the follow-up photometric accuracy is typically not sufficient to trace the \reds \ slope for high-redshift clusters, one has to aim at an average \reds \ color which will be naturally hinged at $m$* galaxies as a result of the achievable depth limit. 

Combining the cluster redshift estimate with the X-ray flux measurement, an approximate X-ray luminosity $L_X$ and a preliminary mass proxy can be obtained based on the $L_X$--$M$ scaling relation.   
The relative galaxy overdensity and \reds \ population additionally yield a first clue on the optical richness of the cluster, and the features in the X-ray morphology might hint at the dynamical state of the system. For an extended cluster candidate evaluation, the following cluster features can be considered: (i) the presence of a central dominant galaxy (see Sect.\,\ref{s10_bcg_assembly}), (ii) a systematic BCG offset from the X-ray center (see Fig.\,\ref{f10_bcg_clustercentric_dist}), (iii) a possible AGN contribution to the X-ray emission, or (iv) the potential coincidence of galaxy and X-ray features such as filamentary extensions (see Sect.\,\ref{s9_cl15_analysis}).   

We have finally arrived at the initial objective of the two-band follow-up imaging strategy. As survey {\bf step 3}, the presence of a cluster can be confirmed based on the observed overdensity of red galaxies, and for  {\bf step 4} a redshift estimate is obtained from the color of the red-sequence. Promising high-redshift candidates are now ready for spectroscopic confirmation as discussed in Chap.\,\ref{c8_SpecAnalysis}.




\subsection{Photometric calibration}
\label{s7_NIR_calibration}

\noindent
We will now come back and discuss the photometric calibration procedure, which was left out in the last section. 
Due to the different nature of the optical Z-band and the NIR H-filter, the calibration methods differ in both band-passes and will be introduced separately.

\subsubsection{2MASS NIR calibration}

\noindent
The 2-Micron All-Sky Survey (2MASS) \cite{Cutri2003a} provides full sky coverage in the J, H, and Ks-band to a minimum depth\footnote{See \url{http://www.ipac.caltech.edu/2mass/releases/allsky/doc/sec2_2b.html}. Minimum depth means that this value is achieved for the full survey area, while parts of the sky have a  deeper coverage.}  of J$_\mathrm{min}$(Vega)=15.9, H$_\mathrm{min}$(Vega)=15.0, and Ks$_\mathrm{min}$(Vega)=14.3 for 10-$\sigma$ point source detections. 
The availability of this high quality data set allows a direct and easy cross-calibration to the 2MASS photometric systems via serendipitous  point sources of sufficient brightness in OMEGA2000's large FoV. 



Since the 2MASS calibration stars with known calibrated magnitudes are in the science data and are only extracted after the observations, they have experienced the same atmospheric extinction as all sources in the field. This implies that airmass corrections as for the Z-band are not needed, and that a calibration is even possible under non-photometric observing conditions. Homogenous extinction across the 15\arcmin-FoV is the only assumption that is made for the case of low or varying transparency, \eg \ due to moderate cloud coverage. This assumption is justified in most cases and can be cross-checked by testing for systematic photometric offsets across the FoV. Hence, the H-band calibration has low  requirements on the conditions and allows continuing observations even under challenging circumstances.    


In practice, the calibration process is achieved through the following steps. (i) The raw photometric H-band catalog\footnote{2MASS data products can be downloaded from \url{http://irsa.ipac.caltech.edu/applications/Gator/}.} of the science field is matched to the 2MASS catalog of the bright (H$<$14.5\,mag) point sources in the FoV as was obtained during the astrometric calibration. (ii) The zero point offset of each source is determined by subtracting the instrumental  4\arcsec -aperture magnitudes from the calibrated H-band object magnitudes in the reference catalog. (iii) The average effective zero point (ZP) for the observation and its standard deviation is computed via iterated outlier clipping of all calibration sources. (iv) This ZP is provided to {\em SExtractor} as an additive offset for the second processing which will yield calibrated magnitudes.  

The calibration process for a sample field is illustrated in the upper right panel of Fig.\,\ref{f7_NIR_calibration} which shows the distribution of the individual calibration source zero points. The typically 20--80 serendipitious bright 2MASS stars per field provide a zero point accuracy of approximately $\sigma_{\mathrm{ZP}}\!\simeq$0.02--0.04\,mag. The upper right panel of 
Fig.\,\ref{f7_NIR_calibration} demonstrates that the determined zero points do not exhibit a systematic trend with the object color, \ie \ a color term correction is not necessary.





\subsubsection{Z-band calibration}

\noindent
Unfortunately, a 2MASS-like all-sky survey in the Z-band does not exist yet resulting in a photometric calibration process for this filter which requires more effort. 
The Sloan Digital Sky Survey (SDSS), with 10\,000 square degree multi-band imaging coverage in the Northern hemisphere, overlaps with about 15--20\% of the XDCP survey fields and thus only allows calibration cross-checks but no substitute for designated standard star observations. 

In order to achieve an accurate calibration to the SDSS photometric system, several Z-band standards from the Smith \etal \ catalog \cite*{Smith2002a} were observed at different airmasses over the course of each photometric night. Since a good fraction of the Z-band science observations were performed in non-photometric conditions, we obtained additional short image series for these fields during the photometric nights resulting in a minimum of five available frames per field for the calibration process. This photometric overlap subsequently allows `bootstrapping' of the calibration for the deep science observation as will be discussed.


Since no experience with the OMEGA2000 Z-band calibration existed before the XDCP program, all necessary calibration information had to be derived from the data.
The following photometric transformation equation is considered

\begin{equation}\label{e7_phot_trafos}
    Z_{\mathrm{cal}} = Z_{\mathrm{instr}} + A_0 - A_1 \cdot X + A_2 \cdot (R\!-\!Z) \ ,
\end{equation}

\noindent 
where $A_0$ is the nightly photometric zero point\footnote{The instrumental magnitude is given by 
$m_{\mathrm{instr}}\!\equiv \!-2.5\log(\mathrm{counts \ s^{-1}})$. Hence, the zero point is in effect the apparent magnitude of a source for which the instrument-telescope system detected one count per second.}, 
 $A_1$ is the Z-band atmospheric extinction coefficient per unit airmass, and $A_2$ is a potential color term that could be introduced by a different filter-detector response relative to the SDSS system. The extinction term $-A_1\!\cdot\!X$ corrects the measured magnitudes to the original flux values above the atmosphere, \ie \ the lost flux due to atmospheric extinction is added.



\begin{figure}[b]
\begin{center}
\includegraphics[angle=0,clip,width=0.49\textwidth]{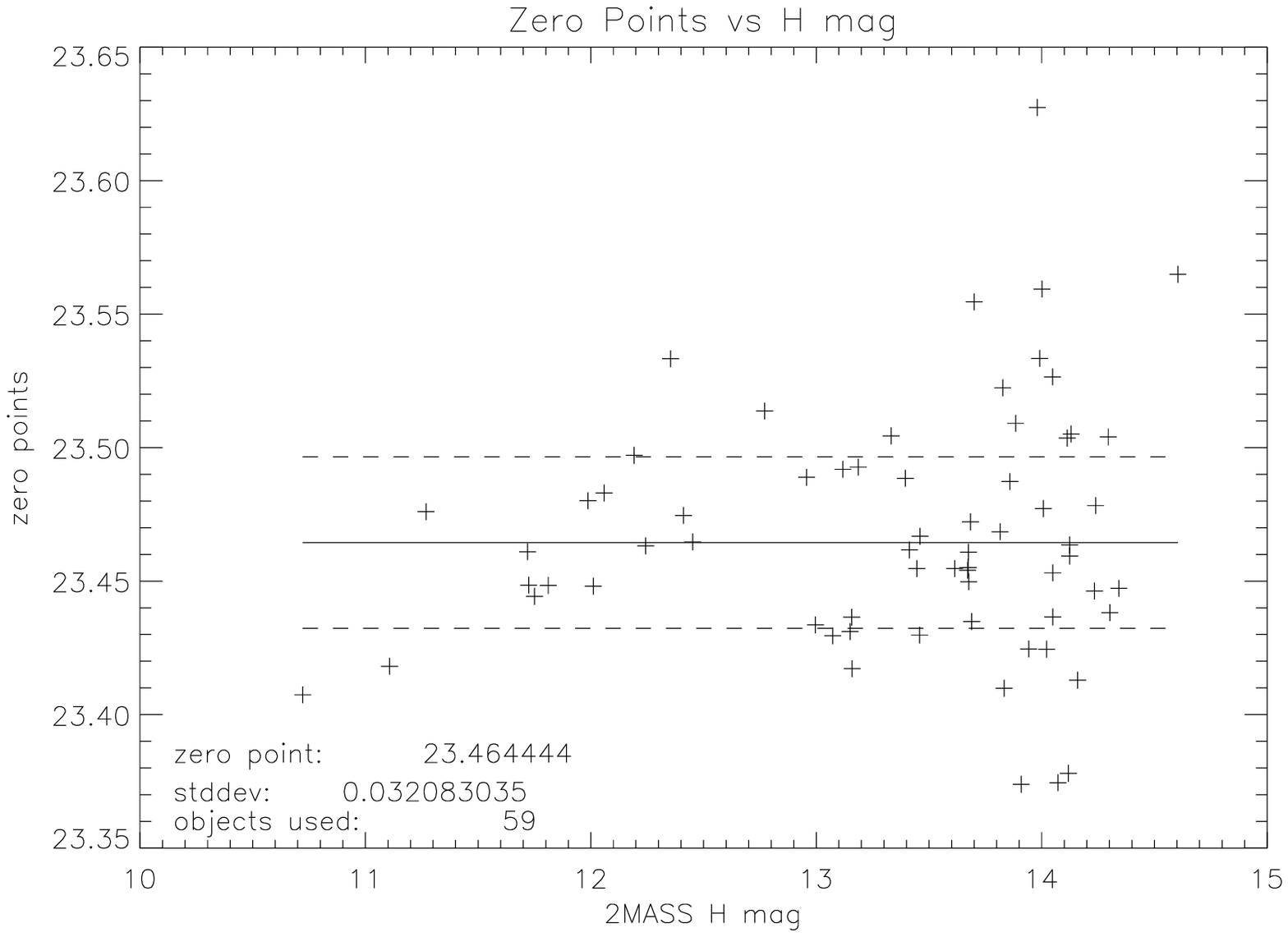}
\includegraphics[angle=0,clip,width=0.49\textwidth]{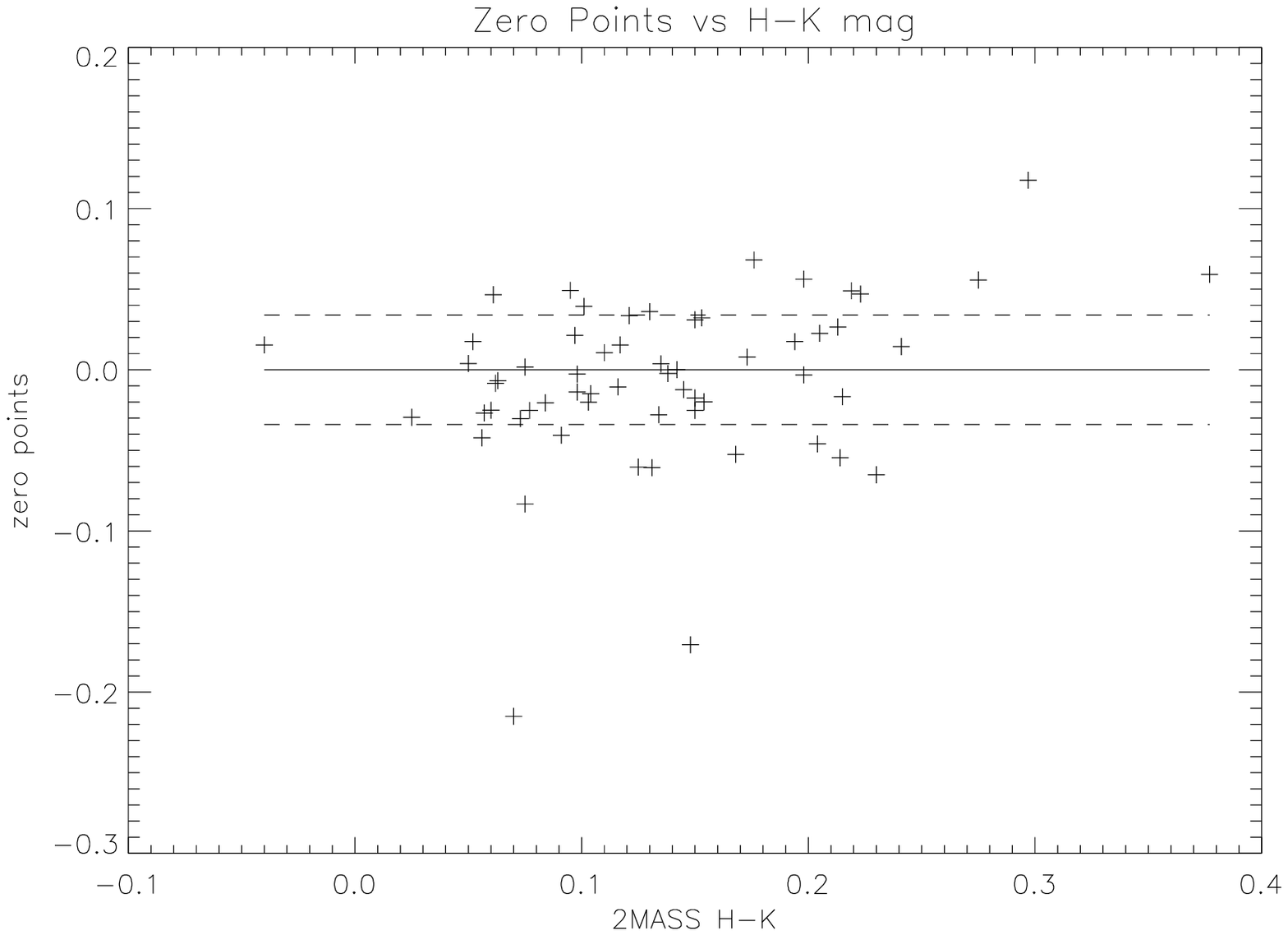}
\includegraphics[angle=0,clip,width=0.49\textwidth]{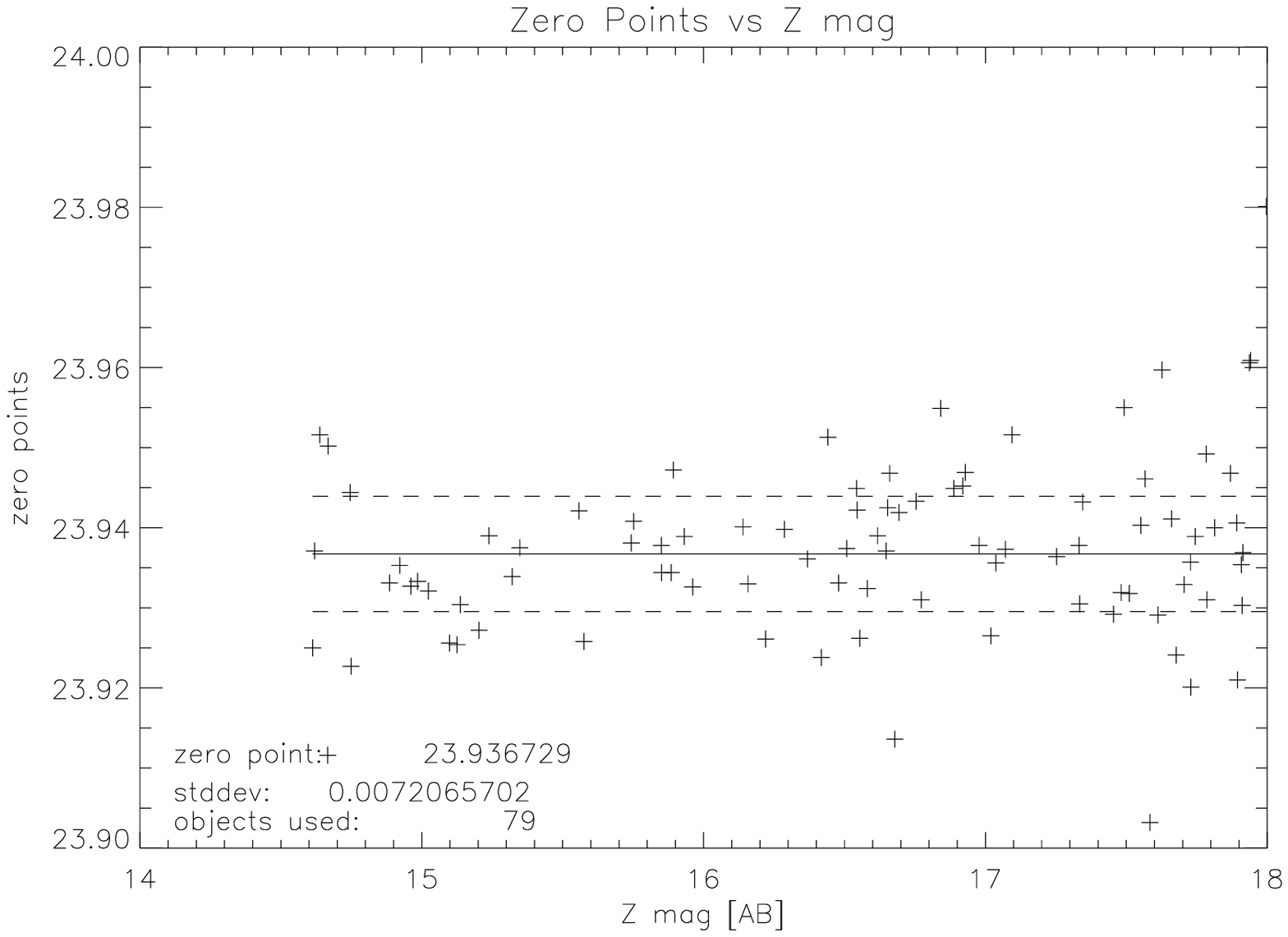}
\includegraphics[angle=0,clip,width=0.49\textwidth]{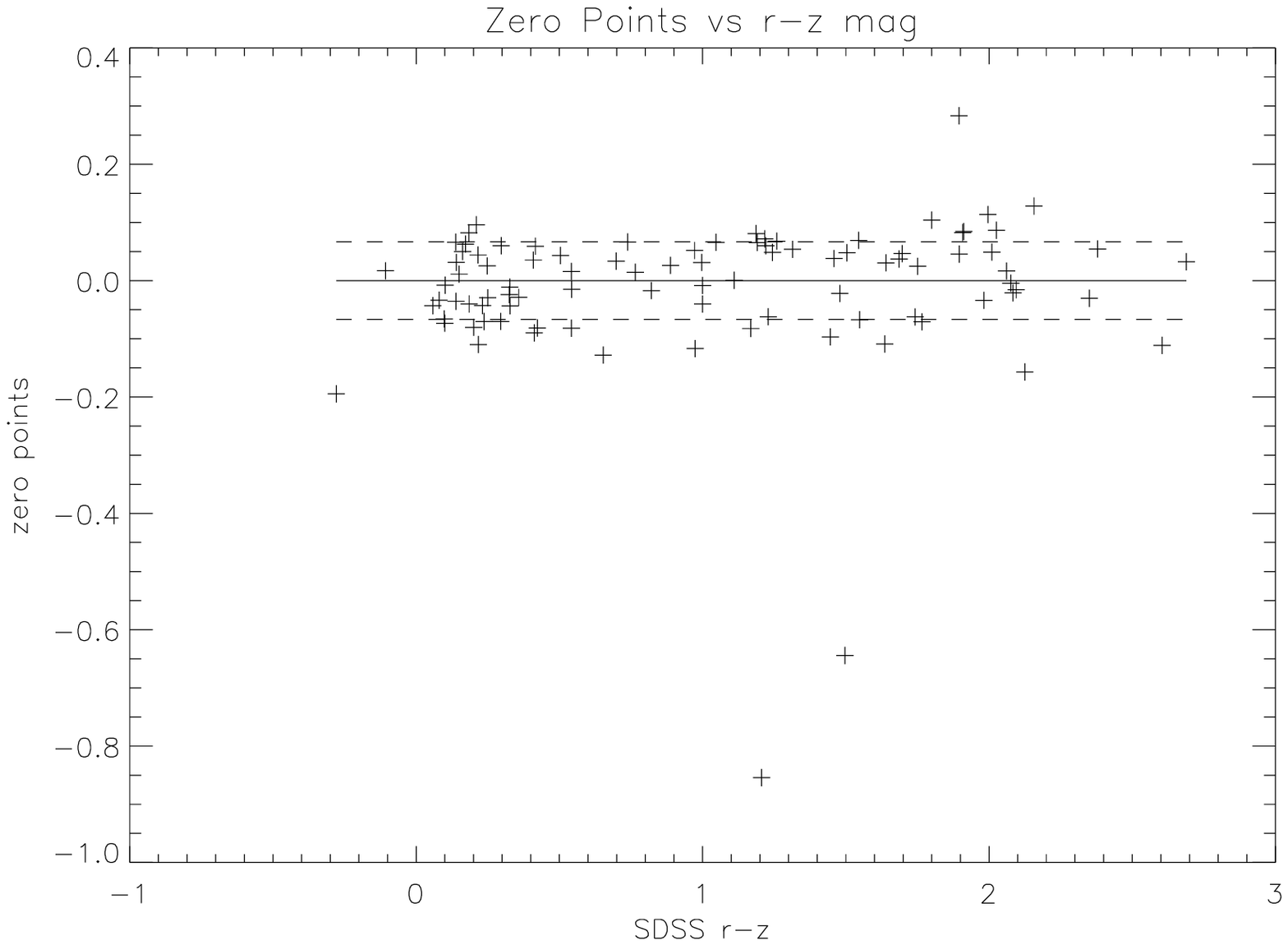}
\end{center}
\vspace{-2ex}
\caption[Photometric Calibration]{Photometric calibration. {\em Top row:} Photometric zero point calibration in the H-band using bright 2MASS reference point sources in the field-of-view ({\em left}). The large FoV usually contains several dozen suitable calibration stars for any pointing and allows a zero point calibration to typically $\sigma_{\mathrm{H}}\!\sim\!0.03$\,mag. The increased scatter towards the faint end  reflects the photometric uncertainties of the shallower 2MASS survey and is the reason for only selecting calibration reference sources with H$<$14.5. 
The right panel shows the zero point offset from the mean as a function of the object H--Ks color  and illustrates that a color term correction with respect to the reference 2MASS system is negligible. {\em Bottom row:} Z-band calibration to the SDSS photometric system. Since most XDCP fields do not have direct SDSS coverage, the nightly zero points have to be obtained from Z-band standard star observations, with which a calibrated reference star catalog can be obtained from short photometric exposures of the science field. The catalog is then used to calibrate the zero point in the deep Z-band field in the same manner as for the H-band ({\em left panel}). The color term check ({\em right panel}) also yields a correction term that is close to zero.} \label{f7_NIR_calibration}
\end{figure}

\clearpage

The available standard star observations were not sufficient in quality and quantity to 
solve the transformation equation simultaneously for the three unknown parameters $A_0$, $A_1$, and $A_2$. 
We thus choose the following approach: (i) Determine the color term coefficient $A_2$ from the best suitable field with SDSS\footnote{SDSS data release\,5 (DR5) archive data products can be retrieved from  \url{http://cas.sdss.or./astodr5/en/tools/search/IQS.asp}.} coverage. (ii) Measure the extinction coefficient $A_1$ with fixed color term  $A_2$ for the photometric night with best standard star coverage and use this value for all nights. (iii) Compute the photometric zero point $A_0$ for each photometric night using the pre-determined values for $A_1$ and $A_2$. 

The bottom right panel of Fig.\,\ref{f7_NIR_calibration} displays the zero point offsets as a function of (R--Z)-color for a SDSS calibration field with about 100 reference stars. The color coefficient $A_2$ is obtained from the slope of the linear regression analysis yielding $A_2~=~0.014~\pm~0.01$ and is thus consistent with zero within the errors. This formal result is good news confirming {\em a posteriori\/} that the Z-band NIR instrument response is very similar to the optical SDSS system. The small color term correction on the percent level thus eases the direct comparison of the OMEGA2000 Z-band magnitudes to SDSS reference values.  


With the color term fixed at the above value, the extinction coefficient $A_1$ is determined from data of the highest quality night (10 November 2006). The instrumental magnitudes $m_{\mathrm{instr}}$ for the observed standard stars are plotted against the airmass under which the data was taken. The average slopes of these relations for each standard star then yield the Z-band atmospheric extinction coefficient  $A_1\!=\!0.045 \pm 0.01$ per unit airmass. This result is consistent with the values obtained at other observatories and implies that the instrumental Z-band magnitudes are only dimmed by $\sim$\,0.05\,mag for observations at airmass two compared to zenith pointings. This moderate extinction, which continues to decrease in the NIR bands, is a manifestation of the atmospheric Rayleigh scattering with cross-section $\sigma_{\mathrm{R}}\!\propto\!\lambda^{-4}$.  The small extinction correction of $A_1\!\cdot\!X\!\sim\!0.05$--0.1 also justifies keeping the coefficient $A_1$ at a fixed value for all nights.


The last unknown parameter for the photometric transformation of Equ.\,\ref{e7_phot_trafos} is the nightly zero point $A_0$. The zero point is obtained by applying Equ.\,\ref{e7_phot_trafos} to all standard star observations of a given night and using the above values for $A_1$ and $A_2$.
Out of the seven photometric nights used for the Z-band calibration, estimated zero point  errors of $\sigma_{\mathrm{ZP}}\!\sim$\,0.02--0.03\,mags were  achieved for five nights; two nights of service observations turned out to be non-photometric with estimated zero point uncertainties of $\sigma_{\mathrm{ZP}}\!\sim$\,0.2\,mags.


We now have all the tools at hand to perform the photometric `bootstrap' calibration of the deep Z-band science fields as  illustrated in Fig.\,\ref{f7_NIR_photometry}. The approach uses the following steps: (i) The astrometric preparatory module is applied to three single photometric snap-shot observations in the same way as the deep Z-band science image, \ie \ the images are aligned and seeing adjusted. (ii) {\em SExtractor} source detection is performed on these calibration images in  dual image mode with the deep detection image as reference. (iii) The magnitudes\footnote{The SDSS {\em model magnitudes} for  $m_{\mathrm{cal}}$ of the calibration stars   are matched to the measured total {\tt MAG\,AUTO\/} magnitudes for $m_{\mathrm{instr}}$.} of the three resulting object catalogs are photometrically calibrated using Equ.\,\ref{e7_phot_trafos} with the appropriate nightly zero point. (iv) Bright but unsaturated stellar sources with $14.5\!\la\!\mathrm{Z}\!\la\!18$ are selected. The stars with a maximum magnitude discrepancy of 0.02\,mag in all three catalogs are chosen as reference objects and saved in a Z-calibration catalog. (v) This reference catalog with calibrated object Z-magnitudes in the science field is now equivalent to the 2MASS input catalog implying that the same calibration procedure can be applied.



In effect, we have used the standard star observations in combination with the photometric snap-shots to bootstrap a calibrated reference catalog for each field. The discussed NIR calibration procedure is then applied as illustrated in the left bottom panel of Fig.\,\ref{f7_NIR_calibration}. The nightly zero point uncertainties of $A_0$ add in quadrature to the scatter of the reference objects resulting in typical achievable Z-band calibration uncertainties of $\sigma_{\mathrm{ZP}}\!\simeq$0.03--0.05\,mag.

\section{Summary}
\label{s7_NIR_summary}

\noindent
After this long and comprehensive chapter on near-infrared imaging, the main achievements within this thesis work are now summarized: 


\begin{itemize}
    \item Development of a new NIR-based Z--H follow-up imaging strategy;
    \item PI of a two-semester 10-night imaging proposal to Calar Alto and award of both observing runs with class A priority;
    \item Led observations at the Calar Alto 3.5\,m telescope with 24 nights of on-site presence;
    \item Co-led a three-night NIR imaging run at the CTIO 4\,m Blanco telescope which complemented  an optical follow-up program with ten nights of  on-site presence (results not included in thesis);  
    \item Development of a versatile NIR science-grade reduction pipeline;
    \item Development of science analysis tools  for \reds \ applications;
    \item Reduction and analysis of the data from all NIR imaging runs.
\end{itemize}





\chapter{Spectroscopic Follow-up}
\label{c8_SpecAnalysis}
\noindent
The spectroscopic confirmation of distant cluster candidates is {\bf step~5} of the survey strategy discussed in Sect.\,\ref{s5_survey_strategy} and constitutes 
the observational bottleneck for cluster studies at high redshift. At \zga1, spectroscopy of cluster galaxies is currently only feasible with very few instruments at 8--10\,m-class telescopes, and at $z\!\ga\!1.5$ any study is extremely challenging with the  available technology. 

The final objectives of the spectroscopic analysis in survey mode are: (i) the confirmation of the cluster as a three-dimensionally bound object, (ii) a precise determination of its redshift, (iii) identification of cluster member galaxies, and (iv) a first estimate of the cluster velocity dispersion. A combination with the available X-ray properties of the cluster enables the derivation of the  X-ray luminosity and thus a mass proxy (see Sect.\,\ref{s2_scaling_relations}). 
These physical  quantities are crucial for linking the observations to theoretical predictions and hence for the use of clusters as cosmological probes. 
Further in-depth investigations for individual selected systems ({\bf step~6}) allow  (v) the study of the cluster dynamics, (vi) a substructure and LSS analysis, and (vii) detailed investigations of the cluster 
galaxy population such as derivations of the star formation history, age, and metallicty of individual galaxies.

Since the spectroscopic work for this thesis has been less extensive than the observational data described in the previous sections, this chapter is intended to provide a compact overview\footnote{A more detailed description of the reduction process can be found in Novak \cite*{Novak2003a}.} of the basic spectroscopic methods and applications.




\section{Basics of Spectroscopy}


Simply speaking, an optical spectrograph consists of (i) a slit aligned to the target in the focal plane, (ii) a collimator, (iii) a light dispersing element, (iv) a camera lens, and (v) a detector. Many modern spectrographs use grisms as dispersing elements, which consist of a combination of a transmission grating and a prism. The latter counteracts the redirection 
from the optical axis and thus allows compact and linear on-axis arrangements.
The principal diffraction maxima  of order $m$ at angle $\theta$ 
for light rays with an incident angle $i$ obey the 
grating equation 

\begin{equation}\label{e8_grating_equ}
m \cdot \lambda_{\mathrm{max}} = d \cdot [ \sin(i) + \sin(\theta)] \ ,
\end{equation}

\noindent
where $d$ is the grating constant (slit distance). Taking the focal length ratio  $f_{\mathrm{cam}}$ of the optical system into account, this  translates into the linear dispersion $\Delta l$ in the detector plane 

\begin{equation}\label{e8_lin_disp}
\frac{\Delta l}{\Delta \lambda} = f_{\mathrm{cam}}  \cdot \frac{m}{d \cdot \cos(\theta)} \ .
\end{equation}

\noindent
The resulting spectral resolving power $R$ of the grating is then given by the expression

\begin{equation}\label{e8_resolving_power}
R = \frac{\lambda}{\Delta \lambda} = m \cdot N ,
\end{equation}

\noindent
with $N$ being the total number of lines. 
Depending on the science application, the grisms can be optimized ({\em blazed}) to yield the maximum throughput at a given central wavelength $\lambda_{\mathrm{cen}}$ for the desired spectral resolution $R$.


The XDCP science goals aim at multi-object spectroscopy of very faint, high-redshift galaxies which requires 
the use of red-sensitive spectrographs at 8\,m-class telescopes. The XDCP instrument-of-choice is the 
{\bf FO}cal {\bf R}educer and low dispersion {\bf S}pectrograph for the Very Large Telescope, or short VLT FORS\,2 \cite{Fors2_Manual}, a first generation VLT multi-purpose instrument which has been refurbished with red-optimized CCDs 
providing an average quantum efficiency of about 65\% in the critical Z-band.
FORS\,2 offers multi-object spectroscopic capabilities with exchangeable laser-cut slit masks (MXU unit) for a simultaneous coverage of about 30 target spectra within the $6.8\arcmin\!\times\!6.8\arcmin$  field-of-view. The GRIS\_300I+21 grism with a central wavelength of $\lambda_{\mathrm{cen}}\!=\!8\,600$\,\AA \ and a peak efficiency of 70\% achieves a wavelength coverage of 6\,000--11\,000\,\AA \ with a resolution of $R\!=\!660$ or 13\,\AA \ (corresponding to four binned pixels). This setup optimizes the signal-to-noise ratio of the target continuum flux, while still resolving the major spectral object features and the sky emission lines. 




The most prominent optical spectral features of galaxies are listed in the left panel of Tab.\,\ref{t8_spec_features}. For galaxies without active nuclei, the spectra reflect the underlying dominant stellar populations 
which depend on the star formation history. As discussed in Sect.\,\ref{s2_galaxy_populations}, the centers of clusters are characterized by the highest density of early-type galaxies, with mostly `old, dead, and red' stellar populations dominated by red giants. Owing to the absence of recent star-formation, elliptical and lenticular galaxies exhibit predominantly absorption features in their spectra. The most  prominent spectral characteristic of early-type galaxies is the D4000 break around 4\,000\,\AA. This break is caused by a complex interplay of a large number of absorption lines (\eg \ Fe\,{\small I}, Mg\,{\small I}, and Balmer lines) in conjunction with the declining continuum of cold stars towards shorter wavelengths. Most notable are the strong H \& K resonance lines of single-ionized Calcium originating in the stellar atmospheres of G and K-giants. The D4000 break requires only a moderate signal-to-noise ratio for identification, shows little contamination from reddening, and does not require absolute fluxes, hence it is the most important redshift indicator for early-type galaxies.        

Spiral galaxies have experienced a more recent star formation epoch and thus have hotter stars contributing to the spectra. Their D4000 breaks are less pronounced than in ellipticals and can exhibit emission lines (\eg \ H$\alpha$, H$\beta$, [O\,{\small II}], [O\,{\small III}]). Also common in clusters are galaxies with active nuclei, \eg \  Seyfert-type galaxies, which show very strong emission lines. Sample spectra of typical low-redshift cluster galaxies with identified features have been shown in Fig.\,\ref{f4_Noras2_spectra}.   


%


\begin{table}[t]    
\begin{center}

\begin{tabular}{|c|c|c|c||c|c|c|}
\hline

\bf{Feature} & \bf{Rest Wavel. [\AA]} & \bf{Type} & \bf{E/S} & \bf{Sky Feature}  & \bf{Wavel. [\AA]} & \bf{Type} \\

\hline\hline

Mg\,{\small II}       & 2\,800  & a  &  E, S  &  O$_2$ B-band  &  6\,867--6\,919 &  a    \\  

[O\,{\small II}]    & 3\,727.0  & e  & S   & O$_2$ A-band  &  7\,607--7\,633  &  a\\
Ca\,{\small II} K   & 3\,933.6  & a  & E, S   & H$_2$O  & 9\,400--9\,9700  & a  \\
Ca\,{\small II} H   & 3\,968.5  & a  & E, S   &  [O\,{\small I}] &  5\,577.3 &  e \\
H$\delta$ & 4\,101.7  & e or a  &  S  &  Na\,{\small I} D &  5\,892.5 &  e \\
CN G-band  & 4\,300.4  & a  & E   & [O\,{\small I}]  & 6\,300.3  & e  \\
H$\gamma$ & 4\,340.5  & e or a  & S  &  [O\,{\small I}]  & 6\,363.8  & e  \\
H$\beta$  & 4\,861.3  & e or a  &  S & OH$^-$  & 6\,820--7\,000  & e \\

[O\,{\small III}]   & 4\,958.9, 5\,006.8  & e  &  S  & OH$^-$ & 7\,240--7\,580  &  e \\   
Mg\,{\small I}      & 5\,175.4 (blend)  & a  & E   &  OH$^-$ & 7\,720--8\,100  &  e \\   
Na\,{\small I} D     & 5\,892.5 (blend)  &  a &  E  &  OH$^-$ & 8\,290--8\,500  &  e \\     
H$\alpha$ & 6\,562.8  &  e &  S  &  OH$^-$  & 8\,760--9\,100  & e  \\  

[S\,{\small II}]    & 6\,717.0, 6\,731.3  & e  & S   & OH$^-$  & 9\,300--10\,200  & e  \\  

\hline
\end{tabular}

\caption[Spectral Galaxy Features]{Optical spectral galaxy and night sky features. {\em Left:} Rest wavelength for the most prominent optical absorption and emission lines for early and late-type galaxies. The line type indicates emission lines (e) or absorption lines (a), E (S) means that the spectral feature is predominantly found in elliptical (spiral) galaxies. {\em Right:} Important telluric absorption lines ({\em top}) and sky emission lines ({\em bottom}) as shown in Fig.\,\ref{f8_sky_features}. Brackets around the element name mark forbidden transition lines. For more details see Cox \cite*{allen2000}.} \label{t8_spec_features}
\end{center}
\end{table}


\section{Challenges of Distant Galaxy Spectroscopy}

\noindent
The spectral galaxy features with rest-frame wavelength $\lambda_{0}$ 
are redshifted by the cosmological expansion to the observed wavelength $\lambda_{\mathrm{obs}}$ given by

\begin{equation}\label{e8_obs_wavelength}
\lambda_{\mathrm{obs}}=(1+z) \cdot \lambda_{0} \ .
\end{equation}

\noindent
For  \zga1 objects this implies that the important D4000 break of early-type galaxies is shifted into the I and Z-band, 
limiting the number of spectral features usable for a secure redshift determination 
to a few lines with  $\lambda_{0}\!<\!4\,500$\,\AA \ (see Tab.\,\ref{t8_spec_features}). 
At $z\!\ga\!1.5$ the D4000 break is completely lost for  optical observations, which marks the onset of the so-called {\em `redshift-desert'}. 

\begin{figure}[t]
\begin{center}
\includegraphics[angle=0,clip,width=8cm]{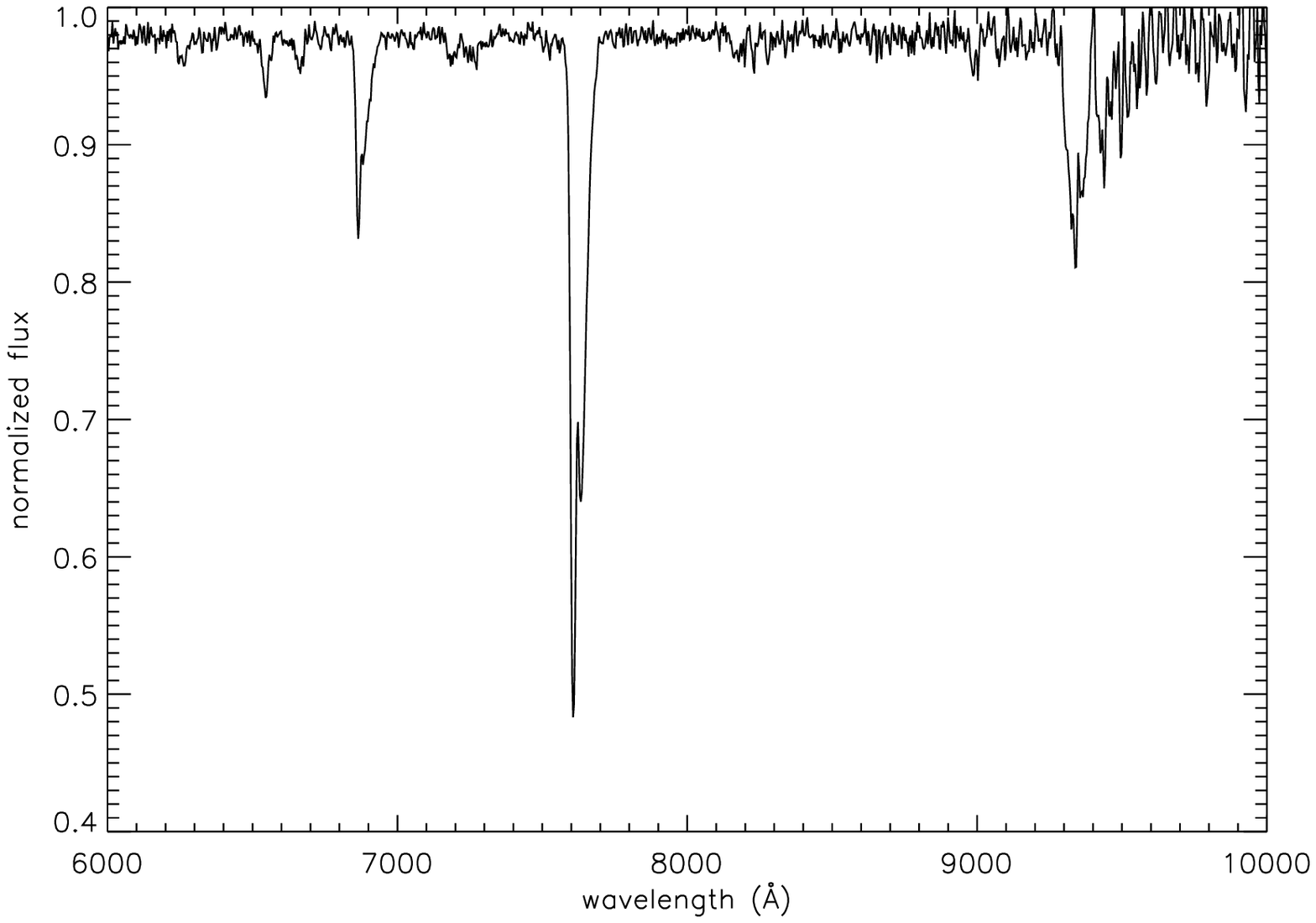}
\includegraphics[angle=0,clip,width=8cm]{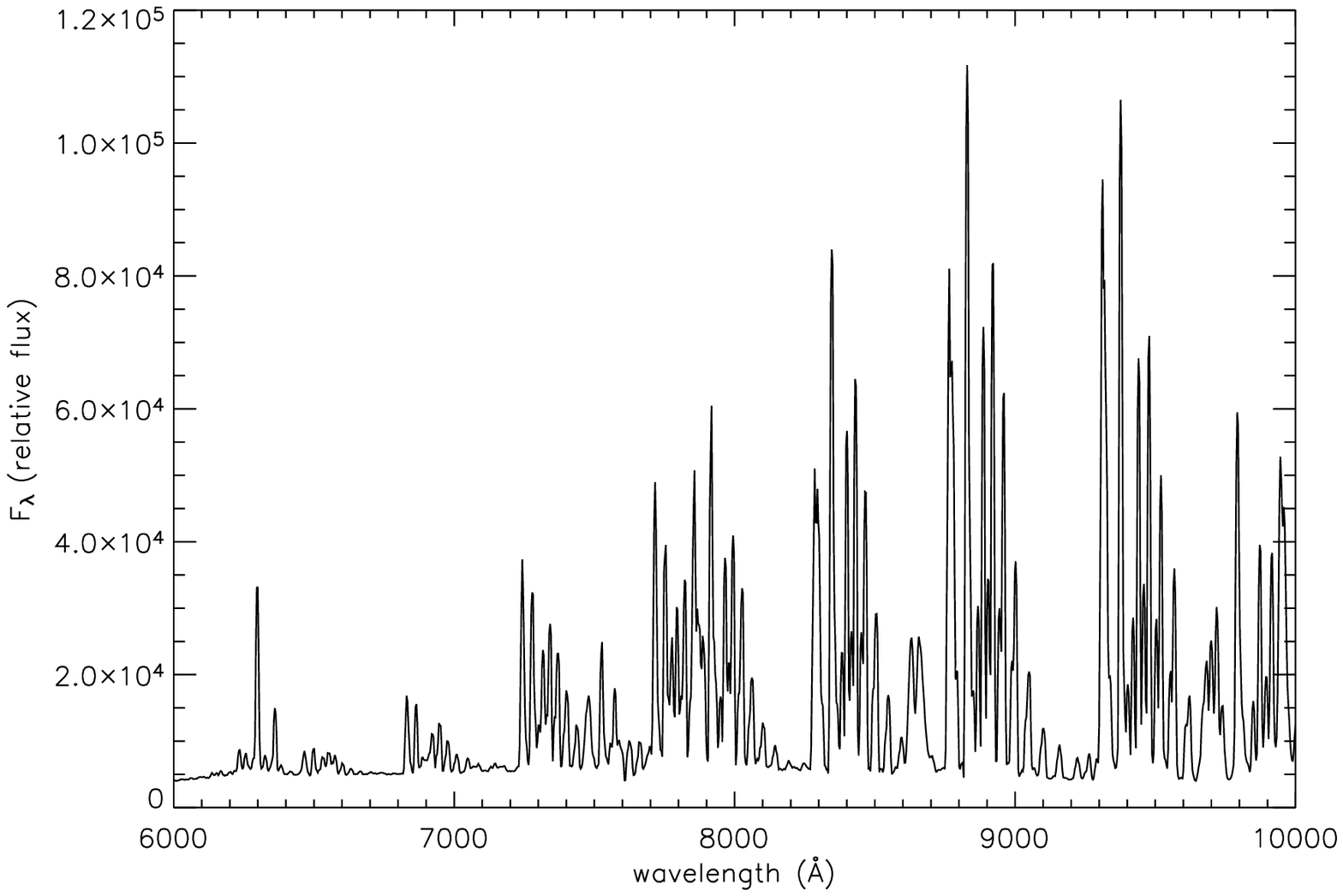}
\end{center}
\caption[Spectral Sky Features]{Spectral absorption and emission features introduced by the night sky. {\em Left:} Normalized telluric absorption lines as determined from the analysis of an intrinsic featureless spectrum of a white dwarf. The strong absorption bands around 6\,900\,\AA \, and 7\,620\,\AA \, are caused by atmospheric O$_2$.  {\em Right:} Night sky emission spectrum dominated by the OH$^-$ radical at wavelength above 6\,800\,\AA. The different molecular {\em Meinel-bands} are easily visible. The apparent weakening towards longer wavelength is caused by the decreasing response of the camera system; the intrinsic line brightness continues to grow drastically towards the NIR regime. The lines above 8\,000\,\AA \, exhibit peak flux levels which are approximately a factor of 1\,000 above the continuum level of typical distant cluster galaxies.} \label{f8_sky_features}
\end{figure}

Besides the decreasing number of accessible spectral features, the second observational challenge is posed by the 
onset of atmospheric contamination in the I-band and in particular in the Z-band. Figure\,\ref{f8_sky_features} illustrates 
the telluric\footnote{{\em Telluric} means in Earth's atmosphere.} absorption (left panel) and the atmospheric emission (right panel) in the wavelength range from 6\,000 to 10\,000\,\AA.
The visible telluric absorption features in the left panel are caused by O$_2$ and H$_2$O. The strong oxygen A-band around  7\,620\,\AA \ and the B-band at  6\,900\,\AA \ can easily mimic galaxy absorption features and have to be carefully excised, in particular if object lines are nearby. Towards the NIR,  water vapor becomes the principal absorber with the first strong absorption band starting at around  9\,400\,\AA. 

The atmospheric emission in the right panel depicts the spectrally resolved reason for the background challenge in the near-infrared, the hydroxyl OH$^-$ emission. These radicals are produced during the day in an excitation reaction of ozone O$_3$ and hydrogen H at an altitude of 85--100\,km and radiate away their excess energy during the night. The airglow emission becomes dominant beyond about 7\,000\,\AA \ and is the limiting factor for ground based observation throughout the NIR regime. The so-called {\em Meinel bands} are easily discernable and correspond to different vibrational molecular transitions, whereas the lines within a band originate from rotational dipole transitions. For comparison, the strong emission lines beyond 8\,000\,\AA \ have peak fluxes roughly 1\,000 times brighter than the continuum of typical \zsim1 galaxies. 
Sky line residuals will thus manifest a noise limit for the identification of galaxy spectral features.



In order to obtain a sufficiently high signal-to-noise ratio for passive cluster galaxies of candidates at redshifts $1\!\la\!z\!\la1.4$, typical exposure times of 2--4\,hours are required even with VLT FORS\,2.
XMMU\,J2235-2557 sample spectra at $z\!\sim\!1.4$ were shown in Fig.\,\ref{f5_survey_strategy}; additional \zsim1 galaxy spectra will be discussed (see Fig.\,\ref{f10_Spectosc_cl15}).


Efficient strategies for the spectroscopic confirmation of the highest redshift cluster candidates at $z\!\ga\!1.5$ have yet to emerge and are currently at the limit of the technological capabilities. In principle, there are two approaches to tackle the problem of $z\!>\!1.5$ redshifts, (i) the classical optical spectroscopy with extremely long exposure times to enable the identification of the weak rest-frame  UV features, or (ii) moving to near-infrared spectroscopy in analogy to the imaging strategy. 

The optical multi-object spectroscopy approach has been used to probe the general galaxy population in the `redshift-desert' for 
the Gemini Deep Deep Survey (GDDS) \cite{Abraham2004a} targeting galaxies with redshifts 1--2. 
Even with typical exposure times of 20--30 hours per field as for the GDDS, a secure redshift determination for passive galaxies is very hard once the D4000 is shifted beyond the probed wavelength range. Owing to the extreme exposure time requirements, the optical spectroscopy approach is not well-suited for routine cluster confirmations of the highest redshift candidates, but is invaluable up to $z\!\sim\!1.5$. 

\pagebreak

Near-infrared spectroscopy could in principle take over once the 4\,000\,\AA \ rest-frame  features are redshifted into the NIR, \ie \ at around $z\!\sim$1.4--1.5. However, several observational and technological issues turn this approach also into a challenging enterprise. (i) The wavelength regimes outside the J, H, and K-band are heavily absorbed by the atmosphere, \ie \ spectra are not homogeneously probed in wavelength. (ii) The sky emission lines within the NIR bands are so strong and numerous that the identification of the galaxy continuum and its absorption features is very difficult. (iii) Current NIR spectroscopy instruments typically exhibit significant trade-offs between wavelength coverage, spectral resolution, and multi-object capabilities, \ie \ for optimal instrument settings single object spectroscopy might be required. 
The technological aspect of the last item will improve significantly over the next few years with several powerful NIR multi-object spectrographs under development or in the commissioning phase. Their performance and application concerning 
high-redshift passive galaxies with spectral absorption features only is yet to be tested carefully.
A possible modified strategy in the NIR regime would be the emphasis on bluer emission line cluster galaxies,  resulting in an 
easier redshift determination at the expense of a less secure object pre-selection, \ie \ a significant fraction of blue objects will be 
foreground sources. 
 
In summary, the spectroscopic confirmation of galaxy clusters at \zg1 \ will always be the observational bottleneck for XDCP-like projects. The best currently available optical spectrographs allow routine identifications in survey mode, \ie \ with reasonable exposure times, out to $z\!\simeq\!1.4$--1.5. Moving beyond this limit into the `redshift desert' is extremely challenging and will likely require the best new NIR multi-object spectrographs of the near future. In any case, the spectroscopic demands posed by the highest redshift clusters are at the technological limit of 8--10\,m-class telescope and will remain challenging for at least another decade.

%

\section{Spectroscopic Data Reduction}

\noindent
In contrast to the reduction software developments in Chaps.\,\ref{c6_XrayAnalysis}\,\&\,\ref{c7_NIRanalysis}, the optical spectroscopic data reduction introduced in this section is standard and did not require an extensive customization beyond the     available {\it IRAF\/}\footnote{For software and documentation see \url{http://iraf.noao.edu}.} reduction routines. Therefore the discussion is kept short with the main aim of providing an overview of the different steps from raw data to the final spectroscopic redshifts.   

\begin{figure}[t]
\begin{center}
\includegraphics[angle=0,clip,width=0.96\textwidth]{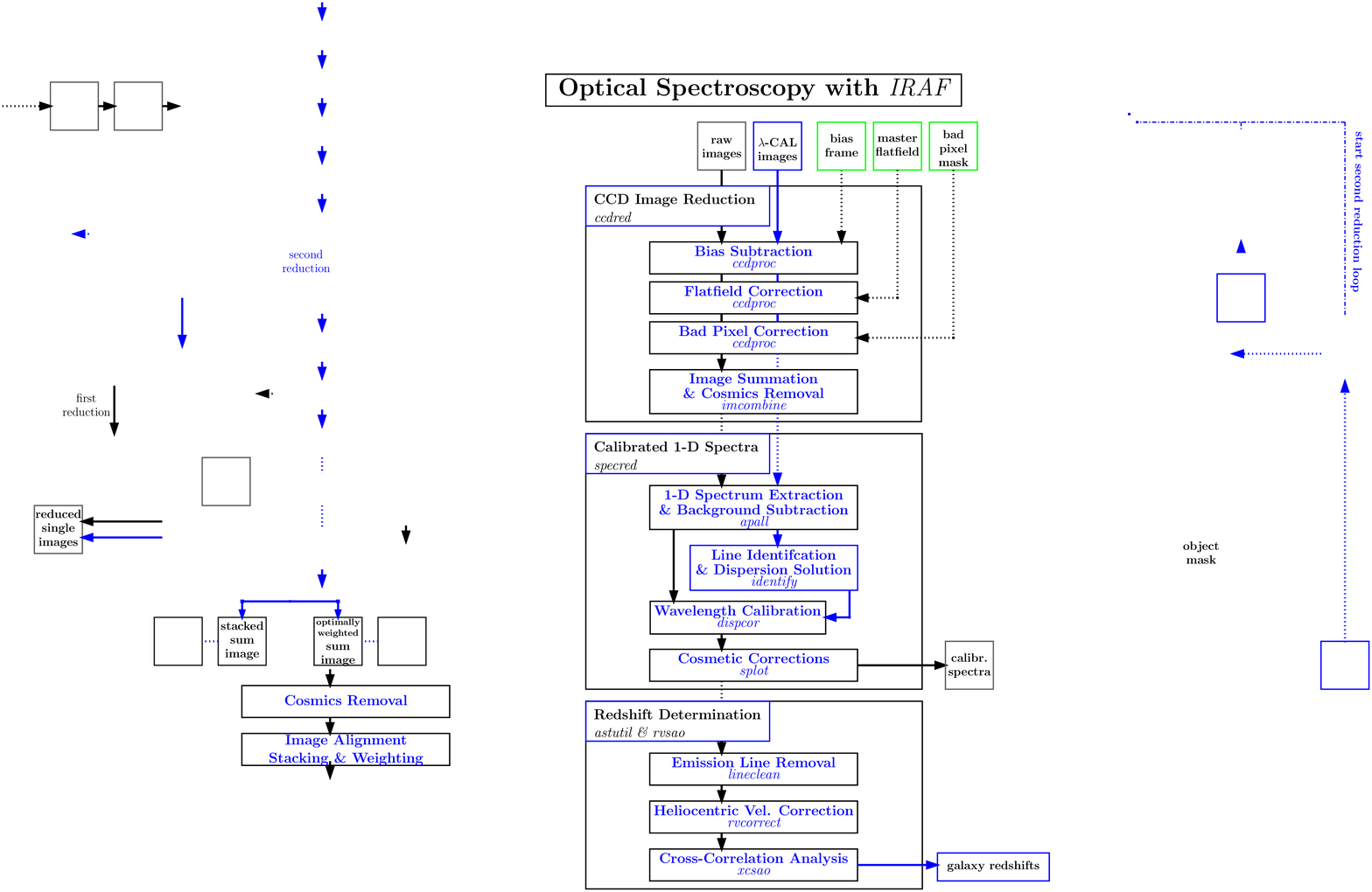}
\end{center}
\vspace{-2ex}
\caption[Spectroscopy Reduction Flow Chart]{Principal data reduction steps for optical spectroscopy with {\it IRAF}. The employed {\it IRAF} packages and main tasks are indicated in the corresponding module boxes.} \label{f8_FC_SpecRed}
\end{figure}

Figure\,\ref{f8_FC_SpecRed} illustrates  the principal reduction steps of optical spectroscopy  in a flow chart with  {\it IRAF\/}  packages and the main routines indicated. The reduction procedure can be subdivided into three main modules: (i) the general CCD image reduction, (ii) the extraction and calibration of the 1-D spectra, and (iii) the redshift analysis.

The CCD image reduction, as the first module, is similar to some of the basic steps discussed for the NIR pipeline in Sect.\,\ref{s7_NIR_pipeline}. As additional preparatory steps for CCD images, the frames have to be first trimmed to extract the actual data section, and then bias subtracted in order to remove the pedestal count level featured by CCDs. The additive applied bias voltage offset can be measured from the overscan CCD regions and designated bias frames taken with zero exposure time. The reconstructed  master bias frame is then subtracted from the data to yield intermediate products which contain the measured signal values. As a second calibration step, the multiplicative flatfield correction is applied by dividing through the fixed-pattern-noise frame. In contrast to the NIR imaging flatfield, which included the global quantum efficiency variations, the spectroscopic flatfield only contains the local pixel-to-pixel variations since the full spectrum is now spectrally dispersed over the CCD length. In practice, the local pixel-to-pixel sensitivity variations are extracted by dividing the global spectroscopic dome flatfields, taken with the proper slit or mask configuration, by a smooth continuum fit along the dispersion axis to remove the global large-scale modulation\footnote{The large-scale variations are caused by the detector QE function and the unknown spectral shape of the flatfield lamp.} and keep the local median levels to unity. In case a proper flux calibration is required (which is often not necessary for a redshift determination) a correction function along the dispersion axis can be constructed from observations of a spectroscopic flux standard star with accurately known spectral properties. The CCD image reduction is completed by applying the bad pixel correction and summing all individual frames per science target (typically three) to a {\em cosmics} cleaned final sum image, in analogy to the NIR imaging procedure.          

\pagebreak

\clearpage

The second module of the reduction procedure in Fig.\,\ref{f8_FC_SpecRed}  has the main objective of obtaining 
wavelength-calibrated 1-D spectra from the reduced 2-D CCD images. As a first task, the object spectra with a corresponding accurate background estimation have to be extracted from the CCD image. For each object an extraction aperture with two adjacent background windows is defined by viewing a CDD cut along the spatial direction for every mask slit. These object apertures are traced along the dispersion axis by following the peak of the spatial source profile. The object flux in the defined apertures is then summed in the spatial direction to recover the full source signal. The same procedure is applied to the designated background pixels, where the sky contribution is scaled to the same area per pixel element as the object aperture, and then subtracted to yield raw 1-D spectra with source signal only. Due to the dominating sky emission lines in the red part of the spectrum (right panel of Fig.\,\,\ref{f8_sky_features}), this last sky subtraction step is critical for the resulting signal-to-noise ratio.    


The extracted 1-D spectra for each object contain the relative source flux values as a function of image coordinate in the dispersion direction. The next task is to find the exact transformation of the image coordinates to physical wavelength units, \ie \ to determine the dispersion solution. This is achieved with the help of wavelength calibration observations using an internal Helium-Argon lamp as a source with numerous sharp and well known emission lines spread over the whole spectral range. The calibration lamp data have been taken with the same slit configuration and ideally at the same telescope position as the science target and are reduced in the same way as the target spectra. For the last extraction step, the  1-D calibration spectra are obtained from  exactly the same aperture traces along the detector as the science objects, which results in one raw 
image-coordinate-matched Helium-Argon calibration spectrum for each object. By identifying several dozen  calibration lines with their physical wavelength over the full spectral range, the dispersion solution is derived for each spectrum by iteratively   fitting a smooth polynomial function to the data points. The final converged functional form for transforming image coordinates to wavelength is now applied to the raw science spectra yielding the final wavelength calibration.      
After some manual cleaning of  sky line residuals, the calibrated 1-D science spectra are ready for the redshift determination of the next section.


\section{Redshift Determination}

\enlargethispage{8ex}

\noindent
In principle, the identification and subsequent wavelength measurement of a single spectral feature determines uniquely the galaxy redshift according to Equ.\,\ref{e8_obs_wavelength}. However, for routine survey applications it is favorable to use a more elaborate cross-correlation analysis 
for several reasons: (i) for passive galaxies the continuum shape and relative position of the spectral features is important; (ii) all spectral absorption features are taken into account and thus a higher accuracy is achieved; (iii) the quality and error of the redshift measurement can be quantified; (iv) the procedure can be  fully automated.

The cross-correlation analysis \cite{Tonry1979a} uses template spectra with high signal-to-noise ratios as a reference. Since our main interest focuses on absorption line spectra of passive galaxies, good templates for a representative selection of  different subclasses of elliptical and lenticular galaxies are of prime importance as a correlation input. 
\pagebreak 
 
The object spectra for the redshift determination need to be prepared for the correlation analysis in three ways. First, emission lines are removed by replacing the spectral sections deviating more than a given threshold above a smooth continuum fit. This measure is performed since emission lines could originate from physically different locations in a galaxy (\eg \ H\,{\small II} regions in spiral disks, and galactic winds) compared to the old stellar populations responsible for the absorption features, and could thus have a slightly different intrinsic redshift. Second, atmospheric spectral features need to be excised or removed to avoid an artificial correlation biasing, in particular with the strong telluric absorption lines (left panel of Fig.\,\ref{f8_sky_features}). Third,  
the {\em heliocentric velocity correction} is to be computed for each observation. With this correction, the stationary reference point for velocity comparisons is placed at the solar system barycenter, \ie \ the center-of-mass, and takes out the seasonal and nightly velocity modulations of up to $\pm$\,30\,km\,s$^{-1}$ due to Earth's motion relative to this reference point.



\enlargethispage{4ex}

The input spectra are now ready for the {\it xcsao} correlation analysis which performs the following simplified steps 
\cite{Kurtz1998a}: (i) Object and template spectra are rebinned to a logarithmic wavelength scale in order to achieve redshift independent separations between spectral features. From Equ.\,\ref{e8_obs_wavelength} we arrive at 
$\ln(\lambda_{\mathrm{obs}})\!=\!\ln(1\!+\!z)\!+\!\ln(\lambda_{0})$ and thus at $z$-independent logarithmic intervals $d\ln(\lambda_{\mathrm{obs}})\!=\!d\ln(\lambda_{0})$. (ii) The spectral continuum is fitted by a smooth function and subsequently subtracted to extract the absorption lines. (iii) The rebinned, continuum-subtracted template and object spectra are Fourier transformed into frequency space, where they are filtered for high and low frequencies beyond the instrumental resolution limits. (iv) The cross-correlation function in wavelength space  $c_{\lambda}(z)$ between object spectrum $o_{\lambda}$ and all template spectra $t_{\lambda}$ is proportional to the convolution of the spectra  $c_{\lambda}(z) \propto o_{\lambda} \bigotimes t_{\lambda}(z)$, \ie \ the correlation signal obtained when shifting the template spectrum to redshift $z$ (with logarithmic binning). In practice, the correlation function is computed in Fourier space, where the convolution simplifies to a multiplication of the object spectrum with the 
template spectra. 
(v) The location of the highest correlation peak in any of the template correlation functions is interpreted as the proper redshift of the observed spectrum. (vi) From the width of the peak a velocity error estimate is obtained, whereas the peak height relative to the average noise peaks yields a measure for the quality and reliability of the determined redshift.


After cross-correlation redshifts have been obtained for all available object spectra, the galaxy cluster redshift is determined as the outlier-clipped average of all cluster members. An object is classified as a cluster galaxy if its redshift deviates less than 3--4 times the expected velocity dispersion from the median redshift, \ie \ within typical 
member selection cuts of about $\pm$\,3\,000\,km\,s$^{-1}$ relative to the average cluster redshift.

\enlargethispage{4ex}

As a last application of this chapter, the proper measurement of the cluster radial velocity dispersion is discussed. 
Following Danese, De Zotti, \& Tullio \cite*{Danes1980a}, the spectroscopic cluster galaxy redshifts can be decomposed in three components: (i) the peculiar velocity of the Milky Way $z_0$, (ii) the proper cosmological redshift  $z_{\mathrm{R}}$ due to the Hubble expansion, and (iii) the galaxy peculiar velocity in the cluster $z_{\mathrm{G}}$. These components have to be multiplied\footnote{The multiplication is intuitive when considering that the cluster velocity dispersion interval is stretched by the cosmological redshift by a factor (1+$z_{\mathrm{R}}$) as all intervals of observables.} (not summed) to yield the observed object redshift $z_{\mathrm{obs}}$ according to 



\begin{equation}\label{e8_dispersion_factors}
z_{\mathrm{obs}}=(1+z_0)(1+z_{\mathrm{R}})(1+z_{\mathrm{G}}) \ .
\end{equation}

\noindent
Neglecting the first term $z_0$  as a $\orderof$($10^{-3}$) effect and approximating the proper cosmological redshift by the average cluster redshift  $z_{\mathrm{R}}\!\approx\!\bar{z}$, we are left with the contribution of the galaxy motion within the cluster. This redshift component is to be interpreted as a special relativistic Doppler shift due to the 
line-of-sight velocity $v_{\|}$ of a particular galaxy in the cluster potential

\begin{equation}\label{e8_Doppler_shift}
1+z_{\mathrm{G}} = 1+\frac{\Delta\!\lambda}{\lambda_0} = \sqrt{\frac{c+v_{\|}}{c-v_{\|}}} \approx 1 +  \frac{v_{\|}}{c} + \orderof\left(\frac{v^2_{\|}}{c^2}\right) \ .
\end{equation}


\noindent
Since $z_{\mathrm{G}}\!\ll\!1$ for the Doppler shifts of galaxies in clusters, only the first order term needs to be taken into account for the last approximation. Plugging this result into Equ.\,\ref{e8_obs_wavelength} and solving for the line-of-sight velocity $v_{\|}$ of a given galaxy yields

\begin{equation}\label{e8_radial_velocity}
v_{\|} = c \left( \frac{z_{\mathrm{obs}}-\bar{z}}{1+\bar{z}} \right) \ .
\end{equation}

\noindent
The final result for the  velocity dispersion of a galaxy cluster with $N$ spectroscopic member galaxies with measured redshifts $z_{\mathrm{obs}}(i)$ is then

\begin{equation}\label{e8_dispersion_factors}
\sigma_{\|}^{2}= \frac{c^2}{N-1} \left[ \sum_{i=1}^{N} \left( \frac{z_{\mathrm{obs}}(i)-\overline{z}}{1+\overline{z}}   \right)^2 \right] - \left( \frac{\delta}{1+\overline{z}} \right)^2 \ ,
\end{equation}

\noindent 
where the last term corrects for the spurious contribution due to the intrinsic average redshift measurement error $\delta$.
For a meaningful estimate of the true velocity dispersion $\sigma_{\|}$ 
a minimum of about ten cluster member redshifts is required.


We have now accomplished the main objectives of {\bf step 5} for the spectroscopic confirmation by measuring an accurate cluster redshift, determining cluster galaxy memberships, and deriving a velocity dispersion estimate.
This also completes the technical part of Chaps.\,\ref{c6_XrayAnalysis}--\ref{c8_SpecAnalysis} with discussions on the applied X-ray, near-infrared, and optical observational methods and strategies for the XDCP survey.




\section{Summary}
\label{s8_spec_summary}

\noindent
The main spectroscopy contributions within this thesis work include the following: 

\begin{samepage}
\enlargethispage{6ex}

\begin{itemize}
    \item Reduction and analysis of two NORAS\,2 spectroscopy runs (see Fig.\,\ref{f4_Noras2_spectra}) with data of 35 low redshift clusters (results not included); 
    \item Reduction and spectroscopic analysis of a \zsim1 \ XDCP cluster candidate obtained through DDT time (see Sect.\,\ref{s9_cl15_analysis});
    \item Led a spectroscopy proposal for the GMOS instrument at the 8\,m Gemini South telescope with 16\,hours of class C time 
    awarded; observations not executed; 
    \item PI of an experimental NIR spectroscopy DDT proposal with Gemini's GNIRS instrument for a high-redshift cluster confirmation; not accepted.  

\end{itemize}    

\end{samepage}



\chapter{XDCP Survey Status and Outlook}
\label{c9_SurveyResults}

\noindent
This chapter summarizes the current XDCP status and provides cluster candidate 
and survey characteristics. In addition, open survey tasks will be briefly discussed.


\section{X-ray Analysis Status}
\label{s9_Xray_Status}

\noindent
The complete detection and identification procedure of serendipitous extended distant cluster candidate sources for the current XDCP survey phase has been discussed in detail in Chap.\,\ref{c6_XrayAnalysis}. In this section, a preliminary assessment of the global survey characteristics is provided.

\subsection{Sky coverage}


\begin{figure}[t]
\begin{center}
\includegraphics[angle=0,clip,width=0.88\textwidth]{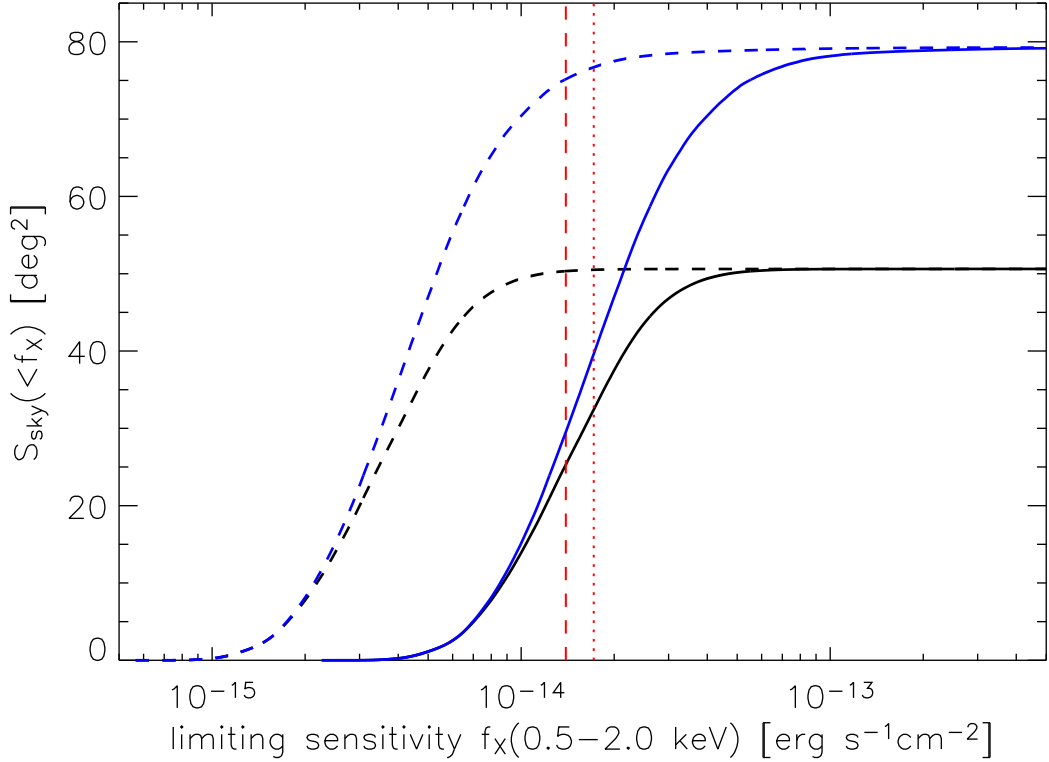}
\end{center}
\vspace{-2ex}
\caption[XDCP Sky Coverage]{XDCP sky coverage. Dashed lines indicate the full survey area ({\em blue}), 79.3 square degrees in total,  and the coverage with a 12\arcmin \ off-axis angle restriction ({\em black}), 50.6 square degrees in total, as a function of limiting {\em point source\/} sensitivity in the 0.5--2.0\,keV band. Solid lines show conservatively estimated {\em extended source\/} sensitivities corresponding to approximately a 4\,$\sigma$ extent significance and 100 source counts in the {\em single\/} 0.35--2.4\,keV detection band. At these levels, to be interpreted as {\em upper limits\/}, the median sensitivity for the full area is $1.7\!\times\!10^{-14}$\,\flux \,({\em dotted red line}), and $1.4\!\times\!10^{-14}$\,\flux \,for the restricted core survey region ({\em dashed red line}).} 
\label{f9_sky_coverage}
\end{figure}

\noindent
An essential ingredient for the evaluation of the survey search volume (Equ.\,\ref{e4_survey_volume_serendip}) is the computation of the survey sky coverage as a function of limiting flux.
The final determination of the XDCP selection function, \ie \ the probability to find a cluster with properties $C=[z, L_{\mathrm{X}}, T_{\mathrm{X}}, r_{\mathrm{c}}, M, ...]$  in the given data set with local characteristics $D=[t_{\mathrm{eff}}, B, \Theta, ...]$, is a challenging task for XMM-Newton surveys and will require extensive simulations  over the next two years (see Sect.\,\ref{s9_Selection_Funct}). At this point, approximate 
global XDCP characteristics can be derived based on some simplifying assumptions. 



The dashed lines in Fig.\,\ref{f9_sky_coverage} show the cumulative sky coverage\footnote{$S_{\mathrm{sky}}(<\!f_{\mathrm{X}})$ denotes the total survey area with a detection sensitivity of at least the given flux $f_{\mathrm{X}}$ or better.} $S_{\mathrm{sky}}(<\!f_{\mathrm{X}})$ as  a function of  limiting 0.5--2.0\,keV {\em point source sensitivity\/} for the full survey area (blue) and the restricted core survey (black) with a maximum off-axis angle of 12\arcmin. The {\em point source sensitivity\/} can be evaluated for every XMM field with the {\tt SAS} task  {\tt esensmap} (see panel 6 of Fig.\,\ref{f6_Xray_products}), which computes the local $3\,\sigma$ point source detection threshold\footnote{Standard ECF values for a power law spectrum with a photon index of 1.7 and an absorbing column of $3\!\times\!10^{20}$\,cm$^{-2}$ are assumed.} ({\tt DET\,ML}) for the combined detectors outside the masked regions from (i) the local background levels $B$, (ii) the effective exposure time $t_{\mathrm{eff}}$  at the pixel position, and (iii) the PSF dependent detection aperture at the given off-axis angle. The determined local {\em point source sensitivities\/} for each pixel element are then merged for the 469 XDCP survey fields. Sky areas with XMM coverage in multiple observations are corrected for by only retaining the most sensitive value of all overlapping pixels.   
The resulting {\em exact\/} total sky coverage for the XDCP survey is $\mathbf{S_{\mathrm{\bf sky}}(\mathrm{\bf tot})\!=\!79.3}$ {\bf square degrees} for the full area and $\mathbf{S_{\mathrm{\bf sky}}(\Theta\!\leq\!12\arcmin)\!=\!50.6}$ {\bf square degrees} for the restricted core survey region. The {\em point source sensitivity\/} varies (10--90\% of $S_{\mathrm{sky}}$) within the {\em full\/} survey (blue) by about a factor of five from  $2$--$10\!\times\!10^{-15}$\,\flux, and by about a factor of 3.5 from  $2$--$7\!\times\!10^{-15}$\,\flux \ for the core survey region (black).

Conservative preliminary estimates of the effective XDCP sky coverage as a function of the {\em extended source sensitivity\/} can be obtained by applying an empirically determined offset factor to the {\em point source sensitivity\/}.
At the detection limit, the average {\em point source\/} contains a total of about 25 counts in the 0.35--2.4\,keV  band. Comparing this value to the identified {\em extended sources\/} in Fig.\,\ref{f6_EXTML_counts} and choosing a conservative extended source count number limit of $\sim$100 photons, corresponding to about a $4\,\sigma$ extent significance, an average offset factor between the point source detection limit and a realistic reliable extended source threshold of a factor of four is obtained. Applying this empirical offset to the {\em point source sensitivity\/} yields a conservative XDCP sky coverage estimate for {\em extended sources\/} illustrated by the solid lines in Fig.\,\ref{f9_sky_coverage}.

The median {\em extended source sensitivity\/} for the full XDCP survey is then approximately 
$f^{\mathrm{med}}_{\mathrm{X}}(\mathrm{tot})\!\simeq\!1.7\!\times\!10^{-14}$\,\flux \ (dotted red line) and $f^{\mathrm{med}}_{\mathrm{X}}(\Theta\!\leq\!12\arcmin)\!\simeq\!1.4\!\times\!10^{-14}$\,\flux \ (dashed red line) for the core survey region. The median XDCP survey sensitivities determined in this way are to be considered as  conservative upper limits since the empirical offset  factor of four compared to the point source threshold is quite large. The question how far the {\em extended source sky coverage\/} curves (solid lines) can be shifted to the left, \ie \ towards lower limiting flux levels, while retaining a high completeness will be answered with simulations in the near future.

  



\subsection{Cluster number counts}
\label{s9_logS_logN}

\begin{figure}[t]
\centering
\includegraphics[angle=0,clip,width=0.94\textwidth]{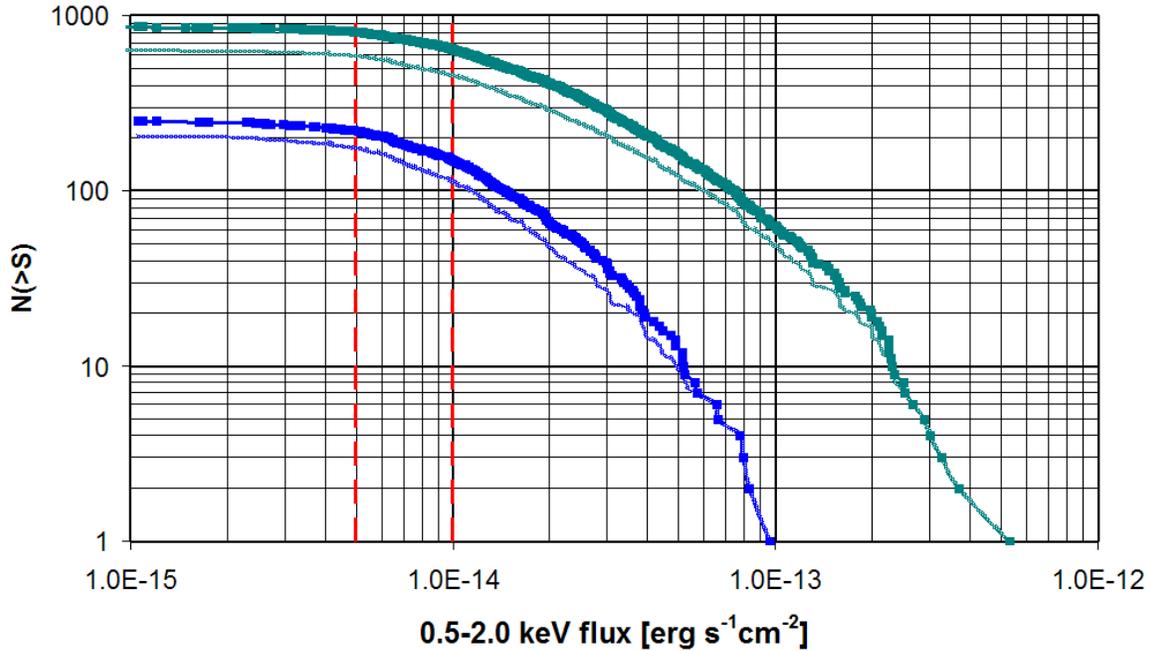}
\caption[Raw $\log N$--$\log S$ of XDCP Clusters]{Raw $\log N$--$\log S$ of XDCP candidates and identified clusters. 
The total number of objects is shown as a function of the 0.5--2.0\,keV flux measured for the PN detector, which is available for about 90\% of all sources. 
The green lines display all identified clusters {\em and\/} distant candidates for the full XMM detector area ({\em top green line}) and the FoV restricted to 12\arcmin \ off-axis angle ({\em lower green line}). 
The blue lines illustrate the cumulative distribution of the distant cluster candidates only.
Vertical red lines indicate the position where the $\log N$--$\log S$ curves start to turn over at approximately  $1\!\times\!10^{-14}$\,\flux and the flux level where the effective area becomes small around $5\!\times\!10^{-15}$\,\flux.  

}
\label{f6_logNlogS}       
\end{figure}

\noindent
The cumulative {\em raw\/} XDCP cluster number counts in the 0.5--2.0\,keV band for {\em distant cluster candidates} (top blue line) and {\em all\/} identified cluster candidate sources (top green line) are shown in Fig.\,\ref{f6_logNlogS}, where the lighter lines indicate the distribution for a core survey restriction with a maximum off-axis angle of 12\arcmin.
This $\log N$--$\log S$ (see Equ.\,\ref{e3_cluster_NumberCounts}) is to be considered as {\em raw\/} in the following sense: (i) the number counts are not normalized to a unit solid angle, (ii) the candidate sample still contains {\em false positives\/}, (iii) the flux determination is based on fiducial values for the Energy Conversion Factor (see Sect.\,\ref{s6_source_detection}) and the $\beta$ parameter (Equ.\,\ref{e2_beta_model}), and (iv)
fluxes are consistently determined for the PN detector, which is not available for about 10\% of the (missing) candidate sources due to a non-imaging operation mode of the PN for this fraction of all observations.   





The overall shape of the cumulative number count distribution yields important diagnostic information for the survey sensitivity characteristics. The flux level where the $\log N$--$\log S$ slope starts deviating from the observed number counts  in deeper surveys (see Fig.\,\ref{f4_ROSAT_logSlogN}) indicates the approximate flux limit of the survey below which the source counts become incomplete. In  Fig.\,\ref{f6_logNlogS} this estimated XDCP completeness level is represented by the red dashed line at  
$f^{\mathrm{com}}_{\mathrm{X}}\!\simeq\!1\!\times\!10^{-14}$\,\flux.
The $\log N$--$\log S$ approaches a constant value, \ie \ no additional sources are found at lower flux levels, at the point where the effective sky coverage of the survey at the given limit drops rapidly towards zero. This approximate minimum survey sensitivity  is indicated  by the  second dashed line at $f^{\mathrm{min}}_{\mathrm{X}}\!\simeq\!5\!\times\!10^{-15}$\,\flux.
At the bright end, the observed cumulative number count distribution appears to be (artificially) truncated for the most luminous clusters. This effect can be attributed to two features of the source detection algorithm which (i) has a fixed upper size limit of  80\arcsec \ for the core radius in the XDCP pipeline setup and (ii) the intrinsic tendency of the {\tt eboxdetect\/} task to split very extended cluster sources into several sub-clumps.


\newpage

With the total number of cluster candidates and the survey area available, we can now estimate the observed XDCP galaxy cluster surface density. In total, the full XDCP sample contains 990 unique cluster candidates, 276 of which are {\em distant   
cluster candidates\/} and 714 are {\em DSS-identified cluster sources}. For the restricted core survey at off-axis angles of less than 12\arcmin, the 226 distant and 526 DSS-identified candidates make up a combined core sample of 752 sources.
Subtracting the approximate {\em false positive\/} fraction of 1/3 for distant and 1/6 for low redshift candidates, the expected sample size of {\em real} galaxy clusters is obtained as $\sim\!780$ for the full region of 79.3 square degrees and $\sim\!590$ objects for the core survey area of 50.6 square degrees. Note that this cluster sample size is comparable to the largest local 
all-sky surveys (see Sect.\,\ref{s4_LocalSurveys}), with the difference that the XDCP survey is not aiming at a full identification of all clusters.

The approximate average XDCP surface density is thus $\sim\!9.8$ clusters per square degree for the overall area and   $\sim\!11.6$  clusters per square degree for the more sensitive  core survey area with $\Theta\!\leq\!12\arcmin$.
From the observed $\log N$--$\log S$ of Fig.\,\ref{f4_ROSAT_logSlogN}, a cluster surface density of 
10 objects per square degree is obtained at a flux limit of $1\!\times\!10^{-14}$\,\flux.
The identified combined XDCP cluster candidate sample is thus fully consistent with an average limiting sensitivity of 
$\bar{f}^{\mathrm{lim}}_{\mathrm{X}}\!\simeq\!1\!\times\!10^{-14}$\,\flux \ for the full survey  area and slightly below this value for the core sample.

The estimated XDCP sensitivity limits from the {\em raw\/} $\log N$--$\log S$ and the total candidate numbers in this section are about 30\% lower than the conservative median sensitivities derived in the previous section. This apparent discrepancy originates from the attempt to characterize an XMM serendipitous surveys with a single number, the flux limit $f^{\mathrm{lim}}_{\mathrm{X}}$. The difficulties arising from this concept of a global survey flux limit will be briefly discussed in the next sub-section.










\begin{figure}
\centering
\includegraphics[angle=0,clip,width=0.94\textwidth]{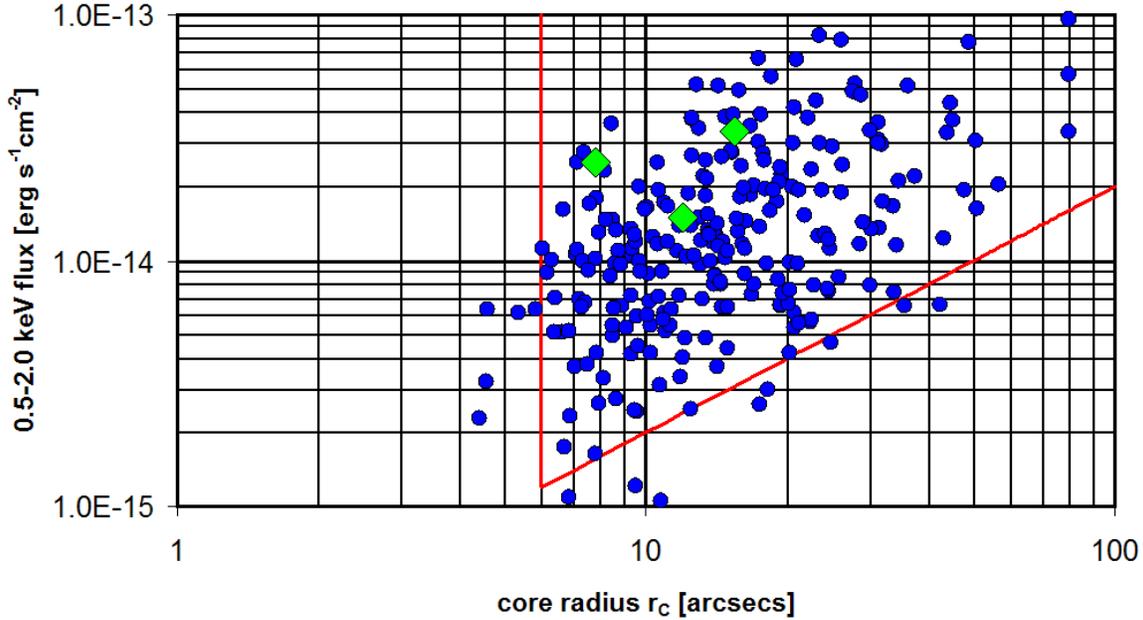}
\vspace{-1ex}
\caption[Extent Properties of XDCP Candidates]{XDCP distant cluster candidates in the 0.5-2.0\,keV flux versus  core radius  plane. The parameter space of  detected candidates is confined by the XMM resolution limit ({\em vertical red line}) at core radii of about 6\arcsec \ 
and the background limit ({\em lower red line}), where the cluster surface brightness drops below the detection threshold. Green diamonds indicate the positions of three most distant spectroscopically confirmed clusters in the Southern hemisphere (using the source parameters of the XDCP X-ray pipeline), from left to right: XMMU~J2235-25 at $z\!=\!1.39$, XCS~J2215-17 at $z\!=\!1.45$, and RDCS\,J1252-29 at $z\!=\!1.24$, see Sect.\,\ref{s9_confirmed_dist_clusters} for references.
}
\label{f9_cands_flux_rc}       
\end{figure}

\subsection{Selection function}
\label{s9_Selection_Funct}

\noindent
In this final section on X-ray properties of XDCP cluster candidates, some important aspects concerning the selection efficiency of serendipitous extended  sources with XMM-Newton are summarized.


Figure\,\ref{f9_cands_flux_rc} displays the extended X-ray sources of the XDCP {\em distant cluster candidate sample} in the 0.5-2.0\,keV flux versus  core radius\footnote{Core radii $r_{\mathrm{c}}$ are determined with {\tt emldetect\/} (see Sect.\,\ref{s6_source_detection}), \ie \ using a fixed $\beta\!=\!2/3$.}  plane (see Equ.\,\ref{e2_beta_model}), the two prime source parameters governing the detection probability.
This empirical plot illustrates nicely the two main XMM limitations that constrain the accessible parameter space of distant cluster candidate sources (see Sect.\,\ref{s6_XMM}). The XMM {\em resolution limit} sets a lower core radius cut-off at about 6\arcsec \ (vertical red line), whereas the {\em background} imposes a limit on the maximum core radius for sources of given flux    (lower red line). Note that the currently three most distant spectroscopically confirmed clusters in the Southern hemisphere (green diamonds) are well away from the sample limits, in particular concerning the accessible minimum flux threshold (vertical offset to the lower red line).


The {\em selection function\/} characterizes a galaxy cluster survey in terms of the extended source detection probability of clusters with given properties in the X-ray data.  
The most important {\em cluster characteristics\/} with prime influence on the detection rate are given in the following:

\begin{description}
    \item[Source Flux:] 
    For an idealized, homogenous X-ray survey, the  flux limit is the {\em single\/} characteristic that determines the detection probability of a source. The received X-ray flux $f_{\mathrm{X}}$ in a given energy band depends on the cluster luminosity  $L_{\mathrm{X}}$, the spectral shape governed by the gas temperature $T_{\mathrm{X}}$, the redshift $z$, and the galactic hydrogen absorption column $N_{\mathrm{H}}$.    
Note that high cluster redshifts lead to severe cosmological surface brightness dimming  (Equ.\,\ref{e3_SB_dim})  and can hence introduce selection biases at faint flux levels, where the surface brightness drops below the signal-to-noise threshold (for a real survey with background).
    
    
    \item[Core Radius:] 
    Once the background surface brightness reaches levels comparable  to that of the source, the physical size of the cluster, characterized by the projected core radius  $r_{\mathrm{c}}$\footnote{To avoid confusion with the off-axis angle $\Theta$, the notation $r_{\mathrm{c}}$ is used for the projected core radius throughout the following discussion.} (see Equ.\,\ref{e2_beta_model}), becomes the second prime property governing the detection rate. For further analytical considerations it is important to note that 
the source flux fraction encircled within the core radius $f_{\mathrm{X}}(r\!<\!r_\mathrm{c})$ is constant for $\beta$-models, \ie \ it does not vary with the size of the source\footnote{This can be shown by integrating Equ.\,\ref{e2_beta_model}.}. For extended sources whose SB-profiles follow the applied $\beta\!=\!2/3$ model, this fraction is   
$f_{\mathrm{X}}(r\!<\!r_\mathrm{c})/{f_{\mathrm{X}}(\mathrm{tot})}\!=\!0.293$, \ie \ about 30\% of the flux originates from the core region. 

            
    
    \item[Cooling Core Activity:] 
    Cooling core activity in clusters (see Sect.\,\ref{s2_CCC}) is connected to very peaked inner SB profiles with smaller core radii than non-cooling core clusters (\eg \ Chen \etal, 2007). \nocite{Chen2007a}
For possible high-redshift CCC, the peaked inner profile can have positive {\em or} negative effects on the detection probability. Moderate cooling core activity, as found for XMMU\,J2235.3$-$2557  at $z\!=\!1.393$ (Santos \etal, submitted), 
\nocite{Santos2007a}
increases the SB contrast and hence improves the signal-to-noise ratio. However, if the core radius of a strong CCC drops below the XMM resolution limit, the extended source detection probability drops and the cluster might be missed.   
    
    \item[Cluster Morphology:] 
For the detection process, cluster sources are  usually  idealized as spherically symmetric. This might not be a good approximation (\eg \ Hashimoto \etal, 2005), in particular at high redshifts, where clusters are either dynamically very young objects or are still in the formation process with a high merger frequency (see Fig.\,\ref{f3_ClusterFormation}). Deviations from spherical symmetry and in particular cluster substructure will lower the detection probability, since the cluster emission might be mistaken for several point sources. 

\nocite{Hashimoto2005a}
    
    \item[AGN Activity:] 
    If a cluster hosts an X-ray bright AGN, the chance for a misidentification as a point source increases. While only 5\% of local clusters have bright AGN close to their center \cite{HxB2004a}, the point source contamination for high-redshift systems can be significant (\eg \ Stanford \etal, 2001). 
\nocite{Stanford2001a}
    
 
\end{description}

\noindent
The relevant {\em XMM instrumental and data characteristics} for the detection of faint extended sources have been discussed in Sect.\,\ref{s6_XMM}. As a function of detector position, \ie \ off-axis angle $\Theta$ and azimuth $\phi$, the key features are (i) the effective, vignetting corrected clean exposure time  $t_{\mathrm{eff}}$($\Theta$), (ii) the PSF FWHM($\Theta$) defining the minimum resolvable cluster angular size $r_{\mathrm{min}}(\Theta)$, (iii) the background surface brightness B($\Theta$, $\phi$), and (iv) detector defects (\eg \ gaps, dead columns, hot pixels).

Close to the detection threshold for extended sources, XMM observations  are, in very good approximation, always  {\em background limited\/} (see Equ.\,\ref{e7_background_limit}).     
Assuming that the detection threshold is defined by a minimum signal-to-noise ratio SNR$_{\mathrm{min}}$ in a given detection aperture, \eg \ the core radius of the source, then the limiting flux threshold scales as $f_{\mathrm{lim}}\!\propto\!\sqrt{B(\Theta,\phi)}/\sqrt{t_{\mathrm{eff}}(\Theta)}$ with the exposure time $t_{\mathrm{eff}}$ and the background level $B$. Similarly, the increased detection aperture of larger extended sources at a given background level has the scaling effect  $f_{\mathrm{lim}}\!\propto\!\sqrt{\pi B r^2_{\mathrm{c}}}\!\propto\!r_{\mathrm{c}}$. This expected linear dependance of the core radius on the detection limit is reflected by the lower red line in Fig.\,\ref{f9_cands_flux_rc}.
{\em A priori\/} unknown effects can be taken into account  by introducing fudge factors  for the cluster AGN and foreground flux contamination of nearby sources $A_{\mathrm{contam}}\!\ga\!1$ and for possible morphological cluster distortions  $B_{\mathrm{morph}}\!\ga\!1$.
By defining the {\em local\/} flux limit as the level where a cluster with resolvable core radius $r_{\mathrm{c}}\!>\!r_{\mathrm{min}}(\Theta)$ is detected with a specified probability, \eg \ 50\%, then the combination of all factors yields 

\begin{equation}\label{e9_flux_limit}
    f_{\mathrm{lim}}(r_{\mathrm{c}}\!>\!r_{\mathrm{min}}) \propto r_{\mathrm{c}}\,\frac{\sqrt{B(\Theta,\phi)}}{\sqrt{t_{\mathrm{eff}}(\Theta)}}\,A_{\mathrm{contam}}\,B_{\mathrm{morph}} \neq const \ .
\end{equation}  

\noindent
The effective {\em local\/}  extended source flux limit is hence a function of the XMM field characteristics ($t_{\mathrm{eff}}$, $B$), the location on the detector ($\Theta$, $\phi$), the sky position for contaminating nearby sources $A_{\mathrm{contam}}$(RA, DEC), and the cluster properties ($r_{\mathrm{c}}$, $B_{\mathrm{morph}}$).
A single {\em global\/} flux limit for background limited XMM serendipitous surveys is hence {\em not well defined}. The variation of the cluster core radius over typical  values in the local Universe of 50-500\,kpc already  introduces a difference in the limiting sensitivity  of one order of magnitude. A stated  flux limit for XMM surveys is only meaningful for a specified cluster size, \eg \ for core radii of $r_{\mathrm{c}}\!=\!15\arcsec$, which can be resolved over the full FoV. All other factors combined cause  sensitivity  variations across the XDCP survey of approximately another factor of ten from the deepest to the shallowest fields.





The currently best modelled selection function of all XMM-Newton surveys has been provided by the XMM-LSS project (see Sect.\,\ref{s4_DeepSurveys}). For the  published five square degrees of data with 29 clusters, Pacaud \etal \ (2006, 2007) 
\nocite{Pacaud2006a}
\nocite{Pacaud2007a}
have determined  the detection probability of  sources as a function of core radius from simulated data assuming symmetric $\beta$-model cluster profiles. As a result, the XMM-LSS survey  has abandoned the concept of a {\em single global flux limit\/}, for the discussed reasons, and replaced it by {\em local likelihood thresholds} for the source extent {\em and} detection significance.    

In close collaboration with a team led by Joe Mohr at the University of Illinois, simulation efforts have started to determine the  XDCP selection function for each of the 469 survey fields. The final goal is to go one step beyond the XMM-LSS modelling and use {\em realistic\/} artificial clusters from high-resolution simulations with different masses, redshifts, morphologies, and dynamical states. These simulated clusters will be placed in mock images that take the  XMM-Newton instrumental characteristics into account. After adding background and a point source population  that follows the observed AGN number counts, the XDCP detection pipeline will be applied to multiple mock realizations of the same field. From the comparison of the detected extended source list with the cluster input catalog, the completeness and contamination functions can be evaluated for each field as a function of input cluster mass $M$, size $r_{\mathrm{c}}$, morphology, and redshift $z$.







\section{Imaging Status}
\label{s9_imaging_status}

\begin{figure}[h]
\begin{center}
\includegraphics[angle=0,clip,width=0.85\textwidth]{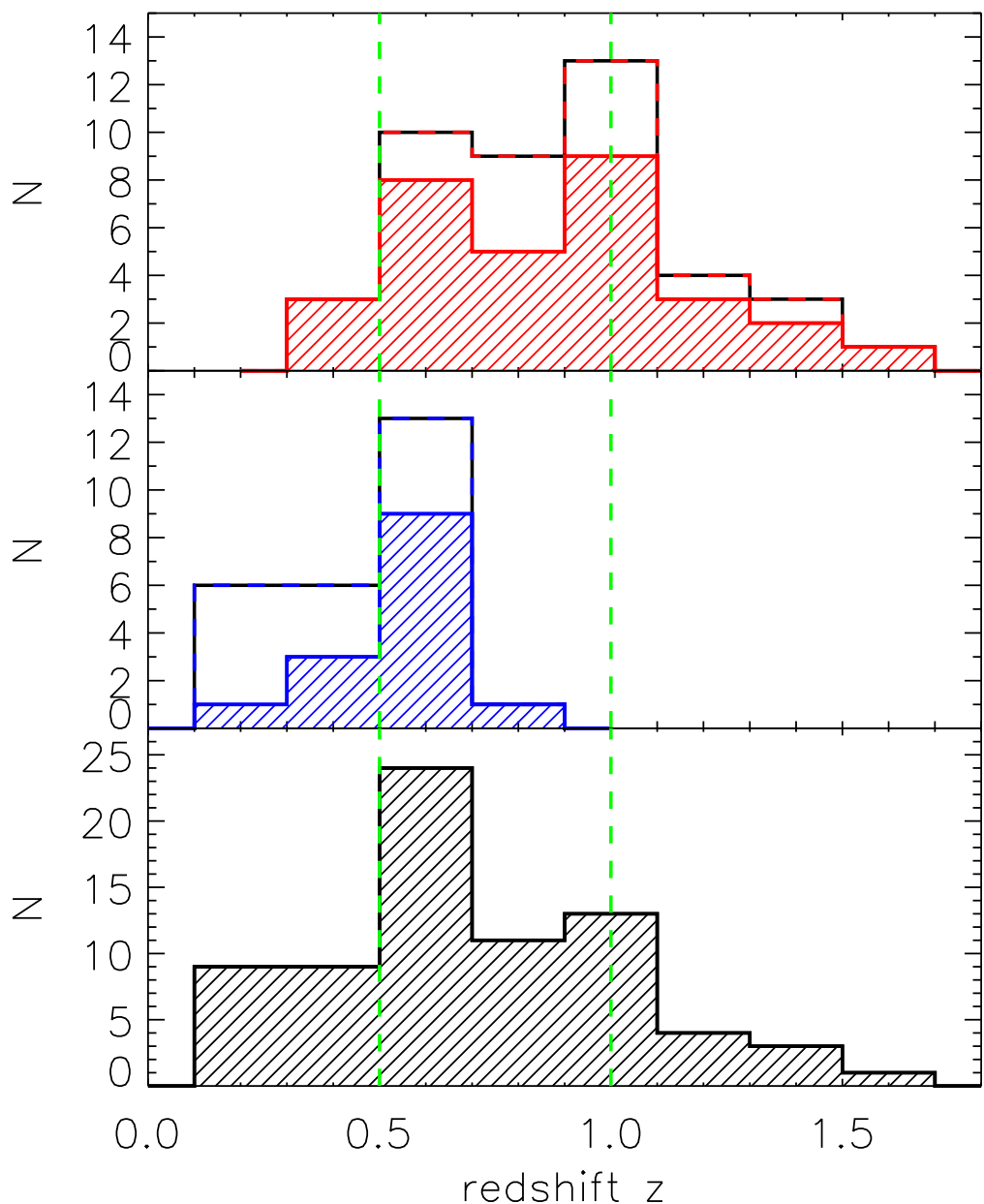}
\end{center}
\vspace{-2ex}
\caption[Calar Alto Imaging Cluster Sample]{Histogram of the final Calar Alto imaging cluster sample based on estimated Z--H \reds redshifts. {\em Top:} Redshift distribution of the targeted {\em distant cluster candidates\/} in $\Delta z\!=\!0.2$ bins with a secure photometric confirmation ({\em red hashed}) and including sources with lower photometric confirmation confidence ({\em top line}). Vertical dashed lines indicate redshifts of 0.5 and 1.0. {\em Center:} Distribution of {\em DSS-identified\/} cluster candidates that were included in the OMEGA\,2000 field-of-view of distant candidates.  {\em Bottom:} Redshift histogram of the combined Calar Alto X-ray cluster sample as the sum of the two upper red and blue curves and five additional XMM target clusters, which are not part of the serendipitous object class. The total Calar Alto sample includes 21 X-ray cluster candidates with photometric \reds redshift estimates of $z\!>\!0.9$, 15 of them with a high confidence level.  
} 
\label{f9_CA_sample}
\end{figure}

\noindent
The results of the two Calar Alto Z--H follow-up imaging campaigns (Chap.\,\ref{c7_NIRanalysis}) are summarized in Fig.\,\ref{f9_CA_sample}. In total, data was obtained for 63 {\em distant cluster candidates\/}. The redshift distribution for 43 systems with detected optical counterparts is shown in the top panel, based on Z--H \reds redshift estimates for most of the sources and some additional available redshifts from the literature (see Sects.\,\ref{s10_formation_epoch}\,\&\,\ref{s9_cl15_analysis}). 
The red hashed region displays secure cluster detections, the top dashed line also includes systems with lower confirmation confidence, usually attributed to sparse, less obvious red-sequences (if real).  
Two sources with central galaxy overdensities exhibit an ambiguous color distribution and are not included in the plot. 
Three objects with counterparts in the H-band still lack Z-band data for a redshift estimate, and 15 X-ray sources were classified as spurious. In combination with a few additional expected non-cluster sources from the lower confirmation confidence sample, an approximate {\em false positive fraction\/} of 1/3 is obtained, as was already discussed in Sect.\,\ref{s6_Source_Screening}.
A total of {\bf 21 cluster candidates} with photometric redshift estimates of $\mathbf{z\!>\!0.9}$ were discovered, the 15 secure systems being high-priority candidates for spectroscopic follow-up.

The center panel of Fig.\,\ref{f9_CA_sample} displays the redshift histogram of {\em DSS-identified low redshift cluster candidates\/}  (see Sect.\,\ref{s6_Source_Screening}) that could be imaged together with distant targets at no extra observational cost. As discussed in Sect.\,\ref{s6_identification_limit}, the blue histogram confirms that a good fraction of clusters out to $z\!\sim\!0.6$ can be successfully identified during the classification procedure, with the positive consequence that the resulting {\em distant cluster candidate\/} distribution (red histogram) is skewed towards a peak at $z\!\sim\!0.9$.
The large fraction of uncertain identifications at low redshifts (top blue line) is to be attributed to low mass groups with only few probed member galaxies and expected Z--H colors in the CMD regions associated with high background object densities. 
The lower black panel illustrates the combined Calar Alto cluster sample obtained  from the sum of the upper red and blue distributions and five additional XMM target clusters that are not part of the XDCP {\em serendipitous\/} cluster catalog.

The observed {\em distant cluster candidates\/} at Calar Alto constitute about 1/4 of the complete XDCP sample. XDCP imaging data for about 2/3 of all candidates have been obtained and will likely be extended to 80\% by the end of 2007. In total, the XMM-Newton Distant Cluster Project has been awarded with ten follow-up imaging runs at four different observatories making use of four different cluster identification techniques (see Sect.\,\ref{s7_NIR_strategy}):

\begin{itemize}
    \item The pilot study (Sect.\,\ref{s5_pilot_study}) started with a two semester VLT FORS\,2 R--Z imaging program in the years 2003/2004;
    \item Additional FORS\,2 R--Z imaging data of cluster candidates was obtained during two recent runs in 2006/2007;
    \item The two Calar Alto Z--H campaigns in 2006 have been discussed in Chap.\,\ref{c7_NIRanalysis}, results will be shown in Chap.\,\ref{c10_HizClusterStudies}; 
    \item The I--H method has been tested during two runs in 2007 using the SOFI and EMMI instruments at ESO's New Technology Telescope (NTT);
    \item A different {\em multi-band photometric redshift method} using  g\,r\,i\,z imaging complemented by H-band observations for very distant cluster candidates has been pursued at the CTIO observatory in  2006 and will be continued in fall 2007. 
\end{itemize}



\noindent
The latter {\em multi-band\/} imaging  program is being conducted in close collaboration with the group of Joe Mohr at the University of Illinois. The scientific aim of this special XDCP sub-project is the optical identification and characterization of {\em all\/} X-ray selected clusters in the SZE survey region of the South Pole Telescope (SPT) (see Sects.\,\ref{s2_SZE}\,\&\,\ref{s11_SZE}). 
The future SZE coverage of the 106 fully analyzed XDCP fields in the SPT region (red symbols in Fig.\,\ref{f6_Sky_Coverage}) with a total effective clean XMM exposure time of almost 2\,Msec will open new possibilities for joint X-ray-optical-SZE cluster studies. The X-ray selected  sample of approximately 200 clusters with  multi-band photometric redshifts over a large baseline will allow a detailed survey cross-calibration between the X-ray-based XDCP and the SZE survey of the South Pole Telescope, a critical pre-requisite for the use of distant clusters as sensitive cosmological probes.







\clearpage



\section{Spectroscopy Status}
\label{s9_spectr_status}

\noindent
The spectroscopic confirmation of \zg1 galaxy cluster candidates is the observational bottleneck of the XDCP survey since it requires the highly competitive use of red-sensitive multi-object spectrographs on 8\,m-class telescopes (see Chap.\,\ref{c8_SpecAnalysis}).
The XDCP status of spectroscopically confirmed high-redshift clusters can be summarized as follows:

\begin{itemize}
    \item XMMU\,J2235.3-2557 at $z\!=\!1.393$ was the first spectroscopically targeted XDCP distant cluster in 2004 (Sect.\,\ref{s5_xmmu_2235});
    \item XMMU\,J0104.4-0630 was the second followed-up pilot study candidate in 2005 and was confirmed at $z\!=\!0.947$ (see Sect.\,\ref{s9_cl15_analysis});
    \item Three cluster candidates with estimated redshifts of $z\!\sim\!0.9$ have been spectroscopically observed at one of the 6\,m-class Magellan telescopes;
    \item Four other XDCP distant cluster candidates have been spectroscopically confirmed through other channels and surveys (see Sect.\,\ref{s9_confirmed_dist_clusters}) at redshifts of 1.45, 0.976, and two at $z\!=\!1.05$; 


    \item Two spectroscopic VLT FORS\,2  XDCP programs are currently being executed targeting a dozen high-redshift objects, completing the pilot study program on one hand and starting the confirmation of Z--H selected Calar Alto candidates on the other;
    \item Spectroscopy time for eight additional FORS\,2 observations has been granted for the upcoming semester.
\end{itemize}


\section{XDCP Outlook}
\label{s9_xdcp_outlook}

\noindent
With the X-ray candidate selection for the current survey phase completed (Chap.\,\ref{c6_XrayAnalysis}) and more than half of all XDCP distant cluster candidates followed-up with imaging observations (Chap.\,\ref{c7_NIRanalysis}), the XDCP project focus will shift towards the spectroscopic confirmation of high-redshift clusters (Chap.\,\ref{c8_SpecAnalysis}) over the next few years.  

The XDCP {\bf X-ray analysis\/} will  change its scope from the search and identification of distant cluster candidates to 
the detailed characterization of the survey sample. The initiated collaborative project of realistic simulated cluster data   (Sect.\,\ref{s9_Selection_Funct}) will be used to evaluate the {\em XDCP selection function\/} in detail. By applying the XDCP X-ray pipeline to simulated cluster fields, the  probability for the  detection of extended sources as a function of various input parameters (\eg \ off-axis angle, exposure time, core radius, background)  will be quantified in conjunction with the fraction of missed cluster sources (\eg \ due to AGN dominated X-ray flux, distorted morphologies, compact cores). With this information at hand, the final strict X-ray selection cuts for the XDCP core survey cluster sample will be defined, which will provide the assessment of the associated cluster search volumes for the upcoming cosmological studies.  

After a full evaluation of all available {\bf follow-up imaging\/}  data of distant cluster candidates and the different identification techniques, the completion of the final quarter of the two-band photometric identification program of XDCP can be prepared in 2008. The available (and scheduled) optical coverage of the full XMM field-of-view for approximately 20\% of the survey overlapping  with the SPT region will additionally allow the  quantitative assessment of the applied screening procedure of Sect.\,\ref{s6_Source_Screening}, in particular the fraction of {\em false negatives\/}, \ie \ detected real distant cluster sources that were not added to  the {\em distant cluster candidate list\/}.   

The completion of the  {\bf spectroscopic follow-up\/} has the longest time line and will likely require 3--4 more years.
The compilation of an X-ray selected,  well characterized and spectroscopically confirmed \zga1 cluster sample over $\sim\,50$ square degrees  within this time span will be highly competitive as a prime cosmological tool and as pathfinder sample for several of the upcoming larger cluster surveys (\eg \ SPT, eROSITA, see Chap.\,\ref{c11_Outlook}).

In principle, the XMM-Newton Distant Cluster Project could be extended in the future on several different scales, provided sufficient man power is available. First, the planned enlargement of the XDCP sky coverage in the SPT  region will add approximately 100 additional XMM archive fields to the survey that have accumulated over the past three years. Second, new archival data could be included for the full Southern hemisphere XDCP sample based on the selection criteria discussed in Sect.\,\ref{s6_survey_definition}, which would increase the number of XMM fields by about 400. The largest sky coverage increase with more than 600 new XMM fields would result from a Northern sky (DEC\,$>+20$\degr) extension of the survey. Suitable efficient follow-up imaging methods applicable for Northern sky observatories have been developed and tested (Chap.\,\ref{c7_NIRanalysis}) and new large facilities, such as the Large Binocular Telescope (LBT), could be involved in the spectroscopic distant cluster work.  In total, XMM archival data for a tripling of the current XDCP survey sky coverage up to $\sim\!150$ square degrees has accumulated, but the related work to {\em fully\/} exploit this unique data set would fill several more PhD theses in the future.


The highest priority of the XMM-Newton Distant Cluster Project is therefore the completion of the ongoing survey phase 
and the scientific exploitation of the {\bf first X-ray selected distant cluster sample with $\mathbf{\ga\!30}$ systems at redshifts of} $\mathbf{z\!\ga\!1}$.
The top three XDCP science drivers to be addressed over the coming years are (i) the determination of the cluster number density evolution out to $z\!\sim\!1.5$ for constraints of prime cosmological parameters (see Sect.\,\ref{s3_cosmo_tests}), (ii) the characterization of evolving cluster properties at \zga1 as a function of redshift and cluster mass 
(see Sects.\,\ref{s2_ICM_form}\,\&\,\ref{s2_scaling_relations}), and (iii) the exploration of the cluster galaxy population evolution at large lookback times (see Sect.\,\ref{s2_galaxy_populations}).   
First XDCP results on galaxy evolution in clusters will be presented in Chap.\,\ref{c10_HizClusterStudies}.






\newpage


\section{Spectroscopically Confirmed $z\!>\!1$ Clusters in 2007}
\label{s9_confirmed_dist_clusters}

\noindent
This chapter on the XDCP survey status is closed with   
 a compilation of all spectroscopically confirmed X-ray luminous galaxy clusters at \zg1 published by August 2007.
Since the beginning of this thesis work the total number of \zg1 X-ray clusters has increased from five to eleven, indicating the advances in the XMM-Newton era on one side but also emphasizing  that these systems are still {\em `rare beasts'}. 

Table\,\ref{t10_Xray_clusters} lists the cluster names, surveys within they were confirmed, the redshift, and basic cluster properties of the eleven known systems. Five out of six confirmed \zg1 clusters in the Southern hemisphere are part of the XDCP distant cluster survey sample indicated by the star in the first column\footnote{RX\,J1252.9-2927 was also detected as a {\em serendipitous\/} source in addition to the targeted observations.}.

Note that several optically selected clusters at high redshift have recently been spectroscopically confirmed, most notable two infrared selected clusters with Spitzer  at redshift $z\!=\!1.41$ \cite{Stanford2005a} and  at $z\!=\!1.51$ 
\cite{McCarthy2007a}. 
These clusters are not yet confirmed in X-rays and are therefore not listed  in Tab.\,\ref{t10_Xray_clusters}.

\enlargethispage{2ex}

The references for clusters with identification number [ID] are: [1] Stanford \etal~\cite*{Stanford2006a},
[2] Mullis \etal~\cite*{Mullis2005a}, [3] Stanford  \etal~\cite*{Stanford1997a}, Ettori  \etal~\cite*{Ettori2004a}, [4] Rosati \etal~\cite*{Rosati1999a}, [5]  Rosati  \etal~\cite*{Rosati2004a}, [6] Bremer \etal~\cite*{Bremer2006a}, [7] Hashimoto \etal~\cite*{Hashimoto2005a},  Hashimoto \etal~\cite*{Hashimoto2002a}, [8]  Stanford \etal~\cite*{Stanford2002a}, [9,10] Andreon \etal~\cite*{Andreon2005a}, Pierre \etal~\cite*{Pierre2006a}, and [11]  Maughan \etal~\cite*{Maughan2006a}, Perlman \etal~\cite*{Perlman2002a}.

\nocite{Lamer2007a}






\begin{table}[t!]    
\begin{center}

\begin{tabular}{|c|c|c|c|c|c|c|c|c|}
\hline


\bf{ID} & \bf{Object Name} & \bf{Survey} & \bf{$\bf z$} & $\bf f_\mathrm{\bf X,0.5-2.0\,keV}$ & 
 \bf{$\bf L_\mathrm{\bf X,bol}$} & $\mathbf{T}_\mathrm{\bf X}$ \\ 

 &  &  &  & erg s$^{-1}$cm$^{-2}$  & $10^{44}$\,erg s$^{-1}$ & keV \\ 

\hline\hline

1* & XMMXCS\,J2215.9$-$1738  & XCS  & 1.45  & $1.2 \times 10^{-14}$  &  4.4  & 7.4  \\
2* & XMMU\,J2235.3$-$2557  & XDCP  & 1.393  & $2.6 \times 10^{-14}$  & 11.1  & 6.0    \\
3 & CIG\,J0848.6+4453  & RDCS  & 1.273  &  $1.0 \times 10^{-14}$    & 1.0  &  2.9  \\
4 & RX\,J0848.9+4453  & RDCS  & 1.261    & $1.8 \times 10^{-14}$  &  2.8  & 5.2   \\
5* & RX\,J1252.9$-$2927  & RDCS  & 1.237    & $2.5 \times 10^{-14}$ & 6.6  & 6.0   \\
6 & XLSS\,J0223.1$-$0436  & XMMLSS  & 1.22  & $6.2 \times 10^{-15}$ &   1.1   &  3.8  \\
7 & RX\,J1053.7+5735  & LH  & 1.134  & $2.0 \times 10^{-14}$  &  2.0  & 4.9   \\
8 & RX\,J0910.8+5422  & RDCS  & 1.106  &  $1.1 \times 10^{-14}$    & 2.5  &  7.2  \\
9* & XLSS\,J0224.1$-$0413  & XMMLSS  & 1.05  & $3.1 \times 10^{-14}$    & 4.8  &  4.1  \\
10* & XLSS\,J0227.1$-$0418  & XMMLSS  & 1.05  & $1.1 \times 10^{-14}$  &  1.5   &  3.7  \\
11& Cl\,J1415.1+3612  & WARPS  & 1.03  &  $1.1 \times 10^{-13}$   &  10.4  &  5.7  \\



\hline
\end{tabular}

\caption[Spectroscopically Confirmed \zg1 X-ray Clusters]{Spectroscopically confirmed \zg1 X-ray luminous galaxy clusters in August 2007. Specified values for the X-ray flux $f_\mathrm{X}$, bolometric luminosity $L_\mathrm{X}$, 
and temperature $T_\mathrm{X}$ are rounded, see references in text for details. Marked objects (*) in the first column are part of the XDCP distant cluster sample.} \label{t10_Xray_clusters}
\end{center}
\end{table}

\vfill

\begin{figure}[h]
\begin{center}
\end{center}
\label{f10_CA_test}
\end{figure}

\begin{figure}[b]
\begin{center}
\end{center}
\label{f10_CA_test}
\end{figure}




\chapter{High-Redshift Cluster Studies: First Results}
\label{c10_HizClusterStudies}


\noindent
The results of the follow-up observations are now used to address some of the scientific questions discussed in the introductory Chaps.\,\ref{c2_cluster_theory}\,\&\,\ref{c3_cosmo_theory}.
The cluster studies in  Chaps.\,\ref{c10_HizClusterStudies}\,\&\,\ref{c10b_science_outlook}  rely 
on photometric data obtained during two imaging campaigns at the Calar Alto Observatory (see Tab.\,\ref{t7_NIR_data_overview} and Chap.\,\ref{c7_NIRanalysis}). 
This chapter will focus on early results based on the available sub-sample of {\em spectroscopically confirmed\/} clusters.

In Sect.\,\ref{s10_formation_epoch}, the new Z--H  method is calibrated with the  spectroscopic sample in order to establish a reference model for the \reds color evolution and to provide first constraints on the formation epoch of the bulk of the stellar populations in early-type galaxies. This model is the basis for the photometric redshift estimates of the full distant cluster candidate sample.  

Section\,\ref{s9_cl15_analysis} presents the spectroscopic, photometric, and X-ray analysis of a newly confirmed galaxy cluster at $z\!=\!0.95$. This example provides a first {\em a posteriori\/} test for the photometric Z--H \reds technique and strengthens its applicability to the cluster identification and redshift estimation.


\section{The Formation Epoch of Red-Sequence Galaxies}
\label{s10_formation_epoch}

\begin{figure}[t]
\begin{center}
\includegraphics[angle=0,clip,width=0.85\textwidth]{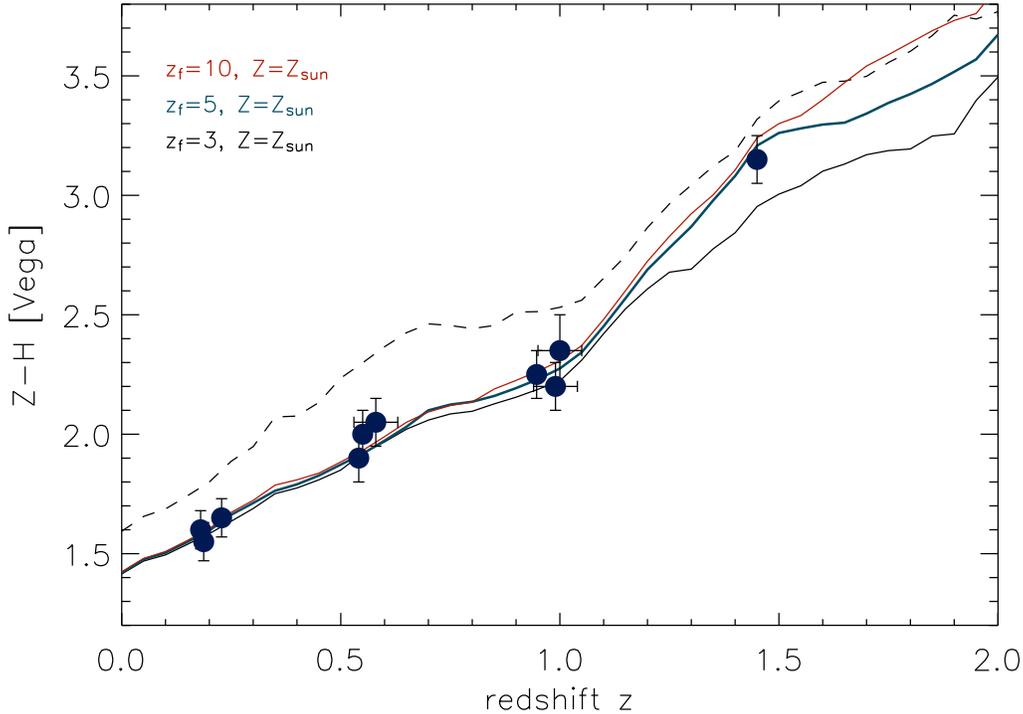}
\end{center}
\vspace{-3ex}
\caption[Z--H Evolution of the Red-Sequence]{Redshift evolution of the Z--H color of \reds galaxies.  Measured (average) \reds \ colors with 1\,$\sigma$ uncertainties for ten galaxy clusters with known redshifts 
in the range $0.18\!\leq\!z\!\leq\!1.45$ ({\em blue dots}) are shown together with different 
synthetic evolutionary tracks set by different models of passively evolving galaxies (\ie \ single burst models).
Solid lines represent solar metallicity models with formation redshifts $z_{\mathrm{f}}\!=\!3$ ({\em black}), $z_{\mathrm{f}}\!=\!5$ ({\em blue}), and $z_{\mathrm{f}}\!=\!10$ ({\em red}). The {\em dashed black line\/} indicates the color offset for a $z_{\mathrm{f}}\!=\!3$ model with three times higher metallicity. 
The blue model ($z_{\mathrm{f}}\!=\!5$, $Z\!=\!Z_{\sun}$) has been used to derive \reds redshift estimates for all observed cluster candidates.

} \label{f10_z_H_calibration}
\end{figure}

\begin{figure}[t]
\begin{center}
\includegraphics[angle=0,clip,width=0.85\textwidth]{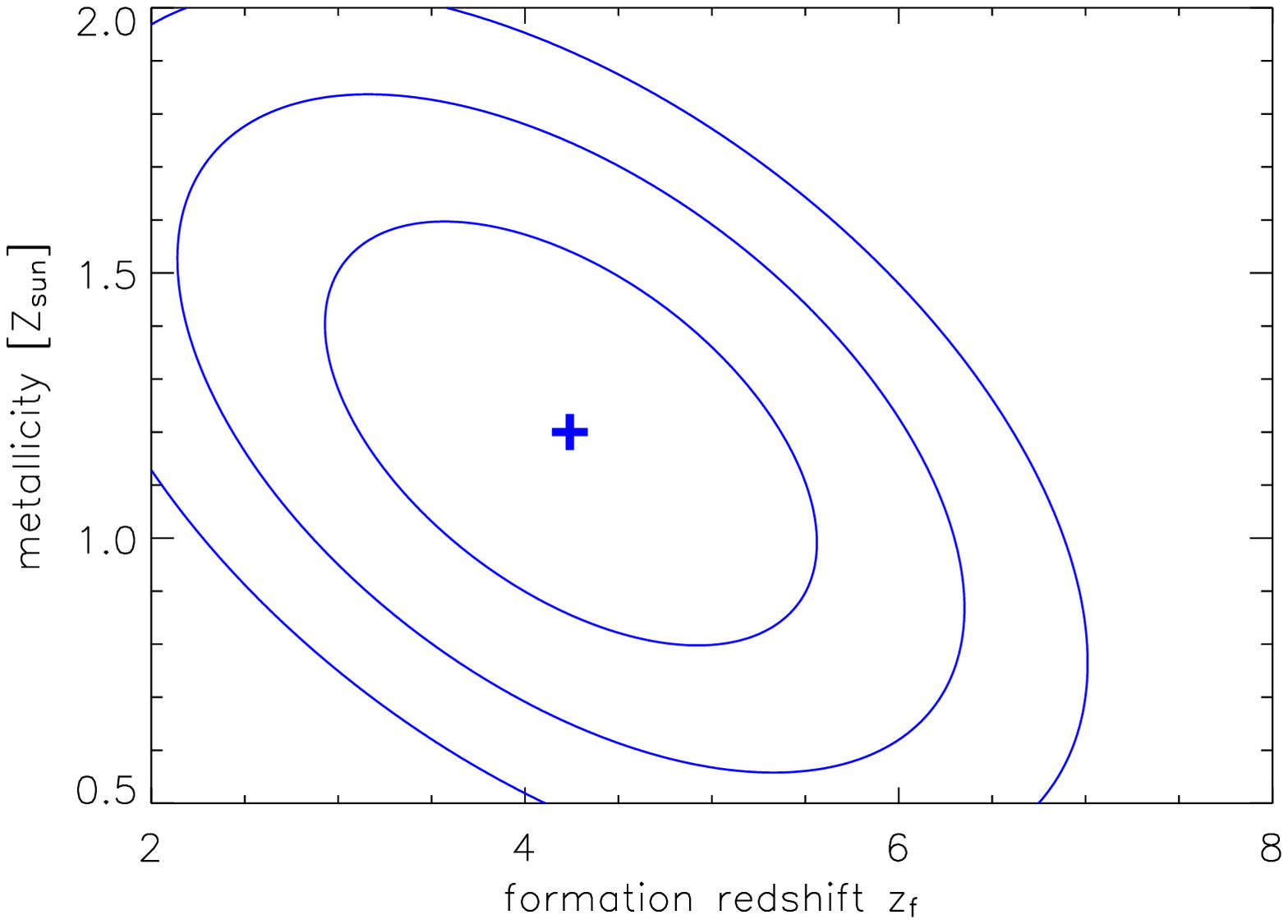}
\end{center}
\vspace{-3ex}
\caption[Formation Redshift of Ellipticals]{Maximum likelihood fit of the formation redshift $z_{\mathrm{f}}$ of passively evolving \reds galaxies {\em and\/} their metallicity $Z$. 
The long redshift baseline allows simultaneous constraints of both parameters yielding a best fit model    
(1\,$\sigma$  errors) of $z_{\mathrm{f}}\!=\!4.24\!\pm\!1.1$ and $Z\!=\!(1.20\!\pm\!0.35)\,Z_{\sun}$.} \label{f10_formation_redshfit}
\end{figure}

\noindent
Formation models for ellipticals and implications of the observed \reds properties of cluster galaxies  have been discussed in Sect.\,\ref{s2_red_sequence}. In this section, we compare observed \reds Z--H colors of ten calibration clusters with known spectroscopic redshifts to the SSP model expectations (see Sect.\,\ref{s7_NIR_strategy}).

\subsection{Calibration of the Z--H method}
\label{s10_Z_H_calibration}



\noindent
Figure\,\ref{f10_z_H_calibration} displays the observed evolution of the average Z--H \reds color as a function of 
 redshift ({\em blue dots}) together with different model expectations ({\em solid and dashed lines}).
In order of increasing redshift, the following clusters have been used 
to calibrate the model for the Z--H color evolution:
WARPS\,J0943+1644  at $z\!=\!0.180$  \cite{Vik1998a},
A\,383   at $z\!=\!0.187$ and
A\,2390  at $z\!=\!0.228$ \cite{Struble1999a},
CL\,0016+1609     at $z\!=\!0.541$ and
RX\,J0018+1617      at $z\!=\!0.55$ \cite{Connolly1996a},
XLSS\,J0227-0411     at $z\!=\!0.58$\footnote{An additional redshift error of $\pm 0.05$ has been added, since the  spectroscopic redshift appears uncertain.}  \cite{Pierre2006a}, 
XMMU\,J0104-0630   at $z\!=\!0.947$ (see Sect.\,\ref{s9_cl15_analysis}), 
XMMU\,J2215-1740            at $z\!=\!0.99 \pm 0.05$\footnote{Photometric redshift estimate based on deep multi-color CFHT photometry.}  (Lamer \etal, in preparation),
XLSS\,J0227-0418    $z\!=\!1.0$\footnote{Spectroscopic redshift, but the presence of a second overlapping foreground structure at $z\!\sim\!0.9$ decreases the reliability (see also Pierre \etal, 2006).} \cite{Valtchanov2004a}, 
and XMMXCS\,J2215-1738   at $z\!=\!1.45$ \cite{Stanford2006a}. 
 \nocite{Pierre2006a}

As can be seen in Fig.\,\ref{f10_z_H_calibration}, the solar metallicity $z_{\mathrm{f}}\!=\!5$ simple stellar population model ({\em blue line}) is fully consistent with all measurements. For this reason, the  ($z_{\mathrm{f}}\!=\!5$, $Z\!=\!Z_{\sun}$)-model was selected as {\em master calibration reference\/} for the photometric \reds redshift estimates of all observed XDCP cluster candidates (see Fig.\,\ref{f9_CA_sample}).  
The visual inspection of Fig.\,\ref{f10_z_H_calibration} reveals that the discrimination power between the model formation epochs rests upon the highest redshift cluster at $z\!=\!1.45$. This single data point currently excludes a $z_{\mathrm{f}}\!=\!3$ model on the 1.5--2\,$\sigma$ level, whereas the high-redshift $z_{\mathrm{f}}\!=\!10$ model is still within the 1\,$\sigma$ uncertainties.

Nevertheless, the unprecedented redshift baseline for X-ray cluster \reds studies in conjunction with the monotonic color-redshift relation of the new  Z--H method allow first reasonable simultaneous constraints of the two main formation model parameters $z_{\mathrm{f}}$ and $Z$.
Formation epoch (\ie \ age) and metallicity are highly degenerate in terms of resulting colors, 
although they are {\em independent\/} model parameters  (see lower right panel of Fig.\,\ref{f7_CMD_models}). Whereas the formation redshift impacts the galaxy colors at high redshifts of \zg1, an increasing metallicity results in an almost constant overall reddening of the color. 
We can hence parameterize the age and metallicity effects as a linear combination of the form


\begin{equation}\label{e10_model_parametr}
    M_{\mathrm{SSP}}[z_{\mathrm{f}}\!=\!(5\!+\!A),Z\!=\!(1\!+\!B)\,Z_{\sun})]\!\approx\!M[5, Z_{\sun}]\!+\! 
A\,\frac{M[5]\!-\!M[3]}{2}\,+\!B\frac{M[3\,Z_{\sun}]\!-\!M[Z_{\sun}]}{2},
\end{equation}

\noindent
where $M_{\mathrm{SSP}}[z_{\mathrm{f}}, Z]$ denotes the simple stellar population model of formation redshift $z_{\mathrm{f}}$ and metallicity $Z$ used for predicting the Z--H colors as a function of redshift. The ($z_{\mathrm{f}}\!=\!5$, $Z\!=\!Z_{\sun}$)-model is used as reference, and the formation redshift parameter $A\!\equiv\!z_{\mathrm{f}}\!-\!5$ and the metallicity offset $B\!\equiv\!(Z/Z_{\sun})-1$  allow continuous model variations based on the linear interpolation of the initially coarse parameter grid.

\subsection{Discussion}
\label{s10_Z_H_discussion}


\noindent
The  maximum likelihood fit result of the parameterized model to the data is displayed in Fig.\,\ref{f10_formation_redshfit}, where the contours define the 1\,$\sigma$, 2\,$\sigma$, and 3\,$\sigma$ confidence regions. The best fit values with 1\,$\sigma$ errors are
$\mathbf{z_{\mathrm{\bf f}}\!=\!4.24\!\pm\!1.1}$ for the formation epoch of the bulk of the stellar populations of \reds galaxies and $\mathbf{Z\!=\!1.20\!\pm\!0.35\,Z_{\sun}}$ for their average metallicity.
The determined direct observational constraint on the formation epoch based on the {\em Z-H color evolution\/} is fully consistent with the lower limits obtained from individual \reds studies at high redshift (\eg \ Mei \etal, 2006a), the age of high-$z$  massive ellipticals (\eg \ Dunlop \etal, 1996), age measurements of local early type galaxies (\eg \ Thomas \etal, 2005), and the latest high resolution simulation results \cite{DeLucia2006a}.  
\nocite{Thomas2005a}
\nocite{Mei2006a}
\nocite{Dunlop1996a}

The current 
constraints 
are based on only ten clusters with known redshifts. By populating the redshift regime $0.9\!\la\!z\!\la\!1.5$ with the XDCP targets currently scheduled for spectroscopic confirmation, we will soon be able to significantly lower the uncertainties on the cosmic formation epoch of early type \reds galaxies in clusters using the Z--H \reds method.

\begin{figure}[h]
\begin{center}
\includegraphics[angle=0,clip,width=13.5cm]{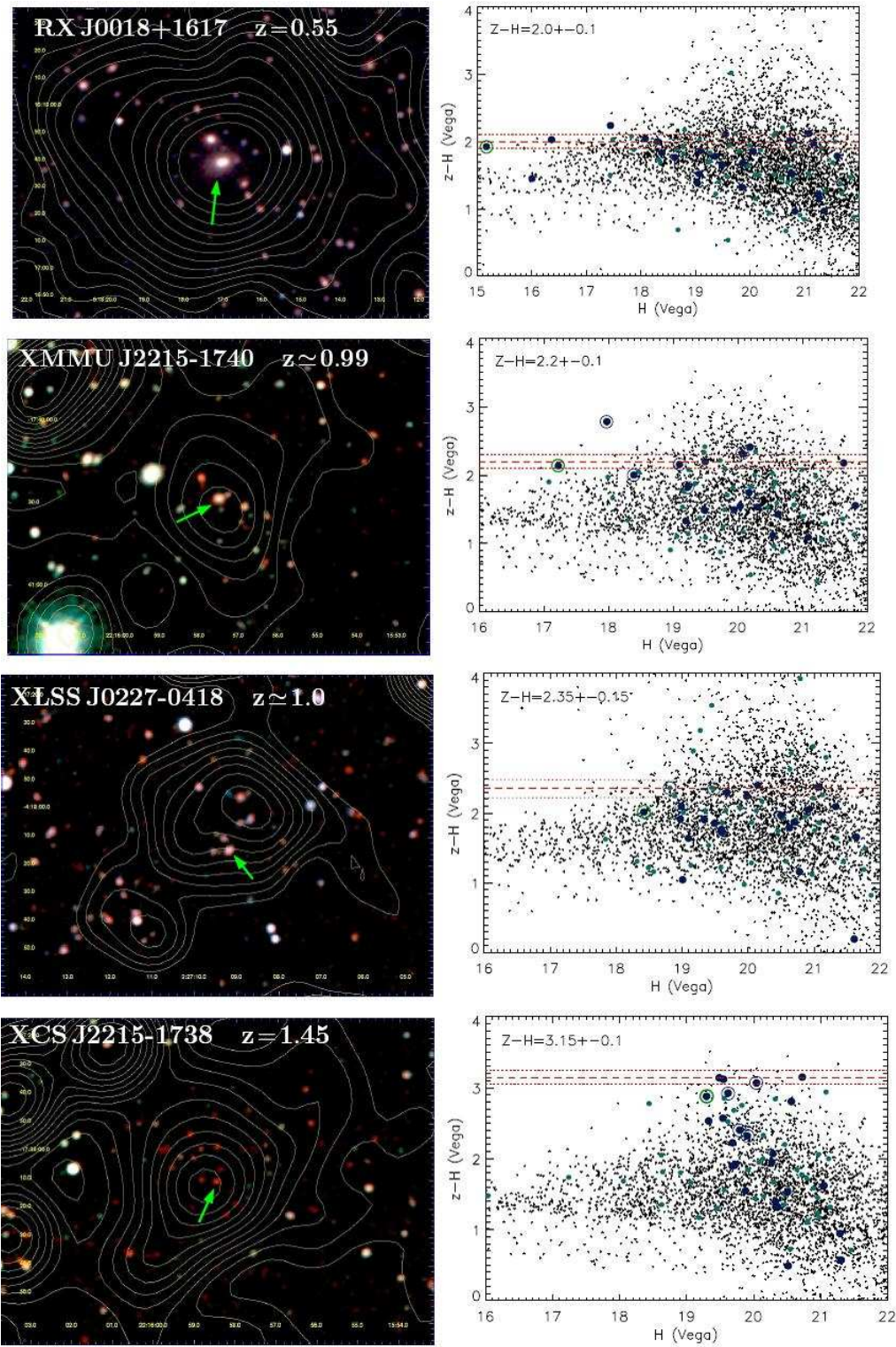}

\end{center}
\vspace{-3ex}
\caption[Calibration Clusters B]{$2\arcmin\!\times\!2\farcm 5$ image cutouts  and Z--H CMDs for the 
high-$z$  calibration clusters with known redshifts. Spectroscopically confirmed cluster galaxies in addition to the BCG are indicated by blue circles for the three most distant systems. In the case of XMMU\,J2215-1740, galaxies with multi-band photometric redshifts in the range $0.93\!\leq\!z\!\leq\!1.05$ are marked.
} \label{f10_calibration_clusters_B}
\end{figure}

Eight\footnote{The low-mass cluster XLSS\,J0227-0411 at $z\!=\!0.58$ has the largest uncertainties, \ie \ the lowest weight, among the calibration objects at $z\!\la\!0.6$ and is therefore not explicitly discussed.} of the calibration clusters over the full redshift baseline together  with their color-magnitude diagrams are shown in Figs.\,\ref{f10_calibration_clusters}\,\&\,\ref{f10_calibration_clusters_B}; XMMU\,J0104-0630 will be presented in Sect.\,\ref{s9_cl15_analysis}. The dashed red horizontal lines in the right column indicate the determined \reds colors based on the data, dotted red lines illustrate  the estimated color uncertainties. Spectroscopically confirmed BCGs are marked by green arrows in images, their corresponding locations in the CMDs are indicated by green circles. Additional confirmed cluster members  are circled in blue in Fig.\,\ref{f10_calibration_clusters_B} for the three calibration clusters with highest redshifts.

In preparation of the preliminary BCG study of Sect.\,\ref{s10_bcg_assembly}, a closer look at the spectroscopic sample   
can provide some observational guidelines for the selection and interpretation of BCGs. 
The BCG identification for all calibration clusters at $z\!\la\!0.6$ is unambiguous based on the central position in the cluster and the total apparent H-band magnitude. At these low and intermediate redshifts, all case study BCGs are located on the \reds within the uncertainties of the color determination, which justifies the assumption that stellar populations of  BCGs are passively evolving. The reduced redshift sensitivity of the Z--H method at $z\!<\!0.9$ is partially compensated by the lower color uncertainties of well populated red-sequences.      

At $z\!\ga\!0.9$, the observational task of locating the cluster \reds for the redshift estimate is more challenging. The red-sequences are often sparsely populated, as shown in the lower three rows of Fig.\,\ref{f10_calibration_clusters_B}. However, the redshift leverage of the Z--H color increases (see Fig.\,\ref{f10_z_H_calibration})  and the  background (black small dots) in the CMD decreases for redder colors. In absence of an obvious sequence, the Z--H color of the red envelope (formed by the filled  symbols) promises to be a reliable redshift proxy, as shown for the cases of  XLSS\,J0227-0418  and XMMXCS\,J2215-1738 in Fig.\,\ref{f10_calibration_clusters_B}. These two case study examples also suggest that BCGs  at high redshifts might not 
sit at
the X-ray center and that their color can be slightly bluer than the upper  envelope. 
For the identification of BCGs of high-$z$ clusters based on photometric data only, the spectroscopic sample points at an outer radius of $\sim\!30\arcsec$ and maximum color tolerances of  $\sim\!0.5$\,mag as reasonable search constraints.






\section{Indications for Large-Scale Structure at $z\!\sim\!1$}
\label{s9_cl15_analysis}

\noindent
In this section, the full XDCP survey analysis ({\bf steps 1--5}) is applied to the environment of  the X-ray cluster XMMU\,J0104.4-0630, the second pilot study selected object with available XDCP  spectroscopic follow-up data.
The relevant background information concerning cluster red-sequences (Sect.\,\ref{s2_red_sequence}) and environmental effects of galaxy evolution (Sect.\,\ref{s2_environmental_effects}) have been discussed in the introductory chapter.
At the cluster redshift of $z\!\simeq\!0.95$, the lookback time is 7.5\,Gyr and one arcsecond corresponds to a physical comoving scale of 7.9\,kpc (see Fig.\,\ref{f3_cosmic_time}\,\&\,\ref{f3_cosmological_distances}).
The results presented here follow closely the  journal letter of Fassbender \etal \ which has been submitted for publication in 
{\em Astronomy\,\&\,Astrophysics}.
\nocite{RF2007a}


\subsection{Observations}
\label{s10_cl15_observations}

\subsubsection{X-ray Data}

\noindent
The galaxy cluster XMMU\,J0104.4-0630 was selected as a serendipitous extended X-ray source in a high-galactic latitude archival field ($b\!=\!+69$\degr ; nominal exposure time: 26.7\,ksec; OBSID: 0112650401; here field 1). A second, partially overlapping XMM-Newton observation (nominal exposure time: 24.9\,ksec; OBSID: 0112650501; here field 2) extends the X-ray coverage to the South, providing moderately deep X-ray data over a field of about $25\arcmin \times 45\arcmin$.

The data were reduced following the procedure discussed in Sect.\,\ref{s6_Xray_Pipeline}. High background periods were excluded by applying a two-step flare cleaning procedure first in the hard (10-14\,keV) and subsequently in the 0.3-10\,keV band following Pratt\,\&\,Arnaud \cite*{Pratt2003a}.
The remaining clean exposure times for field 1 (field 2) are 15.1\,ksec (13.7) for the EPIC PN camera, 23.2\,ksec (19.7) for MOS1, and 23.5\,ksec (21.2) for MOS2, resulting in an effective (sensitivity weighted) clean exposure time of about 18\,ksec throughout most of the covered field.
Images in different bands for the PN and MOS detectors were created from the cleaned event lists and later combined for the overlapping sky region of fields 1 and 2.

\begin{figure}[h]
\begin{center}
\includegraphics[angle=0,clip,width=13.5cm]{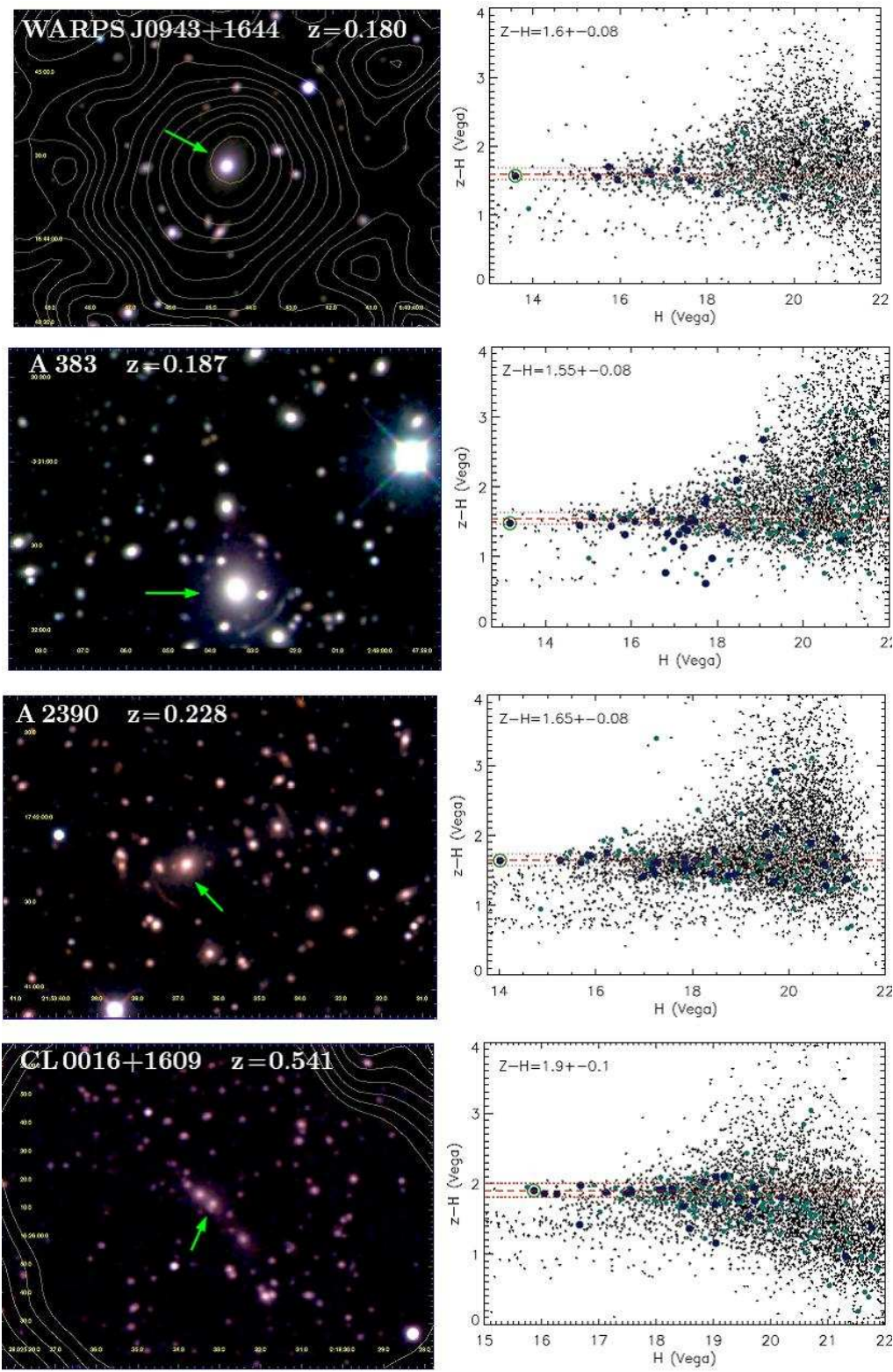}

\end{center}
\vspace{-3ex}
\caption[Calibration Clusters A]{$2\arcmin\!\times\!2\farcm 5$ image cutouts  and Z--H CMDs for the 
low-$z$  calibration clusters. X-ray contours are shown in yellow and BCGs are indicated by green arrows (left column) or a green circle (right column). The \reds color (red dashed line) and the estimated color uncertainties (red dotted lines) are marked.  Dark blue symbols represent objects located within a region of 30\arcsec \ from the  center,  smaller green circles objects within 30--60\arcsec, black dots mark all other objects in the field. 
} \label{f10_calibration_clusters}
\end{figure}

\clearpage

\begin{figure}[t]
\begin{center}
\includegraphics[angle=0,clip,width=\textwidth]{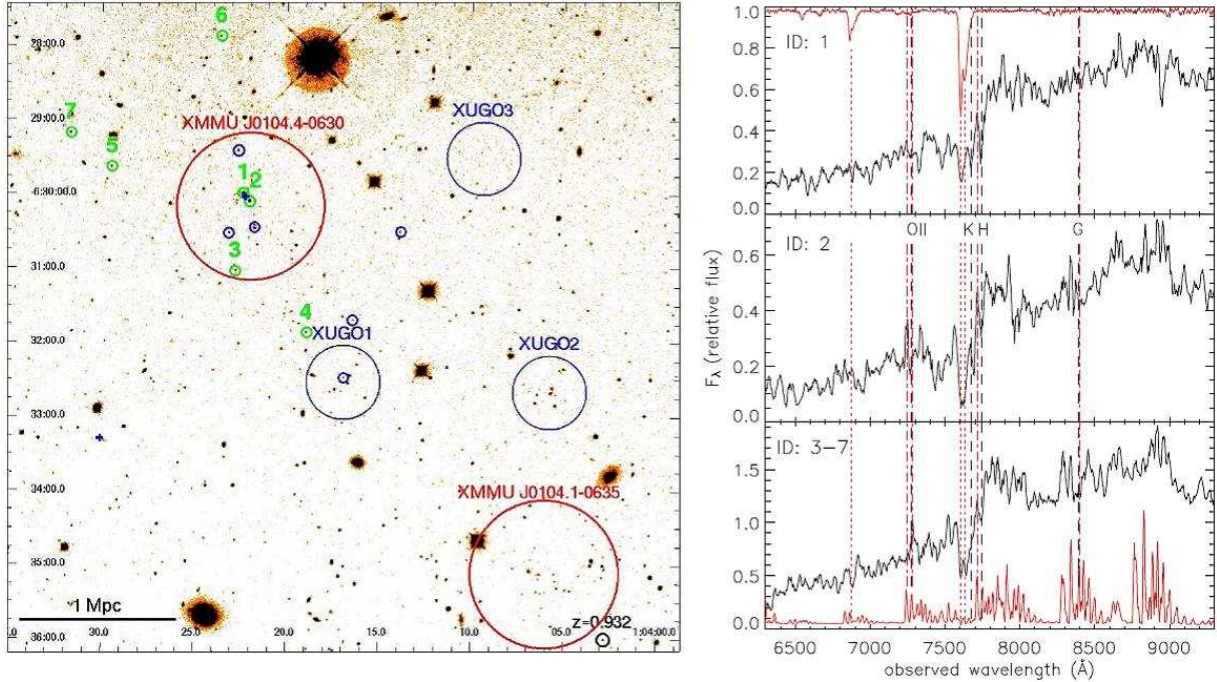}
\end{center}
\vspace{-3ex}
\caption[Spectroscopic Confirmation of XMMU\,J0104.4-0630]{Spectroscopic confirmation of XMMU\,J0104.4-0630.
{\em Left:} $9\arcmin\!\times\!9\arcmin$ H-band image of the environment of XMMU\,J0104.4-0630. Spectroscopically confirmed cluster members are indicated by green circles with identification numbers; tentative cluster members are marked by blue circles, 
 radio sources by crosses, and the consistent literature redshift is shown in black. The cluster radio source is located in between the BCG (ID\,1) and the galaxy coincident with the X-ray center (ID\,2).
{\em Right:} VLT FORS\,2 spectra of the galaxies with secure and concordant redshifts 
giving the cluster redshift of $z\!=\!0.947\!\pm\!0.005$ as labelled in the left panel (smoothed with a 7\,pixel boxcar filter). Flux units are arbitrary; sky emission lines and telluric absorption lines are shown in red at the bottom and top, respectively. The  two spectra from the top belong to the two brightest cluster galaxies in the center, the bottom panel displays the stacked spectrum of the galaxies with IDs 3--7.} 
\label{f10_Spectosc_cl15}
\end{figure}

A total of six extended X-ray sources were detected in the individual fields using the \texttt{SAS} tasks \texttt{eboxdetect} and \texttt{emldetect}. 
This work focusses on two of the sources, the main galaxy cluster XMMU\,J0104.4-0630 (here cluster\,A) and a second system  XMMU\,J0104.1-0635 at a projected distance of 6.4\arcmin \ to the South-West (here cluster\,B). 
Cluster\,A was detected at an off-axis angle of 5.5\arcmin \ in field\,1 with an aperture corrected, unabsorbed flux of $(1.7  \pm 0.3)\times 10^{-14}$\,erg\,s$^{-1}$\,cm$^{-2}$ \ in the 0.5--2.0\,keV energy band, using  $N_{\mathrm{H}}\!=\!5.08\times 10^{20}$\,cm$^{-2}$ from the 21\,cm survey of Dickey\,\&\,Lockman \cite*{Dickey1990a}.
The estimated core radius for a $\beta\!=\!2/3$ model is ($18.8 \pm 0.8$)\arcsec \ corresponding to 150\,kpc at $z=0.95$. The flux of cluster\,B at 7.2\arcmin \ off-axis angle in field\,2  is $(1.2  \pm 0.4)\times 10^{-14}$\,erg\,s$^{-1}$\,cm$^{-2}$, its estimated core radius of ($45 \pm 4$)\arcsec, corresponding to 340\,kpc, is significantly larger.
The main properties of the systems are summarized in Tab.\,10.1.


\subsubsection{Optical and Near-IR follow-up}

\noindent
The sky region around XMMU\,J0104.4-0630 was followed-up with R (1140s) and Z-band (480s) imaging on 18 October 2003 with VLT-FORS\,2 ($6\farcm8\!\times\!6\farcm8$ FoV). The target was reobserved in H (1000s) and Z (1800s) on 30 October 2006 in photometric conditions (1\arcsec \ seeing) using the NIR wide-field camera OMEGA2000 \cite{bailer2000} at the Calar Alto 3.5m telescope with a larger $15.4\arcmin\!\times\!15.4\arcmin$ \ field-of-view, which also covered the second cluster XMMU\,J0104.1-0635 to the South-West. 
The {\em SExtractor\/} 
photometry in this larger field is calibrated to the Vega system using 2MASS point sources \cite{Cutri2003a} in H, and designated SDSS standard star observations \cite{Smith2002a} in Z (see Chap.\,\ref{c7_NIRanalysis}). In both bands the limiting magnitudes (50\% completeness) of H$_\mathrm{lim}\!\sim\!20.7$ and Z$_\mathrm{lim}\!\sim\!22.8$ correspond to $m$*+1.6 for passively evolving galaxies at the cluster redshift 
(see Tab.\,\ref{t7_magnitude_evolution}).
Figure\,\ref{f10_Spectosc_cl15} (left) provides an overview of the field, Fig.\,\ref{f10_CMDs_cl15} (top panels) displays the R+Z+H false-color composite image of XMMU\,J0104.4-0630 (left), the Z+H images of XMMU\,J0104.1-0635 (center), and the region halfway between the clusters (right). The corresponding Z--H color magnitude diagrams  are shown in the bottom panels of Fig.\,\ref{f10_CMDs_cl15}.

Spectroscopic observations were obtained on 2 November 2005 using a VLT-FORS\,2 MXU slit-mask centered on XMMU\,J0104.4-0630 for a total exposure time of 60\,minutes (see Chap.\,\ref{c8_SpecAnalysis}). 
The incomplete execution of the program, originally scheduled for 3\,hours,  and the rather poor seeing conditions (about 2\arcsec)  resulted in  a data quality that allowed a redshift determination for only about half the  targeted sources. However, seven galaxies were found at the same redshift and could thus be identified as 
secure cluster members (see Fig.\,\ref{f10_Spectosc_cl15}, right panel), yielding a cluster redshift for XMMU\,J0104.4-0630 of $z\!=\!0.947\!\pm\!0.005$.  Eight additional galaxies (blue circles in the left panel) are classified as tentative members, with indications of the D4000 break at the expected position but significant contamination from telluric absorption and sky emission lines.


\subsection{Discussion}
\label{s10_cl15_discussion}

\subsubsection{Cluster XMMU\,J0104.4-0630}

\enlargethispage{4ex}

\begin{figure}[b]
\begin{center}


\includegraphics[width=\textwidth]{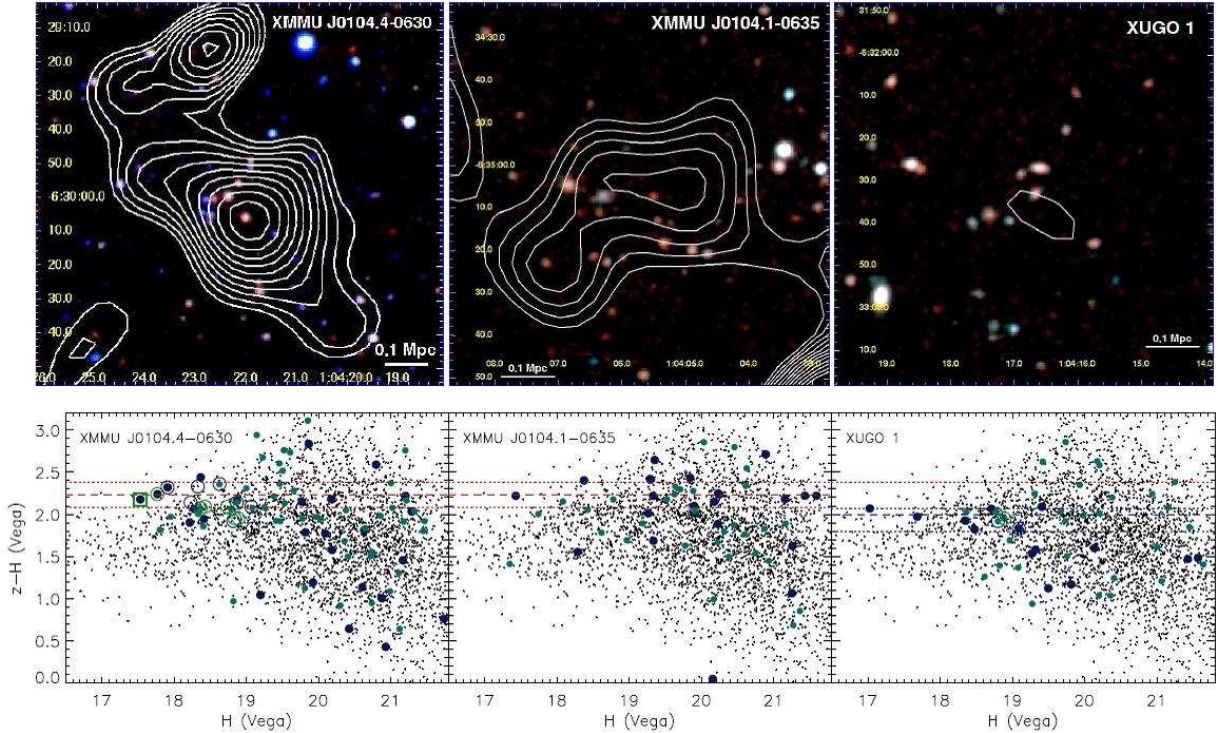}

\end{center}
\vspace{-2ex}
\caption[CMD of XMMU\,J0104.4-0630]{
Color composites and color magnitude diagrams for different target regions of the field.
{\em Top left:} $2\arcmin\!\times\!2\arcmin$ \ R+Z+H color image of XMMU\,J0104.4-0630 at $z\!=\!0.947$ with white X-ray contours overlaid. {\em Top center:} $1.5\arcmin\!\times\!1.5\arcmin$ \ Z+H color composite of the second X-ray selected cluster XMMU\,J0104.1-0635 with the same redshift estimate from its Z--H \reds.
{\em Top right:}
$1.5\arcmin\!\times\!1.5\arcmin$ \ zoom on one of the X-ray undetected overdensities of bluer galaxies in the center of the field. %
{\em Bottom row:}
Z--H CMDs of the three objects shown in the top panels.
Filled dark symbols indicate objects within a 30\arcsec \ radius from the center, filled green symbols are from 0.5\arcmin--1\arcmin, black dots are all objects in the field. {\em Left and center:} CMDs of the two X-ray selected galaxy clusters with concordant red-sequence colors. The dashed horizontal lines indicate the predicted model color for passively evolving galaxies  at $z\!=\!0.95$, the dotted red lines define an interval of $\pm 0.15$ magnitudes around the red-sequence model color represented as the red-galaxy population in Fig.\,\ref{f10_cl15_gal_OD}. Spectroscopically confirmed cluster members of XMMU\,J0104.4-0630 are indicated in the left panel by open symbols. The BCG is marked by the green square, the six additional secure members by green circles, and the eight tentative cluster members by blue open circles.
{\em Right:} CMD of the galaxy overdensity between the two X-ray clusters with a bluer red-sequence color. The three galaxies there  with available spectroscopy are indicated by open circles.  The dashed blue line is shown at the color of the apparent locus, the blue dotted lines define the color cuts for the bluer galaxy population in Fig.\,\ref{f10_cl15_gal_OD}.
}  
\label{f10_CMDs_cl15}
\end{figure}

\noindent
XMMU\,J0104.4-0630 (Fig.\,\ref{f10_CMDs_cl15}, top left)  with a 0.5--2.0\,keV luminosity of $L_\mathrm{x}\!=\!(6.4 \pm 1.3) \times  10^{43}$\,erg\,s$^{-1}$ is an intermediate mass cluster of $M_{500}\!\sim\!1.1 \times 10^{14}\,M_{\sun}$ and $T_{\mathrm{X}}\!\sim\!3$\,keV, based on scaling relations following Finoguenov \etal \ \cite*{Alexis2006a}. 
The system has a compact and fairly regular X-ray appearance with slight extensions to the North-East and South-West.

The cluster core is dominated by early-type red galaxies as in low redshift clusters. The X-ray center coincides with the second brightest galaxy (ID 2 in left panel of Fig.\,\ref{f10_Spectosc_cl15}). The brightest cluster galaxy and the densest part of the cluster core exhibit an offset of about 10\arcsec \ ($\sim\!80$\,kpc) to the North-East. 
The color of the cluster red-sequence of Z--H$\simeq$2.2 (Vega) is fully consistent with model predictions \cite{Fioc1997a} of solar metallicity passively evolving galaxies with formation redshift $z_\mathrm{f}\!\sim\!5$ (see left panel of Fig.\,\ref{f10_CMDs_cl15} and Fig.\,\ref{f7_CMD_models}). 
The  spectroscopic confirmation of  XMMU\,J0104.4-0630 at a redshift very close to the photometric Z--H \reds estimate   
is hence a successful test for the new method. The BCG 
of this cluster is within 0.1\,mag of the expected color for the passive evolution model as indicated by the green square in the CMD of Fig.\,\ref{f10_CMDs_cl15}.
\pagebreak


\begin{table}[t]    
\label{t10_v2_cl15_prop}
\centering
\begin{tabular}{| c | c | c | c | c | c | c | }
\hline 

{\bf ID} & {\bf Name} & $\mathbf{z}$  & $\mathbf{f_{\mathrm{\bf X}}}${\bf \footnotesize{(0.5--2.0\,keV)}} & $\mathbf{r_{\mathrm{\bf c}}}$  &  $\mathbf{L_{\mathrm{\bf X}}}${\bf \footnotesize{(0.5--2.0\,keV)}}  \\ 

 & &    & $10^{-14}$ erg\,s$^{-1}$cm$^{-2}$ & \arcsec &  $10^{43}$ erg\,s$^{-1}$  \\ 
 
\hline\hline 

A & XMMU\,J0104.4-0630 & $0.947\pm 0.005$   & $1.7  \pm 0.3$ &  $18.8 \pm 0.8$  & $6.4\pm 1.3$  \\
B & XMMU\,J0104.1-0635  &  $0.95\pm 0.05$    & $1.2  \pm 0.4$       &  $45 \pm 4$      &  $4.4\pm 1.4$  \\

\hline
\end{tabular}
\caption[X-ray Properties of XMMU\,J0104.4-0630]{Properties of the two newly discovered X-ray clusters.}

\end{table}


\clearpage

\begin{figure}[h]
\begin{center}
\includegraphics[width=\textwidth]{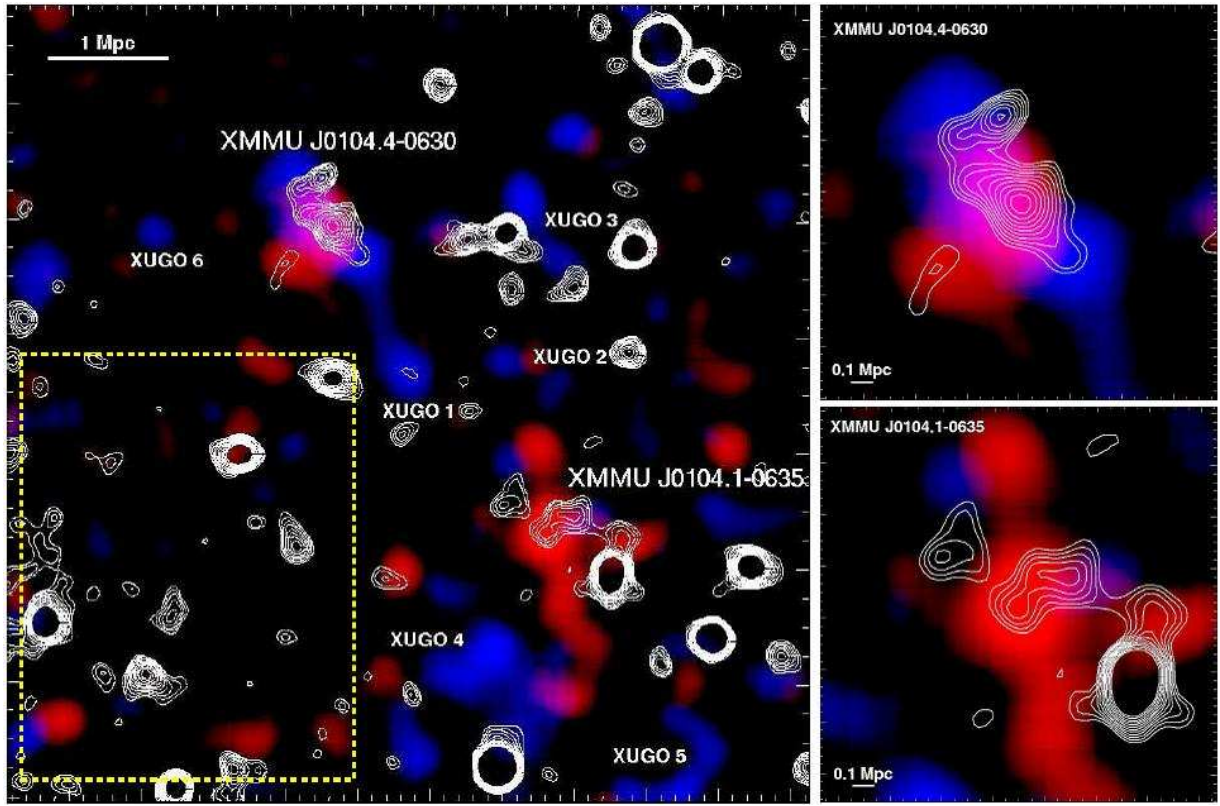}
\end{center}
\caption[Galaxy Overdensities in the Field of XMMU\,J0104.4-0630]{Red and blue galaxy overdensities in the field of XMMU\,J0104.4-0630.
{\em Left:} $14\arcmin\!\times\!14\arcmin$ field-of-view of galaxy overdensities smoothed with a 150\,kpc-kernel  of the {\em red\/} and {\em blue\/} galaxy populations (as defined in Fig.\,\ref{f10_CMDs_cl15}) encoded with the corresponding color; X-ray contours are overlaid in white. The  color cuts for both populations span a 
range of 1.5--5 $\sigma$  relative to the control field in the lower left quadrant (yellow dashed region), which served as the control field. Labels are showing the locations of the two X-ray selected galaxy clusters corresponding to 7-$\sigma$ peaks in the red population, and optically selected 3-6 sigma peaks (XUGO 1--6) in the blue population.   
{\em Right:} $3.5\arcmin\!\times\!3.5\arcmin$  zoom on the two X-ray clusters.

}  
\label{f10_cl15_gal_OD}
\end{figure}

The cluster center has been identified  as the position of a 1.4\,GHz radio source with a flux of $11.9 \pm 1.0$\,mJy in the NVSS survey \cite{Condon1998a}, which is likely to be  attributed to a radio galaxy with a radio power $P_{1.4\,\mathrm{GHz}}\!\sim\!5\!\times\!10^{25}$\,W\,Hz$^{-1}$. 
The positional error circle of a few arcseconds radius includes the two brightest cluster galaxies. The fairly strong radio emission could be an indication of cooling core activity in the cluster center. 
In any case, radio emission of this order in high-$z$ clusters is a prime concern for the cluster selection of upcoming SZE-surveys \cite{Lin2007a} and requires further detailed  studies (see Sect.\,\ref{s2_SZE} \& Sect.\,\ref{s11_SZE}).

\begin{figure}[t]
\begin{center}
\includegraphics[width=\textwidth]{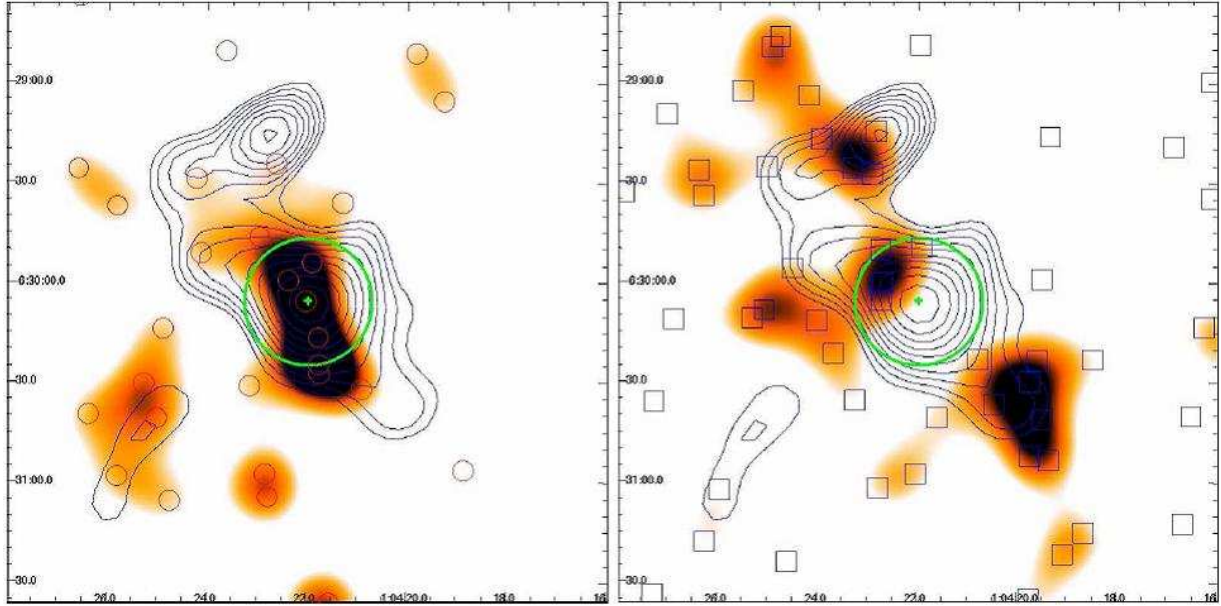}
\end{center}
\caption[Galaxy Populations of XMMU\,J0104.4-0630]{Galaxy populations of XMMU\,J0104.4-0630 at $z\!=\!0.947$.
{\em Left:} $3\arcmin\!\times\!3\arcmin$ view on the {\em red\/} galaxy population smoothed with a 75\,kpc kernel and displayed in 
color cuts spanning overdensities of 2--5 $\sigma$. Red circles indicate the positions of individual galaxies, the green cross marks the X-ray determined cluster center of the blue contours, and the circle represents the cluster core radius of $r_{\mathrm{c}}\!\simeq\!19\arcsec$, corresponding to 150\,kpc. 
{\em Right:} Same FoV and representation for the {\em blue\/} galaxy population.

}  
\label{f10_red_blue_cl15}
\end{figure}

\subsubsection{Detection of the second X-ray cluster XMMU\,J0104.1-0635}


\noindent
The CMD of the second X-ray selected cluster  XMMU\,J0104.1-0635
reveals a \reds color fully consistent 
with the same redshift as
the main cluster A, but with a fainter galaxy population. The overdensity of red galaxies (see Fig.\,\ref{f10_cl15_gal_OD}) selected with a (red) color-cut of $2.23\,\pm\,0.15$ (corresponding to the photometric color uncertainty at H$\sim$20.2) shows a 7\,$\sigma$ peak (6.5\,$\sigma$ for cluster A) relative to the mean density and standard deviation in the 45 square arcminute control field in the lower left quadrant. We thus assign a photometric red-sequence redshift of $z\!=0.95\!\pm\!0.05$ to cluster B. We propose that cluster A and B are physically associated with each other, forming a double cluster with a projected separation in the plane of the sky of 3\,Mpc. 
   A literature redshift of $z\!=\!0.932$ from the XMM-Newton Serendipitous Survey \cite{Barcons2002a} for a BL\,Lac object 75\arcsec \ to the South-West of cluster B
  supports the idea of a large-scale structure filament along the axis of the two systems (see lower right corner of Fig.\,\ref{f10_Spectosc_cl15}).  
  Currently only two X-ray luminous double clusters at higher redshifts are reported in the literature 
(\eg \ Hashimoto \etal, 2005; Nakata \etal, 2005).
\nocite{Hashimoto2005a}
\nocite{Nakata2005a}
The photometric identification of XMMU\,J0104.1-0635 confirms the efficacy of the X-ray detection and selection procedure (Chap.\,\ref{c6_XrayAnalysis}) even for faint extended  sources with large core radii (see Sect.\,\ref{s9_Selection_Funct}).

\subsubsection{Associated large-scale structure environment}

\enlargethispage{-12ex}


The center of the proposed 3\,Mpc-scale cluster-cluster-bridge is marked by an additional significant overdensity of slightly bluer galaxies, with very weak X-ray emission well below the detection threshold for extended sources. 
The color composite and CMD of this X-ray 
Undetected Galaxy Overdensity (XUGO) are shown in the right panels of Fig.\,\ref{f10_CMDs_cl15}  revealing a tight red-sequence at a $\sim\!0.25$\,mag bluer color. 
Without additional spectroscopy, there are two plausible scenarios for this object: (i)  
XUGO\,1 is a foreground group or low-mass cluster at  $z\!\sim\!0.7$ based on the observed Z--H color, or (ii)
this galaxy overdensity is part of the LSS at $z\!\simeq\!0.95$ with a systematically bluer galaxy population. 
This speculative interpretation is supported by three tentative cluster redshifts within 30\arcsec \ (see right lower panel of Fig.\,\ref{f10_CMDs_cl15}) and the suggestive geometric alignment with the two X-ray clusters. 

In order to investigate the nature of this system, we set a second (blue) adjoint color-cut at $1.8\!\le\!\textrm{Z}-\textrm{H}\!<2.08$, corresponding to the blue dotted lines in Fig.\,\ref{f10_CMDs_cl15}. The spatial distribution of  color selected objects was smoothed with a Gaussian kernel of 150\,kpc physical scale (19\arcsec), the approximate core radius of cluster A. The resulting overdensity plot of the $14\arcmin\!\times\!14\arcmin$ \ field-of-view is shown in Fig.\,\ref{f10_cl15_gal_OD}, displayed with  
color cuts ranging from  1.5 to 5 $\sigma$ overdensity relative to a control field in the lower left quadrant. The two brightest peaks in the red galaxy population mark the X-ray selected clusters and are well-centered on the diffuse X-ray emission. On the other hand, the optically selected  system (XUGO\,1) corresponds to a 4.5-$\sigma$ peak of the bluer population. Figure.\,\ref{f10_cl15_gal_OD} (left panel) shows that this population can be traced to the outskirts of the main cluster A, where blue 5-$\sigma$ peaks to the NE and SW enclose  the cluster center along the axis laid out by the X-ray morphology. The drop to 3.5 $\sigma$ in the very center of cluster A in conjunction  with the rising  red overdensity to 6.5 $\sigma$ (see Fig.\,\ref{f10_cl15_gal_OD}, top right) suggests that this is a direct consequence of the proposed scenario of delayed star formation quenching in lower density environments (\eg \ Thomas \etal, 2005; Cucciati \etal, 2006).
\nocite{Thomas2005a}
\nocite{Cucciati2006a}
Under the burst model assumption, the observed  0.15--0.3\,mag  Z--H color shift towards the blue is consistent with an age difference of the stellar populations of 1.2--2.2\,Gyr. 
In Figure\,\ref{f10_red_blue_cl15}, the two different galaxy populations of XMMU\,J0104.4-0630 are displayed in separate panels showing the red cluster core galaxies in the left panel and the proposed filament population on the right.

Systematic blue shifts of the S0 population have been observed in several clusters at lower redshift (\eg \ van Dokkum \etal, 1998; Abraham \etal, 1996) and at $z\!\sim\!1.1$ \cite{Mei2006b}. 
\nocite{VanDokkum1998a}
\nocite{Abraham1996a} 
The observed {\em blue\/} population in Fig.\,\ref{f10_cl15_gal_OD}, which includes the spectroscopic cluster members with IDs 3, 4, 6 and two additional tentative  members, could hence be interpreted as an evolving S0 population.  
However, firm conclusions on physically associated filaments and member galaxies of XMMU\,J0104.4-0630
will only be possible with additional spectroscopic observations of the field.  


\section{Summary and Conclusions}



\begin{itemize}   

\item In Sect.\,\ref{s10_formation_epoch}, the Z--H \reds method was calibrated using ten spectroscopically confirmed clusters.
It was shown that the ($z_{\mathrm{f}}\!=\!5$, $Z\!=\!Z_{\sun}$)-passive evolution model is fully consistent with all observations and was therefore applied as reference model for the photometric redshift estimation of the full cluster candidate sample.

    \item  The observed Z--H  \reds color evolution 
    over the redshift baseline $0.2\!\la\!z\!\la\!1.5$ was used to simultaneously constrain the average metallicity and the  
    formation epoch of \reds galaxies yielding  ${z_{\mathrm{f}}\!=\!4.2\!\pm\!1.1}$ and     $Z\!=\!(1.2\!\pm\!0.4)\,Z_{\sun}$. These early results, based on only ten reference objects, confirm the well-established old age of the stellar populations of \reds galaxies. Using additional spectroscopically confirmed high-redshift clusters of the ongoing XDCP follow-up program, the  Z--H  \reds method has the potential to put stringent constraints on the early-type galaxy formation epoch in the near future.




      \item In Sect.\,\ref{s9_cl15_analysis}, details of the newly discovered X-ray luminous cluster of galaxies         XMMU\,J0104.4-0630 at redshift $z\!=\!0.947\!\pm\!0.005$ were  presented.
       The compact, intermediate mass cluster is found to be in an evolved state and hosts a strong central radio source.
 It was shown that the 
 cluster shows a pronounced stratification of galaxy populations. 
 Whereas the  spatial distribution of the red-sequence population of early type galaxies coincides well with the X-ray emission,  a significantly bluer population dominates beyond 1--2 core radii from the center, suggesting a cluster environment-driven effect of differential galaxy evolution, \eg \  delayed star formation quenching in the outskirts.


\item The Z--H color of the cluster \reds was shown to be in good agreement with the  predictions of the reference model and confirmed the applicability of the method for reliable photometric redshift estimates.

      \item A second  X-ray selected cluster, XMMU\,J0104.1-0635, 6.4\arcmin \ to the South-West could be photometrically identified at $z\!\simeq\!0.95$.
It was speculated that this object is part of the large-scale structure environment of the main cluster, which could include additional optically selected galaxy overdensities.




\end{itemize}      




\chapter{Preliminary Studies and Science Outlook}
\label{c10b_science_outlook}

\noindent
The available Z--H follow-up observations aimed primarily at the cluster {\em identification\/} and {\em confirmation\/} of the principal strategy in terms of limiting depth, photometric precision, and  the choice of filter bands. 
However, the application of the calibrated Z--H-redshift relation to the full distant cluster candidate sample allows a preliminary study of   
the luminosity evolution of the {\em brightest\/} cluster galaxies in the  redshift range $0.2\!\la\!z\!\la\!1.5$, presented  in Sect.\,\ref{s10_bcg_assembly}. 
Section\,\ref{s10_redsequ_population} provides a brief outlook on upcoming investigations of the {\em faint\/} end of the cluster red-sequence. 
The chapter ends with  a gallery of selected, newly discovered high-redshift systems awaiting spectroscopic confirmation.



\section{The Assembly History of Brightest Cluster Galaxies}
\label{s10_bcg_assembly}

\noindent
In this section, the cosmic evolution of the brightest cluster galaxies is investigated. Background information and  the latest simulation results on BCGs have been provided in Sect.\,\ref{s2_BCGs}. The Hubble diagram (see Equ.\,\ref{e3_hubble_diagram} and Sect.\,\ref{s3_cosmo_tests})   will be applied as a diagnostic tool for BCG evolution in a {\em fixed\/} concordance model cosmology. Absolute BCG H-band magnitudes out to $z\!\sim\!1.5$ are derived and compared to predictions of passive evolution models, which have been shown to be very successful for the description of the Z--H color evolution in Sect.\,\ref{s10_formation_epoch}.

\subsection{BCG sample and selection}
\label{s10_BCG_selection}


\noindent
The preliminary study presented here uses the full Calar Alto sample (see Fig.\,\ref{f9_CA_sample}) of X-ray selected clusters, which fulfilled these selection criteria: (i) a significant detection of an optical cluster counterpart  {\em and\/} (ii) a reliable \reds redshift estimate with a maximum accepted uncertainty of $\Delta z\!=\!0.2$.  
The remaining BCG cluster sample contains a total of 63 systems between  $z\!=\!0.18$ and  $z\!\simeq\!1.5$.

\begin{figure}[t]
\begin{center}
\includegraphics[angle=0,clip,width=0.68\textwidth]{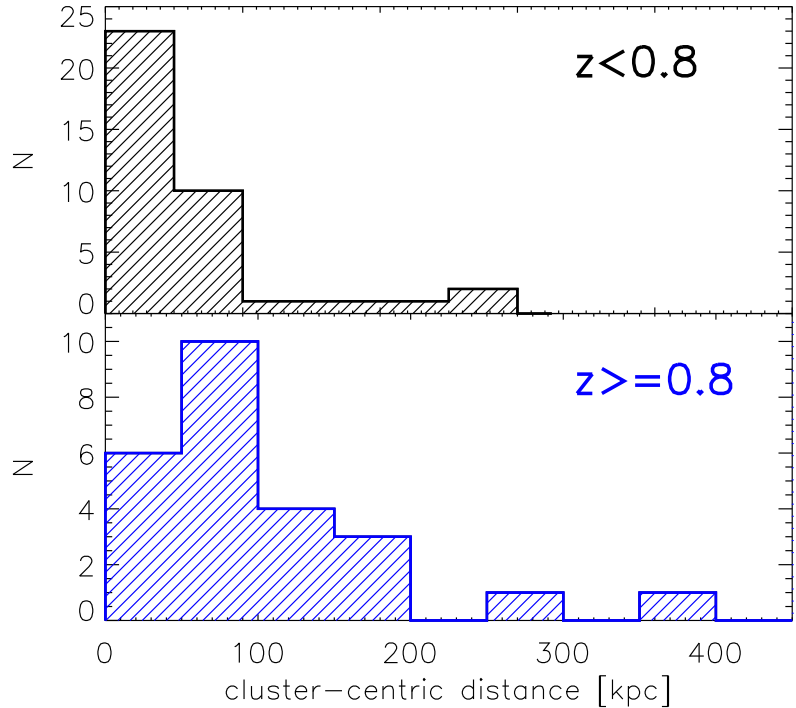}
\end{center}
\vspace{-3ex}
\caption[BCG Selection]{Binned projected cluster-centric distance histogram of the 63 selected BCGs with respect to the X-ray center position. {\em Top:} The BCG locations of the low and intermediate redshift bin with $z\!<\!0.8$ (38 objects) are predominantly coincident with the X-ray marked cluster center. {\em Bottom:} At high redshifts of $z\!\geq\!0.8$ (25 clusters) the distribution is skewed towards slightly larger offsets from the X-ray center. Extreme outliers in both histograms can be mostly attributed to X-ray center biases due to point source flux contaminations.} \label{f10_bcg_clustercentric_dist}
\end{figure}

\enlargethispage{4ex}

The brightest cluster galaxies are selected based on their H-band magnitude, in combination with selection cuts concerning the cluster-centric distance, defined with respect to the X-ray center,  and the Z--H color offset from the red-sequence. The increasing foreground contamination of galaxies with  magnitudes similar  to the BCGs of {\em distant\/} clusters requires tightening identification constraints with increasing cluster-centric distance to ensure a {\em robust} BCG selection. 
In the absence of a spectroscopic identification and  photometric redshifts derived from deep multi-band data, the BCG selection based on cluster-centric distance {\em and\/} reasonable Z--H cuts is currently the best applicable method for the available data. The following selection criteria were applied for the  BCG identification in the 63 X-ray clusters:

\begin{enumerate}
    \item 21 central dominant galaxies within 5\arcsec \ of the X-ray marked center are accepted as BCGs without color cuts;  
    \item 37 identifications are based on galaxies with cluster-centric distances of 5\arcsec--30\arcsec \ {\em and} a color cut of $\pm 0.5$\,mag around the \reds Z--H color;
    \item 5 selected BCGs are located 30\arcsec--60\arcsec \ off-center but are within $\pm 0.3$\,mag of the \reds color.
\end{enumerate}

\noindent
These selection criteria are motivated by the first experiences with the spectroscopically confirmed sub-sample discussed in Sect.\,\ref{s10_Z_H_discussion}. 
However, the robustness of these preliminary selection criteria needs to be investigated and quantified with upcoming
follow-up observations of high-redshift candidates. 

The maximum search radius of 60\arcsec \ corresponds to about 500\,kpc at $z\,\ga\!0.8$. For the majority of {\em X-ray selected\/}  clusters, the identification of BCGs close to the X-ray center is straightforward and fairly unambiguous. The sample contamination level for non-BCG selections is currently expected to be no more than of order 10\%, \ie \  sub-sample averages should be  statistically robust. 
Figure\,\ref{f10_bcg_clustercentric_dist} displays the histogram of binned cluster-centric distances of the identified BCGs for low and intermediate redshift clusters at $z\!<\!0.8$ ({\em top panel}) and for high-redshifts systems at    
$z\!\geq\!0.8$ ({\em bottom panel}). A slight shift towards higher off-center distances can be observed at high redshift as a first tentative indication that BCGs at large lookback times are not yet in their final 
configuration with respect to the cluster center.

\subsection{The BCG H-band Hubble diagram}
\label{s10_BCG_Hubble_Diagram}

\begin{figure}[t]
\begin{center}
\includegraphics[angle=0,clip,width=0.85\textwidth]{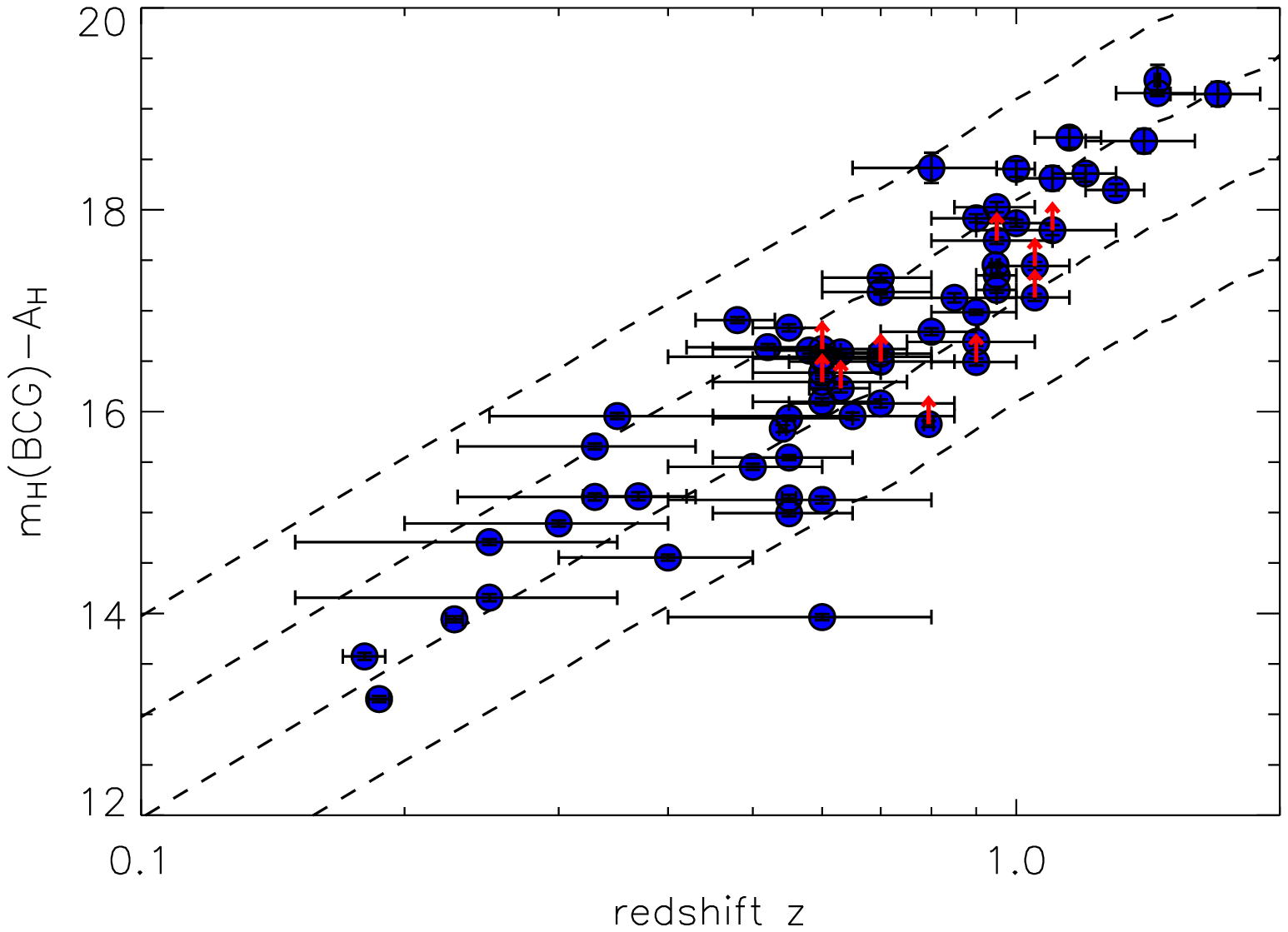}
\includegraphics[angle=0,clip,width=0.85\textwidth]{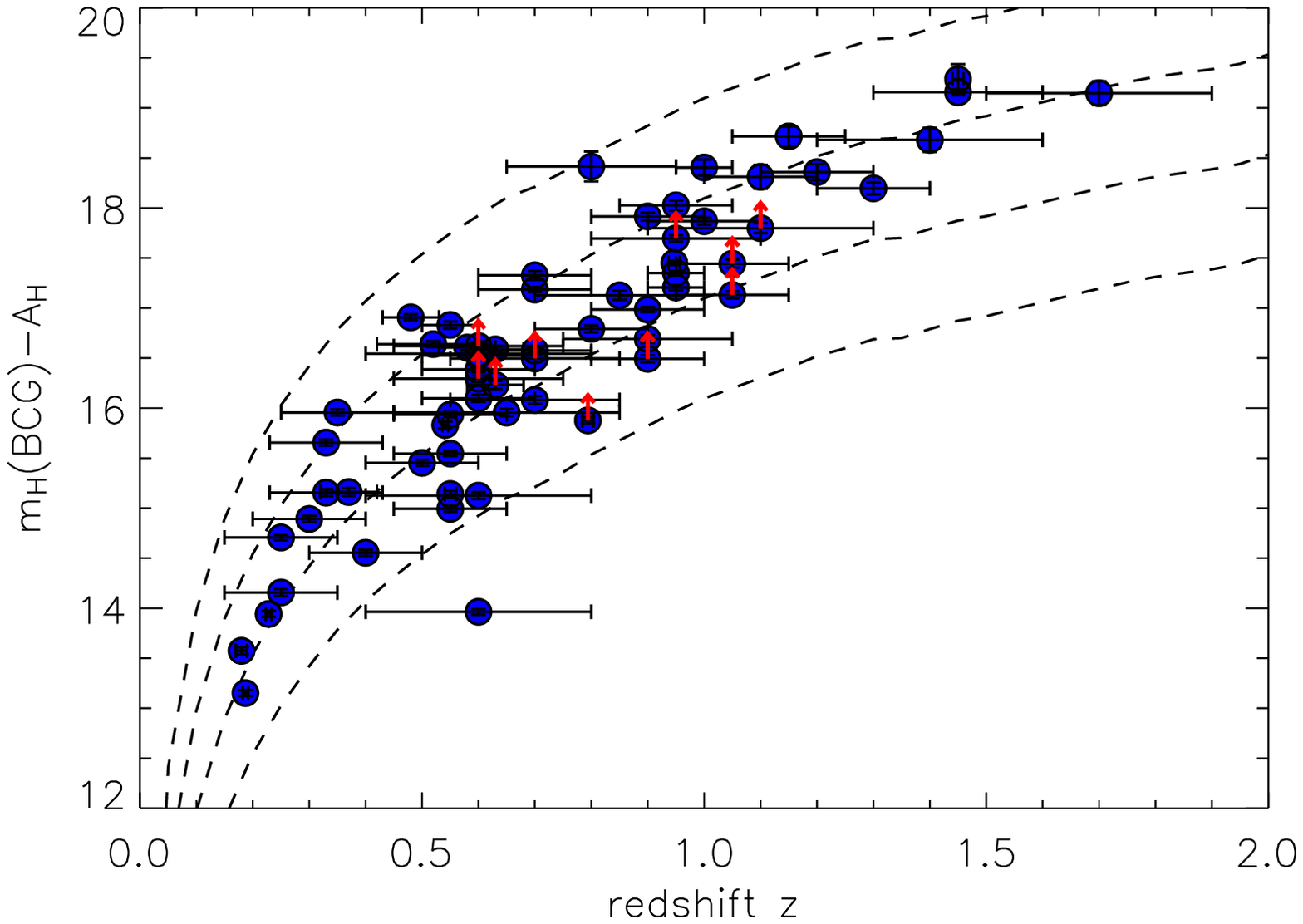}
\end{center}
\vspace{-3ex}
\caption[BCG Hubble diagram]{{\em Top:} H-band Hubble diagram for the  BCGs of 63 X-ray selected clusters between redshifts of $z\!=\!0.18$ and $z\!\simeq\!1.5$ . 
Dashed lines indicate solar metallicity, passive evolution burst model expectations for a formation epoch of $z_{\mathrm{f}}\!=\!5$ and galaxy luminosities of  m*, m*$-$1,  m*$-$2, and  m*$-$3  (top to bottom). BCGs 
with photometry affected by blending 
are marked by red arrows. {\em Bottom:} Same plot with a linear redshift axis.
 } \label{f10_bcg_XDCP_hubble_diagram}
\end{figure}

\noindent
Using the observed apparent {\em total\/} BCG magnitudes and the estimated \reds redshifts, we can construct the H-band BCG Hubble diagram out to $z\!\simeq\!1.5$. H-band galactic extinction corrections  $A_{\mathrm{H}}$ have been applied to  all BCG magnitudes  using the galactic reddening maps\footnote{A galactic extinction calculator is available at \url{http://nedwww.ipac.caltech.edu/forms/calculator.html}.} of Schlegel, Finkbeiner\,\&\,Davis \cite*{Schlegel1998a}.
The final BCG H-band Hubble diagram, \ie \ the $(m_{\mathrm{H}}(\mathrm{BCG})\!-\!A_{\mathrm{H}})$ versus $\log z$ relation, is shown in the top panel of Fig.\,\ref{f10_bcg_XDCP_hubble_diagram}. Black dashed lines indicate the model expectations for solar metallicity, passive evolution burst models with formation epoch $z_{\mathrm{f}}\!=\!5$ for m*, m*$-$1,  m*$-$2, and  m*$-$3 galaxies (top to bottom). Red arrows mark objects with strong signs of object blending of the {\em SExtractor\/} photometry, \ie \ the determined magnitudes are  lower limits (implying upper limits of the luminosity). Photometric errors are typically small, except for blended sources, since all BCGs are at least two magnitudes brighter than the limiting H-band image depth. Errors are hence dominated by the redshift uncertainties of the Z--H \reds method. The lower panel of  Fig.\,\ref{f10_bcg_XDCP_hubble_diagram} shows the same plot with a {\em linear\/} redshift axis, which provides a better illustration of the effects of redshift errors, in particular at low-$z$ where the apparent magnitudes faint rapidly with increasing distance.   

Instead of assuming standard candle properties of the BCGs and applying the Hubble diagram as a cosmological test for parameter constraints (\eg \ Collins\,\&\,Mann, 1998; Arag{\'o}n-Salamanca \etal, 1998), we assume a {\em fixed concordance cosmology\/}  and 
\nocite{Aragon1998a}
\nocite{Collins1998a}
use the Hubble diagram as a tool for BCG evolution studies. Equation\,\ref{e3_hubble_diagram} reveals that variations of the Hubble constant $H_0$ introduce  a  normalization offset, whereas different values for $\Omega_{\mathrm{m}}$,  $\Omega_{\mathrm{\Lambda}}$, and $w$ change the shape of the Hubble relation at high redshifts. However, parameter changes within the concordance model uncertainties introduce variations in the luminosity distance $d_{\mathrm{lum}}$ at the high-redshift end $z\!\sim\!1.5$ of less than 10\%. 




\clearpage

\begin{figure}[t]
\begin{center}
\includegraphics[angle=0,clip,width=0.805\textwidth]{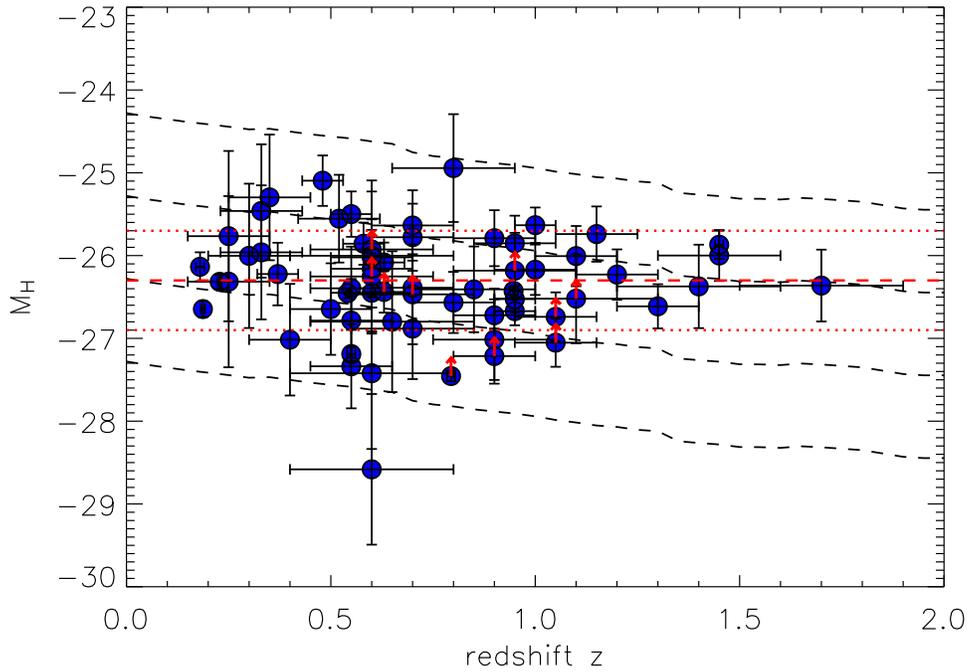}
\end{center}
\vspace{-3.5ex}
\caption[BCG Absolute Magnitudes]{Redshift evolution diagram of the absolute H-band BCG  magnitudes with estimated 1\,$\sigma$ errors. The observations are fully consistent with a non-evolving absolute magnitude of $M_{\mathrm{H}}\!=\!-26.3$\,mag (red dashed line), with an rms scatter of 0.6\,mag (red dotted lines), which is mostly introduced by the photometric redshift uncertainties. Black dashed lines indicate the expected evolution of passive galaxies, which are expected to dim by about one magnitude from redshift $z\!\simeq\!1.5$ to the present epoch. Purely passive BCG luminosity evolution can be excluded at the 2.6\,$\sigma$ level.} \label{f10_bcg_absolute_magnitudes}
\end{figure}

\begin{figure}[b]
\begin{center}
\includegraphics[angle=0,clip,width=0.64\textwidth]{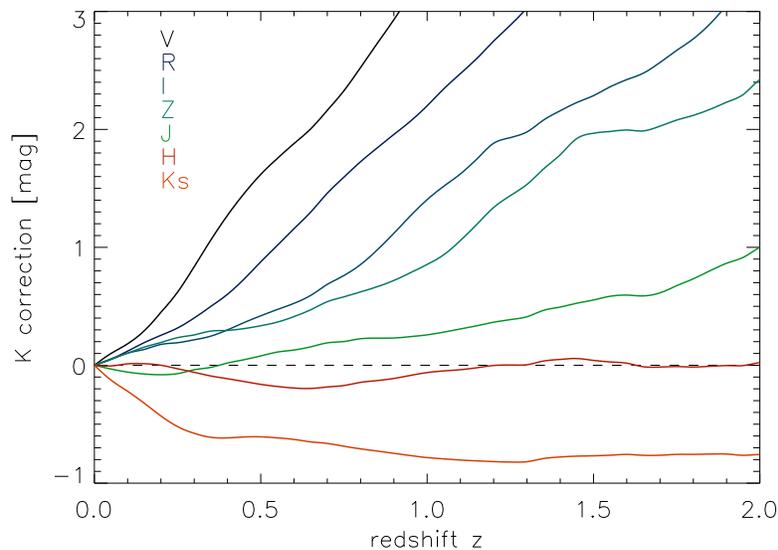}
\end{center}
\vspace{-3.5ex}
\caption[BCG $K$-correction]{
$K$-correction terms as a function of redshift for a $z_{\mathrm{f}}\!=\!5$ passive evolution BCG  model for different bands. The applied H-band correction term (red line) is very small at all relevant redshifts.
} \label{f10_bcg_kcorrection}
\end{figure}

\clearpage

\subsubsection{BCGs as standard candles}

\noindent
As a first application, the absolute H-band magnitudes for the BCGs are computed following Equ.\,\ref{e3_hubble_diagram}

\begin{equation}\label{e10_absolute_MH}
    M_{\mathrm{H}}({\mathrm{BCG}}) =  m_{\mathrm{H}}(\mathrm{BCG})  - A_{\mathrm{H}} - K_{\mathrm{H}}(z) - 25 - 5\log \left(d_{\mathrm{lum}}\, [h^{-1}_{70}\,\mathrm{Mpc}]\right) \ .
\end{equation}

\noindent
The $K$-correction terms\footnote{Models computed by Daniele Pierini.} $K_{\mathrm{H}}$ to transform observed 
$m_{\mathrm{H}}$ magnitudes at redshift $z$ into {\em rest-frame} H-magnitudes have been calculated for the standard $z_{\mathrm{f}}\!=\!5$, solar metallicity passive evolution model and are displayed by the red line in Fig.\,\ref{f10_bcg_kcorrection}.
Note that the H-band observations exhibit the smallest absolute $K$-correction terms of all bands, 
implying that the observed apparent H-band magnitudes are very close to the associated {\em rest-frame\/} values. 


The BCG redshift uncertainties, as the dominant error source for all non-spectroscopically confirmed clusters, enter the error budget of the absolute magnitudes via 
$dM_{\mathrm{H}} = dm_{\mathrm{H}} + 5\,[(d\log d_{\mathrm{lum}})/dz]\,dz$.
In particular at $z\!\la\!0.4$, where the luminosity distance is a steep function of $z$ (see lower panel of Fig.\,\ref{f10_bcg_XDCP_hubble_diagram}), the photometric redshift errors  influence  $dM_{\mathrm{H}}$ in a  magnified way.

The absolute BCG H-band magnitude diagram of Fig.\,\ref{f10_bcg_absolute_magnitudes}, with a full accounting of the error budget included, reveals an astonishing result: the   brightest cluster galaxies of X-ray selected clusters are fully consistent with being true standard candles with constant absolute rest-frame H-band luminosity of $\mathbf{M_{\mathrm{\bf H}}\!\simeq\!-26.3}$\,mag out to ${z\!\simeq\!1.5}$.

This result is statistically very robust. The median {\em and\/} mean values of (i) the {\em full\/} sample of 63 BCGs, (ii) the {\em three lowest redshift\/} clusters (spectroscopically confirmed), and (iii)  the {\em five highest redshift\/} objects all agree within $\pm 0.06$\,mag. The scatter of the full sample around the mean value ${M_{\mathrm{H}}\!\simeq\!-26.3}$\,mag is $\sigma_M\!\simeq\!0.6$\,mag, which is for the largest part to be attributed to the redshift uncertainties.
A linear fit to the full BCG sample yields the formal functional form
$M_{\mathrm{H}} = (-26.23 \pm 0.20) + (-0.10 \pm 0.22) \cdot z$, also fully consistent with a {\em non-evolving\/} absolute H magnitude.

The full BCG sample contains four outliers, 
all of which have associated literature redshifts, \ie \ their estimated   $M_{\mathrm{H}}$ errors are small. The BCGs of the two clusters A\,383  with 
$M_{\mathrm{H}}\!=\!-26.65 \pm 0.1$, and RX\,J0018+1617 with $M_{\mathrm{H}}\!=\!-27.2
 \pm 0.1$ (see Sect.\,\ref{s10_formation_epoch} for references) are likely truly super-luminous systems compared to the sample average.
The remaining two outliers are clusters LCDS\,0504 with spectroscopic redshift $z\!=\!0.794$ and LDCS\,0505 with a photometric literature redshift of $z\!\simeq\!0.48$ from the Las Campanas Distant Cluster Survey \cite{Gonzalez2001a}.
The BCG of LCDS\,0504 with $M_{\mathrm{H}}\!=\!-27.45$ is flagged as a photometric blend 
(see panel 11 of Fig.\,\ref{f10_BCG_gallery_hi}). In the case of LDCS\,0505, with an apparently underluminous BCG of $M_{\mathrm{H}}\!=\!-25.1$, the Z--H \reds color suggests that the published photometric redshift of $z\!\simeq\!0.48$ is underestimated. The latter two outlier objects can thus  be attributed to systematic effects rather than significant intrinsic deviations from the mean absolute BCG  magnitude.  
  
The fact that BCGs are good standard candles has long been established for redshifts of $z\!<\!1$ (see Sect.\,\ref{s2_BCGs}).
The new preliminary result from the sample of this work with the increased redshift baseline out to $z\!\simeq\!1.5$ is the conclusion that the observations are {\bf not consistent with pure passive evolution}. The dashed black lines in Fig.\,\ref{f10_bcg_absolute_magnitudes} indicate the expected fading of the absolute H-band luminosity for passively evolving galaxies by one magnitude from  $z\!\simeq\!1.5$ to $z\!=\!0$, corresponding to an average slope of $dM_{\mathrm{model}}/dz\!\simeq\!-0.67$. 
Comparing this result to the constraints obtained from the linear fit to the observed data of $dM_{\mathrm{H}}/dz\!\simeq\!-0.10 \pm 0.22$, excludes a purely passive BCG luminosity  evolution at the $2.6\,\sigma$~level.




\subsubsection{BCG luminosity evolution}

\begin{figure}[t]
\begin{center}

\includegraphics[angle=0,clip,width=1.0\textwidth]{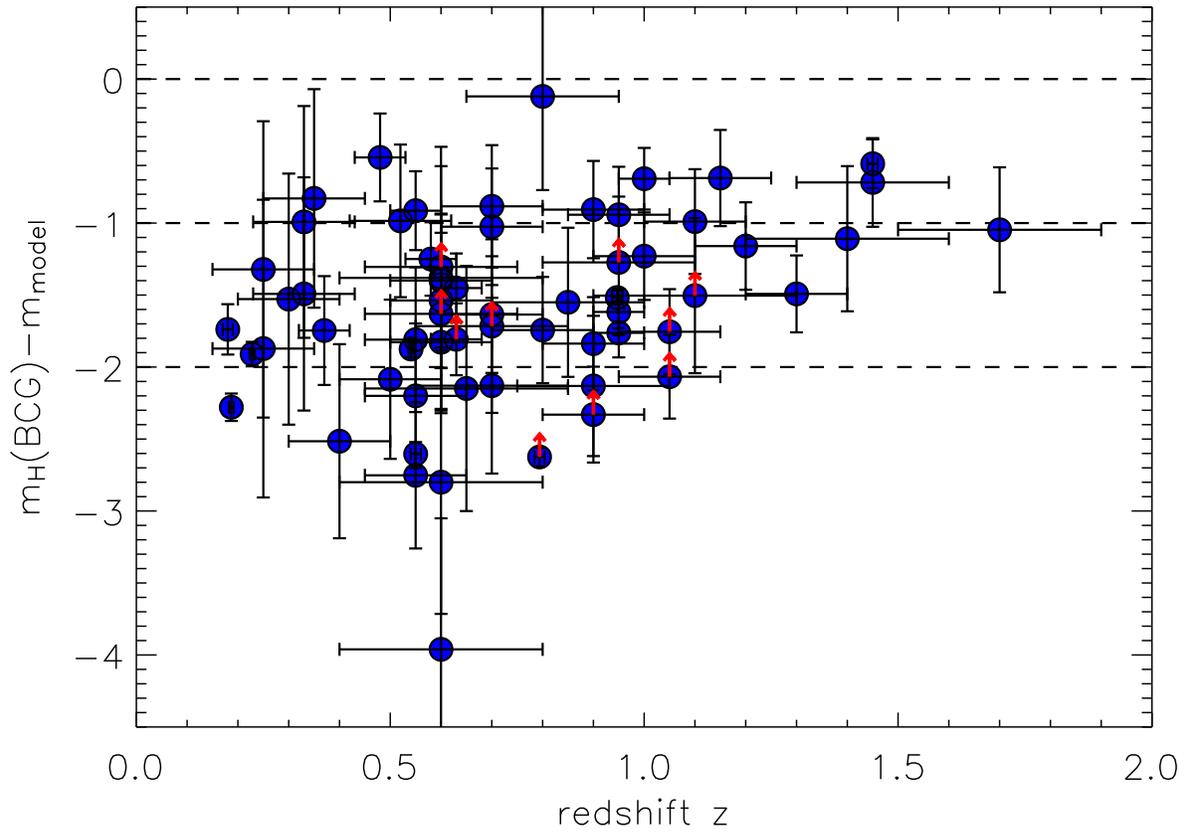}
\end{center}
\vspace{-3ex}
\caption[BCG Luminosity Evolution]{Observed apparent magnitude evolution relative to expectations of passive evolution models.
The expected apparent model magnitudes $m_{\mathrm{mod}}^*(z)$ for an
L$_{\mathrm{mod}}^*$ galaxy  have been subtracted, implying that the two dashed lower lines represent galaxies one (center line) and two (bottom line) magnitudes {\em brighter\/} than M*.} \label{f10_BCG_lum_evolution}
\end{figure}

\noindent
As the next step, the BCG luminosity evolution {\em relative to the passively evolving model\/} is investigated. 
The apparent magnitudes of the standard  $z_{\mathrm{f}}\!=\!5$, solar metallicity  model for an L$_{\mathrm{mod}}^*$ 
galaxy, $m_{\mathrm{model}}$,
are subtracted from the extinction corrected observed magnitudes  $m_{\mathrm{H}}(\mathrm{BCG})\!-\!A_{\mathrm{H}}$. 
The results\footnote{The galactic extinction correction term $A_{\mathrm{H}}$ is omitted in the axis label for better clarity, $m_{\mathrm{H}}(\mathrm{BCG})$ implicitly denotes extinction corrected magnitudes. } are shown in Fig.\,\ref{f10_BCG_lum_evolution}, with the same meaning of the symbols as before. 
In this representation, passively evolving L$_{\mathrm{mod}}^*$ galaxies would be located along the null line (top horizontal dashed line). The bulk of the BCG population is observed in the diagram region in between  galaxies one magnitude (center dashed line) and two magnitudes (lower dashed line) {\em brighter\/} than L$_{\mathrm{mod}}^*$.  
The errors in the model-subtracted magnitudes originate from the redshift uncertainties of the BCGs, $m_{\mathrm{model}}(z\!\pm\Delta z)$, in analogy with Fig.\,\ref{f10_bcg_absolute_magnitudes}.

\begin{figure}[t]
\begin{center}
\includegraphics[angle=0,clip,width=0.85\textwidth]{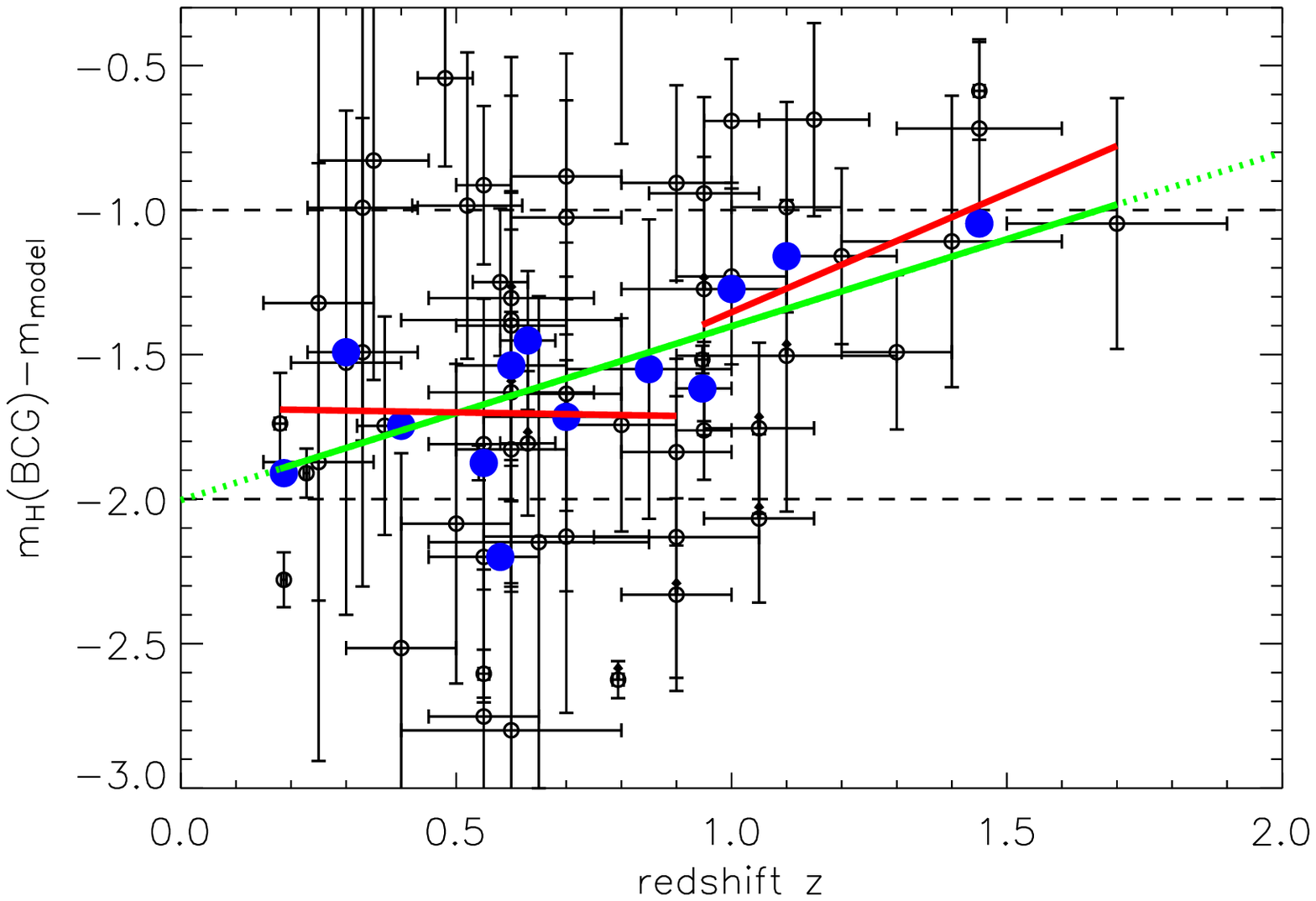}
\includegraphics[angle=0,clip,width=0.85\textwidth]{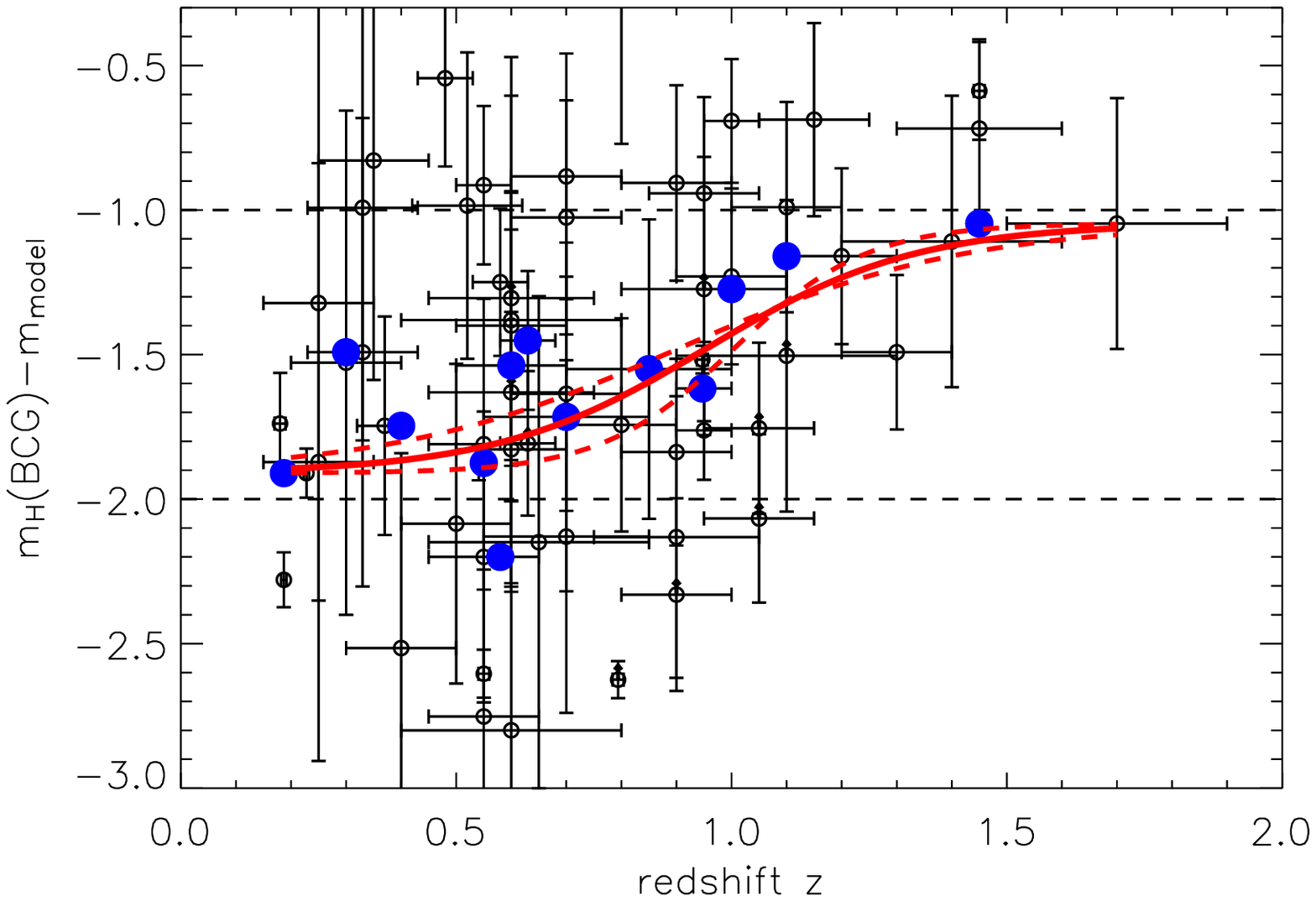}
\end{center}
\vspace{-3ex}
\caption[BCG Assembly Models]{BCG luminosity evolution models compared to observations. {\em Top:} Linear fits to the data over the full sample (green line), and separate linear models for $z\!\leq\!0.9$ and $z\!>\!0.9$ sub-samples (red lines). Black background symbols show the observed data, blue circles represent the medians of independent bins of five objects.
{\em Bottom:} Maximum likelihood model fit for a step function with a softened transition zone given by the red solid line, dashed lines display 1\,$\sigma$ variations of the fit parameters. 
The observations are consistent with a mass  assembly phase of the BCGs centered at $z\!\sim\!0.95$.
Since their colors evolve as for passively evolving stellar populations, the increase in luminosity is driven by an increase in stellar mass.

} \label{f10_BCG_lum_model}
\end{figure}

Since Fig.\,\ref{f10_BCG_lum_evolution} displays the magnitude {\em residuals\/} after accounting for  passive evolution, more negative data points at low redshifts can be interpreted in terms of an {\em increased\/} luminosity, in excess of passive evolution, compared to the high-redshift BCGs.  As a first approach for modelling the residual BCG luminosity evolution, we can assume a linear relationship with redshift of the form 
$m_{\mathrm{H}}(\mathrm{BCG})\!-\!m_{\mathrm{model}}\!=\!A_0\!+\!B\!\cdot\!z$, with fit parameters $A_0$ for the local magnitude difference, and $B$ for the slope. 
A linear fit to the {\em full\/} BCG sample over all redshifts yields the parameter constraints $A_0\!=\!-2.0 \pm 0.2$ and 
$B\!=\!0.60 \pm 0.22$, shown by the green line in the top panel of Fig.\,\ref{f10_BCG_lum_model}. Black symbols with error bars represent the observed BCG data, with one outlier to either side not displayed in the zoomed representation of the central residual magnitude interval of interest. Blue dots symbolize the magnitude and redshift medians of 13 {\em independent\/} bins of five BCGs in order of decreasing redshift (three objects for the lowest $z$ bin). The medians are less affected by outliers compared to averaging, \eg \ due to  blends in the photometry, and should represent a robust  smoothed trend of the observed data as visual guide.
A linear fit to the median smoothed data results in consistent parameters but with decreased uncertainties, ruling out a flat model with passive evolution at the 3.7\,$\sigma$ level, whereas the raw data fit yields a 2.7\,$\sigma$ confidence level. 
 
\newpage

A linear relationship would imply that 2/3 of the luminosity evolution happened at redshifts below unity, at variance with observations (\eg \ Burke \etal, 2000). 
\nocite{Burke2000a}
The apparent discrepancy can be resolved by splitting the BCG sample into a low--intermediate redshift bin at $z\!\leq\!0.9$, approximately corresponding to the range accessible by former BCG analyses, and a high-redshift bin with 18 objects with $z\!>\!0.9$. The linear fits to the two sub-samples yield the parameter constraints      
$A_0\!=\!-1.7 \pm 0.4$, $B\!=\!0.0 \pm 0.5$ for the low redshift bin, and 
$A_0\!=\!-2.2 \pm 0.5$, $B\!=\!-0.8 \pm 0.4$ for the independent high-$z$ sub-sample, displayed by the red lines in the top panel of Fig.\,\ref{f10_BCG_lum_model}. The observed low--intermediate redshift BCGs at $z\!\leq\!0.9$ are hence consistent with a passive evolution model, although with increased errors, in agreement with previous findings. At $z\!>\!0.9$, a significant jump in the residual BCG magnitudes is apparent.   
Note that the non-passive evolution  
signal is present for  the three rightmost blue points, representing the median values of the 15 highest redshift BCGs. The observed effect is hence robust against the removal of individual high-$z$  objects from the sample.   

\newpage

 






\clearpage

The observed discontinuity for the split sample 
suggests the application of 
a step function model with a gradual transition zone,  in analogy  to a Fermi function. The functional form can then be parameterized as 

\begin{equation}\label{e10_fermi_model}
    m_{\mathrm{H}}(\mathrm{BCG})\!-\!m_{\mathrm{model}} = (A_0 + \Delta m) - \frac{\Delta m}{1 + \exp (\frac{z-z_{\mathrm{a}}}{b})} \ ,
\end{equation}

\noindent
where $A_0$ is the local magnitude residual, $\Delta m$ the step height, $z_{\mathrm{a}}$ the central BCG `assembly' redshift, and $b$ a softening parameter that determines the width of the transition zone. Since the data is not sufficient to provide meaningful constraints for a functional fit with {\em four\/} free parameters, the end points are fixed at $A_0\!=\!-1.91$ (at $z\!\simeq\!0.2$) and $\Delta m\!=\!0.86$, consistent with the results of a constant absolute luminosity in Fig.\,\ref{f10_bcg_absolute_magnitudes} and  the linear fit in Fig.\,\ref{f10_BCG_lum_model} (green line in top panel).

A maximum likelihood fit to the data\footnote{The magnitude errors of the four outliers identified in the absolute magnitude diagram of Fig.\,\ref{f10_bcg_absolute_magnitudes} have been 'softened' to 1\,$\sigma$ deviations to prevent the likelihood fit being dominated by individual objects (with systematics).} with  two remaining free parameters  yields the constraints $z_{\mathrm{a}}\!=\!0.96\pm 0.05$ for the central `assembly' redshift and $b\!=\!0.19\pm 0.07$ for the transition zone width. The solid red line in the bottom panel of Fig.\,\ref{f10_BCG_lum_model} displays the best fit  model, with the effects of 1\,$\sigma$ parameter variations shown by the dashed red lines. 
The reduced $\chi^2$ value of 1.78 for this softened step function model 
is improved compared to the linear fit (green fit) with a $\chi^2$ of 1.89. 
The  luminosity evolution fit shown in the lower panel of Fig.\,\ref{f10_BCG_lum_evolution} is currently  the best preliminary model for the available BCG data, consistent with only small deviation from passive evolution at low redshifts as reported by former studies.




\subsubsection{Biases and systematics}

\noindent
We will now address the question whether the observed deviations from  passive BCG luminosity evolution models could be an artefact introduced by systematic effects and selection biases. The main areas of concern are briefly discussed in the following:

\begin{description}
    \item[Non-Representative Cluster Sample:] 
    Using a heterogeneous cluster sample with varying object properties as a function of redshift can introduce biases that mimic cosmic evolution. In fact, Burke, Collins\,\&\,Mann \cite*{Burke2000a} showed that the initial claim of a non-passive BCG luminosity evolution of  Aragon-Salamanca, Baugh\,\&\,Kauffmann \cite*{Aragon1998a} could be attributed to the bias of mixing an optically selected high-$z$ sample and an X-ray selected low-redshift sample. Owing to the observed weak correlation of cluster mass and BCG  luminosity  $L_{\mathrm{BCG}}\!\propto\!M^{\sim\!1/4}_{200}$ (see Sect.\,\ref{s2_BCGs}), the target selection from high-redshift clusters with systematically lower mass, and hence lower X-ray luminosity, introduces the {\em selection artefact\/} of intrinsically fainter distant BCGs.  
    
The cluster sample presented here is X-ray selected at all redshifts, introducing a general selection bias with the opposite sign, \ie \ an effective `{\em anti-bias}'. At a given approximate flux limit, the minimum cluster luminosity is increasing with redshift according to $L_{\mathrm{X}}\!\ga\!4\pi f_{\mathrm{X,lim}}\,d^2_{\mathrm{lum}}$, implying an increasing minimum cluster mass for the 58 serendipitously detected objects. The five additional XMM target clusters (see Sect.\,\ref{s9_imaging_status}) at low and intermediate redshifts are high X-ray luminosity, massive clusters, of which three were identified  as BCG outliers on the bright end. The X-ray selection bias of finding predominantly low mass clusters and groups at lower redshifts (\eg \ Pacaud \etal, 2007)  contributes to the observed increased scatter at low-$z$ in Fig.\,\ref{f10_bcg_absolute_magnitudes}. The detected non-passive BCG luminosity evolution of $\Delta m\!\simeq\!0.9$\,mag is hence to be interpreted as a {\em lower limit\/}, with a {\em larger\/} effect to be expected using a representative cluster sample.    
\nocite{Pacaud2007a} 
In other words, we find distant BCGs to be {\em fainter\/} than expected from passive evolution, 
despite the fact that
the high-redshift objects have an X-ray selection bias towards {\em intrinsically brighter\/} absolute magnitudes, \ie \
we observe a {\em positive\/} slope in Fig.\,\ref{f10_BCG_lum_model} but would expect a {\em negative} correlation. The X-ray selection bias  effectively increases the confidence in the detection of non-passive BCG evolution, but has to be taken into account for the comparison with unbiased predictions from simulations.



    \item[BCG Selection:]
 The BCG selection has been discussed in Sect.\,\ref{s10_BCG_selection}. While individual misclassifications based on the two-band data are possible, the sample statistics should be robust. A significant bias is only expected if a large fraction of distant BCGs is either located at cluster-centric distances of $>$500\,kpc, or  the BCG color is more than 0.5\,mag bluer than the red-sequence; both scenarios seem unlikely based on known cluster populations. Since the BCG identification is more difficult at high redshift, the current approach needs to be justified once additional  data is available leading to possibly refined selection criteria. 


 \item[Cluster Redshift Estimates:]
To be consistent with passive evolution models, the redshifts of the high-$z$ clusters would have to be significantly {\em underestimated\/} (see lower panel of Fig.\,\ref{f10_bcg_XDCP_hubble_diagram}). This would be nice from the distant X-ray cluster discovery point-of-view, but is very unlikely given the high redshift-sensitivity of the Z--H technique at $z\!\ga\!0.9$ (see Sect.\,\ref{s7_performance_expectations}). 
Using the `red-envelope' in the color-magnitude diagram as distance indicator for high-$z$ clusters (see Sect.\,\ref{s10_Z_H_discussion}) could possibly lead to a slight overestimation of the redshift for some distant clusters, which would again lead to an {\em anti-bias\/} in the sense that the distant BCGs are intrinsically even fainter than measured. 
Note that the spectroscopically confirmed cluster XMMXCS\,J2215.9$-$1738 at $z\!=\!1.45$ exhibits the faintest BCG of the whole sample.
     
    \item[Cosmological Model:] 
    As discussed, a modified Hubble parameter $H_0$ only changes the normalization of the Hubble diagram, but not its shape (see Equ.\,\ref{e3_hubble_diagram}). Varying $\Omega_{\mathrm{m}}$ within reasonable limits of $0.2\!\la\!\Omega_{\mathrm{m}}\!\la\!0.4$ changes the luminosity distance of a flat cosmology $d_{\mathrm{lum}}$ by less than 10\% at $z\!\sim\!1.5$, which would result in apparent magnitude differences at the highest redshifts of $\Delta m\!\la\!0.2$\,mag (see Equ.\,\ref{e7_flux_approximation}), \ie \ less than 1/4 of the observed magnitude offset. 
    
    \newpage
        
    \item[Cosmological Surface Brightness Dimming:] 
    The severe cosmological surface brightness dimming (Equ.\,\ref{e3_SB_dim}) can introduce significant selection biases for very distant ($z\!\ga\!2$) galaxy studies close to the detection limit \cite{Lanzetta2002a}. However, all galaxies of the BCG sample are at least two magnitudes above the H-band completeness limit, implying that surface dimming should not play a significant role.     
    \item[Biased Photometry:] 
    The H-band BCG data is taken from a homogeneous data set, observed with the same instrument and reduced in a standardized way. Great care has been taken to optimize the NIR sky subtraction to avoid background  biases, in particular for faint, low surface brightness objects  (see Sect.\,\ref{s7_NIR_pipeline}). The {\em SExtractor\/} photometry software is well tested and robust at  the BCG magnitudes well above the detection threshold. {\em SExtractor\/} systematics due to source blending effects have been discussed and are marked in diagrams by arrows to indicate lower limits. {\em Without\/} source blending, the affected BCGs in the range $0.5\!\ga\!z\ga\!1.1$ could be intrinsically slighter fainter, with minor effects on the overall form of the luminosity evolution.       
    
    \item[Passive Evolution Models:] 
BCGs typically exhibit red, old stellar populations, and are in most cases the top ranked galaxies on  the \reds (see Sect.\,\ref{f10_calibration_clusters}). The assumption that the stellar populations of BCGs evolve passively seems therefore to be well justified. In Sect.\,\ref{s10_formation_epoch}, we could show that the \reds Z--H color evolution out to the highest accessible redshifts is fully consistent with passive evolution models with an early formation epoch. In addition, the X-ray selection process {\em disfavors\/} clusters with central (strongly non-passive) AGN activity with their associated point-like X-ray flux contribution.  
For the BCGs in our {\em X-ray selected, \reds confirmed cluster sample\/}  passive evolution is a good description for the colors of their stellar populations. This means that recent star formation episodes or moderately strong active nuclei are expected to play a minor role.

For the comparison to passive evolution expectations, the {\em standard}  $z_{\mathrm{f}}\!=\!5$, solar metallicity model is applied, which was shown to be close to the best fit parameters of Fig.\,\ref{f10_formation_redshfit}. Variations of the formation age result in a slightly different  $m_{\mathrm{H}}$ evolution over the observed redshift range from $z\!\sim\!1.5$ to  $z\!\sim\!0.2$. At a fixed local apparent H-band magnitude, a $z_{\mathrm{f}}\!=\!3$ passive galaxy is expected to be about 0.2\,mag brighter at $z\!\sim\!1.5$ compared to the standard model, whereas   formation redshift $z_{\mathrm{f}}\!=\!10$ objects are  0.2\,mag fainter. This implies that subtracting passive evolution models with $z_{\mathrm{f}}\!<\!5$, as derived in Sect.\,\ref{s10_formation_epoch}, results in an increased magnitude offset at high redshifts, since the BCG magnitudes are compared to brighter model galaxies. The observed non-passive BCG luminosity evolution will hence be {\em enhanced}, if the formation redshift is lowered.   


    
    
    
\end{description}



%





\noindent
Although a number of possible  biases and selection effects  could be identified, all {\em individual\/} sources of error can at most, under reasonable assumptions, account for 1/4 of the observed non-passive component shown in Fig.\,\ref{f10_BCG_lum_model}. 
The main aim of this preliminary study at this point is to provide  conclusive direct observational indications,
 rather than  final exact values, that BCGs are indeed 
increasing their mass
as expected from hierarchical galaxy formation models, whereas their stellar populations evolve passively.
For an accurate quantitative measurement of the BCGs evolution, all of the discussed possible biases and systematic effects need to be carefully taken into account for detailed follow-up studies. 
 The preliminary total observed non-passive luminosity evolution component of $\Delta m\!\simeq\!0.9$
is likely a {\em lower limit\/}, due to the discussed {\em anti-bias\/} of the X-ray cluster selection and the use of a  
$z_{\mathrm{f}}\!=\!5$ model for comparison.

\subsection{Observing the BCG assembly epoch}
\label{s10_mass_assembly}

\noindent 
Under the assumption that BCGs  
are assembled from progenitors with passively evolving stellar populations, \ie \ associating the   luminosity gain to additional mass, the detected evolution  should be accompanied by an increased merging activity at $z\!\sim\!1$.
The following sub-section will take a closer look at this scenario.

\subsubsection{Merging activity}

\noindent
Before addressing the question of BCG merging events  at high redshifts, we need to approximate the occurrence rate of chance projected alignments of galaxies in clusters. Assuming that the BCG is located close to the center and that the galaxy density in the inner core region is approximately constant,  then the probability that a second detected {\em cluster\/} galaxy is within the projected radius  $r_{\mathrm{proj}}$ of the BCG can be estimated as

\begin{equation}\label{e10_projection_chance}
    p_{\mathrm{proj}} \simeq N_{\mathrm{gal}}\cdot \frac{V_{\mathrm{proj}}}{V_{\mathrm{tot}}} \simeq 1.5\,N_{\mathrm{gal}}\left(\frac{r_{\mathrm{proj}}}{r_{\mathrm{tot}}}\right)^2 \la  0.1 \ ,
\end{equation}

\noindent
where $V_{\mathrm{proj}}\!\simeq\!2\pi\,r^2_{\mathrm{proj}}r_{\mathrm{tot}}$ is the projected cluster volume within $r_{\mathrm{proj}}$ along the cluster diameter $2\,r_{\mathrm{tot}}$,
$V_{\mathrm{tot}}$ is the approximate total volume of the inner cluster region, 
and $N_{\mathrm{gal}}$ is the number of detected cluster galaxies.
For distant objects, observed at seeing limited $\sim\!1\arcsec$ resolution, we can set the radius for apparent projected encounters to $r_{\mathrm{proj}}\!\simeq\!3\arcsec$, \ie \ 2--3 times the seeing, 
the cluster radius for the denser core to  
$r_{\mathrm{tot}}\!\simeq\!0.5\,\mathrm{Mpc}\!\simeq\!60\arcsec$, and the number of detected galaxies at the given image depth to  $N_{\mathrm{gal}}\!\simeq\!20$, yielding a {\em projected encounter chance\/} of about 8\%. 
This value is a first crude estimate of the occurrence  rate of close galaxy pairs due to chance alignments  in cluster cores at the observed image depth, but does not include additional foreground and background objects yet.

For analytic approximations of chance alignments, values for the considered outer cluster radius  $r_{\mathrm{tot}}$ and the total number of detected objects $N_{\mathrm{gal}}$ have to be assumed, which are not accurately known. As an alternative method without strong assumptions, the {\em observed\/} occurrence rate of close pairs for non-BCGs galaxies in the cluster core can be determined and compared to the BCG values. In $1\arcmin\!\times\!1\arcmin$ environments surrounding the BCGs at $0.9\!\la\!z\!\la\!1.2$, about 15\% of the non-BCG galaxies are found as close pairs with separations of $\la\!3\arcsec$. Since this method also accounts for real mergers, the value can be considered as a conservative reference point.       
In summary, an occurrence rate for close galaxy pairs of 10--15\% can be accounted for by chance alignments at the image depth and resolution limit of the available data.






Figures\,\ref{f10_BCG_gallery_hi}\,\&\,\ref{f10_BCG_gallery_low} display a gallery of 24 selected BCGs (marked by green arrows).
The ordering of panels 1--24 is a {\em speculative attempt\/} to place the observed BCG appearance along an evolutionary sequence, \ie \ to provide `typical BCG snap-shots' over the last 9\,Gyr of cosmic time.   
At the earliest observationally accessible epochs (\eg \ panels 1--3), the BCGs appear to be  ordinary galaxies very similar to their neighbors. At redshifts of $z\!\la\!1.3$, almost 40\% of all BCGs exhibit indications for possible ongoing merging events (\eg \ panels 4--15), \ie \ 2--3 times the value that can be accounted for by projection effects. 
While a few of these observed close galaxy encounters are expected to be chance alignments (\eg \ panel 9), the majority is likely to be attributed to real BCGs merging events, in particular for the cases with more than two distinguishable objects (\eg \ panels, 5, 6, 10-15). The apparent merger fraction remains high down to redshifts of $z\!\sim\!0.5$, but the physical mechanism responsible for the close encounters appears to change, as radially infalling  subgroups reach the center (\eg \ panels 16--19). In the final phase of the apparent BCG evolution, the central dominant giant cD galaxies appear on the cosmic stage at redshifts $z\!\la\!0.6$ (\eg \ panels 20--24 in Fig.\,\ref{f10_BCG_gallery_low}).

%

\subsubsection{The four phases of BCG assembly}

\noindent


Since the BCGs constitute {\em the\/} most homogeneous class of  galaxies  in the local Universe (see Sect.\,\ref{s2_BCGs}), it can be expected that their formation history is very similar. 
Based on the findings discussed in  this section, the following {\bf  speculative, observationally motivated formation scenario for BCGs\/} is proposed, complemented by results of numerical simulations (see Sect.\,\ref{s2_galaxy_populations}). 



\begin{figure}[t]
\begin{center}

\includegraphics[angle=0,clip,width=1.05\textwidth]{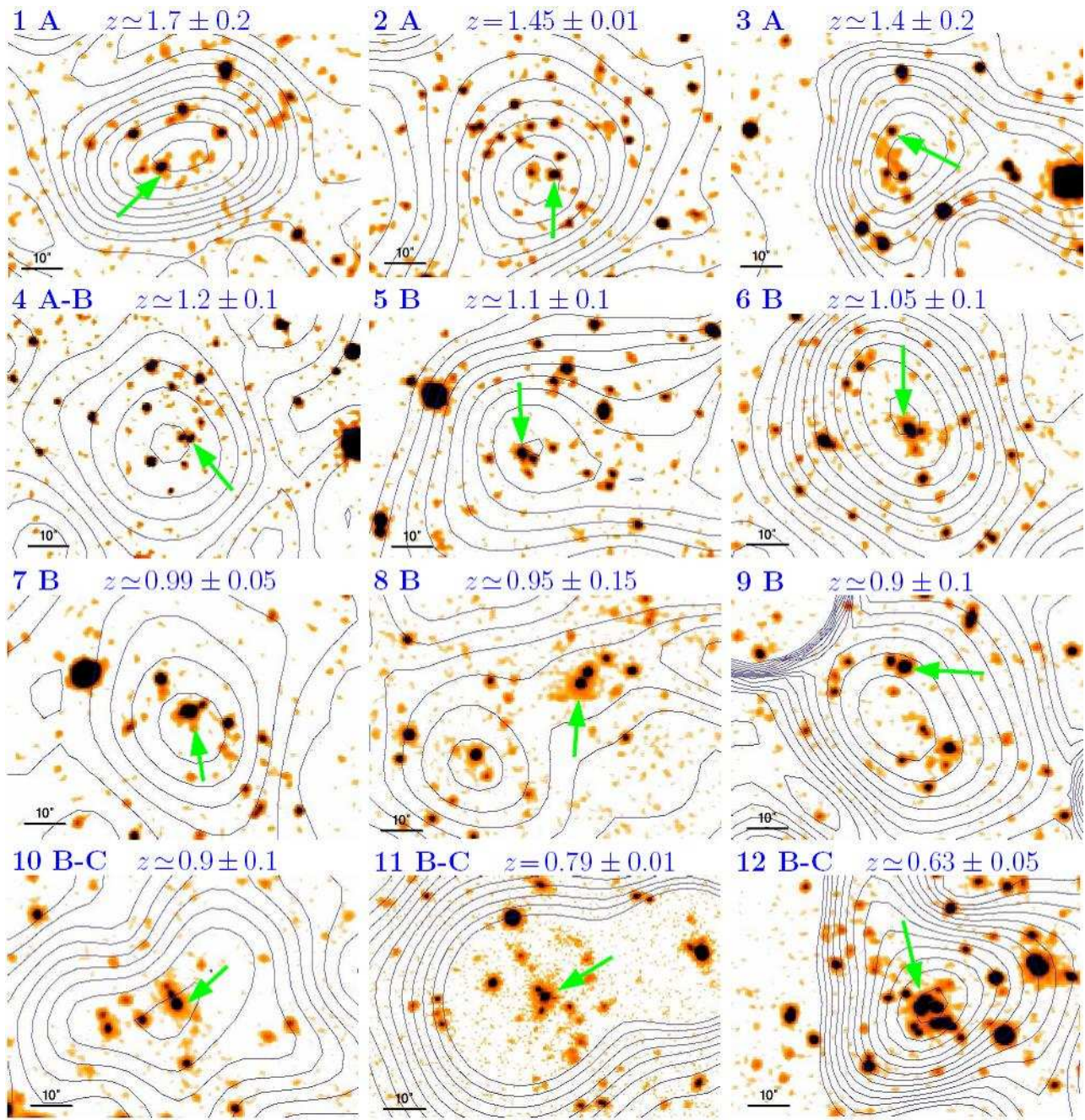}
\end{center}
\caption[High-redshift BCG gallery.]{High-redshift BCG gallery. The panel number, the apparent BCG evolutionary phase indicated by the capital letter (or an intermediate stage), and estimated redshifts are given in the upper left. BCGs are marked by green arrows, X-ray contours are given in blue. All images show H-band or combined Z+H data with identical angular scales.} \label{f10_BCG_gallery_hi}
\end{figure}

\begin{figure}[t]
\begin{center}
\includegraphics[angle=0,clip,width=1.05\textwidth]{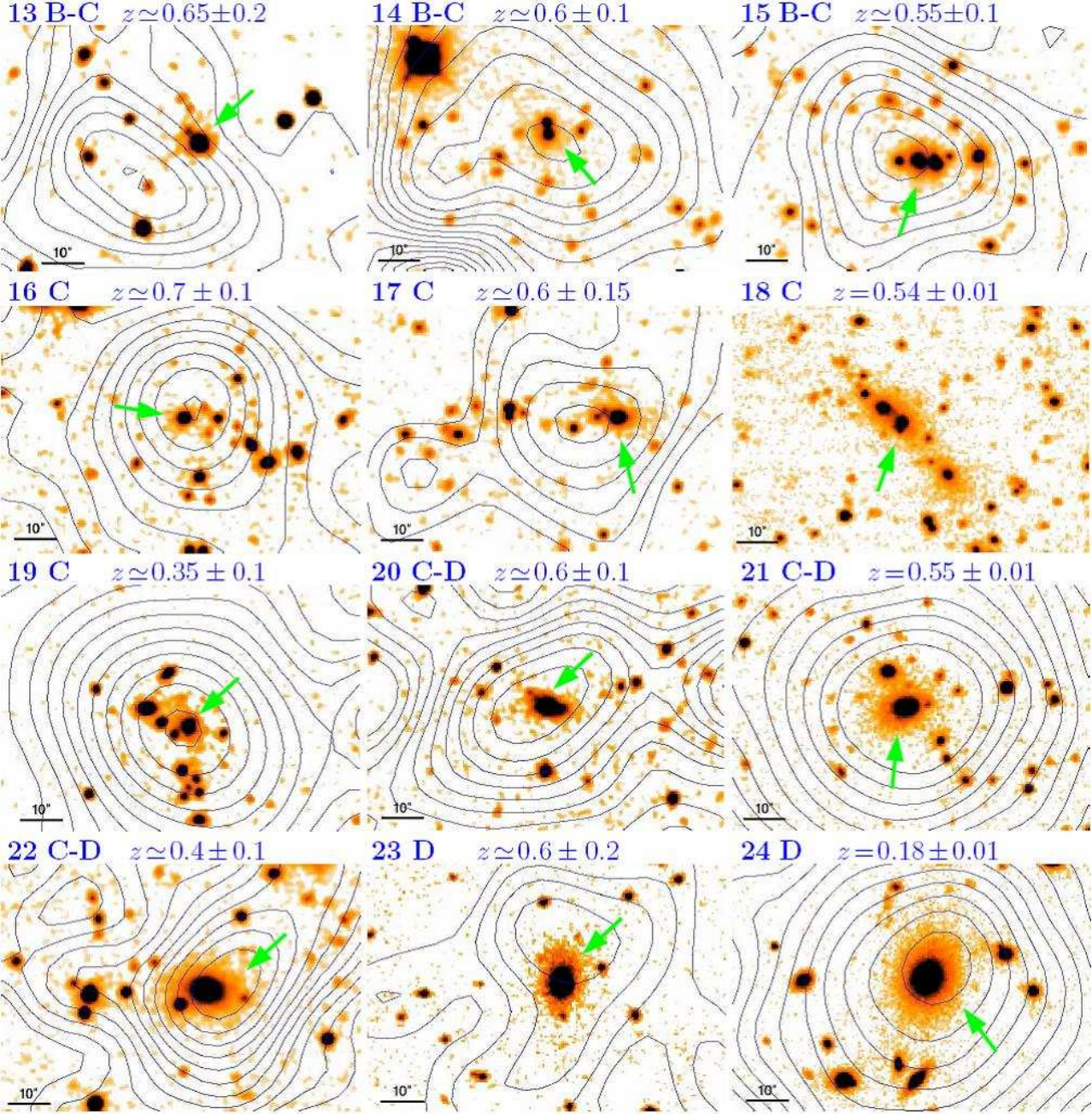}
\end{center}
\caption[Low-redshift BCG gallery.]{Low-redshift BCG gallery. Labels and data are the same as in Fig.\,\ref{f10_BCG_gallery_hi}.} \label{f10_BCG_gallery_low}
\end{figure}

\begin{description}
    \item[Phase A -- Normal Galaxies at High Redshift:] 
    At the presently observationally highest accessible redshifts of $\mathbf{z\!\ga\!1.3}$, BCGs (5 in the current sample, \eg \ panels 1--3 in Fig.\,\ref{f10_BCG_gallery_hi}) seem to be ordinary galaxies with H-band  magnitudes of approximately m*$-$1, consistent with a top ranked galaxy in the cluster luminosity function. Together with other massive cluster galaxies, BCGs migrate towards the bottom of the cluster potential well via the {\em dynamic friction\/} mechanism (see Sect.\,\ref{s2_transformation_mechanisms}).

    \item[Phase B -- Main Merging and Mass Assembly Epoch:] 
    During the cosmic epoch  $\mathbf{1.3\,\ga\!z\!\ga\!0.6}${\bf --}$\mathbf{0.7}$, corresponding to lookback times between about 9 and 6\,Gyr, the main BCG luminosity and mass assembly takes place through  dynamic friction driven dry major merging events  close to the cluster center. During this phase, the BCG roughly doubles in mass and luminosity (\eg \ panels 4--12)  and becomes the  dominant cluster galaxy.

    \item[Phase C -- Radial Infall of Groups and High Speed Encounters:] \ \ \ \\ 
The central BCG is further fed by radially infalling groups of  galaxies with small impact parameters (\eg \ panels 13--19) in the redshift range $\mathbf{0.8\!\ga\!z\!\ga\!0.3}${\bf --}$\mathbf{0.4}$.
In contrast to the friction induced, slow encounters of phase B, the high impact velocities of the order of the cluster velocity dispersion prohibit direct merging (see Sect.\,\ref{s2_transformation_mechanisms}). It seems  likely that radially impacting galaxies are (partially) disrupted and the debris dispersed throughout the cluster core, from where it can settle back onto the BCG to eventually form the extended cD envelope (\eg \ panels 17--22). The mass growth rate during this phase is slowed down, but the characteristic appearance of local cD galaxies is likely shaped in this evolutionary phase.

   
    \item[Phase D -- Settlement and Equilibrium Configuration:] 
Once the stellar debris has settled back onto the central BCG and groups have coalesced with the main cluster, the BCG can assume its final equilibrium configuration as a giant dominant cD galaxy at redshifts $\mathbf{z\!\la\!0.6}$ (\eg \ panels 23--24).

\end{description}

\noindent
This speculative BCG formation picture provides a testable starting point for further, more detailed studies. In particular, the physical mechanism for the formation of the characteristic cD halo requires additional in-depth investigations. The current BCG sample is based on follow-up observations of about one quarter of the XDCP distant cluster sample, implying that significant progress is possible over the next few years using  extended BCG samples in conjunction with spectroscopic and high-resolution studies.

\subsubsection{Discussion and comparison with simulations}

\noindent
The NIR rest-frame H-band luminosity of galaxies is dominated by old stars  and is therefore a good proxy for the total stellar mass.
Under the assumption of passively evolving BCG stellar populations, we can obtain a first order photometric stellar mass estimate by relating the observed averaged apparent H-band magnitude to the model predictions for an L$_{\mathrm{mod}}^*$  galaxy with a total stellar mass of $M_{\mathrm{tot}}(L*)\!=\!2.84\!\times\!10^{11}\,M_{\sun}$. For the best fit model of Fig.\,\ref{f10_BCG_lum_model} (bottom panel), the averaged BCG magnitudes are m*$-$1.05 at $z\!\sim\!1.5$, m*$-$1.4 at $z\!\sim\!1.0$, m*$-$1.7 at $z\!\sim\!0.5$, and m*$-$1.9 at $z\!\sim\!0.2$. These magnitudes yield the estimated stellar mass evolution for BCGs of $M_{\mathrm{tot}}^{\mathrm{BCG}}\!\simeq\!7.5\!\times\!10^{11}\,M_{\sun}$ at $z\!\sim\!1.5$, increasing to  $M_{\mathrm{tot}}^{\mathrm{BCG}}\!\simeq\!1.0\!\times\!10^{12}\,M_{\sun}$ by $z\!\sim\!1.0$ and $M_{\mathrm{tot}}^{\mathrm{BCG}}\!\simeq\!1.4\!\times\!10^{12}\,M_{\sun}$ at $z\!\sim\!0.5$, with a final mass of $M_{\mathrm{tot}}^{\mathrm{BCG}}\!\simeq\!1.7\!\times\!10^{12}\,M_{\sun}$ at $z\!\sim\!0.2$.
The observed luminosity evolution from redshift $z\!\simeq\!1.5$ to $z\!\simeq\!0.2$
can therefore be translated into 
a {\bf stellar mass increase by a factor 2.2\/} over this cosmic epoch. As discussed, this is to be interpreted as a lower limit for the stellar mass evolution  due the {\em anti-bias} of the X-ray selection. 

\enlargethispage{4ex}

We can now compare our preliminary observational results to the  BCG evolution predicted by semi-analytic hierarchical formations models summarized  in Sect.\,\ref{s2_BCGs}. The left panel of Fig.\,\ref{f2_BCG_assembly} shows that the studied model BCG  
experiences a major mass assembly epoch around $z\!\sim\!1$, roughly coincident with the observationally identified epoch of high BCG merging activity (phase B). For the averaged BCG population in the right panel of  Fig.\,\ref{f2_BCG_assembly}, the simulations predict a more gradual stellar mass increase by a factor of three from $z\!\simeq\!1$ to the present epoch, or 
a factor of four for the observationally probed redshift interval $1.5\!\ga\!z\!\ga\!0.2$.
The estimated lower limits for the BCG {\em mass evolution\/} are thus within a factor of two in agreement with the predictions.
A similar offset is found when the  crudely determined {\em total stellar\/} mass estimates are compared to the median BCG mass  predictions of Fig.\,\ref{f2_BCG_masses}.

\newpage

\clearpage

The predicted BCG K-band evolution in Fig.\,\ref{f2_sim_BCG_HD} can be directly compared to the observed absolute H-band magnitudes in Fig.\,\ref{f10_bcg_absolute_magnitudes} and the model subtracted magnitude evolution in Fig.\,\ref{f10_BCG_lum_model}. If the X-ray cluster selection {\em anti-bias} is taken into account by restricting the  simulated high-redshift sample to the brightest BCGs, \ie \ the lower part of the sample bins in Fig.\,\ref{f2_sim_BCG_HD}, the agreement with the simulations is remarkable and well within the uncertainties. The preliminary BCG sample study presented here hence manifests a significant step forward in the convergence process 
between observed
properties of the most massive galaxies and the advancing predictions of the standard hierarchical galaxy formation paradigm.

\section{Cluster Red-Sequences Studies}
\label{s10_redsequ_population}

\noindent
As discussed in Sect.\,\ref{s2_environmental_effects}, the effects of the `downsizing' scenario have  direct consequences on the \reds galaxy population and would result in a `truncation' of the CMR beyond a redshift dependent observed magnitude. 
The most distant cluster for which a deficit in the faint red galaxy population has been reported is RDCS\,J1252.9-2927 and its surrounding LSS at $z\!=\!1.24$, recently investigated in detail by Tanaka \etal \ \cite*{Tanaka2007a}.
While the main cluster exhibits a clear CMR down to magnitudes of $\mathrm{Ks(AB)}\!\sim\!23$\,mag, the \reds of the identified associated clumps in the LSS seem to be truncated at $\mathrm{Ks(AB)}\!\la\!22$\,mag, consistent with the first reports of this effect at similar redshifts (Kajisawa \etal, 2000; Nakata \etal, 2001).
\nocite{Kajisawa2000a}
\nocite{Nakata2001a} 
The observations support the picture of an {\em environment dependent downsizing evolution\/}, where the star formation activity ceases first for the most massive galaxies in clusters and a  {\em delayed\/}  CMR build-up  in lower density environments.  

At redshift $z\!\la\!0.8$, De~Lucia \etal \ \cite*{DeLucia2007b} find a deficit of faint red cluster galaxies in the luminosity range $0.4\!\ga\!L_{\mathrm{faint}}/L*\!\ga\!0.1$ relative to the fraction of $L_{\mathrm{lum}}\!>0.4\,L*$ objects.
This luminous-to-faint galaxy number ratio,  $N_{\mathrm{lum}}/N_{\mathrm{faint}}$, exhibits a   
 decreasing trend towards lower redshift, implying that the faint end of the CMR becomes increasingly populated with increasing cosmic time. A similar evolution in the build-up of the faint \reds galaxy population is observed by Stott \etal \ \cite*{Stott2007a} at $z\!\la\!0.5$. 


The current limited depth of the available Z and H-band follow-up data does not yet allow a quantitative assessment of the galaxy population at the {\em faint end\/} of the red-sequence.
The detailed analysis of the \reds properties at \zga1 requires deep,  high-quality multi-band data for precision photometry and the selection of cluster members using photometric redshift techniques, currently available for only about half a dozen clusters listed in Tab.\,\ref{t10_Xray_clusters}. 

The detailed investigation of the `downsizing' scenario and the related \reds truncation in galaxy clusters at high redshift will be a main  XDCP science focus in the near future. Using guaranteed observing time at the 2.2\,m telescope at the La Silla Observatory, we will use the multi-channel GROND\footnote{More details can be found at \url{http://www.mpe.mpg.de/~jcg/GROND}.} camera to obtain deep optical and NIR photometric coverage for a sample of $z\!\ga\!0.9$ cluster candidates currently scheduled for spectroscopic confirmation. GROND provides 
the unique possibility of performing simultaneous observations in four optical filters and the  J, H, and Ks NIR bands. This data will be used to determine accurate photometric redshifts  and construct CMDs to completeness limits that are 1--2 magnitudes fainter than the currently available data.

In order to fully understand the build-up of the \reds  and its evolution, the CMRs of distant clusters have  to be studied as a function of redshift {\em and\/} cluster mass. 
Two of the important questions that can be addressed with the upcoming XDCP study are (i) whether the truncation of cluster red-sequences at high redshift is a universal phenomenon and (ii) how the cluster mass influences  `downsizing'.

\section{Gallery of High-Redshift Clusters}
\label{s10_highz_gallery}

\begin{figure}[t]
\begin{center}
\includegraphics[angle=0,clip,width=1.0\textwidth]{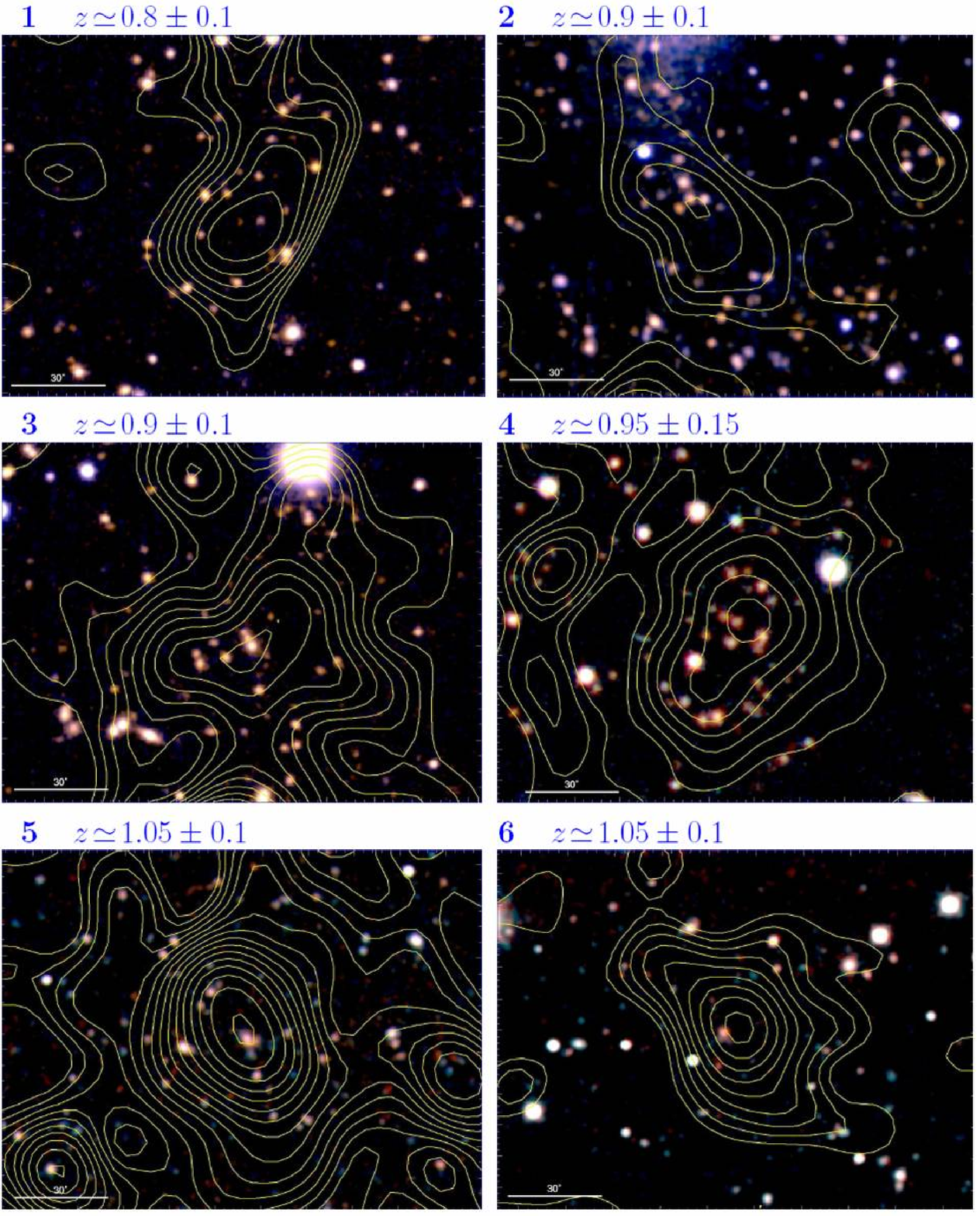}
\end{center}
\caption[High-Redshift Cluster Gallery A.]{High-redshift galaxy cluster gallery A. $2\farcm 5\!\times\!2\arcmin$  Z+H color composites  of a selection of  newly discovered high-redshift clusters with photometric redshift estimates of  $0.8\!\la\!z\!\la\!1.05$ are shown;  X-ray contours are overlaid in yellow.  } \label{f10_hiz_gallery_1}
\end{figure}

\begin{figure}[t]
\begin{center}
\includegraphics[angle=0,clip,width=1.0\textwidth]{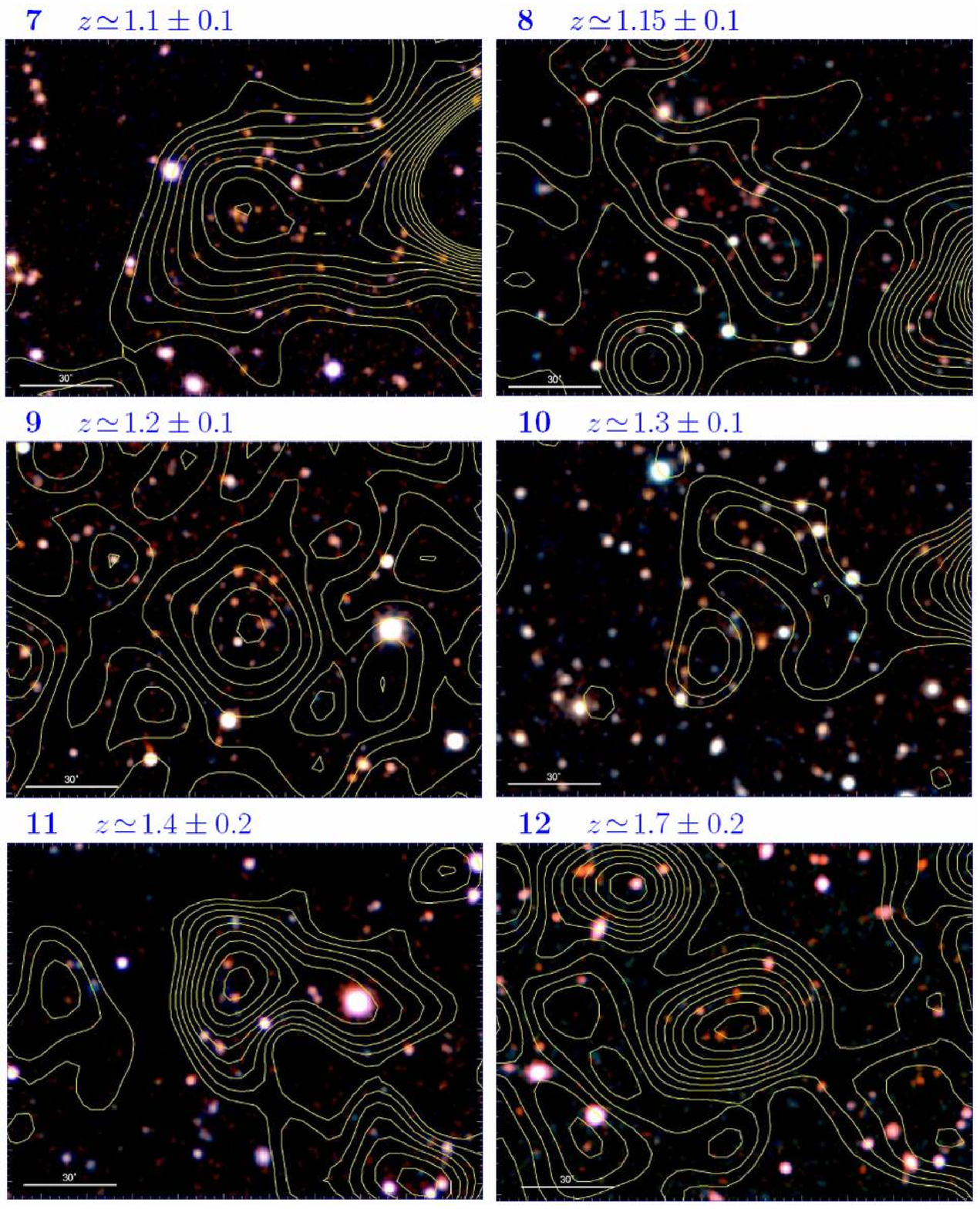}
\end{center}
\caption[High-Redshift Cluster Gallery B.]{High-redshift galaxy cluster gallery B, displaying a selection of the new sample of  candidate clusters with photometric redshift estimates of $z\!\ga\!1.1$.} \label{f10_hiz_gallery_2}
\end{figure}

\noindent
Before concluding, color composite images of a selection of newly discovered high-redshift clusters 
are shown
in Fig.\,\ref{f10_hiz_gallery_1}  for the redshift range $0.8\!\la\!z\!\la\!1.05$ and in Fig.\,\ref{f10_hiz_gallery_2}
for candidate systems at $z\!\ga\!1.1$.
All $2\farcm 5\!\times\!2\arcmin$ \ color images have X-ray contours overlaid in yellow  and are based on Z+H data obtained during the two Calar Alto follow-up campaigns.
The displayed systems are not yet spectroscopically confirmed, but several of the cluster candidates are already scheduled for FORS\,2 spectroscopy. 
The redshift confirmation is
is of particular importance (and a challenge, see Chap.\,\ref{c8_SpecAnalysis}) for candidates at the highest accessible redshifts of $z\!\ga\!1.4$. Three Calar Alto selected candidate systems belong to the latter category, based on typically 4--6 very red objects close to the X-ray center, two of them are displayed in  panels 11\,\&\,12 in Fig.\,\ref{f10_hiz_gallery_2}.







\section{Summary and Conclusions}
\label{s10b_outlook_conclusions}

\begin{itemize}

   
    \item In Sect.\,\ref{s10_bcg_assembly}, the photometric Z--H \reds redshift estimates were used for a preliminary study of the brightest cluster galaxy evolution by constructing the 
 H-band Hubble diagram for BCGs out to $z\!\simeq\!1.5$.

    \item 
It was shown, that the   observed X-ray selected cluster  BCGs   are consistent with being `standard candles'  of absolute H-band magnitude $M_{\mathrm{H}}\!\simeq\!-26.3$\,mag  out to $z\!\simeq\!1.5$.

    \item 
The increased redshift baseline allowed to  exclude 
that the luminosity of
BCGs evolves passively at  the 3\,$\sigma$ confidence level.
Although  a number of possible  selection biases  were identified, which need to be carefully quantified in future studies, none of these  effects could fully account for the observed non-passive evolution component.

    \item 
 The comparison to the expected apparent magnitude evolution of 
 a single model galaxy with passively evolving stellar populations
revealed the  tentative result that BCGs have at least doubled their mass between redshifts of $z\!\simeq\!1.5$ and  $z\!\simeq\!0.2$, whereas their stellar populations evolve passively.

    
    \item It was speculated  that the tentatively identified active BCG assembly phase around $z\!\sim\!1$ is accompanied by an increased observed merger rate at the cluster center, followed by an epoch where radially infalling groups of galaxies seem to be important for the final stages of the BCG evolution.     
    
    \item The preliminary  results suggest that the well-established old age of the stellar populations of BCGs and the expected late mass assembly can be observationally reconciled.  Predictions for the BCG evolution from  the latest simulations and semi-analytic models are in qualitative agreement with our tentative findings.   
        

    \item Upcoming deep multi-band photometric studies of XDCP clusters at high redshift will allow a systematic investigation of the phenomenon of a \reds truncation and the related `downsizing' scenario.
    
        \item A gallery of newly discovered X-ray luminous  high-redshift galaxy cluster candidates was presented. These and other  systems are currently awaiting their spectroscopic redshift confirmation.

    
    
    \item 
    The refinement of the  early results of Chap.\,\ref{c10_HizClusterStudies} and the preliminary BCG evolution study presented in Sect.\,\ref{s10_bcg_assembly} will provide  observational contributions towards an  emerging  consistent picture
of galaxy evolution in clusters.
    
    

\end{itemize}



\chapter{The Future of Galaxy Cluster Surveys}
\label{c11_Outlook}

\noindent 
Before closing this work, a brief outlook of some of the expected major developments in distant galaxy cluster studies over the next decade(s) is presented.


\section{SZE Surveys}
\label{s11_SZE}

\noindent
A new generation of galaxy cluster survey instruments based on the {\em Sunyaev-Zeldovich effect} (SZE) is currently resuming routine operations. 35 years after the theoretical prediction \cite{Sun1972a}, the SZE (see Sect.\,\ref{s2_SZE}) is opening up a new observational survey window to cluster studies.
The nearly redshift independent sensitivity to massive clusters and the complementarity with respect 
to X-ray measurements   have raised high hopes for its impact on cosmological studies. The scientific focus of the large SZE surveys is the determination of the cluster number density evolution out to high redshifts as a means to constrain the properties of {\em Dark Energy\/} (Sect.\,\ref{s3_cosmo_tests}). The cosmological applications are hence similar to the XDCP goals with the main differences that the SZE selected cluster samples (i) follow a shallower, less peaked redshift distribution compared to the X-ray selection (see right panel of Fig.\,\ref{f6_COSMOS_clusters}), and (ii) have a contiguous, wider sky coverage allowing a clustering analysis.    

\begin{figure}[t]
\begin{center}

\includegraphics[angle=0,clip,height=9cm]{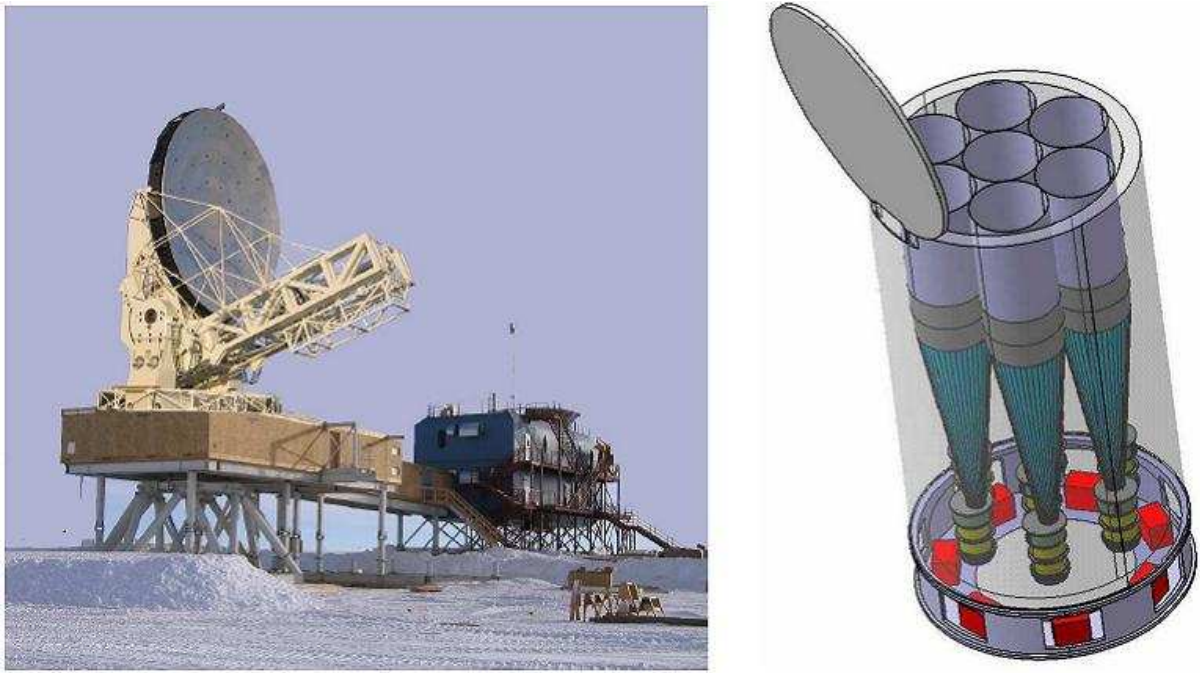}
\end{center}
\caption[South Pole Telescope and eROSITA]{Instruments for upcoming major cluster surveys: {\em Left:}  The 10\,m South Pole Telescope has started SZE observations. Image from press release.\footnotemark {\em Right:} eROSITA with its seven mirror and detector systems will conduct the next generation all-sky X-ray survey starting in 2011. Technical drawing from the mission definition document.\footnotemark} \label{f11_SPT_eROSITA}
\end{figure}

The South Pole Telescope (SPT) survey\footnote{The SPT homepage can be found at \url{http://spt.uchicago.edu/}.} 
\cite{Ruhl2004a}
will cover approximately 4000\,deg$^2$ in the mm-bands centered at 150\,GHz, 219\,GHz, and 274\,GHz. The 10\,m 
off-axis Gregorian telescope, shown in the left panel of Fig.\,\ref{f11_SPT_eROSITA}, has seen first light in February 2007 and is currently preparing for routine survey observations.
The SPT survey  is of special importance for the XMM-Newton Distant Cluster Project because of the substantial overlap of more than 100 XDCP core sample fields with a total nominal exposure time of 3.3\,Msec.  
The ongoing efforts to take full advantage of the  joint SZE-X-ray-optical data synergies have been introduced in Sect.\,\ref{s9_imaging_status}. The future cross-comparison of the X-ray and SZE cluster selection functions
will be of particular significance to understand the different survey systematics, which is crucial for the cosmological 
interpretation of the results.








This last important point for the use of galaxy clusters for precision cosmology
gave rise to the motivation to conduct a designated X-ray cluster survey (PI: H. B\"ohringer) in the joint sky regions  of the three major SZE instruments  SPT, the Atacama Pathfinder EXperiment (APEX-SZ), and the Atacama Cosmology Telescope (ACT). 
The initial 1\,Msec XMM proposal for a 12.5\,deg$^2$ X-ray coverage with typical exposure times of 12\,ksec has been 50\% granted at present.
XMM observations of the first 42 pointings are currently being conducted. The upcoming X-ray analysis of this cluster survey will largely benefit from the gained experience and developed tools of the XDCP survey.



\addtocounter{footnote}{-2}

\addtocounter{footnote}{-1}
\footnotetext{The image can be found at \url{http://www.nsf.gov/news/mmg/media/images/SPT}{\tt \%}\url{20complete}{\tt \%}\url{20s3.JPG}.} 
\addtocounter{footnote}{1}
\footnotetext{The eROSITA mission definition document can be accessed at \url{http://www.mpe.mpg.de/erosita/MDD-6.pdf}.} 



\newpage

\section{eROSITA}
\label{s11_eROSITA}

\noindent
The next generation all-sky  X-ray surveys will be conducted with eROSITA (extended ROentgen Survey with an Imaging Telescope Array) starting in 2011. 
The eROSITA  instrument \cite{Predehl2006a} is fully funded and is currently in an active preparation phase as one of the three main experiments of the Russian Spectrum-RG mission. 

As shown in the right panel of Fig.\,\ref{f11_SPT_eROSITA}, eROSITA will combine seven Wolter-I X-ray mirror modules, each equipped with an improved PN-like detector array with a field-of-view\footnote{The eROSITA detector design has recently been modified and now foresees a FoV diameter of about 60\,arcminutes, which will also increase the stated total grasp.} of about 0.5\,deg$^2$. 
The satellite will be placed in an equatorial Low Earth Orbit with expected good X-ray background properties. 
The total effective area of $\sim\!2\,500$\,cm$^2$ is comparable to the imaging mode capabilities of XMM-Newton, but the wider FoV leads to a total grasp of $g\!\simeq\!700$\,cm$^2$\,deg$^2$, \ie \ a survey efficiency improvement of a factor of 2.5 with respect to the full XMM field or a factor of 3.5 when restricting the XMM FoV to the inner 12 arcminutes.  The design specifications concerning the PSF quality aim at a FWHM resolution of  15\arcsec \ on-axis and a 20\arcsec \ field-of-view average.
The current eROSITA survey plans foresee an all-sky survey with a typical extended source sensitivity of $3\!\times\!10^{-14}$\,\flux \ and a deeper survey region covering a few hundred square degrees to a flux limit of about 
$1\!\times\!10^{-14}$\,\flux.

The challenging main goal of eROSITA will be the identification of 50\,000--100\,000 galaxy clusters out to $z\!\sim\!1.5$.
This unprecedented sample size is required to achieve the prime science objective of tracing the cosmic evolution of the {\em Dark Energy\/} equation-of-state parameter $w$ using the {\em baryonic acoustic oscillation\/} method discussed in Chap.\,\ref{c3_cosmo_theory}.  

XDCP will serve as an important pathfinder sample for eROSITA and its distant cluster cosmology applications. As probably the deepest XMM-Newton cluster survey with a sky coverage of $\gg\!10$\,deg$^2$ and an average sensitivity 3--5 times better than the eROSITA all-sky survey, the XDCP sample will be valuable for quantifying the eROSITA performance and for  calibrating the cluster selection function. In particular, the lower eROSITA resolution will constrain the accessible parameter region for a distant cluster selection based on extent and is hence to be closely investigated (see Fig.\,\ref{f9_cands_flux_rc}).
In any case, eROSITA will increase the known X-ray cluster population by at least one order of magnitude and will open unprecedented possibilities for precision cosmology with galaxy clusters.





\newpage

\section{VADER -- A Concept Study }
\label{s11_VADER}

\enlargethispage{2ex}

\noindent
The first two sections covered new galaxy cluster experiments already existing or in preparation. 
In this final section, a concept study is presented, which is closing in on an {\em ultimate cluster survey}, \ie \ a mission that is able to identify essentially all observable galaxy clusters over a wide sky area.  
The result of this study 
is the Very Ambitious Dark Energy Research mission (VADER), a project that originated at the 2005 ESA summer school in Alpbach, Austria. The objective was a design study for a designated {\em Dark Energy\/} satellite mission within ESA's {\em Horizons 2015--2025} program. The VADER project was continued after the school and led to a publication in the high energy mission section of the SPIE  proceedings \cite{RF2006a}.

\begin{table}[tb]
\begin{center}
\begin{tabular}{|c|c|c|c|c|c|c|c|c|c|}
\hline

 {\bf ID} &  {\bf Method} &  {\bf Test Objects} &  {\bf $\mathbf{\sigma_w}$} &  {\bf $\mathbf{z_{\mathrm{max}}}$ } &  {\bf X} &  {\bf O/IR} &  {\bf Hi\,Res}    \\
\hline\hline

1 & BAOs in galaxy cluster PS  & $1.75\!\times\!10^5$ GC & $<$5\%  &  2  & $\times$  &  $\times$ &      \\
2 & cosmic shear & 10$^9$ G  & $<$5\%  & 2  &   & $\times$  &  $\times$     \\
3 & BAOs in galaxy PS  & 10$^9$ G  & $<$5\%  & 3  &   &  $\times$ &      \\
4 & cluster number counts & $1.75\!\times\!10^5$ GC  & $<$10\%  &  2  & $\times$  & $\times$ &      \\
5 & BAOs in AGN PS & $5\!\times\!10^6$ AGN  & $<$10\%  &  4  & $\times$  &  $\times$ &    \\


\hline
 & all methods combined  &   & $\la$1\%  & & $\times$ & $\times$ & $\times$  \\

\hline
\end{tabular}
\caption[Expected VADER Results]{Expected results and measurement precision on $w$ of a five year VADER survey 
covering 3\,500\,deg$^2$ and probing a comoving volume of 50\,Gpc$^3$ for {\em all\/} {\em Dark Energy\/} tests ($z$$\le 2$). 
From left to right the table lists the scientific priority (ID), the method, the object class with the expected sample size, and the covered redshift range. The last three columns specify the instrumental 
requirements for the given methods distinguishing  X-ray coverage (X), multi-band optical/IR data (O/IR), and diffraction limited high spatial resolution (Hi\,Res).
By combining all independent VADER measurements of $w$ with their different degeneracies, a sub-percent level precision is achievable. 
}

\label{t11_VADER results}
\end{center}
\end{table}


\subsection{Science case}

\noindent
The nature of {\em Dark Energy\/}, as the driving force of the observed acceleration of the Universe, is currently one of the deepest mysteries in astrophysics. 
A precise knowledge of the equation-of-state parameter $w$ (see Equ.\,\ref{e3_density_evol_DE}) and in particular its redshift dependance is required to allow at least a phenomenological understanding of the dominant energy contribution in the Universe. The answer to the fundamental question whether DE is equivalent to Einstein's cosmological constant ($w\!=\!-1$), phantom energy \cite{Caldwell2003a} ($w\!<\!-1$), or time varying as in quintessence models \cite{Wetterich1988a} ($-1\!<\!w\!<\!-1/3$) will not only shed light on the fate of the Universe but will also lead the way to a unified physical theory.


The main science objective of the proposed VADER mission concept is the measurement of the {\em Dark Energy\/}  $w$ parameter with percent-level precision and to accurately constrain its redshift evolution.    
This goal is to be achieved with a single experiment yielding a self-contained, space-quality data set. 
eROSITA, for example, will likely detect $\orderof$(100\,000) extended X-ray sources, but the science applications depend critically on the time-consuming follow-up observation programs (see Chaps.\,\ref{c7_NIRanalysis}\,\&\,\ref{c8_SpecAnalysis}). 



VADER is designed as a multi-wavelength survey mission joining X-ray, optical, and IR instruments for a simultaneous spectral coverage from 4\,$\mu$m (0.3\,eV) to 10\,keV over a field-of-view  of 1 square degree.
This combination enables  several independent {\em Dark Energy\/} tests with low systematics. 
Table\,\ref{t11_VADER results} lists the five DE tests that were identified with highest priority. The associated expected object sample sizes for galaxy clusters (GC), galaxies (G), and Active Galactic Nuclei (AGN) are specified together with conservative estimates of the measurement accuracy, the covered redshift range, and the instrumental requirements.  
As the presumably cleanest {\em Dark Energy\/} test (see Sect.\,\ref{s3_cluster_PS}), a strong focus is placed on the accurate measurement of the {\em baryonic acoustic oscillations\/}  in the power spectrum (PS). This cosmological test (see Tab.\,\ref{t3_cosmo_tests}) effectively determines the Hubble expansion history $H(z)$ and angular size distance $d_{\mathrm{ang}}(z)$ and can be performed independently for the different object classes of galaxy clusters (highest priority), galaxies (medium priority), and AGN (lower priority). Large area measurements of the {\em cosmic shear}, \ie \  the weak lensing effect of the large-scale structure, is a promising alternative DE technique through 
its sensitivity to cosmic distances (see \eg \ Schneider, 2006b). \nocite{Schneider2006b} 
The determination of the number density evolution of galaxy clusters 
(Sect.\,\ref{s3_struct_formation}) completes the main VADER portfolio of {\em Dark Energy\/} tests.   
Note that the BAO test, {\em cosmic shear}, and the cluster number density evolution have later been identified as three of the four most promising routes to {\em Dark Energy\/} research by the NASA Dark Energy Task Force\footnote{\url{http://www.science.doe.gov/hep/DETF-FinalRptJune30,2006.pdf}}.  Supernovae\,Ia studies, as the fourth recommended pillar, apply the Hubble diagram method (see Sect.\,\ref{s3_cosmo_tests}) but require alternative observational techniques.

\subsection{Satellite design concept}

\noindent
In short, the VADER design foresees the combination of two high-throughput XMM-Newton type X-ray imaging telescopes (0.1--10\,keV)  complemented by a 1.5\,m wide-field corrected optical/IR telescope as sketched in the left panel of Fig.\,\ref{f11_VADER}.

The optical/IR beam is dichroic-separated\footnote{The principal design is similar to the GROND camera. More detailed information can be found at \url{http://www.mpe.mpg.de/~jcg/GROND}.}
 into eight bands (U\,G\,R\,Z\,J\,H\,K\,L) allowing the simultaneous coverage from 0.3--4\,$\mu$m. The two X-ray and eight optical/IR imaging cameras all have a field-of-view of 1\,square degree and survey the same patch of sky. 
As indicated in Tab.\,\ref{t11_VADER results}, the weak lensing technique requires 
high spatial resolution and good optical quality for galaxy shape measurements. Diffraction-limited  images are achieved in the Z-band using a 3.7\,Giga pixel optical camera  with a  scale of 0.064\arcsec \ per pixel. Most of the other seven camera systems will  yield undersampled images  with roughly a doubled pixel scale and provide the photometric data for accurate photometric redshifts, the critical pre-requisite of all DE methods.
With the proposed eight photometric bands, the achievable expected redshift uncertainties are about $\Delta z\!\simeq\!0.02\,(1\!+\!z)$ per galaxy and $\Delta z\!\la\!0.01\,(1\!+\!z)$ for the averaged cluster redshifts. 
The near-infrared bands J, H, K, and the additional mid-infrared L-band at 3.6\,\microns \ provide sensitivities which are inaccessible
from the ground and can trace the bulk of the galaxy population out to redshifts of $z\!\sim\!3$ and for AGN even further.   

Additional X-ray imaging data is crucial for the three DE tests using galaxy clusters and AGN as cosmological probes 
(Tab.\,\ref{t11_VADER results}). 
The two VADER X-ray mirror modules are based on the XMM-Newton design in order to combine the total  effective area of 2\,800\,cm$^2$ at 1.5\,keV with an average spatial resolution of about 15\arcsec \ across the 1 square degree FoV. 
This largely {\em extended} \ field-of-view, compared to XMM as discussed in Sect.\,\ref{s6_XMM}, can be achieved with a detector configuration that  follows the curved focal plane of the grazing incident mirrors leading to a significant improvement of the off-axis PSF.
Including vignetting effects, VADER's  total effective X-ray grasp at 1.5\,keV is about  1120\,cm$^2$deg$^2$,
\ie \ a factor of four better than XMM.
\subsection{Survey and results}
\begin{figure}[t]
\begin{center}
\includegraphics[angle=0,clip,height=5.95cm]{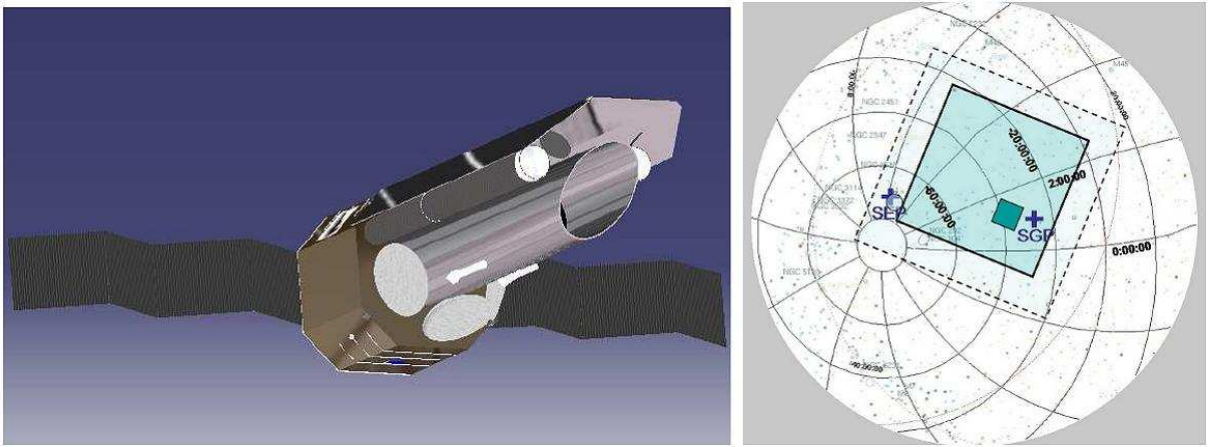}
\end{center}
\caption[VADER Spacecraft and Survey Region]{VADER spacecraft and survey region. {\em Left:} Scheme of the VADER spacecraft with its optical/IR telescope and the two X-ray mirror modules.
{\em Right:} 
The solid square indicates the  3\,500\,deg$^2$ sky area between the South Galactic Pole (SGP) and the South Ecliptic Pole (SEP) covered by a 5 year survey. 
The dashed box  shows the prospects for an extended 10 year survey and the small square  the location of the proposed 100\,deg$^2$ deep survey field. 
}
\label{f11_VADER}       
\end{figure}

\noindent
VADER is designed to perform a very deep extragalactic multi-wavelength survey  over a contiguous sky area of 3\,500 square degrees
during the first five years of the mission.
The selected survey region between the South Galactic Pole (SGP) and the South Ecliptic Pole (SEP) is illustrated in the right panel of Fig.\,\ref{f11_VADER}, the possible survey extension and the location of an ultra-deep field are also indicated.

The satellite will operate in a highly elliptical  72\,hour orbit allowing 60\,hours of consecutive science data acquisition outside the radiation belts.
The orbit was selected to optimize the observational and technical requirements of (i) a low X-ray, optical, and IR background, (ii) a high science data fraction per orbit, (iii) long, pointed observations, (iv) observations of a single, contiguous survey field, (v) a stable thermal environment, and (vi) a high downlink rate for data transmission.

The survey strategy is aiming for a gross exposure time of 10\,h per field, achieved  through an observation pattern of overlapping  single 2\,h exposures sampled over a period of a few weeks in order  to additionally allow time domain astrophysics.
Taking overheads and data losses into account, the average achievable X-ray flux limit will be about $1\!\times\!10^{-15}$\,\flux \ for point sources and  $2\!\times\!10^{-15}$\,\flux \  for extended objects.
With these sensitivities, the expected X-ray selected samples will include approximately $5\!\times\!10^{6}$  AGN out to redshifts of at least $z\!\sim\!4$  and 175\,000 galaxy clusters out to $z\!\sim\!2$. 
These X-ray capabilities, complemented by the deep optical and infrared galaxy photometry, provide accessability to essentially {\em all existing} clusters of galaxies within the survey region with masses of $M\!\ga\!1.5\!\times\!10^{14}\,M\sun$.
The typical achievable limiting AB magnitudes of 26--27\,mag in the optical bands and 23--25\,mag in the infrared bands
will allow the detection, SED characterization, and photometric redshift determination  of about one billion galaxies.

\enlargethispage{3ex}

The special strength of the proposed VADER mission is its ability to provide the most {\em complete and versatile} survey data set for {\em Dark Energy\/} and general astrophysical studies. The independent {\em Dark Energy\/} tests using BAOs, {\em cosmic shear}, and the cluster number density evolution enable critical cross-checks between the methods. 
The final combination of all DE techniques with their different degeneracies brings $w$ constraints with percent-level precision  within reach.    

Concerning the technical feasibility of the mission, the X-ray part is mainly  based on  available technology and  bears low risk. The wide-field optical/IR telescope with its multiplexed imaging capability will require a challenging optical design with substantial additional research and development activities. With the rapidly advancing detector technology, the first Giga pixel camera systems will soon be available and similar sizes in the near-infrared regime will become feasible within the next few years. The project scope for a realization of VADER would be an ESA cornerstone mission  with a possible launch around 2020.

The aim of unveiling the very nature of the dominant {\em Dark Energy\/} component in the Universe has given rise to a hyper-active phase of proposals and developments of a full armada of DE experiments with the result that {\em Dark Energy} has become {\em the buzzword} in current astrophysics.  
However, the theoretical and observational developments in {\em Dark Energy\/} research over the next two decades are difficult to foresee
and critical opinions on the merit of designated DE missions have appeared (\eg \ White, 2007). \nocite{White2007a}
In any case, the solution to the {\em Dark Energy\/} riddle is of fundamental importance 
and the route to its understanding might bear many surprises and dead ends.

\newpage

\section{Concluding Remarks }

\noindent
When speaking of the {\em Golden Age for Astronomy} \ and the dawning 
 {\em Era of  Precision Cosmology}, one can, without exaggeration, also conceive the next 10--15 years as the {\em New Age of Galaxy Cluster Surveys}.   
Here we have only touched on the upcoming developments in X-rays and for the SZ effect. Major survey projects with a strong focus on galaxy clusters are additionally ongoing or in preparation in the optical, near-infrared, and mid-infrared wavebands. Just to name a few, the Dark Energy Survey (DES), the Kilo Degree Survey (KIDS), the Panoramic Survey Telescope and Rapid Response System (PanSTARRS) are optical experiments with expected cluster sample sizes of at least a few thousand systems. At longer wavelength the Visible \& Infrared Survey Telescope for Astronomy\footnote{A list of VISTA surveys can be found at  \url{http://www.eso.org/sci/observing/policies/PublicSurveys/sciencePublicSurveys.html#VISTA}.} (VISTA) or the Spitzer mid-infrared fields\footnote{A list of Spitzer surveys is available at \url{http://ssc.spitzer.caltech.edu/legacy/all.html}.} aim at the compilation of hundreds of objects out to high redshift. 

The common aim of all projects is the identification of systems out to redshifts of \zga1, owing to the recognition 
of galaxy clusters as prime tracers of cosmic evolution.
As a survey in an advanced state, the XMM-Newton Distant Cluster Project is in a good position to take on 
a pathfinder role 
to largely unexplored redshift regimes over the next few years.
In conclusion, the future for distant galaxy cluster studies looks bright. 
Only the tip of the distant cluster iceberg has been found,
 the main body is yet to be unveiled.






\begin{appendix}


\chapter{XDCP Survey Field List}






\begin{table}[!h]
\caption[XDCP Survey Fields]{XDCP survey field summary. For each of the 546 processed XMM-Newton observations the central equatorial (RA, DEC) and galactic coordinates (GL, GB) are given followed by the corresponding galactic hydrogen column density N$_{\mathrm{H}}$ in units of $10^{21}$\,cm$^{-2}$. The last five columns contain the XMM field identifier (OBSID), the nominal exposure time (EXT), the effective clean exposure time (CLT), the processing status (STAT), and the target name of the observation (TARGET). 469 fields have been successfully processed and analyzed (STAT=ok), 48 were highly flared (STAT=flared), and 29 were discarded for other reasons (STAT=discard). }
\label{tA_field_list}
\end{table}

\begin{longtable}{|c|c|c|c|c|c|c|c|c|l|}

 
\hline

{\footnotesize\bf RA} & {\footnotesize\bf DEC} &   {\footnotesize\bf GL}   &   {\footnotesize\bf GB}   & {\footnotesize\bf N$_{\mathbf{H}}$ }  & {\footnotesize\bf OBSID}  &    {\footnotesize\bf EXT}  & {\footnotesize\bf CLT} & {\footnotesize\bf STAT} &   {\footnotesize\bf TARGET} \\ 
\footnotesize h m s & \footnotesize  d m s & \footnotesize d.d   & \footnotesize d.d & \footnotesize $\times 10^{21}$ & & \footnotesize  sec & \footnotesize sec & & \\
 
\hline\hline

\footnotesize 00 00 29.2 & \footnotesize  -25 07 20 & \footnotesize  40.0  & \footnotesize  -78.3 & \footnotesize  0.157 & \footnotesize  0125310101  & \footnotesize  46065 & \footnotesize  16360 & \footnotesize  ok & \footnotesize  Abell 2690 \\ 
\footnotesize 00 02 51.2 & \footnotesize  -29 59 11 & \footnotesize  15.0  & \footnotesize  -78.9 & \footnotesize  0.131 & \footnotesize  0041750101  & \footnotesize  52270 & \footnotesize  46458 & \footnotesize  ok & \footnotesize  BLANCO1 \\
\hline
\end{longtable}


\chapter{List of Technical Terms and Acronyms}

\begin{description}


\item[2MASS:] Two Micron All-Sky Survey

\item[AB Magnitudes:]
    Magnitude system which uses a fiducial source of constant flux $S_{\nu}^{\mathrm{AB}}\!=\!2.89\!\times\!10^{-21}$\,erg \,cm$^{-2}$Hz$^{-1}$ as reference standard \cite{Schneider2006a}. This value is chosen in a way that the V-band magnitudes are equivalent to the Vega system $m_V^{\mathrm{AB}}=m_V^{\mathrm{Vega}}$. Redder filter bands have a positive offset to the Vega system (fainter magnitudes), bluer bands have negative offsets (brighter magnitudes).

\item[ACS:]
    Advanced Camera for Surveys on board the Hubble Space Telescope

\item[ACT:]
    Atacama Cosmology Telescope

\item[AGB:] Asymptotic Giant Branch 

\item[AGN:]  Active Galactic Nuclei 

\item[APEX:]
    Atacama Pathfinder EXperiment 

\item[Baffle:]
    Light shield placed in the optical path to block stray light and
    reduce the thermal background radiation.

\item[BAOs:] Baryon Acoustic Oscillations

\item[BCG:]  Brightest Cluster Galaxy

\item[Bias:]
    An offset voltage applied to all pixels in an array detector.

\item[CA:]
    Calar Alto Observatory in Southern Spain

\item[CAL:] Calibration Access Layer

\item[CCD:]
    Charge-Coupled Device


\item[CDM:] Cold Dark Matter

\item[CFHT:] Canada-France-Hawaii Telescope

\item[Chandra:] 
    American-led X-ray observatory launched in 1999 with an on-axis spatial resolution of 0.5\arcsec. 

\item[CMB:] Cosmic Microwave Background

\item[COSMOS:] Cosmic Evolution Survey


\item[Cosmic:]
    Cosmic ray event in an image. Signal in isolated pixels caused by a
    charged cosmic particle hit.

\item[CTIO:] Cerro Tololo Inter-American Observatory
 
\item[CXB:]  Cosmic X-ray Background 

\item[DDT:] Director's Discretionary Time

\item[DEC:] Declination 

\item[DES:] Dark Energy Survey

\item[Descriptor:]
    Data buffer with a name that contains information about an associated image.

\item[DETF:] Dark Energy Task Force

\item[DET\,ML:]  Detection Maximum Likelihood

\item[Dithering:]
    Common observation technique in infrared astronomy. Multiple images
    of an object are taken with slight telescope offsets in between the
    images. By overlaying several images and using a median process, the
    local sky for the data reduction can be extracted from the science images.

\item[ECF:]  Energy Conversion Factor 

\item[EMSS:]
    Einstein Medium Sensitivity Survey 

\item[EMMI:] The ESO Multi Mode Instrument at the NTT telescope in La Silla.

\item[EoS:] Equation-of-State relating density and pressure

\item[EPIC:] European Photon Imaging Cameras  on XMM-Newton 

\item[eROSITA:] extended R\"ontgen Survey with an Imaging Telescope Array

\item[ESA:] European Space Agency

\item[ESO:]
    European Southern Observatory, with its headquarters in Garching,
    Germany.

\item[EXT\,ML:]  Extent Maximum Likelihood

\item[FITS:]
    Flexible Image Transport System. Standard format for astronomical
    images.

\item[Flatfield:]
    Image of a uniformly-illuminated area used for the correction of
    sensitivity variations across the detector.

\item[FORS\,2:]
    The FOcal Reducer and low dispersion Spectrograph for the Very Large Telescope.

\item[FoV:]
    Field-of-View. Sky area that can be covered with one image of a
    particular instrument.

\item[FP:]   False Positive 

\item[FPN:]
    Fixed Pattern Noise. Pixel-to-pixel sensitivity variations that are
    corrected by dividing through a normalized flatfield.

\item[Fringing:]
    Interference patterns produced in CCD layers by strong monochromatic atmospheric emission lines at long optical wavelengths.

\item[FWHM:]
    Full-Width Half-Maximum,  a measure of
    the width of an object in an image. The FWHM is a well-defined number
    obtained by fitting a Gaussian curve of the form
    $f(x)\!=\!1 / (\sigma \sqrt{2 \pi}) \times exp(-\frac{x^2}{2 \sigma ^2})$
    to the intensity profile of an
    object. The standard deviation $\sigma$ and the full width at
    half the peak intensity of the profile only differ by a constant factor of
    $\sigma\!=\!0.424 \cdot$FWHM \cite{stoecker1998}.

\item[Gain:]
    Conversion factor from detected electrons to digital counts.

\item[GDDS:]
    Gemini Deep Deep Survey

\item[GISSEL:]
    Galaxy Isochrone Synthesis Spectral Evolution Library

\item[GMOS:]
    Gemini Multi-Object Spectrograph at the Gemini South 8\,m telescope.
    
\item[GNIRS:]
    Gemini NIR Spectrograph at the Gemini South 8\,m telescope.    

\item[GR:] General Relativity 

\item[GROND:]  Gamma-Ray Burst Optical Near-IR Detector

\item[GTI:]  Good Time Intervals 

\item[HEW:] Half Energy Width

\item[HgCdTe:]
    Mercury-cadmium-telluride

\item[HIFLUGCS:]
    HIghest X-ray FLUx Galaxy Cluster Sample 

\item[HIROCS:]
    Heidelberg InfraRed / Optical Cluster Survey

\item[HR:]  Hardness Ratio 

\item[HR:]
    Hertzsprung-Russell diagram 

\item[HST:]
    Hubble Space Telescope

\item[ICM:]
    Intracluster Medium

\item[Image Catalog:]
    List of image names that can be used within the MIDAS environment to
    apply a set of commands to all images in the list.


\item[IMF:]
    Stellar Initial Mass Function

\item[IRAC:]
    InfraRed Array Camera on the Spitzer observatory

\item[IRAF:]
    Image Reduction and Analysis Facility. Multi-purpose astronomical software for image reduction and analysis.

\item[I/O:]
    Input/Output

\item[IR:]
    InfraRed

\item[ISAAC:] Infrared Spectrometer And Array Camera at the VLT

\item[Keyword:]
    Variable of a specified data type within the MIDAS environment.

\item[LF:]
    Luminosity Function

\item[LH:]
    Lockman Hole

\item[LM:]
    Limiting Magnitude. Apparent magnitude of objects with a
    signal-to-noise ratio~of~5.

\item[LSS:] Large-Scale Structure

\item[M67:]
    Open star cluster in the constellation Cancer at $\alpha\!=\!8^h51^m$
    and $\delta\!=\!11.\degr 82$.


\item[Median:]
    Central value in an ordered sequence of numbers.

\item[MEKAL:]
    Plasma emission code from MEwe, KAastra, and Liedahl

\item[MH:]  Mexican Hat wavelet 

\item[MIDAS:]
    Munich Image Data Analysis System. Astronomical software package
    developed and maintained by the European Southern Observatory.

\item[ML:]  Maximum Likelihood 

\item[MOS:] Metal Oxid Semi-conductor instrument on XMM-Newton

\item[NED:]  NASA Extragalactic Data Base 

\item[NEP:]
    North Ecliptic Pole survey 



\item[NICMOS:]
    Near Infrared Camera and Multi-Object Spectrometer on the Hubble Space Telescope

\item[NIR:]
    Near-InfraRed. In this thesis, usually used for the spectral region
    1--2.5\,\microns, unless otherwise noted.

\item[NORAS:]
    NOrthern ROSAT All-Sky survey

\item[NTT:]
    The 3.5\,m New Technology Telescope at La Silla Observatory in Chile.

\item[OBSID:] 
Unique identification number for each XMM observation. 

\item[ODF:] Observation Data File

\item[OMEGA2000:] The NIR wide-field imager at the 3.5\,m Calar Alto telescope.

\item[OoT:]  Out-of-Time events 

\item[PanSTARRS:] Panoramic Survey Telescope \& Rapid Response System

\item[PI:]
    Principal Investigator

\item[Pipeline:]
    Automated data reduction software requiring only minimal user
    interaction.

\item[Pixel Coordinates:]
    Coordinate system based on the pixel row and column numbers on the detector.

\item[PN:] Imaging camera on XMM-Newton

\item[Pointing:]
    One principal telescope position. The sum of all images with a
    specific target in the field of view.

\item[PSF:]
    Point-Spread Function. Intensity distribution of a point-like
    light source in an astronomical image.

\item[PSPC:] 
    Position Sensitive Proportional Counter, main instrument of ROSAT. 

\item[QE:]
    Quantum Efficiency.

\item[QSO:]
    Quasi-Stellar Object

\item[Seeing:]
    ``Blurring'' of point-like astronomical objects caused by atmospheric
    turbulence. Without adaptive optics, the seeing limits the highest
    possible resolution in ground based observations. Typical seeing
    values for the Calar Alto observatory are somewhat better than
    1\,arcsec. 

\item[RA:]  Right Ascension 

\item[RASS:]
    ROSAT All-Sky Survey 

\item[RCS:] Red-Sequence Cluster Survey

\item[RDCS:] 
    ROSAT Deep Cluster Survey 

\item[REFLEX:]
    ROSAT-ESO Flux-Limited X-ray cluster survey 

\item[RGS:] Reflecting Grating Spectrometer on XMM-Newton

\item[ROSAT:] ROentgen SATellite, German X-ray survey mission from 1990 to 1999.

\item[RWM:] Robertson-Walker-Metric 

\item[SAM:]
Semi-Analytic Model

\item[SAS:] XMM-Newton Science Analysis Software

\item[SDSS:] Sloan Digital Sky Survey

\item[SED:]
    Spectral Energy Distribution

\item[SEP:]
    South Ecliptic Pole

\item[SFH:]
Star Formation History

\item[SFR:]
Star Formation Rate

\item[SGP:]
    South Galactic Pole

\item[SMF:]  Spectral Matched Filter Scheme 

\item[SNR:]
    Signal-to-Noise-Ratio.

\item[SPIE:]
Society of Photo-Optical Instrumentation Engineers

\item[Spitzer:] Mid-infrared space telescope launched in 2003.

\item[SPT:] South Pole Telescope

\item[SRT:] Special Relativity Theory

\item[SSP:]
    Simple Stellar Population models. Galaxy evolution models assuming a single formation redshift of the stellar population       and only passive evolution afterwards. 

\item[STD:]  Standard Scheme 

\item[SZE:] Sunyaev-Zeldovich Effect

\item[UT:]
    Universal time.

\item[VADER:]
    Very Ambitious Dark Energy Research mission

\item[Vega Magnitudes:]
    Magnitude system calibrated to the A0 star Vega for which all filter bands have by definition a magnitude of 0. Since Vega's   spectrum has decreasing flux towards the red optical and NIR filters, these bands have magnitude values which are smaller (\ie \ brighter) than the corresponding AB magnitudes. The Vega system is still the standard photometric system for NIR observations.

\item[Vignetting:]
    Obscuration of parts of the primary mirror as seen by a detector pixel.

\item[VISTA:]
    Visible and Infrared Survey Telescope for Astronomy in Chile

\item[VLT:]
    Very Large Telescope in Chile

\item[VST:] VLT Survey Telescope

\item[WARPS:]
    Wide Angle ROSAT Pointed Survey

\item[WMAP:] Wilkinson Microwave Anisotropy Probe 

\item[World Coordinates:]
    Coordinate system that is based on absolute positions on the
    celestial sphere.

\item[XCS:]
    XMM Cluster Survey

\item[XDCP:]
    XMM-Newton Distant Cluster Project

\item[XLF:] X-ray Luminosity Function

\item[XMM-LSS:]
    XMM Large-Scale Structure survey  

\item[XMM-Newton:] X-ray Spectroscopy Multi-Mirror Mission

\item[XSA:] XMM-Newton Science Archive

\end{description}

\end{appendix}



\newpage

\thispagestyle{empty}

\bibliographystyle{astron}
\bibliography{astrings,RENE_PhD_BIB_publ_dec07}

\cleardoublepage


\listoffigures 
\listoftables

\cleardoublepage


\pagestyle{empty}

\huge \bfseries Acknowledgements \\ [5ex] \normalsize \mdseries

\normalsize
\noindent 
It is my pleasure to thank 
Hans B\"ohringer for being a superb supervisor, for giving me valuable advice when  needed, for encouraging new projects and initiatives, for supporting new ideas and approaches, for sponsoring frequent travel, and for always spreading a positive, open
attitude.   

\vspace{3ex}

\noindent 
This thesis is dedicated to the late Peter Schuecker, my mentor and advisor, whose enthusiasm and love for the beauty of the Universe has been truly inspiring and will continue to live on.

\vspace{3ex}

\noindent 
I am indebted to many other people who contributed to this thesis either scientifically,  financially, with moral support,
or just by making MPE a great place to work.     
I would like to thank Prof.\,Gregor Morfill for his support over the years, his bottles of wine for late work sessions, and
for being a humorous office neighbor. 
I thank Wolfgang Voges for being a great thesis committee member and for his encouragement. 

\vspace{3ex}

\noindent 
Furthermore, I am grateful to the whole XDCP team. Special thanks go to Chris Mullis for getting me started, Georg Lamer for 
his X-ray help, and Joana Santos for helping with the observations.
I am also indebted to the Calar Alto staff for their  support, with additional thanks to  Aurora Simionescu and Filiberto Braglia for their  participation. I have also enjoyed the CTIO observing run and the ongoing  work  with Joe Mohr.
Special thanks go to Hermann-Josef R\"oser for the  support of OMEGA2000, the active exchange of ideas and codes, and his advice.
Thanks also to Harald Baumgartner and Joachim Paul for their essential technical support over the years.

\vspace{3ex}

\noindent 
My office mate Gabriel Pratt deserves special thanks for putting up with me, for the convivial work atmosphere, his X-ray support, and all the proofreading. I would also like to thank Alexis Finoguenov for sharing his expertise, data, and money with me.
Very warm final thanks go to Daniele Pierini for providing all the models, galaxy expertise, guidance, and proofs, and to   
Martin M\"uhlegger for his careful reading and cheerful personality.

\cleardoublepage





\end{document}